\documentclass[
	12pt,					
	openright,				
	oneside,					
	a4paper,					
	english,					
    oldfontcommands
	]{abntex2}

\usepackage{lmodern}				
\usepackage[T1]{fontenc}			
\usepackage[utf8]{inputenc}		
\usepackage{lastpage}			
\usepackage{indentfirst}			
\usepackage{color}				
\usepackage{graphicx,graphicx}	
\usepackage{subcaption}
\usepackage{graphicx}
\usepackage{caption}

\usepackage{subcaption} 
\usepackage{microtype} 			
		
\usepackage{lipsum}						
\usepackage{amsmath,amssymb,mathrsfs}	
\usepackage{setspace}  					
\usepackage{verbatim}					
\usepackage{tabularx} 					
\usepackage{longtable} 				
\usepackage{afterpage} 					
\usepackage{url} 	
\usepackage{physics}
\usepackage{float}
\usepackage{caption}
\usepackage{subcaption}
\usepackage{makecell}

\usepackage{tensor}
\usepackage{epigraph}   
\usepackage{pdfpages}   

\newif\ifarxiv
\arxivfalse

\usepackage{titlesec}
\usepackage{xcolor} 

\usepackage{lmodern} 
\usepackage{titletoc} 

\usepackage{cite} 

\usepackage{hyperref}

\renewcommand{\thechapter}{\arabic{chapter}}

\titleformat{\chapter}[block]
  {\normalfont\fontsize{24}{28}\bfseries\fontfamily{lmroman}\selectfont\color{black}} 
  {\normalfont{\color{blue}\fontsize{60}{70}\selectfont\thechapter\hspace{0.1em}|}}
  {1em}
  {}

\usepackage[brazil]{babel}
\addto\captionsbrazil{%
\def\bibname{References}%

\renewcommand{\listfigurename}{List of Figures}
\renewcommand{\listtablename}{List of Tables}
\renewcommand{\contentsname}{Contents}

}
\newcommand{\bes}{\begin{subequations}}
\newcommand{\ees}{\end{subequations}}

\ifdefined\backrefpagesname
	\renewcommand{\backrefpagesname}{Cited on page(s):~}
	\renewcommand*{\backrefalt}[4]{
		\ifcase #1 %
		No citation in the text.%
		\or
		Cited on page #2.%
		\else
	    Cited #1 times on pages #2.%
		\fi}%
\fi

\titulo{Quantum-Corrected and Hairy Black Holes via Analogue Gravity}
\autor{Christian Noberto de Souza}
\local{Santo André - SP}
\data{2026}
\orientador{Prof. Dr. Roldão da Rocha Junior}
\instituicao{%
  Federal University of ABC -- UFABC
  \par
  \par
  Graduate Program in Physics}
\tipotrabalho{Thesis (Doctor of Philosophy)}
\preambulo{\textbf{Doctoral Thesis} conducted under the supervision of Prof. Dr. Roldão da Rocha Junior, presented to the Graduate Program in Physics of the Federal University of ABC (UFABC), as part of the requirements for obtaining the title of Doctor of Philosophy in Physics.}

\definecolor{blue}{RGB}{127,0,255}

\makeatletter
\hypersetup{
		pdftitle={\@title}, 
		pdfauthor={\@author},
    		pdfsubject={\imprimirpreambulo},
	    pdfcreator={LaTeX with abnTeX2},
		pdfkeywords={abnt}{latex}{abntex}{abntex2}{trabalho acadêmico}, 
		colorlinks=true,       		
    		linkcolor=blue,          	
    		citecolor=blue,        		
    		filecolor=blue,      		
		urlcolor=blue,
		bookmarksdepth=4
} 
\makeatother

\makeindex

\begin{document}

\frenchspacing

\pretextual

  \begin{capa}%
    \begin{figure}[h!]%
        \centering%
        \includegraphics[scale=1.2]{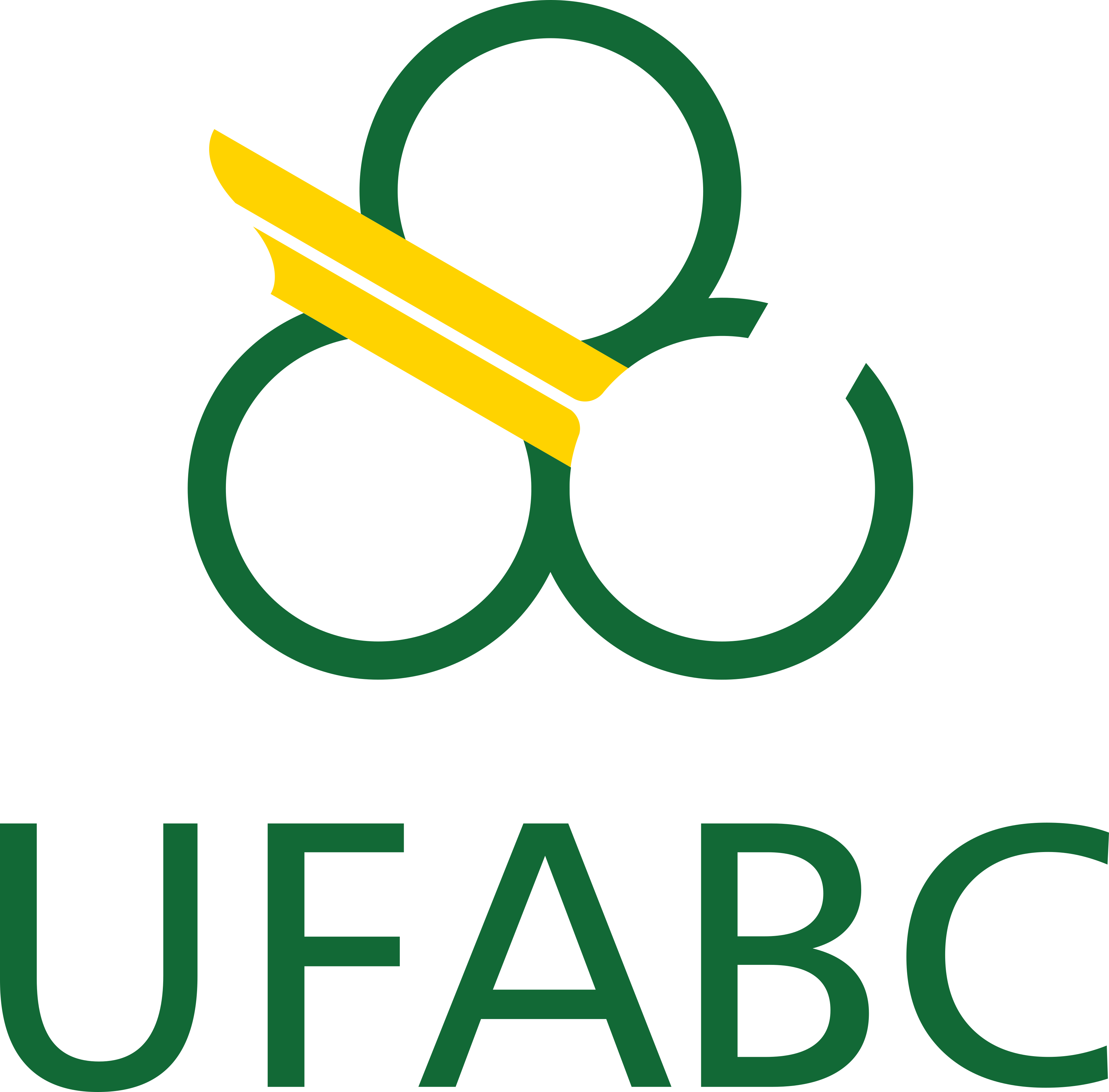}%
      \end{figure}%
    \center
	\ABNTEXchapterfont\large{Federal University of ABC  \\ Graduate Program in Physics}

    \vfill
    \ABNTEXchapterfont\bfseries\LARGE\imprimirtitulo
    \vfill

	\ABNTEXchapterfont\large\imprimirautor
	\vfill
%
	
    \large\imprimirlocal, \large\imprimirdata

    \vspace*{1cm}
  \end{capa}

\clearpage
\imprimirfolhaderosto*

%
\begin{fichacatalografica}
	\vspace*{\fill}					
	\hrule							
	\begin{center}					
	\begin{minipage}[c]{12.5cm}		
	
	\imprimirautor
	
	\hspace{0.5cm} \imprimirtitulo \ / \imprimirautor. --
	\imprimirlocal, \imprimirdata \ --
	
	\hspace{0.5cm} \pageref{LastPage} p. : il. (some colors) ; 30 cm.\\
	
	\hspace{0.5cm} \imprimirorientadorRotulo~\imprimirorientador\\
	
	\hspace{0.5cm}
	\parbox[t]{\textwidth}{\imprimirtipotrabalho~--~\imprimirinstituicao,
	\imprimirdata.}\\
	
	\hspace{0.5cm}
		1. Exotic black holes.
		2. Gravitational decoupling.
        3. Effective field theory of gravity.
		4. Quantum corrections.
		5. Analogue gravity.
		6. Quasinormal modes.
        7. de Laval nozzle.
		I. Rocha Junior, Roldão da, supervisor.
		II. Universidade Federal do ABC.
		III. Quantum-Corrected and Hairy Black Holes via Analogue Gravity\\

	\end{minipage}
	\end{center}
	\hrule
\end{fichacatalografica}

\ifarxiv
\else
  \includepdf[pages=1-2]{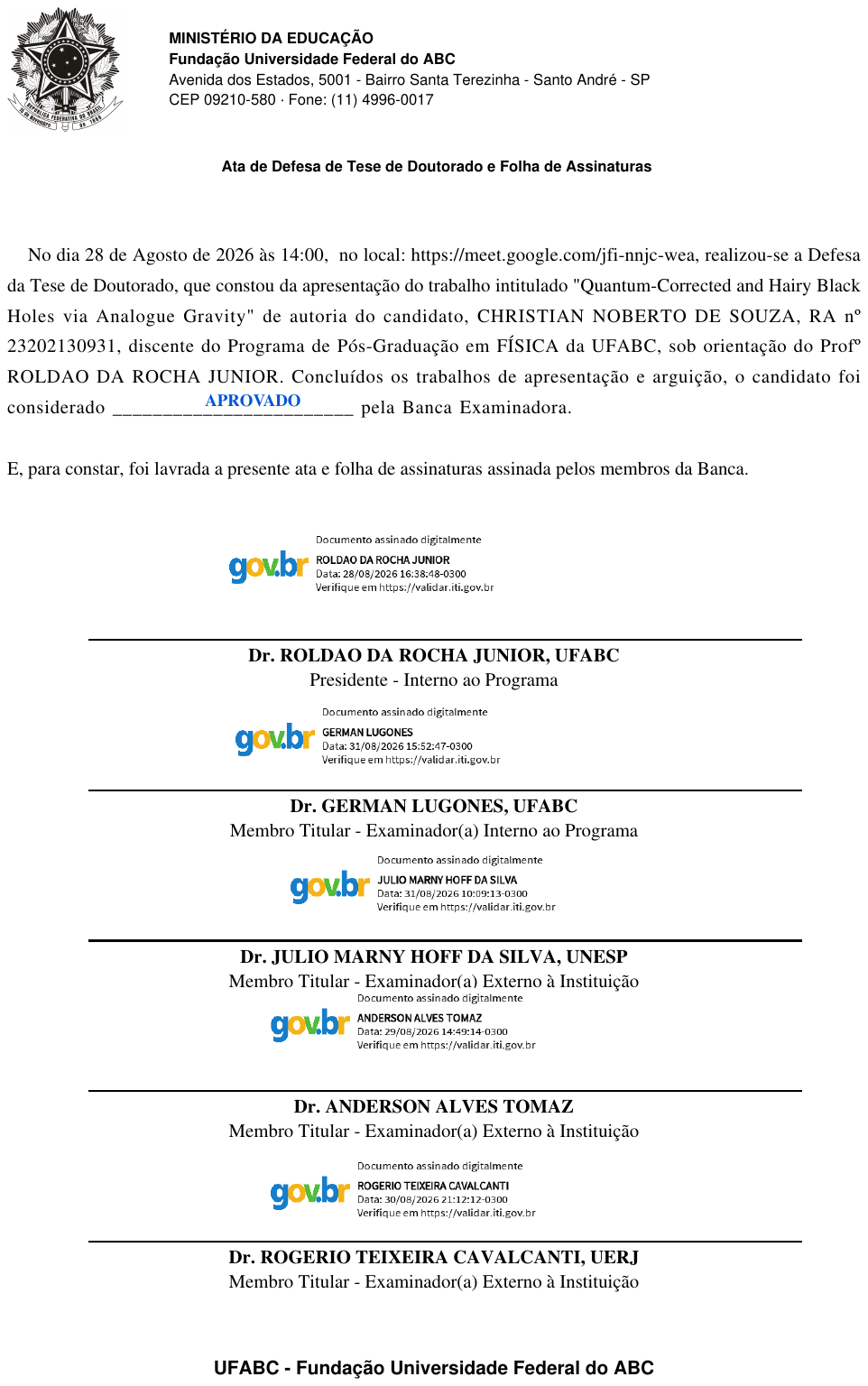}
\fi

\begin{dedicatoria}
   \vspace*{\fill}
   \begin{center}
      \textbf{Declaração de Atendimento às Observações da Banca}
   \end{center}
   \vspace{1em}
   \noindent
   Este exemplar foi revisado e alterado em relação à versão original, de acordo com as observações levantadas pela banca examinadora no dia da defesa, sob responsabilidade única do autor e com a anuência do orientador.
   \vspace*{\fill}
\end{dedicatoria}

\begin{dedicatoria}
   \vspace*{\fill}
   \centering
   \noindent
   \textit{This study was financed in part by the Coordenação de Aperfeiçoamento de Pessoal de Nível Superior -- Brasil (CAPES) -- Finance Code 001.}
   \vspace*{\fill}
\end{dedicatoria}

\begin{dedicatoria}
   \vspace*{\fill}
   \centering
   \noindent
   \textit{To My Family.\\To Bigode, Tonhão, Geralda, Maria, Adolfo, Jamil and Jeremias (my cats).} \vspace*{\fill}
\end{dedicatoria}


\begin{resumo}[Acknowledgements]

I am grateful to the Federal University of ABC. Its interdisciplinary structure is the reason I found physics at all. I arrived intending to become ``only'' an aerospace engineer, and the programme's scope widened my horizons until a different set of questions became impossible to ignore. The path that led to this thesis started there.

I thank my supervisor, Prof. Dr. Roldão da Rocha, who has always been willing to help. His guidance accompanied me from my first undergraduate research project through to the doctorate, and it was never restricted to physics, also extending to the professional decisions I had to make along the way. I am grateful for his availability, for his patience with the pace imposed by a doctorate carried out alongside full-time work, and for the standard of rigour he sets.

I thank my beloved wife, Vitória, who has been my foundation throughout. She always has the right words at hand, and her presence has a way of making even the heaviest weeks feel manageable. She kept me steady through the years this thesis took.

I thank my father and Ale, and my mom, for supporting the decisions I took along the way, and for trusting them even when the reasons were not obvious from the outside.

Finally, I thank my cats, Bigode, Tonhão, Geralda, Maria, Adolfo, Jamil, and Jeremias. Without their company, their fluffy fur, and their remarkable talent for demanding attention at the worst possible moment, I doubt I would have made it through the undergraduate years, a full-time career in credit strategy and data science, first in traditional banks and now in the technology sector, a pandemic, and finally a doctorate.

\end{resumo}

\begin{epigrafe}
    \vspace*{\fill}
	\begin{flushright}
		\textit{``We'll ride the spiral to the end and may just go where no one's been. \\ Spiral out, keep going.''\\
		        Tool, ``Lateralus'' (2001)}
	\end{flushright}
\end{epigrafe}

%

\setlength{\absparsep}{18pt} 
\begin{resumo}[Abstract]

This thesis maps two exotic black-hole geometries onto the transonic flow of a de Laval nozzle and reads their perturbative response in aerodynamic variables. The first is the hairy black hole of gravitational decoupling, constrained by the dominant energy condition and by the existence of a horizon. The second is the Calmet--Kuipers metric, short-ranged and governed by the single Wilson coefficient of the dimension-six operator that the two-loop Goroff--Sagnotti divergence identifies. Perturbations of spin $0$, $1/2$, $1$ and $2$ reduce to a Schr\"odinger-type wave equation, whose spectrum the Mashhoon method returns as a whole overtone tower in closed form, every frequency damped. The two corrections are distinguished by which channel responds and by how strongly all of them do, the quantum coupling moving the quality factor by up to $37\%$ in the scalar and Dirac channels and by about $10\%$ in the other two, while the hair acts on every channel alike. Identifying the two tortoise coordinates turns the equality of the potentials into a Sturm--Liouville problem, so that the prescribed barrier fixes the wall profile and with it the flow variables and the thrust coefficient. The hair is felt across the whole divergent section, whereas the quantum correction registers only near the throat and grows as the inverse fourth power of the mass, which confines it to primordial black holes. The analogy is kinematical, so the duct reproduces the scattering of perturbations off these geometries without reproducing the field equations behind them.

\vspace{\onelineskip}

 Effective field theory of gravity. Quantum corrections. Analogue gravity. Quasinormal modes. de Laval nozzle.

\end{resumo}

\begin{resumo}[Resumo]

Esta tese mapeia duas geometrias exóticas de buraco negro no escoamento transônico de uma tubeira de Laval e lê a resposta perturbativa de ambas em variáveis aerodinâmicas. A primeira é o buraco negro cabeludo do desacoplamento gravitacional, restringido pela condição de energia dominante e pela existência de horizonte. A segunda é a métrica de Calmet--Kuipers, de curto alcance e regida pelo único coeficiente de Wilson do operador de dimensão seis que a divergência de dois laços de Goroff--Sagnotti identifica. Perturbações de spin $0$, $1/2$, $1$ e $2$ se reduzem a uma equação de onda do tipo Schr\"odinger, cujo espectro o método de Mashhoon devolve como uma torre inteira de sobretons em forma fechada, com todas as frequências amortecidas. As duas correções se distinguem por qual canal responde e por quão forte todos respondem, com o acoplamento quântico deslocando o fator de qualidade em até $37\%$ nos canais escalar e de Dirac e em cerca de $10\%$ nos outros dois, enquanto o cabelo age igualmente sobre todos os canais. Identificar as duas coordenadas tartaruga transforma a igualdade dos potenciais num problema de Sturm--Liouville, e assim a barreira prescrita fixa o perfil da parede e, com ele, as variáveis do escoamento e o coeficiente de empuxo. O cabelo é sentido ao longo de toda a seção divergente, enquanto a correção quântica se registra só perto da garganta e cresce com a quarta potência inversa da massa, o que a confina a buracos negros primordiais. A analogia é cinemática, portanto o duto reproduz o espalhamento de perturbações nessas geometrias sem reproduzir as equações de campo por trás delas.

\vspace{\onelineskip}

\noindent
\textbf{Palavras-chave}: Buracos negros exóticos. Desacoplamento gravitacional. Teoria de campos efetiva da gravitação. Correções quânticas. Gravidade análoga. Modos quasinormais. Tubeira de Laval.

\end{resumo}

\begin{resumo}[Publication List]


This thesis has resulted in the following publications:

\begin{enumerate}
\item[{\cite{Casadio:2024uwj}}] Casadio R, Noberto Souza C and da Rocha R 2024 \textit{Eur. Phys. J. C} \textbf{84} 767 (\textit{Preprint} \href{https://arxiv.org/pdf/2402.04682}{2402.04682})
\item[{\cite{Casadio:2025lkn}}] Casadio R, Noberto Souza C and da Rocha R 2026 \textit{Eur. Phys. J. C} \textbf{86} 15 (\textit{Preprint} \href{https://arxiv.org/pdf/2506.10847}{2506.10847})
\end{enumerate}

\end{resumo} 

\pdfbookmark[0]{\listfigurename}{lof}
\listoffigures*
\cleardoublepage

\pdfbookmark[0]{\listtablename}{lot}
\listoftables*
\cleardoublepage

\begin{resumo}[List of abbreviations, acronyms and symbols]

\noindent\textbf{Conventions and acronyms}
\begin{longtable}{@{}c@{\quad}p{0.78\textwidth}@{}}
  $\qty(-,+,+,+)$ & Spacetime signature used throughout the thesis \\
  $c = \hbar = 1$ & Natural units used throughout the thesis \\
  $\mu,\nu,\rho,\sigma,\ldots$ & Greek indices: spacetime components ($0,1,2,3$) \\
  EFE & Einstein's Field Equations \\
  GR & General Relativity \\
  GD & Gravitational Decoupling \\
  MGD & Minimal Geometric Deformation \\
  MGDe & Extended Minimal Geometric Deformation \\
  DEC & Dominant Energy Condition \\
  EFT & Effective Field Theory \\
  WKB & Wentzel--Kramers--Brillouin (semianalytic approximation) \\
  PT & P\"oschl--Teller (potential barrier used in the Mashhoon method)
\end{longtable}


\noindent\textbf{Geometry and gravity}
\begin{longtable}{@{}c@{\quad}p{0.78\textwidth}@{}}
  $g_{\mu\nu}$ & Spacetime metric tensor \\
  $h_{\mu\nu}$ & Metric (graviton) perturbation, $g_{\mu\nu}=\bar g_{\mu\nu}+\kappa h_{\mu\nu}$ (Chapter~\ref{chap:EFT}) \\
  $B(r),\,A(r)$ & Metric functions of the static, spherically symmetric line element, $B=-g_{tt}$, $A=g_{rr}$ \\
  $\theta,\,\phi$ & Polar and azimuthal angles of the spherical coordinates, $\dd\Omega^2=\dd\theta^2+\sin^2\theta\,\dd\phi^2$ \\
  $\xi(r),\,\zeta(r)$ & Temporal and radial exponents of a generic seed metric, $\dd s^2=-e^{\xi}\dd t^2+e^{\zeta}\dd r^2+r^2\dd\Omega^2$ (Chapter~\ref{chap:GD}) \\
  $\nu(r),\,\lambda(r)$ & Temporal and radial exponents of the deformed metric, $B=e^{\nu}$, $A=e^{\lambda}$ (Chapter~\ref{chap:GD}) \\
  $R,\,R_{\mu\nu},\,R_{\mu\nu\rho\sigma}$ & Ricci scalar, Ricci tensor, Riemann tensor \\
  $\Box \equiv \bar g^{\mu\nu}\nabla_\mu\nabla_\nu$ & Covariant d'Alembertian \\
  $G_\text{N}$ & Newton's gravitational constant \\
  $\kappa^2 = 8\pi G_\text{N}$ & Gravitational coupling constant \\
  $M_{\text{Pl}}$ & Planck mass, $M_{\text{Pl}} = (8\pi G_\text{N})^{-1/2}$ \\
  $\Lambda$ & Cosmological constant \\
  $M$ & Black-hole mass \\
  $r_\text{EH}$ & Event-horizon radius (unified notation across the thesis) \\
  $T_\text{H}$ & Hawking temperature \\
  $\kappa_s$ & Surface gravity \\
  $k_\text{B}$ & Boltzmann constant, restored together with $\hbar$ only where a dimensional estimate is made
\end{longtable}


\noindent\textbf{Gravitational decoupling / hairy black hole}
\begin{longtable}{@{}c@{\quad}p{0.78\textwidth}@{}}
  $\theta_{\mu\nu}$ & Energy-momentum tensor of the decoupling source (hair) \\
  $\tilde{T}_{\mu\nu}$ & Effective (seed $+$ source) energy-momentum tensor \\
  $\varrho$ & Energy density; $\varrho^{\text{seed}}$ for the seed source and $\tilde\varrho$ for the effective one (variant symbol, since $\rho$ is the fluid mass density of Part~\ref{part: analogue models}) \\
  $p_{\text{rad}},\,p_{\text{tan}}$ & Radial and tangential pressures of the anisotropic effective source \\
  $f(r),\,g(r)$ & Radial and temporal geometric deformations of the MGD/MGDe transformation \\
  $\alpha$ & Gravitational-decoupling (hair) coupling parameter \\
  $\ell_h$ & Primary-hair length scale (distinct from the multipole number $\ell$) \\
  $Q$ & Charge-like hair parameter of the hairy black hole \\
  $\mathcal{M}$ & Hair-shifted effective mass, $\mathcal{M}=M+\alpha\ell_h/2$ 
\end{longtable}

\noindent\textbf{Effective field theory / quantum correction}
\begin{longtable}{@{}c@{\quad}p{0.78\textwidth}@{}}
  $c_1,c_2,c_3$ & Local (curvature-squared) Wilson coefficients \\
  $\tilde\alpha,\tilde\beta,\tilde\gamma$ & Non-local Wilson coefficients (renamed to free $\alpha$ and $\gamma$) \\
  $c_6$ & Dimensionless dimension-six (Riemann-cubed) coupling \\
  $\mu$ & Renormalisation scale \\
  $S_{\text{Wald}}$ & Wald entropy 
\end{longtable}

\noindent\textbf{Perturbations and quasinormal modes}
\begin{longtable}{@{}c@{\quad}p{0.78\textwidth}@{}}
  $s$ & Spin of the perturbing field \\
  $\ell$ & Multipole (angular) number \\
  $n$ & Overtone number \\
  $r_\star$ & Tortoise coordinate, $\dd r_\star=\dd r/\sqrt{B/A}$ \\
  $V_{\text{eff}}$ & Effective (Schr\"odinger-type) potential \\
  $V_{\text{eff}}^{\max}$ & Height of the effective potential at its peak \\
  $V_{\text{PT}}$ & P\"oschl--Teller barrier fitted to $V_{\text{eff}}$ in the Mashhoon method \\
  $\varpi$ & Scalar gauge function of the vector perturbation, $A_\mu \mapsto A_\mu + \partial_\mu \varpi$ (Chapter~\ref{chap:conexaoentrebnsebocais}) \\
  $\omega = \omega_R + i\,\omega_I$ & Complex quasinormal frequency of the black hole \\
  $q_n$ & Quality factor of the $n$-th mode \\
  $\varkappa$ & P\"oschl--Teller inverse width
\end{longtable}

\noindent\textbf{Fluid mechanics and acoustic analogue}
\begin{longtable}{@{}c@{\quad}p{0.78\textwidth}@{}}
  $\mathcal{N}$ & Generic density of an extensive quantity in the balance law (frees $\mu$) \\
  $\rho$ & Fluid mass density ($\rho_0$: stagnation value; $\rho_*$: critical value), distinct from the energy density $\varrho$ \\
  $\rho_{\text{rel}}$ & Relative density $\rho/\rho_0$, used from Chapter~\ref{chap:acoustic laval} onward \\
  $p$ & Pressure \\
  $c_s$ & Local speed of sound ($c_{s0}$: stagnation value; $c_{s*}$: critical value at the throat) \\
  $\gamma$ & Adiabatic index (ratio of specific heats $c_p/c_v$) \\
  $\mathrm{M}$ & Mach number, $\mathrm{M}=v/c_s$ (upright, distinct from mass $M$) \\
  $\psi$ & Velocity potential of the flow \\
  $\phi_g$ & Gravitational potential (fluid mechanics) \\
  $\mathcal{Q}$ & Volumetric source/sink rate in the conservation law (frees $Q$) \\
  $\mathcal{K}$ & Constant of the canonical acoustic-BH velocity field (frees $\kappa$) \\
  $t,\,\tilde{t}$ & Laboratory (Painlev\'e--Gullstrand) time of the flow; static time of the acoustic metric \\
  $\mathcal{A},\,\mathcal{A}_*$ & Nozzle cross-sectional area; throat area (frees $A$ for the metric function $A(r)$) \\
  $\mathcal{A}_{\text{rel}}$ & Relative area $\mathcal{A}/(\kappa_\gamma\mathcal{A}_*)$, used from Chapter~\ref{chap:acoustic laval} onward \\
  $\kappa_\gamma$ & Constant of the gas in the area normalisation, $0.259$ for air \\
  $R_g,\,\bar{m}$ & Universal gas constant; molecular mass (free $R$ and $\mu$) \\
  $T,\,T_0$ & Static and stagnation temperature \\
  $c^{*}$ & Characteristic velocity \\
  $C_F$ & Thrust coefficient \\
  $\dot{m}$ & Mass-flow rate 
\end{longtable}

\noindent\textbf{Gravity-acoustics bridge}
\begin{longtable}{@{}c@{\quad}p{0.78\textwidth}@{}}
  $g_c = \rho \mathcal{A}/c_s$ & Gravitational-acoustic mapping function, connecting nozzle and black hole \\
  $\omega_{\text{ac}}$ & Acoustic frequency of the duct, related to the black-hole frequency by $\omega_{\text{ac}} = c_{s0}\,\omega$ \\
  $\mathrm{F}(r) = \sqrt{B/A}$ & Metric factor entering the tortoise coordinate \\
  $\chi(r)$ & Frobenius amplitude, $g_c=\chi^2$ \\
  $x,\,x_\star$ & Longitudinal nozzle coordinate and its tortoise version 
\end{longtable}

\end{resumo}

\setcounter{table}{0}

\pdfbookmark[0]{\contentsname}{toc}
\tableofcontents*
\cleardoublepage

\textual
\newcounter{paginaspretextuais}
\setcounter{paginaspretextuais}{\value{page}}
\pagenumbering{arabic}
\setcounter{page}{\value{paginaspretextuais}}
\addtocounter{page}{1}


%
%

\chapter{Introduction}
\label{chap:introduction}

\epigraph{``Somewhere, something incredible is waiting to be known.''}{Carl Sagan}

The first direct detection of gravitational waves from a pair of merging black holes, GW150914~\cite{LIGOScientific:2016aoc}, opened the strong-field regime of gravity to observation. It was followed by the first joint gravitational-wave and electromagnetic observation of two merging neutron stars~\cite{LIGOScientific:2017vwq} and by the growing catalogues of the LIGO--Virgo--KAGRA network~\cite{VIRGO:2014yos,LIGOScientific:2020tif,KAGRA:2021vkt}. A merger has three stages, the inspiral, the merger itself, and the ringdown. Of the three it is the ringdown that says most about the object left behind, because it is a sum of damped oscillations, the quasinormal modes, whose frequencies are fixed by the background geometry and by the spin of the field being perturbed, and not by whatever set the perturbation off~\cite{Kokkotas:1999bd,Nollert:1999ji,Berti:2009kk,Konoplya:2011qq}. The spectrum is therefore a signature of the object alone, which is what black-hole spectroscopy sets out to use~\cite{Dreyer:2003bv,Berti:2005ys,Berti:2015itd}, by identifying more than one mode in a measured ringdown and asking whether they agree with each other.

Current catalogues already support tests of General Relativity in the ringdown~\cite{LIGOScientific:2016lio,LIGOScientific:2020tif,KAGRA:2021vkt}, and reading the overtones has been proposed as a way of starting the programme with a single event~\cite{Giesler:2019uxc,Isi:2019aib}. Third-generation ground detectors such as the Einstein Telescope~\cite{Punturo:2010zz,ET:2019dnz} and the space mission LISA~\cite{LISA:2017pwj,Barausse:2020rsu} are expected to make multi-mode measurements routine~\cite{Berti:2015itd,Cardoso:2018ptl}.

The theory behind this is much older than the measurements. Regge and Wheeler studied the stability of the Schwarzschild solution and found the potential that governs axial perturbations~\cite{Regge:1957td}, Zerilli obtained the polar counterpart~\cite{Zerilli:1970se,Zerilli:1974ai}, and Vishveshwara recognised the damped ringing in the scattering of radiation off a Schwarzschild black hole~\cite{Vishveshwara:1970zz}. The electromagnetic case was treated in the same years~\cite{Ruffini:1972pw} and the Dirac field earlier still~\cite{Brill:1957fx}, its partner potentials being given a modern treatment later~\cite{Cho:2003qe,Dubinsky:2024mwd}. Chandrasekhar and Detweiler then computed the Schwarzschild spectrum in a systematic way~\cite{Chandrasekhar:1975zza}, and a set of methods grew around the problem, from the continued fractions of Leaver~\cite{Leaver:1985ax} to the semianalytic Wentzel--Kramers--Brillouin (WKB) treatment~\cite{Schutz:1985km,Iyer:1986np,Iyer:1986nq} and to the P\"oschl--Teller substitution of Ferrari and Mashhoon~\cite{Ferrari:1984zz,Ferrari:2007dd}, which is the one used in this thesis. The late-time tails found by Price~\cite{Price:1971fb,Price:1972pw} and the correspondence between the lowest modes and the unstable circular photon orbit~\cite{Cardoso:2008bp} complete a picture that the standard reviews collect~\cite{Kokkotas:1999bd,Nollert:1999ji,Berti:2009kk,Konoplya:2011qq}. Because the frequencies are set by the geometry, they offer a clean way of testing any departure from Schwarzschild and Kerr, which is where the ringdown differs from the inspiral and what has driven the search for echoes and other signals from close to the horizon~\cite{Cardoso:2016rao,Abedi:2016hgu,Maggio:2019zyv,Buonanno:2006ui}.
There are good reasons to expect departures. The uniqueness theorems say that a stationary, asymptotically flat black hole in electro-vacuum is fixed by its mass, its angular momentum and its charge~\cite{Israel:1967wq,Carter:1971zc,Robinson:1975bv}, and further results rule out broad classes of extra scalar fields~\cite{Bekenstein:1972ny}. These theorems hold under their own hypotheses. Relaxing them, by adding matter or by widening the gravitational sector, gives black holes that carry extra parameters, of which the Kerr solutions with scalar hair are a well-known case~\cite{Herdeiro:2014goa,Herdeiro:2015waa}. Testing the no-hair hypothesis against data is what the overtone analyses attempt~\cite{Isi:2019aib,Giesler:2019uxc}, and hairy solutions give controlled alternatives against which such tests can be calibrated~\cite{Guimaraes:2025jsh}.
Building those alternatives is not simple, because the Einstein equations are non-linear and adding a source to a known solution usually means solving the problem again from the start. The gravitational-decoupling programme was developed to avoid exactly that. It grew out of work on compact stars in the Randall--Sundrum brane-world~\cite{Randall:1999ee,Randall:1999vf,Shiromizu:1999wj,Maartens:2010ar,daRocha:2012pt,daRocha:2009gb}, where the bulk Weyl fluid deforms the radial part of a general-relativistic seed~\cite{Ovalle:2010zc,Ovalle:2013xla,Casadio:2012rf}, an approach that was also compared against strong-deflection lensing~\cite{Cavalcanti:2016mbe} and whose minimal geometric deformation is reviewed in Ref.~\cite{Ovalle:2016pwp}.

The step that gave the programme its present form was the realisation that the deformation needs no higher-dimensional origin at all. Ovalle showed that a spherically symmetric seed sourced by a perfect fluid can be deformed by an arbitrary extra source, with the field equations splitting into the Einstein system for the seed and a similar system for the deformation alone~\cite{Ovalle:2017fgl,Ovalle:2017wqi}. The extended case, where both metric functions are deformed and the two sources may exchange energy, came soon after~\cite{Ovalle:2018gic}, along with a way of generating acceptable anisotropic solutions~\cite{Ovalle:2018vmg}. Applied to a vacuum seed the same machinery gives black holes~\cite{Ovalle:2018umz} and, in particular, the asymptotically flat hairy solution that obeys the dominant energy condition outside the horizon~\cite{Ovalle:2020kpd}, which is the metric used in Part~\ref{part: extensions of gr}. The family has since grown to include solutions coupled to a scalar field~\cite{Ovalle:2018ans,Arias:2022jax}, regular hairy black holes built from Minkowski space~\cite{Ovalle:2023ref}, non-singular and Hayward-type geometries~\cite{Misyura:2024fho,Naseer:2025ghn}, anti-de~Sitter configurations~\cite{Zhang:2022niv}, static solutions over a wider parameter range~\cite{Avalos:2023ywb}, and rotating black holes obtained by decoupling in axial symmetry~\cite{Contreras:2021yxe,Mahapatra:2022xea,Ovalle:2021jzf}.
The method itself has been carried well beyond its original setting, into $2+1$ dimensions with and without a cosmological term~\cite{Contreras:2018vph,Contreras:2019mhf,Contreras:2019fbk,Contreras:2019iwm,Rincon:2020izv}, into (anti-)de~Sitter backgrounds~\cite{Contreras:2018nfg}, into higher dimensions and pure Lovelock gravity~\cite{Estrada:2018vrl,Estrada:2019aeh}, into higher-order theories~\cite{Sultana:2021cvq}, back into the brane-world as a self-consistent scheme~\cite{Leon:2019abq,Leon:2021qvs}, into cosmology~\cite{LinaresCedeno:2019aul,Sharif:2023opu}, and into Vaidya geometries~\cite{Vertogradov:2022oif}. The inverse problem of recovering the deformation from a given anisotropy was posed in Ref.~\cite{Contreras:2018gzd}, and the isotropic sector and the complete geometric deformation were worked out elsewhere~\cite{Heras:2018cpz,Heras:2021xxz,Maurya:2020gjw,Maurya:2021qye,Maurya:2022yva,Leon:2023nbj}. Thermodynamics has received particular attention, from the first law under decoupling~\cite{Estrada:2020ptc,Estrada:2021kuj} and the near-horizon behaviour of hairy black holes~\cite{Cavalcanti:2022adb} to conformal anomalies~\cite{Meert:2020sqv,Meert:2021khi}, holographic entanglement entropy~\cite{DaRocha:2019fjr} and phase structure~\cite{Ditta:2023arv,Ditta:2024tdo,Abbas:2024pyi}, while the energy exchanged between the two sectors was analysed separately~\cite{Ovalle:2022yjl,Ovalle:2022eqb,Casadio:2019usg}.

A large part of the literature applies the method to stars. Anisotropic interiors have been generated from most of the classical seeds, among them the Tolman, Durgapal, Heintzmann, Finch--Skea, Vaidya--Tikekar and Buchdahl families~\cite{Hensh:2019rtb,Gabbanelli:2018bhs,Gabbanelli:2019txr,Morales:2018urp,Morales:2018nmq,PerezGraterol:2018eut,Singh:2019ktp,Singh:2021iwv,Tello-Ortiz:2019gcl,Tello-Ortiz:2020nuc,Zubair:2020lna,Andrade:2023wux,Pant:2023irv}, with care for the regularity and the causal structure of the anisotropy that results~\cite{Abellan:2020dze,Abellan:2020wjw,Jasim:2021kga,Ramos:2021drk}. Ultracompact and dark-star configurations~\cite{Ovalle:2019lbs,Tello-Ortiz:2020ydf,Carrasco-Hidalgo:2021dyg}, anisotropic neutron stars~\cite{Torres-Sanchez:2019wjv}, polytropes~\cite{Santana:2022vmw,LasHeras:2022pyj}, superfluid and Horndeski hybrid stars~\cite{daRocha:2021aww,daRocha:2021sqd}, Yang--Mills--Dirac configurations~\cite{daRocha:2020jdj}, glueball condensates~\cite{Fernandes-Silva:2018abr} and stars in a cloud of strings~\cite{Panotopoulos:2018law} have all been obtained this way. The complexity factor has served to organise these solutions~\cite{Contreras:2021xkf,Contreras:2022vec,Maurya:2021yhc,Maurya:2022cyv,Maurya:2022pef,Maurya:2023szc,Sharif:2023ecm,Sharif:2023ycm,Naseer:2024jjy,Albalahi:2024vpy}, and the method has been taken into modified gravity as well~\cite{Sharif:2019mjn,Sharif:2020lbt,Sharif:2020vvk,Sharif:2020rlt,Sharif:2021emv,Sharif:2022vvi,Maurya:2019hds,Maurya:2019wsk,Maurya:2019noq,Maurya:2020rny,Maurya:2020ebd,Maurya:2021tca,Maurya:2021zvb,Maurya:2021fuy,Maurya:2022brt,Maurya:2022wwa,Maurya:2022hmw,Maurya:2023dhm,Azmat:2021qig,Azmat:2022bbc,Bamba:2023wok,Tello-Ortiz:2021kxg,Panyasiripan:2024kyu}. Some of these models have been compared with mass-radius data and with the lighter component of GW190814~\cite{Maurya:2023muz,Maurya:2024ylr}.

Closer to the present work is the branch that asks what a decoupled geometry would look like to an observer. Quasinormal modes and greybody factors of hairy black holes have been computed by several groups~\cite{Yang:2022ifo,Cavalcanti:2022cga,Li:2022hkq,Lei:2023wlt,dePaiva:2025eux,Navarro:2026lrf}, and the imprint of the hair on the spectroscopic programme, including whether the axial and polar sectors remain isospectral, was studied in Ref.~\cite{Guimaraes:2025jsh}. Strong lensing~\cite{Mustafa:2024hfq,Zhao:2024hep,Atamurotov:2023rye}, accretion and the resulting appearance~\cite{Rehman:2023eor,Gan:2021xdl,Gan:2021pwu}, geodesics and bound orbits~\cite{Rakhimova:2024hzt,Meng:2024cnq,Li:2023htz}, and energy conditions and stability~\cite{Cavalcanti:2022xet,Ashraf:2025cnz} have all been examined. The link between decoupling and quantum gravity, in which the deformation is produced by higher-curvature operators rather than postulated, was established in Ref.~\cite{Casadio:2024vfh}, with further aspects of the relation between deformed geometries and effective sources treated elsewhere~\cite{Casadio:2017sze,Casadio:2023mgl,daRocha:2020gee,daRocha:2021xwq,daRocha:2024lev,daRocha:2017cxu,daRocha:2017tiz}. In this thesis the hairy geometry is taken as given, and the question asked is whether its perturbative response can be reproduced in a hypothetical laboratory setup.

Decoupling is phenomenological, in the sense that the extra source is postulated rather than derived. A route from first principles to a deformed Schwarzschild geometry is offered instead by the effective field theory of gravity~\cite{Donoghue:1994dn,Burgess:2003jk,Donoghue:2012zc,Donoghue:2015hwa,Donoghue:2017pgk}. General Relativity is not renormalisable as a fundamental quantum field theory, because the gravitational coupling has negative mass dimension and an endless list of counterterms is needed~\cite{Weinberg:1980gg}. Diffeomorphism invariance still allows a tower of higher-curvature operators in the effective action, whose coefficients absorb the divergences order by order, and what the theory predicts at low energy does not depend on how it is completed at high energy~\cite{Bjerrum-Bohr:2002fji,Donoghue:2015hwa}. Explicit calculations bear this out. Pure Einstein gravity is finite at one loop once the background solves the vacuum equations, as 't~Hooft and Veltman showed~\cite{tHooft:1974toh}, while a divergence that cannot be removed appears at two loops through the cubic curvature invariant of Goroff and Sagnotti~\cite{Goroff:1985th}. Integrating out massless fields adds a non-local sector whose coefficients are finite and calculable, and follow from the Barvinsky--Vilkovisky heat-kernel expansion~\cite{Barvinsky:1985an,Barvinsky:1990up,Barvinsky:1994hw,Codello:2012kq,Odintsov:1991nd}.

At second order in the curvature the Schwarzschild metric is left alone, and the corrections show up only in the Wald entropy~\cite{Wald:1993nt,Iyer:1994ys,Iyer:1995kg} and in the pressure that follows from it~\cite{Calmet:2017qqa,Calmet:2018elv,Calmet:2021stu}, whereas matter or an electric charge does deform the exterior geometry already at that order~\cite{Calmet:2019eof,Solodukhin:2019xwx,Perrucci:2024qrr}. For a vacuum black hole the leading correction therefore waits until third order, where the dimension-six Goroff--Sagnotti operator gives the quantum-corrected metric of Calmet and Kuipers~\cite{Calmet:2021lny,Calmet:2023gbw,Calmet:2024pve}. That metric breaks the relation $g_{tt}\,g_{rr}=-1$ and falls off as $r^{-6}$ in $g_{rr}$ and as $r^{-7}$ in $g_{tt}$. Rotating and charged versions have been derived~\cite{Casadio:2023iqt,Casadio:2022ndh}, the non-local logarithmic operator and the actions built from it have been studied~\cite{Kuntz:2019lzq,Kuntz:2019omq,Kuntz:2019gup,Casadio:2021rwj,Casadio:2017cdv,Woodard:2015zca,Rachwal:2021bgb}, as have the status of the higher-derivative truncation~\cite{Simon:1990ic,Sarkar:2017puh,Nastase:2016sji,Antoniadis:1998ig,Delhom:2022xfo} and the regularity of the interiors that result~\cite{Casadio:2024lgw,Casadio:2025pun,Ovalle:2023vvu,Ovalle:2024wtv}. The quasinormal response of higher-derivative and quantum-corrected black holes has been followed in parallel~\cite{Cardoso:2018ptl,Cardoso:2019mqo,Cano:2021myl,Cano:2023jbk,Husken:2026axq,Chen:2025wfi,Konoplya:2022zym,Matyjasek:2020bzc,Bueno:2016lrh,Moreira:2024ewr,Cavalcante:2024kmy,Dubinsky:2024mwd}, and bounds on the single coupling that controls it have been sought through several observables~\cite{Gong:2023ghh,Wang:2025fmz,Marciu:2023hdb,Bukhari:2025fhd,DeFelice:2023vmj,Lambiase:2024vkz}.

Reaching either deformation with a telescope is hard. Parameters are degenerate, the environment matters, and the signals are weak, which limits what even the next generation of detectors can do. The quantum correction is worse off still, since it is negligible for astrophysical masses and becomes appreciable only near the Planck scale, and therefore only for primordial black holes~\cite{Hawking:1971ei,Carr:1974nx,Carr:2020gox}. This is what makes a laboratory approach worth considering, and analogue gravity is the established one~\cite{Barcelo:2005fc}. Its founding observation, due to Unruh, is that sound in a smooth, inviscid flow obeys a wave equation on an effective acoustic metric, so that the point where the flow reaches the local speed of sound acts on phonons as a horizon acts on light~\cite{Unruh:1980cg,Visser:1997ux,Novello:2002qg}. The idea has been realised on several platforms, among them Bose--Einstein condensates, where both stimulated and spontaneous Hawking emission have been reported~\cite{Garay:1999sk,Steinhauer:2015saa,MunozdeNova:2018fxv}, nonlinear optical systems~\cite{Philbin:2007ji} and open-channel water flows~\cite{Weinfurtner:2010nu,Euve:2015vml}. Those experiments aim mostly at the thermal effect, but the same effective potential also governs the scattering resonances of the acoustic horizon, the analogue quasinormal modes~\cite{Cardoso:2016rao,Anacleto:2019rfn,Oliveira:2018oha,Wang:2019rcf}, and the acoustic setting has been used to model a range of exotic geometries~\cite{Mosna:2016nrt,Bazeia:2014xxa,Li:2013fka,Lessa:2023thi,Lessa:2023dbd,deBrito:2020xhy,Abdalla:2009pg,Cuyubamba:2013iwa,Torres:2016iee,Torres:2019sbr,deOliveira:2023qxe}.

Of the available platforms the de~Laval nozzle, which narrows and then widens, is the one adopted here, and two features recommend it. The first is that the two problems already have the same shape. A static, spherically symmetric black hole reduces its perturbation to a single wave equation in one radial variable, and a duct slender enough to be treated as quasi-one-dimensional reduces its own to a single wave equation along the axis, so the two are matched function against function, with nothing averaged over the transverse directions. The second is that the horizon follows from the flow rather than having to be arranged. In an isentropic flow the sonic point can sit only at the throat, where the cross-section is smallest, so a transonic nozzle places the acoustic horizon there by construction, and the isentropic relations then fix every thermodynamic variable from the Mach number alone, which in turn follows from the area ratio. The whole effective geometry is therefore carried by the shape of the wall, one function of one variable, which is what allows a target potential to be prescribed rather than merely observed. The perturbation equation becomes a Schr\"odinger-type problem whose potential is fixed by the shape of the duct and by the flow~\cite{Sakagami:2001ph,Furuhashi:2006dh}. The map can also be run backwards. Given a target potential, one can reconstruct the shape and the thermodynamic profiles that reproduce it, and with them the ringing~\cite{Okuzumi:2007hf,Abdalla:2007dz}. The quantities involved are ordinary aerodynamic ones, measured as a matter of routine in propulsion work~\cite{Anderson:2010,Sutton:2016,Huzel:1992,Hill:1992,Turner:2008}, which is what keeps the correspondence from being merely formal. Acoustic analogues of decoupled geometries have already been built along these lines~\cite{daRocha:2017lqj}.

This thesis puts the two halves together. Two exotic geometries are taken, one from gravitational decoupling, which deforms a vacuum seed by an extra source and gives hairy black holes~\cite{Ovalle:2016pwp,Ovalle:2017fgl,Ovalle:2018gic,Ovalle:2020kpd}, and one from the effective field theory, whose third-order operator gives the quantum-corrected Schwarzschild metric~\cite{Goroff:1985th,Calmet:2021lny}. Their perturbation problems are then mapped onto transonic flows in de~Laval nozzles, and the shape of the duct, the thermodynamic variables, the Mach number and the thrust coefficient are reconstructed from the effective potential, while the quasinormal spectra, overtones included, are computed within controlled approximations. The work follows and systematises what was established with collaborators in Refs.~\cite{Casadio:2024uwj,Casadio:2025lkn}. Because the quantum correction grows as the mass falls towards the Planck scale, the analysis covers a wide range of masses and identifies where the corrections are largest.

\section{Thesis Organisation}
\label{sec:intro organisation}

The thesis has four parts.

Part~\ref{part: extensions of gr} builds the two geometries. Chapter~\ref{chap:GD} presents the gravitational-decoupling method and constructs hairy black holes by minimal geometric deformation, with attention to the hair parameters and to the conditions they must obey. Chapter~\ref{chap:EFT} presents the effective field theory of gravity, the third-order curvature operator and the changes it makes to the Schwarzschild metric functions, together with the Wald entropy and the horizon structure that follow.

Part~\ref{part: quasinormal modes} sets up the perturbation problem. Chapter~\ref{chap:conexaoentrebnsebocais} derives the mapping used throughout, in which both black-hole and nozzle perturbations become Schr\"odinger-type equations, and specialises the master potentials to the two geometries. Chapter~\ref{chap:QNM_spectrum} defines the quasinormal spectrum and its boundary conditions, presents the Mashhoon method, and computes the frequencies and overtones of both families.

Part~\ref{part: analogue models} assembles the acoustic side. Chapter~\ref{chap: analogue gravity} reviews the hydrodynamic assumptions behind the acoustic metric and the appearance of horizons in transonic flow. Chapter~\ref{chap:isentropic equations} collects the isentropic relations and the performance parameters of de~Laval nozzles, and Chapter~\ref{chap:acoustic laval} constructs the acoustic black hole explicitly, along with the procedure that reproduces a prescribed effective potential in a nozzle.

Part~\ref{part: results} applies the mapping to the two families. For each one the shape of the duct and the flow observables are obtained as functions of the black-hole parameters, following the methodology of Refs.~\cite{Casadio:2024uwj,Casadio:2025lkn}, and the final chapter gathers the results and sets out what would still be needed for an experimental reading.

Three appendices carry the longer calculations, so that the chapters keep to the argument. Appendix~\ref{ap: reducao schrodinger} works through the reduction of the acoustic wave equation to Schr\"odinger form, step by step. Appendix~\ref{ap: analytical solution for lambda} solves the Sturm--Liouville problem for the amplitude $\chi(r)$ on which the reconstruction of the duct rests, in closed form on a Schwarzschild background and as a power series for a general spherically symmetric metric. Appendix~\ref{chap:curvature invariants} computes the curvature invariants of every metric used in the thesis, which is what supports the statements made in the main text about where each deformation is felt.

\part{Hairy and Quantum-Corrected Black Hole Metrics} \label{part: extensions of gr}


\chapter{Gravitational Decoupling}\label{chap:GD}

\epigraph{``Knowledge is a process of piling up facts; wisdom lies in their simplification.''}{Martin H. Fischer}

This chapter develops a robust framework for finding new, physically relevant solutions in extended gravity theories, with a particular focus on keeping the new physics sources decoupled from the established general-relativity background. The methods scrutinised here are the Minimal Geometric Deformation (MGD), a technique that decouples Einstein's Field Equations (EFE) through a geometric deformation of the radial term of a spherically symmetric metric while leaving the temporal term intact~\cite{Ovalle:2016pwp,Ovalle:2017fgl,Ovalle:2017wqi}, and its extension, the Extended Minimal Geometric Deformation (MGDe), which deforms both the radial and the temporal terms of the metric~\cite{Ovalle:2018gic}.

The chapter is organised as follows: Sec.~\ref{sec:GRextension} briefly discusses extensions of general relativity (GR) and their implications for anisotropy and coupled physics. Sec.~\ref{sec:GD} introduces the MGD method before turning to the MGDe, developing the tools needed to decouple and solve a modification of GR. Sec.~\ref{sec:HBHbyGD} applies the MGDe directly to construct a hairy black hole whose energy-momentum tensor obeys the Dominant Energy Condition, and maps the domain of hair parameters over which that construction remains admissible. Finally, Sec.~\ref{sec:GD_summary} closes the chapter with a summary of the results and their role in the remainder of the thesis.

\section{First Thoughts on General Relativity Extensions} \label{sec:GRextension}

Deriving EFE via the variational principle requires the specification of an appropriate action. Since the aim is to describe the interaction between gravitational and matter fields in the classical regime, the framework is that of a classical field theory, which makes the Hilbert--Einstein action the natural choice. Considering a background field described by the metric $g_{\mu\nu}$, coupled to a matter field characterised by the Lagrangian density $\mathcal{L}_M$, and admitting a universe with cosmological constant $\Lambda$, the action takes the form,
\begin{equation} \label{eq:HE action}
    S = \int \dd^4 x \sqrt{-g} \qty[\frac{1}{2\kappa^2}\qty(R-2\Lambda)  + \mathcal{L}_M],
\end{equation}
where $\kappa^2 = 8\pi G_\text{N}$, $G_\text{N}$ Newton's gravitational constant and $g=\det \qty(g_{\mu \nu})$.
By imposing the stationarity condition $\delta S = 0$ in Eq.~\eqref{eq:HE action}, one obtains:
\begin{equation}\label{eq:EFE}
    R_{\mu\nu} - \frac{1}{2}R g_{\mu\nu} + \Lambda g_{\mu\nu} = \kappa^2 T_{\mu\nu}.
\end{equation}
This equation encapsulates how a matter field, characterised by the Lagrangian density $\mathcal{L}_M$, sources the curvature of spacetime geometry in a universe with a cosmological constant $\Lambda$.

Extensions to GR typically proceed by modifying the Hilbert--Einstein action through an additional source term. This new gravitational sector is described by a Lagrangian density $\mathcal{L}_\text{correction}$. The modified action becomes:
\begin{equation} \label{eq:HE action modified}
    S = \int \dd^4 x \sqrt{-g} \qty[\frac{1}{2\kappa^2}\qty(R-2\Lambda)  + \mathcal{L}_M] + \alpha \int \dd^4 x \sqrt{-g} \mathcal{L}_\text{correction}.
\end{equation}
Here, the free parameter $\alpha$ quantifies the coupling strength of the correction to the background physics. Applying the variational principle $\delta S=0$ leads to a modified set of field equations:
\begin{equation}\label{eq:modified EFE}
    R_{\mu\nu} - \frac{1}{2}R g_{\mu\nu} + \Lambda g_{\mu\nu} = \kappa^2 T_{\mu\nu} + \alpha \theta_{\mu\nu}^\text{correction},
\end{equation}
where the correction tensor is defined as:
\begin{equation}
    \theta_{\mu\nu}^\text{correction} = \frac{2}{\sqrt{-g}}\frac{\delta \qty(\sqrt{-g} \mathcal{L}_\text{correction})}{\delta g^{\mu\nu}} = 2 \frac{\delta \mathcal{L}_\text{correction}}{\delta g^{\mu\nu}} - g_{\mu\nu} \mathcal{L}_\text{correction}.
\end{equation}
Superficially, the GR limit is recovered by simply setting $\alpha = 0$, suggesting the two theories are trivially consistent. Probing deeper, however, by solving Eq.~\eqref{eq:modified EFE} for a physically motivated source, say a spherically symmetric perfect fluid, reveals potential subtleties. In standard GR, the line element of such a fluid is given by the ansatz:
\begin{equation}\label{eq:spherical ansatz}
    \dd s^2 = - e^{\nu(r)} \dd t^2 + e^{\lambda(r)} \dd r^2 + r^2 \dd \Omega^2,
\end{equation}
where $\dd \Omega^2 = \dd \theta^2 + \sin^2 \theta \dd \phi^2$ and the radial component is parametrised by the Misner--Sharp mass function $m(r)$:
\begin{equation}
e^{-\lambda(r)} = 1 - \frac{2 G_\text{N} m(r)}{r}.    
\end{equation}
When the same perfect fluid is analysed within the extended theory governed by Eq.~\eqref{eq:modified EFE}, the resulting line element generically takes the form:
\begin{equation} \label{eq:grr modified}
e^{-\lambda(r)} = 1 - \frac{2 G_\text{N} m(r)}{r} + \text{(geometric deformation)}.
\end{equation}
The term (geometric deformation) represents the deviation from the standard GR solution induced by the new gravitational sector, and it is precisely this deformation that induces \emph{anisotropy} in the fluid, breaking the equality between radial and tangential effective pressures. The anisotropy $\Pi = {p}_\text{tan} -{p}_\text{rad}$, for a modified action as Eq.~\eqref{eq:HE action modified} generally decomposes linearly in terms of $\alpha$:
\begin{equation}
    \Pi = \Pi_1 + \alpha \Pi_2,
\end{equation}
where $\Pi_1$ contains the natural anisotropy of standard GR and/or the correction applied in the system, and $\Pi_2$ is induced only by the additional term. Correspondingly, the geometric deformation in Eq.~\eqref{eq:grr modified} can be written as:
\begin{equation}
    \text{(geometric deformation)} = X(r) + \alpha Y(r).
\end{equation}
A critical issue arises if $X(r) \neq 0$. Even in the limit $\alpha \to 0$, the solution does not strictly converge to the standard GR result if the correction term modifies the underlying geometry in a parameter-independent manner.

To better illustrate this, consider as an example the Randall--Sundrum brane-world scenario~\cite{Randall:1999ee,Randall:1999vf}\footnote{The development of the Randall--Sundrum brane-world scenario is beyond the scope of this work. It is used here only for illustration. Ref.~\cite{Maartens:2010ar} reviews the subject in full.}, in which gravity propagates in a five-dimensional bulk that influences the (3+1)-dimensional brane where standard model fields reside. This is also the setting in which the geometric deformation was originally introduced, as a means of generating exact interior solutions for brane-world compact stars~\cite{Ovalle:2010zc,Ovalle:2013xla,Casadio:2012rf,Cavalcanti:2016mbe,Ovalle:2016pwp}, and in which variable-tension branes and their black strings were studied~\cite{daRocha:2009gb,daRocha:2012pt,Abdalla:2009pg}. Projecting the five-dimensional field equations onto the brane~\cite{Shiromizu:1999wj}, the effective 4D Einstein equations then contain corrections to the energy-momentum tensor \cite{Ovalle:2016pwp}:
\begin{equation}\label{eq:energymomentumeffectivebulk}
    T_{\mu\nu}^\text{eff} = T_{\mu\nu} + \frac{6}{\sigma}S_{\mu\nu} + \frac{1}{8\pi} \mathcal{E}_{\mu\nu} + \frac{4}{\sigma} \mathcal{F}_{\mu\nu},
\end{equation}
where $\sigma$ relates to the brane tension (with $\sigma^{-1}$ acting as the coupling parameter $\alpha$). $T_{\mu\nu} = (\varrho+p) u_\mu u_\nu + p g_{\mu\nu}$ is the perfect fluid tensor, in which $\varrho$ denotes the energy density, written with the variant symbol throughout this chapter because $\rho$ is reserved for the mass density of the fluid in Part~\ref{part: analogue models}, $S_{\mu\nu}$ encodes high-energy quadratic corrections, $\mathcal{E}_{\mu\nu}$ represents non-local Kaluza--Klein corrections from the bulk Weyl curvature (proportional to $\sigma^{-1}$ as well), and $\mathcal{F}_{\mu\nu}$ accounts for non-standard model fields possibly existing in the bulk (therefore, can be set to zero for a vacuum bulk).

For a static, spherically symmetric spacetime \eqref{eq:spherical ansatz}, the field equations are:
\begin{subequations}\label{eq:EFEsymetricstatic}
\begin{align}
    \kappa^2 {T_0}^0 &= e^{-\lambda}\qty(\frac{1}{r^2} - \frac{\lambda'}{r})-\frac{1}{r^2} \label{eq:EFE00},\\
    \kappa^2 {T_1}^1 &= e^{-\lambda}\qty(\frac{1}{r^2} + \frac{\nu'}{r})-\frac{1}{r^2} \label{eq:EFE11} ,\\
    \kappa^2 {T_2}^2 &= \frac{e^{-\lambda}}{4} \qty(2\nu'' + \nu'^2 - \lambda' \nu' + 2 \frac{\nu'-\lambda'}{r}) \label{eq:EFE22},
\end{align}
\end{subequations}
where primes denote radial derivatives. Substituting $T_{\mu\nu}^\text{eff}$ into Eqs.~\eqref{eq:EFEsymetricstatic} then yields a modified radial metric component:
\begin{equation} \label{eq:grr bulk}
    e^{-\lambda(r)} = 1- \frac{2 G_\text{N} m(r)}{r} + f(r),
\end{equation}
with the geometric deformation $f(r)$ given by \cite{Ovalle:2016pwp}:
\begin{equation} \label{eq:f r example}
    f(r) =   e^{-I(r)}\qty{ \int_0^r \frac{2 \tilde{r} e^{I\qty(\tilde{r})}}{\tilde{r} \nu' + 4} \qty[H\qty(p,\varrho,\nu) + \frac{\kappa^2}{\sigma}\qty(\varrho^2+3\varrho p)] \dd \tilde{r} + \beta(\sigma)}.
\end{equation}
The integrating factor $I(r)$ is defined as:
\begin{equation}
    I(r) \equiv \int_0^r \frac{\tilde{r}^2\qty(2  \nu'' + \nu'^2) + 4\tilde{r} \nu' + 4}{\tilde{r} \qty(\tilde{r} \nu' + 4)} \dd \tilde{r}.
\end{equation}
The term $H\qty(p,\varrho,\nu)$ encapsulates non-local effects:
\begin{equation}
    H\qty(p,\varrho,\nu) \equiv 3\kappa^2 p - \qty[\frac{2 G_\text{N} m}{r}\qty(\frac{1}{r}-\frac{m'}{m}) \qty(\frac{\nu'}{2} + \frac{1}{r}) + \qty(1-\frac{2 G_\text{N} m}{r})\qty(\nu'' + \frac{\nu'^2}{2} + \frac{2\nu'}{r}+\frac{1}{r^2}) - \frac{1}{r^2}],
\end{equation}
and $\beta(\sigma)$ is an integration constant chosen to vanish as $\sigma^{-1} \to 0$.

In particular, the deformation $f(r)$ contains terms that are independent of $\sigma$ (within the integral involving $H$), and if these do not vanish identically, the solution retains a contribution from the new sector even in the limit where the extended theory should reduce to GR. To ensure a consistent limit, one must therefore enforce the constraint:
\begin{equation}
      \lim_{\sigma^{-1} \to 0} \int_0^r \frac{2 \tilde{r} e^{I\qty(\tilde{r})}}{\tilde{r} \nu' + 4} H\qty(p,\varrho,\nu) \dd \tilde{r} = 0.
\end{equation}
The fact that direct resolution of the modified EFE can lead to $\sigma$-independent geometric deformations highlights a potential issue. The new sector may remain coupled to the background physics, complicating the physical interpretation. To resolve this and systematically separate the new physics from the standard background, the gravitational decoupling (GD) method is employed. Its decisive feature is that it does not refer to the five-dimensional origin at all: once the deformation is treated as an ansatz on the metric functions rather than as a projection of a bulk, the brane tension may be replaced by an arbitrary coupling and the additional source by an arbitrary energy-momentum tensor~\cite{Ovalle:2017fgl,Ovalle:2017wqi}. The brane-world example is retained here only because it exhibits the obstruction in a concrete setting, and nothing in the construction that follows depends on it~\cite{Ovalle:2016pwp}.

\section{The Gravitational Decoupling Method}\label{sec:GD}

The GD Method rests on the MGD approach~\cite{Ovalle:2016pwp,Ovalle:2017fgl,Ovalle:2017wqi,Ovalle:2018gic}, whose central idea is to trade a single, fully coupled non-linear problem for two simpler, independent ones. Restricting the analysis to a spherical and static spacetime, with a metric ansatz as in Eq.~\eqref{eq:spherical ansatz}, one writes an effective energy-momentum tensor as the linear combination of a well-behaved, known solution (the seed) and an additional source term (the correction),
\begin{equation} \label{eq:effective energy momentum tensor}
    \tilde{T}_{\mu\nu} = T_{\mu\nu}^{\text{seed}} + \theta_{\mu\nu},
\end{equation}
where $\theta_{\mu\nu}$ collects whatever new physics, such as hair, anisotropy, or a quantum correction, is to be grafted onto the seed. The strategy that gives the method its name is geometric rather than algebraic. Instead of solving directly for $\tilde{T}_{\mu\nu}$, the metric functions themselves are deformed, with the piece of the deformation proportional to the coupling $\alpha$ carrying all the information about $\theta_{\mu\nu}$, while the seed geometry is exactly recovered in the limit $\alpha \to 0$. Substituted back into the EFE, Eq.~\eqref{eq:EFEsymetricstatic}, this deformation splits the problem cleanly in two, one part recovering the ordinary Einstein equations for the seed source alone, and the other a formally analogous but distinct system, the so-called \emph{quasi-Einstein equations}, that determines the deformation sourced purely by $\theta_{\mu\nu}$. Because the two sets decouple, $\theta_{\mu\nu}$ can be prescribed or solved for independently of the seed, which is precisely what allows new physics to be added to a known solution without having to resolve the full non-linear problem from scratch.

Let the seed solution be parametrised by the line element
\begin{equation} \label{eq:knownseed}
    \dd s^2 = - e^{\xi (r)} \dd t^2 + e^{\zeta(r)} \dd r^2 + r^2 \dd \Omega ^2,
\end{equation}
where $e^{-\zeta(r)} = 1-2 G_\text{N} m(r)/r$ also involves the standard Misner--Sharp mass function. Since the metric \eqref{eq:knownseed} is a known solution of the EFE for the source $T_{\mu\nu}^{\text{seed}}$, it can be used as a basis to construct the general solution associated with the full tensor in Eq.~\eqref{eq:effective energy momentum tensor}.

In its original and simplest form, the MGD deforms only the radial metric component, leaving the temporal one untouched. Denoting the seed geometry by $\{\xi, \zeta\}$ and the deformed one by $\{\nu, \lambda\}$, the transformation reads \cite{Ovalle:2016pwp}
\begin{subequations}
\label{eq:mgd_deformation}
\begin{align}
    \xi(r) &\mapsto \nu(r) = \xi(r), \label{eq:mgd_temporal}\\
    e^{-\zeta(r)} &\mapsto e^{-\lambda(r)} = e^{-\zeta(r)} + \alpha f(r), \label{eq:mgd_radial}
\end{align}
\end{subequations}
where $f(r)$ is the radial geometric deformation and $\alpha$ is a coupling parameter tracking the correction strength. The deformation is called \emph{minimal} precisely because it acts on a single metric function, the radial one.

This minimality, however, comes with a warning. Because the temporal function $\xi(r)$ is left untouched by Eq.~\eqref{eq:mgd_temporal}, the relation $g_{tt} = -g_{rr}^{-1}$ that identifies the horizon of many seed solutions no longer holds automatically for the deformed metric. The zero of $g_{tt}$ and the zero of $g_{rr}^{-1}$ can, in general, sit at different radii, leading to a horizon structure that need not resemble the seed's at all. This mismatch can be resolved by the extended version of the method, the MGDe~\cite{Ovalle:2018gic,Ovalle:2018umz}, in which \emph{both} metric functions are deformed,
\begin{subequations}
\label{eq:deformations}
\begin{align}
    \xi(r) &\mapsto \nu(r) = \xi(r) + \alpha g(r), \label{eq:deformation_temporal}\\
    e^{-\zeta(r)} &\mapsto e^{-\lambda(r)} = e^{-\zeta(r)} + \alpha f(r), \label{eq:deformation_radial}
\end{align}
\end{subequations}
with $g(r)$ the temporal geometric deformation. The original MGD is recovered by setting $g=0$. Even in the extended scheme, recovering a single, well-defined horizon is not automatic and must be imposed as an extra physical requirement on $\theta_{\mu\nu}$, as done explicitly in Sec.~\ref{sec:HBHbyGD}.

Substituting the ansatz \eqref{eq:deformations} into the Einstein equations $G_{\mu\nu} = \kappa^2 \tilde{T}_{\mu\nu}$, the system splits into two distinct sets. Modelling the seed source as a perfect fluid with energy density $\varrho^{\text{seed}}$ and radial and tangential pressures $p_{\text{rad}}^{\text{seed}}$, $p_{\text{tan}}^{\text{seed}}$,
\begin{equation}
    {T_\mu}^\nu = \text{diag}\qty(-\varrho^{\text{seed}},\ p_{\text{rad}}^{\text{seed}},\ p_{\text{tan}}^{\text{seed}},\ p_{\text{tan}}^{\text{seed}}),
\end{equation}
the first set corresponds to the standard Einstein equations for the seed source:
\begin{subequations}
\label{eq:seed_equations}
\begin{align}
    \kappa^2 \varrho^{\text{seed}} &= e^{-\zeta}\qty(\frac{1}{r^2} - \frac{\zeta'}{r}) - \frac{1}{r^2}, \\
    \kappa^2 p_{\text{rad}}^{\text{seed}} &=  e^{-\zeta}\qty(\frac{1}{r^2} + \frac{\xi'}{r}) -\frac{1}{r^2}, \\
    \kappa^2 p_{\text{tan}}^{\text{seed}} &= \frac{e^{-\zeta}}{4}\qty(2\xi'' + \xi'^2 - \zeta'\xi' + 2\frac{\xi' - \zeta'}{r}),
\end{align}
\end{subequations}
where the prime, once again, denotes differentiation with respect to $r$. The second set, governing the new source $\theta_{\mu\nu}$, constitutes the quasi-Einstein equations:
\begin{subequations}
\label{eq:quasi_einstein}
\begin{align}
    \kappa^2 {\theta_0}^0 &= \alpha \qty(\frac{f}{r^2} + \frac{f'}{r}), \label{eq:theta00}\\
    \kappa^2 {\theta_1}^1 &= \alpha \qty[ \frac{f}{r^2} + \frac{\nu' f}{r}  + \frac{g'}{r}e^{-\zeta}], \label{eq:theta11}\\
    \kappa^2 {\theta_2}^2 &= \alpha \qty{\frac{f}{4}\qty(2\nu'' + \nu'^2 + \frac{2\nu'}{r}) + \frac{f'}{4}\qty(\nu' + \frac{2}{r}) + \frac{e^{-\zeta}}{2}\qty[ g'' + \frac{g'}{2} \qty(g' + \frac{2}{r} + 2 \xi'  -\zeta' ) ]} . \label{eq:theta22}
\end{align}
\end{subequations}
Note that Eqs.~\eqref{eq:quasi_einstein} are $\alpha$-dependent, so that in the limit $\alpha \to 0$ one recovers GR automatically.

The conservation of the energy-momentum tensor ensures the consistency of this decoupling, $\nabla_\mu \tilde{T}^{\mu\nu} = 0$. Since the seed solution independently satisfies $\nabla_\mu T^{\mu\nu}_{\text{seed}} = 0$, the additional source must satisfy the Bianchi identities in the deformed geometry. Explicitly, the conservation equation for $\theta_{\mu\nu}$ reads:
\begin{equation}
\label{eq:conservation_theta}
    \qty({\theta_1}^1)' - \frac{\nu'}{2}\qty({\theta_0}^0 - {\theta_1}^1) - \frac{2}{r}\qty({\theta_2}^2 - {\theta_1}^1) = 
    \frac{\alpha g'}{2}\qty({T_0}^0 - {T_1}^1),
\end{equation}
which describes the exchange of energy-momentum between the seed and the correction term. When the two sources are non-interacting (only gravitationally coupled), $\nabla_\mu \theta^{\mu\nu} = 0$ must instead hold independently.
The particular case $g=0$, called MGD-decoupling \cite{Ovalle:2017fgl}, realises exactly this non-interacting decoupling. It is worth noting that the same effect can also be enforced, even when the full MGDe is used, through a suitable choice of fluid, namely a barotropic equation of state ${T_0}^0 = {T_1}^1$, or simply a vacuum region $T_{\mu\nu} =0$. The energy exchange described by the right-hand side of Eq.~\eqref{eq:conservation_theta} is not a defect of the extended scheme but a physical channel in its own right, and it has been used to study isotropisation and the change of complexity of a self-gravitating configuration~\cite{Casadio:2019usg,Ovalle:2022yjl}.

This formalism allows for the systematic inclusion of modifications, such as hair, anisotropies, or quantum corrections, into a known general relativity solution while keeping physical relevance. By solving the quasi-Einstein system \eqref{eq:quasi_einstein} for $f(r)$ and $g(r)$ under physically motivated constraints (e.g., the dominant energy condition or specific equations of state), one obtains the complete metric describing the extended system.

The scope of the procedure is considerably wider than the static, spherically symmetric case treated here. Decoupling has been formulated in $2+1$ dimensions, with and without a cosmological term~\cite{Contreras:2018vph,Contreras:2019mhf,Contreras:2019fbk,Contreras:2019iwm,Rincon:2020izv}, in asymptotically (anti-)de~Sitter backgrounds~\cite{Contreras:2018nfg,Zhang:2022niv}, in higher dimensions and in pure Lovelock gravity~\cite{Estrada:2018vrl,Estrada:2019aeh,Estrada:2021kuj}, in higher-order theories of gravity~\cite{Sultana:2021cvq}, for axially symmetric and rotating configurations~\cite{Contreras:2021yxe,Mahapatra:2022xea,Ovalle:2021jzf}, for Vaidya and generalised Vaidya geometries~\cite{Vertogradov:2022oif}, in the brane-world treated as a self-consistent decoupled system~\cite{Leon:2019abq,Leon:2021qvs}, and in cosmology~\cite{LinaresCedeno:2019aul,Sharif:2023opu}. The inverse problem, of reconstructing the deformation from a prescribed anisotropy, was posed in Ref.~\cite{Contreras:2018gzd}, and the complete geometric deformation, in which the two deformation functions are treated on the same footing without the auxiliary constraints used here, was developed in Refs.~\cite{Heras:2018cpz,Heras:2021xxz,Maurya:2020gjw,Maurya:2021qye,Maurya:2022yva,Leon:2023nbj}. The bulk of the applications concerns relativistic stellar structure, where the deformation supplies the anisotropy of the interior~\cite{Ovalle:2018vmg,Estrada:2018zbh,Hensh:2019rtb,Gabbanelli:2018bhs,Gabbanelli:2019txr,Morales:2018urp,Morales:2018nmq,PerezGraterol:2018eut,Singh:2019ktp,Torres-Sanchez:2019wjv,Ovalle:2019lbs,Carrasco-Hidalgo:2021dyg,Santana:2022vmw,daRocha:2021aww,daRocha:2021sqd,daRocha:2020jdj,Fernandes-Silva:2018abr,Panotopoulos:2018law}, often organised through the complexity factor~\cite{Contreras:2021xkf,Contreras:2022vec,Maurya:2021yhc,Maurya:2023szc,Sharif:2023ecm,Naseer:2024jjy} or transplanted into modified gravity~\cite{Sharif:2019mjn,Sharif:2021emv,Maurya:2019hds,Maurya:2020ebd,Maurya:2022wwa,Maurya:2022hmw,Azmat:2021qig,Bamba:2023wok,Panyasiripan:2024kyu}, with observational constraints drawn from strange-star mass-radius relations and from GW190814~\cite{Maurya:2023muz,Maurya:2024ylr}. None of these applications is used below, but they establish that the splitting exploited in this chapter is a general property of the field equations rather than a feature of the particular seed adopted here.

\section{Hairy Black Holes via Gravitational Decoupling}\label{sec:HBHbyGD}

A direct application of the MGDe discussed in Sec.~\ref{sec:GD} is the construction of black hole solutions that evade the no-hair theorem~\cite{Israel:1967wq,Carter:1971zc,Robinson:1975bv,Bekenstein:1972ny}. By introducing a generic source $\theta_{\mu\nu}$, the vacuum Schwarzschild solution can be extended to include a new sector, interpreted here as hair, without disrupting the event horizon structure or introducing pathological singularities~\cite{Ovalle:2018umz,Ovalle:2020kpd}. The construction that follows is the one of Ref.~\cite{Ovalle:2020kpd}, and it is the first member of a family that has since grown to include hairy solutions sourced by a scalar field~\cite{Ovalle:2018ans,Arias:2022jax}, regular hairy black holes obtained by deforming Minkowski space~\cite{Ovalle:2023ref}, non-singular and Hayward-type geometries~\cite{Misyura:2024fho,Naseer:2025ghn}, and solutions with an extended parameter range~\cite{Avalos:2023ywb}.

Consider the Schwarzschild metric as the seed solution, described by the line element \eqref{eq:knownseed} with:
\begin{equation}
    e^{\xi(r)} = e^{-\zeta(r)} = 1 - \frac{2 G_\text{N} M}{r},
\end{equation}
where $M$ is the mass of the black hole, and the corrections introduced by the source $\theta_{\mu\nu}$ deform this geometry via the transformations \eqref{eq:deformations}. To ensure that the resulting solution describes a physically consistent black hole, three fundamental premises are imposed \cite{Ovalle:2020kpd}:

\begin{enumerate}
    \item The solution must represent an extension of the Schwarzschild spacetime, recovering the vacuum solution when the decoupling parameter $\alpha \to 0$;
    \item The solution must have a well-defined event horizon structure. This implies that the deformed metric components should maintain the inverse relationship characteristic of Schwarzschild-like coordinates, i.e., $g_{tt} = - g_{rr}^{-1}$, or simply $e^{\nu(r)} = e^{-\lambda(r)}$;
    \item The effective energy-momentum tensor $\tilde{T}_{\mu\nu}$ must satisfy the Dominant Energy Condition (DEC), ensuring the physical viability of the hair distribution outside the horizon.
\end{enumerate}

The second premise imposes an important constraint on the equation of state. From ${G_0}^0=\kappa^2 {T_0}^0$ and ${G_1}^1=\kappa^2 {T_1}^1$, one obtains the relation:
\begin{equation} \label{eq:state equation}
    \tilde{p}_{\text{rad}} = -\tilde{\varrho},
\end{equation}
where, for Schwarzschild,
\begin{subequations} \label{eq:effective tensor perfect fluid}
\begin{align}
    \tilde{\varrho} &= {T_0}^0 + {\theta_0}^0 = {\theta_0}^0, \\
    \tilde{p}_\text{rad} &= -{T_1}^1 - {\theta_1}^1 = - {\theta_1}^1, \\
    \tilde{p}_\text{tan} &= -{T_2}^2 - {\theta_2}^2 = - {\theta_2}^2.
\end{align}
\end{subequations}
Therefore, only negative radial pressure is obtained for positive density.
The first and second premises together imply that the temporal and radial geometric deformations are coupled:
\begin{equation}\label{eq:deformation_coupling}
    e^{\xi(r)+ \alpha g(r)} = e^{-\zeta(r)}+\alpha f(r) \mapsto \alpha f(r) = \qty(1-\frac{2 G_\text{N} M}{r})\qty(e^{\alpha g(r)} - 1),
\end{equation}
and the final line element becomes
\begin{equation}
    \dd s^2 = -\qty(1-\frac{2 G_\text{N} M}{r})e^{\alpha g(r)} \dd t^2 + \qty(1-\frac{2 G_\text{N} M}{r})^{-1}e^{-\alpha g(r)} \dd r^2 + r^2 \dd \Omega^2.
\end{equation}

The third premise, for a perfect fluid, can be put in terms of two inequalities:
\begin{subequations} \label{eq:dec}
\begin{align}
    \tilde{\varrho} &\geq \qty|\tilde{p}_\text{rad}| \label{eq:dec prad} \\
    \tilde{\varrho} &\geq \qty|\tilde{p}_\text{tan}| \label{eq:dec tan}
\end{align}
\end{subequations}
Because of Eq.~\eqref{eq:state equation}, the inequality \eqref{eq:dec prad} is automatically satisfied and saturated, while the remaining inequality \eqref{eq:dec tan}, combined with Eqs.~\eqref{eq:effective tensor perfect fluid}, gives the following inequations:
\begin{subequations} \label{eq:inequations}
\begin{align}
    {\theta_0}^0 + {\theta_2}^2 &\geq 0 \label{eq:ineq 1}\\
    {\theta_0}^0 - {\theta_2}^2 &\geq 0 \label{eq:ineq 2}.
\end{align}
\end{subequations}
To find $f(r)$ or $g(r)$, the next step is substituting Eqs.~\eqref{eq:quasi_einstein} into Eqs.~\eqref{eq:inequations}, which gives the following differential inequations:
\begin{subequations} \label{eq:dif ineq}
\begin{align} 
I_1 (r) &\equiv -r (r-2 G_\text{N} M) y'' - 4(r-G_\text{N} M) y' - 2y + 2 \geq 0  \label{eq:dif ineq 1}  \\
I_2 (r) &\equiv r (r-2 G_\text{N} M) y'' + 4 G_\text{N} M y' - 2y + 2 \geq 0, \label{eq:dif ineq 2}
\end{align}
\end{subequations}
where $y(r) \equiv e^{\alpha g(r)}$ is introduced to simplify the notation. Saturating Eqs.~\eqref{eq:dif ineq} then gives the solutions:
\begin{subequations}
\begin{align}
    y_1^{\text{saturated}}(r) =& 1 + \qty(1-\frac{2 G_\text{N} M}{r})^{-1} \qty(\frac{c_{11}}{r} - \frac{c_{12}}{r^2}) \\
    y_2^{\text{saturated}}(r) =& 1 + \qty(1-\frac{2 G_\text{N} M}{r})^{-1} \qty(\frac{c_{21}}{r} + c_{22} r^2),
\end{align}
\end{subequations}
where $c_{ij}$ are constants of integration, and these solutions give the following metrics:
\begin{subequations} \label{eq:sol sat} 
\begin{align}
    \text{1:} \ e^{\nu(r)} =  e^{-\lambda(r)} &= 1-\frac{2 G_\text{N} M-c_{11}}{r} - \frac{c_{12}}{r^2} \label{eq:sol numb 1 sat} \\
    \text{2:} \ e^{\nu(r)} =  e^{-\lambda(r)} &= 1-\frac{2 G_\text{N} M-c_{21}}{r} + c_{22} r^2. \label{eq:sol numb 2 sat}
\end{align}
\end{subequations}
Neither solution introduces genuinely new physics. They represent two well-known solutions with a shifted mass~\cite{Ovalle:2018umz,Ovalle:2020kpd}. Solution \#1, Eq.~\eqref{eq:sol numb 1 sat}, represents the Reissner--Nordstr\"om metric, with the mass shifted as $M \mapsto M - c_{11}/(2G_\text{N})$ and the constant $c_{12}$ identified by comparison with that solution as $c_{12} = -Q^2$. Solution \#2, Eq.~\eqref{eq:sol numb 2 sat}, represents the Schwarzschild-de Sitter metric, with the same type of mass shift, $M \mapsto M - c_{21}/(2G_\text{N})$, and $c_{22} = -\Lambda/3$ read off in the same way.
To circumvent saturation and achieve new solutions, the strategy is to saturate only near the boundaries of the outer region, $I_1(r) = 0$ for $r\sim 2 G_\text{N} M$ and $r \gg G_\text{N} M$, so that the solution has a proper horizon and also behaves approximately like the Schwarzschild metric for $r \gg 2 G_\text{N} M$, to meet experimental bounds in the weak field regime \cite{Ovalle:2020kpd}. Imposing these conditions yields a modification of Solution \#1:
\begin{equation}
    y(r) = 1 - \qty(1-\frac{2 G_\text{N} M}{r})^{-1} \qty(\frac{G_\text{N} \alpha \ell_h}{r} + \frac{\alpha G_\text{N} M e^{-r/(G_\text{N} M)}}{r} - \frac{Q^2}{r^2}),
\end{equation}
where the integration constants of Solution~\#1 are identified as $c_{11} \equiv - G_\text{N} \alpha \ell_h$ and $c_{12} \equiv -Q^2$, the signs following from the Reissner--Nordstr\"om reading of Eq.~\eqref{eq:sol numb 1 sat}; the geometric deformation of a genuinely charged background, in which $Q$ is sourced by a gauge field, is a distinct problem and was treated separately in Ref.~\cite{Rincon:2019jal}. The coupling $\alpha$ is dimensionless, since the exponential term of the deformation multiplies the dimensionless combination $G_\text{N}M/r$, so that $\ell_h$ carries dimensions of mass, $G_\text{N}\ell_h$ being the length associated with the primary hair, while $Q$ has dimensions of length. The new metric then becomes
\begin{equation} \label{eq: metric HBH}
    e^{\nu(r)} =  e^{-\lambda(r)} = 1 - \frac{2 G_\text{N} \mathcal{M}}{r} + \frac{Q^2}{r^2} - \frac{\alpha G_\text{N}}{r}\qty(\mathcal{M} - \frac{\alpha \ell_h}{2})e^{-r/G_\text{N}\qty(\mathcal{M}-\frac{\alpha \ell_h}{2})},
\end{equation}
where $\mathcal{M} \equiv M + \alpha \ell_h/2$. It is essential to note that $Q$ does not mean an electric charge, as in the Reissner--Nordström metric, since it is not sourced by any gauge field and, as shown below, is constrained by the correction strength $\alpha$ through the dominant energy condition.

Applying the metric to Eqs.~\eqref{eq:quasi_einstein} gives the effective energy-momentum tensor:
\begin{subequations} \label{eq:effective calc tensor}
\begin{align}
    \tilde{\varrho} = -\tilde{p}_\text{rad} &= \frac{1}{\kappa^2r^2} \qty(\frac{Q^2}{r^2} - \alpha e^{-r/(G_\text{N} M)}), \label{eq:effective calc tensor 00 11} \\
    \tilde{p}_\text{tan} &= \frac{1}{\kappa^2r^2} \qty(\frac{Q^2}{r^2} -  \frac{\alpha r}{2 G_\text{N} M} e^{-r/(G_\text{N} M)}). \label{eq:effective calc tensor 22}
\end{align}
\end{subequations}

An anisotropy arises from Eqs.~\eqref{eq:effective calc tensor}, taking the form
\begin{equation}
    \Pi = \tilde{p}_\text{tan} -\tilde{p}_\text{rad} =  \frac{1}{\kappa^2r^2} \qty[\frac{2Q^2}{r^2} - \qty( 1+ \frac{ r}{2 G_\text{N} M}) \alpha e^{-r/(G_\text{N} M)} ].
\end{equation} 

The curvature of the deformed geometry follows from the same metric, and Appendix~\ref{chap:curvature invariants} computes its invariants in full. Two of the results are worth recording here. The Ricci scalar reduces to
\begin{equation}
\label{eq: ricci hbh}
    R = \frac{\alpha\qty(r-2G_\text{N}M)}{G_\text{N}M\,r^2}\,e^{-r/G_\text{N}M},
\end{equation}
which depends on neither $Q$ nor $\ell_h$, since the charge-like term is traceless and the primary hair only shifts the mass, and which agrees with the trace of Eqs.~\eqref{eq:effective calc tensor}, as it must. The Kretschmann scalar is dominated at large radii by $48G_\text{N}^2\mathcal{M}^2/r^6$, so that $\mathcal{M}$ is the mass registered at infinity. At the centre it diverges as $56Q^4/r^8$ when $Q\neq0$, and as $48G_\text{N}^2\mathcal{M}_\text{eff}^2/r^6$ with $\mathcal{M}_\text{eff} = M + \alpha(M+\ell_h)/2$ when it does not. The hair therefore does not remove the singularity of the seed, it dresses it, and the mass governing the curvature at the centre is not the one measured asymptotically.

With the metric calculated, the next step is to determine the map of accepted parameters and horizon radii.

\subsection{Map of accepted parameters and horizon radii}
\label{subsec: map of accepted}
To ensure that DEC is satisfied, the inequality~\eqref{eq:dec tan} implies that
\begin{equation}
    \tilde{\varrho} - \tilde{p}_\text{tan} = \frac{e^{-r/(G_\text{N} M)}}{2\kappa^2 G_\text{N} M r^2} \alpha \qty(r-2 G_\text{N} M) \geq 0.
\end{equation}
Thus, for $r\geq 2 G_\text{N} M$, $\alpha$ must be greater than or equal to zero.
Since the density is positive, Eq.~\eqref{eq:effective calc tensor 00 11} imposes, for $r\geq 2 G_\text{N} M$, a limit between $Q$, $\alpha$, and $M$:
\begin{equation} \label{eq: limit Q alpha M}
    \tilde{\varrho} \geq 0 \quad \Longrightarrow \quad Q^2 \geq  \frac{4\alpha G_\text{N}^2 M^2}{e^2}.
\end{equation}
The horizon radii $r_\text{EH}$ are given by solving $e^{\nu(r_\text{EH})} = e^{-\lambda(r_\text{EH})} = 0$,
\begin{equation} \label{eq: horizon radii}
    r_\text{EH} - \qty(2  M + \alpha \ell_h) G_\text{N} + \frac{Q^2}{r_\text{EH}} -\alpha G_\text{N} M e^{-r_\text{EH}/(G_\text{N} M)} = 0.
\end{equation}
Evaluating Eq.~\eqref{eq: horizon radii} at $r_\text{EH} = 2 G_\text{N} M$, where the bound~\eqref{eq: limit Q alpha M} is saturated, gives
\begin{equation} \label{eq: Q2 at horizon}
    Q^2 = 2 \alpha G_\text{N}^2 M \qty(\ell_h + \frac{M}{e^2}),
\end{equation}
which, combined with Eq.~\eqref{eq: limit Q alpha M}, yields a lower limit between $\ell_h$ and $M$:
\begin{equation} \label{eq: limit ell M}
    \ell_h \geq \frac{M}{e^2}.
\end{equation}
Equation~\eqref{eq: Q2 at horizon} also shows that $Q^2$ is proportional to $\alpha$, so that
\begin{equation} \label{eq: Q vanishes with alpha}
    \lim_{\alpha \to 0} Q = 0.
\end{equation}
This is not an extra assumption but the first premise of Sec.~\ref{sec:HBHbyGD} at work. The parameter $Q$ enters through the integration constant $c_{12}$ of the deformation~\eqref{eq:deformation_coupling}, which is itself proportional to $\alpha$, so switching off the coupling switches off the charge-like parameter along with the hair, and the Schwarzschild seed is recovered.
To ensure asymptotic flatness, the exponential term in Eq.~\eqref{eq: metric HBH} must diminish at $r\to \infty$, thus
\begin{equation} \label{eq: limit M alpha ell}
    \mathcal{M} > \frac{\alpha \ell_h}{2}
\end{equation}
to keep consistency.

\begin{figure}[htbp]
    \centering
    \includegraphics[width=1\linewidth]{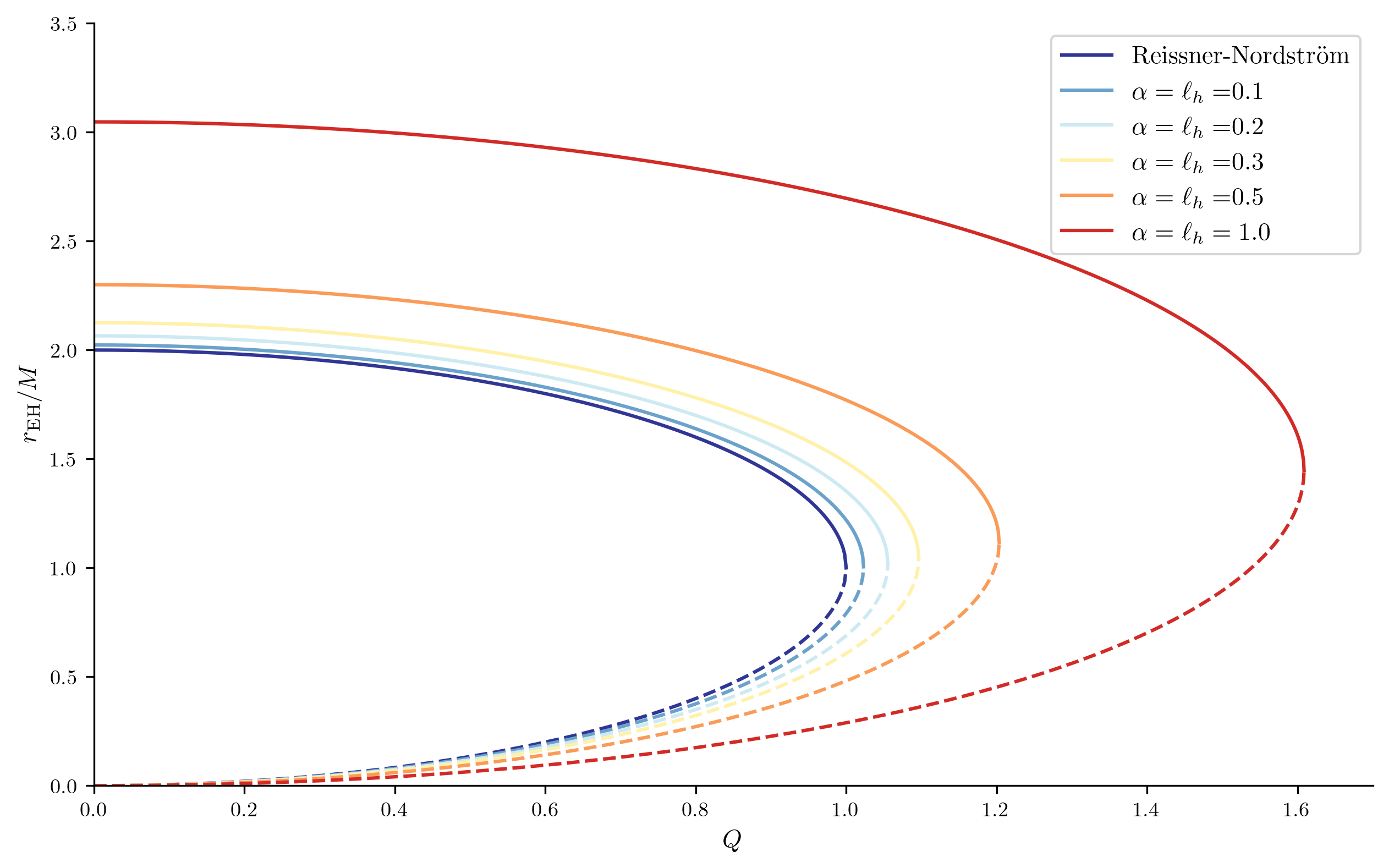}
    \caption[Event and Cauchy horizons of the hairy black hole against $Q$]{Event (solid) and Cauchy (dashed) horizons of the hairy black hole~\eqref{eq: metric HBH} as functions of $Q$, for the values of $\alpha=\ell_h$ given in the legend, with Reissner--Nordstr\"om for comparison. Natural units with $G_\text{N}=M=1$.}
    \label{fig:eventhorizon}
\end{figure}

For the numerical evaluation of Eq.~\eqref{eq: horizon radii} and of the figures below, $G_\text{N}=M=1$ is adopted. The parameter map obtained in this subsection was reported in Ref.~\cite{Casadio:2024uwj}. Numerically solving Eq.~\eqref{eq: horizon radii} then shows that the maximum number of solutions is two, just as for the Reissner--Nordström black hole. Accordingly, the larger root will be called the event horizon and the smaller one the Cauchy horizon. Figure~\ref{fig:eventhorizon} shows the values of the event and Cauchy horizons for different combinations of $\alpha$, $\ell_h$, and $Q$, with solid curves denoting the event horizon, dashed curves the Cauchy horizon, and colour denoting the combination of $\alpha$ and $\ell_h$ used. As with the Reissner--Nordström black hole, there is a saturating value of $Q$ beyond which no horizon exists, and this saturation value grows with both $\alpha$ and $\ell_h$.

\begin{figure}[htbp]
    \centering
    \includegraphics[width=0.8\linewidth]{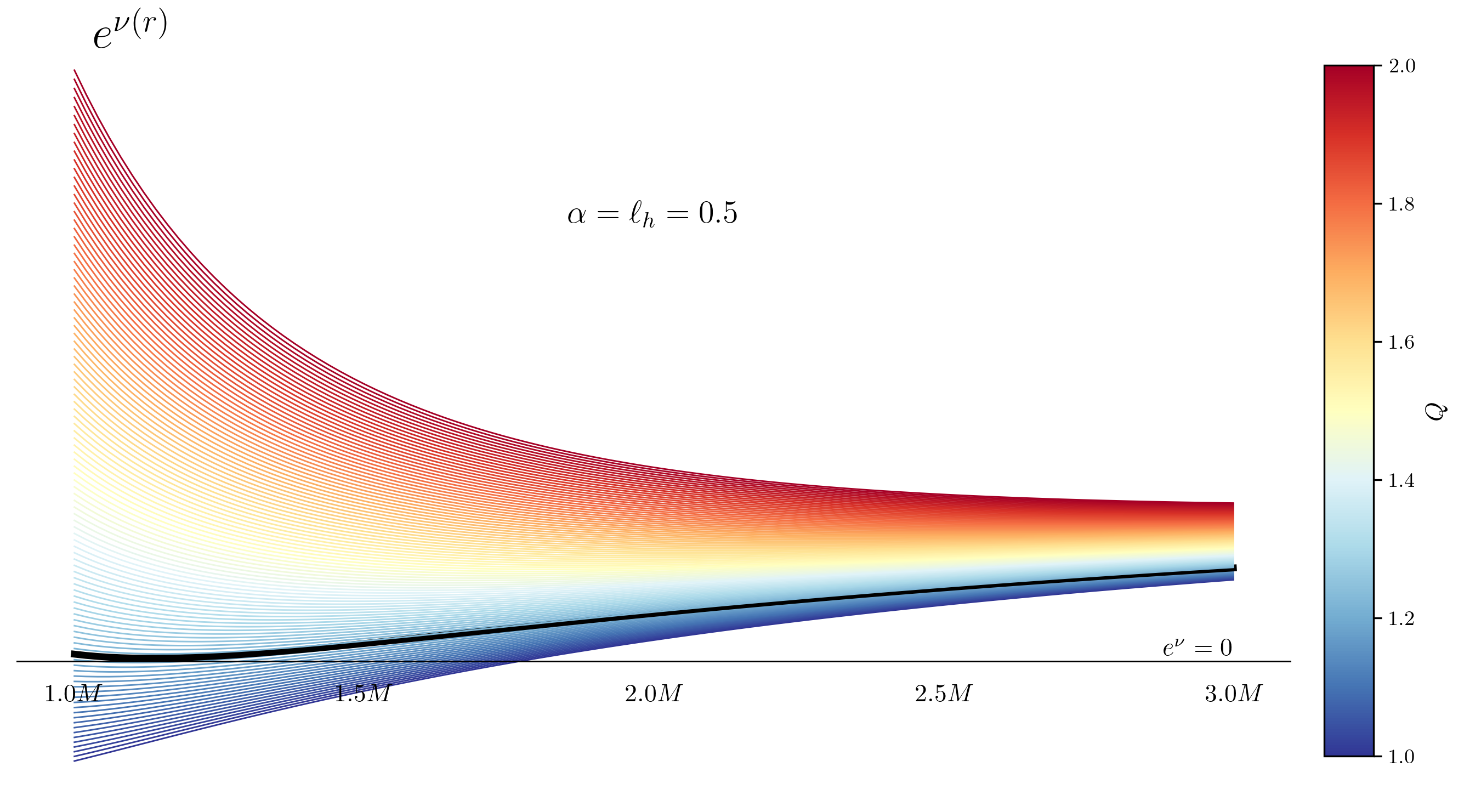}
    \caption[Metric function of the hairy black hole against the radial coordinate]{Metric function $e^{\nu(r)}$ of the hairy black hole~\eqref{eq: metric HBH} against the radial coordinate, for $\alpha = \ell_h = 0.5$ and $Q$ varied over $[1.0, 2.0]$ (colour bar). The black curve marks the saturation, beyond which no horizon exists, and the horizontal line marks $e^{\nu}=0$.}
    \label{fig:expnur}
\end{figure}

Not every $Q$, however, admits a solution for a given pair $(\alpha, \ell_h)$. Fig.~\ref{fig:expnur} plots the metric function for $\alpha = \ell_h = 0.5$ while varying $Q$, and since the horizon radius must remain real, there is an upper limit beyond which no horizon exists. For this pair of parameters the saturation occurs at $Q\approx 1.20$. The restriction applies to the real branch alone. Allowing $Q$ to be imaginary, that is, $Q^2<0$ in Eq.~\eqref{eq: metric HBH}, turns the $Q^2/r^2$ term attractive and restores real roots of Eq.~\eqref{eq: horizon radii} beyond the saturation value, so that further horizon configurations remain available outside the range considered here. Only real $Q$ is retained throughout this thesis, since Eq.~\eqref{eq: limit Q alpha M} requires $Q^2\geq0$ for the effective density to be non-negative.

\begin{figure}[htbp]
    \centering
    \includegraphics[width=0.8\linewidth]{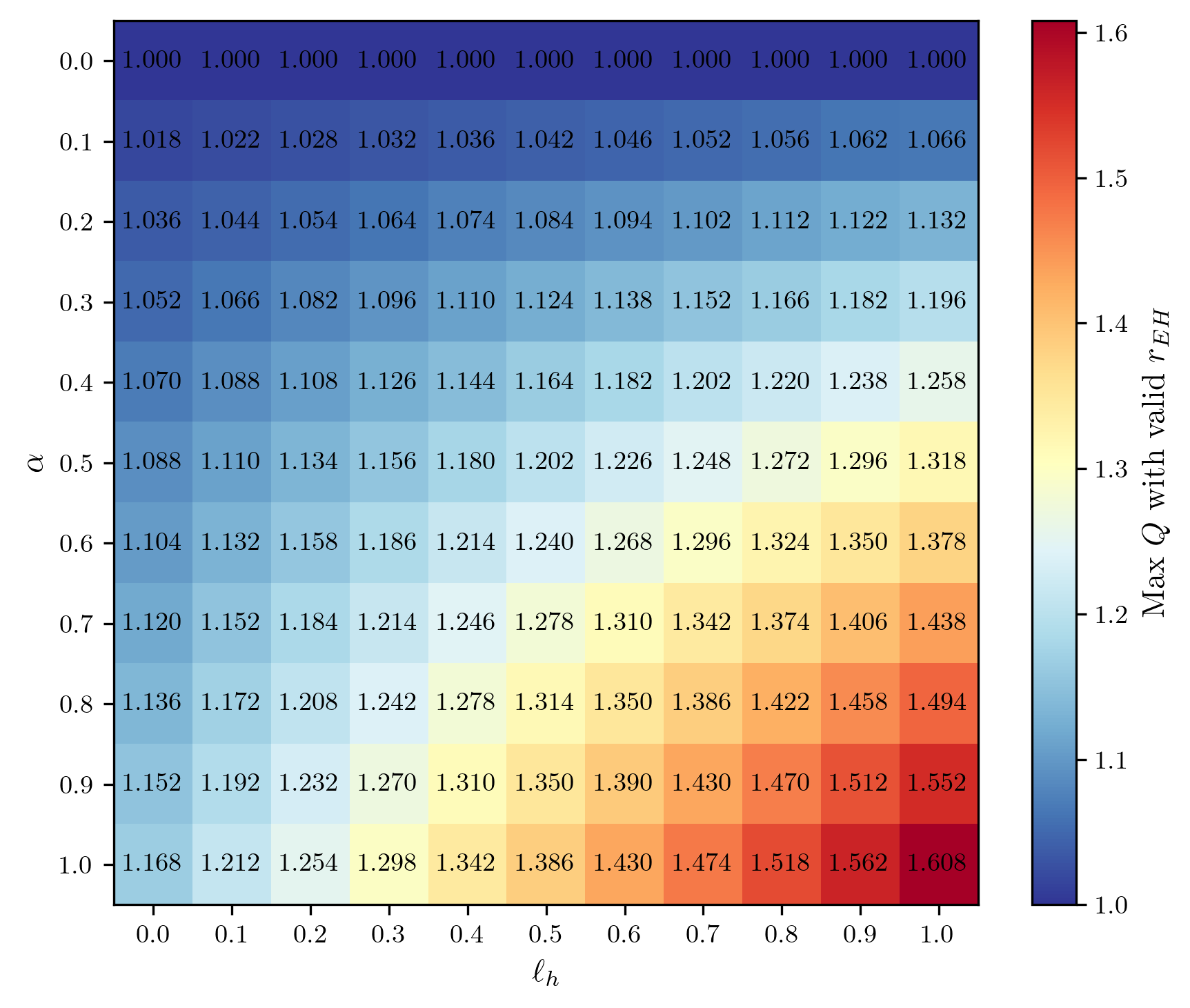}
    \caption[Largest $Q$ admitting a horizon over the $(\alpha,\ell_h)$ grid]{Maximum value $Q$ for which Eq.~\eqref{eq: horizon radii} admits at least one horizon, on a grid in $\alpha$ and $\ell_h$ over $[0,1]$, the value printed in each cell and repeated in the colour scale.}
    \label{fig:maxQwithrEH}
\end{figure}

To map this behaviour systematically, Fig.~\ref{fig:maxQwithrEH} plots the maximum $Q$ for each combination of $\alpha$ and $\ell_h$. Setting $\alpha = 0$ while relaxing Eq.~\eqref{eq: Q vanishes with alpha}, that is, keeping $Q$ finite as the coupling is switched off, recovers the Reissner--Nordström black hole, with $Q_\text{saturation} = 1$. Any combination with $\alpha \neq 0$ instead gives $Q_\text{saturation} > 1$, so that hairy black holes are generally expected to have broader horizon radii than their Reissner--Nordström counterparts.

\begin{figure}[htbp]
    \centering
    \includegraphics[width=0.8\linewidth]{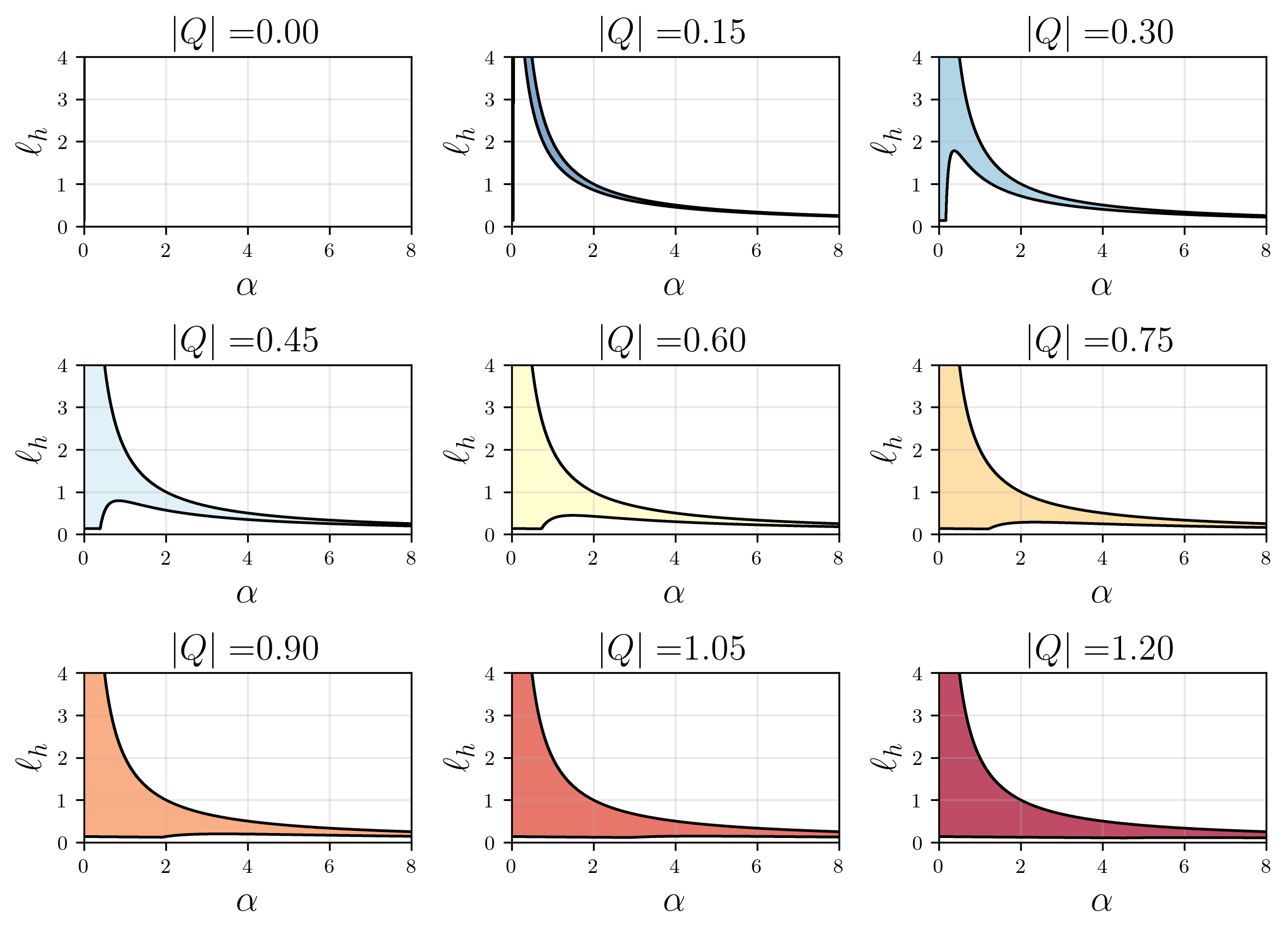}
    \caption[Map of admissible hair parameters, panel by panel in $|Q|$]{Parameter space satisfying the dominant energy condition, asymptotic flatness, and the existence of at least one real horizon. Each panel is a fixed $\abs{Q}$, from $0$ to $1.20$, and the shaded region is the allowed set of $(\alpha,\ell_h)$ over $\alpha \in [0,8]$ and $\ell_h \in [0,4]$.}
    \label{fig:mapadeparametros}
\end{figure}

To summarise all conditions, namely Eqs.~\eqref{eq: limit Q alpha M}, \eqref{eq: limit ell M}, \eqref{eq: limit M alpha ell}, $\alpha \geq 0$ and the results plotted in Fig.~\ref{fig:maxQwithrEH}, everything can be put together in the map of accepted parameters shown in Fig.~\ref{fig:mapadeparametros}.
The graph with $\qty|Q| = 0.00$ shows that the only accepted solution with $\tilde{\varrho}\geq 0$ is Schwarzschild, with $\alpha = \ell_h = Q = 0$.

\section{Summary and Outlook}\label{sec:GD_summary}

This chapter has developed gravitational decoupling as a tool for adding new physics to a known solution without solving the full non-linear problem again. The difficulty it addresses was set out in Sec.~\ref{sec:GRextension}, where the brane-world example showed that solving a modified action directly can leave a geometric deformation that survives the limit in which the new sector is switched off, so that the correction never fully decouples from the background. The method of Sec.~\ref{sec:GD} avoids this by deforming the metric functions rather than the source, Eqs.~\eqref{eq:deformations}, which splits the field equations into the ordinary Einstein equations for the seed and the quasi-Einstein system~\eqref{eq:quasi_einstein} for the new source alone, the latter carrying an overall factor of $\alpha$ and therefore vanishing with it by construction.

Applying the extended method to a Schwarzschild seed in Sec.~\ref{sec:HBHbyGD} produced the hairy black hole~\eqref{eq: metric HBH}. Three requirements fixed it: recovery of Schwarzschild as $\alpha\to0$, a single well-defined horizon, which forces $e^{\nu}=e^{-\lambda}$ and hence the equation of state $\tilde p_\text{rad}=-\tilde\varrho$, and the dominant energy condition on the effective source. The solution carries the dimensionless coupling $\alpha$, the primary hair $\ell_h$ of dimensions of mass, and the charge-like parameter $Q$ of dimensions of length, which despite the Reissner--Nordstr\"om form of the $Q^2/r^2$ term is not sourced by any gauge field and vanishes with $\alpha$ by Eq.~\eqref{eq: Q vanishes with alpha}.

The map of Fig.~\ref{fig:mapadeparametros} delimits the region of $(\alpha,\ell_h,Q)$ over which the hairy solution remains admissible, and it is over that region alone that the effective potentials of Chapter~\ref{chap:conexaoentrebnsebocais} and the quasinormal spectra of Chapter~\ref{chap:QNM_spectrum} are evaluated. The deformation acts over a macroscopic range set by $G_\text{N}M$, which is what makes these configurations comparatively accessible to black-hole spectroscopy, and it is in this respect that they contrast with the short-range quantum correction derived in the next chapter.

That accessibility has been exploited in a substantial body of phenomenological work, which places the present analysis in context. The quasinormal spectrum and the greybody factors of hairy black holes obtained by decoupling have been computed by several groups and by complementary methods~\cite{Yang:2022ifo,Cavalcanti:2022cga,Avalos:2023jeh,Li:2022hkq,Lei:2023wlt,dePaiva:2025eux,Navarro:2026lrf}, and the imprint of the hair on the spectroscopic programme as a whole, including the isospectrality of the axial and polar sectors, was examined in Ref.~\cite{Guimaraes:2025jsh}. Strong gravitational lensing and the deflection of light in the presence of plasma~\cite{Mustafa:2024hfq,Zhao:2024hep,Atamurotov:2023rye}, the optical appearance of accreting configurations and their photon rings~\cite{Rehman:2023eor,Gan:2021xdl,Gan:2021pwu}, geodesic motion, bound orbits, and energy extraction~\cite{Rakhimova:2024hzt,Meng:2024cnq,Li:2023htz,Ashraf:2025cnz}, and the near-horizon thermodynamics, phase structure, and thermal stability of the solutions~\cite{Cavalcanti:2022adb,Ditta:2023arv,Ditta:2024tdo,Abbas:2024pyi,Estrada:2020ptc} have all been analysed for this family. The energy conditions and the stability of the deformed geometries have likewise been scrutinised~\cite{Cavalcanti:2022xet,Meert:2020sqv,Meert:2021khi,DaRocha:2019fjr}. A separate line of work has shown that the deformation need not be postulated at all: the same functional form is generated by the higher-curvature operators of the effective field theory of gravity, so that gravitational decoupling and quantum gravity meet on the same geometry~\cite{Casadio:2024vfh}, which is precisely the pairing exploited in the remainder of this thesis.

In summary, this chapter applied the gravitational-decoupling method, through the extended minimal geometric deformation, to construct a hairy black hole by deforming a Schwarzschild vacuum seed with an additional gravitational source. The resulting metric~\eqref{eq: metric HBH} carries the hair parameters $(\alpha,\ell_h,Q)$, whose physically admissible domain was fixed by requiring the dominant energy condition, asymptotic flatness, and the existence of at least one horizon. The corresponding map of accepted parameters and the event/Cauchy horizon radii was obtained. The hair deforms the geometry at macroscopic scales while preserving the Schwarzschild-like relation $g_{tt} = -g_{rr}^{-1}$. The next chapter develops the complementary, first-principles mechanism, namely the quantum-gravitational corrections predicted by the effective field theory of gravity, which modify the Schwarzschild metric only at short range and, in contrast to the decoupling hair, break that relation.
\chapter{Effective Field Theory and Semiclassical Corrections}
\label{chap:EFT}

\epigraph{``Our mistake is not that we take our theories too seriously, but that we do not take them seriously enough.''}{Steven Weinberg}

In the preceding chapter, exotic black-hole geometries were constructed by supplementing the gravitational sector with an effective source through the GD programme. That construction is phenomenological in nature, since the additional source encodes possible macroscopic departures from vacuum GR without reference to a specific microscopic theory. The present chapter adopts a complementary approach, grounded in first principles, in which the corrections to the Schwarzschild metric are obtained from the effective field theory (EFT) of gravity. The two routes are not disjoint, since the higher-curvature operators derived below can themselves be recast as a decoupled source of the type used in Chapter~\ref{chap:GD}, which is the sense in which gravitational decoupling and quantum gravity describe the same deformation~\cite{Casadio:2024vfh}.

The presentation is organised as follows. Section~\ref{sec:EFT_GR_as_EFT} reviews the sense in which GR is non-renormalisable and the interpretation of this property within the EFT framework valid below the Planck scale~\cite{Donoghue:1994dn,Burgess:2003jk}. Section~\ref{sec:EFT_one_loop} computes the one-loop effective action explicitly, identifying the curvature-squared structure of the divergences and their cancellation on the vacuum equations of motion, together with the divergence that survives at two loops. Section~\ref{sec:EFT_corrections} then determines the corrections to the Schwarzschild geometry order by order, the metric being unchanged at second order in curvature, while at third order a single dimension-six operator yields the quantum-corrected Calmet--Kuipers metric~\cite{Calmet:2021lny}. The second-order corrections are not absent from the physics, only from the metric, since they survive in the Wald entropy and in the black-hole pressure derived from it, both of which are obtained in the same section. The third-order metric provides the theoretical input for the analogue-gravity programme of Parts~\ref{part: analogue models} and~\ref{part: results}. The same section examines the resulting horizon structure and the bound on $c_6$, and verifies that a cosmological constant leaves these conclusions unaffected. Section~\ref{sec:EFT_discussion} closes the chapter with a discussion of the results and their model-independence.

\section{General Relativity as a Low-Energy Effective Field Theory}
\label{sec:EFT_GR_as_EFT}

General Relativity admits two complementary readings. In the geometric formulation originally due to Einstein, the metric $g_{\mu\nu}$ is taken as the dynamical variable and general covariance is imposed as a founding principle. In the particle-physics formulation, one begins instead from the field content and the symmetries of flat-space quantum field theory and asks which interacting theory can describe a single massless spin-$2$ particle, the graviton. At low energies the answer is General Relativity~\cite{Donoghue:1994dn,Burgess:2003jk,Donoghue:2015hwa}, and it is this second reading that makes the EFT treatment of gravity natural.

Both quantum numbers of the graviton are fixed by observation. The spin is $2$ because gravity couples to the energy-momentum tensor $T_{\mu\nu}$, which is symmetric and of rank two, so its mediator must be a symmetric rank-two field. An even spin also makes like sources attract, whereas an odd spin, as in electromagnetism, makes them repel~\cite{Weinberg:1964ew,Weinberg:1965rz,Feynman:1996kb}. The mass vanishes because the interaction is long-ranged. A graviton mass $m$ would replace the Newtonian $1/r$ potential by a Yukawa fall-off $e^{-mr}/r$, and observation bounds it by $m\lesssim 10^{-22}\,\text{eV}$~\cite{LIGOScientific:2016lio}. The massless limit of a massive theory is, moreover, discontinuous, so massive gravity does not reduce smoothly to General Relativity~\cite{vanDam:1970vg,Zakharov:1970cc,deRham:2014zqa}.

Three requirements then fix the low-energy theory almost entirely. The first is that the theory propagate exactly two degrees of freedom, since the unitary irreducible representations of the Poincar\'e group are classified by mass and spin, and a massless particle of spin $s\geq1$ carries only two physical polarisations, the helicity states $\pm s$~\cite{Wigner:1939cj}. A manifestly Lorentz-covariant symmetric tensor $h_{\mu\nu}$ has ten components, so a gauge symmetry is needed to remove the surplus, namely the linearised diffeomorphisms $\delta h_{\mu\nu} = \partial_\mu\xi_\nu + \partial_\nu\xi_\mu$, in which $\xi_\mu(x)$ is an arbitrary infinitesimal vector field, the generator of the coordinate change $x^\mu \mapsto x^\mu - \xi^\mu$. Among the components to be removed are ghosts, that is, modes whose kinetic term enters the action with the wrong sign and which therefore carry negative energy, so that their excitation would destabilise the theory and spoil unitarity~\cite{Woodard:2015zca}. There is exactly one quadratic action for $h_{\mu\nu}$ that carries none of them, known as the Fierz--Pauli action~\cite{Fierz:1939ix}, and it coincides with the Einstein--Hilbert action expanded to quadratic order. That coincidence is what makes the third requirement below land on General Relativity rather than on some other non-linear theory. Second, requiring this field to couple consistently to a conserved source forces the coupling to be universal, so that all forms of energy gravitate alike, which is the statement of the equivalence principle~\cite{Weinberg:1964ew,Weinberg:1965rz}. Third, requiring in addition that the field couple to its own energy-momentum resums that ghost-free theory order by order in the coupling into a fully non-linear one, whose unique two-derivative completion is General Relativity~\cite{Gupta:1954zz,Feynman:1996kb,Deser:1969wk}. Lorentz invariance, the equivalence principle and the restriction to two propagating degrees of freedom therefore single out the Einstein--Hilbert action,
\begin{equation}
    \label{eq:EH_action}
    S_{\text{EH}} = \frac{1}{2\kappa^2}\int \dd^4x \sqrt{-g} R,
    \qquad \kappa^2 \equiv 8\pi G_\text{N}.
\end{equation}
This is Eq.~\eqref{eq:HE action} of Chapter~\ref{chap:GD} with the cosmological constant and the matter Lagrangian switched off. Both are set aside here for reasons of scope rather than of principle. Matter returns in Eq.~\eqref{eq:EFT_action_local} as $\mathcal{L}_{m}$, and the cosmological constant is restored at the end of Subsec.~\ref{subsec:EFT_horizon}, where it is shown to leave the corrected geometry unchanged.

The graviton is the quantum fluctuation of the metric about a fixed background, $g_{\mu\nu} = \bar g_{\mu\nu} + \kappa\, h_{\mu\nu}$, with $\kappa \equiv \sqrt{8\pi G_\text{N}} = M_{\text{Pl}}^{-1}$ the gravitational coupling, $M_{\text{Pl}} = \sqrt{\hbar c/8\pi G_\text{N}}$ denoting the reduced Planck mass, which is the convention used throughout this thesis. Unlike the dimensionless gauge couplings of the Standard Model, this one is dimensionful, $[\kappa] = -1$ in mass units, so the parameter controlling graviton self-interactions carries negative mass dimension, $[\kappa^{2}] = -2$. Because that dimension is fixed, and because every term in the action must be dimensionless overall, counting dimensions is by itself enough to say how many powers of energy have to accompany each power of the coupling. This is what allows power counting to organise the entire ultraviolet behaviour of the theory. The compensation has to occur at every order, and the number of independent divergent structures grows without bound~\cite{Weinberg:1980gg,Burgess:2003jk,Donoghue:2015hwa}.

The counting runs as follows. Adding one loop to a graviton amplitude costs a factor $\kappa^{2}E^{2} = \qty(E/M_{\text{Pl}})^{2}$, where $E$ is the characteristic energy of the process, so that a one-loop contribution is suppressed by that factor relative to the classical, no-loop result and each further loop by one more such factor. Since $[\kappa^{2}]=-2$, the two powers of energy must be supplied by two additional derivatives acting on the background, and two derivatives of the metric are precisely one power of the curvature. The divergent part of the $L$-loop effective action therefore takes the form
\begin{equation}
    \label{eq:tower}
    \Gamma_{\text{div}}^{(L)} \sim
    \frac{\kappa^{2(L-1)}}{\varepsilon}\int \dd^4x\,\sqrt{-g}\;
    \qty(\text{curvature})^{L+1},
\end{equation}
in which $L$ is the number of loops and $\qty(\text{curvature})^{L+1}$ stands for any scalar built from $L+1$ powers of the Riemann tensor and its contractions. The power of $\kappa$ is fixed by dimensions. The integral alone carries mass dimension $2L-2$, since $\dd^4x$ contributes $-4$ and each curvature $+2$, so a factor of dimension $2-2L$ is needed to make $\Gamma$ dimensionless, and $[\kappa]=-1$ gives $\kappa^{2(L-1)}$. The same counting follows from the loop expansion itself, the classical action carrying $\kappa^{-2}$ and each loop one factor of $\kappa^{2}$. The factor $1/\varepsilon$ is the pole in which the divergence resides once the calculation is performed in $d = 4-\varepsilon$ dimensions, the finite part being left implicit. One loop thus calls for curvature-squared counterterms, two loops for curvature-cubed, and so on without end. Since $L$ is unbounded, cancelling every divergence requires an infinite set of independent counterterms, each carrying a coefficient that experiment would have to supply, and General Relativity is non-renormalisable as a fundamental quantum field theory~\cite{Weinberg:1980gg,tHooft:1974toh,Deser:1974cz,Deser:1974xq,Goroff:1985th,vandeVen:1991gw}.
This does not, however, imply a divergence at every order. At one loop the would-be curvature-squared counterterm vanishes once the vacuum equations of motion are imposed, so pure gravity is finite on shell~\cite{tHooft:1974toh}, and the first divergence that genuinely cannot be reabsorbed into the Einstein--Hilbert action arises only at two loops. Both statements are established in Sec.~\ref{sec:EFT_one_loop}, where the two-loop divergence is displayed and its Riemann-cubed structure identified.

Non-renormalisability does not exclude predictive computations. For $E \ll M_{\text{Pl}}$, the ultraviolet completion is not resolved, and one works with the most general diffeomorphism-invariant action ordered in powers of the curvature~\cite{Donoghue:1994dn,Burgess:2003jk,Donoghue:2012zc,Donoghue:2015hwa}. Here a field is regarded as light when its mass is at or below the characteristic infrared scale being probed, so that it remains an active degree of freedom in the low-energy theory. Heavy fields, by contrast, can be integrated out and affect the effective theory mainly through local Wilson coefficients. The action then reads
\begin{multline}
    \label{eq:EFT_action_local}
    S_{\text{EFT}} = \int \dd^4x \sqrt{-g}\,
    \bigg[\, \frac{R}{16\pi G_\text{N}}
        + c_{1}(\mu)\, R^{2}
        + c_{2}(\mu)\, R_{\mu\nu}R^{\mu\nu} + \\
        + c_{3}(\mu)\, R_{\mu\nu\rho\sigma}R^{\mu\nu\rho\sigma}
        + \mathcal{L}_{m}
        + \mathcal{O}\qty(M_{\text{Pl}}^{-2}) \bigg],
\end{multline}
where $\mu$ is the renormalisation scale and the $c_{i}(\mu)$ are dimensionless Wilson coefficients, that is, numbers that the low-energy theory does not itself predict and which either an ultraviolet completion or a measurement has to supply. Operators beyond those displayed are suppressed by further powers of $\qty(E/M_{\text{Pl}})^{2}$, so the expansion is predictive for $E \ll M_{\text{Pl}}$. The universal low-energy quantum corrections are determined by the symmetries and the light field content, whereas the coefficients of local higher-curvature operators encode information about the ultraviolet completion~\cite{Donoghue:2015hwa,Bjerrum-Bohr:2002fji,DaRocha:2020oju,Junqueira:2016hlu}.

This is where the present route parts company with the gravitational decoupling of the preceding chapter. There the departure from Schwarzschild is macroscopic and is carried by an effective source added to the field equations, and it is governed by the three hair parameters. Here no matter is introduced, the departure is purely gravitational and loop-generated, and it is governed by a single Wilson coefficient. The two also differ in reach, since the hair acts at any mass whereas the quantum correction becomes appreciable only for masses approaching the Planck scale. Section~\ref{sec:EFT_corrections} makes the comparison quantitative.

\section{The One-Loop Effective Action}
\label{sec:EFT_one_loop}

The statements of the previous section can be made precise by evaluating the one-loop effective action. In the path-integral quantisation of the metric field, the generating functional is
\begin{equation}
    \label{eq:path_integral}
    Z = \int \mathcal{D}g_{\mu\nu}\; e^{\,iS[g]} \equiv e^{\,i\Gamma[\bar g]}.
\end{equation}
Here $\mathcal{D}g_{\mu\nu}$ is the functional measure, which instructs one to sum over every metric configuration, $S[g]$ is the classical action~\eqref{eq:EH_action} weighting each of them, and $\Gamma[\bar g]$ is the effective action, the quantity that plays for the quantum theory the role the classical action plays for the classical one. Its evaluation proceeds through the background-field method, in which the metric is decomposed as
\begin{equation}
    \label{eq:bfm_split}
    g_{\mu\nu} = \bar g_{\mu\nu} + \kappa\, h_{\mu\nu},
    \qquad \kappa^{2} = 8\pi G_\text{N},
\end{equation}
with $\bar g_{\mu\nu}$ a fixed classical background and $h_{\mu\nu}$ the quantum fluctuation. Since $g_{\mu\nu}$ is dimensionless and $[\kappa]=-1$, the fluctuation carries $[h_{\mu\nu}]=+1$, the mass dimension of a canonically normalised bosonic field in four dimensions. That is the reason for extracting $\kappa$ in Eq.~\eqref{eq:bfm_split} rather than expanding in $g_{\mu\nu}-\bar g_{\mu\nu}$ directly, since it is $h_{\mu\nu}$, and not the bare difference of metrics, that has a canonical kinetic term and therefore a standard propagator. Each quantity in the action is expanded to second order in $h_{\mu\nu}$. For the volume element,
\begin{equation}
    \label{eq:sqrtg_expansion}
    \sqrt{-g} = \sqrt{-\bar g}\qty[
        1 + \frac{\kappa}{2}\,h
        + \frac{\kappa^{2}}{8}\qty(h^{2} - 2\,h_{\mu\nu}h^{\mu\nu})
        + \mathcal{O}(h^{3})],
    \qquad h \equiv \bar g^{\mu\nu}h_{\mu\nu},
\end{equation}
and, for the inverse metric,
\begin{equation}
    \label{eq:ginv_expansion}
    g^{\mu\nu} = \bar g^{\mu\nu} - \kappa\, h^{\mu\nu}
                 + \kappa^{2}\, h^{\mu\alpha}h_{\alpha}{}^{\nu}
                 + \mathcal{O}(h^{3}),
\end{equation}
with a corresponding expansion of the Ricci scalar. Collecting terms by order in $h_{\mu\nu}$,
\begin{equation}
    \label{eq:action_expansion}
    S[\bar g + \kappa h] = S[\bar g] + S^{(1)}[h] + S^{(2)}[h] + \mathcal{O}(h^{3}).
\end{equation}
Substituting Eqs.~\eqref{eq:sqrtg_expansion} and~\eqref{eq:ginv_expansion} into the action~\eqref{eq:EH_action} and retaining the terms with a single power of $h_{\mu\nu}$ gives the linear term,
\begin{equation}
    \label{eq:linear_term}
    S^{(1)}[h] = \int \dd^4x \sqrt{-\bar g}
    \left.\frac{\delta S}{\delta g_{\mu\nu}}\right|_{\bar g}\,h_{\mu\nu},
\end{equation}
whose coefficient is nothing other than the left-hand side of the classical field equations. Varying Eq.~\eqref{eq:EH_action} gives $\delta S/\delta g_{\mu\nu} \propto G^{\mu\nu}$, so demanding that the background solve those equations in vacuum, $\bar R_{\mu\nu}=0$, sets that coefficient to zero and with it the whole linear term. This is the same statement as in ordinary perturbation theory, where expanding about a stationary point removes the term of first order. The leading quantum contribution is then the quadratic term, which may be written as the bilinear form
\begin{equation}
    \label{eq:quadratic_action}
    S^{(2)}[h] = \frac{1}{2}\int \dd^4x \sqrt{-\bar g}
    h^{\mu\nu}\,\mathcal{O}_{\mu\nu\rho\sigma}\,h^{\rho\sigma}.
\end{equation}
The operator $\mathcal{O}_{\mu\nu\rho\sigma}$ is not invertible before the diffeomorphism gauge freedom is fixed, since the action is invariant under $\delta h_{\mu\nu} = \nabla_{\mu}\xi_{\nu} + \nabla_{\nu}\xi_{\mu}$. Introducing a gauge-fixing term for the harmonic (de~Donder) condition,
\begin{equation}
    \label{eq:de_donder}
    \nabla^{\mu} h_{\mu\nu} - \frac{1}{2}\nabla_{\nu} h = 0,
\end{equation}
the kinetic operator separates into a part that carries all the derivatives and a part built algebraically from the background curvature,
\begin{equation}
    \label{eq:hessian}
    \mathcal{O}_{\mu\nu\rho\sigma}
    = -\frac{1}{2}\,\mathcal{C}_{\mu\nu\rho\sigma}\,\Box
      + \mathcal{R}_{\mu\nu\rho\sigma},
    \qquad
    \mathcal{C}_{\mu\nu\rho\sigma} \equiv \bar g_{\mu\rho}\bar g_{\nu\sigma}
      + \bar g_{\mu\sigma}\bar g_{\nu\rho}
      - \bar g_{\mu\nu}\bar g_{\rho\sigma},
\end{equation}
with $\Box \equiv \bar g^{\mu\nu}\nabla_{\mu}\nabla_{\nu}$ and $\mathcal{C}_{\mu\nu\rho\sigma}$ the DeWitt metric on symmetric rank-two tensors. The first part has the same form on every background, and all the curvature dependence sits in the matrix-valued potential $\mathcal{R}_{\mu\nu\rho\sigma}$, which on an arbitrary background reads
\begin{equation}
    \label{eq:R_general}
    \mathcal{R}_{\mu\nu\rho\sigma}
    = -2\,\bar R_{\mu\rho\nu\sigma}
      - 2\,\bar g_{\mu\nu}\bar R_{\rho\sigma}
      - 2\,\bar g_{\rho\sigma}\bar R_{\mu\nu}
      + 2\,\bar g_{\mu\rho}\bar R_{\nu\sigma}
      + 2\,\bar g_{\nu\sigma}\bar R_{\mu\rho}
      + \bar g_{\mu\nu}\bar g_{\rho\sigma}\bar R,
\end{equation}
understood to be symmetrised in $\mu\nu$ and in $\rho\sigma$, since it is contracted with $h^{\mu\nu}$ and $h^{\rho\sigma}$, which are symmetric~\cite{tHooft:1974toh,Christensen:1979iy,Barvinsky:1985an}.

It is worth recording where that curvature dependence comes from, since only part of it originates in the expansions written above. Those of $\sqrt{-g}$ and of $g^{\mu\nu}$ do contribute, through the background $\bar R$ they multiply, but the terms that carry the Riemann tensor arise elsewhere. Expanding $R_{\mu\nu}$ to second order in $h_{\mu\nu}$ produces pairs of covariant derivatives acting on the fluctuation in both orders, and reordering them costs a commutator,
\begin{equation}
    \label{eq:commutator}
    \qty[\nabla_{\alpha},\nabla_{\beta}]h_{\mu\nu}
    = -\bar R^{\rho}{}_{\mu\alpha\beta}h_{\rho\nu}
      - \bar R^{\rho}{}_{\nu\alpha\beta}h_{\mu\rho},
\end{equation}
which converts two derivatives into a curvature. This is what makes $\mathcal{R}_{\mu\nu\rho\sigma}$ contain the Riemann tensor and not merely its contractions.

On a background that solves the vacuum field equations, $\bar R_{\mu\nu}=0$ and every contraction of the Riemann tensor vanishes with it, so that Eq.~\eqref{eq:R_general} collapses to its first term,
\begin{equation}
    \label{eq:W_ricciflat}
    \left.\mathcal{R}_{\mu\nu\rho\sigma}\right|_{\bar R_{\mu\nu}=0}
    = -2\,\bar R_{\mu\rho\nu\sigma},
\end{equation}
and the operator acting on $h_{\mu\nu}$ becomes the Lichnerowicz operator of a Ricci-flat space, which is what the d'Alembertian turns into when it acts on a symmetric rank-two tensor in a curved space. The normalisation can be checked against the linearised theory, since $\mathcal{O}_{\mu\nu\rho\sigma}h^{\rho\sigma}=0$ then returns
\begin{equation}
    \label{eq:linearised_vacuum}
    \Box\, \tilde h_{\mu\nu} + 2\,\bar R_{\mu\rho\nu\sigma}\,\tilde h^{\rho\sigma} = 0,
    \qquad
    \tilde h_{\mu\nu} \equiv h_{\mu\nu} - \tfrac{1}{2}\bar g_{\mu\nu}h,
\end{equation}
which is the linearised vacuum equation in the de~Donder gauge.

The five Ricci terms of Eq.~\eqref{eq:R_general} are carried along rather than discarded, because the divergence of Subsec.~\ref{subsec:EFT_uv} is obtained with the background left arbitrary and only afterwards evaluated on a solution, which is what gives the cancellation found there any content. Leaving the background arbitrary also leaves the linear term~\eqref{eq:linear_term} in place, but it enters the Gaussian integral as a source and not through the operator, so the functional determinant computed below is unaffected. What matters for what follows is the shape of Eq.~\eqref{eq:hessian}, one piece multiplying $\Box$ and one piece with no derivative at all, which is what allows the two to be gathered into the single operator $\mathcal{O}_{\mu\nu\rho\sigma}$.

Fixing the harmonic gauge~\eqref{eq:de_donder} is accompanied by the Faddeev--Popov ghosts~\cite{Faddeev:1967fc}. These are not physical particles but anticommuting vector fields introduced alongside the gauge-fixing term to cancel the overcounting of metric configurations related by a diffeomorphism, and they are unrelated to the negative-energy ghosts of Sec.~\ref{sec:EFT_GR_as_EFT}. Their (vector) determinant contributes with the opposite power and is kept implicit below.

Collecting Eqs.~\eqref{eq:action_expansion} to~\eqref{eq:quadratic_action}, the expansion of the action about the background reads
\begin{equation}
    \label{eq:action_collected}
    S[\bar g + \kappa h]
    = S[\bar g]
      + \frac{1}{2}\int \dd^4x \sqrt{-\bar g}\;
        h^{\mu\nu}\,\mathcal{O}_{\mu\nu\rho\sigma}\,h^{\rho\sigma}
      + \mathcal{O}(h^{3}),
\end{equation}
the linear term having dropped out on shell. The remaining integral over $h_{\mu\nu}$ is Gaussian. In analogy with the finite-dimensional identity $\int \dd^{n}x\; e^{-\frac{1}{2}x^{\mathsf{T}}\!A x} = (2\pi)^{n/2}\qty(\det A)^{-1/2}$, its functional counterpart gives
\begin{equation}
    \label{eq:gaussian}
    Z = e^{\,iS[\bar g]}\int \mathcal{D}h\;
        e^{\,\frac{i}{2}\int h\,\mathcal{O}\,h}
      \;\propto\; e^{\,iS[\bar g]}\,\qty(\det \mathcal{O})^{-1/2}.
\end{equation}
Taking the logarithm of Eq.~\eqref{eq:path_integral} and using the identity $\log\det\mathcal{O} = \Tr\log\mathcal{O}$, the one-loop effective action is~\cite{Parker:2009uva,DeWitt:1967ub,Barvinsky:1985an}
\begin{equation}
    \label{eq:gamma_one_loop}
    \Gamma[\bar g] = S[\bar g] + \frac{i}{2}\log\det\mathcal{O}
                   = S[\bar g] + \frac{i}{2}\Tr\log\mathcal{O}.
\end{equation}
At this order, the quantum corrections are therefore contained in the functional trace of $\log\mathcal{O}$.

\subsection{Ultraviolet structure and the 't~Hooft--Veltman result}
\label{subsec:EFT_uv}

The trace runs over all graviton modes,
\begin{equation}
    \label{eq:trace_modes}
    \Tr\, X = \sum_{\text{pol.}}\int \frac{\dd^d p}{(2\pi)^{d}}\; X(p),
\end{equation}
so its divergences originate in the large-$p$ region. There the kinetic term dominates over the potential $\mathcal{R}_{\mu\nu\rho\sigma}$ of Eq.~\eqref{eq:R_general}, so that for momenta large compared with the background curvature the operator may be written with the kinetic part factored out,
\begin{equation}
    \label{eq:O_largep}
    \mathcal{O} \simeq -p^{2}\qty(1 - \frac{\mathcal{R}}{p^{2}}),
    \qquad p^{2} \gg \abs{\mathcal{R}},
\end{equation}
the indices being suppressed because every term carries the same pair. Expanding the logarithm,
\begin{equation}
    \label{eq:log_expansion}
    \log\mathcal{O}
    = \log(-p^{2}) - \frac{\mathcal{R}}{p^{2}}
      - \frac{1}{2}\frac{\mathcal{R}^{2}}{p^{4}}
      - \mathcal{O}\!\qty(\frac{\mathcal{R}^{3}}{p^{6}}),
\end{equation}
and substituting into Eq.~\eqref{eq:gamma_one_loop} yields
\begin{equation}
    \label{eq:gamma_three_terms}
    \Gamma^{(1)} = \frac{i}{2}\qty[
        \Tr\log(-p^{2})
        - \Tr\!\qty(\frac{\mathcal{R}}{p^{2}})
        - \frac{1}{2}\Tr\!\qty(\frac{\mathcal{R}^{2}}{p^{4}})
        + \cdots ],
\end{equation}
whose three terms have distinct interpretations. The first is independent of the background and renormalises only the vacuum energy (cosmological constant). The second is linear in the curvature, has the structure of the Einstein--Hilbert term, and renormalises $G_\text{N}$, its scaleless integral $\int \dd^d p\, p^{-2}$ vanishing in dimensional regularisation. The third is quadratic in the curvature and is the first term to generate a curvature-squared divergence.

Its momentum integral deserves to be followed slowly, because the divergence is easy to miss. Stripped of the curvature factors, the integral is $\int \dd^d p\; p^{-4}$, which is not the one-dimensional $\int \dd p\, p^{-4}$ and is therefore not convergent. Passing to spherical coordinates in momentum space, $\dd^d p = \Omega_{d-1}p^{d-1}\dd p$ with $\Omega_{d-1}$ the area of the unit $(d-1)$-sphere,
\begin{equation}
    \label{eq:loop_radial}
    \int \frac{\dd^d p}{p^{4}}
    = \Omega_{d-1}\int_{0}^{\infty} \dd p\; p^{\,d-5}
    \;\xrightarrow[\;d=4\;]{}\;
    \Omega_{3}\int^{\Lambda_{\text{UV}}} \frac{\dd p}{p}
    = \Omega_{3}\log\Lambda_{\text{UV}},
\end{equation}
so that in four dimensions the two extra powers of $p$ from the measure cancel two of the four in the denominator and what remains diverges logarithmically. Rather than keeping the cutoff, one continues the dimension to $d=4-\varepsilon$, where the same divergence appears as a pole. Restoring the curvature dependence through an effective mass, the integral to be evaluated is
\begin{equation}
    \label{eq:dimreg_integral}
    \int \frac{\dd^d p}{(2\pi)^{d}}\,\frac{1}{\qty(p^{2}+\Delta)^{2}}
    = \frac{1}{(4\pi)^{d/2}}\,\frac{\Gamma\!\qty(2-\tfrac{d}{2})}{\Gamma(2)}\,
      \Delta^{\,d/2-2},
\end{equation}
where $\Delta$ denotes the curvature-dependent effective mass. Setting $d = 4-\varepsilon$, the divergence is exhibited by the pole of the Gamma function~\cite{tHooft:1972tcz,Parker:2009uva},
\begin{equation}
    \label{eq:gamma_pole}
    \Gamma\!\qty(2-\tfrac{d}{2}) = \Gamma\!\qty(\tfrac{\varepsilon}{2})
    = \frac{2}{\varepsilon} - \gamma_{E} + \mathcal{O}(\varepsilon),
\end{equation}
in which $\gamma_{E}\simeq0.5772$ is the Euler--Mascheroni constant, part of the finite remainder the Gamma function leaves behind once its pole is subtracted. That pole is the dimensional-regularisation counterpart of the logarithmic cutoff dependence, $\log\Lambda_{\text{UV}} \leftrightarrow (4-d)^{-1} = \varepsilon^{-1}$, in which $\Lambda_{\text{UV}}$ is an ultraviolet momentum cutoff and is not to be confused with the cosmological constant $\Lambda$ discussed at the end of Subsec.~\ref{subsec:EFT_horizon}, the convention in which the poles of Eqs.~\eqref{eq:GS_divergence} and~\eqref{eq:tHV} are quoted throughout. Because $\mathcal{R}$ is built from the background curvature tensors, $\Tr(\mathcal{R}^{2})$ reduces to a linear combination of the three quadratic invariants $R_{\mu\nu\rho\sigma}R^{\mu\nu\rho\sigma}$, $R_{\mu\nu}R^{\mu\nu}$, and $R^{2}$, and in four dimensions the Gauss--Bonnet identity eliminates the Riemann-squared term up to a topological invariant~\cite{tHooft:1974toh,Parker:2009uva}. Fixing the coefficients through the heat-kernel (Seeley--DeWitt) expansion gives the 't~Hooft--Veltman result~\cite{tHooft:1974toh},
\begin{equation}
    \label{eq:tHV}
    \Gamma_{\text{div}}^{(1)}
    = \frac{1}{8\pi^{2}\,(4-d)}\int \dd^4x\,\sqrt{-\bar g}
      \qty(\frac{7}{20}\,\bar R_{\mu\nu}\bar R^{\mu\nu}
           + \frac{1}{120}\,\bar R^{2}).
\end{equation}
On the Schwarzschild background the vacuum equations impose $\bar R_{\mu\nu}=0$, and hence $\bar R = 0$, so both terms in Eq.~\eqref{eq:tHV} vanish. Pure Einstein gravity is thus one-loop finite on shell~\cite{tHooft:1974toh}, the phrase meaning that the statement holds only for backgrounds that solve the classical field equations, and not for an arbitrary metric. This on-shell one-loop finiteness should not be confused with perturbative renormalisability, since the two-loop divergence is a genuine independent counterterm.

The cancellation is specific to one loop. By Eq.~\eqref{eq:tower} the next rung of the tower is cubic in the curvature, and a cubic invariant can be built from the Riemann tensor alone, so it is not annihilated by $\bar R_{\mu\nu}=0$. Carrying the two-loop calculation through, Goroff and Sagnotti obtained~\cite{Goroff:1985th}
\begin{equation}
    \label{eq:GS_divergence}
    \Gamma_{\text{div}}^{(2)}
    = \frac{209}{2880\,(4\pi)^{4}}\,\frac{G_\text{N}}{4-d}
      \int \dd^4x\,\sqrt{-g}\;
      \tensor{R}{_{\mu\nu}^{\alpha\sigma}}
      \tensor{R}{_{\alpha\sigma}^{\beta\rho}}
      \tensor{R}{_{\beta\rho}^{\mu\nu}},
\end{equation}
proportional to the unique Riemann-cubed invariant and later confirmed by an independent computation~\cite{vandeVen:1991gw}. This operator is absent from the Einstein--Hilbert action~\eqref{eq:EH_action} and is not reabsorbed by a renormalisation of $G_\text{N}$, which is the precise sense in which vacuum General Relativity fails to be renormalisable. The same invariant returns in Sec.~\ref{sec:EFT_corrections} as the source of the leading correction to the Schwarzschild metric.

\section{Quantum Corrections to the Schwarzschild Geometry}
\label{sec:EFT_corrections}

The corrections to the metric are now determined order by order in the curvature expansion of Eq.~\eqref{eq:EFT_action_local}.

\subsection{Second order: Wald entropy and black-hole pressure}
\label{subsec:EFT_second_order}

Beyond the local action~\eqref{eq:EFT_action_local}, integrating out the massless fields at one loop generates a non-local sector, which cannot be expressed as a polynomial in the curvature~\cite{Barvinsky:1985an,Barvinsky:1990up,Barvinsky:1994hw,Codello:2012kq,Odintsov:1991nd}. Its origin is already visible in Subsec.~\ref{subsec:EFT_uv}. The quadratic term of Eq.~\eqref{eq:gamma_three_terms} was there evaluated only for its pole, through the integral~\eqref{eq:dimreg_integral}, but the same integral carries a finite part alongside the pole, and by Eq.~\eqref{eq:gamma_pole} that finite part is a logarithm of the momentum, $\log(p^{2}/\mu^{2})$. Subtracting the pole removes the divergence and leaves the logarithm behind. Returning to position space, where $p^{2}\to-\Box$, turns it into $\log(\Box/\mu^{2})$ acting between two curvatures, which is a non-local operator because no finite number of derivatives reproduces it. Collecting the three ways two curvatures can be contracted gives
\begin{equation}
    \label{eq:nonlocal_action}
    \Gamma_{NL}^{(2)} = \int \sqrt{-g}\,\dd^4x
    \qty[
        \tilde\alpha\, R \log\!\qty(\frac{\Box}{\mu^{2}}) R
        + \tilde\beta\, R_{\mu\nu} \log\!\qty(\frac{\Box}{\mu^{2}}) R^{\mu\nu}
        + \tilde\gamma\, R_{\mu\nu\rho\sigma} \log\!\qty(\frac{\Box}{\mu^{2}}) R^{\mu\nu\rho\sigma}
    ],
\end{equation}
with coefficients $\tilde\alpha,\tilde\beta,\tilde\gamma$ that are finite, calculable, and universal. They do not come out of the expansions used so far, which only ever produce polynomials in the curvature. What produces them is the same functional trace $\Tr\log\mathcal{O}$, evaluated this time by the heat-kernel expansion carried to second order in the curvature and with its finite part retained rather than discarded, which is the computation of Refs.~\cite{Barvinsky:1990up,Barvinsky:1994hw}. Their being universal follows from their depending only on which massless fields run in the loop, and not on the ultraviolet completion, so that each field species contributes an additive amount fixed by its spin. The values are listed in Table~\ref{tab:wilson_coefficients}.

The operator $\log(\Box/\mu^{2})$ is defined by its action in momentum space and is not a differential operator of finite order, so that evaluating it requires more than the local expansions used so far. The covariant perturbation theory of Barvinsky and Vilkovisky provides the general framework~\cite{Barvinsky:1985an,Barvinsky:1990up,Barvinsky:1994hw}, and its use in the gravitational effective field theory is reviewed in Ref.~\cite{Donoghue:2017pgk}. In outline, that framework does for a curved background what ordinary perturbation theory does for a flat one, except that the expansion parameter is the curvature rather than a coupling, and each order is kept in covariant form. The heat kernel of the fluctuation operator is written as a series in the curvature whose coefficients are themselves functions of $\Box$, and it is by keeping those functions whole, instead of expanding them in powers of derivatives, that the non-local logarithms survive. For the static, spherically symmetric functions of interest here the operator can still be evaluated explicitly. Going to the Fourier representation and using the spherical symmetry to place $\vec{x}=(0,0,r)$,
\begin{equation}
    \label{eq:log_box_radial}
    \log\!\qty(\frac{\Box}{\mu^{2}})f(r)
    = \frac{1}{\pi^{2}r}\int_{0}^{\infty}\dd k\; k
      \log\!\qty(\frac{k^{2}}{\mu^{2}})\sin(kr)\,\hat f(k),
    \qquad
    \hat f_{\nu}(k) = \int_{0}^{\infty} f(r)\,J_{\nu}(kr)\,r\,\dd r,
\end{equation}
the second expression being the Hankel transform of order $\nu$, with $J_{\nu}$ the Bessel function of the first kind, which is what plays in three dimensions the role the Fourier transform plays in one. What that returns is a principal-value integral over the radial coordinate together with a pole term at coincident points, and the calculation is carried out in full in Ref.~\cite{Perrucci:2024qrr}. None of it is needed in what follows, since the non-local sector does not act on a Schwarzschild background, as shown below.

\begin{table}[htbp]
\centering
\caption{Non-local Wilson coefficients $\tilde\alpha$, $\tilde\beta$, $\tilde\gamma$ for massless fields, computed via the Barvinsky--Vilkovisky heat-kernel expansion~\cite{Barvinsky:1990up}. All entries must be multiplied by $\qty(11520\pi^{2})^{-1}$, and $\xi$ denotes the non-minimal curvature coupling of the scalar, minimal for $\xi = 0$ and conformal for $\xi = 1/6$.}
\label{tab:wilson_coefficients}
\begin{tabular}{lccc}
\hline\hline
Field & $11520\pi^{2}\,\tilde\alpha$ & $11520\pi^{2}\,\tilde\beta$ & $11520\pi^{2}\,\tilde\gamma$ \\
\hline
Scalar ($s=0$)    & $5(6\xi-1)^{2}$ & $-2$   & $2$   \\
Fermion ($s=1/2$) & $-5$            & $8$    & $7$   \\
Vector ($s=1$)    & $-50$           & $176$  & $-26$ \\
Graviton ($s=2$)  & $250$           & $-244$ & $424$ \\
\hline\hline
\end{tabular}
\end{table}

Writing $\Gamma^{(2)}_{L}$ for the vacuum part of the local action~\eqref{eq:EFT_action_local} truncated at second order in curvature, and varying it together with the non-local sector, $\delta\qty(\Gamma^{(2)}_{L}+\Gamma_{NL}^{(2)})/\delta g^{\mu\nu}=0$, yields the modified field equations
\begin{equation}
    \label{eq:modified_Einstein}
    G_{\mu\nu} + 16\pi G_\text{N}\qty(H^{L}_{\mu\nu} + H^{NL}_{\mu\nu}) = 0,
\end{equation}
where $H^{L}_{\mu\nu}$ and $H^{NL}_{\mu\nu}$ are the local and non-local correction tensors. It is worth writing the first of them out, since the property that matters here can then be read off directly. Using the Gauss--Bonnet identity to remove the Riemann-squared operator from Eq.~\eqref{eq:EFT_action_local}, at the cost of shifting the remaining two coefficients to $\bar c_{1}=c_{1}-c_{3}$ and $\bar c_{2}=c_{2}+4c_{3}$, the local tensor is~\cite{Perrucci:2024qrr,Calmet:2019eof}
\begin{subequations}
\label{eq:HL_open}
\begin{align}
    H^{L}_{\mu\nu} &= \bar c_{1}\qty(
        2RR_{\mu\nu} - \tfrac{1}{2}g_{\mu\nu}R^{2}
        + 2g_{\mu\nu}\Box R - 2\nabla_{\mu}\nabla_{\nu}R)
    \nonumber\\[2pt]
    &\quad + \bar c_{2}\qty(
        2R^{\alpha}{}_{\mu}R_{\nu\alpha}
        - \tfrac{1}{2}g_{\mu\nu}R_{\alpha\beta}R^{\alpha\beta}
        + \Box R_{\mu\nu} + \tfrac{1}{2}g_{\mu\nu}\Box R
        - \nabla_{\alpha}\nabla_{\mu}R^{\alpha}{}_{\nu}
        - \nabla_{\alpha}\nabla_{\nu}R^{\alpha}{}_{\mu}).
\end{align}
\end{subequations}
Every term carries at least one factor of $R_{\mu\nu}$ or $R$, and no term carries the Riemann tensor on its own, which is what makes $H^{L}_{\mu\nu}$ vanish identically on a Ricci-flat background. The non-local tensor $H^{NL}_{\mu\nu}$ is of the same character in its $\tilde\alpha$ and $\tilde\beta$ parts, both of which act on $R$ and on $R_{\mu\nu}$. Its $\tilde\gamma$ part, however, acts on $R_{\alpha\beta}{}^{\sigma\tau}$, which does not vanish on Schwarzschild, so for that piece the cancellation is not a matter of inspection and rests instead on the result of Ref.~\cite{Calmet:2018elv}. On a vacuum background, where $R_{\mu\nu}=0$,
\begin{equation}
    \label{eq:H_vanishes}
    \left.H^{L}_{\mu\nu}\right|_{\text{Schw.}} = \left.H^{NL}_{\mu\nu}\right|_{\text{Schw.}} = 0,
\end{equation}
no fluctuation $h_{\mu\nu}$ is sourced, and the Schwarzschild metric receives no correction at second order in curvature~\cite{Calmet:2017qqa,Calmet:2018elv}. For a star or a charged black hole, by contrast, the matter source maintains $R_{\mu\nu}\neq0$, and second-order corrections do appear in the exterior metric~\cite{Calmet:2019eof}.

Although the metric is unchanged, the entropy is modified. For a diffeomorphism-invariant Lagrangian $\mathcal{L}$, it is given by the Wald functional~\cite{Wald:1993nt,Iyer:1994ys,Iyer:1995kg}
\begin{equation}
    \label{eq:Wald_entropy}
    S_{\text{Wald}} = -2\pi \int \dd\Sigma\;
    \epsilon_{\mu\nu}\,\epsilon_{\rho\sigma}\;
    \lim_{r\to r_{\text{EH}}}
    \pdv{\mathcal{L}}{R_{\mu\nu\rho\sigma}},
\end{equation}
with $\dd\Sigma = r^{2}\sin\theta\,\dd\theta\,\dd\phi$ the horizon areal element, $\epsilon_{\mu\nu}$ its binormal (normalised to $\epsilon_{\mu\nu}\epsilon^{\mu\nu}=-2$), and $r_{\text{EH}}=2G_\text{N} M$. The derivative $\partial\mathcal{L}/\partial R_{\mu\nu\rho\sigma}$ receives one contribution from each term of the action. For the Einstein--Hilbert term,
\begin{equation}
    \label{eq:dLdR_EH}
    \pdv{}{R_{\mu\nu\rho\sigma}}\qty(\frac{R}{16\pi G_\text{N}})
    = \frac{1}{32\pi G_\text{N}}\qty(g^{\mu\rho}g^{\nu\sigma} - g^{\mu\sigma}g^{\nu\rho}),
\end{equation}
so that, contracting the binormals,
\begin{equation}
    \label{eq:binormal_contraction}
    \epsilon_{\mu\nu}\epsilon_{\rho\sigma}
    \qty(g^{\mu\rho}g^{\nu\sigma} - g^{\mu\sigma}g^{\nu\rho})
    = 2\,\epsilon_{\mu\nu}\epsilon^{\mu\nu} = -4,
\end{equation}
Eq.~\eqref{eq:Wald_entropy} reproduces the Bekenstein--Hawking area law,
\begin{equation}
    \label{eq:bekenstein_hawking}
    S_{0} = \frac{A}{4G_\text{N}},
    \qquad A = 16\pi\qty(G_\text{N} M)^{2}.
\end{equation}
The curvature-squared and non-local terms contribute through derivatives such as
\begin{equation}
    \label{eq:dLdR_riem2}
    \pdv{}{R_{\mu\nu\rho\sigma}}
    \qty(R_{\alpha\beta\gamma\delta}R^{\alpha\beta\gamma\delta})
    = 2R^{\mu\nu\rho\sigma},
\end{equation}
evaluated on the Schwarzschild horizon. Carrying out the contraction and the horizon integral gives~\cite{Calmet:2021lny}
\begin{equation}
    \label{eq:Wald_entropy_2nd}
    S_{\text{Wald}}^{(2)} = \frac{A}{4G_\text{N}}
    + 64\pi^{2}
    \qty{c_{3}(\mu) + \tilde\gamma\qty[\log\qty(4G_\text{N}^{2}M^{2}\mu^{2}) - 2 + 2\gamma_{E}]},
\end{equation}
The entropy cannot depend on $\mu$, which is an arbitrary scale introduced by the renormalisation and not a property of the black hole. In Eq.~\eqref{eq:Wald_entropy_2nd} that scale enters in two places only, through the Wilson coefficient $c_{3}(\mu)$ and through the non-local logarithm, so imposing
\begin{equation}
    \label{eq:mu_independence}
    \mu\,\pdv{S_{\text{Wald}}^{(2)}}{\mu} = 0
\end{equation}
ties the running of the first to the coefficient $\tilde\gamma$ of the second. Since $\mu\,\partial_{\mu}\log\qty(4G_\text{N}^{2}M^{2}\mu^{2})=2$, the result is the renormalisation-group equation
\begin{equation}
    \label{eq:rg_c3}
    \mu\,\pdv{c_{3}}{\mu} = -2\,\tilde\gamma,
\end{equation}
so that the two dependences cancel against each other and $S_{\text{Wald}}^{(2)}$ is a physical quantity. Logarithmic terms of this kind are a general feature of the loop expansion of black-hole entropy, and their fate at higher loop order is examined in Ref.~\cite{Solodukhin:2019xwx}.

Since the metric, and therefore the Hawking temperature $T_{H}=(8\pi G_\text{N} M)^{-1}$, is unchanged while the entropy is not, the first law acquires an additional work term. Differentiating Eq.~\eqref{eq:Wald_entropy_2nd} with respect to $M$ at fixed $\mu$, and using $S_{0}=4\pi G_\text{N} M^{2}$, gives
\begin{equation}
    \label{eq:first_law_corrected}
    T_{H}\,\dd S_{\text{Wald}}
    = \qty(1 + \frac{16\pi\tilde\gamma}{G_\text{N} M^{2}})\dd M,
\end{equation}
the correction arising from the $\log\qty(4G_\text{N}^{2}M^{2}\mu^{2})$ term. Writing the first law with a work term,
\begin{equation}
    \label{eq:first_law_PdV}
    \dd M = T_{H}\,\dd S_{\text{Wald}} + P\,\dd V,
    \qquad\text{so that}\qquad
    P\,\dd V = -\frac{16\pi\tilde\gamma}{G_\text{N} M^{2}}\,\dd M,
\end{equation}
and taking for the thermodynamic volume the one enclosed by the horizon,
\begin{equation}
    \label{eq:thermo_volume}
    V = \frac{4\pi}{3}\,r_\text{EH}^{3} = \frac{32\pi}{3}\,G_\text{N}^{3}M^{3},
    \qquad\text{whence}\qquad
    \dv{V}{M} = 32\pi G_\text{N}^{3}M^{2},
\end{equation}
determines the black-hole pressure
\begin{equation}
    \label{eq:BH_pressure}
    P = -\frac{16\pi\tilde\gamma}{G_\text{N} M^{2}}\,\qty(\dv{V}{M})^{-1}
      = -\frac{16\pi\tilde\gamma}{G_\text{N} M^{2}}\,\frac{1}{32\pi G_\text{N}^{3}M^{2}}
      = -\frac{\tilde\gamma}{2\,G_\text{N}^{4}M^{4}}.
\end{equation}
This is a genuinely quantum feature. A classical Schwarzschild black hole has no pressure at all, its first law being simply $\dd M = T_{H}\,\dd S_{0}$ with no work term to speak of, and the term appears here only because the entropy acquires a correction while the metric, and with it the temperature, does not.

Its sign is set by the field content, with $\tilde\gamma>0$ for spin-$0$, spin-$1/2$, and spin-$2$ fields and $\tilde\gamma<0$ for spin-$1$ fields, cf. Table~\ref{tab:wilson_coefficients}. The two cases differ in how the black hole responds to a change of size. A negative $P$ means that work must be done on the black hole to expand it, so the quantum correction resists growth and holds the horizon together, in the manner of a surface tension. A positive $P$ means the opposite, that the correction pushes the horizon outwards and the black hole would do work in expanding. Which of the two occurs is therefore decided by the spins of the light fields, and not by anything intrinsic to the geometry.

\subsection{Third order: the Calmet--Kuipers metric}
\label{subsec:EFT_third_order}

As the metric is unchanged at second order, the leading correction arises at third order in curvature, corresponding to dimension-six operators suppressed by $M_{\text{Pl}}^{-2}$. In four-dimensional vacuum GR, there is a single independent Riemann-cubed invariant, since any other cubic contraction reduces to it through the Bianchi identity, the on-shell condition $R_{\mu\nu}=0$, and topological identities. The associated Lagrangian is
\begin{equation}
    \label{eq:L3_operator}
    \mathcal{L}^{(3)} = c_{6}\, G_\text{N}\,
    \tensor{R}{_{\mu\nu}^{\alpha\sigma}}
    \tensor{R}{_{\alpha\sigma}^{\beta\rho}}
    \tensor{R}{_{\beta\rho}^{\mu\nu}},
\end{equation}
with $c_{6}$ dimensionless. The bare coupling is split into a pole and a finite, scale-dependent part,
\begin{equation}
    \label{eq:c6_renorm}
    c_{6}(\mu) = c_{6}^{\text{bare}}
               + \frac{209}{2880\,(4\pi)^{4}}\,\frac{1}{\varepsilon} 
               + c_6^\text{finite}\qty(\mu),
\end{equation}
in which the pole absorbs the Goroff--Sagnotti divergence~\eqref{eq:GS_divergence}, $c_6^{\text{finite}}(\mu)$ carries the scheme dependence, and $c_{6}(\mu)$ is the renormalised coupling. Since that divergence is not contained in the Einstein--Hilbert action, $c_{6}$ is a free parameter of the EFT and its determination requires ultraviolet input~\cite{Calmet:2023gbw,Calmet:2024pve}. Following Refs.~\cite{Calmet:2021lny,Casadio:2025lkn}, the accompanying non-local cubic operator is not retained below~\cite{Goroff:1985th}.

A cubic operator raises the field equations above second order, and solving them exactly would bring in extra solutions that do not survive the limit $c_6\to0$. Solving them perturbatively in $c_6$ keeps those solutions out of the physical spectrum and is the standard treatment of such truncations~\cite{Simon:1990ic,Sarkar:2017puh,Nastase:2016sji,Delhom:2022xfo}. Theories that instead retain them as genuine degrees of freedom lie outside the effective description adopted here~\cite{Holdom:2022zzo,Asorey:2024mkb,Rachwal:2021bgb}, as do the hairy solutions of Einstein--Weyl gravity, whose extra structure comes from the quadratic sector~\cite{Sultana:2020pcc}.

The corrected geometry follows from the field equations of $S_{\text{EH}}+\int\sqrt{-g}\,\mathcal{L}^{(3)}$, solved perturbatively in $c_6$. Taking the static, spherically symmetric ansatz~\eqref{eq:spherical ansatz}, with $B \equiv e^{\nu}$ and $A \equiv e^{\lambda}$, one sets $B = B_0 + c_6\,\delta B$ and $A^{-1} = B_0 + c_6\,\delta(A^{-1})$, with $B_0 = 1-2G_\text{N} M/r$ the Schwarzschild function. Denoting by
\begin{equation}
    \label{eq:T3_source}
    \tensor{T}{^{(3)}_{\mu\nu}}
    = \frac{1}{\sqrt{-g}}\,\frac{\delta}{\delta g^{\mu\nu}}
      \int\sqrt{-g}\,\mathcal{L}^{(3)}\,\dd^4x
    \;\Bigg|_{B_0}
\end{equation}
the variation of the cubic operator evaluated on the Schwarzschild background, the field equations reduce to $G_{\mu\nu}\propto\tensor{T}{^{(3)}_{\mu\nu}}$ at first order in $c_6$. Although it sits where an energy-momentum tensor would, $\tensor{T}{^{(3)}_{\mu\nu}}$ is not one. No matter has been introduced anywhere in this chapter, and what the variation returns is a purely gravitational object, the higher-curvature terms rearranged so as to act as an effective source for the Einstein tensor. This is the same rearrangement that lets the quantum correction be read as a decoupled source in the sense of Chapter~\ref{chap:GD}. Their $(tt)$ and $(rr)$ projections are two decoupled linear ordinary differential equations for $\delta B(r)$ and $\delta(A^{-1})(r)$, sourced by the corresponding components of $\tensor{T}{^{(3)}_{\mu\nu}}$. Integrating these equations and fixing the integration constants by asymptotic flatness, $\delta B,\ \delta(A^{-1}) \to 0$ as $r\to\infty$, gives the metric functions~\cite{Calmet:2021lny,Casadio:2025lkn}
\begin{subequations}
\label{eq:CK_metric}
\begin{align}
    \label{eq:B_CK}
    B(r) &= 1 - \frac{2G_\text{N} M}{r}
             + 640\pi\, c_{6}\,\frac{G_\text{N}^{5}M^{3}}{r^{7}}, \\[4pt]
    \label{eq:A_CK}
    A(r) &= \qty[
             1 - \frac{2G_\text{N} M}{r}
             + 3456\pi\, c_{6}\,\frac{G_\text{N}^{4}M^{2}}{r^{6}}
               \qty(1 - \frac{49\,G_\text{N} M}{27\,r})
             ]^{-1},
\end{align}
\end{subequations}
which constitute the Calmet--Kuipers metric. Three properties may be noted. First, the metric violates the relation $A\,B=1$, reflecting the distinct $(tt)$ and $(rr)$ projections of the source~\eqref{eq:T3_source}. Second, the correction is short-ranged, the term in $A^{-1}$ decaying as $r^{-6}$ and the one in $B$ as $r^{-7}$, against the classical $r^{-1}$, so that it is concentrated near the horizon. Third, the classical limit is recovered, since $c_6\to0$ restores $B=A^{-1}=1-2G_\text{N} M/r$.

The curvature invariants of Eqs.~\eqref{eq:CK_metric} are computed in Appendix~\ref{chap:curvature invariants}. To first order in the coupling,
\begin{equation}
\label{eq: invariantes ck}
    R = \frac{768\pi\, c_6\, G_\text{N}^4M^2\qty(45r - 106G_\text{N}M)}{r^9},
    \qquad
    K = \frac{48G_\text{N}^2M^2}{r^6} - \frac{18432\pi\, c_6\, G_\text{N}^5M^3\qty(12r - 25G_\text{N}M)}{r^{12}},
\end{equation}
while $R_{\mu\nu}R^{\mu\nu}$ is of order $c_6^2$, the Ricci tensor itself being of order $c_6$ on a Ricci-flat background. The correction to the Kretschmann scalar falls as $r^{-12}$ against the $r^{-6}$ of the Schwarzschild term, which is the covariant statement of the short range noted above, and it changes sign at $r = 25G_\text{N}M/12$, just outside the uncorrected horizon.

\subsection{Horizon structure and the bound on $c_6$}
\label{subsec:EFT_horizon}

The event horizon is the largest root of $A(r)^{-1}=0$. It is obtained perturbatively by writing $r_\text{EH} = 2G_\text{N} M + \delta$ with $\delta = \mathcal{O}(c_6)$ and expanding Eq.~\eqref{eq:A_CK}. To first order, the Schwarzschild part gives $1 - 2G_\text{N} M/r_\text{EH} \simeq \delta/(2G_\text{N} M)$, while the $c_6$ term, evaluated at $r=2G_\text{N} M$, is
\begin{equation}
    \label{eq:c6_term_at_horizon}
    3456\pi c_6\,\frac{G_\text{N}^{4}M^{2}}{(2G_\text{N} M)^{6}}\qty(1-\frac{49}{54})
    = 3456\pi c_6\,\frac{1}{64\,G_\text{N}^{2}M^{4}}\cdot\frac{5}{54}
    = \frac{5\pi c_6}{G_\text{N}^{2}M^{4}}.
\end{equation}
Requiring the sum to vanish, $\delta/(2G_\text{N} M) + 5\pi c_6/(G_\text{N}^2 M^4) = 0$, gives $\delta = -10\pi c_6/(G_\text{N} M^{3})$, and hence~\cite{Calmet:2021lny,Casadio:2025lkn}
\begin{equation}
    \label{eq:horizon_shift}
    r_{\text{EH}} = 2G_\text{N} M\qty(1 - c_{6}\,\frac{5\pi}{G_\text{N}^{2}M^{4}}).
\end{equation}

Defining the horizon through $A(r)^{-1}=0$ needs a word, because $A\neq B^{-1}$ leaves two distinguished surfaces rather than one, the other being $B(r)=0$, where the redshift becomes infinite. Expanding $B$ about $r=2G_\text{N}M$ as in Eq.~\eqref{eq:c6_term_at_horizon} returns $5\pi c_6/(G_\text{N}^2M^4)$, the same value found there for $A^{-1}$, so the two are displaced by the same fraction~\eqref{eq:horizon_shift} at linear order and separate only at $\order{c_6^2}$, where the metric is not determined anyway.

The two surfaces are not on the same footing, however. Measuring radii in units of the Schwarzschild radius, $x \equiv r/2G_\text{N}M$, and collecting the correction into the single fractional parameter
\begin{equation}
    \label{eq:kappa_def}
    \kappa_{6} \equiv \frac{5\pi c_{6}}{G_\text{N}^{2}M^{4}},
\end{equation}
which is not to be confused with the gravitational coupling $\kappa$ of Eq.~\eqref{eq:EH_action} or with the surface gravity $\kappa_s$ below, the condition $B=0$ reads $x^{7}-x^{6}+\kappa_{6}=0$. The function $x^{7}-x^{6}$ attains its minimum at $x=6/7$, where it equals $-6^{6}/7^{7}$, so a real root exists only for
\begin{equation}
    \label{eq:B_surface_exists}
    \kappa_{6} \leq \frac{6^{6}}{7^{7}} \approx 0.0567
    \qquad\Longleftrightarrow\qquad
    c_{6} \leq \frac{6^{6}}{5\pi\,7^{7}}\,G_\text{N}^{2}M^{4} \approx 3.61\times10^{-3}\,G_\text{N}^{2}M^{4}.
\end{equation}
Above this value $B>0$ everywhere and the surface of infinite redshift disappears, leaving only $A^{-1}=0$, which persists for every $c_6>0$ and tends to $r=49G_\text{N}M/27$. A surface where the radial metric blows up without the redshift diverging is a wormhole throat and not an event horizon, which is the sense in which the truncated metric ceases to describe a black hole, and is why the horizon is defined here through $A^{-1}$, the only one of the two that always exists. On the negative branch both surfaces exist for all $c_6<0$ and the question does not arise.

Requiring $r_\text{EH}>0$ bounds $c_6$ from above, while perturbativity at the horizon, $\abs{\delta r_\text{EH}}/r_\text{EH} = 5\pi\abs{c_6}/(G_\text{N}^{2}M^{4}) < 1$, bounds it on both sides,
\begin{equation}
    \label{eq:c6_bound}
    \abs{c_{6}} < \frac{G_\text{N}^{2}M^{4}}{5\pi}
    \qquad\xrightarrow{\;G_\text{N} M^{2}\sim1\;}\qquad
    \abs{c_{6}} < \frac{1}{5\pi}\approx 0.0636
\end{equation}
for a black hole whose mass is of the order of the Planck mass. For heavier holes the bound is weaker, although the fractional shift $\delta r_\text{EH}/r_\text{EH} \sim c_6\,\qty(G_\text{N}M^{2})^{-2}$ is correspondingly small.

The temperature follows from the surface gravity. Near the horizon, $B(r)\simeq B'(r_\text{EH})(r-r_\text{EH})$ and $A(r)^{-1}\simeq \qty(A^{-1})'(r_\text{EH})(r-r_\text{EH})$, so that
\begin{equation}
    \label{eq:surface_gravity}
    \kappa_s = \frac{B'(r_\text{EH})}{2\sqrt{A(r_\text{EH})B(r_\text{EH})}}
             = \frac{1}{2}\sqrt{B'(r_\text{EH})\,\qty(A^{-1})'(r_\text{EH})},
\end{equation}
and $T_\text{H} = \kappa_s/(2\pi)$. Expanding to first order in $c_6$,
\begin{equation}
    \label{eq:TH_EFT}
    T_{H}^{\text{EFT}} = \frac{1}{8\pi G_\text{N} M}
    \qty(1 + c_{6}\,\frac{2\pi}{G_\text{N}^{2}M^{4}}) + \mathcal{O}(c_{6}^{2}),
\end{equation}
so that $c_6>0$ raises and $c_6<0$ lowers the evaporation temperature, consistently with the horizon shift~\eqref{eq:horizon_shift}. The coefficient does not simply follow the horizon shift, since $T_\text{H}$ cannot be read off as $1/(4\pi r_\text{EH})$ once $A\,B\neq1$, and the two derivatives in Eq.~\eqref{eq:surface_gravity} have to be taken separately.
The range of validity of the truncation should be kept in mind when reading these estimates. On a Schwarzschild horizon the Kretschmann invariant is $K = 3/\qty(4G_\text{N}^{4}M^{4})$, which is of order $G_\text{N}^{-2}$, the Planck curvature scale, once $G_\text{N}M^{2}\sim1$, so the regime in which the bound~\eqref{eq:c6_bound} becomes restrictive is also the one in which operators beyond cubic order cease to be suppressed. In the analogue programme of Parts~\ref{part: analogue models} and~\ref{part: results} this is not an obstacle, since $c_6$ is treated as a control parameter of the effective potential to be scanned in the laboratory.

Finally, reinstating the cosmological constant that Eq.~\eqref{eq:EH_action} switched off changes none of this. The term it adds to $A(r)^{-1}$ is $\Lambda r^{2}/3$, which for a black hole of one kilogram, with $r_\text{EH}\sim10^{-27}\,\text{m}$ and $\Lambda\approx10^{-52}\,\text{m}^{-2}$, is of order $10^{-106}$ against a Schwarzschild term of order unity. Ref.~\cite{Casadio:2025lkn} confirms this numerically, scanning $c_{6}\in[10^{-10},10^{10}]$ against $\Lambda\in[10^{-52},10^{-22}]$ and finding the horizon radius unchanged, so $\Lambda$ is set to zero throughout.

\section{Summary and Outlook}
\label{sec:EFT_discussion}
Treating gravity as an effective field theory, the one-loop divergences~\eqref{eq:tHV} cancel on the vacuum equations of motion, and the first that does not is the two-loop Goroff--Sagnotti term~\eqref{eq:GS_divergence}, which identifies the dimension-six operator used here. At second order in curvature the Schwarzschild metric is untouched and the corrections survive only in the Wald entropy~\eqref{eq:Wald_entropy_2nd} and the pressure~\eqref{eq:BH_pressure} that follows from it. At third order the Riemann-cubed operator deforms the metric into the Calmet--Kuipers geometry~\eqref{eq:CK_metric}, governed by the single dimensionless $c_{6}$, with $\abs{c_6}<1/(5\pi)$ for a hole of about the Planck mass and a horizon that shrinks for $c_6>0$ and grows for $c_6<0$.

What makes the result worth carrying forward is that it is model-independent, following from the low-energy symmetries and field content alone and not from the ultraviolet completion~\cite{Bjerrum-Bohr:2002fji,Donoghue:2015hwa,Donoghue:2017pgk,Donoghue:2018izj}, so that a measurement of $c_{6}$ would probe the Wilson coefficient associated with the leading two-loop counterterm of pure quantum gravity. Bounds have been sought through the shadow, the geodesic structure, and cosmological and astrophysical imprints~\cite{Gong:2023ghh,Wang:2025fmz,Marciu:2023hdb,Bukhari:2025fhd,DeFelice:2023vmj,Lambiase:2024vkz}. Together with the hairy black holes of Chapter~\ref{chap:GD}, this geometry supplies the two backgrounds studied in the rest of the thesis. Three extensions are left to the concluding chapter of Part~\ref{part: results}. Rotating and charged quantum-corrected geometries have already been derived~\cite{Casadio:2023iqt,Casadio:2022ndh}, and the frequency shifts they carry can be larger than in the static case treated here~\cite{Cano:2021myl,Cano:2023jbk,Husken:2026axq,Chen:2025wfi}. For charged black holes the corrections appear already at second order in curvature~\cite{Calmet:2019eof,Perrucci:2024qrr}. The regularity of the resulting interiors is a separate question, which the exterior construction of this thesis does not address~\cite{Casadio:2024lgw,Casadio:2025pun,Ovalle:2023vvu,Ovalle:2024wtv}.

\part{Gravitational Perturbations and Quasinormal Modes} \label{part: quasinormal modes}


\chapter{Propagation of Black Holes through Quasinormal Modes}
\label{chap:conexaoentrebnsebocais}

\epigraph{``The black holes of nature are the most perfect macroscopic objects there are in the universe: the only elements in their construction are our concepts of space and time.''}{Subrahmanyan Chandrasekhar}

Any astrophysical black hole is subject to perturbations, whether from accreting matter, from a companion in a binary system, or from the residual fluctuations that follow a merger. The relaxation of the geometry back to equilibrium is dominated, at intermediate times, by a discrete set of damped oscillations known as quasinormal modes, first identified in the scattering of gravitational radiation off a Schwarzschild background~\cite{Vishveshwara:1970zz,Chandrasekhar:1975zza}. They solve the linearised field equations on a fixed background, subject to two boundary conditions: at the horizon, radiation may only travel inwards, and at spatial infinity only outwards. Their frequencies depend only on the parameters of the background geometry and are independent of the initial data that excited the perturbation, which makes the quasinormal spectrum a characteristic signature of the unperturbed spacetime~\cite{Kokkotas:1999bd,Berti:2009kk,Konoplya:2011qq}.
This property carries direct observational significance since the ringdown stage of a compact binary coalescence is described by the quasinormal spectrum of the remnant, and the identification of two or more modes provides a consistency test of the Kerr hypothesis within GR~\cite{Dreyer:2003bv,Berti:2005ys,LIGOScientific:2016aoc}. Departures from the Schwarzschild and Kerr spectra therefore constitute a quantitative diagnostic of modifications of the vacuum solutions of GR, which is precisely the situation of the geometries constructed in Part~\ref{part: extensions of gr}.

The present chapter derives the equation governing such perturbations, restricting the analysis to the gravitational setting. Sec.~\ref{sec:BH_QNM_propagation} shows that, for a static, spherically symmetric background, linear perturbations of arbitrary spin reduce to a single Schr\"odinger-type wave equation, written in terms of a tortoise coordinate, in which the whole influence of the geometry is carried by an effective potential. Sec.~\ref{sec:veff hairy and qc} then specialises that potential to the two exotic geometries of Part~\ref{part: extensions of gr}, the hairy black hole from gravitational decoupling and the quantum-corrected Schwarzschild metric, in preparation for the quasinormal-mode spectra computed in Chapter~\ref{chap:QNM_spectrum}. Finally, Sec.~\ref{sec:veff comparison} closes the chapter by contrasting the two deformations and their imprint on the barrier.

\section{Black Hole Propagation via Quasinormal Modes}
\label{sec:BH_QNM_propagation}

In the study of four-dimensional stationary black holes, whose vacuum solutions are fixed by the uniqueness theorems~\cite{Israel:1967wq,Carter:1971zc,Robinson:1975bv}, the solutions to the Einstein field equations generally take the form
\begin{align}
    \label{sol gerais da eq einstein}
    \dd s^2 = g_{\mu \nu}\qty(r) \dd x^\mu \dd x^\nu,  &&
    \phi_{(i)}=\phi_{(i)}\qty(r), &&
    \frac{\delta \mathcal{L}_M}{\delta \phi_{(i)}} = 0,
\end{align}
where the metric components in Eqs.~\eqref{sol gerais da eq einstein} are independent of time in the appropriate coordinate system and gauge. Since astrophysical black holes are inevitably perturbed, linear perturbations $\delta g_{\mu \nu} \ll \bar g_{\mu \nu}$ are considered, such that
\begin{align}
       g_{\mu \nu} = \bar g_{\mu \nu} + \delta g_{\mu \nu} + \order{\delta g_{\mu \nu}^2},  &&
       \phi_{(i)} = \bar \phi_{(i)} + \delta \phi_{(i)} + \order{\delta \phi_{(i)}^2},
\end{align}
where $\bar g_{\mu \nu}$ is the background metric of the unperturbed black hole. The dynamical response of a perturbed black hole proceeds through the ringing of its quasinormal modes, which encode the black-hole geometry and are central to the analysis of its stability and dynamics~\cite{Regge:1957td,Vishveshwara:1970zz,Chandrasekhar:1975zza}.

The first case treated is that of a \emph{test field} propagating on the fixed background, the simplest of which is a real, massless scalar $\Psi\qty(x^\mu)$. Two statements are to be kept apart at this point. The linearised Einstein equations, $\delta G_{\mu\nu}=0$, govern the perturbation of the metric itself, and are the subject of the spin-$2$ analysis at the end of this section. A scalar field is instead an external probe, so its dynamics does not follow from the vacuum equations, but from an action of its own. What allows the background to be held fixed while the field evolves on it is the test-field approximation. The energy-momentum tensor of the scalar,
\begin{equation}
    \label{eq:scalar Tmunu}
    T_{\mu\nu} = \partial_\mu \Psi\, \partial_\nu \Psi
    - \frac{1}{2}\, \bar g_{\mu\nu}\, \bar g^{\alpha\beta}\, \partial_\alpha \Psi\, \partial_\beta \Psi,
\end{equation}
is quadratic in $\Psi$, so a perturbation of amplitude comparable to $\delta g_{\mu\nu}$ sources a metric response of order $\delta g_{\mu\nu}^2$, which is the order already discarded in the linearisation. The background therefore stays at $\bar g_{\mu\nu}$ throughout, and is written simply as $g_{\mu\nu}$ from here on.

The dynamics follows from the action of a real, massless scalar field minimally coupled to gravity~\cite{Parker:2009uva},
\begin{equation}
    \label{eq:scalar action}
    S\qty[\Psi] = -\frac{1}{2}\int \dd^4 x\, \sqrt{-g}\;
    g^{\mu\nu}\, \partial_\mu \Psi\, \partial_\nu \Psi,
\end{equation}
in the signature $\qty(-,+,+,+)$ used throughout, for which this sign gives a positive kinetic energy. Varying with respect to $\Psi$, with $\delta \Psi$ vanishing on the boundary of the region of integration, and integrating by parts to move the derivative off $\delta \Psi$,
\begin{equation}
    \label{eq:scalar variation}
    \delta S = -\int \dd^4 x\, \sqrt{-g}\; g^{\mu\nu}\, \partial_\mu \delta \Psi\, \partial_\nu \Psi
    = \int \dd^4 x\; \delta \Psi\, \partial_\mu \qty(\sqrt{-g}\, g^{\mu\nu}\, \partial_\nu \Psi),
\end{equation}
so that stationarity for an arbitrary $\delta \Psi$ requires the bracket to vanish. In covariant form the same condition reads $\Box \Psi \equiv \nabla^\mu \nabla_\mu \Psi = 0$, and the two are one and the same statement. For a scalar the covariant derivative reduces to the partial one, $\nabla_\mu \Psi = \partial_\mu \Psi$, whereas the divergence of a vector carries a single contracted Christoffel symbol,
\begin{equation}
    \label{eq:divergence identity}
    \nabla_\mu V^\mu = \partial_\mu V^\mu + \Gamma^{\mu}_{\ \mu\nu} V^\nu,
    \qquad
    \Gamma^{\mu}_{\ \mu\nu} = \frac{1}{2}\, g^{\alpha\beta}\, \partial_\nu g_{\alpha\beta}
    = \partial_\nu \ln \sqrt{-g},
\end{equation}
the last equality being Jacobi's formula for the derivative of a determinant~\cite{Nakahara:2003nw}. Any divergence therefore collapses to $\nabla_\mu V^\mu = \qty(1/\sqrt{-g})\, \partial_\mu \qty(\sqrt{-g}\, V^\mu)$, and setting $V^\mu = g^{\mu\nu} \partial_\nu \Psi$ turns $\Box \Psi = 0$ into the Klein--Gordon equation in curved spacetime,
\begin{equation}
\label{Klein-Gordon}
    \frac{1}{\sqrt{-g}} \partial_\mu \qty(\sqrt{-g}\, g^{\mu \nu} \partial_\nu) \Psi \qty(x^\mu) = 0,
\end{equation}
in which the whole influence of the geometry is carried by $g^{\mu\nu}$ and $\sqrt{-g}$ alone.
Assuming a stationary, spherically symmetric ansatz, Eq.~\eqref{eq:spherical ansatz}, in which the metric is diagonal, Eq.~\eqref{Klein-Gordon} expands to
\begin{multline}
\label{KG aberta na metrica}
  \bigg[ g^{00} \partial^2_0 + \frac{1}{\sqrt{-g}} \qty(\partial_1 \sqrt{-g} ) g^{11} \partial_1  + \qty(\partial_1 g^{11}) \partial_1 + g^{11} \partial^2_1 + \\
  + g^{22}   \partial^2_2 +  \frac{1}{\sqrt{-g}} \qty(\partial_2 \sqrt{-g}) g^{22} \partial_2 +g^{33} \partial^2_3 \bigg]  \Psi \qty(x^\mu)  = 0.
\end{multline}

\begin{figure}[htbp]
    \centering
    \includegraphics[width=0.7\linewidth]{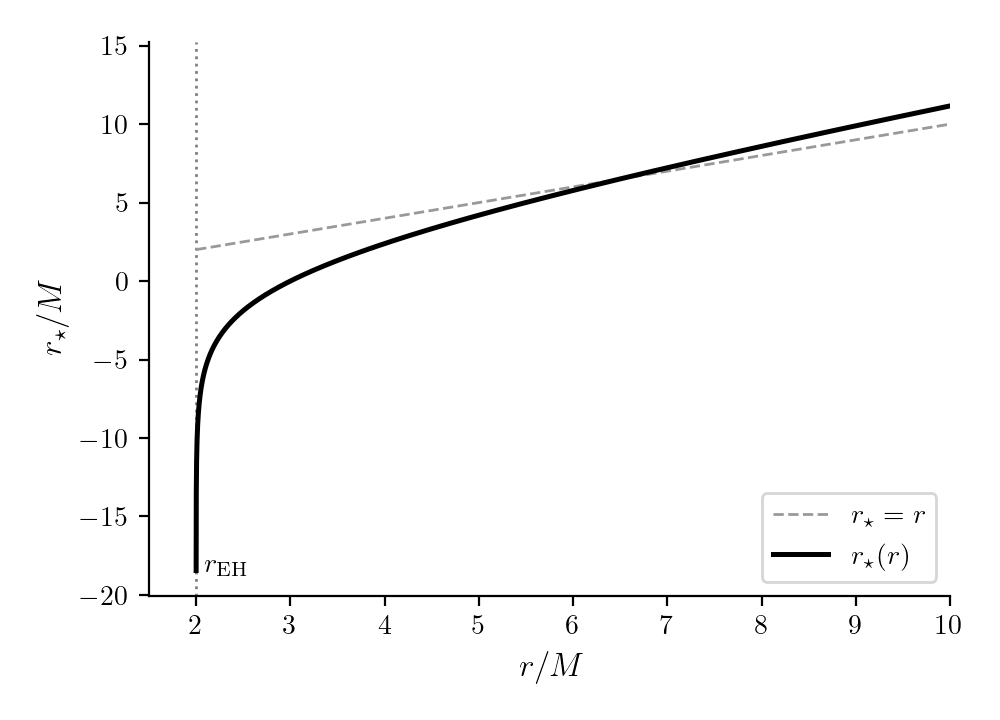}
    \caption[Tortoise coordinate of the Schwarzschild geometry]{Tortoise coordinate $r_\star(r)$ of the Schwarzschild geometry, Eq.~\eqref{eq:tortoise schw}, in natural units. The dashed line is $r_\star=r$, and the integration constant is fixed by $r_\star(3M)=0$.}
    \label{fig:tortoise_schw}
\end{figure}

To simplify the notation, one sets $B(r) \equiv e^{\nu(r)}$ and $A(r) \equiv e^{\lambda(r)}$. Substituting the metric coefficients gives
\begin{multline}
    \label{KG aberta na metrica com valores}
- \frac{1}{B}\frac{\partial^2 \Psi}{\partial t^2} + \frac{1}{A} \qty(\frac{2}{r} + \frac{1}{2B} \frac{\dd B}{\dd r} - \frac{1}{2A} \frac{\dd A}{\dd r}) \pdv{\Psi}{r} + \frac{1}{A} \pdv[2]{\Psi}{r} + \\ +\frac{1}{r^2} \qty[\frac{1}{\sin \theta} \frac{\partial}{\partial \theta}\qty(\sin \theta \pdv{\Psi}{\theta}) + \frac{1}{\sin ^2 \theta} \pdv[2]{\Psi}{\phi}] = 0.
\end{multline}
The angular variables in Eq.~\eqref{KG aberta na metrica com valores} enter only through the final bracket, which is precisely the angular part of the Laplacian on the unit sphere. This structure permits a separation of variables, splitting the equation into one part containing only the derivatives in $t$ and $r$ and another containing only the derivatives in $\theta$ and $\phi$. Since a function of $\qty(t,r)$ can equal a function of $\qty(\theta,\phi)$ only if both are constant, the two parts are set equal to a common separation constant. Requiring the angular solution to be regular and single-valued over the sphere quantises this constant to $\ell\qty(\ell+1)$, with $\ell=0,1,2,\dots$ the multipole (angular-momentum) number and the associated eigenfunctions the spherical harmonics. Because the background is spherically symmetric, the separation constant is independent of the azimuthal number $m$, so the radial-temporal dynamics is degenerate in $m$. Separating the $\qty{t,r}$ and angular components,
\begin{equation}
    \Psi_{\ell m}\qty(t,r,\theta,\phi)=Z\qty(t,r)\, Y_{\ell}^m\qty(\theta,\phi),
\end{equation}
where $Y_{\ell}^{m}\qty(\theta,\phi)$ are the spherical harmonics of degree $\ell$ and order $m$,
\begin{equation}
    Y_{\ell}^{m}\qty(\theta,\phi) = \frac{\qty(\sin \theta)^{|m|}}{2^\ell \ell!} \qty(\dv{}{\cos \theta})^{|m|+\ell} \qty(1-\cos^2 \theta)^{-\ell} \ e^{im\phi}
    \ ,
\end{equation}
and, the radial-temporal function $Z\qty(t,r)$ then satisfies
\begin{equation}
\label{eq gigante}
    -\frac{\partial^2 Z}{\partial t^2} + \frac{B}{A} \qty(\frac{2}{r} + \frac{1}{2B} \frac{\dd B}{\dd r} - \frac{1}{2A} \frac{\dd A}{\dd r}) \pdv{Z}{r} + \frac{B}{A} \pdv[2]{Z}{r} = \frac{\ell\qty(\ell+1)}{r^2} B\,Z.
\end{equation}
To recast Eq.~\eqref{eq gigante} into a standard wave equation, the tortoise coordinate $r = r(r_\star)$ is introduced and, to absorb the first-derivative term into a scaling function, one sets
\begin{align}
    Z\qty(t,r)=N\qty(t,r)\,b\qty(r).
\end{align}
With the notation $(\,)' \equiv \partial_r$, used for radial derivatives throughout this thesis, the dot reserved here for $\dot{(\,)} \equiv \partial_{r_\star}$, and the shorthand $f \equiv B/A$, Eq.~\eqref{eq gigante} becomes
\begin{multline} \label{eq:mama}
    -\pdv[2]{N}{t} + f {r_\star'}^2 \pdv[2]{N}{r_\star} + \pdv{N}{r_\star} \qty[ f r_\star' \dot{r_\star'} + 2 f {r_\star'}^2 \frac{\dot b}{b} + \qty(\frac{2}{r} + \frac{r_\star' \dot f}{2f}) f r_\star'] + \\ +  N \qty[f r_\star' \dot{r_\star'}  \frac{\dot b}{b} + f {r_\star'}^2 \frac{\ddot b}{b} +  \qty(\frac{2}{r} + \frac{r_\star' \dot f}{2f}) f r_\star' \frac{\dot b}{b} ]  = \frac{\ell\qty(\ell+1)}{r^2} B\,N.
\end{multline}
The standard waveform is
\begin{equation}
    \label{equacao de onda}
    {-\pdv[2]{N(t,r_\star)}{t} + \pdv[2]{N(t,r_\star)}{r_\star} = V_{\text{eff}}(r_\star)\, N(t,r_\star)},
\end{equation}
and, working in tortoise coordinates, imposing the equality of Eq.~\eqref{eq:mama} and Eq.~\eqref{equacao de onda} yields the two conditions
\begin{align}
    \label{Termo que vai a zero}
     f r_\star' \dot{r_\star'} + 2 f {r_\star'}^2 \frac{\dot b}{b} + \qty(\frac{2}{r} + \frac{r_\star' \dot f}{2f}) f r_\star' = 0,
     \\[0.25cm]
     \label{Termo que diz a coordenada tartaruga}
     f {r_\star'}^2 = 1.
\end{align}
Eq.~\eqref{Termo que diz a coordenada tartaruga} recovers the definition of the tortoise coordinate,
\begin{equation}
    \label{coordenada tartaruga final}
    {\dd r_\star = \frac{\dd r}{\sqrt{f}} = \frac{\dd r}{\sqrt{B/A}}}.
\end{equation}
This coordinate maps the semi-infinite region from the horizon to infinity onto $\qty(-\infty,+\infty)$, moving the physical (and, in most cases, also the mathematical) singularities out of the region of analysis. For the Schwarzschild geometry, where $A=B^{-1}=\qty(1-2G_\text{N}M/r)^{-1}$, the integration can be carried out in closed form,
\begin{equation}
    \label{eq:tortoise schw}
    r_\star = r + 2G_\text{N}M \ln\abs{\frac{r}{2G_\text{N}M} - 1} + C,
\end{equation}
with $C$ an arbitrary integration constant. The logarithm diverges as $r\to r_{\text{EH}}=2G_\text{N}M$, which sends the horizon to $r_\star\to-\infty$, whereas at large radii the slope tends to unity and $r_\star/r\to1$, the difference $r_\star-r$ growing only logarithmically. Both regimes are shown in Fig.~\ref{fig:tortoise_schw}.

Substituting Eq.~\eqref{coordenada tartaruga final} into Eq.~\eqref{Termo que vai a zero} fixes the scaling function,
\begin{equation}
    \label{b em funcao de r}
    b(r)=\frac{1}{r},
\end{equation}
and the effective potential follows as
\begin{equation}
\label{potencial efetivo final}
    {V_{\text{eff}}(r) = B(r) \frac{\ell\qty(\ell+1)}{r^2} + \frac{1}{2r} \dv{}{r}\qty(\frac{B(r)}{A(r)})}.
\end{equation}
Assuming a periodic time dependence, $N(t,r_\star) = e^{-i \omega t} R(r_\star)$, one arrives at the Schr\"odinger-like equation
\begin{equation}
    \label{equacao dos modos quasinormais}
    \qty[\dv[2]{}{r_\star} + \omega^2 -  V_{\text{eff}}(r_\star)] R(r_\star) = 0.
\end{equation}
Eqs.~\eqref{potencial efetivo final} and~\eqref{equacao dos modos quasinormais} describe the propagation of a spin-$0$ particle in a curved spacetime, and hold throughout the region outside the event horizon. Fields of other spins follow the same route, starting from their respective equations of motion~\cite{Fierz:1939ix,Kokkotas:1999bd,Berti:2009kk}, although each demands further ingredients of its own, such as a gauge choice in the vector and tensor sectors, a decomposition into parity sectors, or a tetrad and spin connection for the Dirac field. In every case, the perturbation is first reduced to a two-dimensional wavelike form in which the angular variables decouple, the tortoise coordinate is then introduced, and the radial and temporal variables are left obeying a wavelike equation that, for stationary backgrounds, is of Schr\"odinger type. Once that reduction is achieved, the spin survives only through the effective potential.

For vector perturbations (spin-$1$ particles), one has the covariant generalisation of the Proca equations~\cite{Proca:1936fbw,Konoplya:2011qq,Kokkotas:1999bd,Berti:2009kk}, which for a massless vector potential $A_\mu$ read
\begin{equation} \label{eq: proca equations}
    \frac{1}{\sqrt{-g}}\partial_\mu \qty(F_{\rho \sigma} g^{\rho\nu} g^{\sigma \mu}\sqrt{-g}) = \frac{1}{\sqrt{-g}}\partial_\mu \qty(
    \qty(\partial_\rho A_\sigma - \partial_\sigma A_\rho)
    g^{\rho\nu} g^{\sigma \mu}\sqrt{-g}) = 0,
\end{equation}
where $F_{\mu \nu}$ is the electromagnetic tensor. Since $A_\mu$ carries a vector index, the angular dependence is now resolved by an expansion in vector spherical harmonics, which splits the perturbation into a polar and an axial sector, and the gauge freedom $A_\mu \mapsto A_\mu + \partial_\mu \varpi$, where $\varpi$ is an arbitrary scalar gauge function, must be fixed before the non-dynamical components are eliminated~\cite{Ruffini:1972pw,Zerilli:1974ai}. Each sector then reduces to a single master function obeying Eq.~\eqref{equacao dos modos quasinormais}, with effective potential
\begin{equation}
\label{potencial efetivo final vector}
    {V_{\text{eff}}(r) = B(r) \frac{\ell\qty(\ell+1)}{r^2}}.
\end{equation}
The derivative term of Eq.~\eqref{potencial efetivo final} is absent here because its coefficient vanishes for $s=1$, which reflects the conformal invariance of the Maxwell field in four dimensions~\cite{Ruffini:1972pw,Konoplya:2011qq}.

For massless Dirac fields (spin-$1/2$ particles), the equations of motion are
\begin{equation}
    \gamma^\alpha {e_\alpha}^\mu \qty(\partial_\mu + \Gamma_\mu) {\Phi} = 0,
\end{equation}
where ${e_\alpha}^\mu$ is the tetrad field connecting $g_{\mu\nu}$ and the Minkowski metric $\eta_{\alpha\beta}$ through
\begin{subequations}
\begin{align}
    g_{\mu\nu}\equiv \eta_{\alpha \beta} {e_\mu}^\alpha {e_\nu}^\beta, &&
g^{\mu\nu}\equiv \eta^{\alpha \beta} {e_\alpha}^\mu {e_\beta}^\nu, \\
{e_\alpha}^\mu {e_\mu}^\beta = {\delta_\alpha}^\beta, &&
{e_\mu}^\alpha {e_\alpha}^\nu = {\delta_\mu}^\nu,
\end{align}
\end{subequations}
$\gamma^\alpha$ are the Dirac matrices obeying
\begin{equation}
\qty{\gamma^\mu,\gamma^\nu} = - 2\eta^{\mu\nu},
\end{equation}
and $\Gamma_\mu$ is the spin connection~\cite{Nakahara:2003nw}
\begin{equation}
    \Gamma_\mu = -\frac{1}{8} \qty[\gamma^{\rho},\gamma^{\sigma}]   \ g_{\alpha\tau} \ e_{\rho}^{\ \alpha} \ \nabla_{\mu}e_{\sigma}^{\ \tau}.
\end{equation}
Here the angular dependence is resolved by spinor spherical harmonics, and the radial functions first obey a coupled first-order system, which decouples into two isospectral Schr\"odinger-like wave functions with supersymmetric partner potentials~\cite{Brill:1957fx,Cho:2003qe,Dubinsky:2024mwd}
\begin{equation}
    V_{\text{eff}}^\pm(r) = \frac{B(r)}{r^2} \qty(\ell+\frac{1}{2})^2 \pm \qty(\ell+\frac{1}{2}) \sqrt{\frac{B(r)}{A(r)}} \dv{}{r}\left(\frac{\sqrt{B\qty(r)}}{r}\right).
    \label{final effective potential half aux}
\end{equation}
Through the Darboux transformation
\begin{equation}
    \label{eq:darboux}
    R^+(r_\star) \propto \qty[\qty(\ell+\frac12)\frac{\sqrt{B(r)}}{r} + \dv{}{r_\star}]  R^-(r_\star),
\end{equation}
the two isospectral wave functions are mapped into one another~\cite{Chandrasekhar:1975zza,Dubinsky:2024mwd}, so that only one is needed for the quasinormal-mode analysis. The potential $V_{\text{eff}}^+(r)$ is therefore retained.

The spin-$2$ (gravitational) case governs the tensor perturbations produced in a black-hole merger~\cite{LIGOScientific:2016aoc,LIGOScientific:2016lio} and is used repeatedly in the following chapters, so its main steps are recalled here. The perturbation is now of the metric itself,
\begin{equation}
    \label{eq:metric perturbation}
    g_{\mu\nu} \mapsto \bar g_{\mu\nu} + h_{\mu\nu},
    \qquad \abs{h_{\mu\nu}} \ll \abs{\bar g_{\mu\nu}},
\end{equation}
with $\bar g_{\mu\nu}$ the background~\eqref{eq:spherical ansatz}. The reduction of the ten components of $h_{\mu\nu}$ to a single wave equation follows three standard steps. Spherical symmetry allows $h_{\mu\nu}$ to be expanded in tensor spherical harmonics, so that each multipole $(\ell,m)$ evolves independently. Parity then splits the harmonics into a polar (even) and an axial (odd) sector, which do not mix at linear order and may be treated separately. Finally, the gauge freedom of an infinitesimal change of coordinates removes the remaining redundant components, the Regge--Wheeler gauge being the usual choice. Inserting the axial ansatz into the linearised vacuum equations, $\delta R_{\mu\nu}=0$, and rescaling the surviving radial function leave a single master function obeying Eq.~\eqref{equacao dos modos quasinormais}. The construction is classical and is developed in full in Refs.~\cite{Regge:1957td,Zerilli:1970se,Vishveshwara:1970zz,Chandrasekhar:1975zza,Kokkotas:1999bd,Berti:2009kk}. The resulting Regge--Wheeler potential~\cite{Regge:1957td} reads
\begin{multline}
\label{potencial efetivo final graviton}
    V_{\text{eff}}(r) = B(r) \frac{\ell\qty(\ell+1)}{r^2} - \frac{3}{2r} \dv{}{r}\qty(\frac{B(r)}{A(r)}) \\
    \Longrightarrow \quad \left. \qty(1-\frac{2G_\text{N} M}{r}) \qty(\frac{\ell(\ell+1)}{r^2}- \frac{6G_\text{N} M}{r^3}) \right|_{\text{Schwarzschild}}.
\end{multline}
This coincides with the $s=2$ member of the integer-spin master family~\eqref{final effective potential int}, the additional term $-\tfrac{3}{2r}\,\dv{}{r}\qty(B/A)$ being the signature of the spin of the graviton. For the Schwarzschild background, it reduces to the classical Regge--Wheeler result shown on the right-hand side of Eq.~\eqref{potencial efetivo final graviton}, whose $-6G_\text{N} M/r^3$ curvature term is precisely what distinguishes gravitational from scalar perturbations. The polar (even-parity) sector is instead governed by the Zerilli equation~\cite{Zerilli:1970se}. Since the two sectors are isospectral~\cite{Chandrasekhar:1975zza}, the axial (Regge--Wheeler) analysis already fixes the gravitational quasinormal spectrum, as exploited in the hairy black-hole spectroscopy of Ref.~\cite{Guimaraes:2025jsh}.

In summary, the quasinormal-mode analysis employs the tortoise coordinate~\eqref{coordenada tartaruga final}, the Schr\"odinger-like wave equation~\eqref{equacao dos modos quasinormais}, and the effective potentials
\begin{subequations}
\label{eq:final effective potentials}
\begin{align}
    V_{\text{eff}}(r) &= B(r) \frac{\ell\qty(\ell+1)}{r^2}
    + \frac{1-s^2}{2r} \dv{}{r}\qty(\frac{B\qty(r)}{A\qty(r)}),
    && \text{for} \;\; s\in\mathbb{Z}, \label{final effective potential int}
    \\[4pt]
    V_{\text{eff}}^{+}(r) &= \frac{B(r)}{r^2} \qty(\ell+\frac{1}{2})^2
    + \qty(\ell+\frac{1}{2}) \sqrt{\frac{B(r)}{A(r)}} \dv{}{r}\qty(\frac{\sqrt{B\qty(r)}}{r}),
    && \text{for}\;\; s=\frac12, \label{final effective potential half}
\end{align}
\end{subequations}
where only the partner potential with the positive sign is retained in the spin-$1/2$ case, the negative one being recovered from it through the Darboux transformation~\eqref{eq:darboux} and yielding the same spectrum. Eqs.~\eqref{final effective potential int} and~\eqref{final effective potential half} are determined entirely by the two metric functions $B(r)$ and $A(r)$, the spin $s$, and the multipole number $\ell$. Once the master potential is known, characterising the ringdown of a given static, spherically symmetric black hole reduces to inserting its metric functions, an observation that is the basis of the parametrised treatments of non-rotating ringdown~\cite{Cardoso:2019mqo,Konoplya:2011qq}.

\section{Effective Potential for Hairy and Quantum-Corrected Black Holes}
\label{sec:veff hairy and qc}

The master potentials~\eqref{final effective potential int} and~\eqref{final effective potential half} were obtained without reference to any particular solution, applying to any static, spherically symmetric line element of the form~\eqref{eq:spherical ansatz}. They are now evaluated on the two metrics constructed in Part~\ref{part: extensions of gr}. The two cases illustrate distinct ways in which a geometry may depart from Schwarzschild, since the hairy black hole retains the Schwarzschild-like relation $A(r)=B(r)^{-1}$, whereas the quantum correction breaks it.

\subsection{Hairy black holes from gravitational decoupling}
\label{subsec:veff hairy}

\begin{figure}[htbp]
    \centering
    \includegraphics[width=0.7\linewidth]{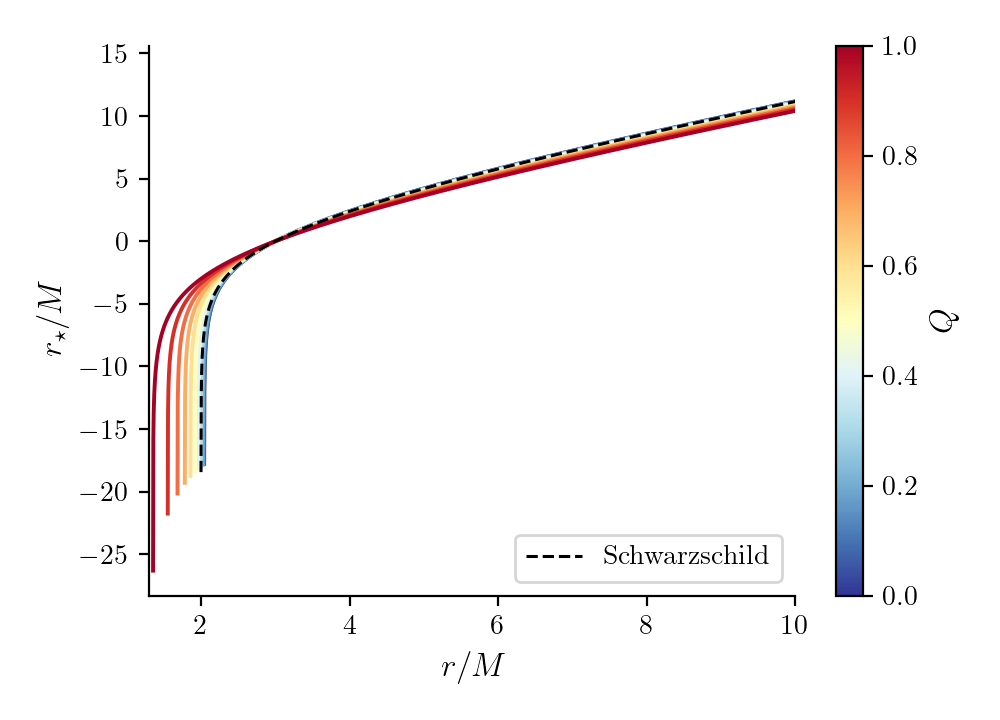}
    \caption[Tortoise coordinate of the hairy black hole]{Tortoise coordinate $r_\star(r)$ of the hairy black hole~\eqref{eq:hbh metric qnm} for $\alpha=\ell_h=0.5$, as the charge $Q$ is varied (colour bar), with Schwarzschild (dashed) for comparison. The charge is given in units of $G_\text{N}M$, with the \emph{bare} mass $M=1$, so that the total mass $\mathcal{M}=M+\alpha\ell_h/2$ grows with the hair, and the integration constant of each curve is fixed by $r_\star(3M)=0$.}
    \label{fig:tortoise_hbh}
\end{figure}

The gravitational-decoupling construction of Part~\ref{part: extensions of gr} yields the static, spherically symmetric geometry
\begin{equation}
    \label{eq:hbh metric qnm}
    B(r) = \frac{1}{A(r)} = 1 - \frac{2G_\text{N}\mathcal{M}}{r} + \frac{Q^2}{r^2}
    - \frac{\alpha G_\text{N}}{r}\qty(\mathcal{M} - \frac{\alpha \ell_h}{2})\,e^{-r/G_\text{N}\qty(\mathcal{M}-\frac{\alpha \ell_h}{2})},
    \qquad \mathcal{M} \equiv M + \frac{\alpha \ell_h}{2},
\end{equation} 
where $\alpha$ is the dimensionless decoupling parameter, $\ell_h$ the primary-hair parameter associated with the additional gravitational sector, carrying dimensions of mass as fixed in Chapter~\ref{chap:GD}, so that $G_\text{N}\ell_h$ is the corresponding length (the subscript distinguishes it from the multipole number $\ell$), and $Q$ a parameter of dimensions of length. Despite the Reissner--Nordstr\"om form of the term $Q^2/r^2$, $Q$ is not an electric charge, nor any other kind of charge, since it is not sourced by a gauge field. It is an integration constant of the decoupling procedure, constrained by $\alpha$ through the dominant energy condition, as established in Chapter~\ref{chap:GD}. Since $A=B^{-1}$, this solution reduces the tortoise coordinate~\eqref{coordenada tartaruga final} to its Schwarzschild form $\dd r_\star = \dd r/B(r)$, shown in Fig.~\ref{fig:tortoise_hbh} for a family of charge-like parameters $Q$ at fixed $\alpha=\ell_h=0.5$\footnote{This choice is arbitrary. Figure~\ref{fig:eventhorizon} shows that the metric~\eqref{eq:hbh metric qnm} depends monotonically on $\alpha$ and on $\ell_h$, so no value of either parameter is qualitatively distinguished, and a single representative pair suffices to display the behaviour. The dependence on both is examined later in this section.}. Increasing $Q$ draws the horizon inwards, as in Reissner--Nordstr\"om, since the term $Q^2/r^2$ raises $B(r)$ and displaces its outermost root to smaller radii. 
Two remarks make the resulting potentials compact. First, $\dv{}{r}\qty(B/A) = \dv{}{r}\qty(B^2) = 2BB'$, with the prime denoting $\dv*{}{r}$ as in Eq.~\eqref{eq:mama}. Second, $\mathcal{M}-\alpha\ell_h/2 = M$ by the definition in Eq.~\eqref{eq:hbh metric qnm}, so that the range of the Yukawa term is set by the bare mass $M$, while the Newtonian term is governed by the shifted mass $\mathcal{M}$. The master potentials~\eqref{final effective potential int}-\eqref{final effective potential half} can then be written with the blackening factor in evidence. For integer spin,
\begin{multline}
    \label{eq:hbh veff int explicit}
    V_{\text{eff}}(r) = \qty[1 - \frac{2G_\text{N}\mathcal{M}}{r} + \frac{Q^2}{r^2}
    - \frac{\alpha G_\text{N}}{r}\qty(\mathcal{M} - \frac{\alpha \ell_h}{2})\,e^{-r/G_\text{N}\qty(\mathcal{M}-\frac{\alpha \ell_h}{2})} ] \\ \times \qty{\frac{\ell\qty(\ell+1)}{r^2}
    + \frac{1-s^2}{r}\qty[\frac{2G_\text{N}\mathcal{M}}{r^2} - \frac{2Q^2}{r^3}
    + \frac{\alpha}{r}\qty(1+\frac{G_\text{N}M}{r})e^{-r/G_\text{N}M}]},
\end{multline}
and for spin-$1/2$,
\begin{multline}
    \label{eq:hbh veff half explicit}
    V_{\text{eff}}^{+}(r) = \frac{1}{r^2}\qty(\ell+\frac12)\Bigg\{\qty(\ell+\frac12)
    \qty[1 - \frac{2G_\text{N}\mathcal{M}}{r} + \frac{Q^2}{r^2}
    - \frac{\alpha G_\text{N}}{r}\qty(\mathcal{M} - \tfrac{\alpha \ell_h}{2})e^{-r/G_\text{N}\qty(\mathcal{M}-\frac{\alpha \ell_h}{2})}] \\
    - \sqrt{1 - \frac{2G_\text{N}\mathcal{M}}{r} + \frac{Q^2}{r^2}
    - \frac{\alpha G_\text{N}}{r}\qty(\mathcal{M} - \tfrac{\alpha \ell_h}{2})e^{-r/G_\text{N}\qty(\mathcal{M}-\frac{\alpha \ell_h}{2})}} \\
    \times \qty[1 - \frac{3G_\text{N}\mathcal{M}}{r} + \frac{2Q^2}{r^2}
    - \frac{\alpha}{2}\qty(1+\frac{3G_\text{N}M}{r})e^{-r/G_\text{N}M}]\Bigg\}.
\end{multline}

Two limits confirm the consistency of Eqs.~\eqref{eq:hbh veff int explicit}-\eqref{eq:hbh veff half explicit}. For $\alpha=0$ and $Q=0$ one has $B(r)=1-2G_\text{N} M/r$, and Eq.~\eqref{eq:hbh veff int explicit} reduces to the Schwarzschild potentials, recovering the Regge--Wheeler potential~\eqref{potencial efetivo final graviton} for $s=2$. For $\alpha=0$ and relaxing $Q=0$ one has $B(r)=1-2G_\text{N} M/r+Q^2/r^2$, the Reissner--Nordstr\"om blackening factor, with the corresponding charged-black-hole potentials.

The deviation from these limits is carried by the Yukawa-type term $\propto\alpha\,e^{-r/G_\text{N}\qty(\mathcal{M}-\alpha\ell_h/2)}$ in Eq.~\eqref{eq:hbh metric qnm}. Unlike the quantum correction that will be discussed below, this term modifies the barrier over a macroscopic range set by the length scale $G_\text{N}M$. Its effect is to lower[raise] the peak as $\alpha$ or $\ell_h$ increases[decreases], the two hair parameters acting in the same sense and in opposition to the charge, which raises[lowers] the barrier as $Q$ increases[decreases]. The hair therefore reshapes the potential in the very region that controls the ringdown, and does so at scales far larger than the short-range quantum term of the next subsection. This is what makes hairy configurations comparatively promising targets for black-hole spectroscopy, although sustaining such hair requires evading the classical no-hair theorems~\cite{Bekenstein:1972ny,Herdeiro:2014goa,Herdeiro:2015waa,Guimaraes:2025jsh}. The quasinormal response of this family has been examined independently, for scalar and gravitational perturbations, for rotating configurations, and through the greybody factors, with results that agree on the sign and the order of magnitude of the effect reported here~\cite{Yang:2022ifo,Cavalcanti:2022cga,Li:2022hkq,Lei:2023wlt,dePaiva:2025eux,Navarro:2026lrf}.

\begin{figure}[htbp]
    \centering
    \begin{subfigure}{0.49\textwidth}
        \centering\includegraphics[width=\linewidth]{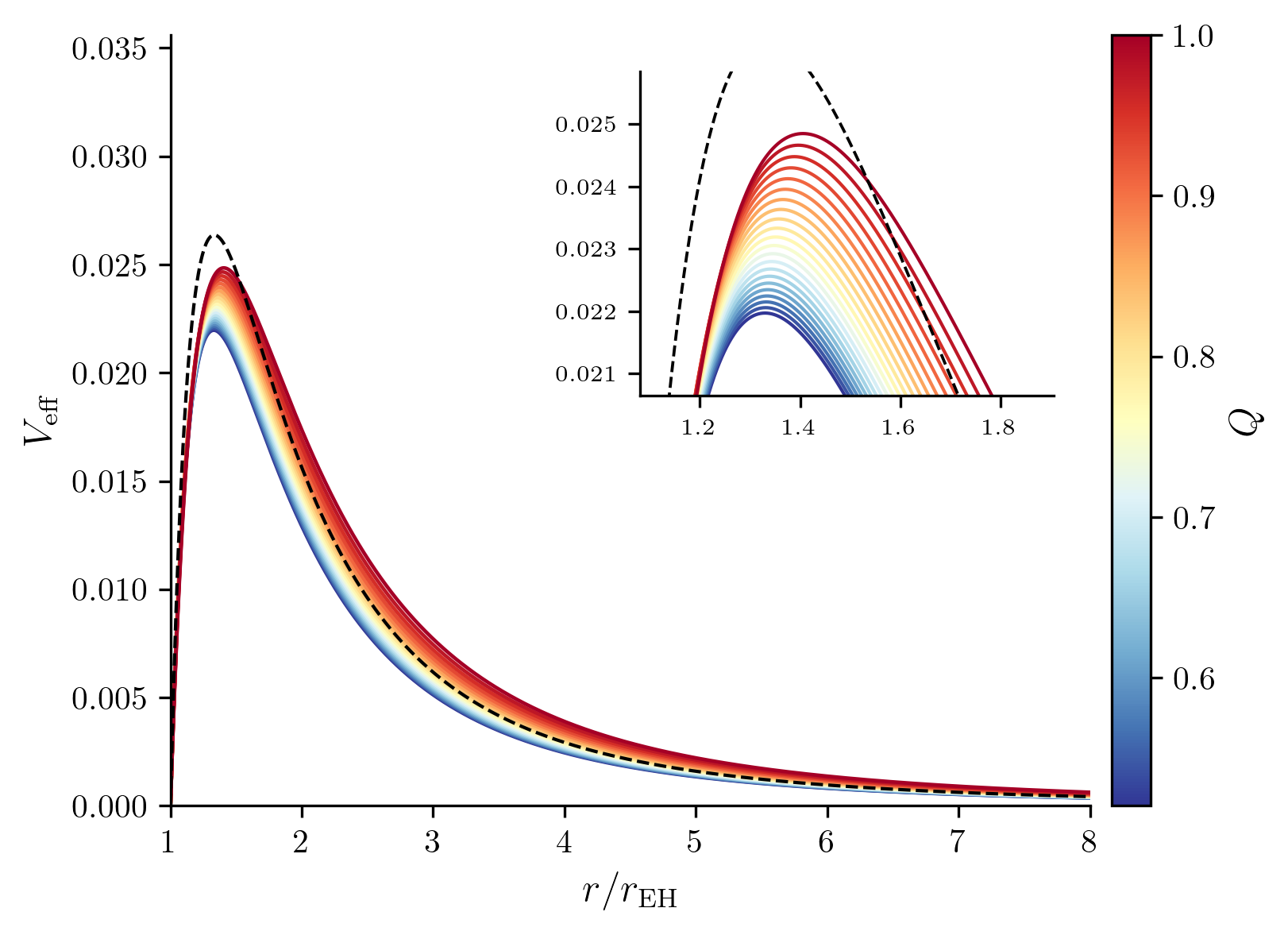}
        \caption{Scalar, $s=\ell=0$.}\label{fig:hbh veff s0}
    \end{subfigure}\hfill
    \begin{subfigure}{0.49\textwidth}
        \centering\includegraphics[width=\linewidth]{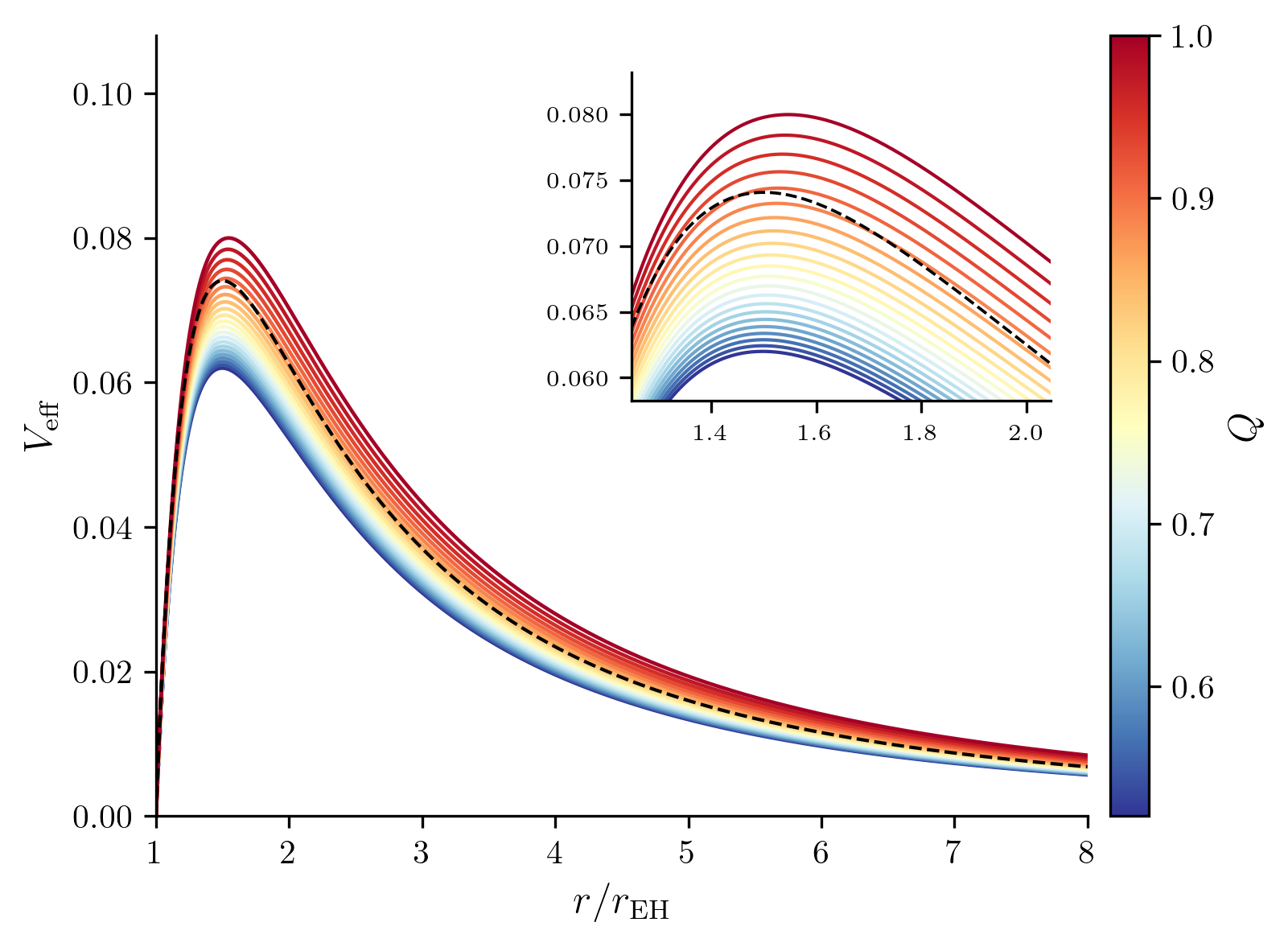}
        \caption{Electromagnetic, $s=\ell=1$.}\label{fig:hbh veff s1}
    \end{subfigure}

    \begin{subfigure}{0.49\textwidth}
        \centering\includegraphics[width=\linewidth]{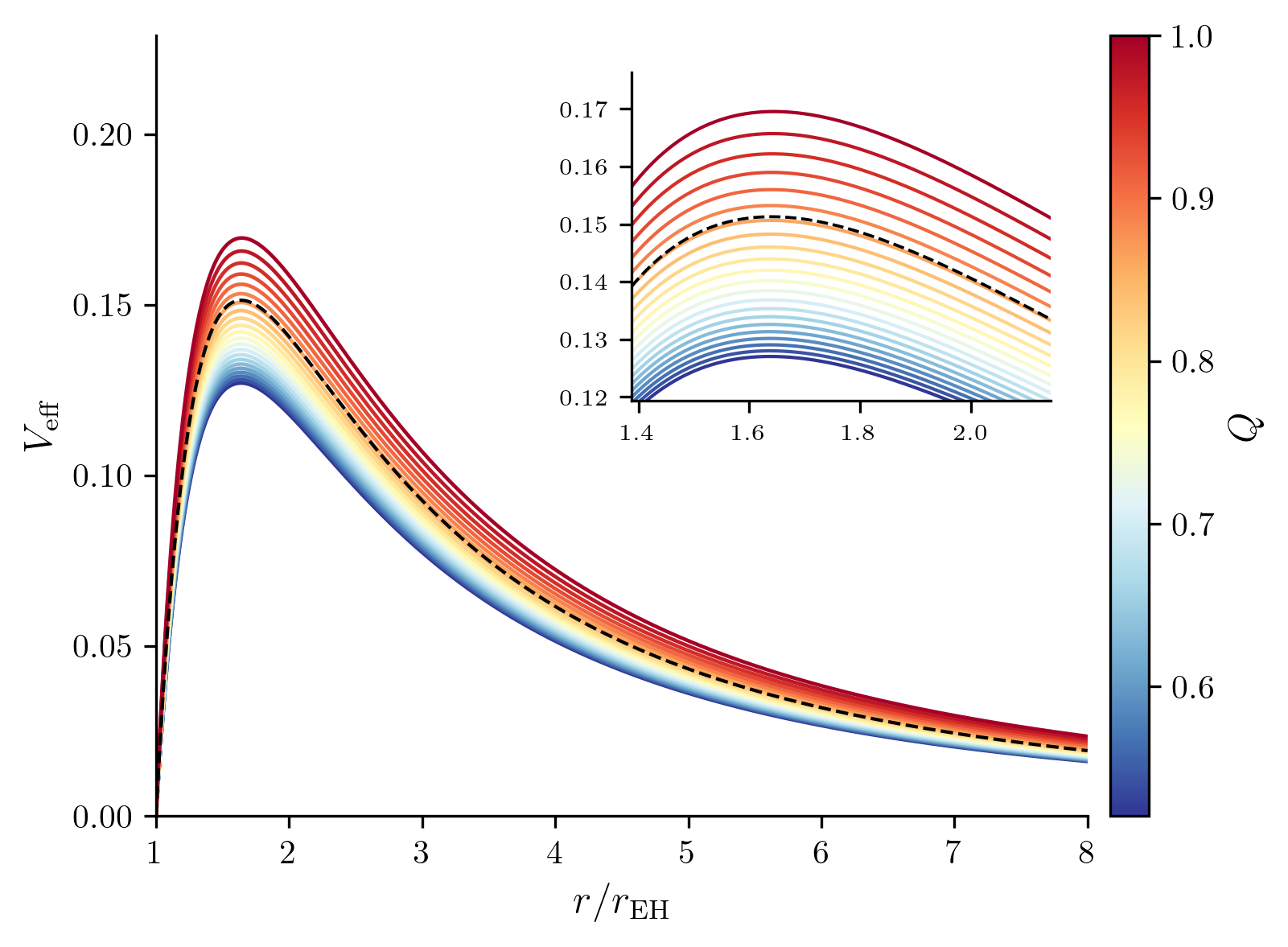}
        \caption{Gravitational, $s=\ell=2$.}\label{fig:hbh veff s2}
    \end{subfigure}\hfill
    \begin{subfigure}{0.49\textwidth}
        \centering\includegraphics[width=\linewidth]{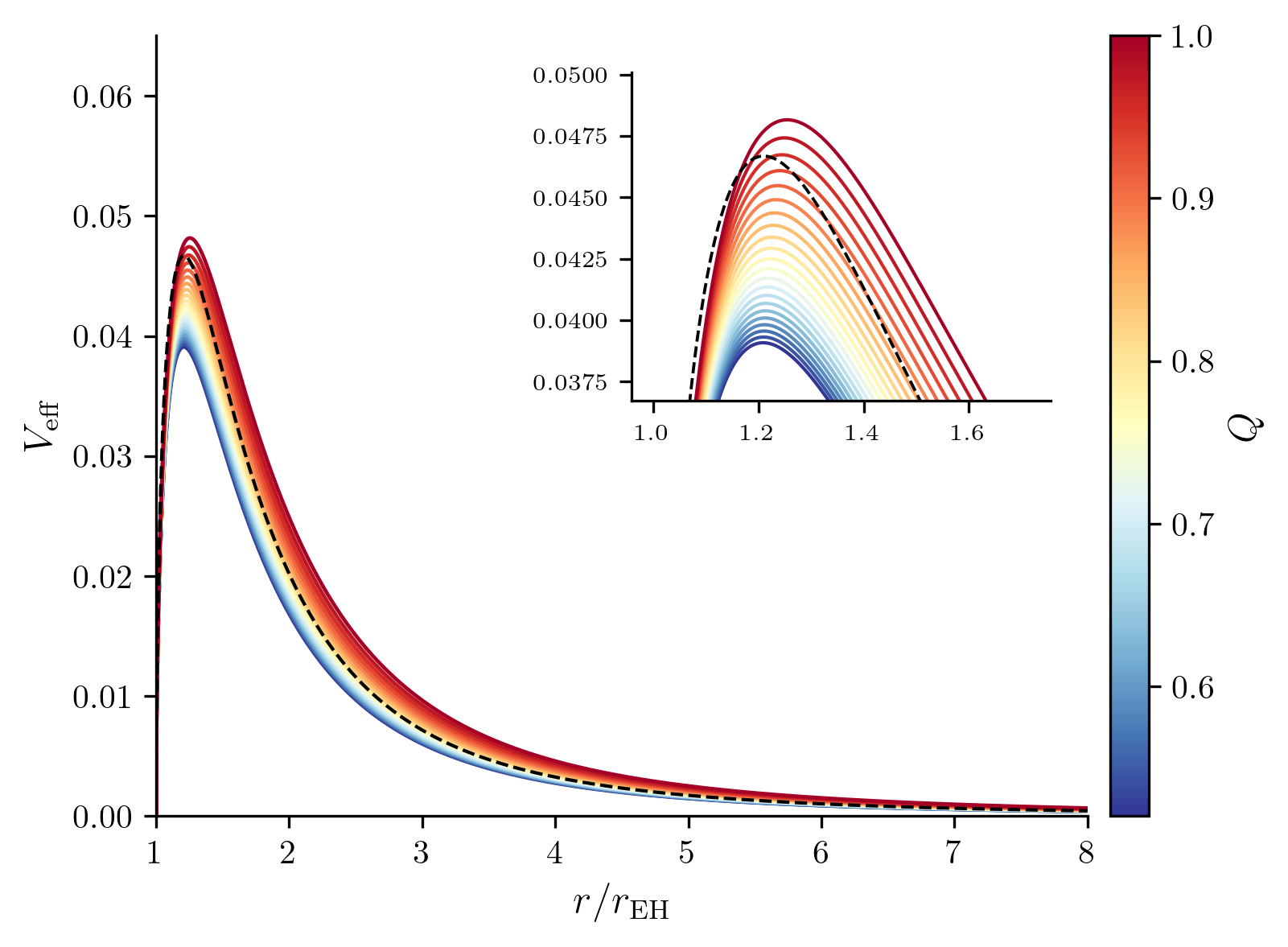}
        \caption{Dirac, $s=\ell=1/2$.}\label{fig:hbh veff s12}
    \end{subfigure}
    \caption[Effective potential of the hairy black hole, four channels]{Effective potential of the hairy black hole, Eqs.~\eqref{eq:hbh veff int explicit} and~\eqref{eq:hbh veff half explicit}, as a function of $r/r_{\text{EH}}$, at fixed $\alpha=\ell_h=0.5$ and with the charge varied (colour bar). The dashed line is the Schwarzschild potential of the same channel, and the inset magnifies the maximum of the barrier. Natural units with the \emph{bare} mass $M=1$, so that the total mass $\mathcal{M}=M+\alpha\ell_h/2$ grows with the hair.}
    \label{fig:hbh veff}
\end{figure}

The domain of $(\alpha,\ell_h,Q)$ over which Eq.~\eqref{eq:hbh metric qnm} describes an admissible black hole (satisfying the dominant energy condition, asymptotic flatness, and the existence of a horizon) was mapped in Part~\ref{part: extensions of gr}, following Refs.~\cite{Ovalle:2020kpd,Casadio:2024uwj}, and the potentials~\eqref{eq:hbh veff int explicit}-\eqref{eq:hbh veff half explicit} are to be evaluated over that domain. In particular, the charge-like parameter is not free once $\alpha$ is fixed, since the dominant energy condition requires $Q^2 \geq 4\alpha G_\text{N}^2M^2/e^2$, so that $\alpha=\ell_h=0.5$ admits only $Q\gtrsim0.52\,G_\text{N}M$, and the curves are drawn over that range alone. The resulting barriers are shown in Fig.~\ref{fig:hbh veff}, for the four channels used throughout this thesis.

\begin{figure}[htbp]
    \centering
    \begin{subfigure}{0.49\textwidth}
        \centering\includegraphics[width=\linewidth]{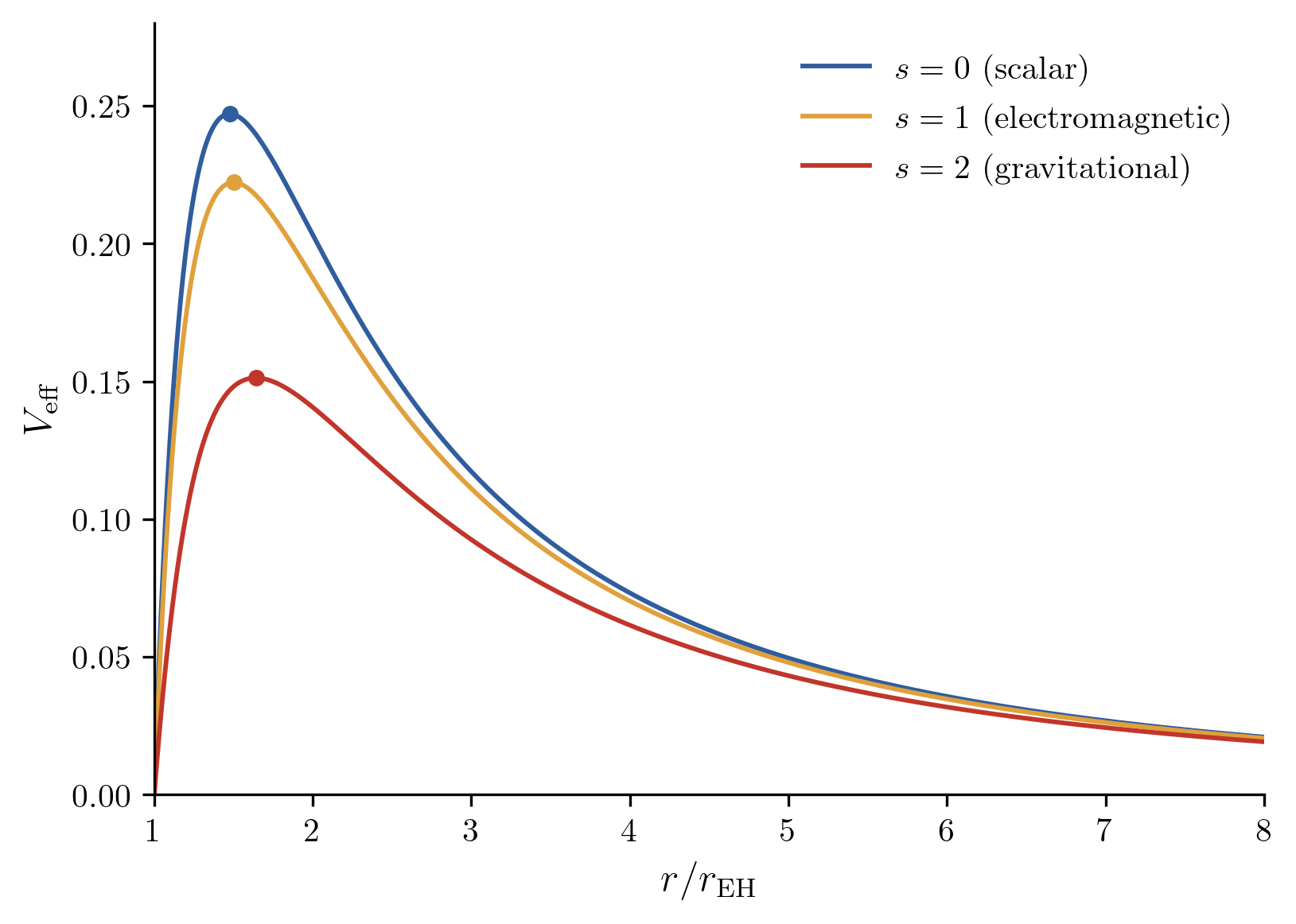}
        \caption{Spin varied at fixed $\ell=2$.}\label{fig:schw spin s}
    \end{subfigure}\hfill
    \begin{subfigure}{0.49\textwidth}
        \centering\includegraphics[width=\linewidth]{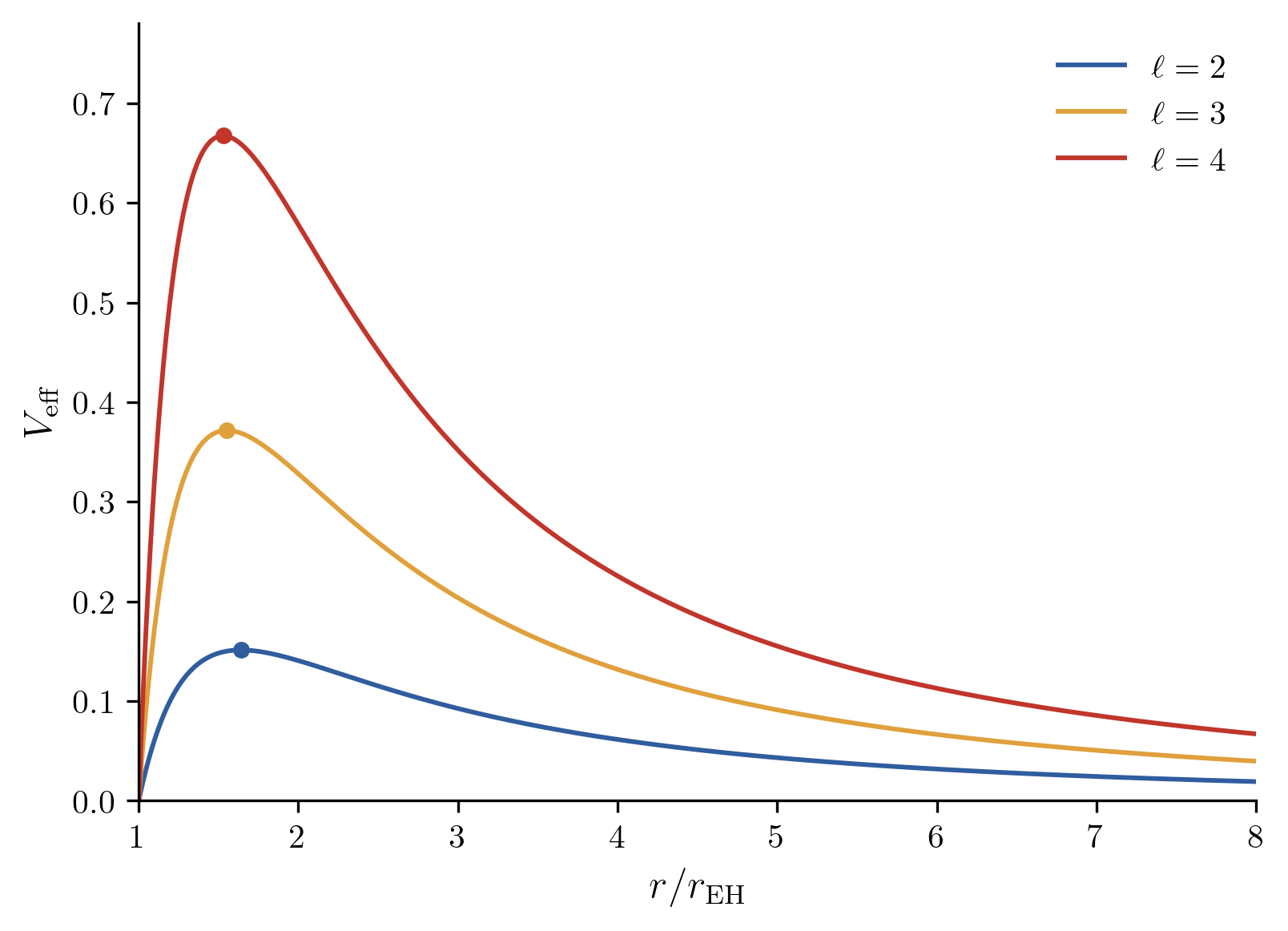}
        \caption{Multipole varied at fixed $s=2$.}\label{fig:schw spin l}
    \end{subfigure}
    \caption[Schwarzschild effective potential under spin and multipole]{Schwarzschild effective potential under the two movements that set the ordering of the peaks. Panel~\subref{fig:schw spin s} fixes the multipole and varies the spin; panel~\subref{fig:schw spin l} fixes the spin and varies the multipole, for which only $\ell\geq s$ is admissible. The markers indicate the maxima, and note the difference in vertical scale between the two panels. Natural units, $M=1$.}
    \label{fig:schw spin}
\end{figure}

Two features are common to the four panels of Fig.~\ref{fig:hbh veff}. The charge raises the barrier monotonically, exactly as in Reissner--Nordstr\"om, whereas the position of the peak in units of the horizon radius is almost unaffected, as quantified below. The ordering of the peaks across the panels is then set by the multipole rather than by the spin, which follows directly from Eq.~\eqref{eq:hbh veff int explicit}. At fixed $\ell$ the spin enters only through the factor $1-s^2$ multiplying $B'(r)/r$, which is the bracket written out in the second line of Eq.~\eqref{eq:hbh veff int explicit}, and $B'(r)>0$ outside the horizon, so that this term is positive for $s=0$, vanishes for $s=1$, and turns negative for $s\geq2$. The barrier therefore decreases with increasing spin at fixed $\ell$, as Fig.~\ref{fig:schw spin s} shows for Schwarzschild at $\ell=2$, where the maximum falls from $0.247$ in the scalar channel to $0.222$ in the electromagnetic one and $0.151$ in the gravitational one. The multipole acts in the opposite sense and far more strongly, through the centrifugal term $\ell(\ell+1)/r^2$. Fixing $s=2$ instead, as in Fig.~\ref{fig:schw spin l}, the maximum climbs from $0.151$ at $\ell=2$ to $0.372$ at $\ell=3$ and $0.667$ at $\ell=4$, so that one unit of $\ell$ multiplies the barrier by roughly two, whereas the whole range of integer spins at fixed $\ell=2$ changes it by less than forty per cent. Each panel of Fig.~\ref{fig:hbh veff} displays the lowest multipole of its channel, $\ell=s$, so that the growth of the peak along the sequence $\ell=s=0,\tfrac12,1,2$ reflects the accompanying growth of $\ell$, which dominates over the spin term.

\begin{figure}[htbp]
    \centering
    \begin{subfigure}{0.49\textwidth}
        \centering\includegraphics[width=\linewidth]{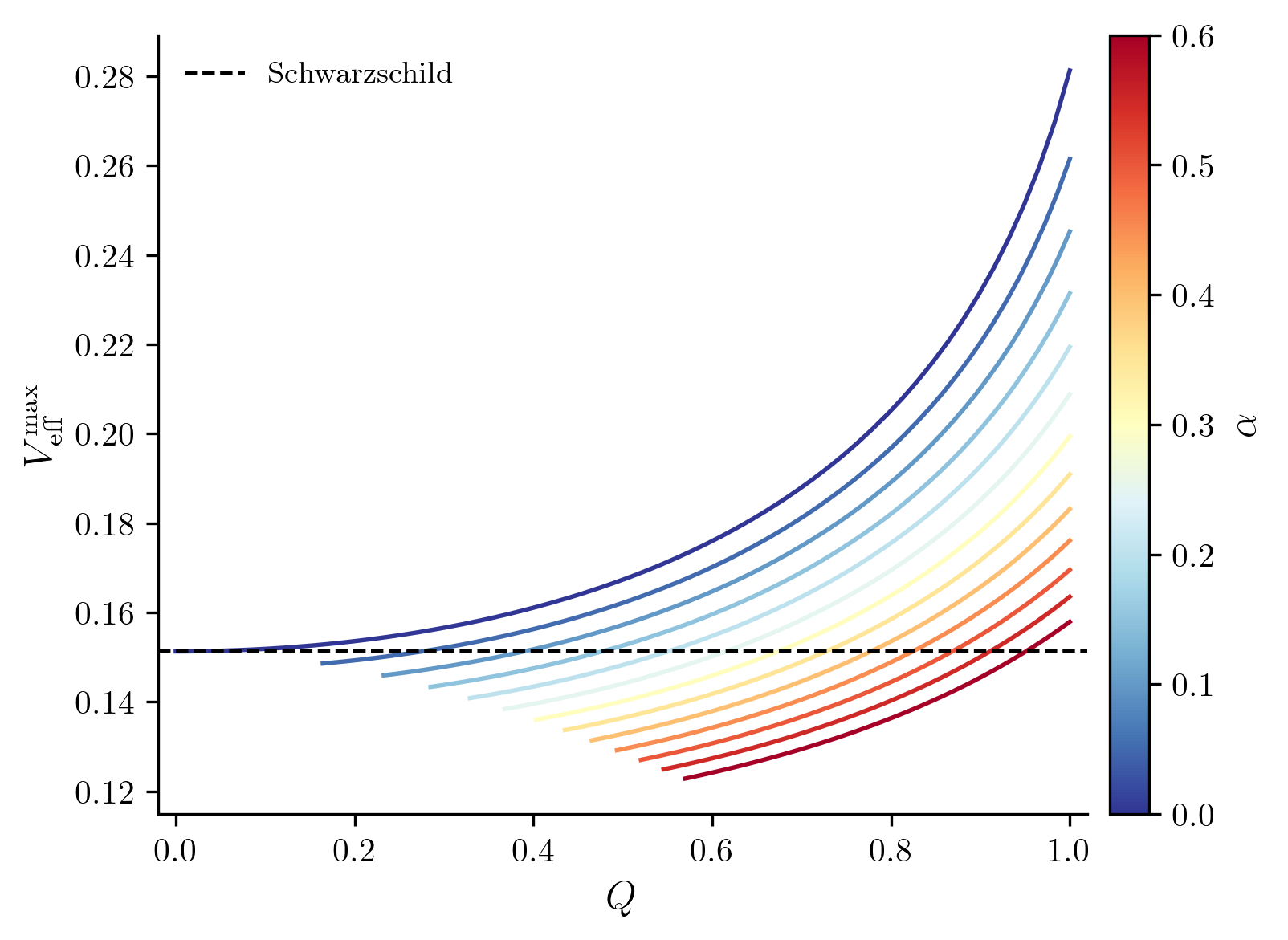}
        \caption{Height of the barrier.}\label{fig:hbh peak height}
    \end{subfigure}\hfill
    \begin{subfigure}{0.49\textwidth}
        \centering\includegraphics[width=\linewidth]{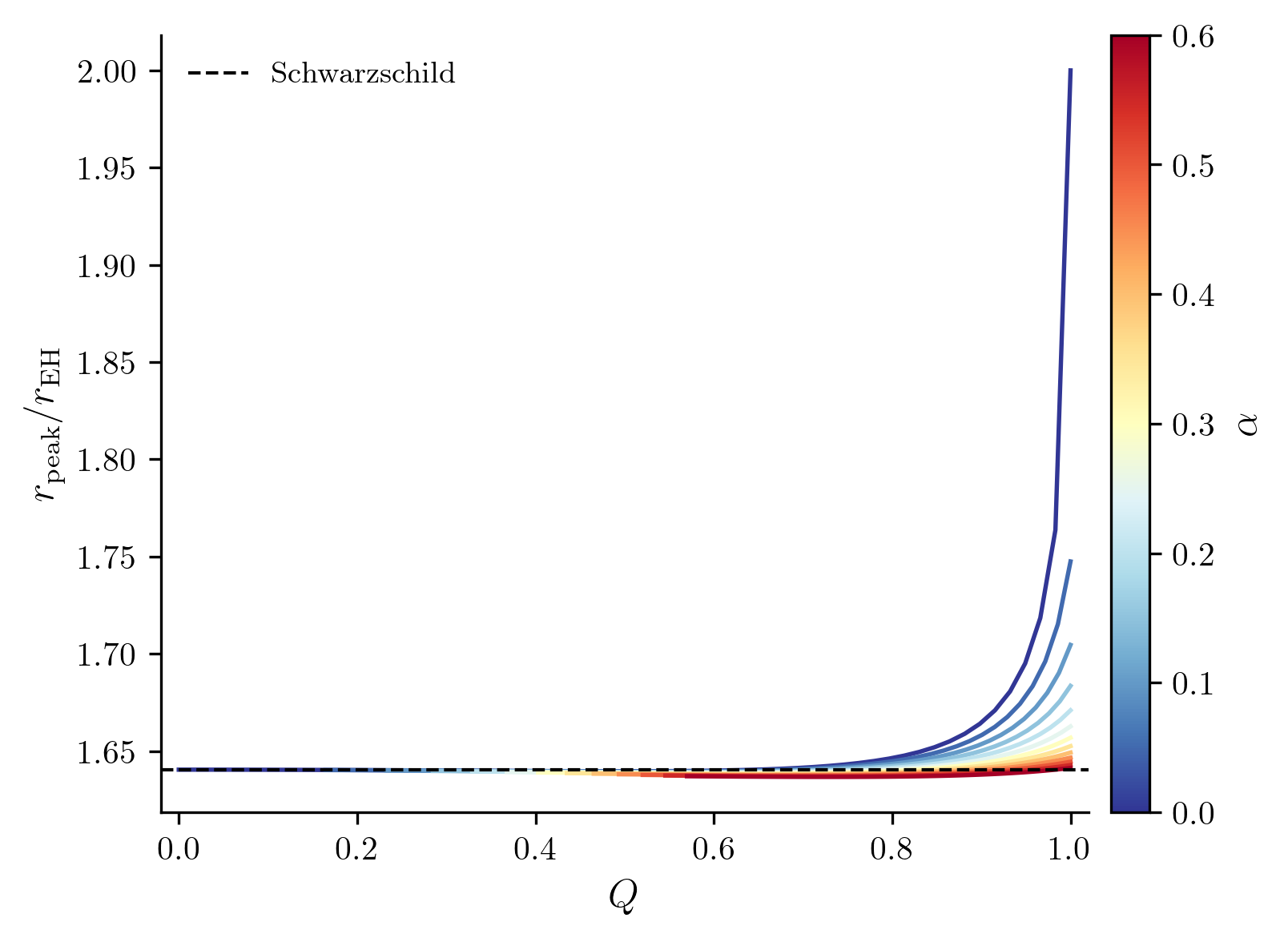}
        \caption{Position of the maximum.}\label{fig:hbh peak radius}
    \end{subfigure}
    \caption[Height and position of the hairy barrier over the admissible domain]{Maximum $V_{\text{eff}}^{\max}$ of the effective potential and its position $r_{\text{peak}}$, in units of the horizon radius, for the gravitational channel $s=\ell=2$ at $\ell_h=0.5$, as functions of the charge $Q$ and of the decoupling parameter $\alpha$ (colour bar). The dashed line marks the Schwarzschild value. Natural units with the \emph{bare} mass $M=1$, so that the total mass $\mathcal{M}=M+\alpha\ell_h/2$ grows with the hair.}
    \label{fig:hbh peak}
\end{figure}

The quasinormal frequencies of a barrier of this type are controlled by two numbers~\cite{Schutz:1985km,Iyer:1986np}, the height of the peak and the curvature there,
\begin{equation}
    \label{eq:peak data}
    V_{\text{eff}}^{\max} = V_{\text{eff}}\qty(r_{\text{peak}})
    \qquad\text{and}\qquad
    \left.\dv[2]{V_{\text{eff}}}{r_\star}\right|_{r_{\text{peak}}},
\end{equation}
where $r_{\text{peak}}$ is the radius at which the potential is largest and the second derivative is taken with respect to the tortoise coordinate~\eqref{coordenada tartaruga final}. The height sets the frequency of the oscillation and the curvature its damping rate, as Chapter~\ref{chap:QNM_spectrum} makes explicit. It is therefore useful to isolate the peak itself. Figure~\ref{fig:hbh peak} tracks the height $V_{\text{eff}}^{\max}$ and the position $r_{\text{peak}}$ of the maximum over the admissible domain, for the gravitational channel. The staggered onset of the curves is the dominant energy condition at work, each value of $\alpha$ being allowed only above its own threshold in $Q$. The two parameters act in opposition, the charge raising the barrier and the hair lowering it, so that the Schwarzschild height is recovered along a line in the $(\alpha,Q)$ plane rather than at a point. The position of the peak, by contrast, is remarkably rigid, remaining within one per cent of $r_{\text{peak}}\simeq1.64\,r_{\text{EH}}$ over most of the domain and moving outwards only in the near-extremal corner $Q\to1$. The hair therefore modulates mainly the height of the barrier, leaving its location essentially fixed.

\subsection{Quantum-corrected Schwarzschild metric}
\label{subsec:veff quantum}

The analysis of this subsection is organised differently from the previous one. The hairy solution carries three parameters, whereas the quantum correction carries the single coupling $c_6$, which allows the dependence of the barrier on it to be characterised in closed form.

\begin{figure}[htbp]
    \centering
    \includegraphics[width=0.7\linewidth]{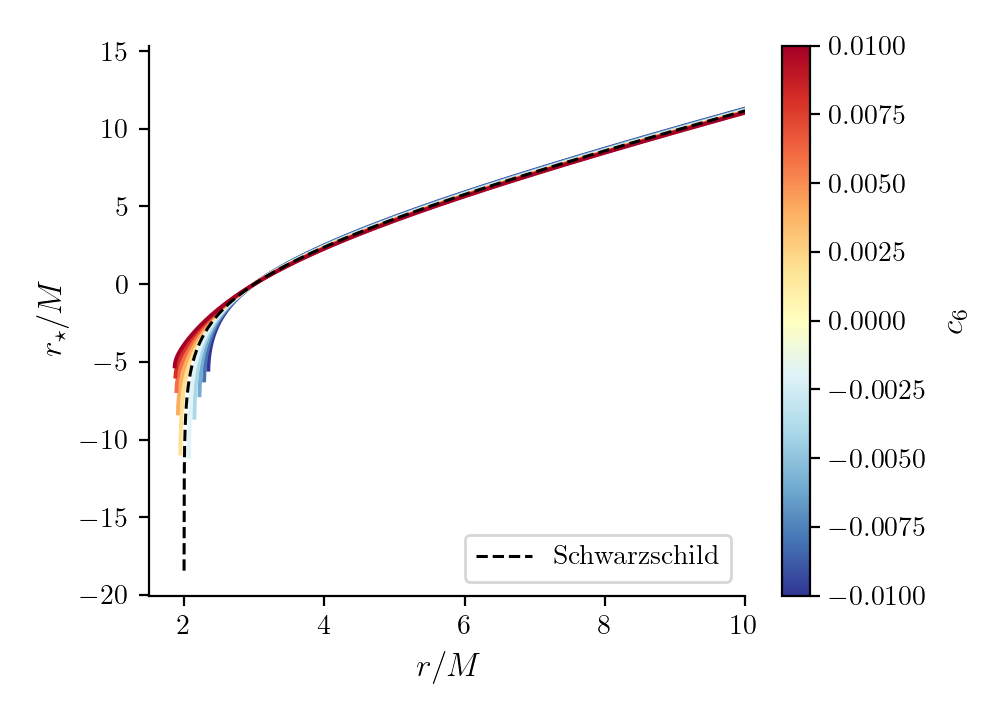}
    \caption[Tortoise coordinate of the quantum-corrected metric]{Tortoise coordinate $r_\star(r)$ of the quantum-corrected metric~\eqref{eq:qc B}-\eqref{eq:qc A} for varying $c_6$ (colour bar), with Schwarzschild (dashed) for comparison. Natural units, with the integration constant of each curve fixed by $r_\star(3M)=0$.}
    \label{fig:tortoise_qc}
\end{figure}

The Calmet--Kuipers metric derived in Chapter~\ref{chap:EFT} reads~\cite{Calmet:2019eof,Calmet:2021lny,Casadio:2025lkn}
\begin{subequations}
\label{eq:qc metric qnm}
\begin{align}
    \label{eq:qc B}
    B(r) &= 1 - \frac{2G_\text{N} M}{r} + 640\pi\, c_6\,\frac{G_\text{N}^{5} M^{3}}{r^{7}}, \\[4pt]
    \label{eq:qc A}
    A(r)^{-1} &= 1 - \frac{2G_\text{N} M}{r}
        + 3456\pi\, c_6\,\frac{G_\text{N}^{4} M^{2}}{r^{6}}\qty(1 - \frac{49\,G_\text{N} M}{27\,r}),
\end{align}
\end{subequations}
where $B(r) = -g_{tt}$ and $A(r) = g_{rr}$. In contrast to the hairy solution, here
\begin{equation}
    \label{eq:qc AB broken}
    A(r) \neq \frac{1}{B(r)}.
\end{equation}
Since Eq.~\eqref{eq:qc A} gives the inverse metric function directly, tortoise coordinates are
\begin{equation}
    \label{eq:qc f}
    \dd r_\star = \frac{\dd r}{\sqrt{B(r)\,A(r)^{-1}}}.
\end{equation}
Because $A\neq B^{-1}$, the surfaces $B=0$ and $A^{-1}=0$ do not coincide, so $1/\sqrt{B A^{-1}}$ has only an integrable singularity at the outermost surface and $r_\star$ stays finite there, instead of diverging as for Schwarzschild (Fig.~\ref{fig:tortoise_qc}).

This feature deserves comment, since it bears on the boundary condition to be imposed at the inner end of the domain. Truncated at the order at which it is derived, the metric~\eqref{eq:qc B}-\eqref{eq:qc A} describes a wormhole rather than a black hole, the outermost surface being a throat of finite proper distance instead of a horizon, and a genuine horizon is recovered only by resumming the correction to higher orders~\cite{Dubinsky:2024mwd}. This is the geometric counterpart of the statement, made in Subsec.~\ref{subsec:EFT_third_order}, that a higher-derivative truncation is to be solved perturbatively rather than exactly~\cite{Simon:1990ic,Sarkar:2017puh}. The structure of the two surfaces was worked out in Subsec.~\ref{subsec:EFT_horizon}, and two of its consequences matter here. The first is that they are displaced by the same fraction at linear order and separate only at $\order{c_6^2}$. The metric is derived only to first order in $c_6$, so it does not predict how large that gap is. The second is that both remain within a fifth of $r=2G_\text{N}M$ over the range $\abs{c_6}\leq0.01$ used for the potentials below, whereas the maximum of the barrier sits near the light ring at $r\simeq3G_\text{N}M$. The barrier that fixes the spectrum is therefore far outside the region where the distinction between throat and horizon arises, and the ingoing boundary condition is imposed at the outermost surface of the truncated geometry.

\begin{figure}[htbp]
    \centering
    \begin{subfigure}{0.49\textwidth}
        \centering\includegraphics[width=\linewidth]{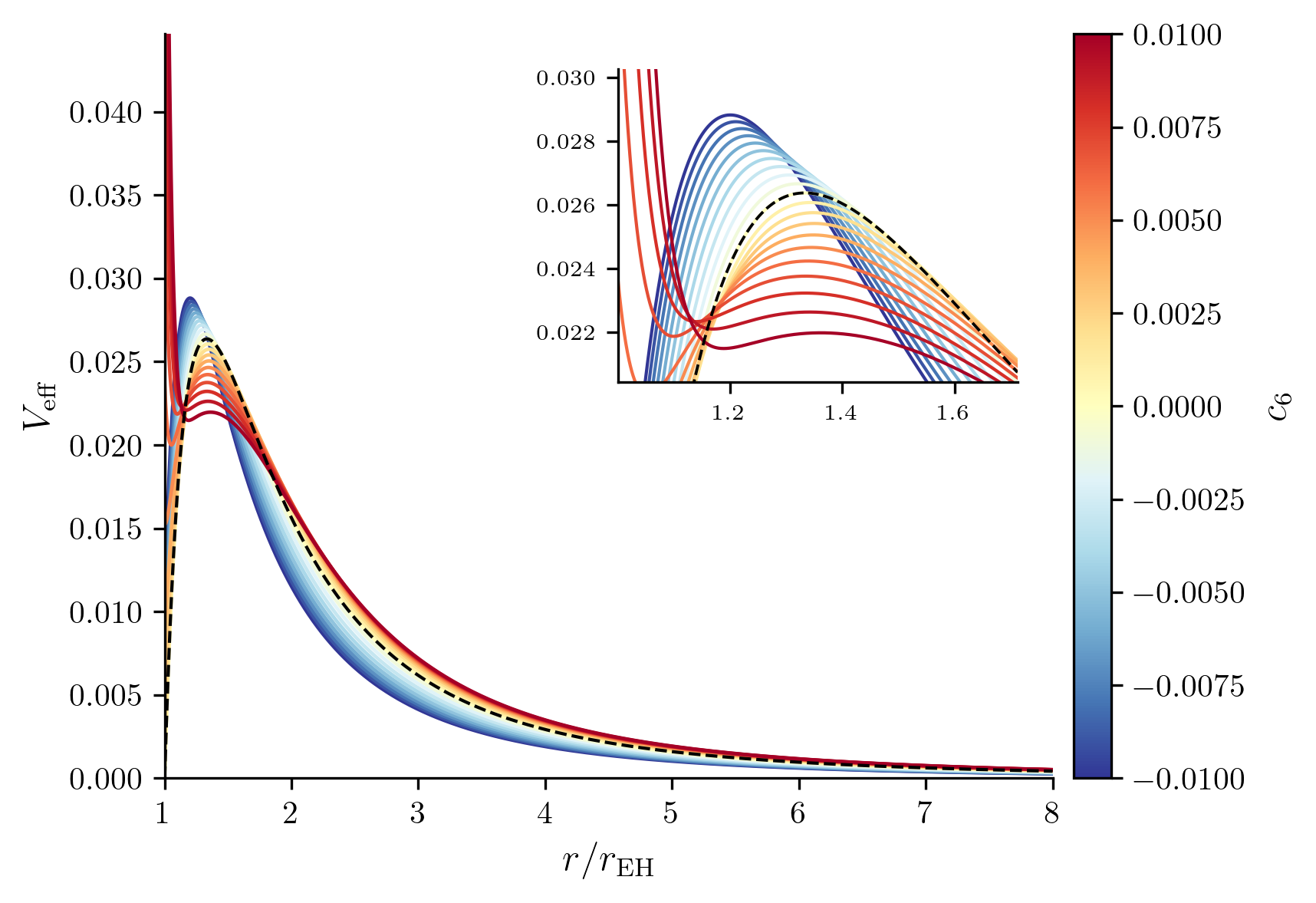}
        \caption{Scalar, $s=\ell=0$.}\label{fig:qc veff s0}
    \end{subfigure}\hfill
    \begin{subfigure}{0.49\textwidth}
        \centering\includegraphics[width=\linewidth]{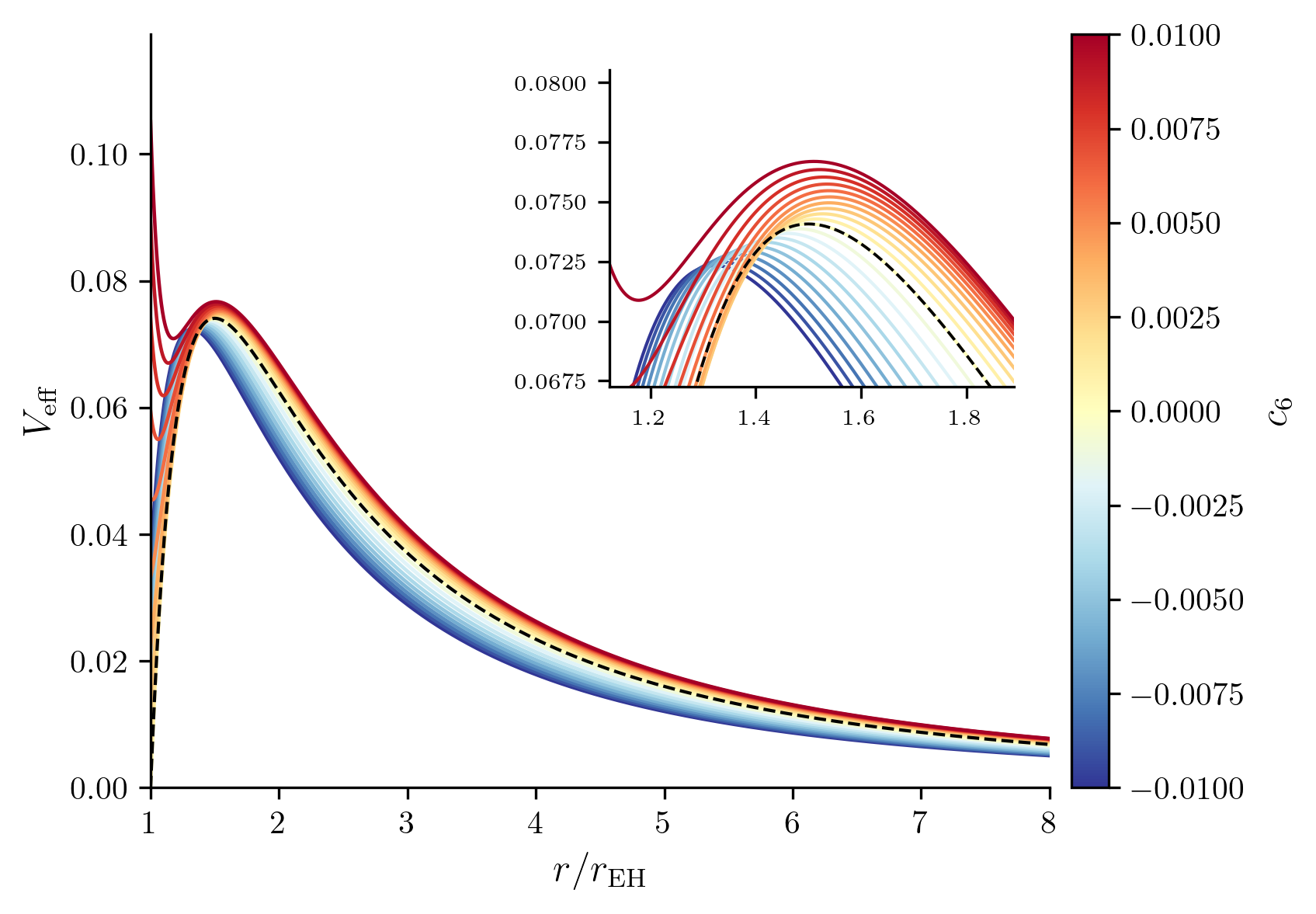}
        \caption{Electromagnetic, $s=\ell=1$.}\label{fig:qc veff s1}
    \end{subfigure}

    \begin{subfigure}{0.49\textwidth}
        \centering\includegraphics[width=\linewidth]{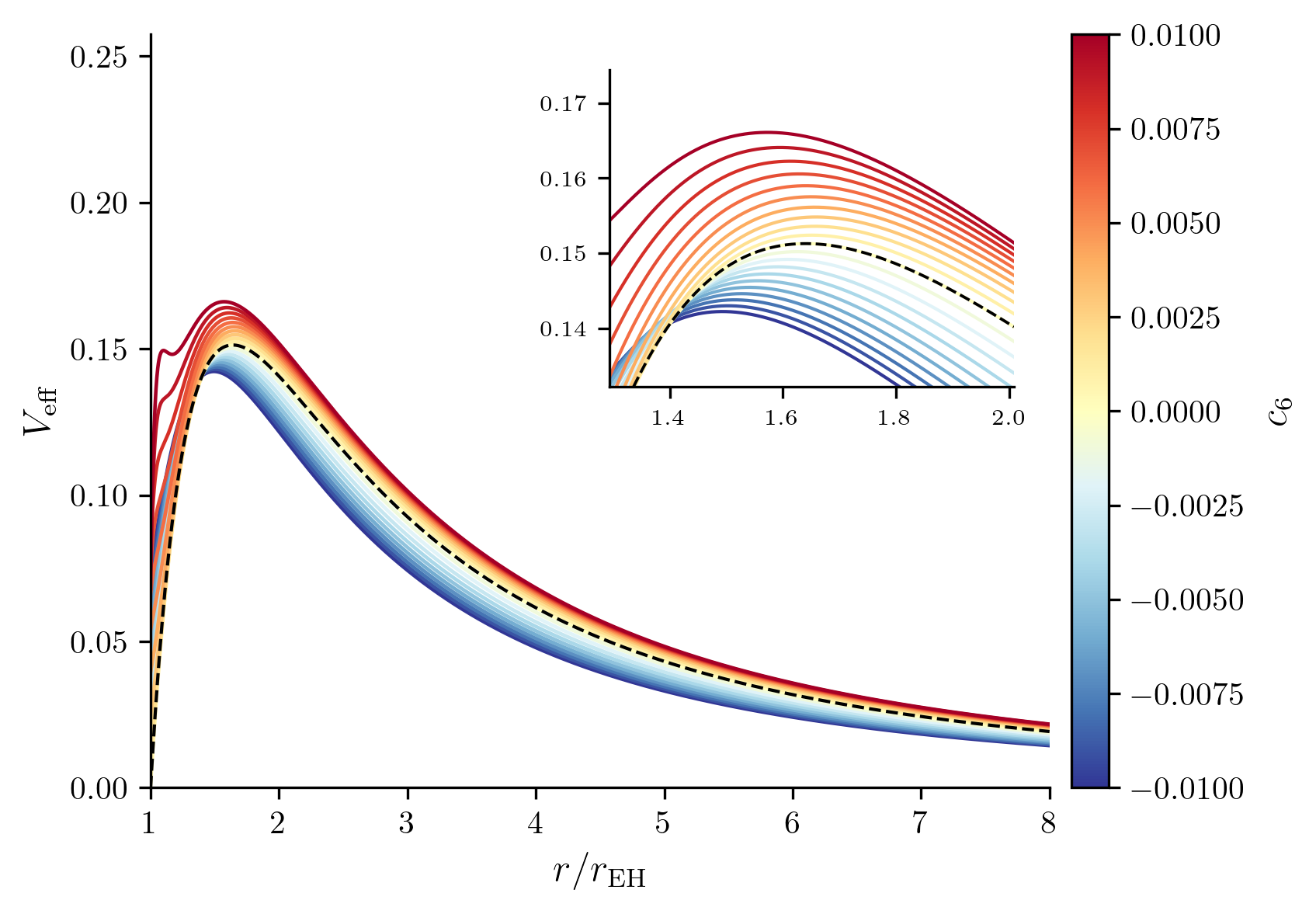}
        \caption{Gravitational, $s=\ell=2$.}\label{fig:qc veff s2}
    \end{subfigure}\hfill
    \begin{subfigure}{0.49\textwidth}
        \centering\includegraphics[width=\linewidth]{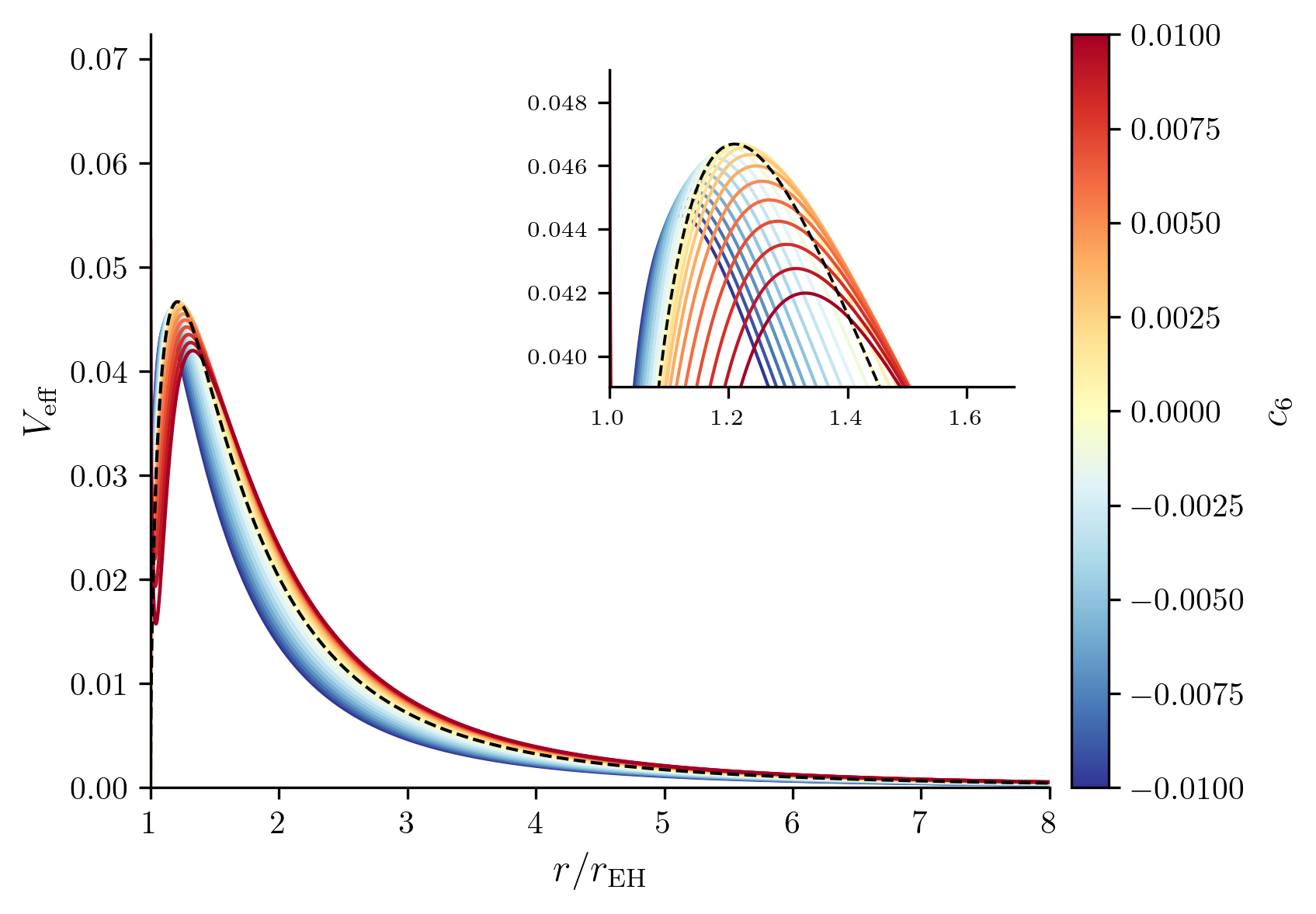}
        \caption{Dirac, $s=\ell=1/2$.}\label{fig:qc veff s12}
    \end{subfigure}
    \caption[Effective potential of the quantum-corrected black hole, four channels]{Effective potential of the quantum-corrected black hole, Eqs.~\eqref{eq:qc V expansion} and~\eqref{eq:qc fermionic}, as a function of $r/r_{\text{EH}}$, with the coupling varied (colour bar). The dashed line is the Schwarzschild potential of the same channel and the inset magnifies the maximum of the barrier. The radius is normalised to the outermost surface at which $B$ or $A^{-1}$ vanishes. Natural units, with mass $M=1$.}
    \label{fig:qc veff}
\end{figure}

One consequence is visible in the panels that follow. Above $c_6\simeq3.6\times10^{-3}G_\text{N}^2M^4$, the threshold~\eqref{eq:B_surface_exists}, the surface $B=0$ ceases to exist and only the throat at $A^{-1}=0$ remains. The blackening factor is then strictly positive at the outermost surface, so the effective potential no longer vanishes there and instead rises steeply as that surface is approached, as seen at the left edge of Fig.~\ref{fig:qc veff} for the largest positive couplings.

Because the metric functions~\eqref{eq:qc B}-\eqref{eq:qc A} are polynomials in $c_6$, so is the integer-spin potential~\eqref{final effective potential int}, which can be written exactly as~\cite{Casadio:2025lkn}
\begin{equation}
    \label{eq:qc V expansion}
    V_{\text{eff}}^{\text{bosonic}}(r) = V_0(r) + V_1(r)\,c_6 + V_2(r)\,c_6^2,
\end{equation}
with $V_0(r)$ the Schwarzschild potential,
\begin{equation}
    \label{eq:qc V0}
    V_0(r) = \qty(1-\frac{2G_\text{N} M}{r})
    \qty[\frac{\ell(\ell+1)}{r^2} + \frac{2G_\text{N} M}{r^3}\qty(1-s^2)],
\end{equation}
the first correction
\begin{equation}
    \label{eq:qc V1}
    V_1(r) = \frac{\pi G_\text{N}^4 M^2}{r^8}
    \qty[\frac{640\,G_\text{N} M}{r}\,\ell(\ell+1)
    + \qty(10368 - \frac{43904\,G_\text{N} M}{r} + \frac{45056\,G_\text{N}^2 M^2}{r^2})(s^2-1)],
\end{equation}
and the quadratic coefficient
\begin{equation}
    \label{eq:qc V2}
    V_2(r) = \frac{\pi^2 G_\text{N}^9 M^5}{r^{15}}\,(s^2-1)
    \qty(14376960 - \frac{28098560\,G_\text{N} M}{r}).
\end{equation}
The quadratic structure of Eq.~\eqref{eq:qc V expansion} shows that integer-spin fields sense the quantum correction only up to $\order{c_6^2}$. Requiring $\partial V_{\text{eff}}/\partial c_6 = 0$ then identifies, at each radius, the coupling that extremises the barrier,
\begin{equation}
    \label{eq:qc saturation}
    c_6^{\text{sat}}(r) = \left.c_6(r)\right|_{V_{\text{eff}} = V_{\text{eff}}^{\max}}  = -\left.\frac{1}{2}\,\frac{V_1(r)}{V_2(r)}\right|_\text{Integer-spin},
\end{equation}
which is called the saturation value. For spin-$1/2$, the situation is qualitatively different. Substituting the metric~\eqref{eq:qc B}-\eqref{eq:qc A} into the fermionic branch, Eq.~\eqref{final effective potential half}, yields
\begin{multline}
    \label{eq:qc fermionic}
    V_{\text{eff}}^{+}(r) = \frac{1}{r^2}\qty(\ell+\tfrac12)^2
    \qty(1 - \frac{2G_\text{N} M}{r} + 640\pi\, c_6\,\frac{G_\text{N}^5 M^3}{r^7}) \\
    - \frac{1}{r^2}\qty(\ell+\tfrac12)
    \sqrt{1 - \frac{2G_\text{N} M}{r} + 3456\pi\, c_6\,\frac{G_\text{N}^4 M^2}{r^6}\qty(1-\frac{49\,G_\text{N} M}{27\,r})} \\ \times
    \qty(1 - \frac{3G_\text{N} M}{r} + 2880\pi\, c_6\,\frac{G_\text{N}^5 M^3}{r^7}).
\end{multline}
Because of the square root, Eq.~\eqref{eq:qc fermionic} is \emph{not} a finite polynomial in $c_6$, so the Dirac sector responds differently (and, perhaps, more sensitively) to the quantum correction than the bosonic one~\cite{Casadio:2025lkn,Dubinsky:2024mwd}.

The magnitude of $c_6$ is not a free parameter either. The horizon shift~\eqref{eq:horizon_shift} bounds it on both sides, and most tightly for the lightest black hole to which the effective description still applies, that is, at the Planck scale, where the curvature at the horizon reaches the cut-off of the effective field theory~\cite{Burgess:2003jk}. Evaluated there the bound reads $\abs{c_6}<1/5\pi\approx0.0636$, Eq.~\eqref{eq:c6_bound}, comfortably within the perturbative requirement $\abs{c_6}\lesssim\order{1}$~\cite{Calmet:2021lny,Casadio:2025lkn}.

The resulting barriers are shown in Fig.~\ref{fig:qc veff}, for the same four channels used for the hairy solution, with the scan restricted to $\abs{c_6}\leq0.01$, some six times inside the bound~\eqref{eq:c6_bound}. 

Comparing Fig.~\ref{fig:qc veff} with Fig.~\ref{fig:hbh veff} shows how differently the two deformations act on the barrier. The quantum correction is short-ranged, its slowest-decaying part falling as $r^{-6}$ against the classical $r^{-1}$ term, so the curves spread out mainly around the peak, where the standard deviation of the family in the gravitational channel, measured against the Schwarzschild barrier height, reaches $5$ per cent. Further out the two families are in fact comparable, the dispersion at $r=8\,r_{\text{EH}}$ being $1.7$ per cent for the quantum family against $1.4$ per cent for the hairy one, so that the impression left by the panels, that the quantum curves merge with Schwarzschild while the hairy ones remain separated, is an effect of the vertical scale rather than a genuine difference in range. What survives at large radius is in any case largely an artefact of the normalisation, since $r_{\text{EH}}$ depends on the parameter being varied, so that curves plotted at equal $r/r_{\text{EH}}$ are evaluated at different physical radii, where the common asymptotic form $V_{\text{eff}}\to\ell(\ell+1)/r^2$ takes slightly different values. In the bosonic channels the response to $c_6$ is monotonic, a positive coupling lowering the barrier and a negative one raising it. The Dirac channel behaves differently, both signs of $c_6$ lowering the peak with respect to Schwarzschild.

\begin{figure}[htbp]
    \centering
    \includegraphics[width=0.8\linewidth]{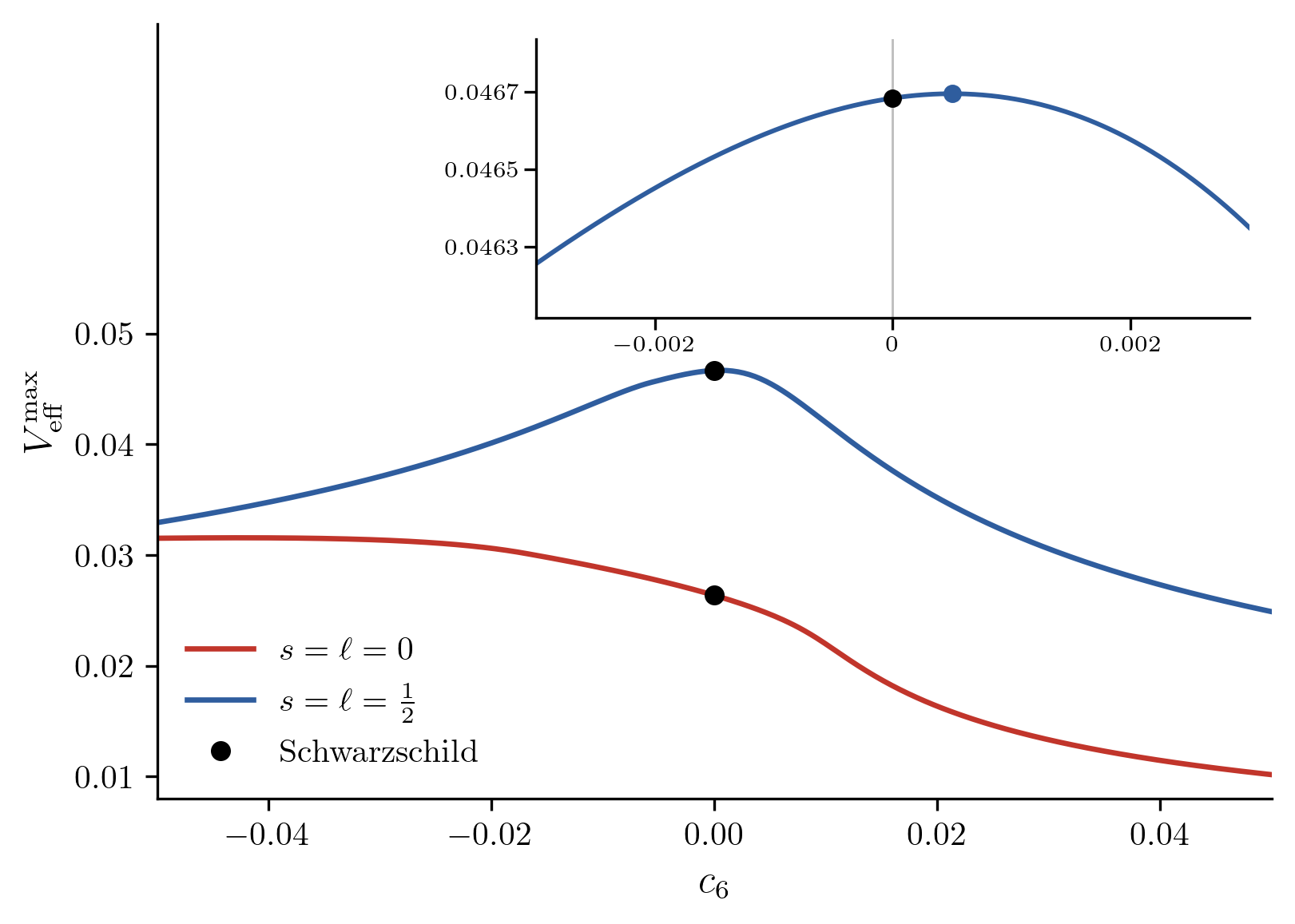}
    \caption[Barrier height against the coupling, scalar and Dirac channels]{Height $V_{\text{eff}}^{\max}$ of the barrier near the light ring against the coupling $c_6$, for the scalar ($s=\ell=0$) and Dirac ($s=\ell=1/2$) channels. The black markers are the Schwarzschild values, both at $c_6=0$ but at different heights, $0.0264$ in the scalar channel and $0.0467$ in the Dirac one. The inset magnifies $\abs{c_6}\leq3\times10^{-3}$, where the Dirac maximum separates from its marker, exceeding the Schwarzschild value by $1.2\times10^{-5}$. Natural units, $M=1$.}
    \label{fig:qc satVeff}
\end{figure}

\begin{figure}[htbp]
    \centering
    \includegraphics[width=0.8\linewidth]{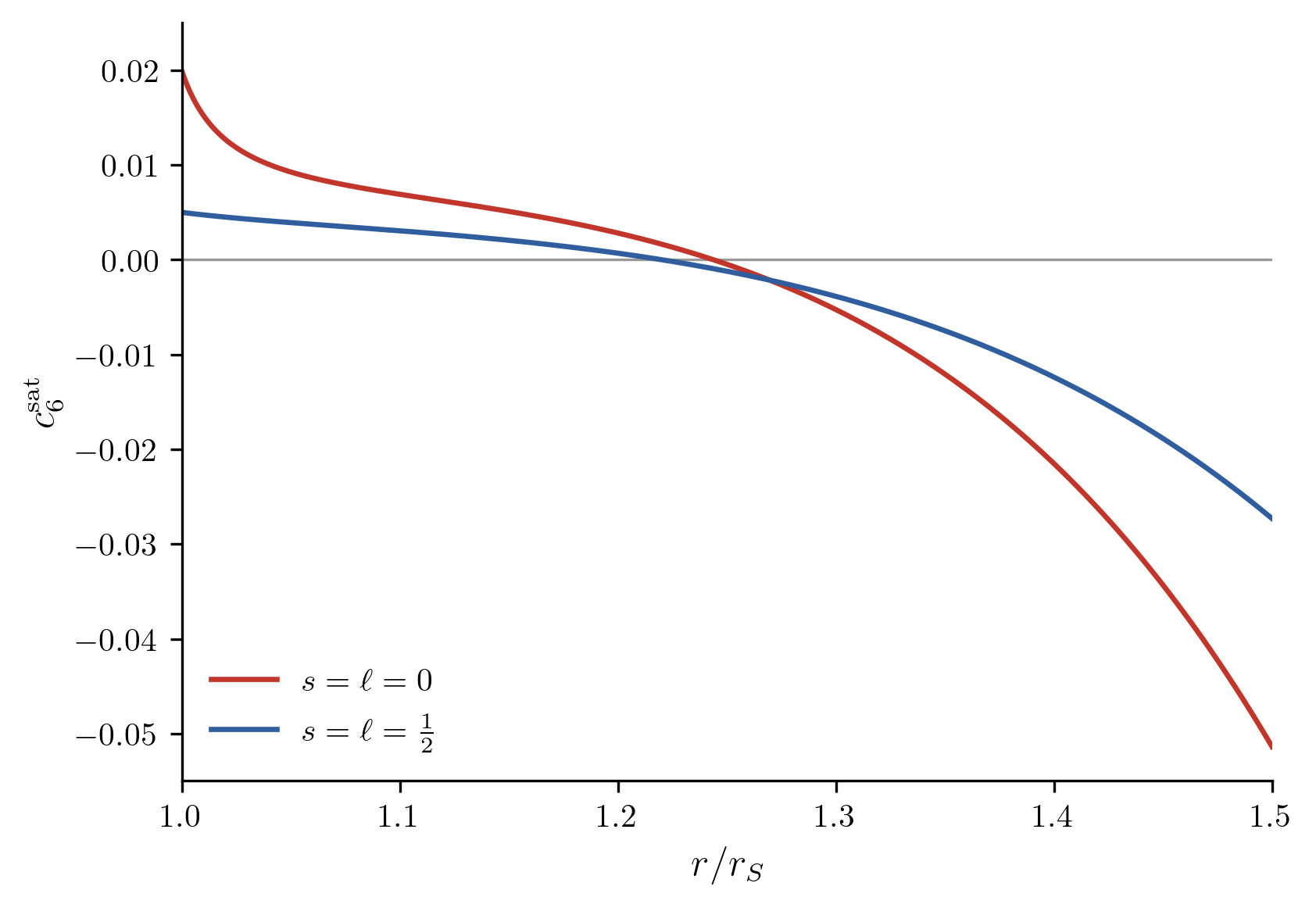}
    \caption[Saturation value of the coupling against the radius]{Saturation value $c_6^{\text{sat}}$ of Eq.~\eqref{eq:qc saturation}, that is, the coupling that extremises the effective potential at each radius, as a function of $r/r_S$ with $r_S = 2G_\text{N} M$, for the scalar ($s=\ell=0$) and Dirac ($s=\ell=1/2$) channels.}
    \label{fig:qc satc6}
\end{figure}

Two quantities summarise how the barrier responds to the coupling. The first is its height $V_{\text{eff}}^{\max}$, Eq.~\eqref{eq:peak data}, which together with the curvature defined there governs the transmission of a wave through the barrier, and hence the damped oscillations it supports~\cite{Schutz:1985km,Iyer:1986np}. The second is the saturation value $c_6^{\text{sat}}(r)$, Eq.~\eqref{eq:qc saturation}, the coupling that extremises the potential at a given radius. Figure~\ref{fig:qc satVeff} plots the height against $c_6$ for the scalar and Dirac channels, and Fig.~\ref{fig:qc satc6} the saturation value against $r/r_S$ for the same two, with $r_S = 2G_\text{N} M$ the Schwarzschild radius. Both scan $c_6$ over $[-0.05,\,0.05]$, in natural units with $M=1$~\cite{Casadio:2025lkn}.

The integer-spin (bosonic) branch is monotonic in $c_6$, in the sense that $V_{\text{eff}}^{\max}$ decreases as $c_6$ increases, with a minimum global slope of $\approx -0.69$ near $c_6\approx0.011$. Over the admissible range it develops no well-defined global extremum, one appearing only for values of $c_6$ far outside the bound~\eqref{eq:c6_bound}. The spin-$1/2$ (fermionic) branch is instead non-monotonic, since $V_{\text{eff}}^{\max}$ increases for $c_6\lesssim0$ and decreases for $c_6\gtrsim0$, more steeply than in the bosonic case, with extreme slopes $\approx -0.78$ at $c_6\approx0.010$ and $\approx 0.42$ at $c_6\approx-0.011$. Its global maximum sits at $c_6\approx5\times10^{-4}$, only marginally displaced from Schwarzschild. Spin-$1/2$ particles therefore experience the maximum attraction essentially at $c_6=0$, and respond to a small quantum perturbation more rapidly than bosons do~\cite{Casadio:2025lkn}.

Figure~\ref{fig:qc satc6} adds two points. Both branches change sign of $c_6^{\text{sat}}$ just inside the light ring, the Dirac one at $r/r_S\approx1.22$ and the scalar one at $r/r_S\approx1.24$, and they cross shortly after, at $r/r_S\approx1.27$. Except within that narrow window, the Dirac saturation value is the smaller of the two in magnitude, by a factor of two to four, so that a weaker coupling suffices to extremise the fermionic barrier. A small positive $c_6$ raises the potential inside the sign change and lowers it outside, the barrier itself sitting on the outer side.

Eq.~\eqref{eq:horizon_shift} also makes the mass dependence of the correction explicit, since the fractional shift of the horizon, and hence of the barrier that sits just outside it, scales as $M^{-4}$. The quantum term is thus fully negligible for astrophysical black holes and matters only near the Planck mass, so that any observable imprint of $c_6$ on the ringdown would be confined to primordial black holes~\cite{Hawking:1971ei,Carr:1974nx,Carr:2020gox}, the only population whose masses can reach that regime. A comparable suppression is found in other higher-derivative extensions of GR, where the corrections to the quasinormal spectrum are likewise controlled by the ratio between the scale of the new operators and the curvature scale of the horizon~\cite{Cardoso:2018ptl,Cano:2021myl,Cano:2023jbk,Husken:2026axq}.

\section{Summary and Outlook}
\label{sec:veff comparison}

The two backgrounds deform Schwarzschild in complementary ways, and this difference propagates into the effective potential. For the hairy black hole the relation $A=B^{-1}$ is preserved, so the tortoise coordinate keeps its Schwarzschild form $\dd r_\star=\dd r/B$ and the potentials~\eqref{eq:hbh veff int explicit}-\eqref{eq:hbh veff half explicit} depend on $B(r)$ alone. The resulting deformation is macroscopic, controlled by $\alpha$, $\ell_h$, and $Q$. For the quantum-corrected black hole the relation $A=B^{-1}$ is broken, so that $B$ and $A^{-1}$ enter the potential independently, and the deformation is instead short-ranged and governed by the single parameter $c_6$. Both geometries are carried into Parts~\ref{part: analogue models} and~\ref{part: results}, where the effective potentials derived here are reproduced by the acoustic response of a de~Laval nozzle~\cite{Visser:1997ux,Barcelo:2005fc,daRocha:2017lqj,Casadio:2024uwj,Casadio:2025lkn}. The quasinormal frequencies and overtones associated with these potentials are computed in the next chapter.

\chapter{Quasinormal modes, eigenfrequencies, and overtones}
\label{chap:QNM_spectrum}

\epigraph{``Never send to know for whom the bell tolls, it tolls for thee.''}{\emph{Devotions upon Emergent Occasions} \\ John Donne}

Chapter~\ref{chap:conexaoentrebnsebocais} reduced the problem of a black-hole perturbation to a single Schr\"odinger-type wave equation, governed by an effective-potential barrier. What is actually observed in a ringdown, however, is not the barrier itself but the frequencies at which the geometry oscillates in response to it. These form a discrete set of complex numbers, and they are fixed by two conditions. At the horizon the radiation can only go in, and at infinity it can only go out. These are the quasinormal frequencies, and finding them is a non-Hermitian eigenvalue problem for which no closed-form solution exists for the barriers considered here~\cite{Kokkotas:1999bd,Berti:2009kk,Konoplya:2011qq}. The spectrum is instead approximated throughout this chapter by the Mashhoon method~\cite{Ferrari:1984zz}, presented in Sec.~\ref{sec:mashhoon}.

The method is then applied to the two backgrounds constructed in Part~\ref{part: extensions of gr}, the hairy black hole from gravitational decoupling in Sec.~\ref{sec:QNM_hairy} and the quantum-corrected Schwarzschild black hole in Sec.~\ref{sec:QNM_QC}. Particular attention is paid to the higher overtones, since they are the most sensitive probes of the shape of the barrier, and therefore of the underlying geometry. Sec.~\ref{sec:QNM_boundary} opens the chapter by defining the quasinormal spectrum and the boundary conditions that select it, and Sec.~\ref{sec:QNM_discussion} closes it by comparing the two spectra obtained.

\section{The quasinormal spectrum and its boundary conditions}
\label{sec:QNM_boundary}

As established in Chapter~\ref{chap:conexaoentrebnsebocais}, a field perturbation of spin $s$ and multipole number $\ell$ on a background $\dd s^2 = -B(r)\dd t^2 + A(r)\dd r^2 + r^2\dd\Omega^2$ obeys, after separation of variables and the introduction of the tortoise coordinate~\eqref{coordenada tartaruga final}, the master equation
\begin{equation}
    \label{eq:qnm_master}
    \qty[\dv[2]{}{r_\star} + \omega^2 - V_{\text{eff}}(r)]R(r_\star) = 0,
\end{equation}
with the time dependence taken as $e^{-i\omega t}$. The effective potential is a barrier, vanishing at the horizon ($r_\star\to-\infty$) and at spatial infinity ($r_\star\to+\infty$), with a single maximum in between. Quasinormal modes are the solutions of Eq.~\eqref{eq:qnm_master} that are purely ingoing at the horizon and purely outgoing at infinity,
\begin{equation}
    \label{eq:qnm_bc}
    R(r_\star) \sim
    \begin{cases}
        e^{-i\omega r_\star}, & r_\star\to-\infty \quad(\text{horizon}),\\[4pt]
        e^{+i\omega r_\star}, & r_\star\to+\infty \quad(\text{infinity}).
    \end{cases}
\end{equation}
These conditions, no radiation entering from infinity and nothing emerging from the horizon, turn Eq.~\eqref{eq:qnm_master} into a non-Hermitian eigenvalue problem, whose spectrum is a discrete, infinite set of complex frequencies~\cite{Kokkotas:1999bd,Berti:2009kk,Konoplya:2011qq}. Following the convention of Refs.~\cite{Casadio:2024uwj,Casadio:2025lkn}, they are written as
\begin{equation}
    \label{eq:qnm_freq}
    \omega_n = \omega_{R,n} + i\,\omega_{I,n},
    \qquad n = 0,1,2,\ldots,
\end{equation}
where the integer $n$ is the overtone number, ordered by increasing $\omega_{I,n}$, with $n=0$ the fundamental mode. The real part $\omega_{R}$ is the oscillation frequency. The imaginary part is reported throughout as a damping rate $\omega_{I}$, so that the amplitude of the mode falls off as $e^{-\omega_{I}t}$ and the damping time is $\tau = 1/\omega_{I}$. Its sign carries physical content, since a mode with $\omega_{I}<0$ would grow without bound and the background would be unstable against the very perturbation that revealed it. The condition $\omega_{I}>0$ is therefore the statement that the geometry is stable at linear order, and every frequency obtained in this chapter satisfies it, in both backgrounds and over the whole admissible range of their parameters. A convenient dimensionless measure of how many cycles the signal completes before it decays is the quality factor
\begin{equation}
    \label{eq:quality_factor}
    q_n \equiv \frac{\omega_{R,n}}{\omega_{I,n}}.
\end{equation}
The fundamental mode dominates the late-time ringdown, but the early part of the signal (the interval immediately after the merger, before the fundamental mode sets in) is controlled by the tower of overtones~\cite{Buonanno:2006ui,Giesler:2019uxc,Casadio:2025lkn}. A complete characterisation of the ringdown therefore requires the overtones, and it is precisely there that the departures from the Schwarzschild spectrum are most pronounced, which is also what makes them the natural target of single-event spectroscopy~\cite{Isi:2019aib,Berti:2015itd}.

\section{The Mashhoon method}
\label{sec:mashhoon}

Because the quasinormal boundary conditions~\eqref{eq:qnm_bc} render the problem non-Hermitian, the eigenfrequencies cannot be obtained by elementary means, and a range of techniques has been developed, direct integration, continued fractions~\cite{Leaver:1985ax}, and the WKB expansion~\cite{Schutz:1985km,Iyer:1986np}, among others~\cite{Kokkotas:1999bd,Nollert:1999ji,Berti:2009kk,Konoplya:2011qq}. A recurring obstacle is the slow decay of the effective potential at large $r$, which introduces a branch cut and produces backscattered tails, the late-time power-law relaxation first identified in Refs.~\cite{Price:1971fb,Price:1972pw}. The approach adopted in Refs.~\cite{Casadio:2024uwj,Casadio:2025lkn}, and followed here, is the Mashhoon method, which sidesteps that difficulty by working only with the local shape of the barrier at its peak.

The physical motivation for doing so traces back to the light ring itself. A photon perturbed off the unstable circular orbit at the peak of $V_{\text{eff}}$ does not return to it, diverging away exponentially in time, at a rate fixed by the curvature of the potential at the peak, the Lyapunov exponent that governs the instability of the orbit and, in the eikonal limit, the quasinormal frequencies themselves~\cite{Cardoso:2008bp}. The wave problem shares the same local structure, since Eq.~\eqref{eq:qnm_master} is a stationary Schr\"odinger-type equation with $V_{\text{eff}}$ playing the role of the barrier. A wave packet briefly held near the peak leaks away towards the horizon and towards infinity, at a rate again set by that same curvature, and it oscillates at a frequency set by the height of the barrier. It is therefore the local, quadratic behaviour of the barrier at its maximum, located at $r_{\star,\text{peak}}$ in the tortoise coordinate,
\begin{equation}
    \label{eq:peak_quadratic}
    V_{\text{eff}}(r_\star) \approx V_{\text{eff}}^{\max} + \frac{1}{2}\left.\dv[2]{V_{\text{eff}}}{r_\star}\right|_{r_\star = r_{\star,\text{peak}}}(r_\star - r_{\star,\text{peak}})^2,
\end{equation}
rather than the detailed shape of the barrier far from the peak, that controls the least-damped part of the spectrum, the modes probed here. A bare inverted parabola, however, cannot itself be substituted into the boundary-value problem, since it diverges rather than vanishing as $r_\star\to\pm\infty$, as required by Eq.~\eqref{eq:qnm_bc}. What is needed is a trial potential that reproduces the same peak and curvature while still decaying to zero on both sides, and that remains exactly solvable, so that its spectrum can be written down rather than computed again for every choice of parameters. The P\"oschl--Teller profile,
\begin{equation}
    \label{eq:PT_potential}
    V_{\text{PT}}(r_\star) = \frac{V_{\text{eff}}^{\max}}{\cosh^2\!\qty[\varkappa\,(r_\star - r_{\star,\text{peak}})]},
\end{equation}
meets both requirements. Expanded about $r_{\star,\text{peak}}$ it reduces to exactly the parabola of Eq.~\eqref{eq:peak_quadratic}, while asymptotically it falls off exponentially like the true barrier, and it is one of the few potentials for which the wave equation~\eqref{eq:qnm_master} is solvable in closed form~\cite{Kokkotas:1999bd,Berti:2009kk,Konoplya:2011qq}. This is the substitution proposed by Ferrari and Mashhoon~\cite{Ferrari:1984zz,Ferrari:2007dd}, whose accuracy against exact methods has been assessed for several backgrounds~\cite{Castello-Branco:2004rzk,Cuyubamba:2013iwa}, and the construction proceeds in three steps. First, the single maximum $r_{\star,\text{peak}}$ of $V_{\text{eff}}(r_\star)$ is located and the barrier is Taylor-expanded there to quadratic order, Eq.~\eqref{eq:peak_quadratic}, the linear term vanishing by definition of a maximum. Second, the trial profile~\eqref{eq:PT_potential} is expanded in the same way, and demanding that the two quadratic expansions agree fixes its two free parameters, the height $V_{\text{eff}}^{\max}$ and the inverse width
\begin{equation}
    \label{eq:PT_kappa}
    \varkappa = \sqrt{-\,\frac{1}{2V_{\text{eff}}^{\max}}\,
        \left.\dv[2]{V_{\text{eff}}}{r_\star}\right|_{r_\star = r_{\star,\text{peak}}}},
\end{equation}
both defined in Eq.~\eqref{eq:peak data}, with the second derivative negative at a maximum so that $\varkappa$ is real. Third, because $V_{\text{PT}}$ is exactly solvable, its quasinormal spectrum is known in closed form, and evaluating it at the $V_{\text{eff}}^{\max}$ and $\varkappa$ just fixed gives the eigenfrequencies of the true barrier without any further calculation~\cite{Casadio:2024uwj,Casadio:2025lkn},
\begin{equation}
    \label{eq:PT_omega}
    \omega_n = \pm\sqrt{V_{\text{eff}}^{\max} - \frac{\varkappa^2}{4}} + i\,\varkappa\qty(n + \frac{1}{2}),
    \qquad n = 0,1,2,\ldots.
\end{equation}
Two features of Eq.~\eqref{eq:PT_omega} are relevant. First, the real part is common to all overtones at fixed $(s,\ell)$, being set by $V_{\text{eff}}^{\max}$ and $\varkappa$, whereas the imaginary part grows linearly in $n$, so the higher overtones are increasingly short-lived and increasingly sensitive to the shape of the barrier. Second, the quality factor~\eqref{eq:quality_factor} follows in closed form,
\begin{equation}
    \label{eq:PT_quality}
    q_n = \frac{\operatorname{Re}(\omega_n)}{\operatorname{Im}(\omega_n)}
        = \frac{1}{2n+1}\sqrt{\frac{4V_{\text{eff}}^{\max}}{\varkappa^2} - 1},
\end{equation}
which decreases with the overtone number. The method is exact when the barrier is genuinely of P\"oschl--Teller form and is accurate for sharply peaked potentials, and its reliability for the fundamental mode and the low overtones was checked in Ref.~\cite{Casadio:2025lkn}.

\begin{figure}[htbp]
    \centering
    \begin{subfigure}{0.49\textwidth}
        \centering\includegraphics[width=\linewidth]{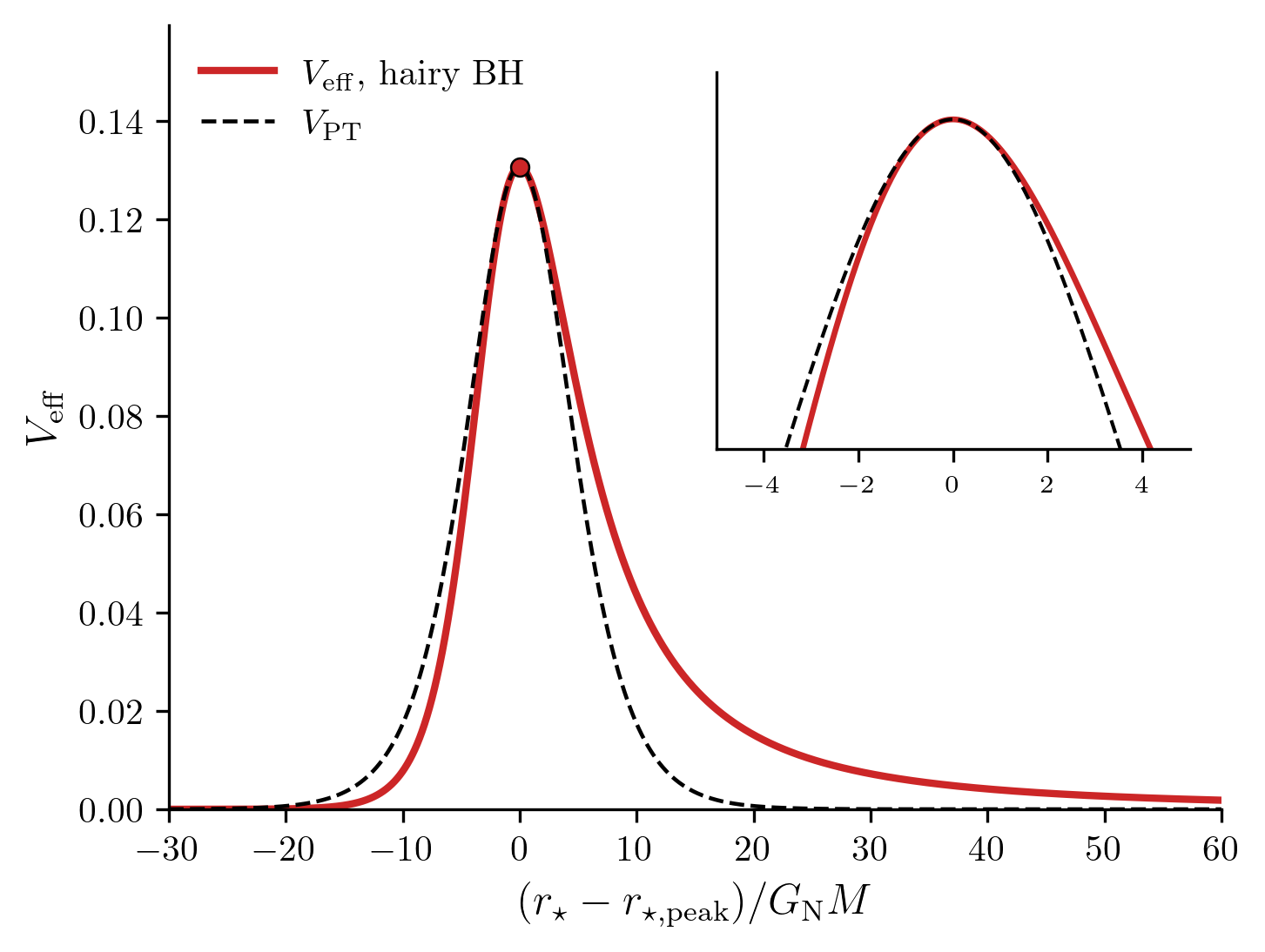}
        \caption{Eq.~\eqref{eq:hbh metric qnm}, $\alpha=\ell_h=0.5$ and $Q=0.6$.}\label{fig:veff_vs_pt_hbh}
    \end{subfigure}\hfill
    \begin{subfigure}{0.49\textwidth}
        \centering\includegraphics[width=\linewidth]{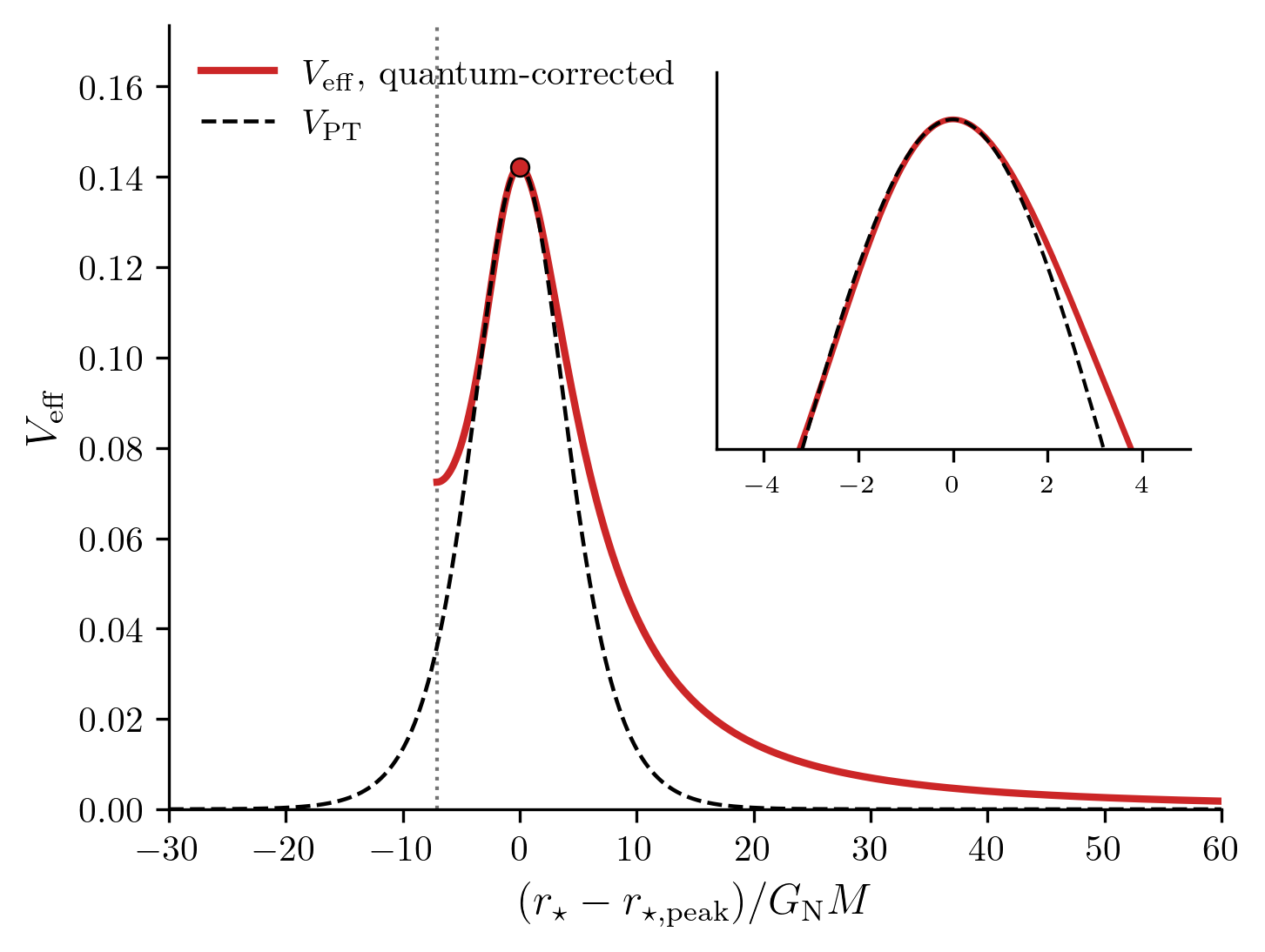}
        \caption{Eq.~\eqref{eq:qc metric qnm}, $c_6=-0.01$.}\label{fig:veff_vs_pt_qc}
    \end{subfigure}
    \caption[Exact barrier against the P\"oschl--Teller fit]{Exact effective potential (solid) and the P\"oschl--Teller profile~\eqref{eq:PT_potential} fitted to it (dashed), in the gravitational channel $s=\ell=2$, against the tortoise coordinate measured from the peak. Panel~\subref{fig:veff_vs_pt_hbh} is the hairy metric~\eqref{eq:hbh metric qnm} at $\alpha=\ell_h=0.5$ and $Q=0.6$; panel~\subref{fig:veff_vs_pt_qc} the quantum-corrected metric~\eqref{eq:qc metric qnm} at $c_6=-0.01$, the dotted line marking the outermost surface of the truncated geometry. The markers indicate the maxima and the insets magnify them.}
    \label{fig:veff_vs_pt}
\end{figure}

Figure~\ref{fig:veff_vs_pt} shows what the substitution amounts to for the two backgrounds treated in this chapter. Each panel draws the exact barrier against the P\"oschl--Teller profile fitted to it through Eqs.~\eqref{eq:PT_potential} and~\eqref{eq:PT_kappa}, at a single representative choice of parameters, since the purpose here is to display the quality of the fit rather than its dependence on the background. The insets confirm that the two curves are indistinguishable over the region around the maximum, which is what the construction imposes and what fixes the spectrum through Eq.~\eqref{eq:PT_omega}. Further out they separate, the exact barrier decaying as a power law towards infinity while the fitted profile decays exponentially. The discrepancy is therefore confined to the region responsible for the late-time tails, and not to the one that sets the modes computed here. In the quantum-corrected panel the exact potential stops at the dotted line, the outermost surface of the truncated geometry, where the tortoise coordinate stays finite and the barrier does not vanish because $B$ and $A^{-1}$ have different roots, as discussed in Subsec.~\ref{subsec:veff quantum}.

The main advantage of the Mashhoon method in the present context is its closed-form spectrum rather than its absolute accuracy. Leaver's continued-fraction method~\cite{Leaver:1985ax} and direct integration are formally exact, but return one frequency at a time, so every point of a parameter sweep such as Fig.~\ref{fig:qnm_hairy} demands a fresh calculation. Direct integration is also unstable at large $n$. The WKB expansion is cheaper, but its accuracy falls with the overtone number~\cite{Schutz:1985km,Iyer:1986np}, the opposite of what the overtones studied here require. Equation~\eqref{eq:PT_omega} instead gives every overtone algebraically, anywhere in the admissible domain, at the cost of a few per cent on the fundamental mode~\cite{Casadio:2025lkn}. Neither background is read perturbatively here for the first time, which matters for the same reason. The quasinormal spectrum of hairy black holes from gravitational decoupling has been computed by several groups~\cite{Yang:2022ifo,Cavalcanti:2022cga,dePaiva:2025eux,Navarro:2026lrf}, and the quantum-corrected Schwarzschild metric has been treated in the same way~\cite{Dubinsky:2024mwd}, so the frequencies tabulated below can be read against an existing body of results obtained by other routes.

The size of the error can be stated for the grid used here. Measured against Leaver's continued fraction on a Schwarzschild background, the fundamental mode of Eq.~\eqref{eq:PT_omega} lies within $4\%$ of the exact real part and within $10\%$ of the exact imaginary part at every integer spin and multipole of Tables~\ref{tab:QNM_GD} and~\ref{tab:QNM_QC}. The first overtone costs between $5\%$ and $33\%$ of the real part, and the second between $13\%$ and $52\%$, the departure growing with $n$ and falling as $\ell$ rises. The reason is visible in Eq.~\eqref{eq:PT_omega}, whose real part is the same for every overtone while the exact one decreases with $n$, so the error accumulates overtone by overtone and the approximation holds where $n<\ell$, the condition also quoted for the WKB expansion~\cite{Schutz:1985km,Iyer:1986np} and consistent with the assessments of Refs.~\cite{Castello-Branco:2004rzk,Cuyubamba:2013iwa}. Appendix~\ref{chap:qnm methods} extends that measurement to the two exotic backgrounds, comparing the Mashhoon method against direct integration and against the WKB expansion on both, and reports where the three cease to agree. What this chapter reports, however, is not the frequency itself but its shift away from the Schwarzschild value at fixed $(s,\ell,n)$, computed by the same method on both backgrounds, so the figures quoted above bound the absolute frequencies rather than those shifts. Both families are anchored to that same reference geometry. The quantum-corrected metric returns to Schwarzschild as $c_6\to0$, and the hairy metric, with the coupling switched off at finite $Q$, returns to the Reissner--Nordstr\"om solution~\eqref{eq:sol numb 1 sat}, whose exact spectrum is known~\cite{Leaver:1990zz} and which is itself Schwarzschild at $Q=0$.

\section{Hairy black hole from gravitational decoupling}
\label{sec:QNM_hairy}

For the hairy black hole of the gravitational-decoupling chapter, Eq.~\eqref{eq: metric HBH}, the Schwarzschild-like relation $A = B^{-1}$ holds, so the tortoise coordinate reduces to $\dd r_\star = \dd r/B$ and the effective potential depends on $B(r)$ alone, as in Chapter~\ref{chap:conexaoentrebnsebocais}. (The hair parameter is written as $\ell_{h}$ here to avoid confusion with the multipole number $\ell$. Recall that $\ell_h$ itself carries dimensions of mass, $G_\text{N}\ell_h$ being the associated length.) The quasinormal frequencies again follow from Eqs.~\eqref{eq:PT_kappa}-\eqref{eq:PT_omega}, now over the admissible domain of the hair parameters $(\alpha,\ell_{h},Q)$ mapped in Chapter~\ref{chap:GD}. Table~\ref{tab:QNM_GD} lists the spectrum, computed with $G_\text{N}=M=1$, for two representative hair configurations, $\alpha=\ell_h=0.1$ and $\alpha=\ell_h=1$, each evaluated at the two ends of the admissible charge range, namely the dominant-energy-condition threshold, Eq.~\eqref{eq: limit Q alpha M}, and the horizon-existence bound $Q_\text{max}$ read off Fig.~\ref{fig:maxQwithrEH}. The Schwarzschild column is the $\alpha=\ell_h=Q=0$ limit.

The trends in Table~\ref{tab:QNM_GD} follow those reported in Ref.~\cite{Casadio:2024uwj}. At fixed multipole $\ell$ and overtone $n$, increasing the decoupling parameter $\alpha$ lowers both the real and the imaginary parts of the frequency, a slower oscillation with a longer damping time. At fixed hair, the real part grows with $\ell$ and the imaginary part grows with the overtone $n$, exactly as for the quantum-corrected case below.

\begin{table}[htbp]
\centering
\caption{Quasinormal-mode frequencies $\omega_n$ for the hairy black hole from gravitational decoupling~\eqref{eq: metric HBH}, for $\alpha=\ell_h=0.1$ and $\alpha=\ell_h=1$, each at the two ends of the admissible charge range, namely the dominant-energy-condition threshold $Q_{\rm DEC}=2\sqrt{\alpha}\,G_\text{N}M/e$~\eqref{eq: limit Q alpha M} and the horizon-existence bound $Q_{\rm max}$ of Fig.~\ref{fig:maxQwithrEH}. The Schwarzschild column is the $\alpha=\ell_h=Q=0$ limit. Computed with $G_\text{N}=M=1$ by the Mashhoon method. Overtones marked with $\dagger$ have $n\geq\ell$ and therefore lie outside the range in which the method is accurate, quantified in Sec.~\ref{sec:mashhoon}. The fundamental mode stays within $4\%$ of the exact real part at every integer spin and multipole.}
\label{tab:QNM_GD}
\footnotesize
\setlength{\tabcolsep}{3.5pt}
\renewcommand{\arraystretch}{1.2}
\begin{tabular}{||c|c|c||c|c|c|c|c||}
\hline\hline
 & & & & \multicolumn{2}{c|}{$\alpha=\ell_h=0.1$} & \multicolumn{2}{c||}{$\alpha=\ell_h=1$} \\
\cline{5-8}
$s$ & $\ell$ & \, $n$ \, & Schwarzschild & $Q_{\rm DEC}\approx0.2327$ & $Q_{\rm max}\approx1.0231$ & $Q_{\rm DEC}\approx0.7358$ & $Q_{\rm max}\approx1.6086$ \\
\hline\hline
$0$ & $0$ & $0$ & $0.1148+0.1148i$ & $0.1148+0.1151i$ & $0.1396+0.1089i$ & $0.0790+0.0788i$ & $0.0963+0.0752i$ \\ \hline
$0$ & $0$ & $1^{\dagger}$ & $0.1148+0.3445i$ & $0.1148+0.3453i$ & $0.1396+0.3268i$ & $0.0790+0.2365i$ & $0.0963+0.2256i$ \\ \hline
$0$ & $0$ & $2^{\dagger}$ & $0.1148+0.5741i$ & $0.1148+0.5755i$ & $0.1396+0.5447i$ & $0.0790+0.3941i$ & $0.0963+0.3759i$ \\ \hline
$1/2$ & $1/2$ & $0$ & $0.1890+0.1048i$ & $0.1892+0.1050i$ & $0.2458+0.0935i$ & $0.1310+0.0721i$ & $0.1711+0.0652i$ \\ \hline
$1/2$ & $1/2$ & $1^{\dagger}$ & $0.1890+0.3143i$ & $0.1892+0.3149i$ & $0.2458+0.2806i$ & $0.1310+0.2163i$ & $0.1711+0.1957i$ \\ \hline
$1/2$ & $1/2$ & $2^{\dagger}$ & $0.1890+0.5238i$ & $0.1892+0.5249i$ & $0.2458+0.4677i$ & $0.1310+0.3605i$ & $0.1711+0.3261i$ \\ \hline
$0$ & $1$ & $0$ & $0.2985+0.1006i$ & $0.2990+0.1008i$ & $0.3842+0.0932i$ & $0.2065+0.0690i$ & $0.2669+0.0647i$ \\ \hline
$0$ & $1$ & $1^{\dagger}$ & $0.2985+0.3019i$ & $0.2990+0.3025i$ & $0.3842+0.2797i$ & $0.2065+0.2071i$ & $0.2669+0.1940i$ \\ \hline
$0$ & $1$ & $2^{\dagger}$ & $0.2985+0.5032i$ & $0.2990+0.5042i$ & $0.3842+0.4662i$ & $0.2065+0.3452i$ & $0.2669+0.3234i$ \\ \hline
$1$ & $1$ & $0$ & $0.2546+0.0962i$ & $0.2550+0.0964i$ & $0.3422+0.0889i$ & $0.1765+0.0661i$ & $0.2380+0.0618i$ \\ \hline
$1$ & $1$ & $1^{\dagger}$ & $0.2546+0.2887i$ & $0.2550+0.2892i$ & $0.3422+0.2666i$ & $0.1765+0.1983i$ & $0.2380+0.1854i$ \\ \hline
$1$ & $1$ & $2^{\dagger}$ & $0.2546+0.4811i$ & $0.2550+0.4820i$ & $0.3422+0.4444i$ & $0.1765+0.3306i$ & $0.2380+0.3091i$ \\ \hline
$1/2$ & $3/2$ & $0$ & $0.3855+0.0991i$ & $0.3861+0.0993i$ & $0.4992+0.0906i$ & $0.2670+0.0681i$ & $0.3474+0.0631i$ \\ \hline
$1/2$ & $3/2$ & $1$ & $0.3855+0.2972i$ & $0.3861+0.2978i$ & $0.4992+0.2719i$ & $0.2670+0.2042i$ & $0.3474+0.1892i$ \\ \hline
$1/2$ & $3/2$ & $2^{\dagger}$ & $0.3855+0.4954i$ & $0.3861+0.4963i$ & $0.4992+0.4532i$ & $0.2670+0.3404i$ & $0.3474+0.3154i$ \\ \hline
$0$ & $2$ & $0$ & $0.4874+0.0979i$ & $0.4881+0.0981i$ & $0.6309+0.0906i$ & $0.3374+0.0672i$ & $0.4387+0.0629i$ \\ \hline
$0$ & $2$ & $1$ & $0.4874+0.2937i$ & $0.4881+0.2942i$ & $0.6309+0.2717i$ & $0.3374+0.2016i$ & $0.4387+0.1888i$ \\ \hline
$0$ & $2$ & $2^{\dagger}$ & $0.4874+0.4895i$ & $0.4881+0.4904i$ & $0.6309+0.4529i$ & $0.3374+0.3360i$ & $0.4387+0.3146i$ \\ \hline
$1$ & $2$ & $0$ & $0.4615+0.0962i$ & $0.4622+0.0964i$ & $0.6058+0.0889i$ & $0.3197+0.0661i$ & $0.4215+0.0618i$ \\ \hline
$1$ & $2$ & $1$ & $0.4615+0.2887i$ & $0.4622+0.2892i$ & $0.6058+0.2666i$ & $0.3197+0.1983i$ & $0.4215+0.1854i$ \\ \hline
$1$ & $2$ & $2^{\dagger}$ & $0.4615+0.4811i$ & $0.4622+0.4820i$ & $0.6058+0.4444i$ & $0.3197+0.3306i$ & $0.4215+0.3091i$ \\ \hline
$2$ & $2$ & $0$ & $0.3783+0.0905i$ & $0.3790+0.0907i$ & $0.5237+0.0811i$ & $0.2630+0.0625i$ & $0.3653+0.0568i$ \\ \hline
$2$ & $2$ & $1$ & $0.3783+0.2716i$ & $0.3790+0.2722i$ & $0.5237+0.2432i$ & $0.2630+0.1874i$ & $0.3653+0.1703i$ \\ \hline
$2$ & $2$ & $2^{\dagger}$ & $0.3783+0.4527i$ & $0.3790+0.4536i$ & $0.5237+0.4053i$ & $0.2630+0.3124i$ & $0.3653+0.2839i$ \\ \hline
$1/2$ & $5/2$ & $0$ & $0.5779+0.0975i$ & $0.5788+0.0976i$ & $0.7497+0.0897i$ & $0.4002+0.0670i$ & $0.5216+0.0624i$ \\ \hline
$1/2$ & $5/2$ & $1$ & $0.5779+0.2924i$ & $0.5788+0.2929i$ & $0.7497+0.2692i$ & $0.4002+0.2009i$ & $0.5216+0.1873i$ \\ \hline
$1/2$ & $5/2$ & $2$ & $0.5779+0.4873i$ & $0.5788+0.4882i$ & $0.7497+0.4487i$ & $0.4002+0.3348i$ & $0.5216+0.3121i$ \\ \hline
\hline
\end{tabular}
\end{table}

The hair leaves a larger imprint on the spectrum than a first estimate would suggest. In Eq.~\eqref{eq: metric HBH} the exponential term decays over the length $G_\text{N}M$ fixed by the bare mass $M=\mathcal{M}-\alpha\ell_h/2$, and not by the total mass $\mathcal{M}$, whatever the values of $\alpha$ and $\ell_h$. That length is comparable to the light-ring radius, which is where the barrier peaks and where the spectrum is fixed, so the exponential suppression of the hair at the peak is only mild ($e^{-r_\text{peak}/G_\text{N}M} \sim e^{-3} = 0.05$). Table~\ref{tab:QNM_GD} makes the consequence explicit. At $\alpha=\ell_h=0.1$, with the charge held at $Q=Q_\text{DEC}$, the hair changes $\omega_R$ by less than one per cent relative to Schwarzschild. At $\alpha=\ell_h=1$ it lowers $\omega_R$ by close to $30\%$ already at $Q=Q_\text{DEC}$, and the suppression is strong enough to keep $\omega_R$ below its Schwarzschild value over the whole admissible range, including at $Q_\text{max}$, in every channel listed.

\begin{figure}[htbp]
    \centering
    \begin{subfigure}{0.48\textwidth}
        \centering\includegraphics[width=\linewidth]{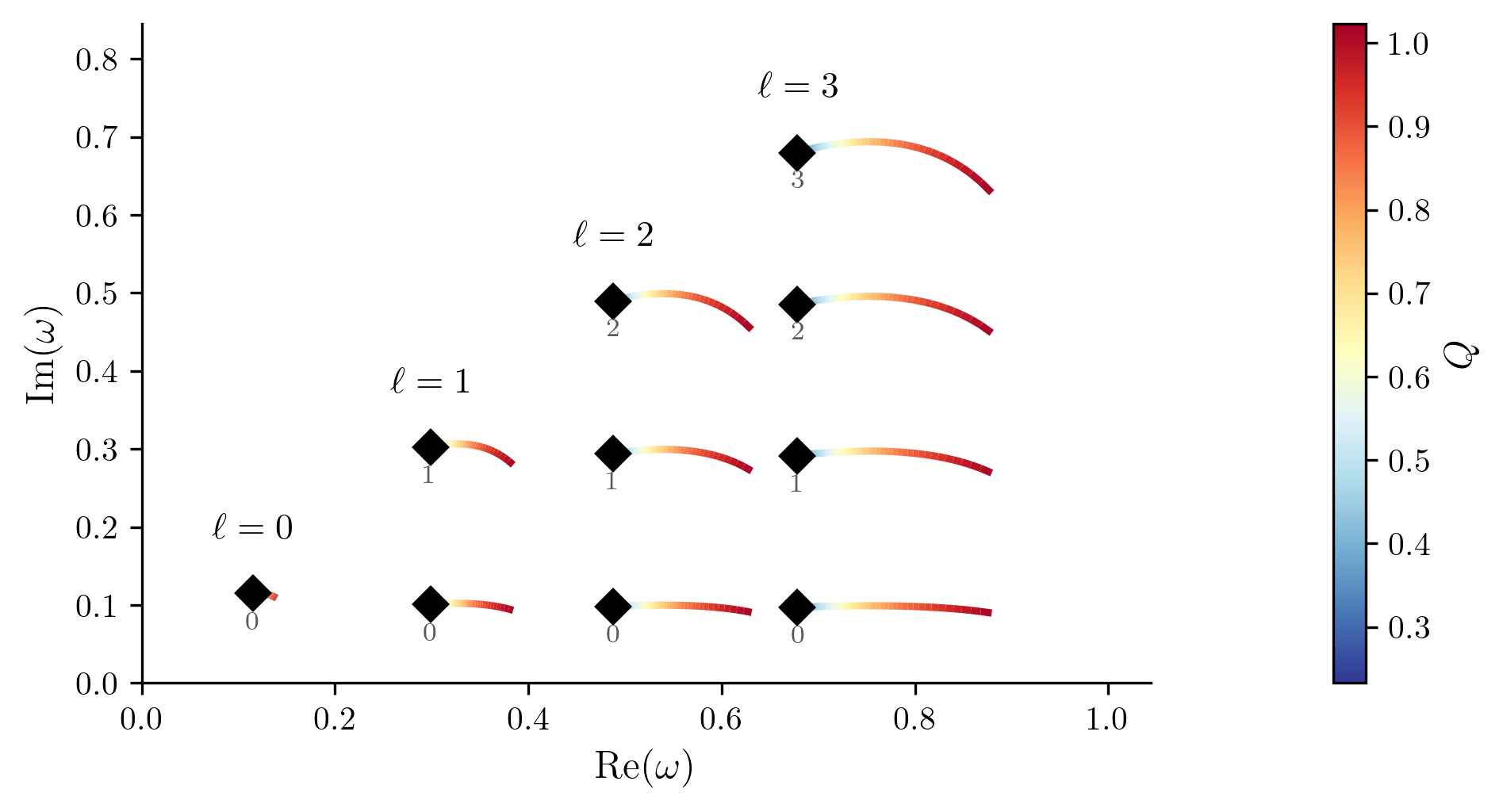}
        \caption{$s=0$, $\alpha=\ell_h=0.1$}\label{fig:qnm_hbh_s0a}
    \end{subfigure}\hfill
    \begin{subfigure}{0.48\textwidth}
        \centering\includegraphics[width=\linewidth]{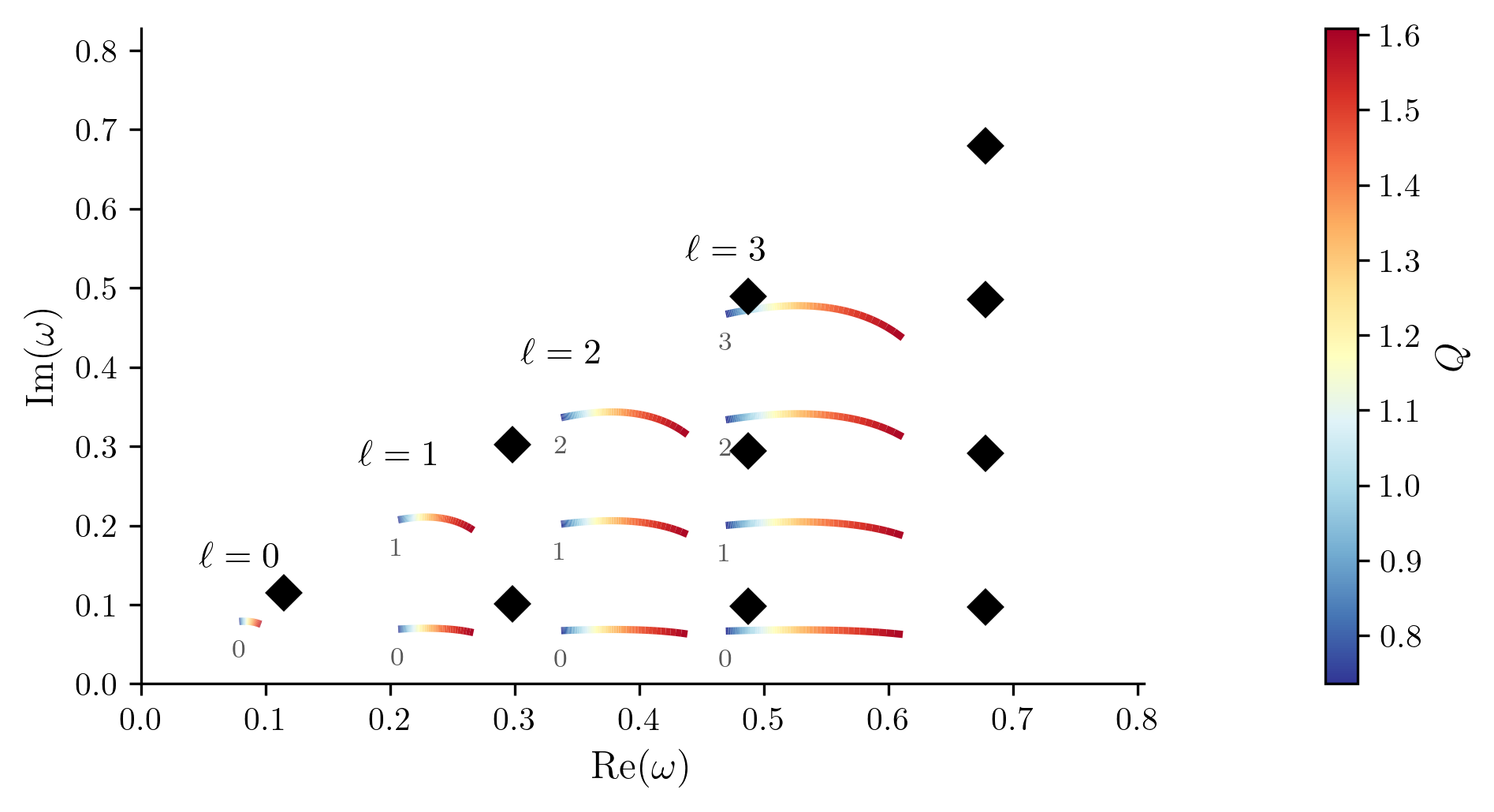}
        \caption{$s=0$, $\alpha=\ell_h=1$}\label{fig:qnm_hbh_s0b}
    \end{subfigure}

    \begin{subfigure}{0.48\textwidth}
        \centering\includegraphics[width=\linewidth]{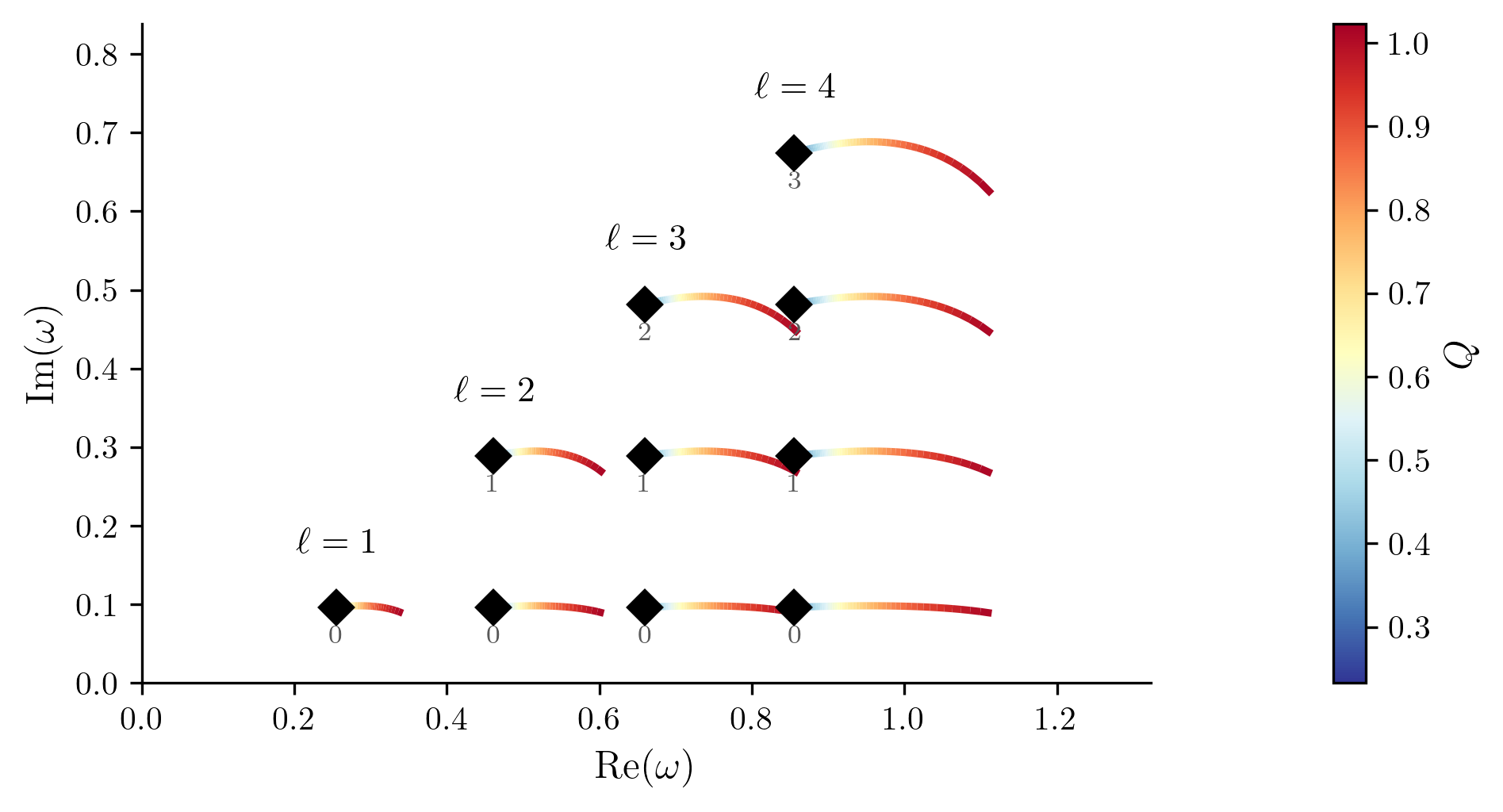}
        \caption{$s=1$, $\alpha=\ell_h=0.1$}\label{fig:qnm_hbh_s1a}
    \end{subfigure}\hfill
    \begin{subfigure}{0.48\textwidth}
        \centering\includegraphics[width=\linewidth]{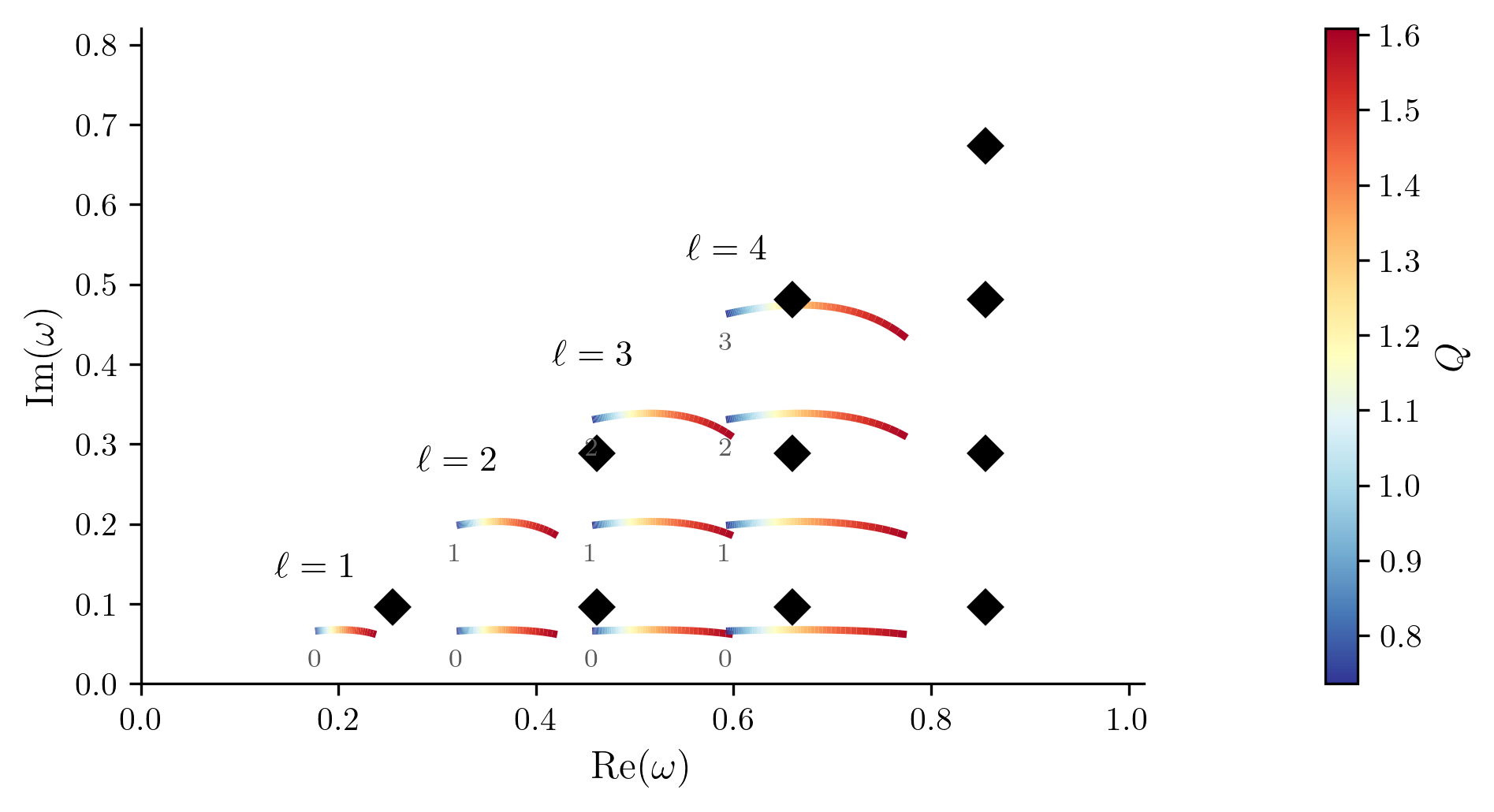}
        \caption{$s=1$, $\alpha=\ell_h=1$}\label{fig:qnm_hbh_s1b}
    \end{subfigure}

    \begin{subfigure}{0.48\textwidth}
        \centering\includegraphics[width=\linewidth]{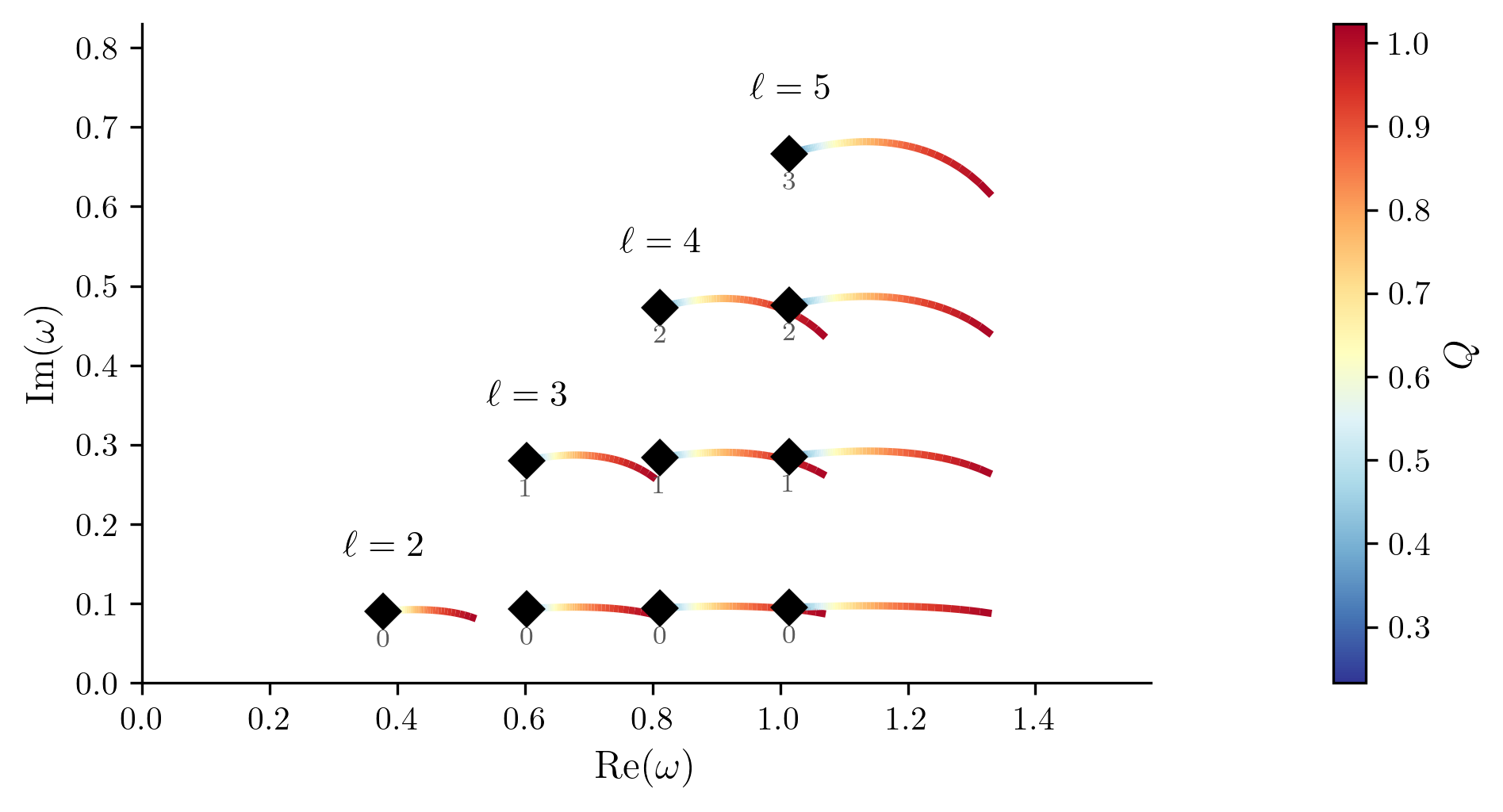}
        \caption{$s=2$, $\alpha=\ell_h=0.1$}\label{fig:qnm_hbh_s2a}
    \end{subfigure}\hfill
    \begin{subfigure}{0.48\textwidth}
        \centering\includegraphics[width=\linewidth]{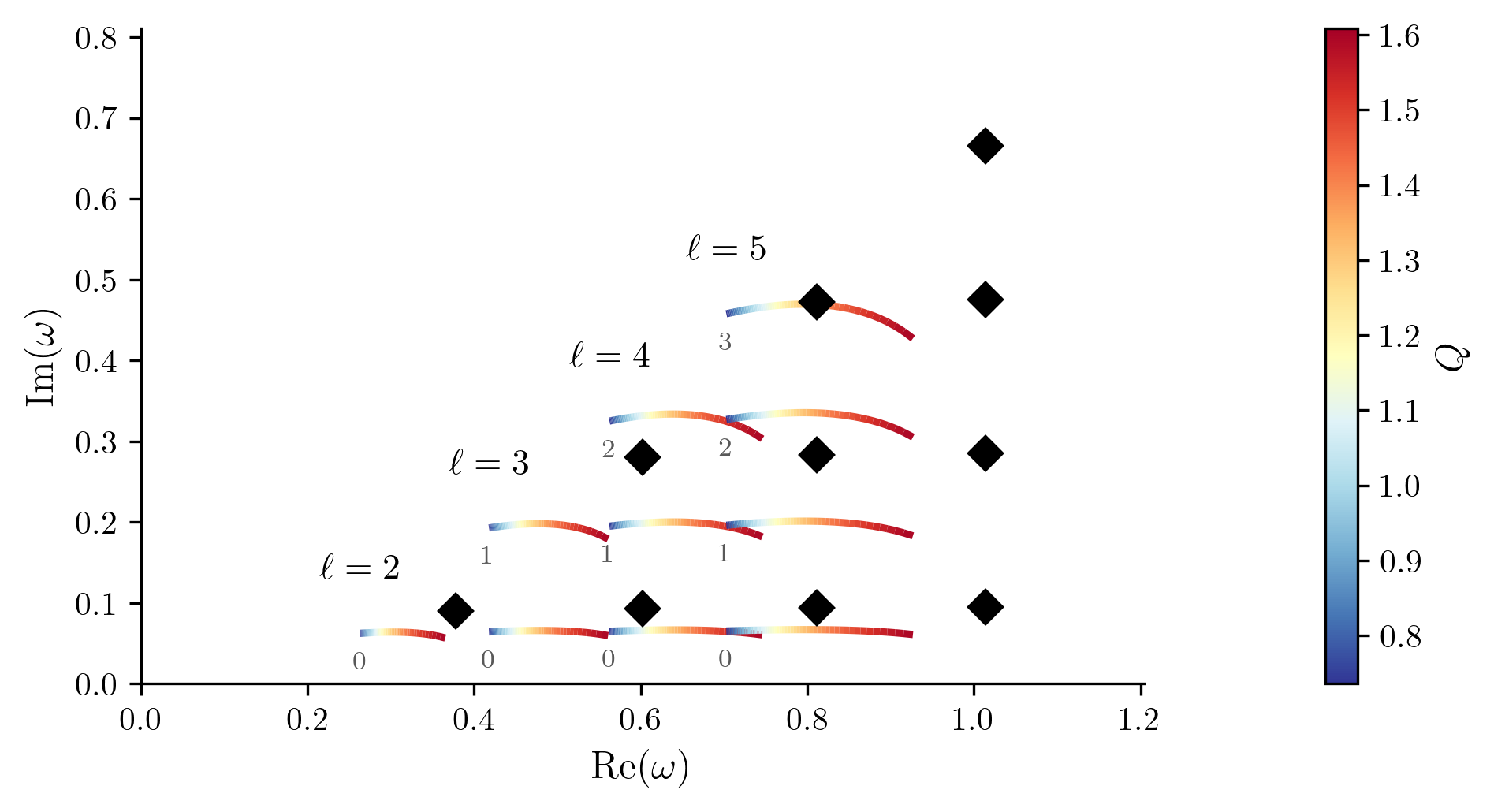}
        \caption{$s=2$, $\alpha=\ell_h=1$}\label{fig:qnm_hbh_s2b}
    \end{subfigure}

    \begin{subfigure}{0.48\textwidth}
        \centering\includegraphics[width=\linewidth]{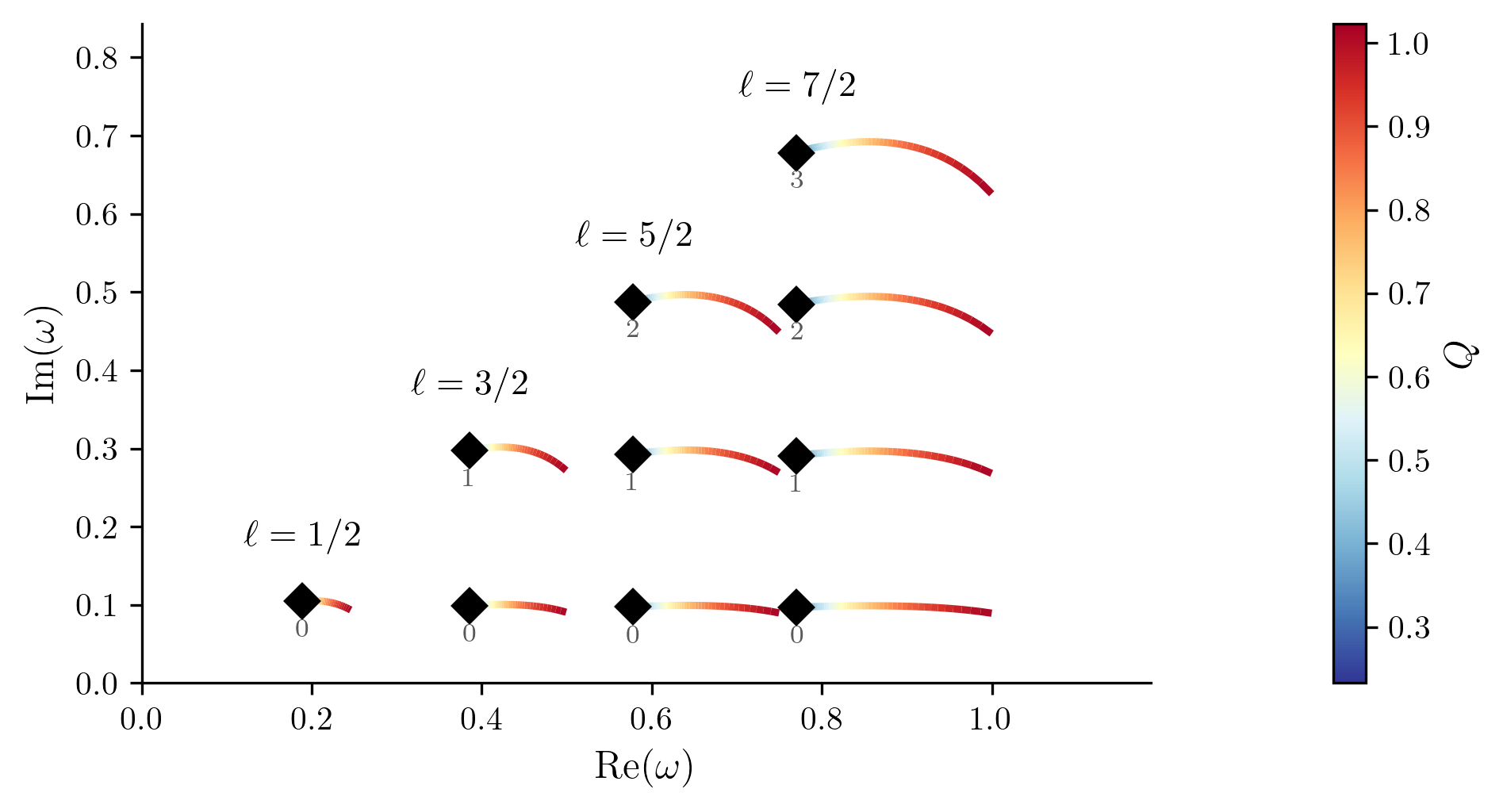}
        \caption{$s=1/2$, $\alpha=\ell_h=0.1$}\label{fig:qnm_hbh_fa}
    \end{subfigure}\hfill
    \begin{subfigure}{0.48\textwidth}
        \centering\includegraphics[width=\linewidth]{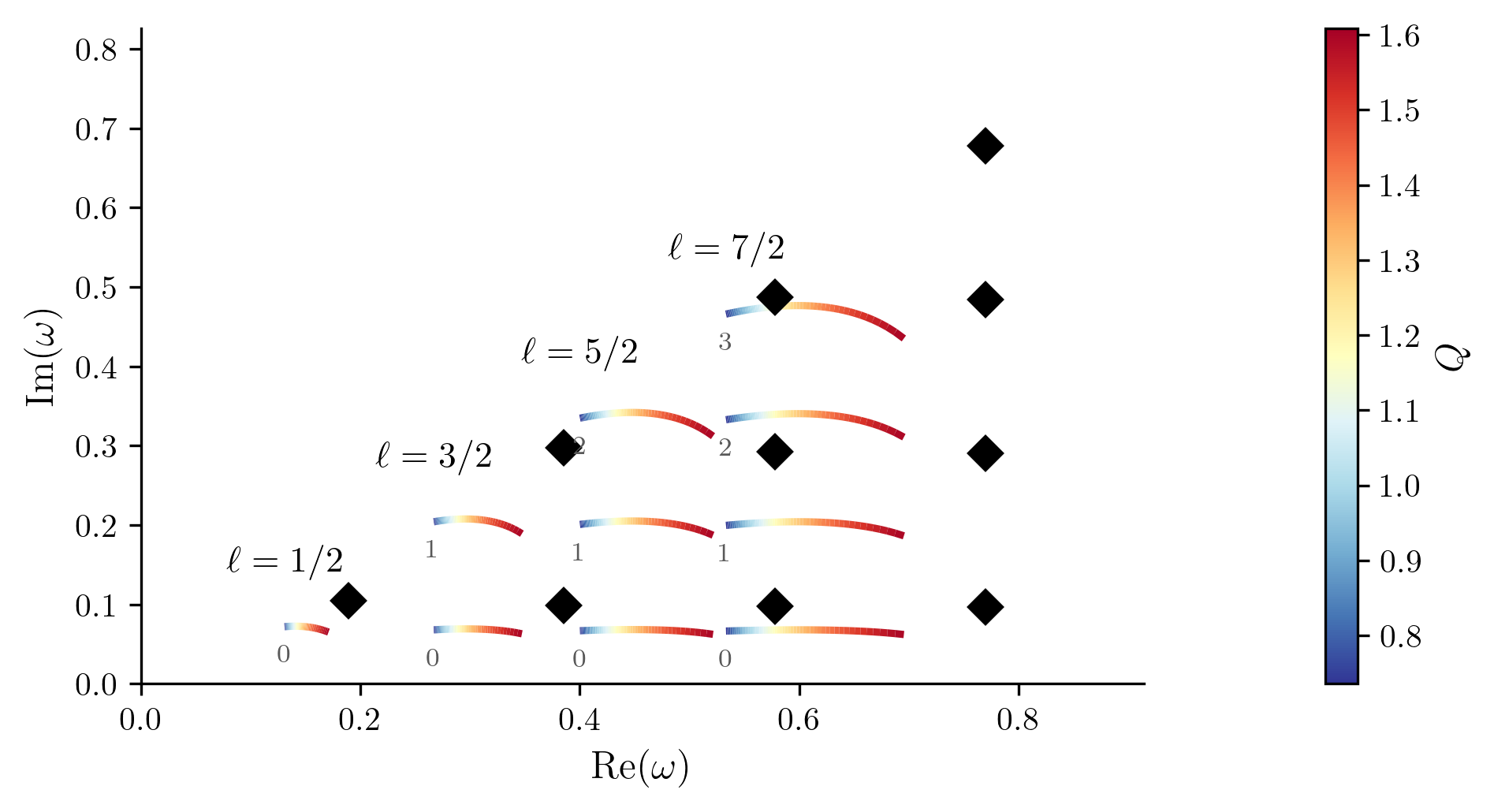}
        \caption{$s=1/2$, $\alpha=\ell_h=1$}\label{fig:qnm_hbh_fb}
    \end{subfigure}
    \caption[Quasinormal spectra of the hairy black hole]{Quasinormal spectra of the hairy black hole~\eqref{eq: metric HBH} for bosonic fields ($s=0,1,2$, top three rows) and the Dirac field ($s=1/2$, bottom row), computed by the Mashhoon method with $G_\text{N}=M=1$. Left column $\alpha=\ell_h=0.1$, right column $\alpha=\ell_h=1$, with the charge swept (colour bar). Black diamonds mark the Schwarzschild values and the numeral next to each curve is its overtone $n$.}
    \label{fig:qnm_hairy}
\end{figure}

The charge-like parameter $Q$ acts in the opposite direction. It enters through a $Q^2/r^2$ term that is not exponentially suppressed, so it raises the barrier, and with it $\omega_R$, over the entire range. The form of that term is the one familiar from the Reissner--Nordstr\"om solution, but the analogy stops there, because $Q$ cannot be switched on independently of the hair. Equation~\eqref{eq: Q vanishes with alpha} shows that $Q$ vanishes with $\alpha$, both being carried by the same integration constant of the deformation, so a configuration with $Q\neq0$ at $\alpha=0$ falls outside the family constructed in Chapter~\ref{chap:GD} and would require relaxing the conditions imposed there. Within the admissible domain the two effects compete, and Fig.~\ref{fig:qnm_hairy} shows how the competition is resolved.

The quasinormal spectra computed here by the Mashhoon method are shown in Fig.~\ref{fig:qnm_hairy} (bosonic fields, $s=0,1,2$, and the Dirac field, $s=1/2$), for the same two hair configurations of Table~\ref{tab:QNM_GD}, with the charge-like parameter swept over the admissible range $Q\in[Q_\text{DEC},\,Q_\text{max}]$ (colour bar), for four multipoles $\ell=s,\ldots,s+3$ and, for each $\ell$, overtones $n=0,\ldots,\min(3,\ell-s)$. The black diamonds mark the Schwarzschild values and the small numeral next to the start of each curve is its overtone $n$. Increasing $Q$ displaces the modes towards larger $\omega_R$, as the $Q^2/r^2$ term alone would do. For $\alpha=\ell_h=0.1$ the curves stay close to their Schwarzschild diamond. For $\alpha=\ell_h=1$ the whole tower is instead pulled to the left and downwards, towards smaller $\omega_R$ and smaller $\omega_I$, by enough that curves belonging to different $(\ell,n)$ pairs cross and overlap.
Table~\ref{tab:QNM_GD} puts numbers on that displacement. In the gravitational channel $s=\ell=2$ at $\alpha=\ell_h=1$ and $Q=Q_\text{DEC}$, the fundamental mode moves from the Schwarzschild value $0.3783+0.0905i$ to $0.2630+0.0625i$, a reduction of $30.5\%$ in $\omega_R$ and of $30.9\%$ in $\omega_I$. Equation~\eqref{eq:PT_omega} makes $\omega_R$ independent of $n$ and $\omega_I$ proportional to $n+1/2$, so the same fractional reduction applies to every overtone and the absolute displacement grows linearly with $n$. The $n=2$ mode of that channel falls from $0.3783+0.4527i$ to $0.2630+0.3124i$, moving five times further down the imaginary axis than the fundamental mode does. This is why the high-overtone sector is the sharpest probe of the hair once the coupling approaches order unity.

\begin{figure}[htbp]
    \centering
    \begin{subfigure}{0.48\textwidth}
        \centering\includegraphics[width=\linewidth]{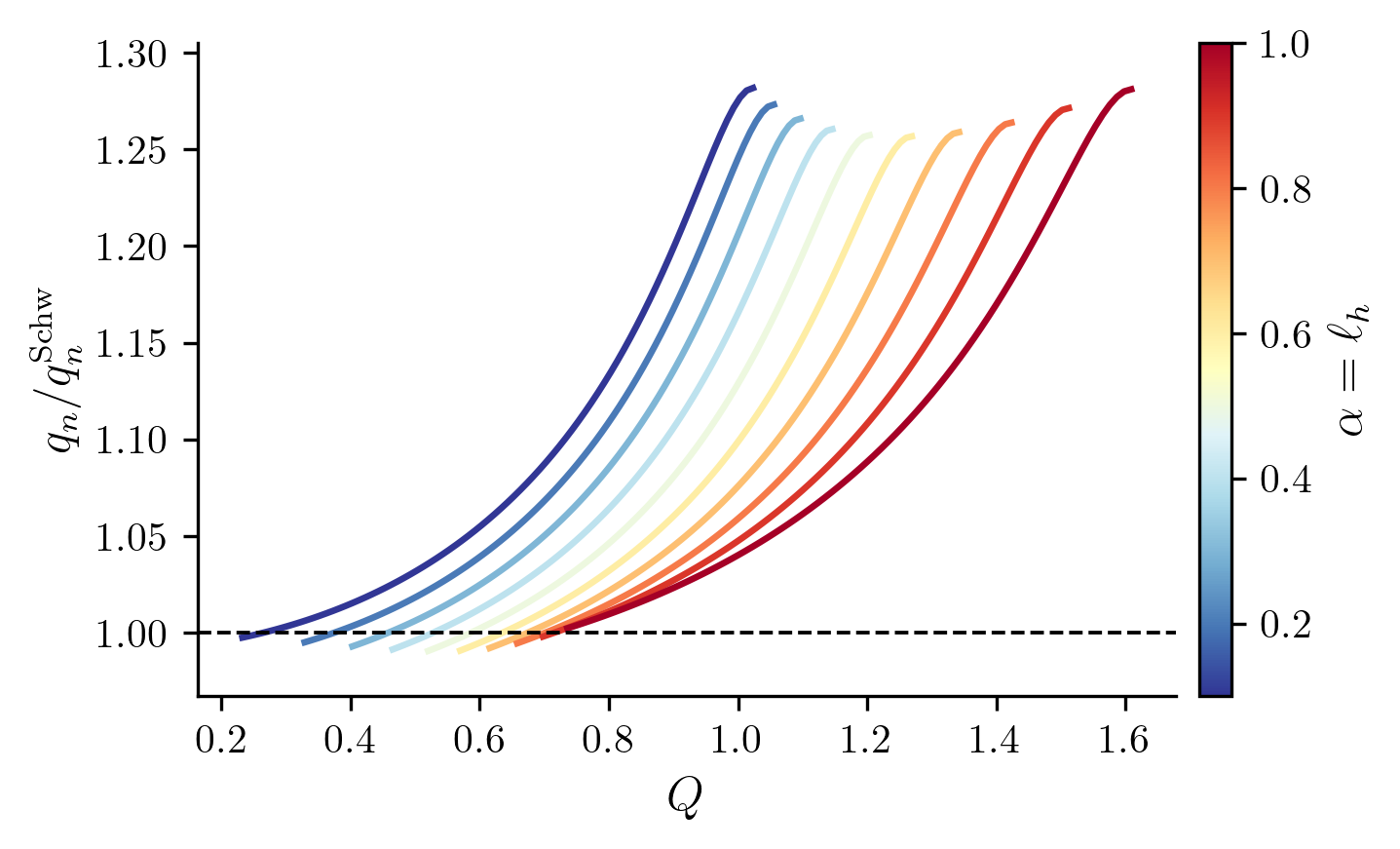}
        \caption{$s=\ell=0$}\label{fig:qf_gd_s0}
    \end{subfigure}\hfill
    \begin{subfigure}{0.48\textwidth}
        \centering\includegraphics[width=\linewidth]{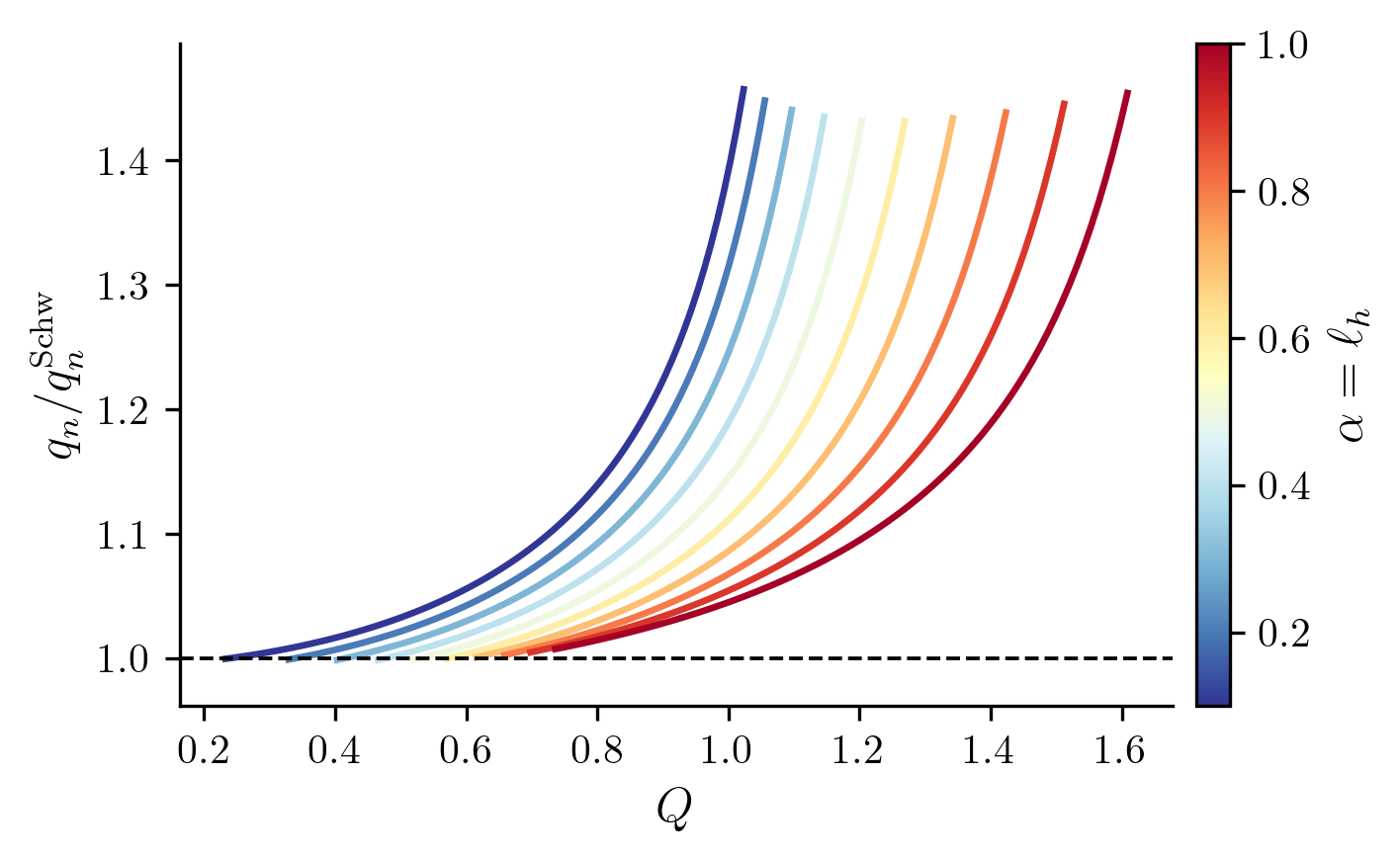}
        \caption{$s=\ell=1/2$}\label{fig:qf_gd_s12}
    \end{subfigure}

    \begin{subfigure}{0.48\textwidth}
        \centering\includegraphics[width=\linewidth]{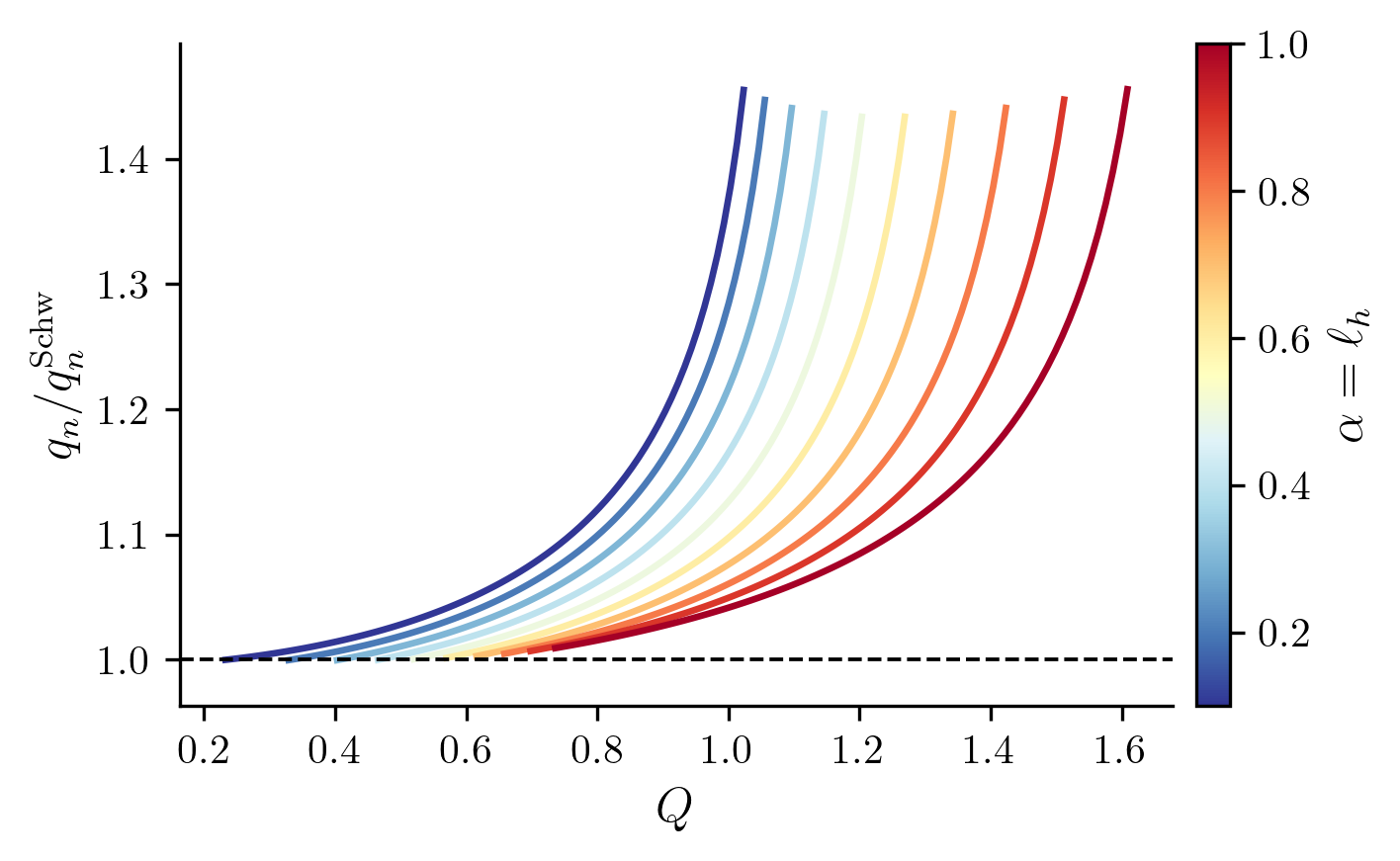}
        \caption{$s=\ell=1$}\label{fig:qf_gd_s1}
    \end{subfigure}\hfill
    \begin{subfigure}{0.48\textwidth}
        \centering\includegraphics[width=\linewidth]{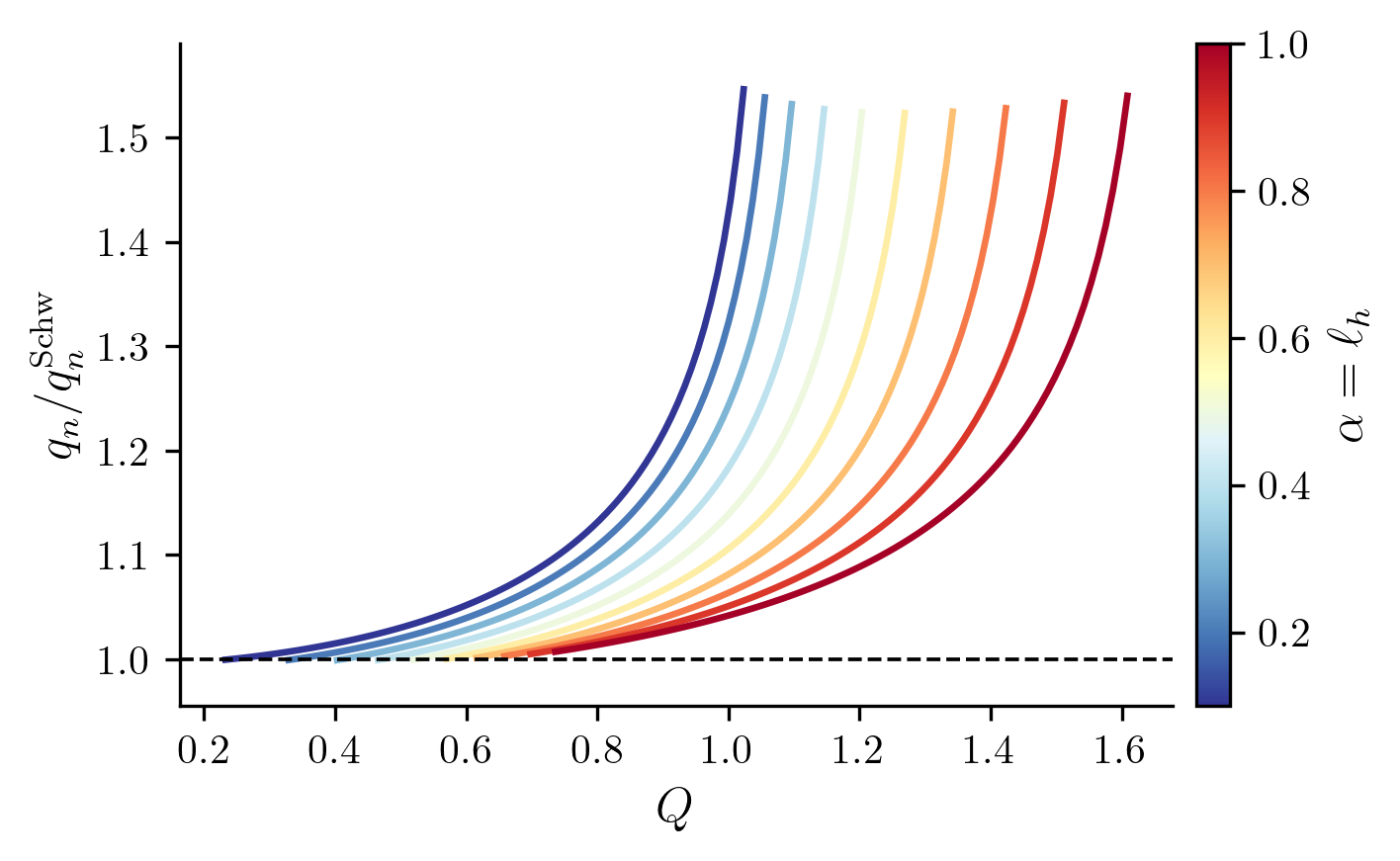}
        \caption{$s=\ell=2$}\label{fig:qf_gd_s2}
    \end{subfigure}
    \caption[Quality-factor ratio of the hairy black hole against $Q$]{Quality-factor ratio $q_n/q_n^{\text{Schw}}$ of the hairy black hole~\eqref{eq: metric HBH} against the charge $Q$, for the lowest multipole of each channel. The hair parameters are varied together over $\alpha=\ell_h\in[0.1,1]$ (colour bar), each curve spanning its own admissible range in $Q$, and the dashed line is the Schwarzschild reference. Computed with $G_\text{N}=M=1$ by the Mashhoon method.}
    \label{fig:quality_gd}
\end{figure}

A further consequence of Eq.~\eqref{eq:PT_quality} is that the fractional deviation of the quality factor from its Schwarzschild value is independent of the overtone number, since the factor $1/(2n+1)$ common to $q_n$ at fixed $(s,\ell)$ cancels in the ratio $q_n/q_n^{\text{Schw}}$. Whatever enhancement or suppression the hair produces at the fundamental mode therefore persists, unchanged in relative terms, through the whole overtone tower, even though $q_n$ itself decreases with $n$ for every background alike, Eq.~\eqref{eq:PT_quality}, and grows with $\ell$ at fixed $n$, as reported in Ref.~\cite{Casadio:2024uwj} and seen in Fig.~\ref{fig:qnm_hairy}. A single curve per background then describes the entire tower, and Fig.~\ref{fig:quality_gd} plots that ratio against $Q$ for the lowest multipole of each channel, $\ell=s$, with the hair parameters varied together over $\alpha=\ell_h\in[0.1,1]$ on the colour bar.

In every channel the ratio starts at unity (or close) on the dominant-energy-condition threshold, rises monotonically with $Q$, and steepens towards the extremal bound $Q_\text{max}$, where the two horizons of Fig.~\ref{fig:eventhorizon} merge. The size of the enhancement attained there is governed by the spin rather than by the hair, reaching, when compared to Schwarzschild, about $27\%$ in the scalar channel, $45\%$ in the Dirac and electromagnetic ones, and $54\%$ in the gravitational one, in each case to within a couple of per cent across the whole range of $\alpha=\ell_h$. What the hair controls is the location of the admissible window in $Q$, since both $Q_\text{DEC}$ and $Q_\text{max}$ grow with $\alpha$, which is why the curves of Fig.~\ref{fig:quality_gd} are displaced to the right as the colour bar is ascended. The ringdown of a strongly hairy configuration is thus not more resonant than that of a weakly hairy one. It merely requires a larger charge to reach the same degree of resonance.

\section{Quantum-corrected Schwarzschild black hole}
\label{sec:QNM_QC}

For the Calmet--Kuipers metric of Chapter~\ref{chap:EFT}, Eq.~\eqref{eq:CK_metric}, the effective potential is the one derived in Chapter~\ref{chap:conexaoentrebnsebocais}, Eq.~\eqref{eq:qc V expansion} for integer spin and Eq.~\eqref{eq:qc fermionic} for the Dirac field. 
Inserting $V_{\text{eff}}$ into Eqs.~\eqref{eq:PT_kappa} and~\eqref{eq:PT_omega} gives the quasinormal frequencies as functions of $c_6$. The effective field theory bounds the coupling by $\abs{c_6}<1/(5\pi)\approx0.0636$, Eq.~\eqref{eq:c6_bound}, and the whole negative half of that interval is accessible. The positive half is not, because for $c_6\gtrsim0.01$ the barrier flattens until $V_{\text{eff}}^{\max}<\varkappa^2/4$ and Eq.~\eqref{eq:PT_omega} stops returning an oscillating mode, so no sweep below goes past $c_6\approx0.005$. Table~\ref{tab:QNM_QC} lists the spectrum, computed with $G_\text{N} = M = 1$, for the fundamental mode and the first two overtones of scalar ($s=0$), Dirac ($s=1/2$), electromagnetic ($s=1$), and axial gravitational ($s=2$) perturbations.
Table~\ref{tab:QNM_QC} also exhibits three regularities that hold across the spectrum~\cite{Casadio:2025lkn}. At fixed $c_6$ the real part grows with the multipole $\ell$, and at fixed $c_6$ and $\ell$ it decreases with increasing spin $s$. At fixed $s$ and $\ell$, the imaginary part increases with the overtone number $n$, in accordance with Eq.~\eqref{eq:PT_omega}. The deviations from the Schwarzschild values ($c_6=0$) grow with $n$, so the higher overtones are the more sensitive probes of the quantum correction.

\begin{figure}[htbp]
    \centering
    \begin{subfigure}{0.49\textwidth}\centering
        \includegraphics[width=\linewidth]{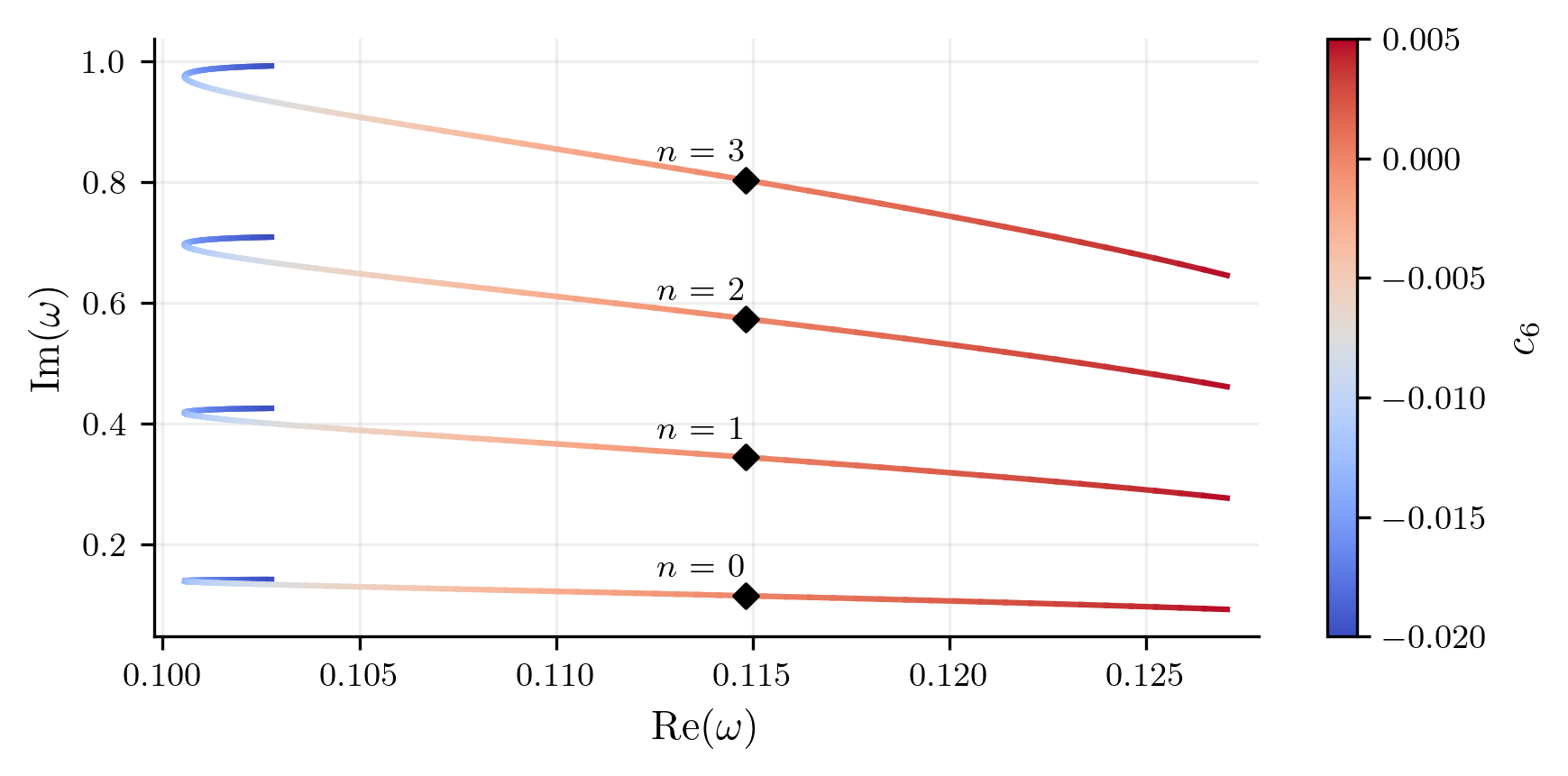}
        \caption{$s=\ell=0$}\label{fig:qnm_b_s0l0}
    \end{subfigure}\hfill
    \begin{subfigure}{0.49\textwidth}\centering
        \includegraphics[width=\linewidth]{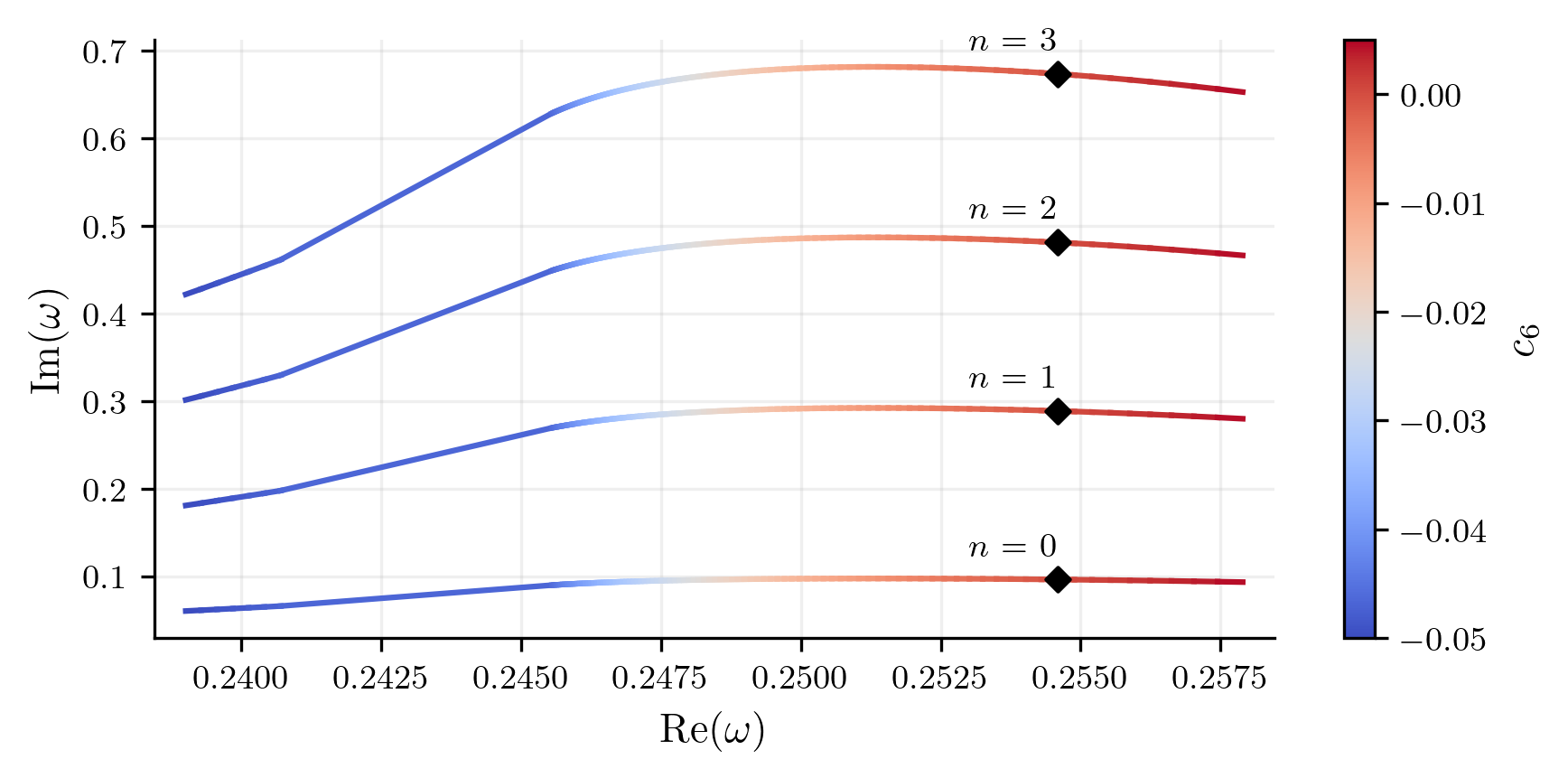}
        \caption{$s=\ell=1$}\label{fig:qnm_b_s1l1}
    \end{subfigure}

    \begin{subfigure}{0.49\textwidth}\centering
        \includegraphics[width=\linewidth]{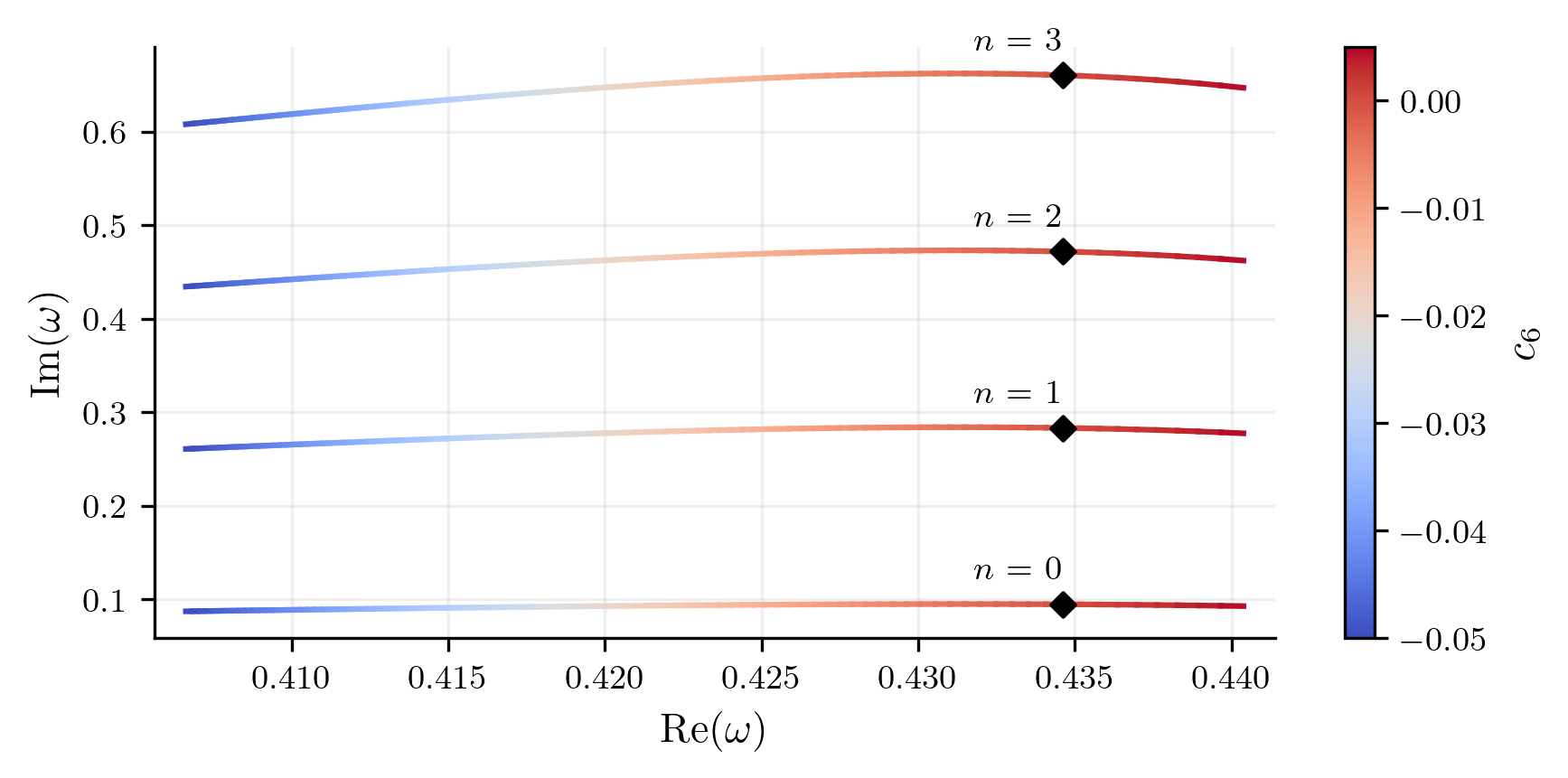}
        \caption{$s=\ell=2$}\label{fig:qnm_b_s2l2}
    \end{subfigure}\hfill 
    \begin{subfigure}{0.49\textwidth}\centering
        \includegraphics[width=\linewidth]{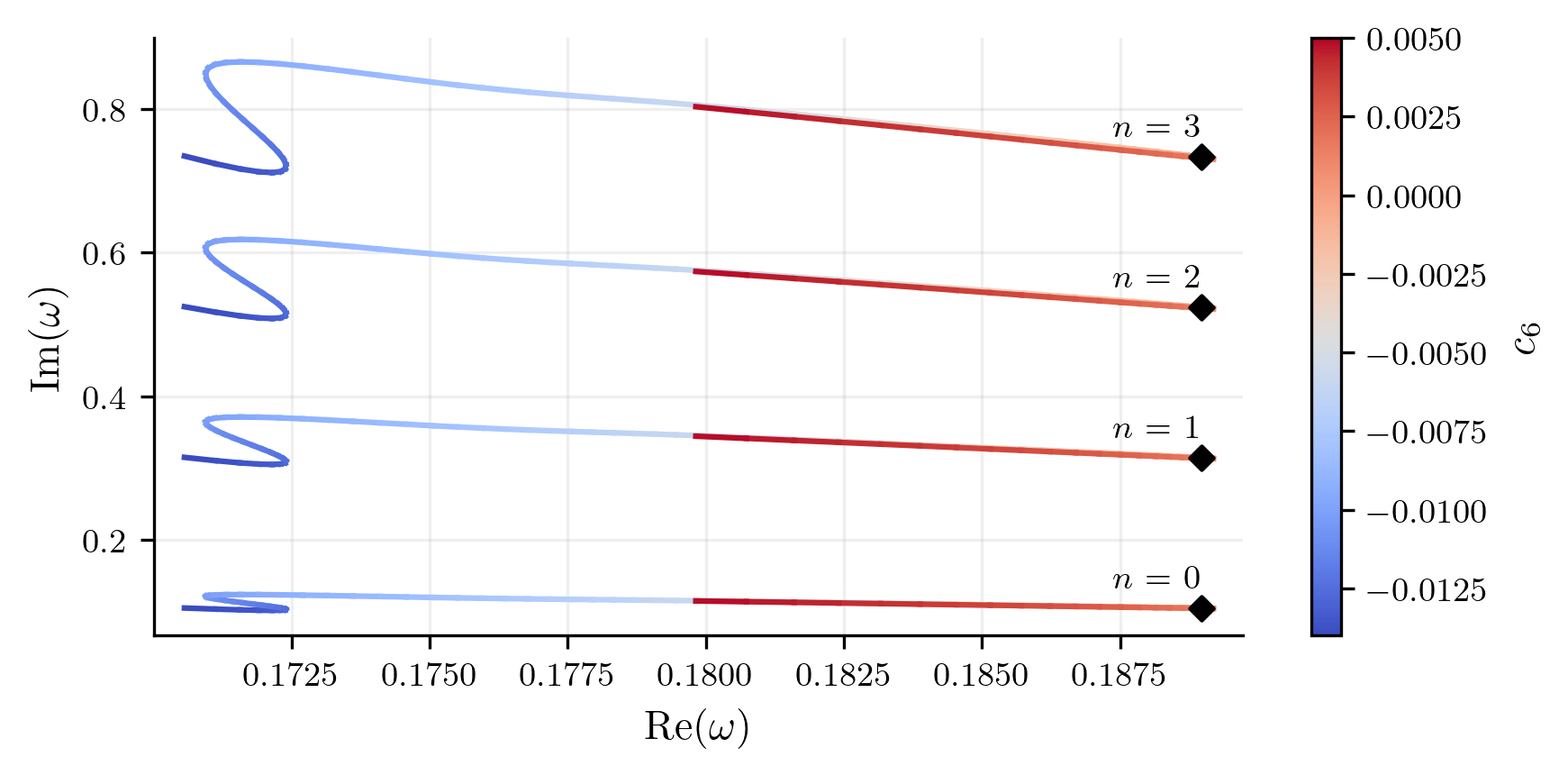}
        \caption{$s=\ell=1/2$}\label{fig:qnm_f_l05}
    \end{subfigure}
    \caption[Quasinormal frequencies of the quantum-corrected black hole]{Quasinormal-mode frequencies $\omega_n$ in the complex plane for bosonic ($s=0,1,2$) and fermionic ($s=1/2$) perturbations, with overtones up to $n=3$ and varying $c_6$, for the quantum-corrected Schwarzschild metric. The black dot marks Schwarzschild ($c_6\to0$); $G_\text{N}=M=1$. First published in Ref.~\cite{Casadio:2025lkn}.}
    \label{fig:qnm_bosonic}
\end{figure}

\begin{figure}[htbp]
    \centering
    \begin{subfigure}{0.49\textwidth}\centering
        \includegraphics[width=\linewidth]{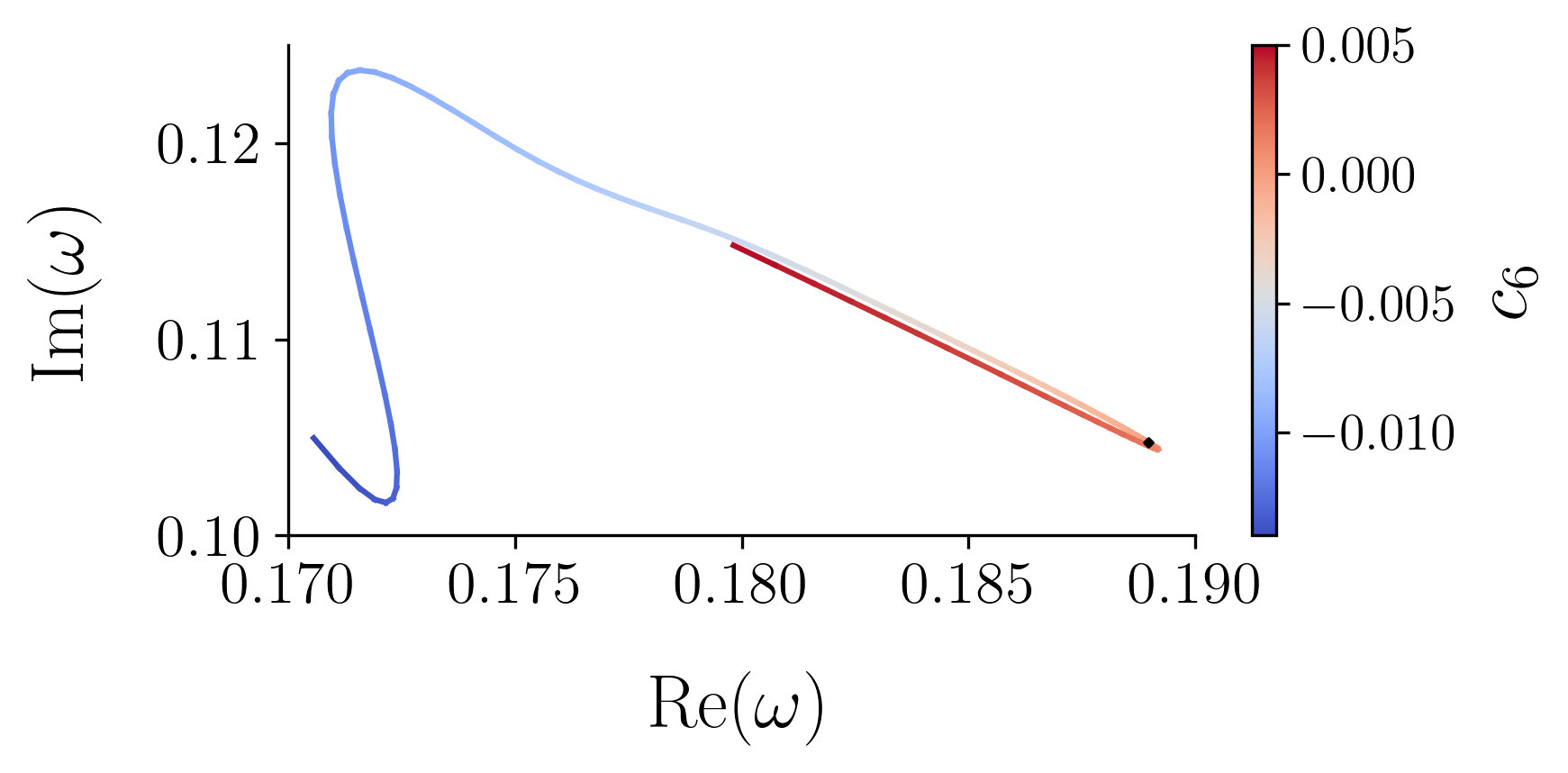}
        \caption{Focus on $s=\ell=1/2$}\label{fig:qnm_f_foc05}
    \end{subfigure}\hfill
    \begin{subfigure}{0.49\textwidth}\centering
        \includegraphics[width=\linewidth]{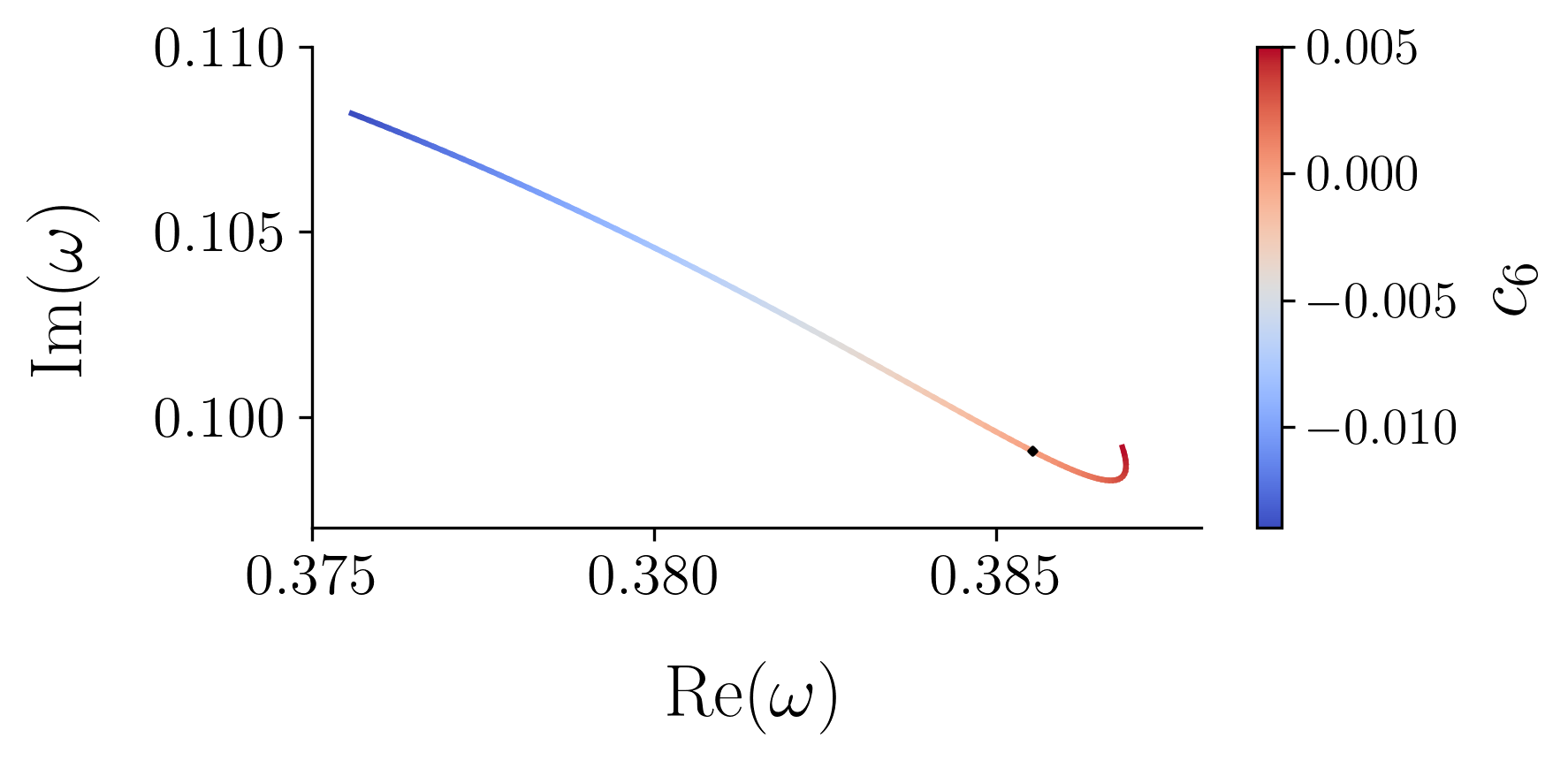}
        \caption{Focus on $s=1/2,\ell=3/2$}\label{fig:qnm_f_foc15}
    \end{subfigure}
    \caption[Fermionic quasinormal frequencies of the quantum-corrected black hole]{Quasinormal-mode frequencies $\omega_n$ in the complex plane for fermionic perturbations ($s=1/2$), focusing on $n=0$ and varying $c_6$, with the shallow maximum of $\operatorname{Re}(\omega)$ near $c_6\approx0.0008$ for $\ell=1/2$ and $c_6\approx0.0043$ for $\ell=3/2$, where the trajectory doubles back. The black dot marks Schwarzschild; $G_\text{N}=M=1$. First published in Ref.~\cite{Casadio:2025lkn}.}
    \label{fig:qnm_fermionic}
\end{figure}

The distribution of the modes and their overtones in the complex-frequency plane is displayed in Fig.~\ref{fig:qnm_bosonic}, for overtones up to $n=3$ and varying $c_6$, with Fig.~\ref{fig:qnm_fermionic} focusing on the fermionic trajectories. The black dots mark the Schwarzschild values ($c_6\to0$), and the deviations grow with the overtone number.
The real part is not monotonic in $c_6$ in the scalar and Dirac channels, and only in those two. The scalar channel passes through a minimum, at $c_6\approx-0.0135$ for $\ell=0$ and at $c_6\approx-0.0122$ for $\ell=1$, and $\omega_R$ climbs again for couplings more negative than that, which is the doubling back seen at the left edge of Fig.~\ref{fig:qnm_bosonic}. The Dirac channel turns the other way and much closer to Schwarzschild, reaching a maximum at $c_6\approx0.0008$ for $\ell=1/2$ and at $c_6\approx0.0043$ for $\ell=3/2$, in both cases exceeding the Schwarzschild value by a few tenths of a per cent before turning over, as Fig.~\ref{fig:qnm_fermionic} resolves~\cite{Casadio:2025lkn}. In both channels the feature disappears at the highest multipole computed, $\ell=2$ for the scalar field and $\ell=5/2$ for the Dirac one, and it is absent from the electromagnetic and gravitational channels over the whole range accessible to the method.

\begin{table}[htbp]
\centering
\caption{Quasinormal-mode frequencies $\omega_n$ for the quantum-corrected Schwarzschild (Calmet--Kuipers) metric~\eqref{eq:CK_metric}, for varying $c_6$, spin $s$, multipole $\ell$, and overtones $n = 0,1,2$, computed with $G_\text{N} = M = 1$ by the Mashhoon method from the effective potentials of Chapter~\ref{chap:conexaoentrebnsebocais}, with no truncation in $c_6$. The Schwarzschild column is the $c_6=0$ limit and agrees with Table~\ref{tab:QNM_GD} mode by mode. Overtones marked with $\dagger$ have $n\geq\ell$ and therefore lie outside the range in which the method is accurate, quantified in Sec.~\ref{sec:mashhoon}. The fundamental mode stays within $4\%$ of the exact real part at every integer spin and multipole.}
\label{tab:QNM_QC}
\footnotesize
\setlength{\tabcolsep}{3.5pt}
\renewcommand{\arraystretch}{1.2}
\begin{tabular}
{||c|c|c||c|c|c|c|c||}
\hline\hline
$s$ & $\ell$ & \, $n$  \, & $c_{6}=-0.0014$ & $c_{6}=-0.006$ & Schwarzschild & $c_{6}=0.003$ & $c_{6}=0.005$ \\
\hline\hline 
$0$ & $0$ & $0$ & $0.1118+0.1194i$ & $0.1046+0.1304i$ & $0.1148+0.1148i$ & $0.1221+0.1024i$ & $0.1271+0.0923i$ \\ \hline
$0$ & $0$ & $1^{\dagger}$ & $0.1118+0.3582i$ & $0.1046+0.3911i$ & $0.1148+0.3445i$ & $0.1221+0.3073i$ & $0.1271+0.2768i$ \\ \hline
$0$ & $0$ & $2^{\dagger}$ & $0.1118+0.5971i$ & $0.1046+0.6518i$ & $0.1148+0.5741i$ & $0.1221+0.5122i$ & $0.1271+0.4613i$ \\ \hline
$1/2$ & $1/2$ & $0$ & $0.1877+0.1065i$ & $0.1793+0.1156i$ & $0.1890+0.1048i$ & $0.1869+0.1069i$ & $0.1798+0.1148i$ \\ \hline
$1/2$ & $1/2$ & $1^{\dagger}$ & $0.1877+0.3194i$ & $0.1793+0.3467i$ & $0.1890+0.3143i$ & $0.1869+0.3207i$ & $0.1798+0.3444i$ \\ \hline
$1/2$ & $1/2$ & $2^{\dagger}$ & $0.1877+0.5323i$ & $0.1793+0.5778i$ & $0.1890+0.5238i$ & $0.1869+0.5344i$ & $0.1798+0.5740i$ \\ \hline
$0$ & $1$ & $0$ & $0.2983+0.1020i$ & $0.2977+0.1052i$ & $0.2985+0.1006i$ & $0.2993+0.0972i$ & $0.2999+0.0942i$ \\ \hline
$0$ & $1$ & $1^{\dagger}$ & $0.2983+0.3059i$ & $0.2977+0.3157i$ & $0.2985+0.3019i$ & $0.2993+0.2915i$ & $0.2999+0.2827i$ \\ \hline
$0$ & $1$ & $2^{\dagger}$ & $0.2983+0.5098i$ & $0.2977+0.5261i$ & $0.2985+0.5032i$ & $0.2993+0.4859i$ & $0.2999+0.4712i$ \\ \hline
$1$ & $1$ & $0$ & $0.2539+0.0966i$ & $0.2520+0.0973i$ & $0.2546+0.0962i$ & $0.2564+0.0948i$ & $0.2579+0.0933i$ \\ \hline
$1$ & $1$ & $1^{\dagger}$ & $0.2539+0.2899i$ & $0.2520+0.2919i$ & $0.2546+0.2887i$ & $0.2564+0.2844i$ & $0.2579+0.2798i$ \\ \hline
$1$ & $1$ & $2^{\dagger}$ & $0.2539+0.4832i$ & $0.2520+0.4866i$ & $0.2546+0.4811i$ & $0.2564+0.4740i$ & $0.2579+0.4663i$ \\ \hline
$1/2$ & $3/2$ & $0$ & $0.3847+0.0999i$ & $0.3814+0.1032i$ & $0.3855+0.0991i$ & $0.3868+0.0983i$ & $0.3868+0.0992i$ \\ \hline
$1/2$ & $3/2$ & $1$ & $0.3847+0.2997i$ & $0.3814+0.3097i$ & $0.3855+0.2972i$ & $0.3868+0.2949i$ & $0.3868+0.2976i$ \\ \hline
$1/2$ & $3/2$ & $2^{\dagger}$ & $0.3847+0.4996i$ & $0.3814+0.5162i$ & $0.3855+0.4954i$ & $0.3868+0.4915i$ & $0.3868+0.4959i$ \\ \hline
$0$ & $2$ & $0$ & $0.4867+0.0986i$ & $0.4850+0.1002i$ & $0.4874+0.0979i$ & $0.4889+0.0958i$ & $0.4902+0.0938i$ \\ \hline
$0$ & $2$ & $1$ & $0.4867+0.2959i$ & $0.4850+0.3005i$ & $0.4874+0.2937i$ & $0.4889+0.2874i$ & $0.4902+0.2813i$ \\ \hline
$0$ & $2$ & $2^{\dagger}$ & $0.4867+0.4931i$ & $0.4850+0.5009i$ & $0.4874+0.4895i$ & $0.4889+0.4789i$ & $0.4902+0.4688i$ \\ \hline
$1$ & $2$ & $0$ & $0.4605+0.0966i$ & $0.4576+0.0973i$ & $0.4615+0.0962i$ & $0.4639+0.0948i$ & $0.4658+0.0933i$ \\ \hline
$1$ & $2$ & $1$ & $0.4605+0.2899i$ & $0.4576+0.2919i$ & $0.4615+0.2887i$ & $0.4639+0.2844i$ & $0.4658+0.2798i$ \\ \hline
$1$ & $2$ & $2^{\dagger}$ & $0.4605+0.4832i$ & $0.4576+0.4866i$ & $0.4615+0.4811i$ & $0.4639+0.4740i$ & $0.4658+0.4663i$ \\ \hline
$2$ & $2$ & $0$ & $0.3762+0.0908i$ & $0.3702+0.0918i$ & $0.3783+0.0905i$ & $0.3831+0.0898i$ & $0.3867+0.0891i$ \\ \hline
$2$ & $2$ & $1$ & $0.3762+0.2725i$ & $0.3702+0.2753i$ & $0.3783+0.2716i$ & $0.3831+0.2695i$ & $0.3867+0.2674i$ \\ \hline
$2$ & $2$ & $2^{\dagger}$ & $0.3762+0.4541i$ & $0.3702+0.4588i$ & $0.3783+0.4527i$ & $0.3831+0.4491i$ & $0.3867+0.4457i$ \\ \hline
$1/2$ & $5/2$ & $0$ & $0.5768+0.0980i$ & $0.5732+0.0998i$ & $0.5779+0.0975i$ & $0.5803+0.0962i$ & $0.5818+0.0956i$ \\ \hline
$1/2$ & $5/2$ & $1$ & $0.5768+0.2941i$ & $0.5732+0.2994i$ & $0.5779+0.2924i$ & $0.5803+0.2887i$ & $0.5818+0.2868i$ \\ \hline
$1/2$ & $5/2$ & $2$ & $0.5768+0.4902i$ & $0.5732+0.4991i$ & $0.5779+0.4873i$ & $0.5803+0.4812i$ & $0.5818+0.4780i$ \\
\hline\hline
\end{tabular}
\end{table}

\begin{figure}[htbp]
    \centering
    \includegraphics[width=0.72\textwidth]{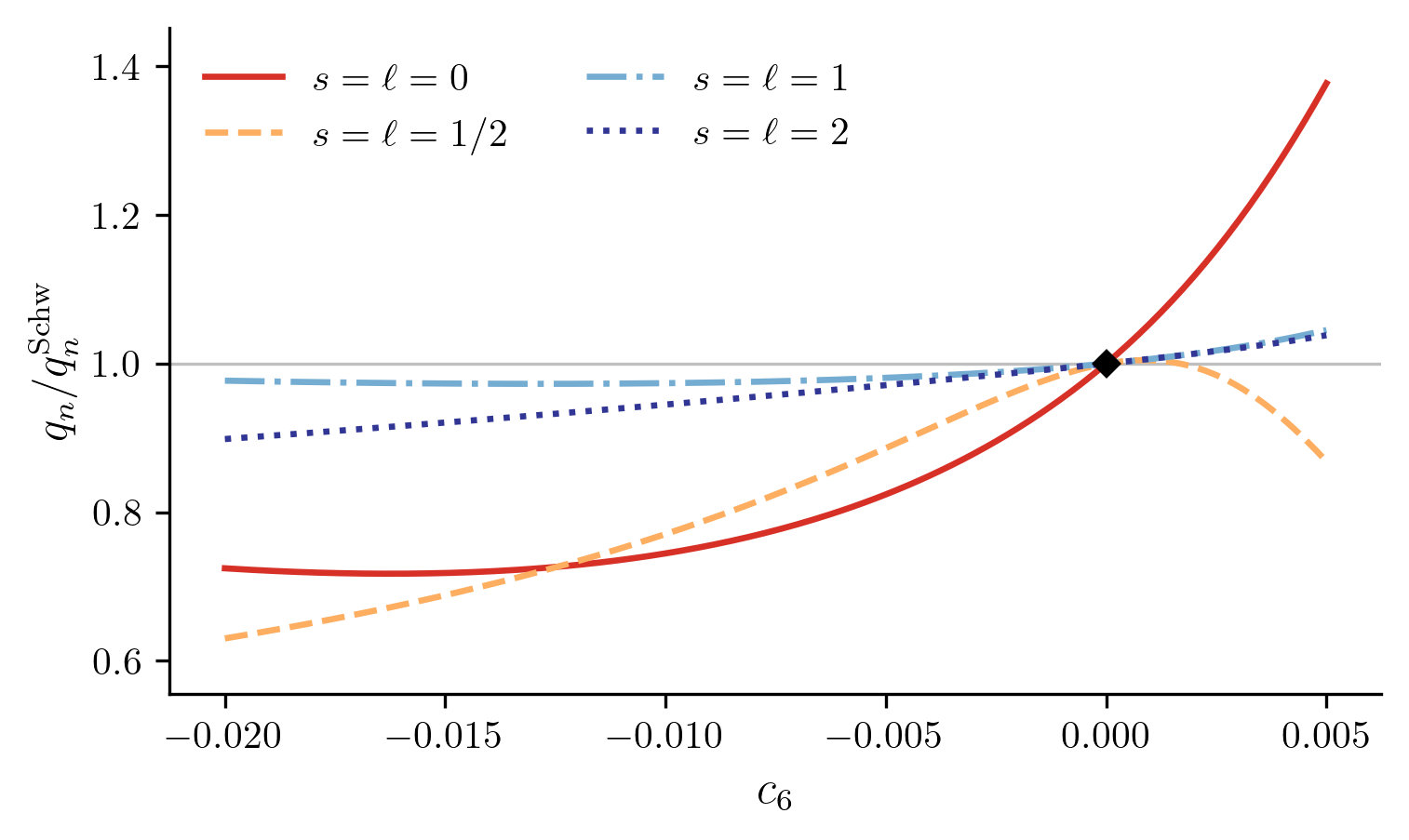}
    \caption[Quality-factor ratio of the quantum-corrected black hole against $c_6$]{Quality-factor ratio $q_n/q_n^{\text{Schw}}$ of the quantum-corrected Schwarzschild metric~\eqref{eq:CK_metric} against the coupling $c_6$, for the lowest multipole of each channel. The black diamond marks the Schwarzschild point $c_6=0$. Computed with $G_\text{N}=M=1$ by the Mashhoon method.}
    \label{fig:quality_qc}
\end{figure}

The quality factor sharpens the comparison between channels. The same cancellation of the factor $1/(2n+1)$ that was exploited in Sec.~\ref{sec:QNM_hairy} applies here, so one curve per channel again describes the whole overtone tower, and the four channels fit side by side in the single panel of Fig.~\ref{fig:quality_qc}, which plots the ratio $q_n/q_n^{\text{Schw}}$ against $c_6$ for the lowest multipole of each channel, $\ell=s$.

All four curves of Fig.~\ref{fig:quality_qc} are anchored at the black diamond, where $c_6=0$ recovers the Schwarzschild value, and they fan out from it in an order set by the spin. The scalar curve is the steepest, rising to $1.38$ at $c_6=0.005$ and falling to $0.72$ at the far end of the window, where it flattens into a shallow minimum near $c_6\approx-0.016$ and turns marginally upwards again, the same doubling back that Fig.~\ref{fig:qnm_bosonic} shows in $\omega_R$. The Dirac curve descends faster still, reaching $0.63$ at $c_6=-0.02$, a deviation of $37\%$ against the $10\%$ of the gravitational curve at the same coupling, and it crosses the scalar one at $c_6\approx-0.0125$, so which of the two channels is the more strongly affected depends on how negative the coupling is. On the positive side the Dirac curve behaves unlike any other, topping unity by half a per cent near $c_6\approx0.0009$, returning below it at $c_6\approx0.0017$ and falling to $0.87$ by the end of the sweep, which is the quality-factor counterpart of the maximum of $\omega_R$ resolved in Fig.~\ref{fig:qnm_fermionic}. The electromagnetic curve, by contrast, is almost flat across the whole window, confined between $0.97$ and $1.05$, and the axial gravitational one runs between the two regimes, reaching $0.90$ at the negative end while staying indistinguishable from the electromagnetic curve for $c_6>0$.

One difference from the hairy case of Fig.~\ref{fig:quality_gd} stands out. There the ratio never drops below unity, whereas here every channel falls below it for $c_6<0$, and the three bosonic ones rise above it for $c_6>0$, the Dirac channel doing so only over the narrow interval noted above. The sign of the coupling therefore decides whether the quantum correction makes the ringdown more or less resonant than Schwarzschild, which the hair, restricted to a single sign by the dominant energy condition, cannot do. Taken together, these results place the scalar and Dirac channels, read through the quality factor rather than through the frequency or the damping rate alone, as the most promising observational signatures of the quantum correction~\cite{Casadio:2025lkn}.

\section{Summary and Outlook}
\label{sec:QNM_discussion}

This chapter has treated the quasinormal spectrum as a non-Hermitian eigenvalue problem, selected by the requirement that radiation only enters at the horizon and only leaves at infinity, and has computed it by the Mashhoon method for the two backgrounds of Part~\ref{part: extensions of gr}. Because the eigenfrequencies follow from the height and the curvature of the barrier alone, Eqs.~\eqref{eq:PT_kappa} and~\eqref{eq:PT_omega} return the whole overtone tower in closed form, and every frequency so obtained has $\omega_I>0$, so neither geometry is unstable at linear order. The two corrections then separate by the way they act. The Calmet--Kuipers coupling is selective, moving the quality factor by up to $37\%$ in the scalar and Dirac channels at their lowest multipole while the remaining two stay within about $10\%$. The hair acts on every channel alike, lowering $\omega_R$ by close to $30\%$ once $\alpha=\ell_h$ approaches unity, with its charge-like partner $Q$ raising the quality factor at extremality by between $27\%$ and $54\%$ according to the spin. One correction is identified by which channel responds, the other by how strongly all of them do.

In both cases the fractional deviation of the quality factor is common to the whole overtone tower, so a single curve characterises a background completely, whereas the absolute displacement of a mode grows linearly with $n$, which makes the higher overtones the sharpest probes of the geometry and the ones that govern the early ringdown. That the spectrum is fixed by the shape of the barrier alone is also the point of departure for the rest of the thesis, since any system whose perturbations obey a barrier of the same form rings at the same frequencies. Parts~\ref{part: analogue models} and~\ref{part: results} make that correspondence concrete, mapping the barrier onto the geometry of a de~Laval nozzle and recovering the black-hole ringdown as the acoustic response of a transonic flow~\cite{Casadio:2024uwj,Casadio:2025lkn}.

\part{Analogue Gravity and Acoustic Systems} \label{part: analogue models}


\chapter{Analogue Gravity and Acoustic Black Holes}
\label{chap: analogue gravity}

\begingroup
\setlength{\epigraphwidth}{0.45\textwidth}
\epigraph{``The echo of a distant time \\
Comes willowing across the sand \\
And everything is green and submarine''}{\emph{``Echoes'', Meddle, 1971} \\ Pink Floyd}
\endgroup

This chapter opens Part~\ref{part: analogue models} and introduces the analogue-gravity programme that underlies the acoustic side of the correspondence developed in the remainder of this thesis. The programme originates in the observation, made by Unruh in 1981, that sound propagating in a transonic fluid flow is trapped beyond the sonic point in the same way that light is trapped beyond the horizon of a black hole, so that the Hawking process admits a laboratory counterpart~\cite{Unruh:1980cg}. The formalism has since been developed for a wide range of media and is reviewed in Refs.~\cite{Visser:1997ux,Barcelo:2005fc,Novello:2002qg}. Starting from the equations of motion of an ideal, irrotational, and isentropic fluid, it is shown that linear acoustic perturbations obey a wave equation formally identical to that of a scalar field on a curved Lorentzian background, the so-called acoustic metric, whose horizon coincides with the sonic point of a transonic flow. 

The chapter is organised as follows. Sec.~\ref{sec:fluid_primer} reviews the conservation laws and the Bernoulli equation governing an ideal fluid. Sec.~\ref{sec:linear perturbations in fluids} then linearises these equations to obtain the wave equation for the propagation of sound. Sec.~\ref{sec: acoustic metric} recasts this wave equation as that of a scalar field on an effective Lorentzian geometry, establishing the acoustic analogue of a black-hole horizon at the sonic point. Sec.~\ref{sec:other analogues} surveys the other media in which an analogue horizon has been realised, and sets out why a transonic gas flow is the system adopted here. Finally, Sec.~\ref{sec:analogue summary} closes the chapter with a summary of the results and their role in the remainder of the thesis.

\section{A Primer on Fluid Mechanics}
\label{sec:fluid_primer}

Consider an arbitrary physical density $\mathcal{N} = \dv*{f}{V}$, which may represent charge, mass, energy, entropy, momentum, etc., within a fluid medium \cite{Landau:1987,Cattani:2005}. Focusing on a fixed control surface $S_0$ enclosing a volume $V_0$, the total amount of the quantity $f$ at a given instant $t$ is expressed as
\begin{equation}
    \int_{V_0} \mathcal{N} \dd V.
\end{equation}
The temporal variation of this quantity within the volume equals the net rate at which it is created or destroyed ($\mathcal{Q}$, with units of $f/(\text{m}^3\,\text{s})$) minus the net outward flux ($\mathcal{N} \vb{v}\cdot \vb{A}$) across the boundary, leading to the balance equation:
\begin{equation}
\pdv{}{t} \int_{V_0} \mathcal{N} \dd V = - \oint_{S_0} \mathcal{N} \vb{v}\cdot\dd \vb{A} + \int_{V_0} \mathcal{Q} \dd V.
\end{equation}
Applying the divergence theorem to the surface integral and rearranging the terms yields
\begin{equation}
    \int_{V_0} \qty[\pdv{\mathcal{N}}{t} + \div{(\mathcal{N} \vb{v})} - \mathcal{Q}]\dd V = 0.
\end{equation}
Since the control volume $V_0$ is arbitrary, the general local conservation law for $\mathcal{N}$ follows:
\begin{equation} \label{eq:general conservation}
    \pdv{\mathcal{N}}{t} + \div{(\mathcal{N} \vb{v})} = \mathcal{Q}.
\end{equation}
Specifically, when describing the fluid mass, the identification $\mathcal{N} = \rho$ applies, where $\rho$ is the mass density. Assuming mass is neither created nor destroyed ($\mathcal{Q}=0$), the standard continuity equation follows:
\begin{equation}
\label{eq da continuidade}
    \pdv{\rho}{t} + \div{(\rho \vb{v})} = 0.
\end{equation}
In an incompressible flow the density of each fluid element is constant along its motion. If it is in addition uniform, $\rho = \text{const}$, Eq.~\eqref{eq da continuidade} reduces to the solenoidal condition:
\begin{equation}
\label{cond ideal}
    \div{\vb{v}}=0.
\end{equation}
Another important quantity is the vorticity, defined by the curl of $\vb{v}$~\cite{Landau:1987},
\begin{equation} \label{eq: vorticity}
    \vb*{\omega} = \curl{\vb{v}},
\end{equation}
half of which is the local angular velocity of each volume element. It is introduced here because it is what the whole construction of this chapter has to exclude. A velocity field admits a scalar potential, $\vb{v}=-\grad\psi$, if and only if $\vb*{\omega}=0$, and it is that potential which obeys the wave equation of Sec.~\ref{sec:linear perturbations in fluids} and from which the acoustic metric of Sec.~\ref{sec: acoustic metric} is read off. A flow carrying vorticity has no such potential, so the perturbations are not described by a single scalar field and no effective metric follows. Vorticity is therefore the obstruction to the analogy, not an incidental quantity, and the irrotationality assumed from Sec.~\ref{sec:fluid_primer} onwards is the statement that it vanishes.

\subsection{Euler's Equation}

Consider a fluid element with volume $\delta V$. Newton's second law governs the dynamics of this element. In the absence of external forces other than gravity, the motion is driven solely by the gravitational field and internal pressure gradients.

The gravitational force can be expressed via a potential $\phi_g$, such that its action on $\delta V$ reads $\delta \vb{F}_g = - \rho \delta V \grad{\phi_g}$. The pressure acts inwards on each face of the element, so that the face at $x$ is pushed along $+\vu{x}$ with a force $p(x)\,\delta y \delta z$ and the opposite face at $x + \delta x$ along $-\vu{x}$ with a force $p(x+\delta x)\,\delta y \delta z$. The net force along $\vu{x}$ is therefore the difference between the two,
\begin{align}
    \delta F_x = \qty[p(x) - p(x+\delta x)] \delta y \delta z = - \pdv{p}{x} \delta x \delta y \delta z = - \pdv{p}{x} \delta V,
\end{align}
and the same holds along $\vu{y}$ and $\vu{z}$, or, in vector notation:
\begin{equation}
     \delta \vb{F}_p = - \grad{p} \delta V.
\end{equation}
Applying Newton's second law and expanding the material derivative leads to Euler's equation for inviscid fluid dynamics:
\begin{equation}
\label{eq euler}
    \pdv{\vb{v}}{t} +\qty(\vb{v}\cdot \grad) \vb{v} = -\frac{\grad{p}}{\rho} - \grad{\phi_g}.
\end{equation}
It is crucial to note that Euler's equation assumes an inviscid flow, neglecting viscous dissipation terms. The equation that retains them, the Navier--Stokes equation, is displayed next for technical completeness, since it is the general equation of motion of a viscous fluid and places Eq.~\eqref{eq euler} in context. It is not the equation used in this thesis: the analogy requires an inviscid flow, and what is employed throughout is Euler's equation, that is, the Navier--Stokes equation with the viscous terms removed~\cite{Landau:1987}:
\begin{equation}
\label{eq:navier stokes}
    \rho \qty[\pdv{\vb{v}}{t} +\qty(\vb{v}\cdot \grad) \vb{v}] = -\grad{p} - \rho \grad{\phi_g} + \eta \grad^2 \vb{v} + \qty(\xi + \frac{1}{3}\eta) \nabla \qty(\div{\vb{v}}),
\end{equation}
where $\eta$ is the dynamic shear viscosity and $\xi$ represents the bulk viscosity. Setting $\eta=\xi=0$ recovers Eq.~\eqref{eq euler}. The omission is not a mere convenience: a viscous term would damp the acoustic perturbations, so that the wave equation derived in Sec.~\ref{sec:linear perturbations in fluids} would no longer be that of a field propagating freely on an effective geometry, and the correspondence with a black-hole background would be lost. Viscosity therefore sets a practical limit on how faithfully a real nozzle can reproduce the analogue spectrum, rather than being a term that could simply be carried along.

Since the analogy is made using ideal isentropic flows, Eq.~\eqref{eq euler} will be employed. Further simplification is possible by employing the vector identity:
\begin{equation}
    (\vb{v}\cdot \grad)\vb{v} = \grad{\qty( \frac{1}{2} \vb{v}\cdot \vb{v})} - \vb{v}\cross (\curl{\vb{v}}).
\end{equation}
By Kelvin's circulation theorem, an ideal barotropic flow that is irrotational at some initial instant remains irrotational at all later times, so that the vorticity~\eqref{eq: vorticity} vanishes identically, $\vb*{\omega} = \curl{\vb{v}} = 0$, and the last term of the identity above drops out~\cite{Landau:1987,Visser:1997ux}. Then Eq.~\eqref{eq euler} becomes:
\begin{equation}
    \pdv{\vb{v}}{t} = - \frac{\grad{p}}{\rho} - \grad{\qty(\frac{1}{2}v^2 + \phi_g)},
\end{equation}
where $\vb{v}\cdot \vb{v} = v^2$.

Because the flow is irrotational, the velocity field can be derived from a scalar potential $\psi$, such that $\vb{v} = -\grad{\psi}$. The flow is further assumed to be isentropic, so that pressure and density obey the polytropic relation
\begin{equation}
\label{eq: politropica}
    p \propto \rho^\gamma,
\end{equation}
where $\gamma$ is the adiabatic index of the medium. Three reasons motivate this assumption. It closes the system with a single free parameter, $\gamma$, in place of an independent energy equation, so that every thermodynamic variable of the flow follows from one of them. It is also the regime in which the de~Laval nozzle of Chapter~\ref{chap:isentropic equations} operates, the flow through a nozzle being fast enough for the heat exchanged with the walls to be negligible and smooth enough for the entropy produced by friction and shocks to be small, which is the standard working hypothesis of quasi-one-dimensional nozzle theory. Finally, it renders the fluid barotropic, $\rho = \rho(p)$, which is the property the acoustic metric of Sec.~\ref{sec: acoustic metric} rests upon. With Eq.~\eqref{eq: politropica} in force, the specific enthalpy is defined as:
\begin{align}
    h(p)=\int_0^p \frac{\dd p'}{\rho(p')} && \Longrightarrow
    && \grad{h} = \frac{\grad{p}}{\rho}.
\end{align}
Substituting these relations gives:
\begin{equation}
    \pdv{}{t} \qty( - \grad \psi ) =-\grad \qty(h + \frac{1}{2} v^2 + \phi_g).
\end{equation}
Commuting the gradient and time derivative operators shows that the gradient of a single bracket vanishes. That bracket is therefore independent of position, and integrating in space leaves an arbitrary function of time,
\begin{equation}
\label{eq bernoulli com constante}
    -\pdv{\psi}{t} + h + \frac{1}{2} (\grad \psi)^2 + \phi_g = \mathcal{C}(t),
\end{equation}
which is the generalised Bernoulli equation in the presence of gravitational and pressure forces. It states that the sum of the enthalpy, the kinetic term and the gravitational potential is conserved along the motion of the fluid. Because the flow is irrotational, the statement is in fact stronger than the usual one, the sum taking the same value on every streamline at a given instant, and not merely along each of them separately.

The function $\mathcal{C}(t)$ is not physical. The velocity potential is defined only up to an additive function of time, since $\vb{v} = -\grad\psi$ is unchanged by $\psi \mapsto \psi - \int^t \mathcal{C}(t')\dd t'$, and that freedom may be used to absorb it. In this gauge Eq.~\eqref{eq bernoulli com constante} reads
\begin{equation}
\label{eq bernoulli}
    -\pdv{\psi}{t} + h + \frac{1}{2} (\grad \psi)^2 + \phi_g = 0,
\end{equation}
which is the form used throughout. The following section discusses the propagation of linear perturbations within this framework.

\section{Linear Perturbations in Ideal Flows}
\label{sec:linear perturbations in fluids}

Three equations of Sec.~\ref{sec:fluid_primer} determine the instantaneous state of the fluid, namely the continuity equation~\eqref{eq da continuidade}, the Bernoulli equation~\eqref{eq bernoulli}, and the barotropic equation of state $\rho = \rho(p)$. They are used in that form throughout this section, with no cross-sectional area attached to the density. The duct enters only at the end of the section, where the resulting wave equation is rewritten for a flow confined to a channel of varying section, which is the form Chapter~\ref{chap:acoustic laval} takes up.

The assumptions of ideal, irrotational, and isentropic flow reduce the system from five variables to three, rendering it determinate: irrotationality replaces the three components of $\vb{v}$ by the single potential $\psi$, leaving $\rho$, $p$, and $\psi$. The focus now turns to first-order perturbations of this ideal fluid, examining the propagation of sound waves within a perfect isentropic medium. The derivation that follows is the standard one of the analogue-gravity literature~\cite{Unruh:1980cg,Visser:1997ux,Barcelo:2005fc}.

The equation of state is used only in the barotropic form $\rho = \rho(p)$, no polytropic law being needed at this stage, so that the results below hold for any medium in which the density is a function of the pressure alone. The density then carries no explicit dependence on position or time, its whole variation across the flow coming through that of the pressure, which is what allows it to be eliminated from the perturbed equations rather than evolved alongside them. Linearising this relation yields:
\begin{equation}
    \rho = \rho(p_0)+\qty(\dv{\rho}{p})_{p=p_0} \qty(p-p_0) + \mathcal{O}(\delta^2),
\end{equation}
where $p_0$ is the background pressure and $\mathcal{O}(\delta^2)$ collects the terms of second order in the departure from it. Writing $\delta\rho = \rho - \rho(p_0)$ and $\delta p = p - p_0$, the first-order relation reads
\begin{align}
 \delta\rho = \qty(\dv{\rho}{p})_{p=p_0}  \delta p.
\end{align}
A dimensional analysis of the quantity $\qty(\dv{\rho}{p})_{p=p_0}$ reveals this quantity corresponds to the inverse square of a velocity. As demonstrated later, its reciprocal is the square of the speed of sound. For now, it is defined as:
\begin{equation}
\label{velocidade do som}
    c_s^{-2} \equiv \qty(\dv{\rho}{p})_{p=p_0}.
\end{equation}

To simplify the derivation, the flow is assumed to be horizontal. Consequently, the gravitational potential $\phi_g$ is constant and vanishes from the Bernoulli equation~\eqref{eq bernoulli}. The first-order perturbation of that equation gives:
\begin{align}
    \delta \qty[-\pdv{\psi}{t} + h + \frac{1}{2} (\grad \psi)^2] =-\pdv{(\delta \psi)}{t} + \frac{\delta p}{\rho} - \vb{v}\cdot( \grad \delta \psi)= 0.
\end{align}
Applying the chain rule to express pressure variations in terms of density, isolating $\delta \rho$ and utilising Eq.~\eqref{velocidade do som} leads to:
\begin{equation}
\label{var rho}
    \delta \rho = \frac{\rho}{c_s^2}\qty[\pdv{(\delta \psi)}{t} +\vb{v}\cdot \qty(\grad \delta \psi)].
\end{equation}

Next, the continuity equation~\eqref{eq da continuidade} is perturbed to first order:
\begin{align}
    \delta \qty[\pdv{\rho}{t} + \div{(\rho \vb{v})}] =
    \pdv{(\delta \rho)}{t} + \div{ \qty[ \delta \rho \, \vb{v} -  \rho \grad \delta \psi]} = 0.
\end{align}
Substituting Eq.~\eqref{var rho} into the expression above yields:

\begin{equation}
\label{eq propag som}
    -\pdv{}{t}\qty{\frac{\rho}{c_s^2}\qty[\pdv{(\delta \psi)}{t} +\vb{v}\cdot \qty(\grad \delta \psi)]}
    +  \div{
    \qty{
     \rho \grad \delta \psi
     - \frac{\rho}{c_s^2}\qty[\pdv{(\delta \psi)}{t} +\vb{v}\cdot \qty(\grad \delta \psi)] \vb{v}
    }} = 0.
\end{equation}

Subject to specific boundary conditions, this equation determines the velocity potential associated with linear perturbation propagation in an ideal, isentropic, and irrotational fluid.

A simple limit checks the result. Take an incompressible medium of uniform density, $\rho=\text{const}$, at rest in the laboratory, so that it carries no velocity profile at all, $\vb{v}=0$. Equation~\eqref{eq propag som} then reduces to
\begin{equation}
\label{eq onda}
    \laplacian (\delta\psi) = \frac{1}{c_s^2} \pdv[2]{(\delta\psi)}{t},
\end{equation}
which is the standard wave equation for fluids at rest. This confirms the physical interpretation of Eq.~\eqref{velocidade do som} as the speed of sound.

The derivation above describes a fluid that fills space. A flow confined to a duct of cross-sectional area $\mathcal{A}$ is covered by the same equations under a single replacement. It is enough to take the conserved density of Eq.~\eqref{eq:general conservation} to be the mass per unit length of the duct, $\mathcal{N} \equiv \rho \mathcal{A}$, which amounts to $\rho \mapsto \rho\mathcal{A}$ wherever the density appears. The continuity equation then reads
\begin{equation}
\label{eq da continuidade area}
    \pdv{(\rho \mathcal{A})}{t} + \div{(\rho \mathcal{A}\vb{v})} = 0,
\end{equation}
and the same substitution in Eq.~\eqref{eq propag som} gives
\begin{equation}
\label{eq propag som area}
    -\pdv{}{t}\qty{\frac{\rho \mathcal{A}}{c_s^2}\qty[\pdv{(\delta \psi)}{t} +\vb{v}\cdot \qty(\grad \delta \psi)]}
    +  \div{
    \qty{
     \rho \mathcal{A} \grad \delta \psi
     - \frac{\rho \mathcal{A}}{c_s^2}\qty[\pdv{(\delta \psi)}{t} +\vb{v}\cdot \qty(\grad \delta \psi)] \vb{v}
    }} = 0.
\end{equation}
The area plays no role in the present chapter, where it is constant and cancels. Equation~\eqref{eq propag som area} is recorded here because Chapter~\ref{chap:acoustic laval} works in the quasi-one-dimensional regime of a de~Laval nozzle, in which $\mathcal{A}$ varies along the flow and carries the whole of the geometry.

Sec.~\ref{sec: acoustic metric} explicitly establishes the connection between this formalism and the description of photons in General Relativity.

\section{Acoustic Metric} \label{sec: acoustic metric}

The reason why Eq.~\eqref{eq propag som} can be expected to hide a spacetime geometry is its structure. It is a conservation law, a time derivative of one quantity balanced against the divergence of another, and both quantities are linear in the first derivatives of $\delta\psi$, with coefficients built from the background alone. A massless scalar field on a curved background obeys an equation of exactly that structure, since the Klein--Gordon equation $\partial_\mu(\sqrt{-g}\,g^{\mu\nu}\partial_\nu \Psi)=0$ of Chapter~\ref{chap:conexaoentrebnsebocais} is the statement that the current $\sqrt{-g}\,g^{\mu\nu}\partial_\nu\Psi$ is divergence free. Two equations of the same structure differ only in their coefficients, so that comparing the two currents term by term is enough to decide whether a metric exists that reproduces the acoustic problem, and to read it off when it does. No further input is required, because the whole of the geometry sits in the coefficient matrix of that current.

Following Ref.~\cite{Novello:2002qg}, the acoustic current is written as a four-vector $V^\mu = (V^t,\vb{V})$ built from the perturbation and the background quantities, so that Eq.~\eqref{eq propag som} becomes $\partial_\mu V^\mu = 0$. Lowering the temporal index with a Minkowski signature, $V_t = -V^t$, this reads
\begin{equation}
\label{quadrivetor vel}
    -\pdv{V_t}{t} + \div{\vb{V}} = 0,
\end{equation}
which accounts for the relative sign between the two terms. Matching Eq.~\eqref{quadrivetor vel} term by term with Eq.~\eqref{eq propag som} fixes the components as:
\begin{align}
\label{vt}
    V_t =     \frac{\rho}{c_s^2}\qty[\pdv{(\delta \psi)}{t} +\vb{v}\cdot \qty(\grad \delta \psi)]
\end{align}
\begin{align}
\label{vr}
    \vb{V} =    \rho \grad \delta \psi
     - \frac{\rho}{c_s^2}\qty[\pdv{(\delta \psi)}{t} +\vb{v}\cdot \qty(\grad \delta \psi)] \vb{v}.
\end{align}

Only the first derivatives of $\delta\psi$ appear in Eqs.~\eqref{vt} and~\eqref{vr}, and the background enters through $\rho$, $c_s$ and $\vb{v}$ alone. That is the structure of the Klein--Gordon current, and the coefficients are now compared with it directly. The perturbation keeps the label $\delta\psi$ throughout, here and in Chapter~\ref{chap:acoustic laval}, so that it is never confused with the background potential $\psi$ of Sec.~\ref{sec:fluid_primer}.

What remains is to read off the metric. Consider first an equation of the form
\begin{equation}
\label{eq f mu nu}
    \partial_\mu\qty(f^{\mu\nu}\partial_\nu \delta\psi)=0,
\end{equation}
where $\partial_\mu \equiv \pdv*{}{x^\mu}$, from which the metric is derived. Comparing Eq.~\eqref{eq f mu nu} with the Klein--Gordon equation yields:
\begin{equation}
   f^{\mu\nu} = \sqrt{-g}g^{\mu\nu}.
\end{equation}
This implies:
\begin{equation}
    \det(f^{\mu\nu})=\det(\sqrt{-g}g^{\mu\nu})=(\sqrt{-g})^4g^{-1} = g.
\end{equation}

Grouping terms of Eq.~\eqref{eq propag som} in index notation, with the coefficients kept inside the derivatives as required by Eq.~\eqref{eq f mu nu}, leads to:
\begin{equation}
\label{eq propag som indices}
   -\partial_t\qty{\frac{\rho}{c_s^2}\qty[\partial_t + v_j \partial_j]\delta\psi}
   + \partial_i\qty{\rho\,\partial_i \delta\psi - \frac{\rho}{c_s^2} v_i \qty[\partial_t + v_j \partial_j]\delta\psi} = 0.
\end{equation}
Reading off the coefficient of each pair $\partial_\mu \cdots \partial_\nu$ in Eq.~\eqref{eq propag som indices} allows the four-dimensional matrix to be constructed:
\begin{equation}
\label{matriz f}
   f^{\mu\nu} = -\frac{\rho}{c_s^2} \left[
    \begin{array}{cccc}
       1  & v_x & v_y & v_z  \\
        v_x & v_x^2 - c_s^2 & v_x v_y & v_x v_z \\
        v_y & v_y v_x & v_y^2 - c_s^2 & v_y v_z \\
        v_z & v_z v_x & v_z v_y & v_z^2 - c_s^2 
    \end{array}
    \right].
\end{equation}

The matrix \eqref{matriz f} has the determinant
\begin{equation}
    g = \qty(\frac{\rho}{c_s^2})^4 \qty[\qty(v^2 - c_s^2)- v^2]\cdot c_s^2 \cdot c_s^2 = - \frac{\rho^4}{c_s^2},
\end{equation}
which is negative as required for a Lorentzian spacetime signature \cite{Nakahara:2003nw}. Thus,
\begin{equation}
    \sqrt{-g} = \frac{\rho^2}{c_s}.
\end{equation}

Consequently, $g^{\mu\nu}$ becomes:
\begin{equation}
\label{matriz g contra}
   g^{\mu\nu} = -\frac{1}{\rho c_s} \left[
    \begin{array}{cccc}
       1  & v_x & v_y & v_z  \\
        v_x & v_x^2 - c_s^2 & v_x v_y & v_x v_z \\
        v_y & v_y v_x & v_y^2 - c_s^2 & v_y v_z \\
        v_z & v_z v_x & v_z v_y & v_z^2 - c_s^2 
    \end{array}
    \right], 
\end{equation}
and its inverse $g_{\mu\nu}$ is given by
\begin{equation}
\label{matriz g cova}
   g_{\mu\nu} = -\frac{\rho}{c_s} \left[
    \begin{array}{cccc}
       c_s^2 - v^2 & v_x & v_y & v_z  \\
        v_x & -1 & 0 & 0 \\
        v_y & 0 & -1 & 0 \\
        v_z & 0 & 0 & -1 
    \end{array}
    \right], 
\end{equation}
bringing Eq.~\eqref{quadrivetor vel} into the form of the Klein--Gordon equation. Equation~\eqref{matriz g cova} is the acoustic metric, obtained by Unruh and developed by Visser~\cite{Unruh:1980cg,Visser:1997ux}, and reviewed in Ref.~\cite{Barcelo:2005fc}. It is recovered above by the route of Ref.~\cite{Novello:2002qg}, in which the wave equation is first written as the vanishing divergence of a four-vector.

Thus, the propagation of linear perturbations in an ideal, isentropic, and irrotational fluid obeys the same equation as a massless Klein--Gordon field on a curved Lorentzian background, in the way that light does. The correspondence established here is one of propagation alone. The acoustic metric is built entirely from the density and the velocity field of the flow, and no Einstein equation has been used or is implied, a point returned to in Sec.~\ref{sec:analogue summary}.

The line element associated with the metric tensor \eqref{matriz g cova} is
\begin{equation}
\label{elem linha geral}
    \dd s^2 = -\frac{\rho}{c_s}\qty[(c_s^2-v^2)\dd t^2 + 2 \vb{v}\cdot \vb{\dd r}\dd t - \vb{\dd r}\cdot\vb{\dd r} ];
\end{equation}
hence, the acoustic spacetime is fully determined by three quantities of the flow, namely the density, the velocity field, and the local speed of sound. The three are not independent, since $c_s$ is fixed by the density through the equation of state~\eqref{velocidade do som}, so that a solution of the fluid equations determines the geometry completely. Here $t$ is the laboratory time in which the fluid equations of Secs.~\ref{sec:fluid_primer} and~\ref{sec:linear perturbations in fluids} were written, and it remains the laboratory time throughout. The cross term shows that these coordinates are of Painlev\'e--Gullstrand type, regular across the sonic point, and a distinct, static time coordinate is introduced below and denoted $\tilde t$.

The horizon of this geometry can be located before any particular flow is chosen. Sound rays are the radial null curves of the line element~\eqref{elem linha geral}, obtained by setting $\dd s^2 = 0$ at fixed angle for a radial velocity field $\vb{v} = v_r \vu{r}$, which gives
\begin{equation}
\label{eq: raios sonoros}
    \dv{r}{t} = v_r \pm c_s.
\end{equation}
The two signs are the ray emitted outwards and the ray emitted inwards, each carried along by the flow that transports it. For an inward flow, $v_r = -v$ with $v>0$, the outgoing ray advances at $c_s - v$. It recedes from the centre where the flow is subsonic, is held at a fixed radius where $v = c_s$, and is dragged inwards wherever $v > c_s$, alongside the ingoing ray. Both branches of Eq.~\eqref{eq: raios sonoros} therefore point towards the centre throughout the supersonic region, which is trapped: no sound emitted inside it reaches the surface $v = c_s$, whatever the direction of emission.

Two features of this result deserve to be stated. The trapping surface is set by the velocity profile and not by any property of the region of space it occupies, so that the same point of the laboratory lies inside or outside the horizon according to how fast the fluid moves through it. And for a profile in which the Mach number $\mathrm{M} = v/c_s$ increases monotonically towards the centre, the condition $\mathrm{M}=1$ has a single root, so the geometry has one horizon and no inner counterpart. Event horizon and Cauchy horizon coincide, and the trapped region extends from the sonic surface inwards. A transonic flow of this kind is therefore an acoustic black hole.

\subsection{Canonical Acoustic Black Hole Metric}

As an application, consider the simplest acoustic black hole model, the canonical one of Ref.~\cite{Visser:1997ux}, which is spherically symmetric and static. The velocity is $\vb{v} = v_r(r) \vu{r}$, and the continuity equation \eqref{eq da continuidade} imposes the constraint:
\begin{equation}
\label{vinculo}
    \frac{1}{r^2} \pdv{}{r} \qty(r^2 \rho v_r) =0.
\end{equation}

Equation~\eqref{vinculo} gives $r^2\rho\, v_r = \text{const}$, so that $v_r \propto 1/(\rho r^2)$. The canonical model further takes the flow to be incompressible, $\rho = \text{const}$, which is the condition~\eqref{cond ideal} and is what makes the conformal factor $\rho/c_s$ in Eq.~\eqref{elem linha geral} constant as well. Spherical symmetry and stationarity leave $r$ as the only variable the density could depend on, so that here constancy in $r$ and $t$ amounts to a genuine constant. Under that assumption the velocity field reads
\begin{equation}
    \vb{v} = -\frac{\mathcal{K}}{r^2} \vu{r},
\end{equation}
where $\mathcal{K}$ is a constant.
Assuming the flow becomes supersonic at a radius $r = r_\text{EH}$, which by the argument of Eq.~\eqref{eq: raios sonoros} is where the outgoing sound ray is held fixed and beyond which every ray is dragged inwards, hence the subscript, the constant becomes
\begin{equation}
    \mathcal{K} = c_s \cdot r_\text{EH}^2,
\end{equation}
and the velocity profile is
\begin{equation}
\label{eq: campo velocidades}
    \vb{v}=-c_s\frac{r_\text{EH}^2}{r^2}\vu{r}.
\end{equation}
Consequently, the line element \eqref{elem linha geral} becomes
\begin{equation}
\label{eq:metrica bac pg}
    \dd s^2 = \frac{\rho}{c_s}\qty[-c_s^2 \qty(1- \frac{r_\text{EH}^4}{r^4}) \dd t^2 + 2c_s\frac{r_\text{EH}^2}{r^2}\dd r \dd t + \dd r^2 + r^2 \dd \Omega^2],
\end{equation}
where $\dd \Omega^2 = \dd \theta^2 + \sin^2 \theta \dd \phi^2$ is the solid angle element.

Performing a coordinate transformation on the time coordinate \cite{Novello:2002qg}, which trades the laboratory time $t$ for a static time $\tilde t$,
\begin{equation}
    \dd \tilde t = \dd t \pm \frac{\frac{r_\text{EH}^2}{r^2}}{c_s\qty(1-\frac{r_\text{EH}^4}{r^4})} \dd r,
\end{equation}
yields the following line element:
\begin{equation}
\label{metrica bac}
    \dd s^2 = \frac{\rho}{c_s} \qty[-\qty(1-\frac{r_\text{EH}^4}{r^4}) c_s^2\dd \tilde t^2 + \qty(1 - \frac{r_\text{EH}^4}{r^4})^{-1}\dd r^2 + r^2 \dd \Omega^2].
\end{equation}
This is the canonical acoustic black hole in static form, a Schwarzschild-like line element whose blackening function is $1-r_\text{EH}^4/r^4$, vanishing at the sonic radius and tending to unity far from it, where the fluid is at rest with respect to sound.

Note that by introducing the Mach number $\mathrm{M} = v/c_s$ with radial velocity (i.e., $\vb{v} = -v(r) \vu{r}$) into Eq.~\eqref{elem linha geral}, rather than substituting the specific velocity field \eqref{eq: campo velocidades}, gives the line element:
\begin{equation}
    \dd s^2 = \frac{\rho}{c_s}\qty[-c_s^2 \qty(1- \mathrm{M}^2) \dd t^2 + 2c_s \mathrm{M} \dd r \dd t + \dd r^2 + r^2 \dd \Omega^2].
\end{equation}
Applying a similar time coordinate transformation:
\begin{equation}
    \dd \tilde t = \dd t \pm \frac{\mathrm{M}}{c_s\qty(1-\mathrm{M}^2)} \dd r,
\end{equation}
yields the metric
\begin{equation}
\label{eq:metrica bac mais geral}
    \dd s^2 = \frac{\rho}{c_s} \qty[-\qty(1-\mathrm{M}^2) c_s^2\dd \tilde t^2 + \frac{1}{(1 - \mathrm{M}^2)}\dd r^2 + r^2 \dd \Omega^2].
\end{equation}
Equation~\eqref{eq:metrica bac mais geral} is the static acoustic metric expressed solely in terms of the Mach number and therefore applies to any stationary spherically symmetric radial flow, whereas Eq.~\eqref{metrica bac} corresponds to the particular canonical profile $\mathrm{M}=(r_\text{EH}/r)^2$.

\begin{figure}[htbp]
    \centering
    \includegraphics[width=0.7\linewidth]{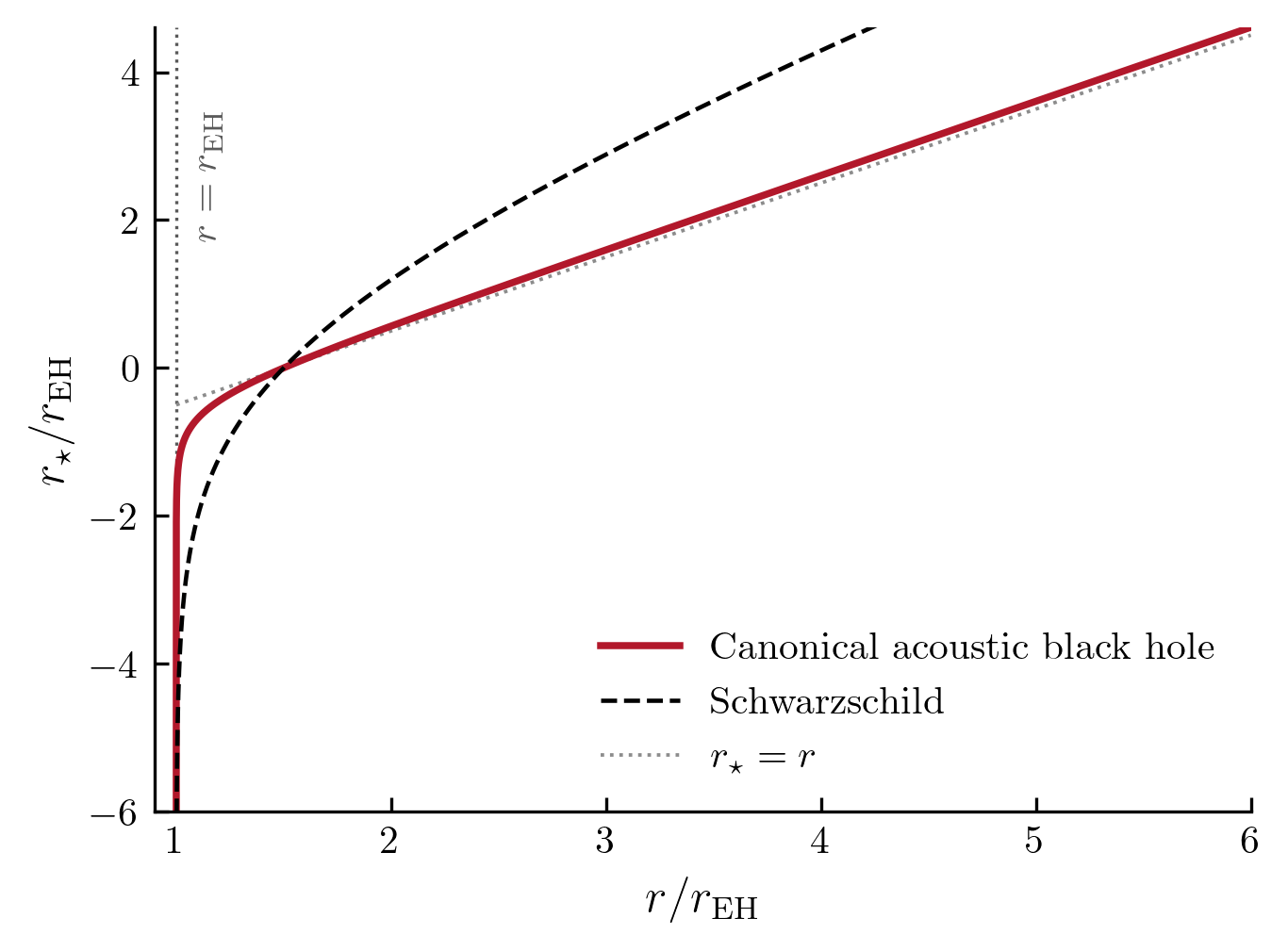}
    \caption[Tortoise coordinate of the canonical acoustic black hole]{Tortoise coordinate $r_\star(r)$ of the canonical acoustic black hole, Eq.~\eqref{eq: tartaruga acustica canonica}, compared with the Schwarzschild result~\eqref{eq:tortoise schw}, both in units of the respective horizon radius and with $c_s=1$. The dashed line is $r_\star=r$. The integration constant of each curve is fixed by $r_\star(1.5\,r_\text{EH})=0$, which for Schwarzschild is the choice $r_\star(3M)=0$ used in Fig.~\ref{fig:tortoise_schw}.}
    \label{fig:tortoise_acoustic}
\end{figure}

Written in the form~\eqref{eq:metrica bac mais geral}, the acoustic geometry is a static, spherically symmetric line element of exactly the type treated in Chapter~\ref{chap:conexaoentrebnsebocais}, once the time coordinate is measured in units of length, $\tau \equiv c_{s0} \tilde t$, where $c_{s0}$ is a constant reference value of the sound speed, taken in Chapter~\ref{chap:isentropic equations} as its stagnation value. The speed of sound is thereby carried by the time coordinate, and the metric functions read $B = (1-\mathrm{M}^2)\,c_s^2/c_{s0}^2$ and $A = (1-\mathrm{M}^2)^{-1}$. In the canonical model the sound speed is uniform, $c_s = c_{s0}$, so that $B = (1-\mathrm{M}^2)$ and $B/A = (1-\mathrm{M}^2)^2$. This identification suppresses the overall factor $\rho/c_s$, which is legitimate here because the canonical model is incompressible, so that $c_s$ is uniform as well and the factor amounts to a constant rescaling of the line element. It is not legitimate in general. In a de~Laval nozzle both $\rho$ and $c_s$ vary along the flow, and the factor then carries physical content: it is precisely what reappears in Chapter~\ref{chap:acoustic laval} as the gravitational-acoustic mapping function $g_c = \rho\mathcal{A}/c_s$, whose derivatives build the acoustic effective potential, and as the ratio $c_{s0}/c_s$ that rescales the acoustic tortoise coordinate. The present subsection should therefore be read as the constant-$\rho/c_s$ case of that construction. The tortoise coordinate~\eqref{coordenada tartaruga final} therefore reads
\begin{equation}
\label{eq: tartaruga acustica}
    \dd r_\star = \frac{c_{s0}}{c_s}\frac{\dd r}{\qty(1-\mathrm{M}^2)},
\end{equation}
which is regular wherever the flow is strictly subsonic or strictly supersonic and diverges at the sonic point, precisely as the Schwarzschild tortoise coordinate diverges at the horizon. For the canonical profile $\mathrm{M} = \qty(r_\text{EH}/r)^2$ the integration can be carried out in closed form,
\begin{equation}
\label{eq: tartaruga acustica canonica}
    r_\star = \frac{c_{s0}}{c_s}\qty[ r
    + \frac{r_\text{EH}}{4}\ln\abs{\frac{r-r_\text{EH}}{r+r_\text{EH}}}
    - \frac{r_\text{EH}}{2}\arctan\!\qty(\frac{r}{r_\text{EH}})] + C,
\end{equation}
with $C$ an arbitrary integration constant. The structure mirrors that of the Schwarzschild result~\eqref{eq:tortoise schw}: the logarithm sends the acoustic horizon to $r_\star\to-\infty$, while at large radii the slope tends to $c_{s0}/c_s$ and the difference $r_\star - r c_{s0}/c_s$ approaches a constant. Both regimes are shown in Fig.~\ref{fig:tortoise_acoustic}, together with the Schwarzschild curve for comparison. The coefficient of the logarithm differs between the two, being $r_\text{EH}/4$ for the canonical acoustic black hole against $r_\text{EH}$ for Schwarzschild in the same units, so that the acoustic horizon is approached more slowly in $r_\star$. That coefficient is the inverse of twice the surface gravity, which for an acoustic horizon reads~\cite{Visser:1997ux}
\begin{equation}
\label{eq: gravidade superficial acustica}
    \kappa_s = \frac{1}{2c_s}\eval{\dv{\qty(c_s^2-v^2)}{r}}_{r_\text{EH}},
\end{equation}
and evaluates to $\kappa_s = 2c_s/r_\text{EH}$ for the canonical profile. Defined in this way $\kappa_s$ has dimensions of inverse time, since it is referred to the laboratory time. Referred instead to the rescaled coordinate $\tau = c_{s0}\tilde t$ it becomes $\kappa_s/c_{s0}$, an inverse length, which is the form directly comparable with the Schwarzschild value.

The same quantity fixes the analogue Hawking temperature. Applying the standard relation to the acoustic horizon gives~\cite{Visser:1997ux}
\begin{equation}
\label{eq: temperatura hawking acustica}
    T_\text{H} = \frac{\hbar\kappa_s}{2\pi k_\text{B}} = \frac{\hbar}{\pi k_\text{B}}\frac{c_s}{r_\text{EH}},
\end{equation}
where $\hbar$ and $k_\text{B}$ have been restored to display the dimensions. Since $\kappa_s$ is an inverse time, the combination $\hbar\kappa_s/k_\text{B}$ already carries the dimensions of a temperature, and no further factor of a velocity is needed. This is the result anticipated by Unruh~\cite{Unruh:1980cg}: the trapping of sound at the sonic point is accompanied by a thermal emission of phonons, at a temperature set by the gradient of the flow at the horizon rather than by the mass of a compact object. It is worth stressing that the analogue temperature is a property of the effective geometry alone, and that no assumption about the field equations of gravity enters its derivation. The Hawking process is not pursued further in this thesis, whose observable of interest is the quasinormal spectrum, but Eq.~\eqref{eq: temperatura hawking acustica} makes explicit that the acoustic horizon reproduces the kinematical structure on which that process rests.

\begin{figure}[htbp]
    \centering
    \includegraphics[width=0.7\linewidth]{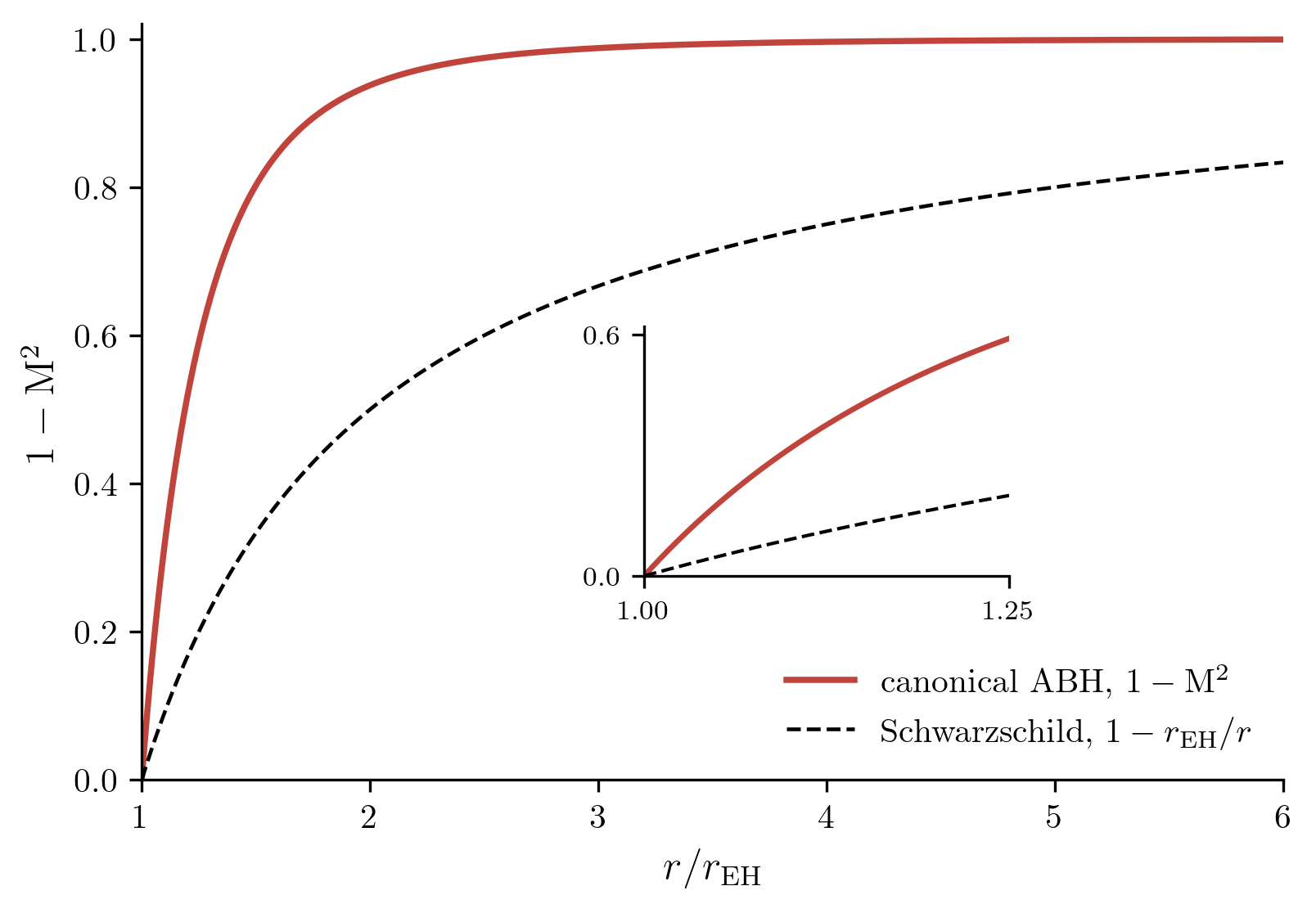}
    \caption[Acoustic analogue of the blackening function]{Acoustic analogue of the blackening function, $1-\mathrm{M}^2$ with $\mathrm{M}=(r_\text{EH}/r)^2$, for the canonical acoustic black hole, compared with the Schwarzschild function $1-r_\text{EH}/r$ (dashed), both in units of the respective horizon radius. The inset shows the neighbourhood of the horizon, where the two slopes are read off.}
    \label{fig:blackening_acoustic}
\end{figure}

The contrast in surface gravity is visible directly in the acoustic analogue of the blackening function. Figure~\ref{fig:blackening_acoustic} shows $1-\mathrm{M}^2$ for the canonical profile against the Schwarzschild function $1-r_\text{EH}/r$. Both vanish at $r_\text{EH}$ and tend to unity at large radii, but the acoustic curve rises four times more steeply at the horizon, which is the same factor of four that separates the two logarithmic coefficients in Fig.~\ref{fig:tortoise_acoustic} and the two surface gravities, $\kappa_s/c_{s0} = 2/r_\text{EH}$ for the canonical acoustic black hole against $\kappa/c = 1/(2 r_\text{EH})$ for Schwarzschild, both referred to a time coordinate of length units.

As previously stated, these spacetimes are static and spherically symmetric. The velocity field~\eqref{eq: campo velocidades} diverges at $r=0$, which is where the geometry becomes genuinely singular, whereas the surface $r=r_\text{EH}$, equivalently $\mathrm{M}=1$, is the acoustic horizon, acting as a trapping boundary for sound waves. The two must not be confused: the divergence of the metric coefficients at $r_\text{EH}$ is an artefact of the static coordinates $(\tilde t,r)$, and is removed by returning to the laboratory coordinates $(t,r)$ of Eq.~\eqref{eq:metrica bac pg}, in which the line element is regular there. The curvature invariants of this geometry are collected in Appendix~\ref{chap:curvature invariants}, alongside those of the black holes of Part~\ref{part: extensions of gr}.

Mathematically, Eq.~\eqref{metrica bac} indicates that at the acoustic horizon the coefficient of $\dd \tilde t^2$ vanishes while that of $\dd r^2$ diverges. Two statements follow from that, and they say different things. A wave falling in from the exterior takes an infinite coordinate time $\tilde t$ to reach $r_\text{EH}$, as registered by an observer at rest far from the horizon, although it crosses in finite time in the laboratory coordinates of Eq.~\eqref{eq:metrica bac pg}, which are regular there. Escape from the interior, by contrast, is not merely slow but impossible: for $r<r_\text{EH}$ the flow is supersonic, $v>c_s$, so that every phonon is swept inwards whatever its direction of emission, and no outgoing signal reaches the horizon at all. This is the same pair of statements that holds for photons in the Schwarzschild metric, completing the analogy between fluid mechanics and gravity.

\section{Other Analogue Systems}
\label{sec:other analogues}

Nothing in the construction of Sec.~\ref{sec: acoustic metric} was specific to sound in air. The derivation used only two ingredients: a medium whose linear excitations obey a wave equation, and a background flow whose velocity enters that equation through the combination $c^2-v^2$, where $c$ is the propagation speed of the excitation in the rest frame of the medium. Wherever those two ingredients are present, a surface on which the flow speed reaches $c$ behaves as a horizon for the excitation, and an effective Lorentzian metric of the form~\eqref{elem linha geral} follows. This is why the analogue-gravity programme is not tied to any single laboratory system, and several distinct platforms have been realised experimentally. Table~\ref{tab:analogue platforms} collects the main ones.

\begin{table}[htbp]
\centering
\caption{Laboratory platforms realising an analogue horizon. In each case the horizon is the surface on which the background flow speed reaches the propagation speed of the excitation in the medium. The last row is the system adopted in this thesis.}
\label{tab:analogue platforms}
\footnotesize
\setlength{\tabcolsep}{4pt}
\renewcommand{\arraystretch}{1.3}
\begin{tabular}{||>{\raggedright\arraybackslash}p{0.20\textwidth}|>{\raggedright\arraybackslash}p{0.16\textwidth}|>{\raggedright\arraybackslash}p{0.18\textwidth}|>{\raggedright\arraybackslash}p{0.28\textwidth}||}
\hline\hline
Medium & Excitation & Horizon condition & Reported observations \\
\hline\hline
Dilute Bose--Einstein condensate & Bogoliubov phonons & Flow exceeds the condensate sound speed & Stimulated emission, then spontaneous thermal radiation with its temperature and entanglement measured~\cite{Garay:1999sk,Steinhauer:2015saa,MunozdeNova:2018fxv} \\ \hline
Nonlinear optical fibre & Light, near a co-propagating pulse & Probe group velocity matches the pulse velocity & Optical analogue of a horizon in a fibre~\cite{Philbin:2007ji} \\ \hline
Open-channel water flow & Long surface gravity waves & Flow exceeds the shallow-water wave speed & Stimulated Hawking emission and noise correlated by the effect~\cite{Weinfurtner:2010nu,Euve:2015vml} \\ \hline
Transonic gas flow in a de~Laval nozzle & Sound & $\mathrm{M}=1$, reached at the throat & Quasinormal ringing computed and simulated numerically~\cite{Okuzumi:2007hf,Abdalla:2007dz} \\
\hline\hline
\end{tabular}
\end{table}

All these platforms share the kinematical character discussed above and differ in the regime each one reaches. Condensates work in the quantum regime, where spontaneous emission from the horizon can be resolved, and it is there that the analogue Hawking temperature of Eq.~\eqref{eq: temperatura hawking acustica} has been measured. Their horizons are small, cold, and short-lived. Water channels and gas flows work in the opposite regime, with horizons of laboratory size, at room temperature and stable in time, at the price of a thermal emission buried under the noise of the flow~\cite{Barcelo:2005fc}. What they give access to is the scattering of sound off the effective geometry, that is, transmission, reflection, and resonance~\cite{Anacleto:2019rfn,Oliveira:2018oha,Wang:2019rcf,Cardoso:2016rao}, and the quasinormal ringing of acoustic geometries has been computed for a range of such configurations~\cite{Cuyubamba:2013iwa,Mosna:2016nrt,Bazeia:2014xxa,Li:2013fka,Lessa:2023thi,Lessa:2023dbd,deBrito:2020xhy,Abdalla:2009pg,Torres:2016iee,Torres:2019sbr,deOliveira:2023qxe}, as has the question of how faithfully an analogue reproduces the temperature and the entropy of the geometry it emulates~\cite{Yang:2024fql}.

That difference is what selects the system used in the remainder of this thesis. The observable pursued here is the quasinormal spectrum, which is fixed by the barrier of the effective potential and not by the horizon alone, as Chapters~\ref{chap:conexaoentrebnsebocais} and~\ref{chap:QNM_spectrum} established, so that a platform is useful here to the extent that its effective potential can be prescribed rather than merely observed. A de~Laval nozzle allows exactly that, because its acoustic potential is fixed by the thermodynamic profiles along the duct, namely the density, the temperature and the local speed of sound, which the isentropic relations of Chapter~\ref{chap:isentropic equations} tie to the Mach number and hence to the shape of the duct. Those are standard aerodynamic quantities, measured routinely, which is what makes the correspondence quantitative and allows the thrust coefficient of Chapter~\ref{chap:isentropic equations} to serve as a diagnostic of the black-hole parameters in Part~\ref{part: results}. Acoustic analogues of geometries deformed by gravitational decoupling have already been constructed in a nozzle along these lines~\cite{daRocha:2017lqj}. It should be said at the outset that what follows maps a black-hole geometry onto a nozzle flow, matching effective potentials, and does not build a fluid system that reproduces the gravitational dynamics.

\section{Summary and Outlook}
\label{sec:analogue summary}

This chapter established the acoustic side of the correspondence developed in the remainder of the thesis. Starting from the continuity and Euler equations of an ideal, irrotational, and isentropic fluid, it was shown that the velocity potential of a linear acoustic perturbation obeys a massless Klein--Gordon equation on an effective Lorentzian geometry, the acoustic metric~\eqref{elem linha geral}, which is fixed entirely by the density, the velocity field, and the local speed of sound of the background flow. The radial null curves of that metric locate the horizon at the surface where the flow turns supersonic, so that it is the velocity profile, and not any feature of the region of space occupied by the fluid, that defines it. The construction is exact at linear order and makes no reference to the Einstein equations, the analogy being kinematical rather than dynamical~\cite{Barcelo:2005fc,Visser:1997ux}, so that the acoustic geometry reproduces the propagation of perturbations on a black-hole background without reproducing the field equations that would generate it.

Specialising to a steady, purely radial flow gave the canonical acoustic black hole~\eqref{metrica bac}, whose horizon sits at the sonic point, and the more general form~\eqref{eq:metrica bac mais geral}, written in terms of the Mach number alone and therefore independent of any particular velocity profile. That last expression is the one carried forward. Since the effective potential governing acoustic perturbations is fixed by the Mach-number profile, prescribing a target potential amounts to prescribing a flow. Chapter~\ref{chap:isentropic equations} assembles the isentropic-flow relations of a de~Laval nozzle, the laboratory system in which such a transonic profile is realised, and Chapter~\ref{chap:acoustic laval} constructs the acoustic black hole explicitly in that geometry, reinstating there the factor $\rho/c_s$ that the canonical model holds constant.

\chapter{Isentropic Flow and Performance Parameters of de Laval Nozzles}
\label{chap:isentropic equations}

\epigraph{``The Earth is the cradle of humanity, but one cannot remain in the cradle forever.''}{Konstantin Tsiolkovsky}

The acoustic black-hole analogy introduced in the previous chapter rests on the existence of a transonic flow whose sonic point plays the role of an event horizon. Its most direct laboratory realisation is the de~Laval nozzle, in which a compressible gas is accelerated from subsonic to supersonic speeds across a converging-diverging throat, the sonic point being reached at the throat. This chapter assembles the thermodynamic and hydrodynamic relations governing an ideal, isentropic flow of this type, and derives the performance parameters, the characteristic velocity and the thrust coefficient, that quantify the nozzle. These relations are the working tools for the construction of the de~Laval acoustic black hole in Chapter~\ref{chap:acoustic laval}, and for the analogue realisations of the exotic geometries of Part~\ref{part: extensions of gr} developed in Part~\ref{part: results}~\cite{Casadio:2024uwj,Casadio:2025lkn}. 

The chapter is organised as follows. Sec.~\ref{sec:isentropic relations} assembles the isentropic flow relations for an ideal compressible fluid, and establishes that the sonic point of such a flow can only sit at the throat. Sec.~\ref{sec:throat horizon} then states the conditions under which the sonic point is actually reached, writes the acoustic surface gravity of Chapter~\ref{chap: analogue gravity} in terms of the flow, and records where a real duct departs from the ideal description. Sec.~\ref{sec:performance parameters} derives the performance parameters, the characteristic velocity and the thrust coefficient, that will be used to characterise the nozzle in later chapters. Finally, Sec.~\ref{sec:isentropic summary} closes the chapter with a summary of the results and their role in the remainder of the thesis.

\section{Isentropic Flow Relations}
\label{sec:isentropic relations}

\begin{figure}[htbp]
    \centering
    \begin{subfigure}[b]{0.49\textwidth}
        \centering
        \includegraphics[width=\linewidth]{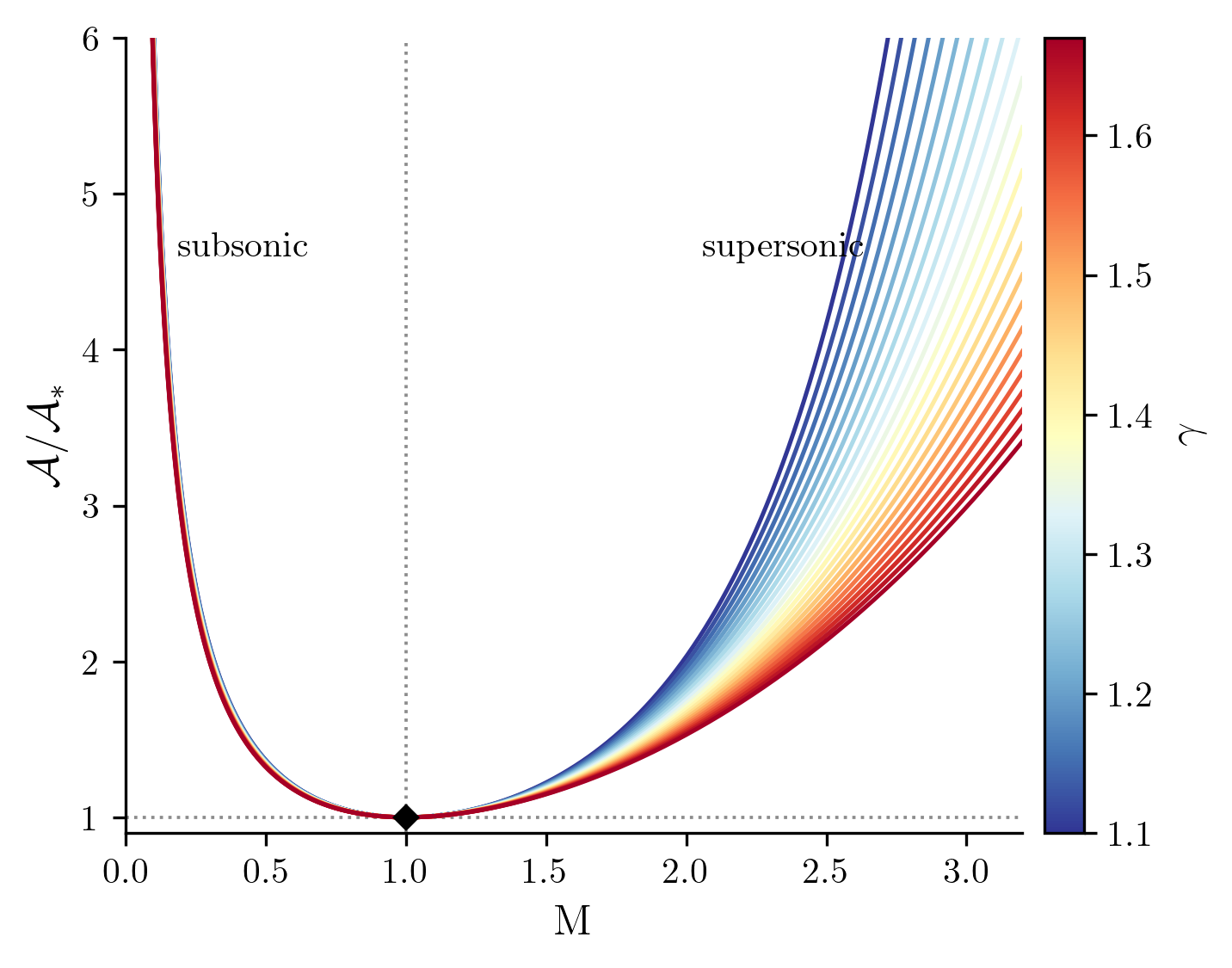}
        \caption{}
        \label{fig:area mach}
    \end{subfigure}
    \hfill
    \begin{subfigure}[b]{0.49\textwidth}
        \centering
        \includegraphics[width=\linewidth]{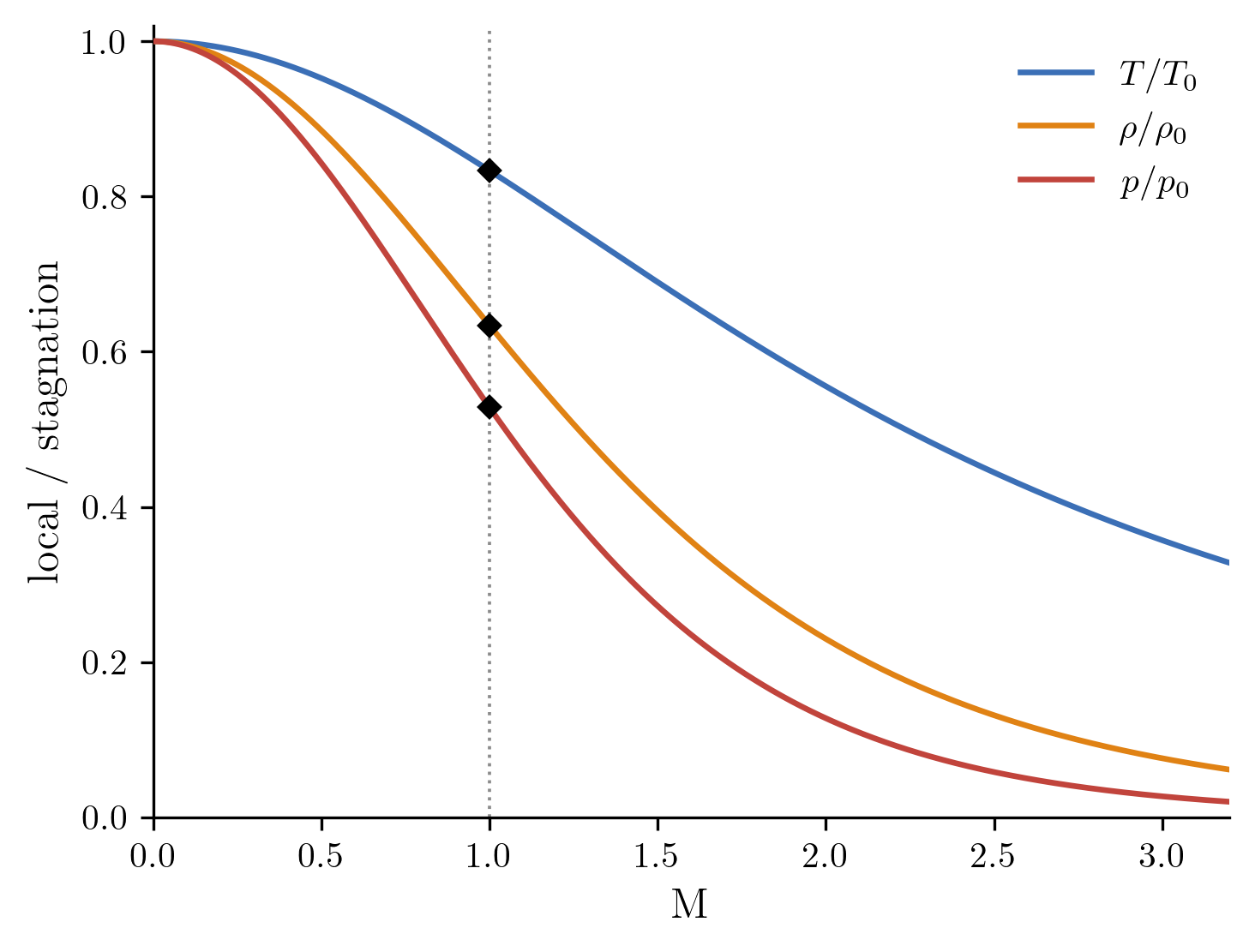}
        \caption{}
        \label{fig:isentropic ratios}
    \end{subfigure}
    \caption[Isentropic flow relations of a de~Laval nozzle]{Isentropic flow relations of a de~Laval nozzle. (a) Area-Mach-number relation~\eqref{eq: area em relacao ao mach} as the adiabatic index is varied (colour bar), every curve sharing the single minimum $\mathcal{A}/\mathcal{A}_*=1$ at $\mathrm{M}=1$ (diamond), the two branches separating more strongly for lighter gases. (b) Local-to-stagnation ratios of temperature, density, and pressure, Eqs.~\eqref{eq: rel local total}, for air ($\gamma=1.4$), with the critical values~\eqref{eq: rel critico total} marked (diamonds): $T_*/T_0=0.833$, $\rho_*/\rho_0=0.634$, and $p_*/p_0=0.528$.}
    \label{fig:isentropic}
\end{figure}

Since the system under consideration is an ideal fluid undergoing isentropic flow, the standard thermodynamic relations are employed~\cite{Anderson:2010}. The \emph{stagnation} (or total) conditions are introduced first. Consider a fluid element with local static properties: pressure $p$, temperature $T$, density $\rho$, Mach number $\mathrm{M}$, and velocity $v$. If this element is adiabatically decelerated to rest, its state variables shift to values designated as stagnation properties; in particular, the temperature attained at zero velocity is the stagnation temperature $T_0$. For the steady isentropic flow of an ideal fluid, the stagnation enthalpy $h_0 = c_p T_0$ is conserved along a streamline~\cite{Anderson:2010} and relates to the local properties through
\begin{align}
\label{eq: entalpia total}
    h_0 = h + \frac{v^2}{2}
    \quad \Longrightarrow \quad
    \frac{T_0}{T} = 1 + \frac{v^2}{2h}.
\end{align}
The aim is to express the thermodynamic variables as functions of the flow velocity. Equation~\eqref{eq: entalpia total} depends explicitly on the local enthalpy $h$. For isentropic flows the polytropic relation~\eqref{eq: politropica} holds, so the speed of sound may be written as
\begin{equation}
\label{eq: vel gas perfeito}
    c^2_s = \frac{\gamma p}{\rho} = \frac{\gamma R_g T}{\bar{m}},
\end{equation}
where the ideal-gas equation of state $p/\rho = R_g T/\bar{m}$ has been used. Recalling that the specific gas constant is $R_g/\bar{m} = c_p - c_v$ and that the adiabatic index is $\gamma = c_p/c_v$, it follows that
\begin{align}
\label{eq: entalpia aa}
    c^2_s = (\gamma - 1)c_p T
    \quad \Longrightarrow \quad
    h = \frac{c^2_s}{\gamma -1}.
\end{align}
Substituting Eq.~\eqref{eq: entalpia aa} into Eq.~\eqref{eq: entalpia total} yields the temperature ratio in terms of the Mach number $\mathrm{M}=v/c_s$, and the isentropic relations for an ideal gas then supply the pressure and density ratios as well, so that all three read
\begin{align}
\label{eq: rel local total}
    \frac{T_0}{T} = 1 + \frac{\gamma -1 }{2} \mathrm{M}^2, \quad
    \frac{p_0}{p} = \left(1 + \frac{\gamma -1 }{2} \mathrm{M}^2\right)^{\frac{\gamma}{\gamma -1}}, \quad
    \frac{\rho_0}{\rho} = \left(1 + \frac{\gamma -1 }{2} \mathrm{M}^2\right)^{\frac{1}{\gamma -1}}.
\end{align}
A second reference state is the \emph{critical condition}, reached when the fluid velocity equals the local speed of sound ($\mathrm{M}=1$); thermodynamic quantities at this state are denoted by an asterisk ($*$). The speed of sound evaluated at the stagnation state is written $c_{s0}$, and it is this constant that Chapter~\ref{chap: analogue gravity} used to give the acoustic tortoise coordinate its dimensions of length. Using the conservation of the total enthalpy~\eqref{eq: entalpia total}, the critical parameters relate to the stagnation values by
\begin{align}
\label{eq: rel crit}
    \frac{1}{2}c^2_{s*} + h_* = h_0
    \quad \Longrightarrow \quad
    \frac{c^2_{s*}}{c^2_{s0}} = \frac{T_*}{T_0} = \frac{2}{\gamma +1},
\end{align}
and, combining Eq.~\eqref{eq: rel crit} with the isentropic relations, the critical ratios read
\begin{align}
\label{eq: rel critico total}
    \frac{T_*}{T_0} = \frac{2}{\gamma +1}, \quad
    \frac{p_*}{p_0} = \left(\frac{2}{\gamma+1}\right)^{\frac{\gamma}{\gamma -1}}, \quad
    \frac{\rho_*}{\rho_0} = \left(\frac{2}{\gamma+1}\right)^{\frac{1}{\gamma -1}}.
\end{align}
Equations~\eqref{eq: rel local total} and~\eqref{eq: rel critico total} are plotted in Fig.~\ref{fig:isentropic ratios} for air, the three ratios falling monotonically with the Mach number and taking their critical values at the sonic point. It is often convenient to express the local velocity in terms of the critical speed of sound,
\begin{equation}
\label{eq: vel som critico}
    \frac{v^2}{c^2_{s*}}=\frac{\gamma+1}{2} \frac{\mathrm{M}^2}{\left(1+\frac{\gamma-1}{2}\mathrm{M}^2\right)}.
\end{equation}

\begin{figure}[htbp]
    \centering
    \begin{subfigure}[b]{0.49\textwidth}
        \centering
        \includegraphics[width=\linewidth]{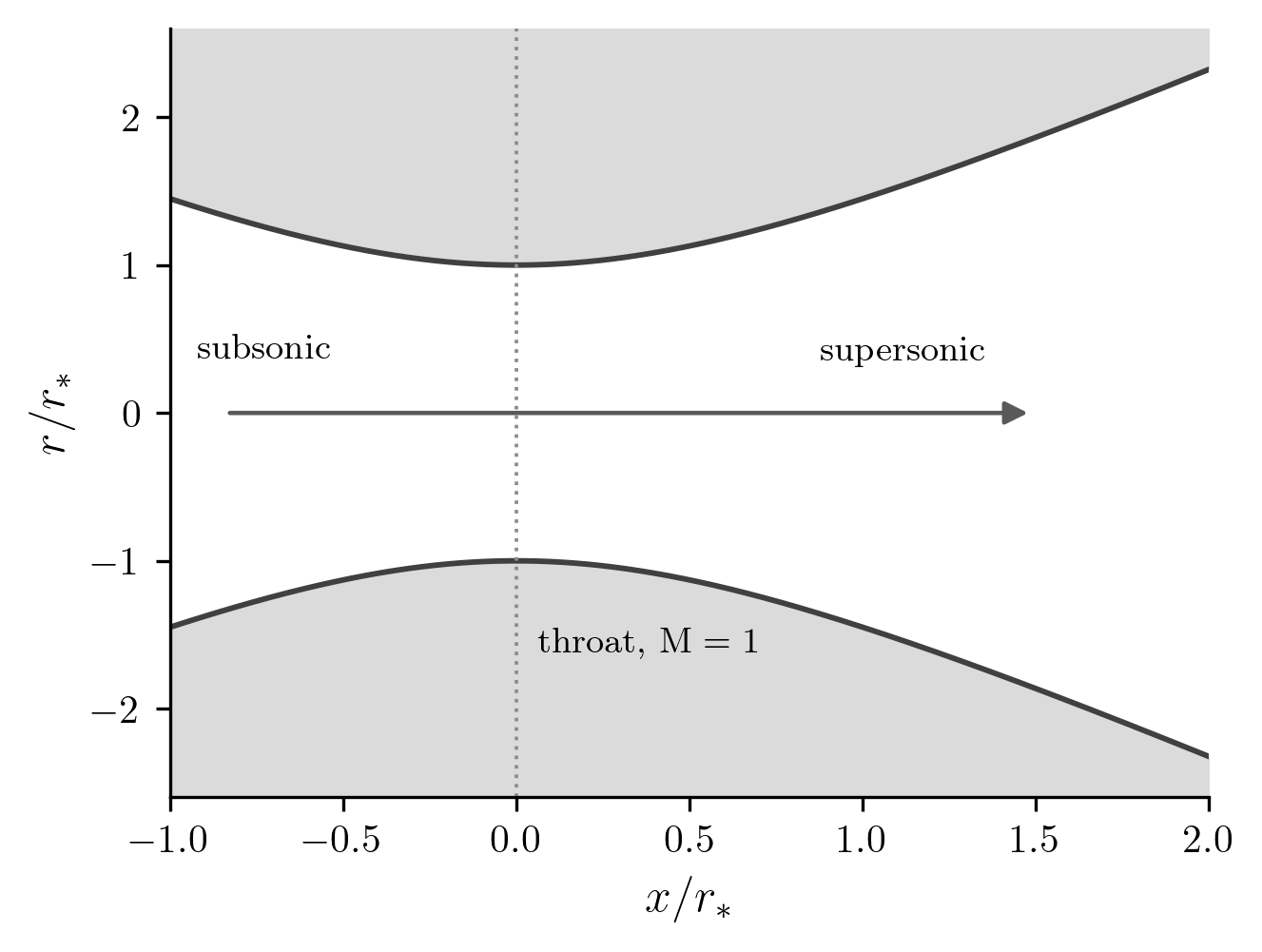}
        \caption{}
        \label{fig:laval wall}
    \end{subfigure}
    \hfill
    \begin{subfigure}[b]{0.49\textwidth}
        \centering
        \includegraphics[width=\linewidth]{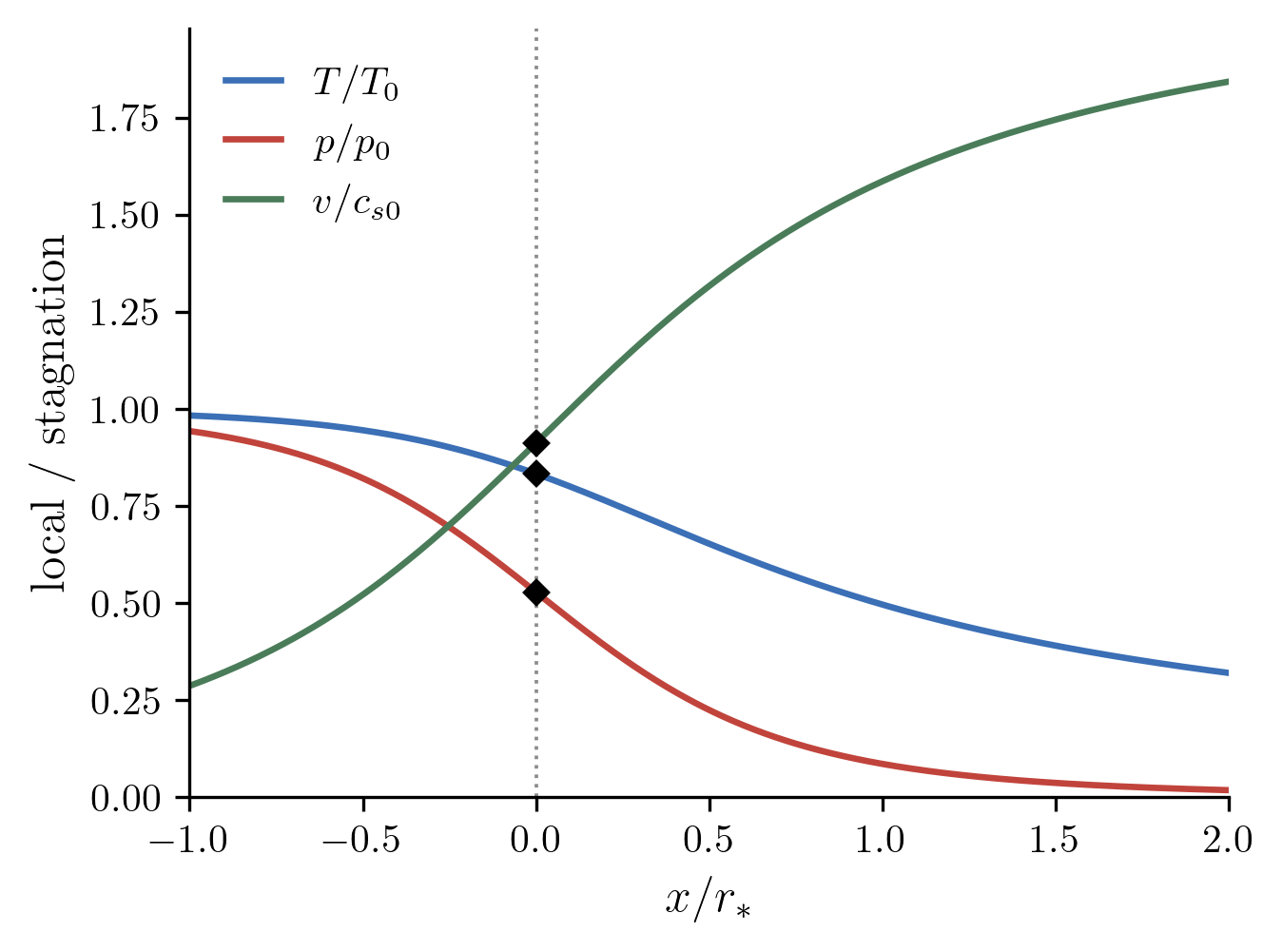}
        \caption{}
        \label{fig:laval profiles}
    \end{subfigure}
    \caption[A de~Laval nozzle in the transonic regime]{\label{fig:bocal laval}A de~Laval nozzle in the transonic regime. (a) The wall of the duct, in units of the throat radius, for the area distribution $\mathcal{A}/\mathcal{A}_*=1+1.1\,(x/r_*)^2$. (b) The temperature, the pressure, and the velocity along the same axis, from the area-Mach-number relation~\eqref{eq: area em relacao ao mach} inverted on the appropriate branch at each station, for air ($\gamma=1.4$), with the critical values~\eqref{eq: rel critico total} at the throat marked by diamonds.}
\end{figure}

Following Refs.~\cite{Okuzumi:2007hf, Abdalla:2007dz}, transonic air flow through a de~Laval nozzle is taken as an accessible model for acoustic black holes. A de~Laval nozzle has a central throat, giving it an asymmetric hourglass shape (see Fig.~\ref{fig:bocal laval}). Upstream of the throat its operation is the familiar Venturi one, the flow accelerating and the pressure dropping as the duct narrows~\cite{Anderson:2010}. Downstream of the throat that picture no longer holds, since the flow keeps accelerating while the duct widens, and the reason is set out below.

The area distribution of Fig.~\ref{fig:bocal laval} is chosen rather than derived, and a quadratic one is the generic behaviour of any smooth throat, since a wall of curvature radius $R_c$ there gives $\mathcal{A}/\mathcal{A}_*\simeq 1+x^2/(R_c r_*)$. The coefficient used corresponds to $R_c \simeq 0.9\,r_*$, within the usual range of nozzle design. The shape is illustrative and unrelated to the ducts reconstructed in Part~\ref{part: results}, which follow instead from the effective potential of a black hole. What the figure displays is the behaviour the relations above already contain: the pressure and the temperature fall monotonically along the duct while the velocity rises, with the sonic point at the minimum cross-section, which is what the remainder of this section establishes.

That the sonic point must sit at the throat, and nowhere else, follows from the equations already assembled. The first of them is the conservation of mass, taken in the form~\eqref{eq da continuidade area}, in which the cross-sectional area is carried along with the density. The description is quasi-one-dimensional throughout this chapter and the next two: the duct is taken to be slender enough for its cross-section $\mathcal{A}(x)$ to vary slowly along the axis $x$, so that the flow variables are uniform over each cross-section and depend on the single coordinate $x$, which runs along the duct with the throat as its origin. A steady flow removes the time derivative, and this reduces the divergence to an ordinary derivative,
\begin{equation}
\label{eq: continuidade unidimensional}
    \dv{}{x}\qty(\rho v \mathcal{A}) = 0.
\end{equation}
What vanishes is the variation of the mass flux along the duct, and not the flux itself. Integrating once, the mass-flow rate
\begin{equation}
\label{eq: fluxo de massa constante}
    \dot{m} = \rho v \mathcal{A} = \text{const}
\end{equation}
is a constant of the flow rather than a local quantity, taking the same value at every station. The content of the statement is that the wall of the duct is impermeable and that nothing accumulates in a steady state, so that whatever crosses one section crosses every other. Differentiating Eq.~\eqref{eq: fluxo de massa constante} gives
\begin{equation}
\label{eq: fluxo de massa diferencial}
    \rho v \mathcal{A} = \text{const}
    \quad \Longrightarrow \quad
    \frac{\dd\rho}{\rho} + \frac{\dd v}{v} + \frac{\dd\mathcal{A}}{\mathcal{A}} = 0.
\end{equation}
The second is Euler's equation~\eqref{eq euler}, which for a steady, horizontal flow loses its time derivative and its gravitational term and reduces to $v\dd v = -\dd p/\rho$. Writing $\dd p = c_s^2 \dd\rho$, by the definition~\eqref{velocidade do som} of the speed of sound, this becomes $\dd\rho/\rho = -\mathrm{M}^2\,\dd v/v$. Eliminating the density between Eq.~\eqref{eq: fluxo de massa diferencial} and this last relation yields the differential area-Mach-number relation~\cite{Anderson:2010},
\begin{equation}
\label{eq: area mach diferencial}
    \frac{\dd \mathcal{A}}{\mathcal{A}} = \qty(\mathrm{M}^2 - 1)\frac{\dd v}{v}.
\end{equation}
Its consequences are immediate. In subsonic flow, $\mathrm{M}<1$, the bracket is negative and the flow accelerates only where the duct converges, as ordinary intuition suggests. In supersonic flow, $\mathrm{M}>1$, the bracket is positive and the flow accelerates where the duct diverges instead. A flow that is to pass continuously from one regime to the other must therefore do so at a point where $\dd\mathcal{A}=0$, that is, at the throat, since $\mathrm{M}=1$ makes the right-hand side vanish for any finite acceleration. This is the reason a converging-diverging duct is required at all, and it is what places the acoustic horizon at the minimum cross-section.

The physical origin of the supersonic branch is worth examining, since it runs against everyday intuition. What is held fixed along the duct is the mass flow $\rho v \mathcal{A}$ and not the volume flow, and the relation $\dd\rho/\rho = -\mathrm{M}^2\,\dd v/v$ shows that beyond $\mathrm{M}=1$ the density falls faster, in relative terms, than the velocity rises. The gas thins out more quickly than it speeds up, so that the same mass per unit time comes to occupy a larger volume, and the duct has to open in order to accommodate it. Nothing dissipative is at work in this, the flow remaining isentropic throughout. Shock waves are a separate phenomenon, arising when the pressure imposed at the exit is incompatible with the supersonic solution, and their effect is the opposite one, returning the flow abruptly to the subsonic branch with a jump in entropy.

Invoking the conservation of mass~\eqref{eq: fluxo de massa constante} and using the critical conditions at the throat as reference,
\begin{align}
    \frac{\mathcal{A}}{\mathcal{A}_*} =
    \frac{\rho_*}{\rho_0}
    \frac{\rho_0}{\rho}
    \frac{c_{s*}}{v},
\end{align}
and substituting Eqs.~\eqref{eq: rel local total}, \eqref{eq: rel critico total}, and~\eqref{eq: vel som critico}, one obtains the area-Mach-number relation
\begin{equation}
\label{eq: area em relacao ao mach}
        \frac{\mathcal{A}}{\mathcal{A}_*} = \frac{1}{\mathrm{M}} \left[\frac{2}{\gamma+1}\left(1+\frac{\gamma-1}{2}\mathrm{M}^2\right)\right]^{\frac{\gamma+1}{2(\gamma-1)}}.
\end{equation}
This relation is plotted in Fig.~\ref{fig:area mach}. It has a single minimum, $\mathcal{A}/\mathcal{A}_*=1$ at $\mathrm{M}=1$, in agreement with Eq.~\eqref{eq: area mach diferencial}, and is therefore not invertible: every area ratio above unity is attained at two distinct Mach numbers, one subsonic and one supersonic. The nozzle geometry alone does not select between them, and the branch actually realised is fixed by the pressure imposed at the exit. The transonic solution, which follows the subsonic branch up to the throat and the supersonic branch beyond it, is the one carrying an acoustic horizon, and it is the solution adopted throughout this thesis.

Using Eq.~\eqref{eq: rel local total}, the dependence on the pressure or the density follows as
\begin{equation}
\label{eq: area em funcao da pressao ou densidade}
    \frac{\mathcal{A}}{\mathcal{A}_*} =
    \frac{\left(\frac{\gamma-1}{2}\right)^{1/2} \left(\frac{2}{\gamma+1}\right)^{\frac{\gamma+1}{2(\gamma-1)}}}{\left(\frac{p}{p_0}\right)^{1/\gamma} \left[1 - \left(\frac{p}{p_0}\right)^{\frac{\gamma-1}{\gamma}}\right]^{1/2}}
    =
    \frac{\left(\frac{\gamma-1}{2}\right)^{1/2} \left(\frac{2}{\gamma+1}\right)^{\frac{\gamma+1}{2(\gamma-1)}}}{\left(\frac{\rho}{\rho_0}\right)\left[1 - \left(\frac{\rho}{\rho_0}\right)^{\gamma-1}\right]^{1/2}}.
\end{equation}

\section{The Throat as an Acoustic Horizon}
\label{sec:throat horizon}

The relations of Sec.~\ref{sec:isentropic relations} describe any isentropic duct flow. What turns a particular one into the analogue system of Chapter~\ref{chap: analogue gravity} is that the sonic point is actually reached, which is a condition on how the nozzle is operated and not on how it is shaped. Acoustic black holes in a de~Laval nozzle were proposed in this form by Sakagami and Ohashi~\cite{Sakagami:2001ph} and simulated numerically by Furuhashi, Nambu and Saida~\cite{Furuhashi:2006dh}, and the quasinormal ringing of the resulting geometry was computed in Refs.~\cite{Okuzumi:2007hf,Abdalla:2007dz}, which are the works this thesis builds upon.

Sonic conditions at the throat require the pressure there to have fallen to the critical value~\eqref{eq: rel critico total}, that is, to $p_*/p_0=0.528$ for air. The back pressure at which that first happens is fixed by the expansion ratio of the duct through the subsonic branch of Eq.~\eqref{eq: area em relacao ao mach} or Eq.~\eqref{eq: area em funcao da pressao ou densidade}. For $\mathcal{A}_e/\mathcal{A}_*=4$ it lies at $p_e/p_0 = 0.985$, so that the throat chokes as soon as the exit pressure falls a per cent or so below the reservoir pressure. Once it does, the mass flow saturates at Eq.~\eqref{eq: fluxo de massa} and no further reduction of the exit pressure changes it, which is what makes $\dot m$ a property of the reservoir and of the throat alone.

Choking is necessary but not sufficient. Between that back pressure and the design value at which the flow remains supersonic all the way to the exit, a normal shock stands somewhere in the divergent section, returning the flow to the subsonic branch and destroying the acoustic horizon downstream of it~\cite{Anderson:2010}. The design condition is fixed by the same area ratio on the supersonic branch, and the pressure it demands is far lower. For $\mathcal{A}_e/\mathcal{A}_*=4$ it gives $\mathrm{M}_e = 2.94$ and $p_e/p_0 = 0.030$, a pressure ratio of about $1/0.030 \approx 34$ across the nozzle, against $10.6$ for $\mathcal{A}_e/\mathcal{A}_*=2$. A laboratory realisation therefore either discharges into an evacuated tank or works with a modest expansion ratio, and this, rather than the shape of the duct, is what limits how far into the supersonic branch an experiment can reach.

Because the throat is the acoustic horizon, the surface gravity~\eqref{eq: gravidade superficial acustica} can be written in terms of the flow variables. Conservation of the stagnation enthalpy~\eqref{eq: entalpia total} with Eq.~\eqref{eq: entalpia aa} gives $c_s^2 = c_{s0}^2 - \frac{\gamma-1}{2}v^2$, so that
\begin{equation}
\label{eq: c2 menos v2 bocal}
    c_s^2 - v^2 = c_{s0}^2 - \frac{\gamma+1}{2}v^2,
\end{equation}
and differentiating at the throat, where $v = c_{s*}$, yields
\begin{equation}
\label{eq: gravidade superficial bocal}
    \kappa_s = \frac{\gamma+1}{2}\eval{\dv{v}{x}}_{*}
    = c_{s*}\eval{\dv{\mathrm{M}}{x}}_{*}.
\end{equation}
The analogue surface gravity is thus the velocity gradient at the throat, up to the factor $(\gamma+1)/2$, or equivalently the critical speed of sound times the gradient of the Mach number. The second form is the useful one here, since it is the Mach-number profile that the reconstruction of Part~\ref{part: results} delivers, so that the analogue temperature can be read off the same curves that carry the black-hole parameters. Substituting Eq.~\eqref{eq: gravidade superficial bocal} into Eq.~\eqref{eq: temperatura hawking acustica} gives
\begin{equation}
\label{eq: temperatura hawking bocal}
    T_\text{H} = \frac{\hbar\, c_{s*}}{2\pi k_\text{B}}\eval{\dv{\mathrm{M}}{x}}_{*}.
\end{equation}
For air at room temperature, $c_{s0} = 347$~m/s and $c_{s*} = 317$~m/s, so a Mach-number gradient of the order of one per centimetre gives $\kappa_s \sim 3\times10^{4}\,\text{s}^{-1}$ and $T_\text{H} \sim 40$~nK. The estimate makes concrete the statement of Sec.~\ref{sec:other analogues}: the thermal emission of a classical analogue lies far below the noise of the flow, which is why the observable pursued in this thesis is the scattering spectrum and not the emission.

Finally, it is worth recording where a real duct departs from the description above, since every relation of this chapter assumes an ideal flow. Viscosity, neglected along with the Navier--Stokes terms in Sec.~\ref{sec:fluid_primer}, builds a boundary layer along the wall, so that the section available to the flow is slightly smaller than the geometric one and the measured mass flow falls a per cent or so below the theoretical maximum, which the isentropic relations fix in the next section, Eq.~\eqref{eq: fluxo de massa}~\cite{Anderson:2010}. The same viscosity damps the acoustic perturbation, which is the practical limit anticipated in Chapter~\ref{chap: analogue gravity} on how faithfully a duct reproduces the analogue spectrum. Heat exchanged with the walls, condensation of water vapour in the rapid expansion, and turbulence downstream of the throat act in the same direction. None of these is carried in what follows, and the profiles reconstructed in Part~\ref{part: results} are to be read as the ideal flow that a real nozzle approximates.

The relations assembled so far fix the state of the gas at every station of the duct once the Mach number is known, and they fix the shape of the duct along with it. What they do not provide is a measure of how well a given nozzle performs, which is what a laboratory reports and what allows two nozzles to be told apart. The next section derives the two standard parameters used for that purpose, the characteristic velocity, which reports the reservoir feeding the duct, and the thrust coefficient, which reports the expansion through it.

Although the present nozzle is not intended for propulsion, the same isentropic relations that define the acoustic horizon also determine the engineering performance of the flow. These quantities provide experimentally accessible observables that will later be correlated with the reconstructed acoustic geometries.

\section{Performance Parameters}
\label{sec:performance parameters}

The parameters derived in this section come from the propulsion literature, where they characterise a thrust chamber. They are the standard way of reducing a nozzle flow to a small number of measured quantities, and the relations are the same whether the duct carries the combustion products of an engine or the air of an analogue experiment. It is the thrust coefficient that is evaluated along the reconstructed nozzles of Part~\ref{part: results}, where it carries the imprint of the black-hole parameters.

\subsection{Characteristic Velocity}
\label{subsec:characteristic velocity}

The efficiency of the combustion and of the propellants in releasing thermal energy from chemical energy is measured by the characteristic velocity $c^*$, which despite the notation is not the speed of sound at the critical point, $c_{s*}$, and is defined as~\cite{Sutton:2016, Huzel:1992, Hill:1992, Turner:2008}
\begin{equation}
\label{eq: cstar definicao}
    c^* = \frac{p_0 \mathcal{A}_*}{\dot{m}},
\end{equation}
where $p_0$ is the chamber (stagnation) pressure, $\mathcal{A}_*$ the throat area, and $\dot{m}$ the mass-flow rate of Eq.~\eqref{eq: fluxo de massa constante}.

The reason for this particular combination is that it isolates the reservoir from everything downstream of the throat. A given stagnation pressure drives a definite mass flow through a throat of a given area, and the ratio between the two measures how much pressure the gas delivers per unit of mass expended, which is what a hotter or lighter gas does better. Since the flow is choked at the throat, the mass flow no longer depends on what happens in the divergent section, so that $c^*$ is blind to the shape of the nozzle and reports the reservoir alone.

Since the mass flow is constant along the nozzle, $\dot{m}$ can be evaluated at the throat, where the velocity is the critical speed of sound, $v_* = c_{s*} = c_{s0}\sqrt{2/(\gamma+1)}$ by Eq.~\eqref{eq: rel crit}, and the density is $\rho_*$ of Eq.~\eqref{eq: rel critico total},
\begin{equation}
\label{eq: fluxo de massa}
    \dot{m} = \rho_* c_{s*} \mathcal{A}_*
    = \rho_0\, c_{s0} \mathcal{A}_* \left(\frac{2}{\gamma+1}\right)^{\frac{\gamma+1}{2(\gamma -1)}}.
\end{equation}
Substituting this into the definition of the characteristic velocity gives
\begin{align}
    c^* = \frac{p_0 \mathcal{A}_*}{\rho_0\, c_{s0} \mathcal{A}_* \left(\frac{2}{\gamma+1}\right)^{\frac{\gamma+1}{2(\gamma -1)}}}  
    = \frac{p_0}{\rho_0} \frac{1}{c_{s0}} \left(\frac{\gamma+1}{2}\right)^{\frac{\gamma+1}{2(\gamma -1)}}, 
\end{align}
and, using $p_0/\rho_0 = R_g T_0/\bar{m} = c_{s0}^2/\gamma$ from Eq.~\eqref{eq: vel gas perfeito},
\begin{equation}
\label{eq: cstar final}
    c^* = \frac{c_{s0}^2}{c_{s0}} \frac{1}{\gamma} \left(\frac{\gamma+1}{2}\right)^{\frac{\gamma+1}{2(\gamma -1)}} = \frac{c_{s0}}{\gamma} \left(\frac{\gamma+1}{2}\right)^{\frac{\gamma+1}{2(\gamma -1)}}.
\end{equation}
For air, $\gamma \approx 1.4$ gives $c^* \simeq 1.234\,c_{s0}$. The sense in which $c^*$ is constant is worth making precise, since it is not a universal constant and nothing in the isentropic relations is being held fixed by hand. It is uniform along the duct, taking the same value at every station, because the two quantities it depends on are themselves uniform, namely the adiabatic index $\gamma$ of the gas and the stagnation sound speed $c_{s0}$, which the flow preserves precisely because it is isentropic and adiabatic. Local quantities such as $p$, $\rho$, $T$ and $c_s$ do vary from station to station, and $c^*$ is built out of stagnation values instead. A different reservoir temperature, or a different gas, gives a different $c^*$. When normalised by the \emph{local} speed of sound $c_s$, which varies along the nozzle as $c_s = c_{s0}\sqrt{T/T_0}$, it therefore acquires an explicit position dependence,
\begin{equation}
\label{eq: cstar adimensional}
    \frac{c^*}{c_s} = \sqrt{\frac{T_0}{T}}\, \frac{1}{\gamma} \left(\frac{\gamma+1}{2}\right)^{\frac{\gamma+1}{2(\gamma -1)}}.
\end{equation}

\subsection{Thrust Coefficient}
\label{subsec:thrust coefficient}

The efficiency of the nozzle in converting thermal energy into kinetic energy is measured by the thrust coefficient $C_F$, defined as
\begin{equation}
\label{eq: definicao coeficiente de empuxo}
C_F = \frac{{F_{\text{thrust}}}}{{\dot{m} c^*}} = \frac{{F_{\text{thrust}}}}{p_0 \mathcal{A}_*}.
\end{equation}
This coefficient quantifies the amplification of thrust produced by the expansion of the gas through the divergent section, relative to the thrust that would be generated were the chamber connected only to the convergent section and the throat. The normalisation by $p_0\mathcal{A}_*$, which is the same combination that defines $c^*$, is what makes the two parameters complementary, since it removes from $C_F$ everything the reservoir contributes and leaves the expansion alone.

The thrust itself follows from a momentum balance. A nozzle discharging a mass flow $\dot m$ at an exit velocity $v_e$ delivers a momentum flux $\dot m\, v_e$, to which is added the force exerted on the exit plane whenever the exhaust does not leave at the ambient pressure,
\begin{equation}
\label{eq: empuxo momento}
    F_{\text{thrust}} = \dot{m}\, v_e + \mathcal{A}_e\qty(p_e - p_{\text{amb}}),
\end{equation}
where $p_e$ and $\mathcal{A}_e$ are the pressure and the cross-sectional area at the exit plane of the nozzle, and $p_{\text{amb}}$ is the ambient pressure. The exit velocity is fixed by the conservation of the stagnation enthalpy~\eqref{eq: entalpia total}, which with Eq.~\eqref{eq: entalpia aa} and the pressure ratio of Eqs.~\eqref{eq: rel local total} gives
\begin{equation}
\label{eq: velocidade de saida}
    v_e^2 = 2\qty(h_0 - h_e) = \frac{2c_{s0}^2}{\gamma-1}\qty[1-\qty(\frac{p_e}{p_0})^{\frac{\gamma-1}{\gamma}}],
\end{equation}
so that the exhaust is fastest when the gas is expanded to the lowest exit pressure. Substituting Eq.~\eqref{eq: velocidade de saida} and the mass flow~\eqref{eq: fluxo de massa} into Eq.~\eqref{eq: empuxo momento}, and using $\rho_0 c_{s0}^2 = \gamma p_0$ from Eq.~\eqref{eq: vel gas perfeito} to collect the stagnation quantities, yields the standard expression~\cite{Sutton:2016,Huzel:1992,Hill:1992,Turner:2008}
\begin{equation}
\label{eq: empuxo}
    F_{\text{thrust}} = p_0\mathcal{A}_* \sqrt{\frac{2\gamma^2}{\gamma-1}\left(\frac{2}{\gamma+1}\right)^{\frac{\gamma+1}{\gamma-1}}\left[1-\left(\frac{p_e}{p_0}\right)^\frac{\gamma-1}{\gamma}\right]} + \mathcal{A}_e (p_e - p_{\text{amb}}),
\end{equation}
in which the first term is the momentum flux of the exhaust and the second the residual pressure force on the exit plane. Dividing by $p_0 \mathcal{A}_*$ gives the coefficient
\begin{equation}
\label{eq: coeficiente de empuxo}
    C_F = \sqrt{\frac{2\gamma^2}{\gamma-1}\left(\frac{2}{\gamma+1}\right)^{\frac{\gamma+1}{\gamma-1}}\left[1-\left(\frac{p_e}{p_0}\right)^\frac{\gamma-1}{\gamma}\right]} + \frac{\mathcal{A}_e}{\mathcal{A}_*} \left(\frac{p_e}{p_0} - \frac{p_{\text{amb}}}{p_0}\right).
\end{equation}
Thus $C_F$ depends on $\gamma$, on the pressure ratio across the nozzle, on the area expansion ratio, and on the ambient pressure. For a local analysis in a vacuum ($p_{\text{amb}}=0$), the coefficient can be written as a function of the longitudinal position $x$, by treating each station along the duct as if it were the exit plane,
\begin{equation}
\label{eq: coeficiente de empuxo local}
    C_F(x) = \sqrt{\frac{2\gamma^2}{\gamma-1}\left(\frac{2}{\gamma+1}\right)^{\frac{\gamma+1}{\gamma-1}}\left[1-\left(\frac{p_x}{p_0}\right)^\frac{\gamma-1}{\gamma}\right]} + \frac{\mathcal{A}_x}{\mathcal{A}_*} \frac{p_x}{p_0},
\end{equation}
where $p_x$ and $\mathcal{A}_x$ are the pressure and area at position $x$. It is Eq.~\eqref{eq: coeficiente de empuxo local} that is evaluated along the reconstructed nozzles of Part~\ref{part: results}. The theoretical maximum is attained for perfect expansion into a vacuum, $p_e = 0$,
\begin{equation}
\label{eq: coeficiente de empuxo maximo}
    C_F^{\max} = \sqrt{\frac{2\gamma^2}{\gamma-1}\left(\frac{2}{\gamma+1}\right)^{\frac{\gamma+1}{\gamma-1}}},
\end{equation}
which depends on the adiabatic index alone and therefore sets, for a given gas, the most that any bell can deliver. For air, $\gamma=1.4$, it gives $C_F^{\max} \simeq 1.812$, against $C_F^{\max} \simeq 2.247$ for the $\gamma\simeq1.2$ typical of hot combustion products.

\begin{figure}[htbp]
    \centering
    \includegraphics[width=0.72\linewidth]{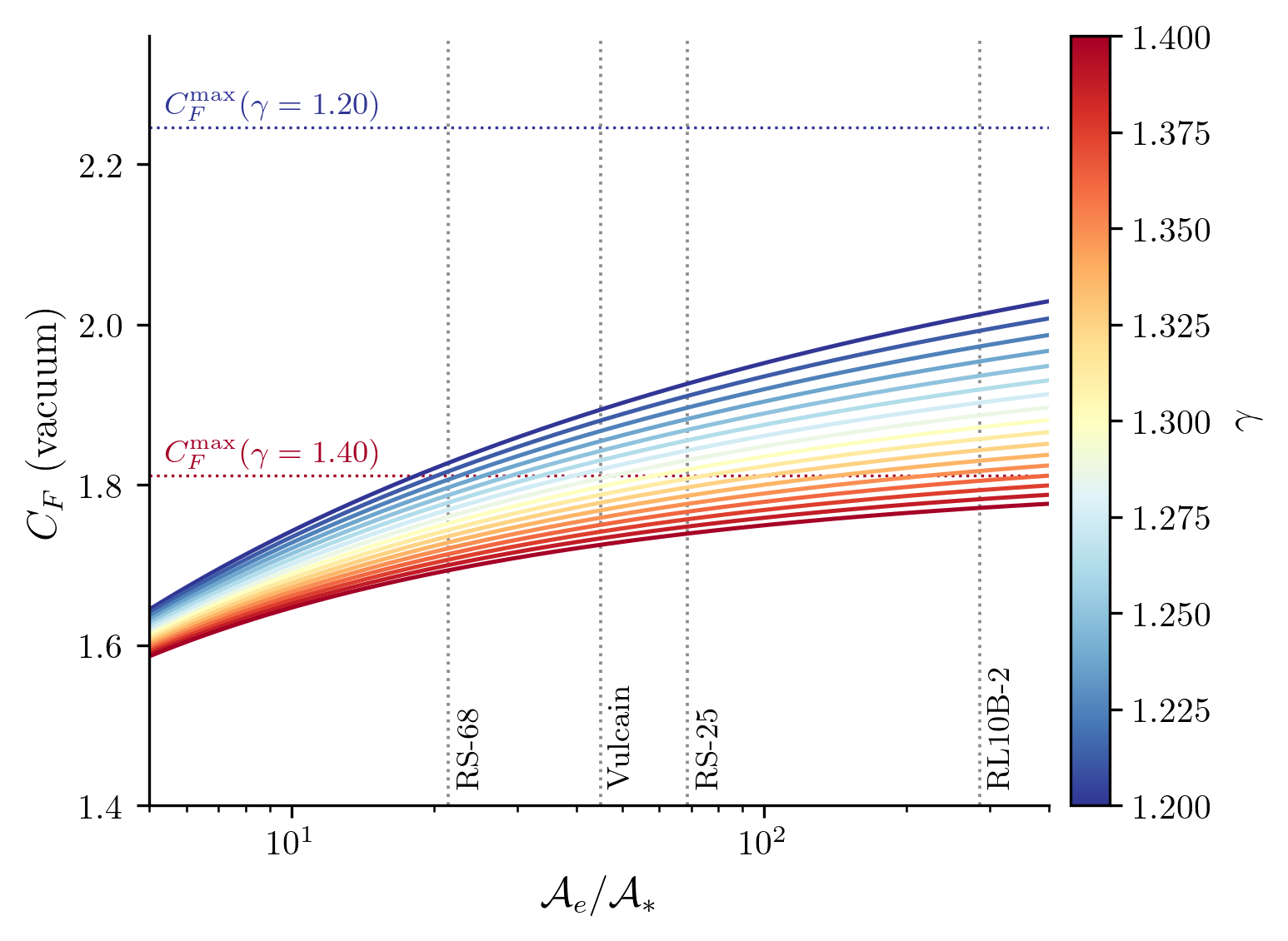}
    \caption[Thrust coefficient in vacuum against the expansion ratio]{Thrust coefficient in vacuum, Eq.~\eqref{eq: coeficiente de empuxo} with $p_{\text{amb}}=0$, against the expansion ratio, for adiabatic indices between the $\gamma\simeq1.2$ of hot combustion products and the $\gamma=1.4$ of air (colour bar). The dotted lines are the perfect-expansion limits~\eqref{eq: coeficiente de empuxo maximo} of the two extremes, and the vertical lines mark the expansion ratios of four engines, from the ground-firing RS-68 to the vacuum-only RL10B-2, as tabulated in Ref.~\cite{Sutton:2016}.}
    \label{fig:cf expansao}
\end{figure}

These are the numbers by which real engines are specified, and it is worth seeing where they fall. Expansion ratios of flight engines span more than an order of magnitude~\cite{Sutton:2016}, from $21.5$ for the RS-68 of the Delta, which fires from the ground, through $45$ for the Vulcain of the Ariane~5 and $68.8$ for the RS-25 of the Space Shuttle, up to $285$ for an upper-stage engine such as the RL10B-2, which fires only in vacuum and can therefore carry an extendible bell far wider than anything useful at sea level. Figure~\ref{fig:cf expansao} places those ratios on the curve that Eq.~\eqref{eq: coeficiente de empuxo} gives in vacuum. The coefficient climbs steeply up to $\mathcal{A}_e/\mathcal{A}_*\sim20$ and then flattens against the limit~\eqref{eq: coeficiente de empuxo maximo}, so that the last factor of four in area, from the RS-25 to the RL10B-2, buys some four per cent of thrust. That is why a bell is always truncated well before the theoretical limit, and measured coefficients fall a few per cent below the ideal curve, since the exhaust is not perfectly axial at the exit plane and the boundary layer of Sec.~\ref{sec:throat horizon} is not accounted for~\cite{Sutton:2016}.

The characteristic velocity is a catalogue quantity in the same way, $1.69$~km/s for the kerosene and oxygen of the RS-27 and $2.31$~km/s for the hydrogen and oxygen of the RL10B-2~\cite{Sutton:2016}, against $c^* \simeq 0.43$~km/s for the room-temperature air of the analogue nozzle. The whole of that difference is stagnation temperature and molecular mass, through $c_{s0}$ in Eq.~\eqref{eq: cstar final}, and none of it is geometry. The analogue duct therefore obeys the same relations as a rocket nozzle and differs from one only in the reservoir that feeds it, which is what makes the aerodynamic literature directly applicable to the profiles reconstructed in Part~\ref{part: results}.

\subsection{Thrust Decomposition}
\label{subsec:thrust decomposition}

Combining the definitions above, the thrust can be written in terms of the three parameters as
\begin{equation}
    F_{\text{thrust}} = \dot{m}\, c^*\, C_F,
\end{equation}
so that it decomposes into three distinct contributions:
\begin{itemize}
    \item $\dot{m}$: the mass-flow rate, reflecting the efficiency of the injectors;
    \item $c^*$: the characteristic velocity, a function of the combustion-chamber parameters and propellant properties;
    \item $C_F$: the thrust coefficient, a function of the nozzle expansion parameters, representing the thermodynamic limit.
\end{itemize}
The decomposition also settles which of the three is the useful observable here. The mass-flow rate and the characteristic velocity are properties of the reservoir and of the throat, and are therefore insensitive to the profile of the divergent section, which is exactly where the black-hole parameters leave their imprint. It is the thrust coefficient that is evaluated along the nozzle reproducing a given black-hole effective potential, for the quantum-corrected and hairy analogues of Part~\ref{part: results}~\cite{Casadio:2024uwj,Casadio:2025lkn}, alongside the Mach-number and thermodynamic profiles.

\section{Summary and Outlook}
\label{sec:isentropic summary}

This chapter assembled the working relations of the laboratory system in which the acoustic horizon of Chapter~\ref{chap: analogue gravity} is realised. For an ideal gas in steady isentropic flow, the conservation of the stagnation enthalpy expresses every thermodynamic ratio as a function of the Mach number alone, Eqs.~\eqref{eq: rel local total}, and mass conservation fixes the cross-sectional area in the same way, Eq.~\eqref{eq: area em relacao ao mach}, the sonic condition being attained at the minimum area, Eq.~\eqref{eq: area mach diferencial}. The throat of a de~Laval nozzle is therefore the acoustic horizon, provided the duct is operated so that the flow reaches it, which demands a pressure ratio of about $34$ across an expansion ratio of four, a shock standing in the divergent section for anything less. Referred to the flow, the acoustic surface gravity is the gradient of the Mach number at the throat, Eq.~\eqref{eq: gravidade superficial bocal}, and the analogue temperature that follows is of the order of tens of nanokelvin for air, which is why the observable pursued here is the scattering spectrum and not the emission.

Two performance parameters were then derived. The characteristic velocity~\eqref{eq: cstar final} is uniform along the duct and measures the reservoir rather than the nozzle, whereas the thrust coefficient~\eqref{eq: coeficiente de empuxo} measures the expansion through the divergent section and therefore depends on the geometry, which is what makes it the observable carried into Part~\ref{part: results}, in the local form~\eqref{eq: coeficiente de empuxo local}. Chapter~\ref{chap:acoustic laval} takes the construction further, recasting both sides in terms of the single variable $g_c=\rho\mathcal{A}/c_s$.

The central result of this chapter is that the Mach-number profile simultaneously determines the acoustic geometry, the nozzle shape, and the thermodynamic state of the flow. This one-to-one correspondence is what allows black-hole effective potentials to be translated into experimentally realisable nozzle geometries.

\chapter{Acoustic Black Holes in de Laval Nozzles}
\label{chap:acoustic laval}

\epigraph{``The only way of discovering the limits of the possible is to venture a little way past them into the impossible.''}{\emph{Profiles of the Future} \\ Arthur C. Clarke}

Building on the acoustic-metric formalism of Chapter~\ref{chap: analogue gravity} and the isentropic flow relations of Chapter~\ref{chap:isentropic equations}, this chapter specialises the propagation of linear perturbations to the quasi-one-dimensional geometry of a de~Laval nozzle, recasting it as a Schr\"odinger-type wave equation with an explicit effective potential, and sets out the computational procedure used to reconstruct a nozzle profile from a prescribed potential. 

The chapter is organised as follows. Sec.~\ref{sec:schrodinger_acoustic_wave} derives the Schr\"odinger-type acoustic wave equation and shows that its effective potential is fixed by the single combination $g_c=\rho\mathcal{A}/c_s$ of the density, the cross-sectional area, and the local speed of sound. That combination is not an aerodynamic observable in its own right, but the variable in which the acoustic and the gravitational problems take the same form, and it is called here the gravitational-acoustic mapping function. Sec.~\ref{sec:construcaometodocomp} then writes the nozzle relations in that same variable, providing the working tool used in Part~\ref{part: results} to construct the acoustic analogues of the exotic black holes of Part~\ref{part: extensions of gr}. Finally, Sec.~\ref{sec:laval summary} closes the chapter with a summary of the results and their role in the remainder of the thesis.

\section{Schr\"odinger-type Acoustic Wave Equation}
\label{sec:schrodinger_acoustic_wave}

Returning to the equations for linear perturbations in ideal flows from Sec.~\ref{sec:linear perturbations in fluids}, the quasi-one-dimensional approximation is adopted, assuming the nozzle radius $r(x)$ varies slowly along the axis $x$, so that only longitudinal modes are retained and the transverse acoustic modes of the duct are neglected. Equation~\eqref{eq propag som area}, the form in which the cross-sectional area is retained, then reduces to a single spatial coordinate, with $\delta \psi = \delta\psi (t,x)$ and $v_x \equiv v$,
\begin{align}
\label{eq:eq onda unidimensional}
    \qty[\qty(\pdv{}{t} + \pdv{}{x}\,v) \frac{\rho \mathcal{A}}{c^2_s} \qty(\pdv{}{t} + v\pdv{}{x}) - \pdv{}{x} \qty(\rho \mathcal{A} \pdv{}{x})] \delta\psi = 0.
\end{align}
In the leftmost bracket, $\pdv*{}{x}\,v$ is the operator that multiplies by $v$ and then differentiates, not the derivative of $v$. The asymmetry between that bracket and the rightmost one is inherited from Eq.~\eqref{eq propag som area}, where the coefficients sit inside the outer derivatives and outside the inner ones.

Assuming stationary solutions via the Fourier transform 
\begin{equation}
    \delta\psi(x,t) = \frac{1}{2\pi} \int \dd \omega_{\text{ac}} e^{-i\omega_{\text{ac}} t} \delta\psi_\omega(x),
\end{equation}
the time-independent differential equation follows:
\begin{multline}
\label{eq: eq onda diferencial independente do tempo}
\frac{1}{2\pi} \int \dd \omega_{\text{ac}}  e^{-i \omega_{\text{ac}} t}\left\{\rho \mathcal{A}\qty(1-\frac{v^{2}}{c_{s}^{2}}) \dv[2]{}{x} +\qty[\dv{(\rho \mathcal{A})}{x}+ 2 i \omega_{\text{ac}} \frac{\rho \mathcal{A} v}{c_{s}^{2}}-\dv{}{x} \qty(\frac{\rho \mathcal{A} v^{2}}{c_{s}^{2}})] \dv{}{x}\right. \\
\left.+\qty[\omega_{\text{ac}}^{2} \frac{\rho \mathcal{A}}{c_{s}^{2}}+i \omega_{\text{ac}} \dv{}{x}\qty(\frac{\rho \mathcal{A} v}{c_{s}^{2}})] \right\}\delta\psi_{\omega} =0.
\end{multline}
Since this must hold at every instant, the expression in braces vanishes for each frequency separately, leaving an ordinary differential equation in $x$ for $\delta\psi_\omega$. It carries a first-derivative term, part of it independent of $\omega_{\text{ac}}$ and part linear in it, and the next step removes both.

Following \cite{Okuzumi:2007hf, Abdalla:2007dz}, a transfer function is introduced, defined as:
\begin{equation}
\label{eq:funcao transferencia}
    H_\omega (x) = \sqrt{g_c} \int \dd t\, e^{i\omega_{\text{ac}} \qty[t + f(x)]} \delta\psi(t,x),
\end{equation}
where
\begin{align}
\label{eq: fase acustica}
    f(x) = \int^{x} \frac{\abs{v(x')}}{c^2_s(x') - v^2(x')}\,\dd x',
\end{align}
and
\begin{equation}
\label{eq: definicao gc}
        g_c \equiv \frac{\rho \mathcal{A}}{c_s}
\end{equation}
is the \emph{gravitational-acoustic mapping function}, which connects the thermodynamic and aerodynamic quantities of the nozzle to the geometric ones of the black hole. It is not a measured property of the flow, nor a quantity with a standing meaning in aerodynamics, but rather the combination in which the acoustic and gravitational descriptions become formally identical, and it is introduced here for that purpose alone. The two factors in Eq.~\eqref{eq:funcao transferencia} play distinct roles: the phase $f(x)$ removes the term linear in $\omega_{\text{ac}}$, and the amplitude $\sqrt{g_c}$ removes the first derivative in $x$, which is the standard reduction of a second-order equation to normal form. Appendix~\ref{ap: reducao schrodinger} carries that reduction out term by term, and only its outcome is needed here.

The phase is fixed rather than chosen, since demanding the cancellation of the term linear in $\omega_{\text{ac}}$ is what gives Eq.~\eqref{eq: fase acustica}, and the same choice removes every remaining odd power of $\omega_{\text{ac}}$. Writing $\delta\psi_\omega = e^{-i\omega_{\text{ac}} f}u$, the equation is left in self-adjoint form,
\begin{equation}
\label{eq: forma autoadjunta}
    \dv{}{x}\qty[\rho\mathcal{A}\qty(1-\mathrm{M}^2)\dv{u}{x}]
    + \frac{\omega_{\text{ac}}^2}{c_s^2}\,\frac{\rho\mathcal{A}}{1-\mathrm{M}^2}\,u = 0.
\end{equation}

The acoustic tortoise coordinate of Chapter~\ref{chap: analogue gravity}, Eq.~\eqref{eq: tartaruga acustica}, is then carried over to the axis of the duct, the ratio of the stagnation sound speed $c_{s0}$ to the local sound speed $c_s$ giving $x_\star$ dimensions of length,
\begin{equation}
\label{eq: coordenada tartaruga bna}
    \dd x_\star = \frac{c_{s0}}{c_s} \frac{\dd x}{(1 - \mathrm{M}^2)}.
\end{equation}
Its integrand diverges at $\mathrm{M}=1$, so the throat lies at infinite $x_\star$, which is what makes it a horizon for sound. The change of variable is also what brings $g_c$ out, since $\rho\mathcal{A}\qty(1-\mathrm{M}^2)\dv*{x_\star}{x} = c_{s0}\,g_c$, so that Eq.~\eqref{eq: forma autoadjunta} becomes
\begin{equation}
\label{eq: sturm liouville acustico}
    \dv{}{x_\star}\qty(g_c \dv{u}{x_\star}) + \frac{\omega_{\text{ac}}^2}{c_{s0}^2}\,g_c\,u = 0,
\end{equation}
a Sturm--Liouville problem in which $g_c$ enters twice as the same weight. That is the precise sense in which it is the natural variable of the problem rather than a convenient shorthand.

The amplitude closes the reduction. Substituting $u = g_c^{-1/2} H_\omega$ in Eq.~\eqref{eq: sturm liouville acustico} cancels the first-derivative term and leaves the Schr\"odinger form:
\begin{equation}
    \label{eq:equacao schrodinger bna}
    \qty[\dv[2]{}{x_\star} + \frac{\omega_{\text{ac}}^2}{c_{s0}^2} -  V_{\text{eff}}(x_\star)] H_\omega(x_\star) = 0,
\end{equation}
with the effective acoustic potential:
\begin{equation}
    \label{eq: pot eff bna}
    V_{\text{eff}}(x_\star) = \frac{1}{2g_c} \dv[2]{g_c}{x_\star} - \frac{1}{4g_c^2}\qty(\dv{g_c}{x_\star})^2.
\end{equation}

This potential $V_{\text{eff}}(x_\star)$ (in units of $\text{m}^{-2}$) governs the scattering of sound waves in the acoustic black hole, completing the analogy with the gravitational case. Nothing in the reduction used a property of the flow, the density, the area, the sound speed and the velocity having entered as arbitrary functions of $x$, so the passage from Eq.~\eqref{eq: eq onda diferencial independente do tempo} to Eq.~\eqref{eq:equacao schrodinger bna} is an exact rearrangement rather than an approximation, and the quasi-one-dimensional assumption of Eq.~\eqref{eq:eq onda unidimensional} remains the only one made. The isentropic relations enter only in the next section, where $g_c$ is tied to the shape of the duct.

\begin{figure}[htbp]
    \centering
    \includegraphics[width=\linewidth]{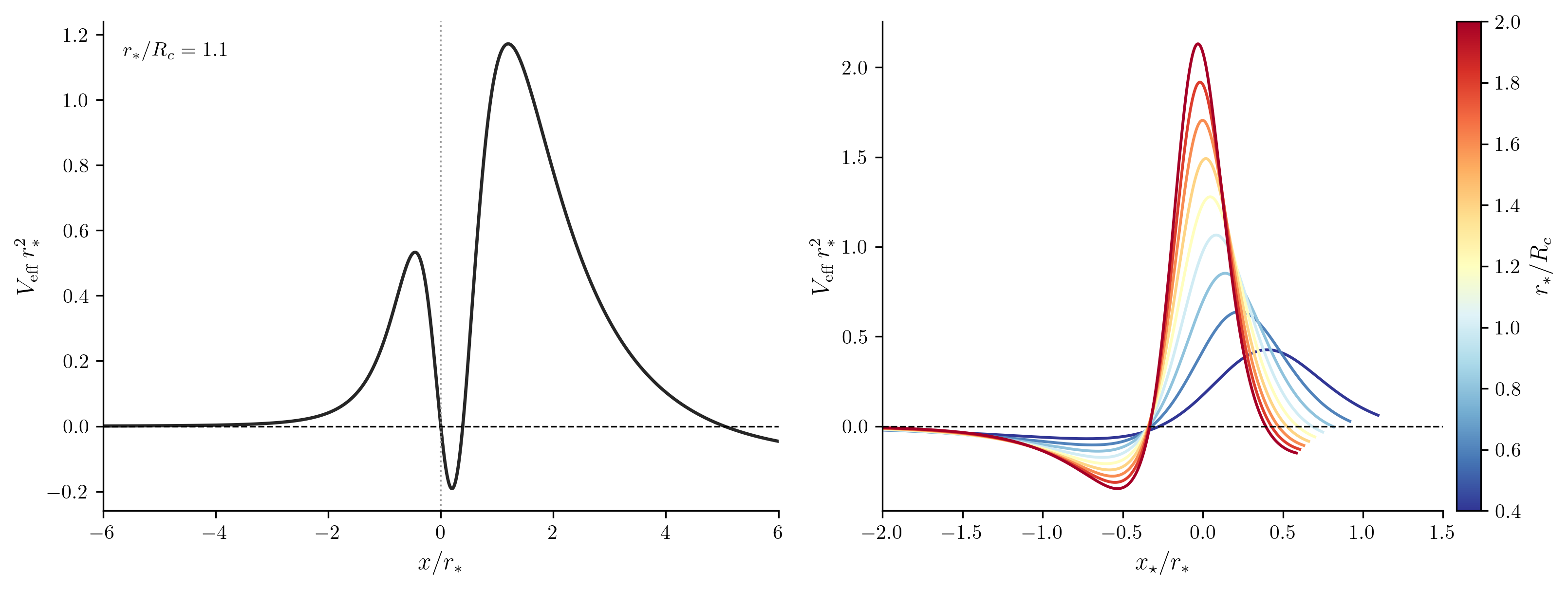}
    \caption[The acoustic effective potential of a de~Laval nozzle]{Effective acoustic potential~\eqref{eq: pot eff bna} of a de~Laval nozzle, in units of the throat radius $r_*$, for the area distribution $\mathcal{A}/\mathcal{A}_*=1+(r_*/R_c)(x/r_*)^2$ of Fig.~\ref{fig:bocal laval} and air ($\gamma=1.4$). (a) Along the axis of the duct, for $r_*/R_c=1.1$, with the throat dotted. (b) Along the acoustic tortoise coordinate~\eqref{eq: coordenada tartaruga bna} on the divergent branch, as the curvature of the throat is varied (colour bar). The additive constant of $x_\star$ is fixed by $x_\star=0$ at $x=r_*$, and its sign is chosen so that the throat lies at $x_\star\to-\infty$.}
    \label{fig: potencial acustico}
\end{figure}

Being an explicit function of the flow, the potential can be drawn for a given duct without any reference to a black hole. Figure~\ref{fig: potencial acustico} does so for the illustrative nozzle of Fig.~\ref{fig:bocal laval}, and the outcome is a barrier of the same type as the gravitational ones of Chapter~\ref{chap:conexaoentrebnsebocais}, sharper for a tighter throat. Matching the two is what Part~\ref{part: results} carries out.

The nature of that analogy is worth stating plainly before the construction proceeds. What the remainder of this chapter does is map a black-hole geometry onto a nozzle flow, by requiring the two effective potentials to agree, and the nozzle so obtained reproduces the scattering of perturbations off that geometry. It is not a fluid system that reproduces the gravitational dynamics, no field equation being imposed on the flow, and the parameters of the black hole survive in the nozzle only as features of its shape and of its thermodynamic profiles.

\section{The de Laval Nozzle Equations in Terms of $g_c$}
\label{sec:construcaometodocomp}

The equations describing the acoustic behaviour of the nozzle must now be written in the same variable as the effective potential~\eqref{eq: pot eff bna}, that is, in terms of the gravitational-acoustic mapping function $g_c$ of Eq.~\eqref{eq: definicao gc}, following Ref.~\cite{Abdalla:2007dz}. The section proceeds in two steps. The cross-sectional area, the density, and the Mach number of the nozzle flow are first written as functions of $g_c$ alone, which is what places the duct and the black hole in a common variable, and the sonic condition at the throat then fixes the boundary value required by the integration. What remains for the correspondence to be complete is the relation between the radial coordinate of the black hole and the longitudinal coordinate of the duct, which is derived at the start of Chapter~\ref{chap:original results}, where it is put to use.

Since Eqs.~\eqref{eq:equacao schrodinger bna} and~\eqref{eq: pot eff bna} are formulated in terms of $g_c$, it is convenient to express the nozzle area equation in the same variable. From this point onward, and for the remainder of the thesis, the density and the cross-sectional area enter only in relative form, the density in units of its stagnation value and the area in units of the throat area, up to the constant of the gas
\begin{equation}
\label{eq: kappa gamma}
    \kappa_\gamma = \sqrt{\frac{\gamma-1}{2}}\qty(\frac{2}{\gamma+1})^{\frac{\gamma+1}{2(\gamma-1)}},
\end{equation}
which is the numerator of Eq.~\eqref{eq: area em funcao da pressao ou densidade} and equals $0.259$ for air. The two relative variables are written with a subscript throughout,
\begin{equation}
\label{eq: variaveis relativas}
    \rho_{\text{rel}} \equiv \frac{\rho}{\rho_0},
    \qquad
    \mathcal{A}_{\text{rel}} \equiv \frac{\mathcal{A}}{\kappa_\gamma \mathcal{A}_*},
\end{equation}
so that $\rho$ and $\mathcal{A}$ keep their dimensional meaning wherever they appear, as in the definition~\eqref{eq: definicao gc} of $g_c$ itself.

One point of the area normalisation deserves care. The constant in Eq.~\eqref{eq: variaveis relativas} is fixed by convenience rather than by the sonic condition, so that $\mathcal{A}_{\text{rel}}$ takes the value $1/\kappa_\gamma \simeq 3.86$ at the throat rather than unity, and it differs from the ratio $\mathcal{A}/\mathcal{A}_*$ used in Chapter~\ref{chap:isentropic equations} by that same factor. Since only the shape of the duct carries physical content, a constant rescaling of the area is immaterial, but the distinction has to be kept in mind when absolute values are read off the profiles.

Using the definition of $g_c$ alongside the isentropic gas relations and the area relation in terms of pressure or density, detailed in Chapter~\ref{chap:isentropic equations}, Eq.~\eqref{eq: area em funcao da pressao ou densidade}, one obtains:
\begin{align}
\label{eq: g e a propto}
    g_c \propto \frac{\rho_{\text{rel}} \mathcal{A}_{\text{rel}}}{\rho_{\text{rel}}^{\frac{\gamma -1}{2}}}, \qquad
    \mathcal{A}_{\text{rel}}^{-1} \propto \rho_{\text{rel}} \qty[1-\rho_{\text{rel}}^{(\gamma-1)}]^{1/2}.
\end{align}
Equations~\eqref{eq:equacao schrodinger bna} and~\eqref{eq: pot eff bna} remain invariant under any scaling of the factor $g_c$, so the proportionality constants in Eq.~\eqref{eq: g e a propto} can be fixed arbitrarily as:
\begin{align}
\label{eq: g e a igual}
    g_c = \frac{\rho_{\text{rel}} \mathcal{A}_{\text{rel}}}{2\rho_{\text{rel}}^{\frac{\gamma -1}{2}}}, \qquad
    \mathcal{A}_{\text{rel}}^{-1} = \rho_{\text{rel}} \qty[1-\rho_{\text{rel}}^{(\gamma-1)}]^{1/2}.
\end{align}
It follows that:
\begin{align}
\label{eq: rho em funcao de g}
      g_c = \frac{ \rho_{\text{rel}} ^{1-\gamma} }
    {2 \qty[ \rho_{\text{rel}} ^{(1-\gamma)} -1] ^{1/2}}
    \quad \Longrightarrow \quad
    \rho_{\text{rel}}^{1-\gamma} = 2g_c^2\qty(1-\sqrt{1-g_c^{-2}}).
\end{align}
Substituting Eq.~\eqref{eq: rho em funcao de g} into Eq.~\eqref{eq: g e a igual}, the cross-sectional area can be expressed as a function of $g_c$:
\begin{equation}
\label{eq: area em funcao de g}
    \mathcal{A}_{\text{rel}} = \frac{\sqrt{2}\qty[2g_c^2\qty(1-\sqrt{1-g_c^{-2}})]^{\frac{1}{\gamma-1}}}{\qty(1-\sqrt{1-g_c^{-2}})^{1/2}}.
\end{equation}
Using Eq.~\eqref{eq: rel local total} together with Eq.~\eqref{eq: rho em funcao de g}, the Mach number can be determined in terms of $g_c$:
\begin{align}
\label{eq: mach em funcao de g}
    \rho_{\text{rel}}^{1-\gamma}=1+\frac{\gamma-1}{2}\mathrm{M}^2
    \quad \Longrightarrow \quad
    \mathrm{M}^2 = \frac{2}{\gamma-1}\qty[2g_c^2\qty(1-\sqrt{1-g_c^{-2}})-1].
\end{align}
Solving for $g_c$:
\begin{equation}
\label{eq: g em funcao de mach}
g_c = \frac{1}{\mathrm{M}}\qty(1+\frac{\gamma-1}{2}\mathrm{M}^2)\frac{1}{\sqrt{2}\sqrt{\gamma-1}}.
\end{equation}

Equation~\eqref{eq: g em funcao de mach} is not monotonic, and that carries a consequence for the inversion. Read as a quadratic in $\mathrm{M}$, it is solved by
\begin{equation}
\label{eq: mach duas raizes}
    \mathrm{M}_\pm = \mathrm{M}_*\qty(g_c \pm \sqrt{g_c^2-1}),
    \qquad
    \mathrm{M}_+ \mathrm{M}_- = \mathrm{M}_*^2,
    \qquad
    \mathrm{M}_* \equiv \sqrt{\frac{2}{\gamma-1}},
\end{equation}
so that each value of $g_c$ corresponds to two Mach numbers, one on either side of $\mathrm{M}_*$, which is $\sqrt5 \simeq 2.24$ for air. The two roots meet at $g_c=1$, where both equal $\mathrm{M}_*$ and where the temperature has fallen to half its stagnation value. Equation~\eqref{eq: g em funcao de mach} therefore gives $g_c \geq 1$ for every Mach number, with equality only at $\mathrm{M}=\mathrm{M}_*$, which is what makes the square roots in Eqs.~\eqref{eq: rho em funcao de g} and~\eqref{eq: mach duas raizes} real. The root displayed in Eq.~\eqref{eq: mach em funcao de g} is $\mathrm{M}_-$ and covers only $\mathrm{M}\leq\mathrm{M}_*$. The divergent section of the nozzles reconstructed in Part~\ref{part: results} runs well past that value, the exit Mach number of the Schwarzschild analogue being $6.23$, so that the branch carried there is $\mathrm{M}_+$. Which of the two applies is settled by the station along the duct and not by the equation itself, exactly as for the area-Mach relation~\eqref{eq: area em relacao ao mach} of Chapter~\ref{chap:isentropic equations}.

\begin{figure}[htbp]
    \centering
    \begin{subfigure}[b]{0.49\textwidth}
        \centering
        \includegraphics[width=\linewidth]{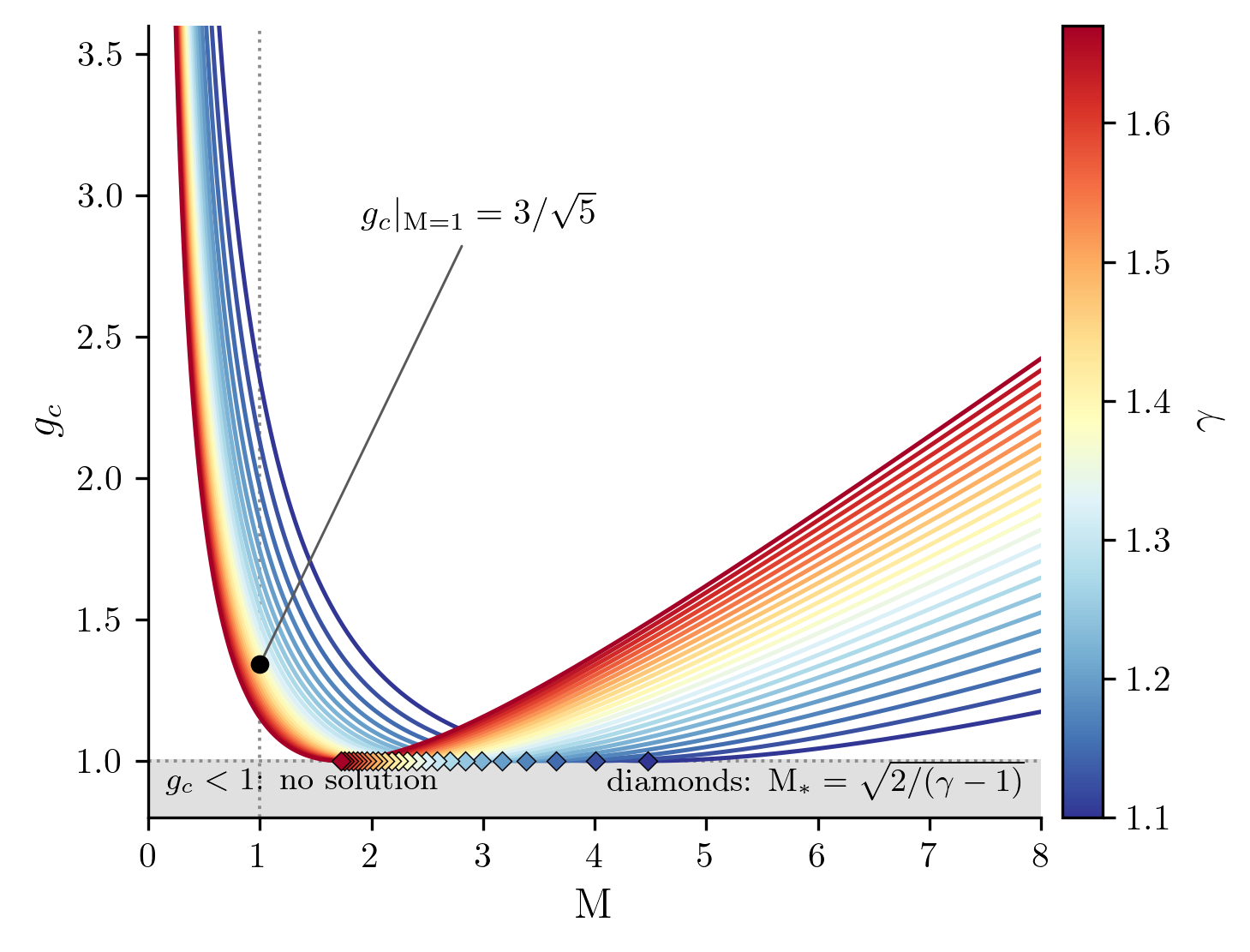}
        \caption{}
        \label{fig: gc mach}
    \end{subfigure}
    \hfill
    \begin{subfigure}[b]{0.49\textwidth}
        \centering
        \includegraphics[width=\linewidth]{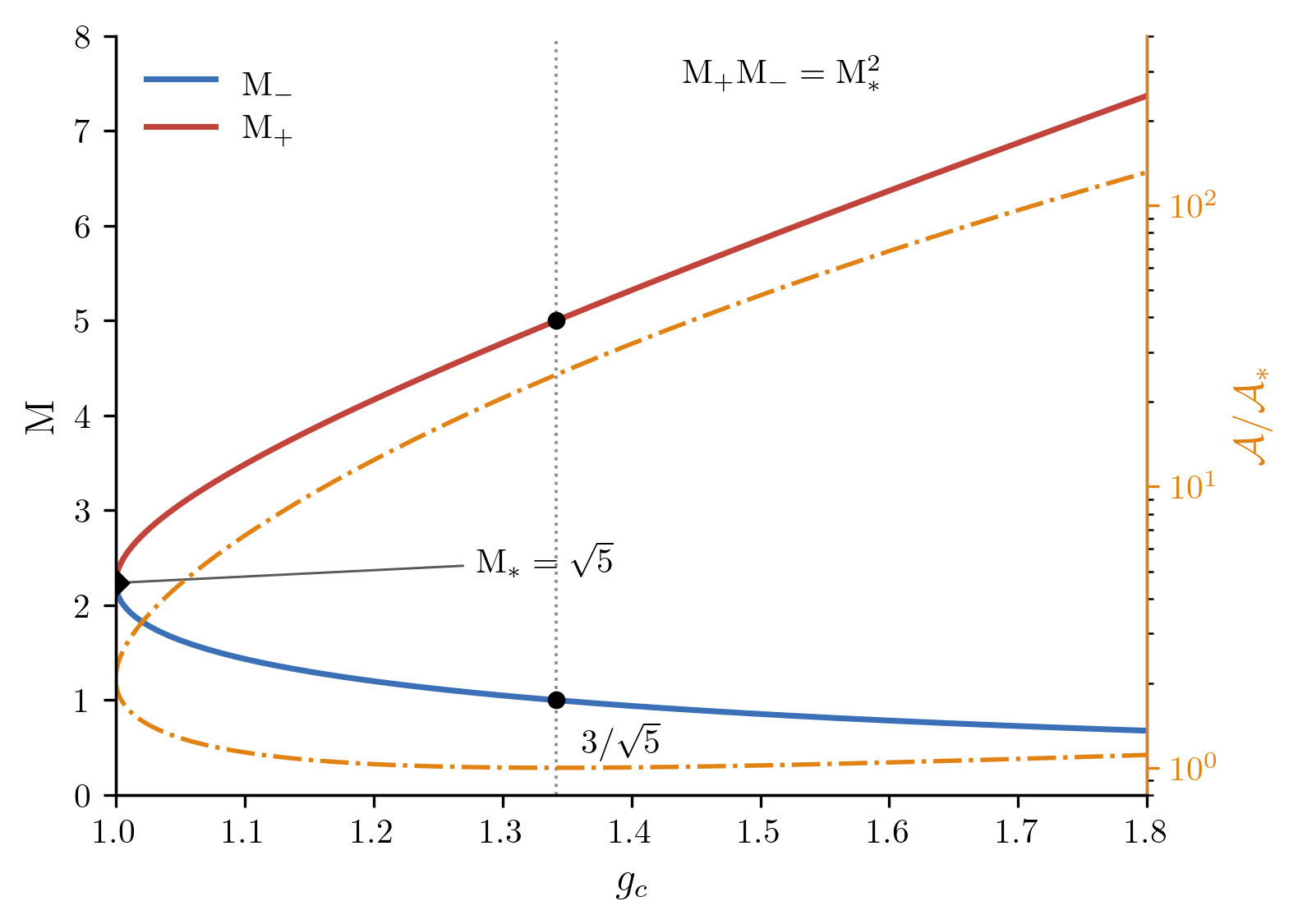}
        \caption{}
        \label{fig: gc inversao}
    \end{subfigure}
    \caption[The gravitational-acoustic mapping function and its inversion]{The gravitational-acoustic mapping function $g_c=\rho\mathcal{A}/c_s$ as a dictionary between the nozzle flow and the acoustic geometry. (a) Relation~\eqref{eq: g em funcao de mach} between $g_c$ and the Mach number as the adiabatic index is varied (colour bar). Every curve has the same minimum $g_c=1$ at $\mathrm{M}_*=\sqrt{2/(\gamma-1)}$ (diamonds), the shaded band is the region no flow can reach, and the circle marks the throat value~\eqref{eq: cond contorno g} for air. (b) The inversion~\eqref{eq: mach duas raizes} for air, its two roots meeting at the branch point $g_c=1$ (diamond), with the boundary value $3/\sqrt5$ dotted (circles). The right-hand axis gives the area ratio $\mathcal{A}/\mathcal{A}_*$ along each branch, that is, $\kappa_\gamma \mathcal{A}_{\text{rel}}$ from Eqs.~\eqref{eq: variaveis relativas} and~\eqref{eq: area em funcao de g}.}
    \label{fig: gc dictionary}
\end{figure}

At the throat the sonic condition $\mathrm{M}=1$ fixes the value of $g_c$ there, which for air ($\gamma \approx 1.4$) reads
\begin{equation}
\label{eq: cond contorno g}
    \left.g_c\right|_{\mathrm{M}=1} = \frac{\gamma+1}{2\sqrt{2}\sqrt{\gamma-1}}=\frac{3}{\sqrt{5}}>1,
\end{equation}
and serves as the boundary condition for the subsequent integrations. Both features are displayed in Fig.~\ref{fig: gc dictionary}, which also shows that the same value $3/\sqrt5$ is attained again at $\mathrm{M}=5$ on the supersonic branch, as Eq.~\eqref{eq: mach duas raizes} requires.

\section{Summary and Outlook}
\label{sec:laval summary}

This chapter completed the acoustic side of the correspondence by moving from the idealised geometry of Chapter~\ref{chap: analogue gravity} to a system that can be built. In the quasi-one-dimensional regime, the acoustic perturbation of a nozzle flow obeys the Schr\"odinger-type equation~\eqref{eq:equacao schrodinger bna}, with the effective potential~\eqref{eq: pot eff bna} determined by a single variable, the gravitational-acoustic mapping function $g_c = \rho\mathcal{A}/c_s$. This is the point at which the factor $\rho/c_s$ set aside in the canonical model of Chapter~\ref{chap: analogue gravity} returns, no longer as a constant conformal rescaling but as the object whose derivatives with respect to the acoustic tortoise coordinate~\eqref{eq: coordenada tartaruga bna} build the barrier itself.

The isentropic relations of Chapter~\ref{chap:isentropic equations} then close the system, since they express the area, the density, and the Mach number as functions of $g_c$ alone, Eqs.~\eqref{eq: area em funcao de g} and~\eqref{eq: mach em funcao de g}, with the sonic condition at the throat fixing the boundary value~\eqref{eq: cond contorno g}. The correspondence between $g_c$ and the Mach number is two-valued, Eq.~\eqref{eq: mach duas raizes}, the branch being selected by the station along the duct, and Fig.~\ref{fig: gc dictionary} collects both statements. Two features of the potential~\eqref{eq: pot eff bna} deserve emphasis. It is invariant under any rescaling of $g_c$, which is what allowed the proportionality constants in Eq.~\eqref{eq: g e a propto} to be fixed arbitrarily, and it involves only $g_c$ and its derivatives, so that a prescribed potential determines $g_c$ by integration and the nozzle profile follows. The procedure therefore runs in the direction required by Part~\ref{part: results}: given the effective potential of a black hole from Part~\ref{part: extensions of gr}, computed in Chapter~\ref{chap:conexaoentrebnsebocais}, the nozzle that reproduces it can be reconstructed, together with the thermodynamic and performance quantities of Chapter~\ref{chap:isentropic equations}. The constraint equations that make this inversion explicit are derived at the start of Chapter~\ref{chap:original results}.

\part{Bridging Gravity and Acoustics} \label{part: results}


\chapter{Acoustic Analogues for Exotic Black Holes}
\label{chap:original results}

\epigraph{``We have found a strange foot-print on the shores of the unknown.''}{\emph{Space, Time and Gravitation} \\ Arthur S. Eddington}

This chapter presents the central original results of the thesis, namely the explicit construction of the acoustic analogues of the two exotic black-hole geometries developed in Part~\ref{part: extensions of gr}, the hairy black hole obtained by gravitational decoupling and the quantum-corrected Schwarzschild metric from the effective field theory of gravity. Building on the correspondence established in Part~\ref{part: quasinormal modes} between the black-hole and acoustic wave equations, the effective potential of each geometry is mapped onto a de~Laval nozzle profile, from which the thermodynamic variables, the Mach number, and the thrust coefficient are reconstructed as functions of the black-hole parameters. The results follow and systematise those of Refs.~\cite{Casadio:2024uwj,Casadio:2025lkn}, the reconstruction procedure being the one introduced for the Schwarzschild and minimally deformed cases in Refs.~\cite{Okuzumi:2007hf,Abdalla:2007dz,daRocha:2017lqj}.

The chapter is organised as follows. Sec.~\ref{sec:constraint_equations} derives the constraint equations that identify the tortoise coordinates of the two systems and determine the function $g_c(r)$ specifying the nozzle profile. Sec.~\ref{sec:schwarzschild analogue} reads the undeformed duct, which is the reference both halves of the chapter are measured against, and shows how it already depends on the channel. Sec.~\ref{sec:results HBH by GD} presents the results for the hairy black hole, weighing the three hair parameters against one another, fixing how far the charge-like one can be taken, and then sweeping it, and Sec.~\ref{sec: results quantum correction} the corresponding results for the quantum-corrected metric as the coupling $c_6$ is swept. Finally, Sec.~\ref{sec:results summary} closes the chapter with a summary of the results and their role in the remainder of the thesis.

\section{Derivation of Constraint Equations}
\label{sec:constraint_equations}

Both systems are now described by a Schr\"odinger-type equation built from an effective potential and a tortoise coordinate, the gravitational one being Eq.~\eqref{equacao de onda} and the acoustic one Eq.~\eqref{eq:equacao schrodinger bna}. Making them the same equation requires two conditions rather than one, and it is worth separating them before either is imposed. The first is that the two effective potentials agree. The second is that they agree as functions of the same variable, since the same potential written in a different coordinate is a different scattering problem, with different transmission through the barrier and a different spectrum. The construction of this chapter therefore imposes both at once, identifying the tortoise coordinates and then demanding that the nozzle reproduce the prescribed black-hole potential in that common coordinate.

Identifying the coordinates does more than fix the map $x=x(r)$. The gravitational tortoise coordinate~\eqref{coordenada tartaruga final} diverges at the event horizon and the acoustic one~\eqref{eq: coordenada tartaruga bna} at the throat, so $\dd r_\star=\dd x_\star$ sends one to the other necessarily, and not by a choice of integration constant. Integrating outwards from there carries the exterior $r>r_\text{EH}$ monotonically onto the divergent section, station by station, while what lies inside the horizon has no counterpart at all. The correspondence is thus between the flow and the exterior of the black hole, which is also the only region a scattering experiment probes.

What that correspondence transfers is wave propagation, not thermodynamics. A field of spin $s$ and multipole $\ell$ scattering off the black hole and a sound wave travelling along the duct obey the same equation, so the barrier, the transmitted and reflected fractions, and the quasinormal frequencies are common to both, up to the scale relating $\omega_{\text{ac}}$ to $\omega$ recorded at the end of this section. The pressure, temperature, and density are not: they are those of the gas in the duct, and the temperature plotted along the nozzle is the static temperature of the flow, unrelated to any temperature attributed to the black hole. The one thermal quantity that does cross over is the analogue Hawking temperature~\eqref{eq: temperatura hawking bocal}, and only formally, since the phonons the duct would emit are quanta of the sound field of a laboratory gas and the estimate of Chapter~\ref{chap:isentropic equations} puts the value at tens of nanokelvin, far below the thermal noise of any real flow.

\subsection{Constraint on the tortoise coordinates}
From Eqs.~\eqref{eq: coordenada tartaruga bna}, \eqref{eq: rho em funcao de g}, and~\eqref{eq: mach em funcao de g}, the first condition reads
\begin{equation}
\label{eq: coordenadas iguais}
    \dd r_\star = \dd x_\star = 
    \frac{c_{s0}}{c_s} \frac{\dd x}{(1 - \mathrm{M}^2)} =
    \frac{\rho_{\text{rel}}^{\frac{1-\gamma}{2}}}{1-\mathrm{M}^2}\dd x =
    \frac{\qty[2g_c^2\qty(1-\sqrt{1-g_c^{-2}})]^{1/2}}{1- \frac{2}{\gamma-1}\qty[2g_c^2\qty(1-\sqrt{1-g_c^{-2}})-1]} \dd x.
\end{equation}
The coordinate relation $r=r(x)$ can be made explicit using Eq.~\eqref{coordenada tartaruga final}. Defining $\mathrm{F}(r) \equiv \sqrt{B/A}$ to simplify calculations and fixing the event horizon radius at $r=r_\text{EH}$, Eq.~\eqref{eq: coordenadas iguais} becomes:
\begin{equation}
\label{eq:xemfuncaoder}
    x = \int_{r_\text{EH}}^{r}{\frac{\qty[\gamma+1-4g_c(r)^2\qty(1-\sqrt{1-g_c(r)^{-2}})]}
    {\mathrm{F}(r)\qty(\gamma-1)\qty[2g_c(r)^2 \qty(1-\sqrt{1-g_c(r)^{-2}})]^{1/2}}}\dd r,
\end{equation}
or, equivalently, in differential form:
\begin{equation}
\label{eq:xemfuncaoder123}
    \dv{x}{r} = \frac{\qty[\gamma+1-4g_c(r)^2\qty(1-\sqrt{1-g_c(r)^{-2}})]}
    {\mathrm{F}(r)\qty(\gamma-1)\qty[2g_c(r)^2 \qty(1-\sqrt{1-g_c(r)^{-2}})]^{1/2}}.
\end{equation}
The integration constant is fixed by the lower limit already written in Eq.~\eqref{eq:xemfuncaoder}, that is, by taking the duct to begin at $x=0$ where $r=r_\text{EH}$, so that the throat of the nozzle is placed at the analogue horizon. Apart from the profile $g_c(r)$ itself, the right-hand side of Eq.~\eqref{eq:xemfuncaoder123} carries only two ingredients, the black hole through $\mathrm{F}=\sqrt{B/A}$ and the working fluid through the adiabatic index $\gamma$. The map $x=x(r)$ is therefore determined once the metric, the gas, and $g_c(r)$ are given, and it is the second condition that supplies the last of the three.

\subsection{Constraint on the potential}
The second condition is the equality of the potentials, and it is what determines $g_c(r)$. With the coordinates already identified, the acoustic potential~\eqref{eq: pot eff bna} can be read as a function of the radial coordinate of the black hole, making the $r$-dependence explicit via Eq.~\eqref{coordenada tartaruga final}:
\begin{align*}
    V_{\text{eff}}(r_\star) &= \frac{1}{2g_c} \dv[2]{g_c}{r_\star} - \frac{1}{4g_c^2}\qty(\dv{g_c}{r_\star})^2 \\
    &=\frac{1}{2g_c} \dv{}{r_\star} \qty[\dv{g_c}{r} \dv{r}{r_\star}] - \frac{1}{4g_c^2} \qty(\dv{g_c}{r} \dv{r}{r_\star})^2 \\
    &=\frac{1}{2g_c} \dv{}{r_\star} \qty[g_c' \mathrm{F}] - \frac{1}{4g_c^2} \qty(g_c' \mathrm{F})^2 \\
    &=\frac{1}{2g_c} \qty[g_c''\dv{r}{r_\star}\mathrm{F}+\mathrm{F}'\dv{r}{r_\star}g_c' ]- \frac{1}{4g_c^2} \qty(g_c' \mathrm{F})^2, 
\end{align*}
yielding:
\begin{equation}
    \label{eq: pot em funcao de r}
    \frac{g''_c(r)}{g_c(r)} + \frac{g'_c(r)}{g_c(r)} \qty(\frac{\mathrm{F}'(r)}{\mathrm{F}(r)} - \frac{1}{2}\frac{g'_c(r)}{g_c(r)}) = \frac{2 V_\text{eff}(r)}{\mathrm{F}^2(r)}.
\end{equation}
The potential on the right of Eq.~\eqref{eq: pot em funcao de r} is now the gravitational one, computed in Chapter~\ref{chap:conexaoentrebnsebocais} for the geometry in question, so that the equation is no longer a definition of $V_\text{eff}$ but an equation for $g_c$ given a prescribed barrier. This is the point at which the correspondence changes direction. Up to Chapter~\ref{chap:acoustic laval} the flow determined the potential, whereas here the potential determines the flow, and with it the shape of the duct.

Since the nozzle carries no angular sector, each profile reproduces the effective potential of a single pair $(s,\ell)$. A given nozzle is therefore the analogue of one multipole of the black-hole spectrum rather than of the geometry as a whole.

Being of second order, Eq.~\eqref{eq: pot em funcao de r} needs two boundary conditions to single out a solution, and both are imposed at the throat, which is where the integration starts. The first is the sonic condition~\eqref{eq: cond contorno g}, which fixes the value of $g_c$ at the station where $\mathrm{M}=1$. The second is the requirement that the solution remain finite and smooth there, which discards the branch that is singular at the throat and fixes the derivative. Their explicit form is given in Eq.~\eqref{eq:condinilamb} below, once the substitution of the next subsection has been made.

\subsection{Solution for the mapping function}
Singularities in solving Eq.~\eqref{eq: pot em funcao de r} are resolved via the substitution $g_c\qty(r) \equiv \chi^2\qty(r)$, yielding a Sturm--Liouville type of problem:
\begin{align}\label{eq: lambda}
\dv{}{r}\qty(\mathrm{F}(r) \dv{\chi(r)}{r}) = \frac{V_\text{eff}(r)}{\mathrm{F}(r)} \chi(r).
\end{align}
For the Schwarzschild metric and bosonic perturbations, Eq.~\eqref{eq: lambda} admits a closed-form solution,
\begin{equation}
\label{gc12}
g_c
=
\frac{\gamma+1}{2\sqrt{2}\sqrt{\gamma-1}}
\qty[\frac{\Gamma(1+\ell+{s}){}_2F_1\qty({s}-\ell,{s}+\ell+1,1+2{s},\dfrac{r}{2G_\text{N} M})}{(2{s})!(\ell-{s})!}\qty(\frac{r}{2G_\text{N} M})^{{s}+1}]^2,
\end{equation}
consistent with prior results~\cite{Abdalla:2007dz}. In the variable $z\equiv r/2G_\text{N}M$, Eq.~\eqref{eq: lambda} has regular singular points at $z=0$, $z=1$ and $z=\infty$, and is thus of hypergeometric type. The horizon sits at $z=1$, and of the two branches there, finiteness selects the one regular at the horizon, whose normalisation the sonic condition~\eqref{eq: cond contorno g} then fixes.

For the deformed metrics no such solution is available and Eq.~\eqref{eq: lambda} has to be integrated numerically, which requires a starting point. Written in normal form,
\begin{align}\label{eq: lambda 2}
\chi''(r) + \frac{\mathrm{F}'(r)}{\mathrm{F}(r)}\chi'(r) - \frac{V_\text{eff}(r)}{\mathrm{F}^2(r)} \chi(r) = 0,
\end{align}
the equation has the appropriate form for the Frobenius method. The horizon is a regular singular point, since $\mathrm{F}'/\mathrm{F}$ and $V_\text{eff}/\mathrm{F}^2$ have at most simple poles there. The two indicial exponents coincide, so that one solution is an ordinary power series, whereas the second generally contains a logarithmic term, and the requirement of finiteness at the horizon selects the first,
\begin{equation}
    \chi\qty(r) = \sum_{k=0}^\infty a_k \qty(r-r_\text{EH})^{k}, \label{eq:frob1}
\end{equation}
the two boundary conditions stated above, the sonic condition~\eqref{eq: cond contorno g} and smoothness at the throat, fixing its first two coefficients,
\begin{equation}
\label{eq:condinilamb}
a_0 = \sqrt{\frac{\gamma+1}{2\sqrt{2}\sqrt{\gamma-1}}}, \quad a_1 = 0,
\end{equation}
and substituting Eq.~\eqref{eq:frob1} into Eq.~\eqref{eq: lambda 2} generates the recurrence relation
\begin{equation}\label{eq: recurrence eq}
    \qty(k +2)\qty(k+1) a_{k+2} + \qty(k+1)a_{k+1} \frac{\mathrm{F}'(r)}{\mathrm{F}(r)} - a_k\frac{V_\text{eff}(r)}{\mathrm{F}^2(r)} =0,
\end{equation}
which leads to
\begin{eqnarray}
\label{eq: serie truncada}
    g_c\qty(r) = {\frac{\gamma+1}{2\sqrt{2}\sqrt{\gamma-1}}} \qty[1 +\frac{1}{2} \frac{
    V_\text{eff}
    (r)}{\mathrm{F}^2(r)} \qty(r-r_\text{EH})^2 - \frac{1}{6} \frac{V_\text{eff}(r)}{\mathrm{F}^2(r)} \frac{\mathrm{F}'(r)}{\mathrm{F}(r)} \qty(r-r_\text{EH})^3 + \cdots]^2.
\end{eqnarray}
The truncation is a local estimate, valid near the horizon. Appendix~\ref{ap: analytical solution for lambda} collects the coefficients to arbitrary order and reports both solutions (closed-form Schwarzschild and power series for spherically symmetric metric).

Equation~\eqref{eq: serie truncada} supplies the initial data slightly outside the horizon, and a Runge--Kutta scheme then integrates Eq.~\eqref{eq: lambda} over $r > r_\text{EH}$. That determines $g_c(r)$, which is subsequently mapped to the nozzle longitudinal coordinate $x$ using Eq.~\eqref{eq:xemfuncaoder123}. The procedure applies unchanged to the hairy metric~\eqref{eq: metric HBH} and to the quantum-corrected metric~\eqref{eq:CK_metric}. One consequence deserves to be recorded here. Carrying the same $V_\text{eff}$ in the same tortoise coordinate, Eqs.~\eqref{eq:equacao schrodinger bna} and~\eqref{equacao de onda} differ only in the eigenvalue term, $\omega_{\text{ac}}^2/c_{s0}^2$ against $\omega^2$, so that $\omega_{\text{ac}} = c_{s0}\,\omega$ and the quasinormal spectrum of the duct is the one computed in Chapter~\ref{chap:QNM_spectrum} for the same channel, rescaled by the stagnation sound speed, and is not recomputed here. This is exact for the potential, and holds for the spectrum so far as the duct, finite by construction, approximates the asymptotic conditions in which those frequencies are defined. Finally, with $g_c$ and $x(r)$ calculated, Eqs.~\eqref{eq: area em funcao de g} and~\eqref{eq: g em funcao de mach} together with \eqref{eq: area em funcao da pressao ou densidade} and \eqref{eq: coeficiente de empuxo local} are used to complete the aerodynamic profile of the analogue.

With the numerical method now fixed, the rest of the chapter applies it. Sec.~\ref{sec:schwarzschild analogue} reads the undeformed duct first, Sec.~\ref{sec:results HBH by GD} takes the hairy black hole obtained by gravitational decoupling, and Sec.~\ref{sec: results quantum correction} the quantum-corrected metric. The conventions are common to the three, so that the two sets of results can be read against each other and against the same Schwarzschild reference. All nozzle geometries are calculated to be symmetric about $x=0$, therefore, the flow is analogue only in the divergent section. The convergent section is displayed and calculated with the Mach-number profile lowest root~\eqref{eq: mach duas raizes}.

A duct machined to any of these profiles would not reproduce the correspondence to the precision the figures display, and the results of this chapter should not be read as the specification of a feasible experiment. Two obstacles are enough to settle the point. The walls dilate under the temperature the flow imposes on them, by an amount that varies from station to station because the static temperature does, so that the area ratio the reconstruction prescribes is not the area ratio the running duct has. Beyond that, the stagnation conditions and the back pressure would have to be held inside a narrow window for the divergent section to remain free of shocks, and a shock anywhere in it destroys the isentropic relations~\eqref{eq: rel local total} on which the whole map rests. What the construction delivers is therefore numerical rather than experimental, in that it recasts the quasinormal problem in the variables of a laboratory flow, where the barrier becomes a wall profile and the ringdown a scattering of sound, without requiring that flow to be realised.

\section{The Schwarzschild Analogue}
\label{sec:schwarzschild analogue}

Both halves of this chapter measure a deformed duct against an undeformed one, and it is the same undeformed duct in the two of them: setting $\alpha=\ell_h=Q=0$ in the hairy metric~\eqref{eq: metric HBH} and $c_6=0$ in the quantum-corrected one~\eqref{eq:CK_metric} returns Schwarzschild in both cases. What is left once the deformation is switched off is the channel, through the pair $(s,\ell)$ that the effective potential carries, and it is worth reading that dependence before any hair or coupling is introduced, since every difference reported below is taken against it.
Figure~\ref{fig:sensitivity_HBHGD} shows the Schwarzschild aerodynamic profile into four panels separated by the barrier each channel carries. Along the diagonal $\ell=s$ the ordering is set by the multipole rather than by the spin, as Chapter~\ref{chap:conexaoentrebnsebocais} establishes from the centrifugal term $\ell(\ell+1)/r^2$, so the barrier grows from the scalar channel to the gravitational one. A taller barrier has to be reproduced by a wider expansion, and the panels show exactly that. Every figure of this chapter is drawn over $\abs{x}\leq20$, so that the exit of the duct is the station $x_e=20$, and it is there that all the values quoted below are read. The Mach number at the exit climbs with the spin, from $5.45$ to $7.35$, $12.35$ and $26.08$, and the wall climbs with it, $\sqrt{\mathcal{A}_{\text{rel}}}$ at $\abs{x}=20$ being $11.7$, $22.3$, $75.2$ and $469.6$. What the spin buys is therefore not a wider throat, which is common to all four, but a steeper divergent section.

\begin{figure}[htbp]
    \centering
    \begin{subfigure}{0.48\textwidth}
        \centering
        \includegraphics[width=\linewidth]{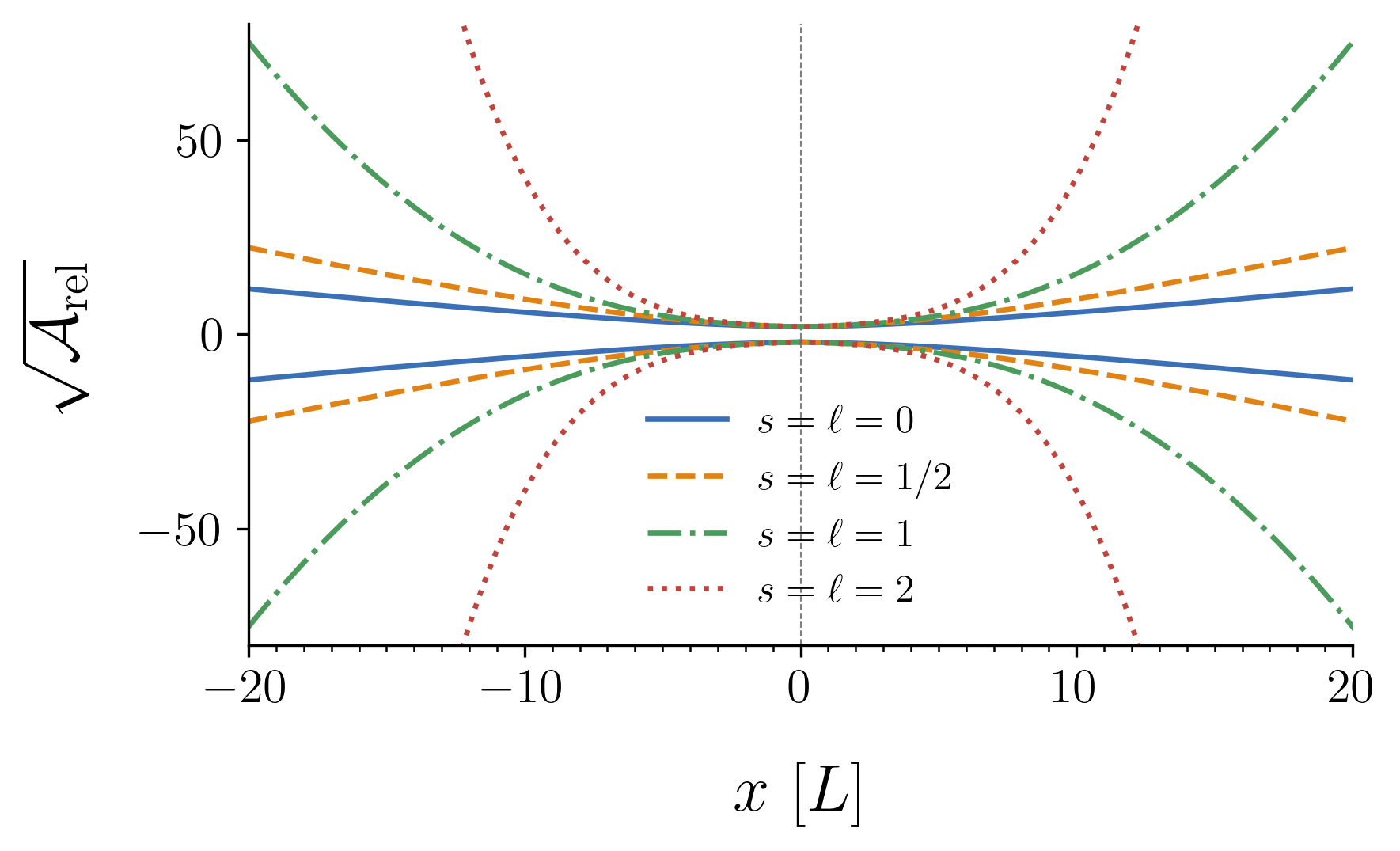}
        \caption{Nozzle shape.}
        \label{fig:schw_nozzle_shape}
    \end{subfigure}
    \hfill
    \begin{subfigure}{0.48\textwidth}
        \centering
        \includegraphics[width=\linewidth]{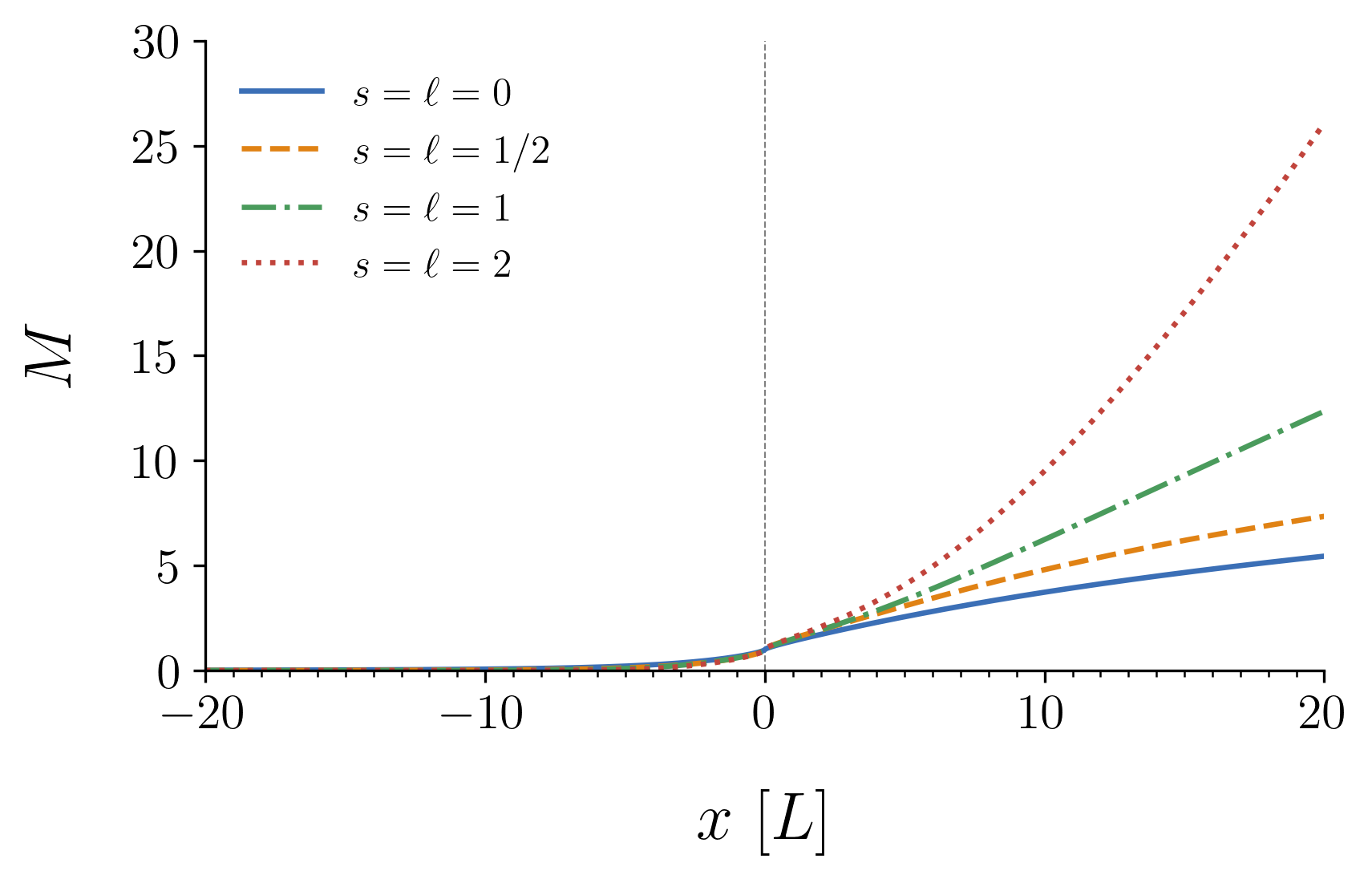}
        \caption{Mach number.}
        \label{fig:schw_Mach}
    \end{subfigure}

    \begin{subfigure}{0.48\textwidth}
        \centering
        \includegraphics[width=\linewidth]{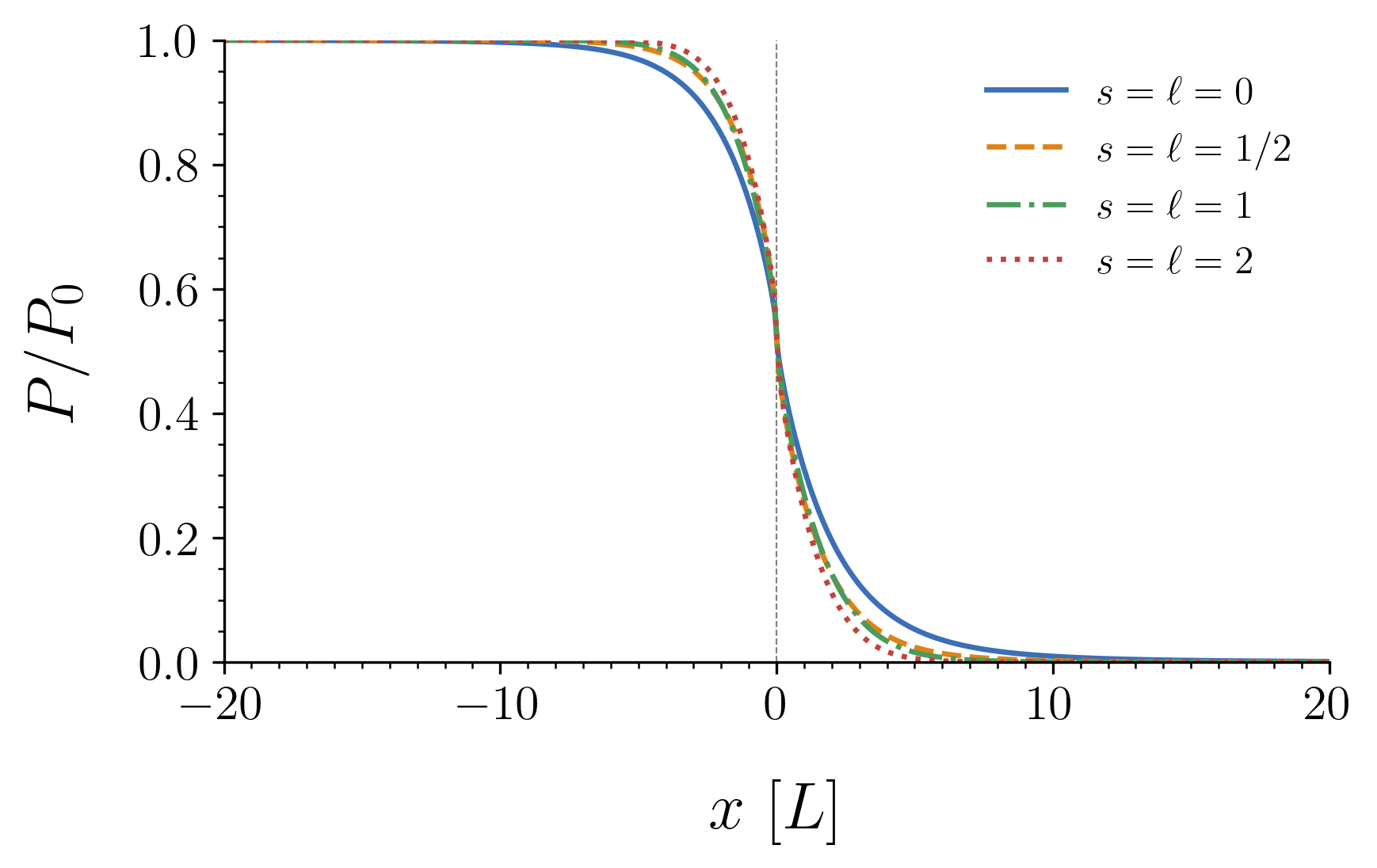}
        \caption{Pressure.}
        \label{fig:schw_Pressure}
    \end{subfigure}
    \hfill
    \begin{subfigure}{0.48\textwidth}
        \centering
        \includegraphics[width=\linewidth]{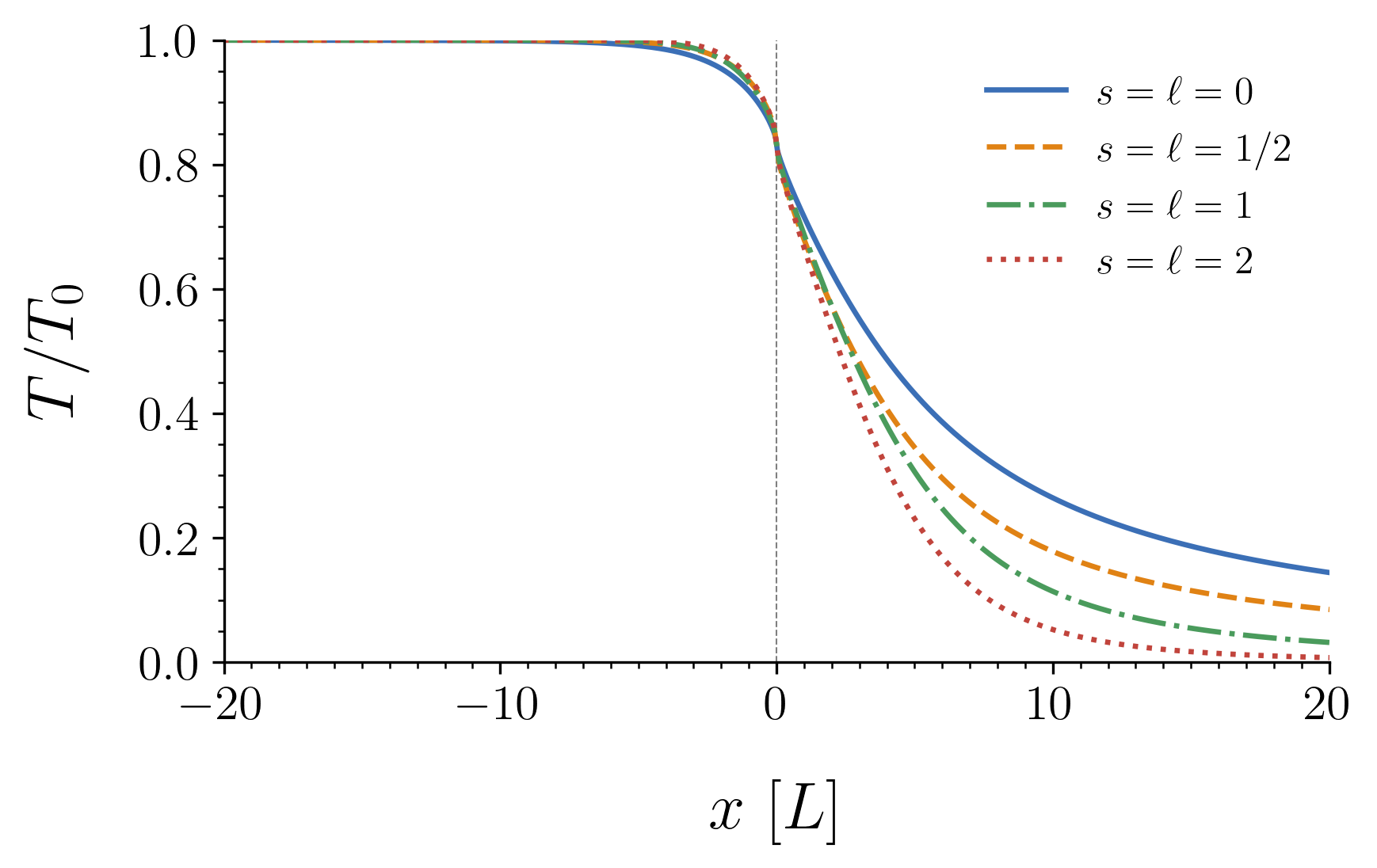}
        \caption{Temperature.}
        \label{fig:schw_Temperature}
    \end{subfigure}

    \begin{subfigure}{0.48\textwidth}
        \centering
        \includegraphics[width=\linewidth]{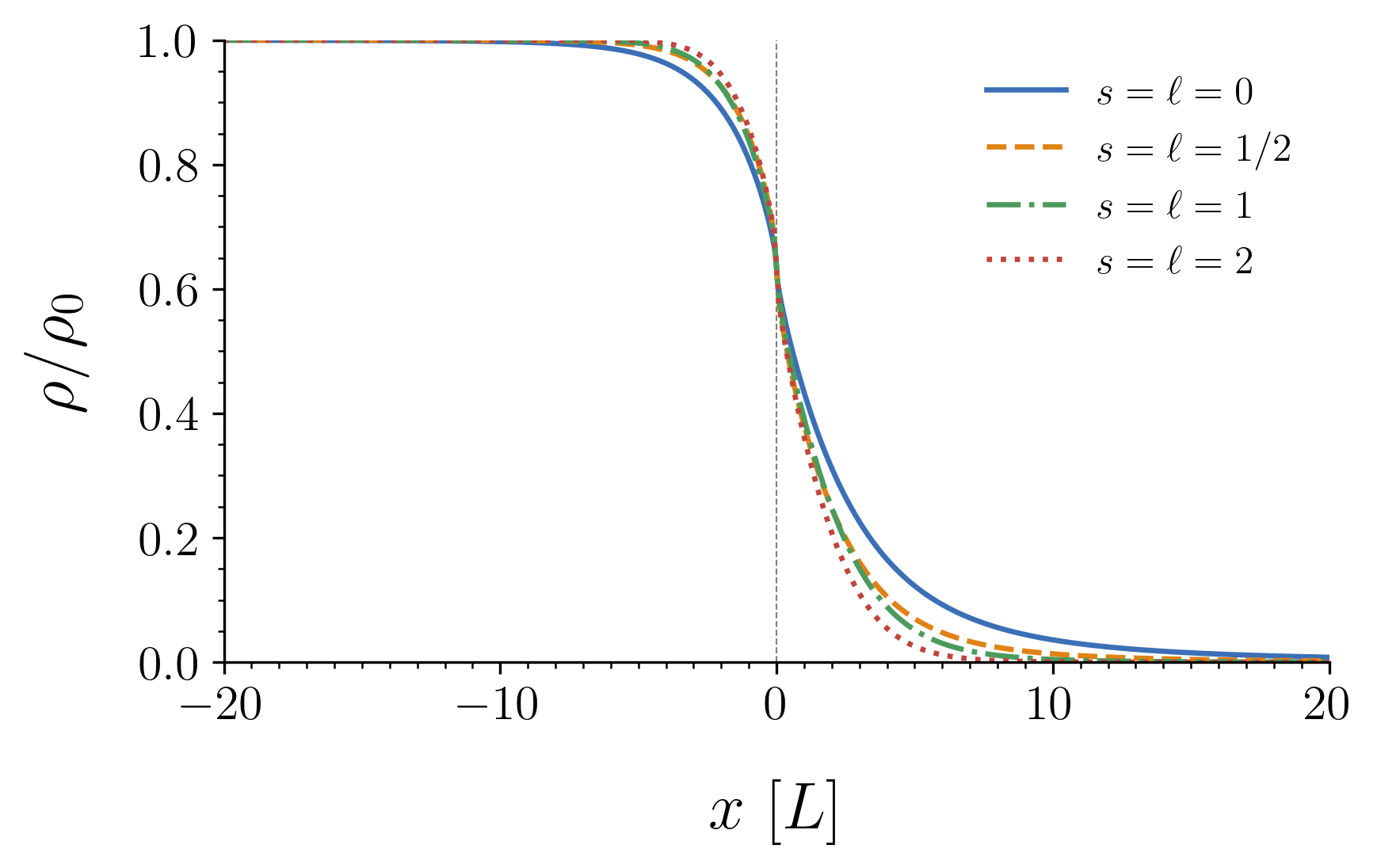}
        \caption{Density.}
        \label{fig:schw_Density}
    \end{subfigure}
    \hfill
    \begin{subfigure}{0.48\textwidth}
        \centering
        \includegraphics[width=\linewidth]{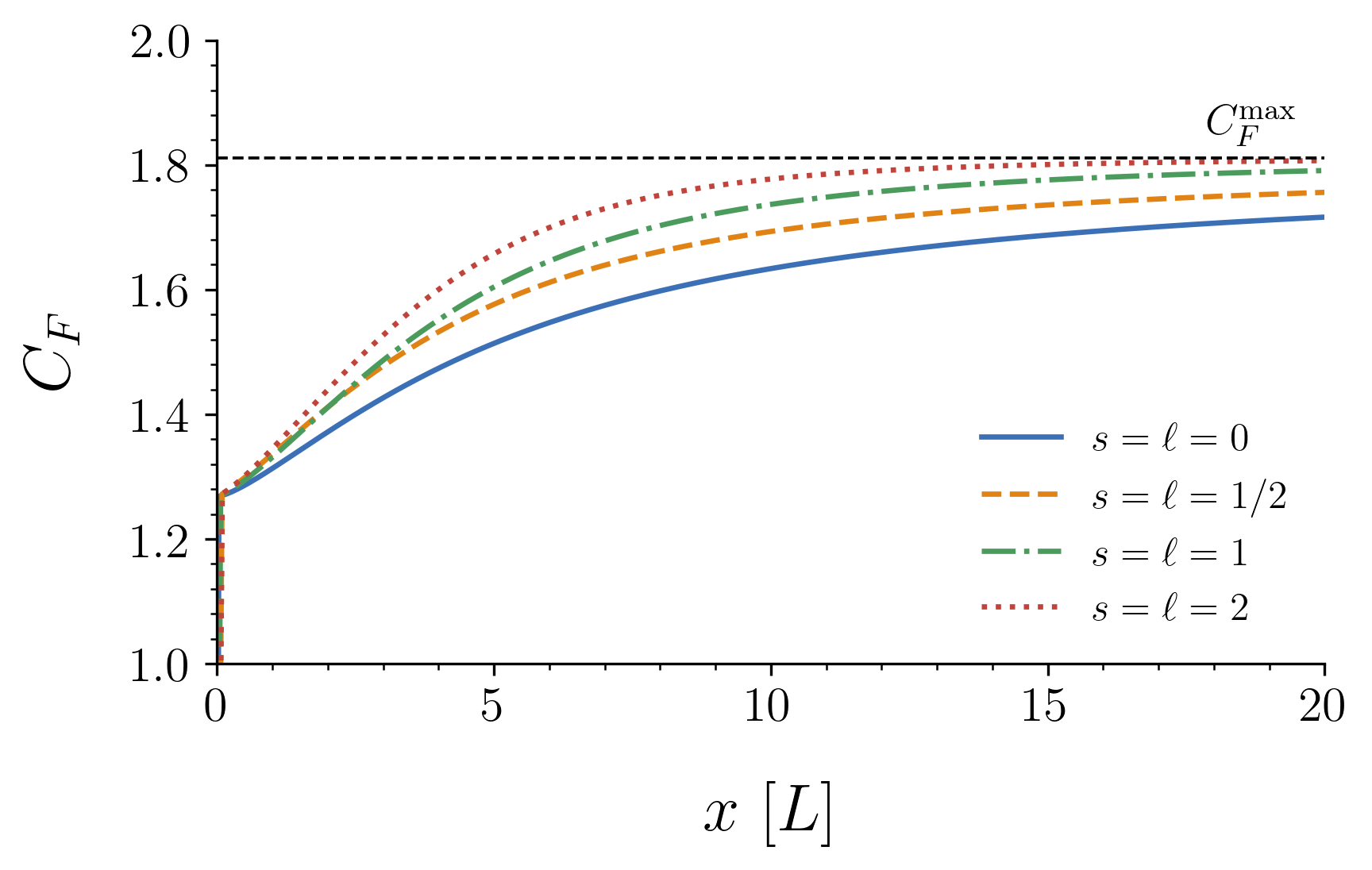}
        \caption{Thrust coefficient.}
        \label{fig:schw_C_F}
    \end{subfigure}
    \caption[The Schwarzschild analogue, by channel]{The Schwarzschild analogue of the four channels, which is the reference every deformed case of this chapter is measured against. Each panel carries the scalar, Dirac, electromagnetic, and gravitational channels at their lowest admissible multipole, $\ell=s$. The duct is the same object in the two halves of the chapter, so these curves are the $\alpha=\ell_h=Q=0$ limit of Sec.~\ref{sec:results HBH by GD} and the $c_6=0$ limit of Sec.~\ref{sec: results quantum correction} alike. The dashed line in panel~(f) is the vacuum limit~\eqref{eq: coeficiente de empuxo maximo} of the thrust coefficient, $C_F^{\max}\simeq1.812$ for air. Natural units with $G_\text{N}=M=1$.}
    \label{fig:schw_reference}
\end{figure}

The remaining panels follow from the Mach number through the isentropic relations~\eqref{eq: rel local total}, and are what an hypothetical experiment reads. All three of pressure, temperature and density leave the stagnation value, pass through the critical values $0.528$, $0.833$ and $0.634$ at the throat, and decay along the divergent section, the faster the higher the spin. At the exit the temperature has fallen to $0.144$ of stagnation in the scalar channel and to $0.007$ in the gravitational one, the density to $0.008$ and to less than $10^{-4}$, and the pressure is below $10^{-3}$ in both. The thrust coefficient runs the other way, rising towards its vacuum limit~\eqref{eq: coeficiente de empuxo maximo}, drawn as the dashed line in panel~\ref{fig:schw_C_F}, which for air is $C_F^{\max}\simeq1.812$ and depends on the gas alone. The four channels reach $1.716$, $1.756$, $1.791$ and $1.807$ at the exit, from $94.7\%$ to $99.7\%$ of the limit, so that the gravitational duct is within three parts in a thousand of the most any bell nozzle could deliver while the scalar one still has five per cent to give. The ordering is the same throughout, the gravitational channel is the extreme case of every statement made in the rest of the chapter, and the scalar one the mildest.

\section{Acoustic Analogue of the Hairy Black Hole}
\label{sec:results HBH by GD}

\begin{figure}[H]
    \centering
    \begin{subfigure}{0.48\textwidth}
        \centering
        \includegraphics[width=\linewidth]{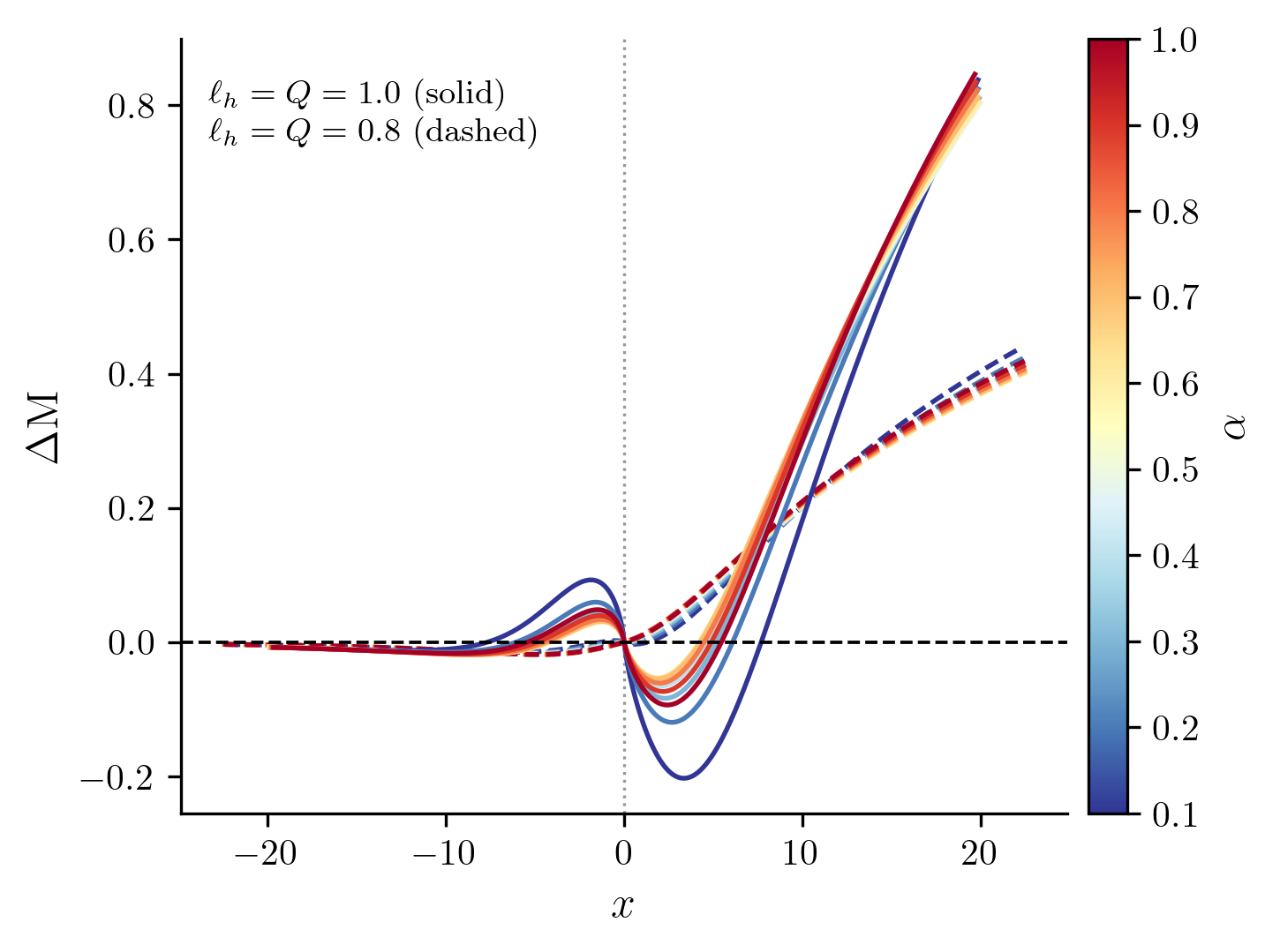}
        \caption{Varying $\alpha$.}
        \label{fig:sens_s0_a}
    \end{subfigure}
    \hfill
    \begin{subfigure}{0.48\textwidth}
        \centering
        \includegraphics[width=\linewidth]{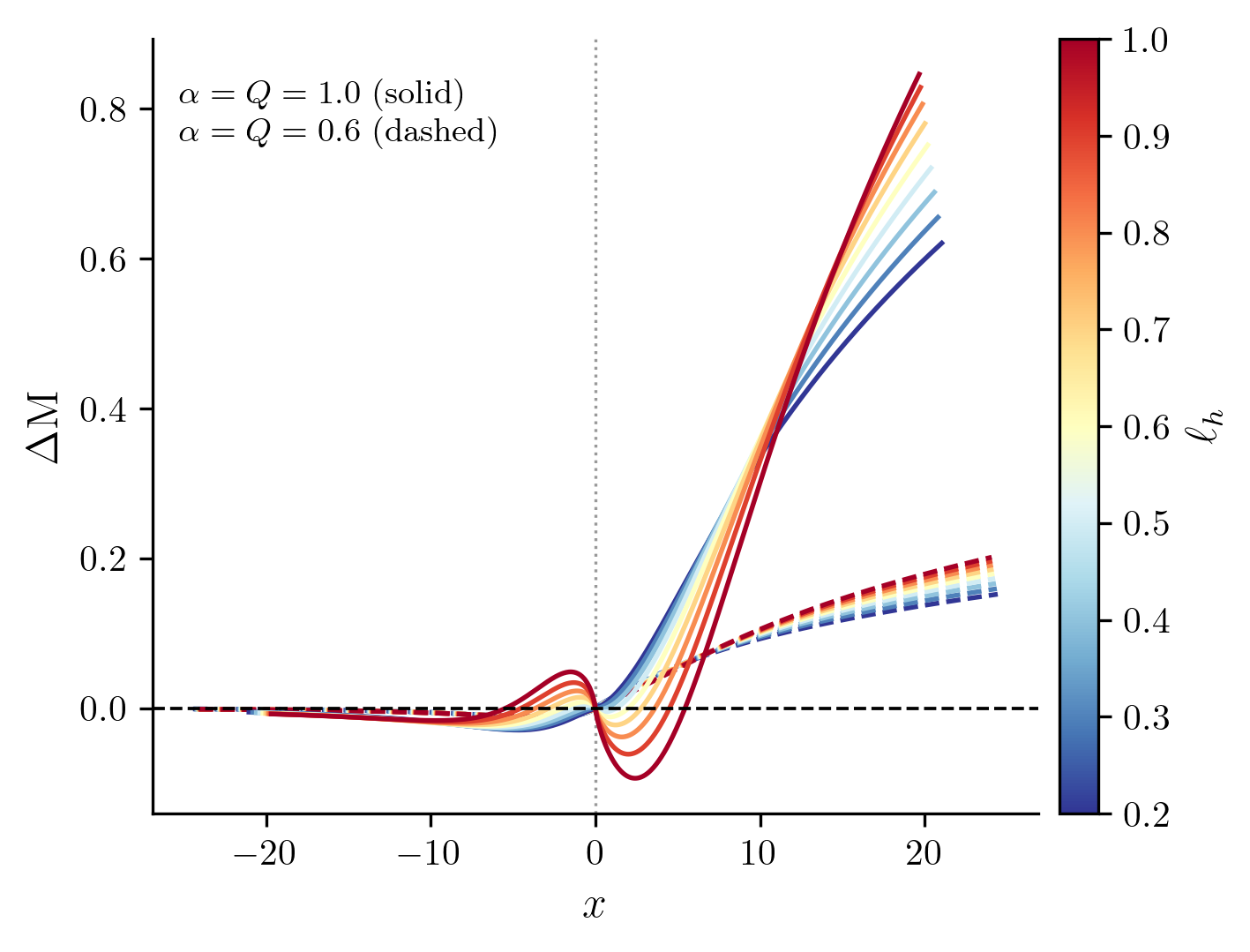}
        \caption{Varying $\ell_h$.}
        \label{fig:sens_s0_b}
    \end{subfigure}

    \begin{subfigure}{0.48\textwidth}
        \centering
        \includegraphics[width=\linewidth]{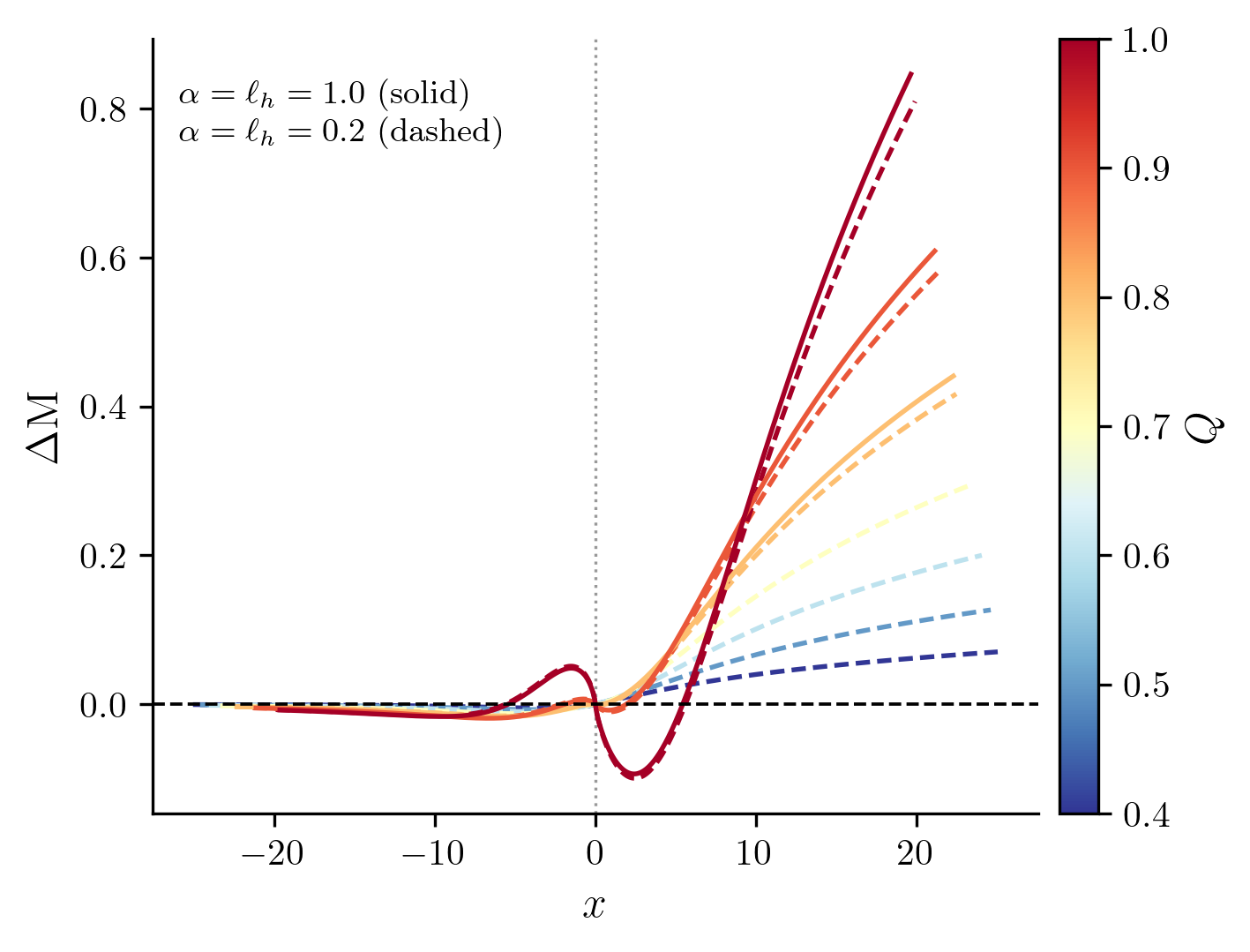}
        \caption{Varying $Q$.}
        \label{fig:sens_s0_c}
    \end{subfigure}
    \hfill
    \begin{subfigure}{0.48\textwidth}
        \centering
        \includegraphics[width=\linewidth]{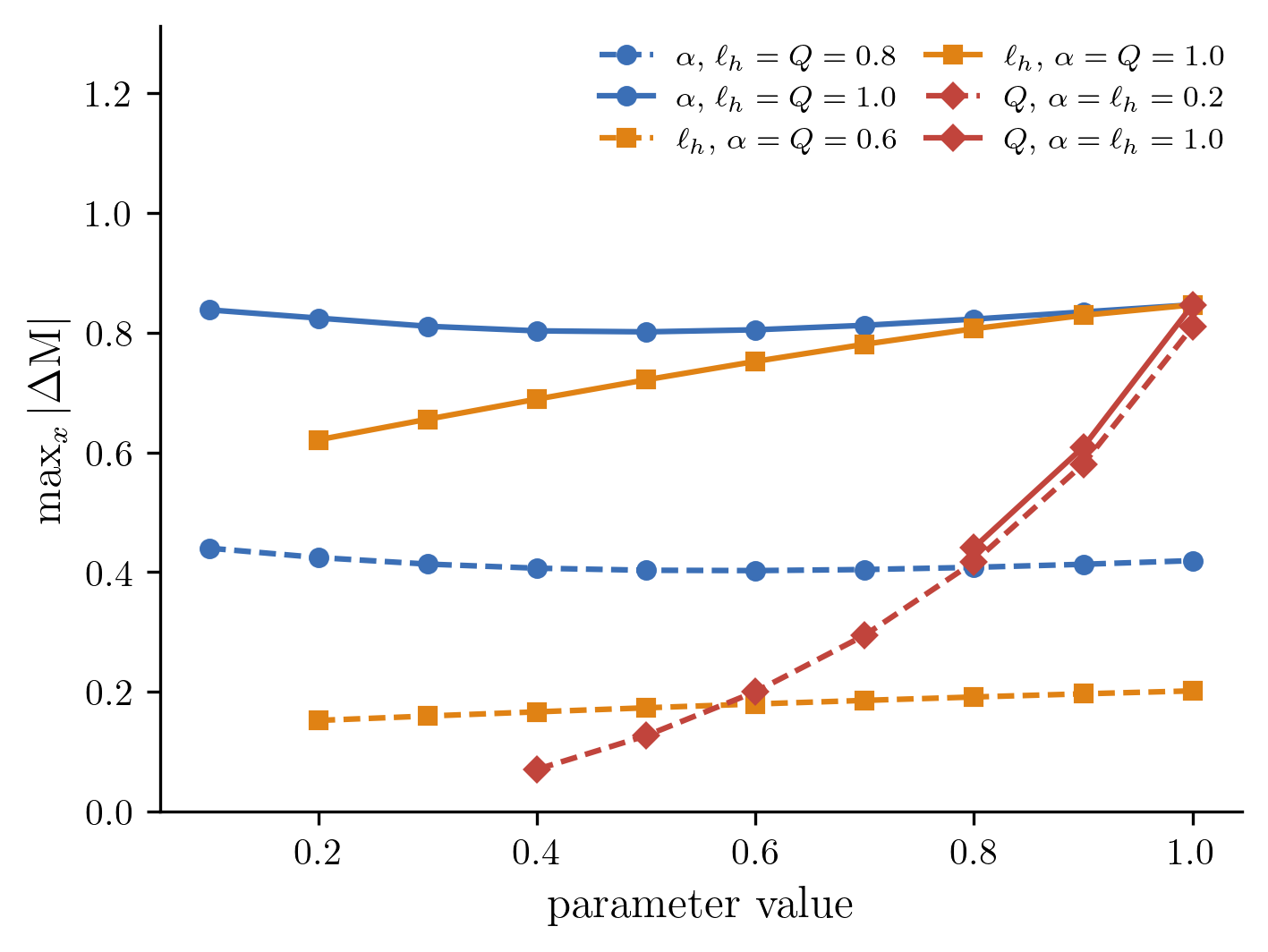}
        \caption{Penetrating power of each parameter.}
        \label{fig:sens_s0_d}
    \end{subfigure}
    \caption[Sensitivity of the acoustic analogue to the three hair parameters, scalar channel]{Sensitivity of the acoustic analogue to the three hair parameters, for the scalar channel $s=\ell=0$. Panels (a), (b), and (c) give the deviation of the Mach profile from the Schwarzschild analogue, $\Delta\mathrm{M}=\mathrm{M}-\mathrm{M}^\text{Schw}$ read station by station along the duct, as $\alpha$, $\ell_h$, and $Q$ are swept in turn (colour bar). Each panel carries two configurations of the two parameters held fixed, the smallest and the largest values of the grid admitted by the dominant energy condition~\eqref{eq: limit Q alpha M}, the larger drawn solid and the smaller dashed, as annotated in each panel. Panel (d) collects the largest deviation reached along the duct against the value of the parameter swept. Natural units with $G_\text{N}=M=1$.}
    \label{fig:sensitivity_HBHGD}
\end{figure}

The three hair parameters $(\alpha,\ell_h,Q)$ of the metric~\eqref{eq: metric HBH} do not act with equal strength on the nozzle, and the difference is large enough to organise the rest of the section. Figure~\ref{fig:sensitivity_HBHGD} makes it quantitative, holding two of them fixed and sweeping the third, and reporting the deviation of the Mach profile from Schwarzschild, $\Delta\mathrm{M}=\mathrm{M}-\mathrm{M}^\text{Schw}$, read station by station along the duct.
Panel~\ref{fig:sens_s0_d} orders the three. The charge-like parameter $Q$ is the only one that acts as a dial. Over the range admitted by the dominant energy condition its imprint grows from $\Delta\mathrm{M}=0.070$ to $0.810$ at $\alpha=\ell_h=0.2$, a factor of eleven, and at a given $Q$ the two configurations of the parameters held fixed give almost the same answer, $0.846$ against $0.810$ at $Q=1$, so that the charge sets the deviation nearly on its own. The decoupling parameter $\alpha$ behaves instead as a switch. Once the hair is present, moving $\alpha$ across its whole range shifts the imprint only between $0.838$ and $0.846$, with a shallow minimum of $0.801$ near $\alpha=0.5$, which is five per cent of its own value. The primary-hair scale $\ell_h$ falls between the two and is derivative of the coupling, rising from $0.621$ to $0.846$ at $\alpha=Q=1$ but only from $0.152$ to $0.202$ at $\alpha=Q=0.6$.

Where each of them is felt follows from the term it enters. The charge-like parameter acts through $Q^2/r^2$, the only term of the metric~\eqref{eq: metric HBH} that falls faster than the Newtonian $1/r$, so that raising $Q$ makes that term progressively more present and concentrates its weight at small $r$, which is to say at the horizon. The reconstruction carries this straight into the duct, the deviation in the flow being concentrated around $x=0$ and growing with $Q$, whereas $\alpha$ and $\ell_h$, which enter through the effective mass and the exponential, are felt over the outer part of the duct instead.
That last statement can be made exact since $\mathcal{M}-\alpha\ell_h/2=M$, the exponential term of Eq.~\eqref{eq: metric HBH} does not contain $\ell_h$ at all, which therefore survives only in $-2G_\text{N}\mathcal{M}/r$, so that its whole effect is the mass shift 
\begin{equation}
\label{eq: ell_h como deslocamento de massa}
    M \mapsto M + \frac{\alpha \ell_h}{2},
\end{equation}
and its influence vanishes with the coupling. The bound $\ell_h\geq M/e^2$ of Eq.~\eqref{eq: limit ell M} keeps the shift from being made arbitrarily small at fixed $\alpha$, but nothing keeps it from being switched off by $\alpha$ itself. An experiment able to resolve a deviation of a few per cent in the Mach profile would consequently constrain the product $\alpha\ell_h$ rather than either factor separately.

The rest of the section therefore sweeps $Q$ alone, at fixed $(\alpha,\ell_h)$. Neither of the other two is inert, but neither separates the profiles across its own range in the way $Q$ does, and what they do change is already contained in Fig.~\ref{fig:sensitivity_HBHGD}. This is the analogue counterpart of the statement made in Chapter~\ref{chap:conexaoentrebnsebocais}, that the hair modulates the height of the barrier while leaving its position essentially fixed, the height responding to $Q$ and the nozzle inheriting that response.

Two things bound the sweep from below and from above, and the upper one is worth stating carefully, because it is also what limits the numerics. From below, the dominant energy condition of Chapter~\ref{chap:GD} requires $Q \geq 2\sqrt{\alpha}\,G_\text{N}M/e$, which at the $\alpha=\ell_h=0.2$ used below reads $Q\geq0.329$.
From above, the limit is the existence of the horizon itself, and the mechanism is the one already identified: $Q^2/r^2$ is the only term of the metric falling faster than $1/r$, so it is the term that governs the small-$r$ region where the roots of Eq.~\eqref{eq: horizon radii} live. Raising $Q$ therefore pushes the inner root outwards and the outer root inwards until the two meet. At $\alpha=\ell_h=0.2$ the event and Cauchy horizons sit at $1.987$ and $0.074$ for $Q=0.4$, still $1.91$ apart, and at $1.352$ and $0.689$ for $Q=1$, only $0.66$ apart, and beyond $Q=1$ the equation has no real root at all, and thus there is no black hole to reconstruct. This was made clear in Subsec.~\ref{subsec: map of accepted}.
The reconstruction feels that approach before the horizon disappears. The integration of Eq.~\eqref{eq: lambda} starts from the series~\eqref{eq: serie truncada} about the horizon, whose coefficients carry $\mathrm{F}'(r_\text{EH})$ in the denominator, and $\mathrm{F}'(r_\text{EH})$ is what vanishes when the two roots merge. Over the sweep it falls from $0.987$ at $Q=0.4$ to $0.711$ at $Q=0.9$ and $0.505$ at $Q=1$, so that the leading Frobenius coefficient, which goes as $\mathrm{F}'^{-2}$, has almost quadrupled by the top of the grid. The signature is: the exit Mach number grows in steps of a fairly constant ratio up to $Q=0.9$, and then jumps by between two and four times the previous increment at $Q=1$, in all four channels at once. That last point is the truncation degrading, not the flow changing, and the sweeps below therefore stop at $Q=0.9$.

\begin{figure}[htbp]
    \centering
    \begin{subfigure}{0.48\textwidth}
        \centering
        \includegraphics[width=\linewidth]{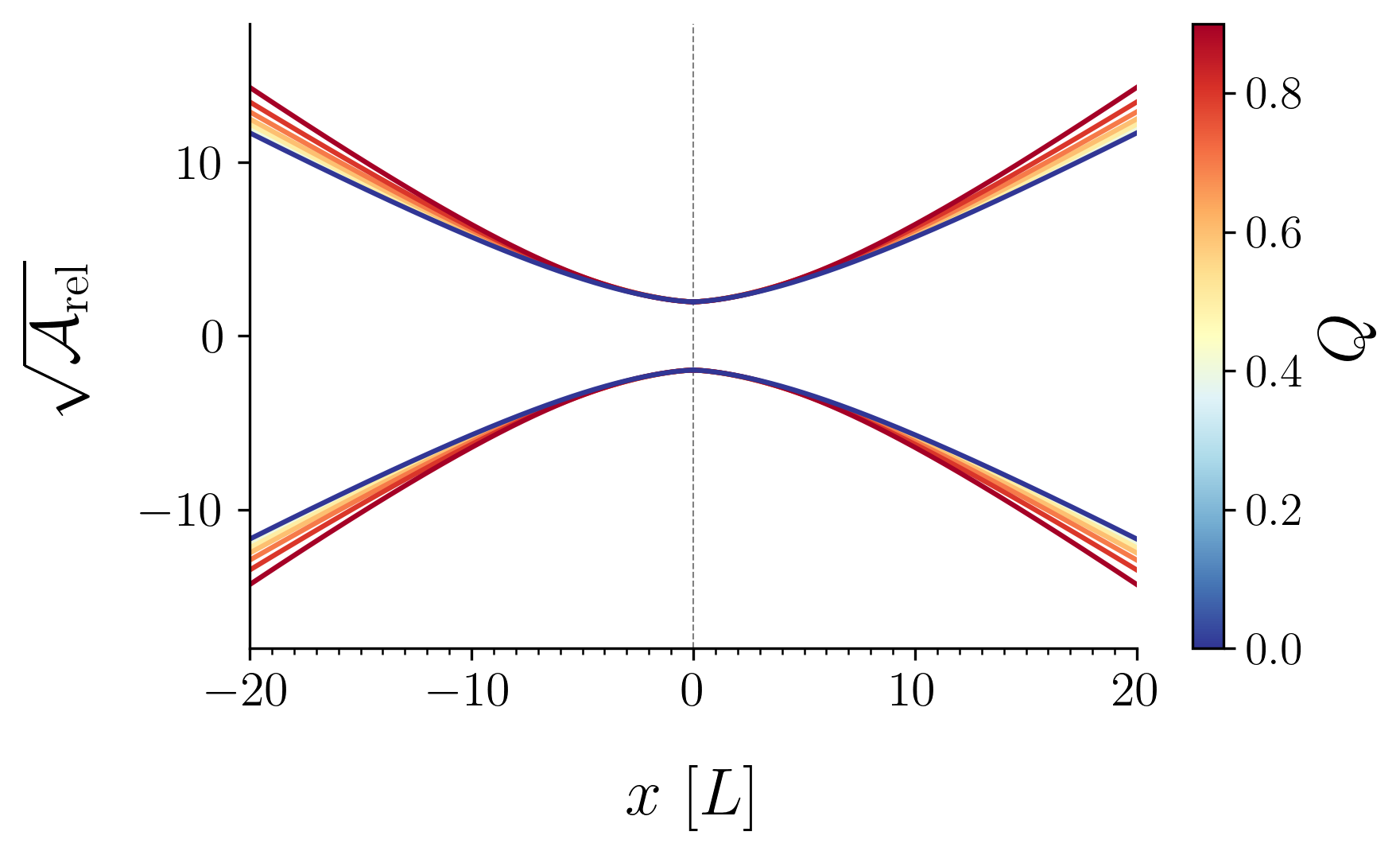}
        \caption{$s=\ell=0$.}
        \label{fig:varQ_nozzle_shape_HBHGD_a}
    \end{subfigure}
    \hfill
    \begin{subfigure}{0.48\textwidth}
        \centering
        \includegraphics[width=\linewidth]{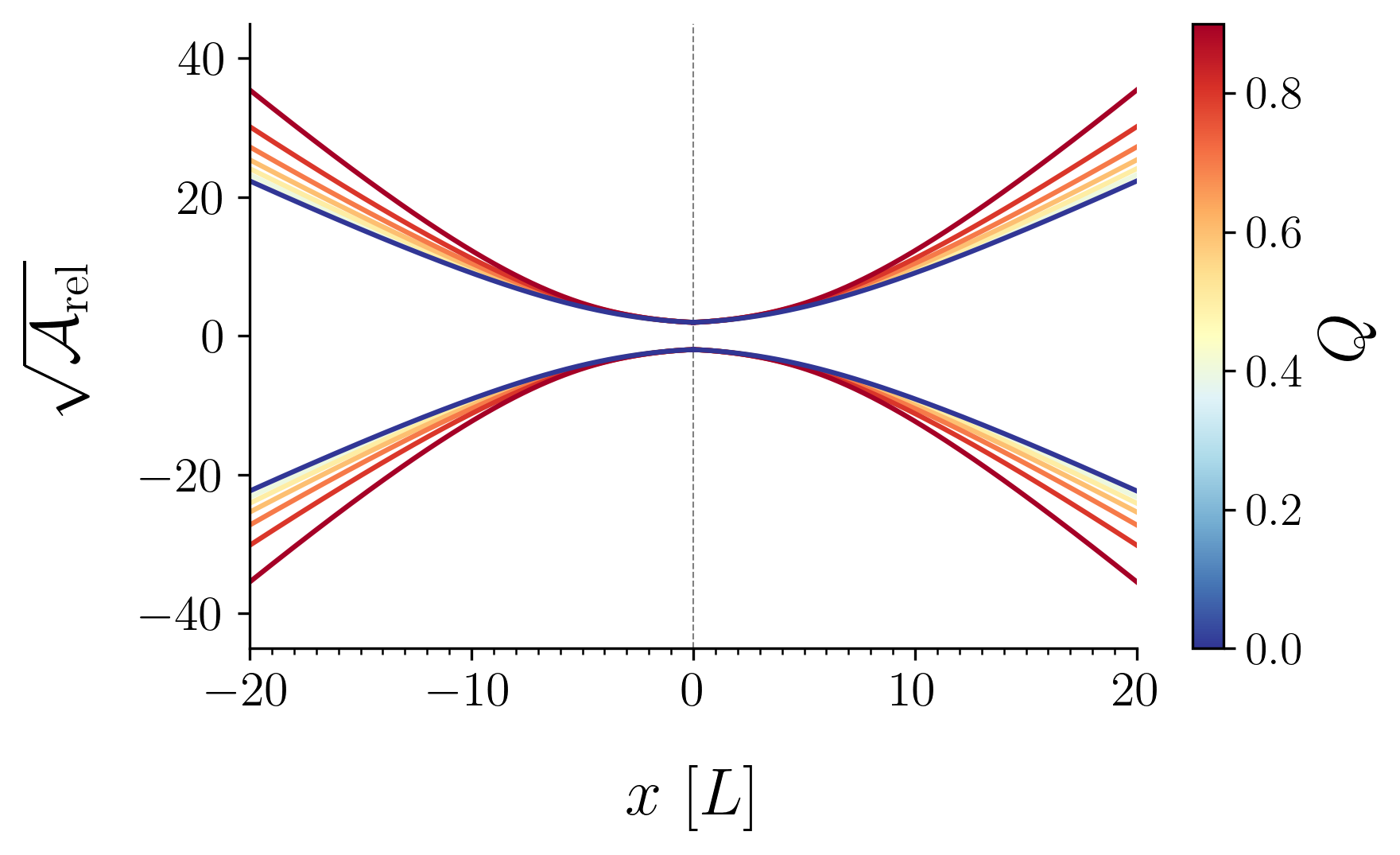}
        \caption{$s=\ell=1/2$.}
        \label{fig:varQ_nozzle_shape_HBHGD_b}
    \end{subfigure}

    \begin{subfigure}{0.48\textwidth}
        \centering
        \includegraphics[width=\linewidth]{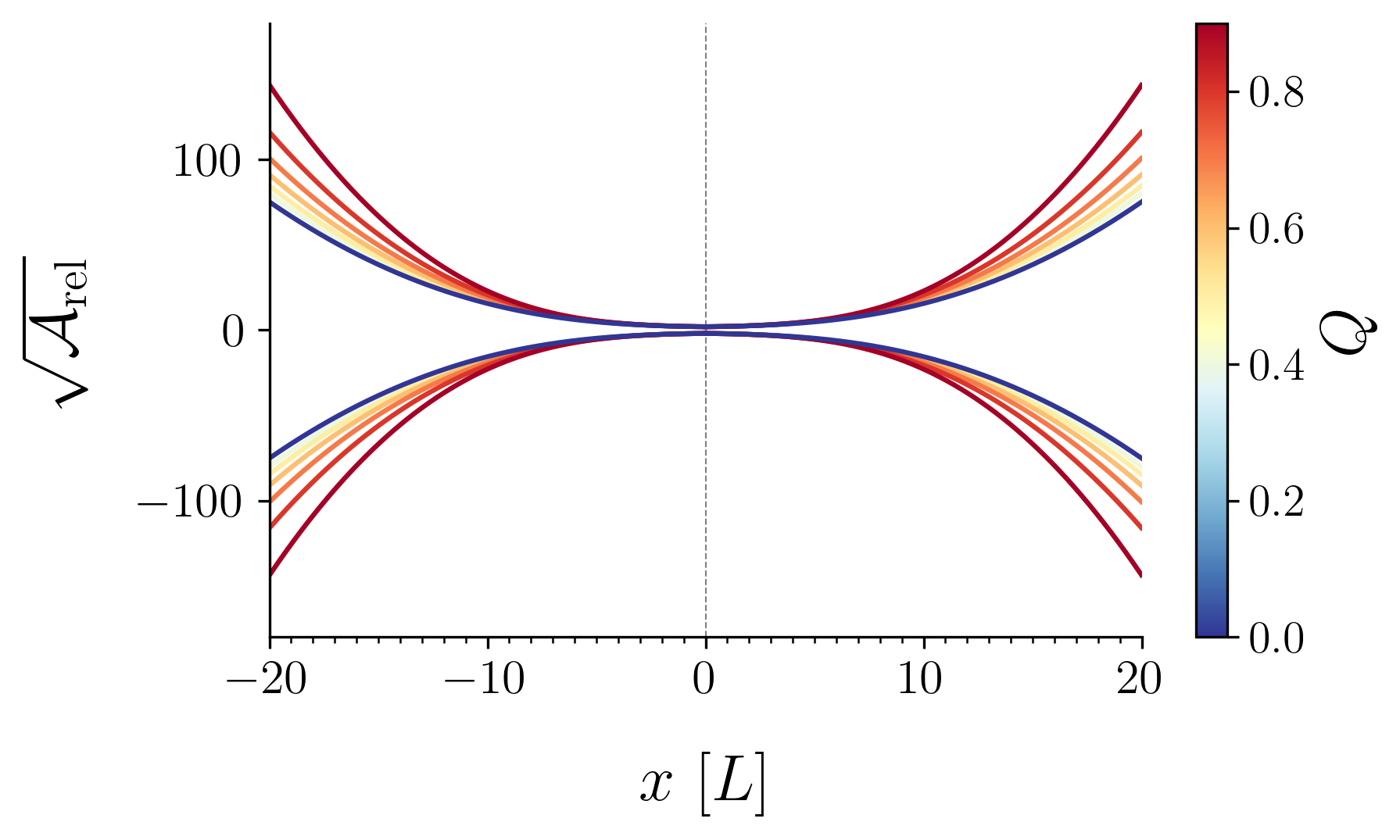}
        \caption{$s=\ell=1$.}
        \label{fig:varQ_nozzle_shape_HBHGD_c}
    \end{subfigure}
    \hfill
    \begin{subfigure}{0.48\textwidth}
        \centering
        \includegraphics[width=\linewidth]{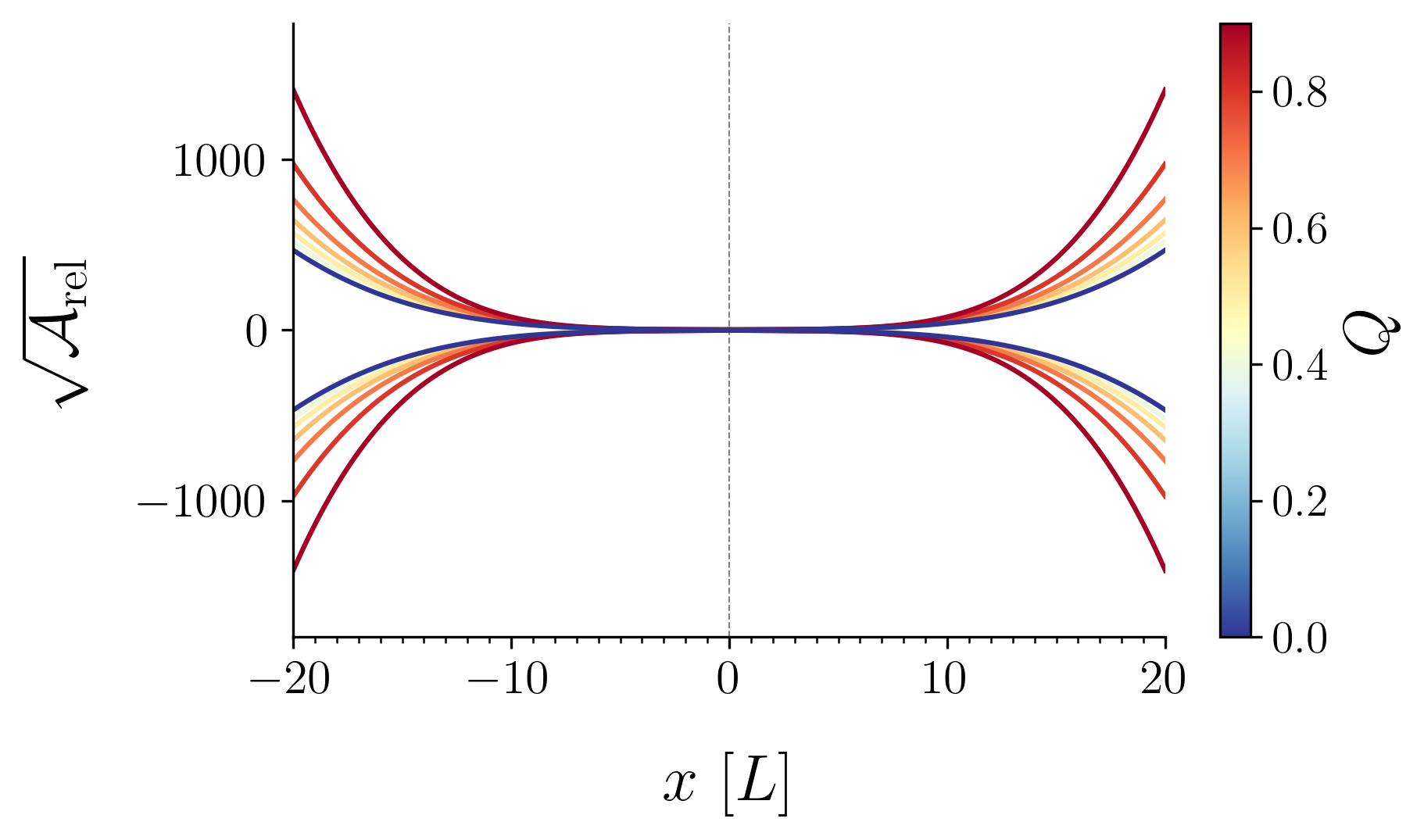}
        \caption{$s=\ell=2$.}
        \label{fig:varQ_nozzle_shape_HBHGD_d}
    \end{subfigure}
    \caption[Nozzle shape of the hairy analogue, varying $Q$]{Nozzle shape geometry along the de~Laval nozzle, varying $Q$, for the hairy black hole~\eqref{eq: metric HBH} at $\alpha=\ell_h=0.2$, one panel per channel, the dashed curve being the Schwarzschild analogue. The colour bar spans the sweep, which the dominant energy condition~\eqref{eq: limit Q alpha M} bounds below at $Q\geq0.329$ for $\alpha=\ell_h=0.2$ and which is stopped above at $Q=0.9$. Natural units with $G_\text{N}=M=1$. First published in Ref.~\cite{Casadio:2024uwj}.}
    \label{fig:varQ_nozzle_shape_HBHGD}
\end{figure}
\begin{figure}[H]
    \centering
    \begin{subfigure}{0.48\textwidth}
        \centering
        \includegraphics[width=\linewidth]{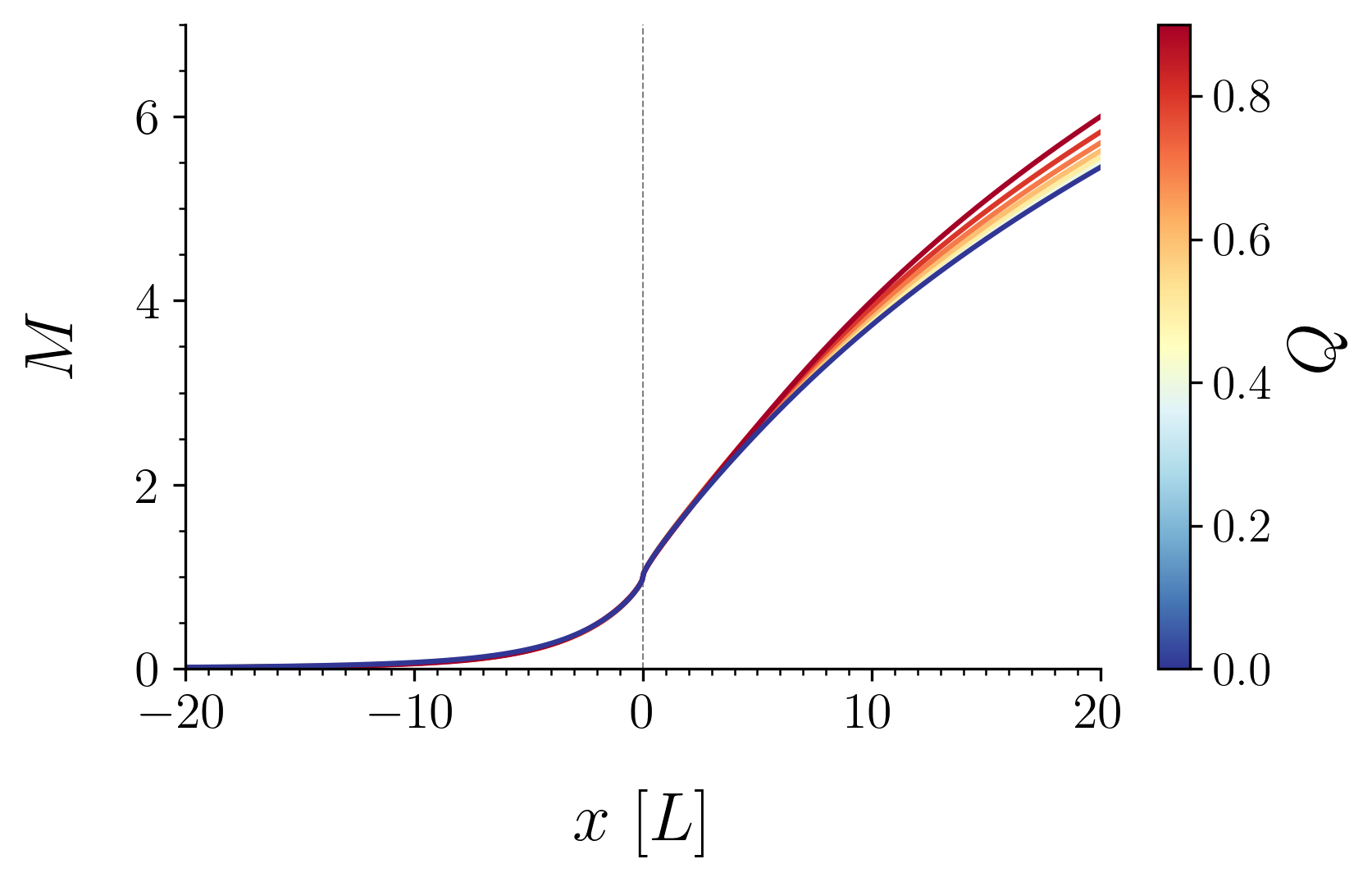}
        \caption{$s=\ell=0$.}
        \label{fig:varQ_Mach_HBHGD_a}
    \end{subfigure}
    \hfill
    \begin{subfigure}{0.48\textwidth}
        \centering
        \includegraphics[width=\linewidth]{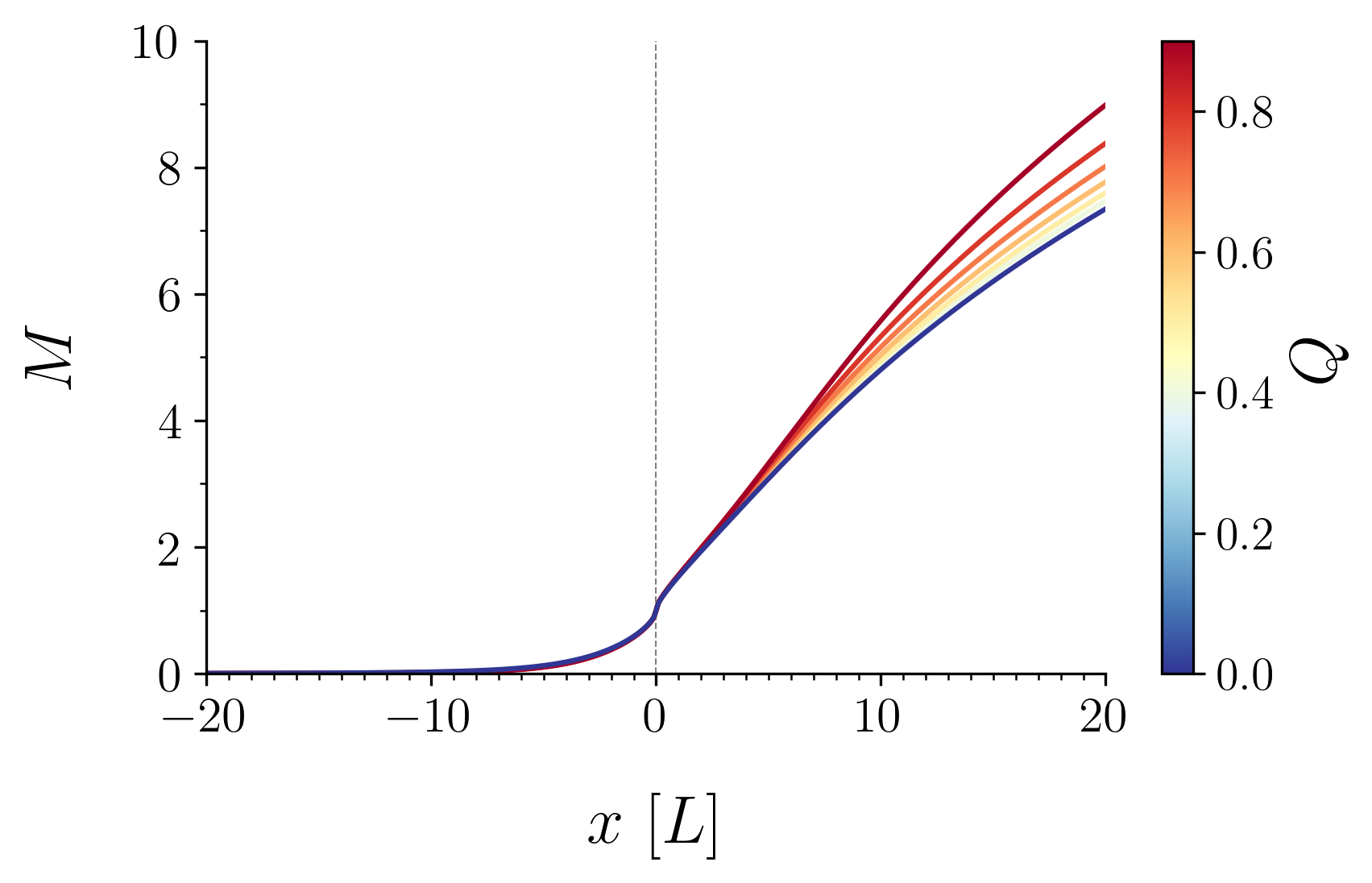}
        \caption{$s=\ell=1/2$.}
        \label{fig:varQ_Mach_HBHGD_b}
    \end{subfigure}

    \begin{subfigure}{0.48\textwidth}
        \centering
        \includegraphics[width=\linewidth]{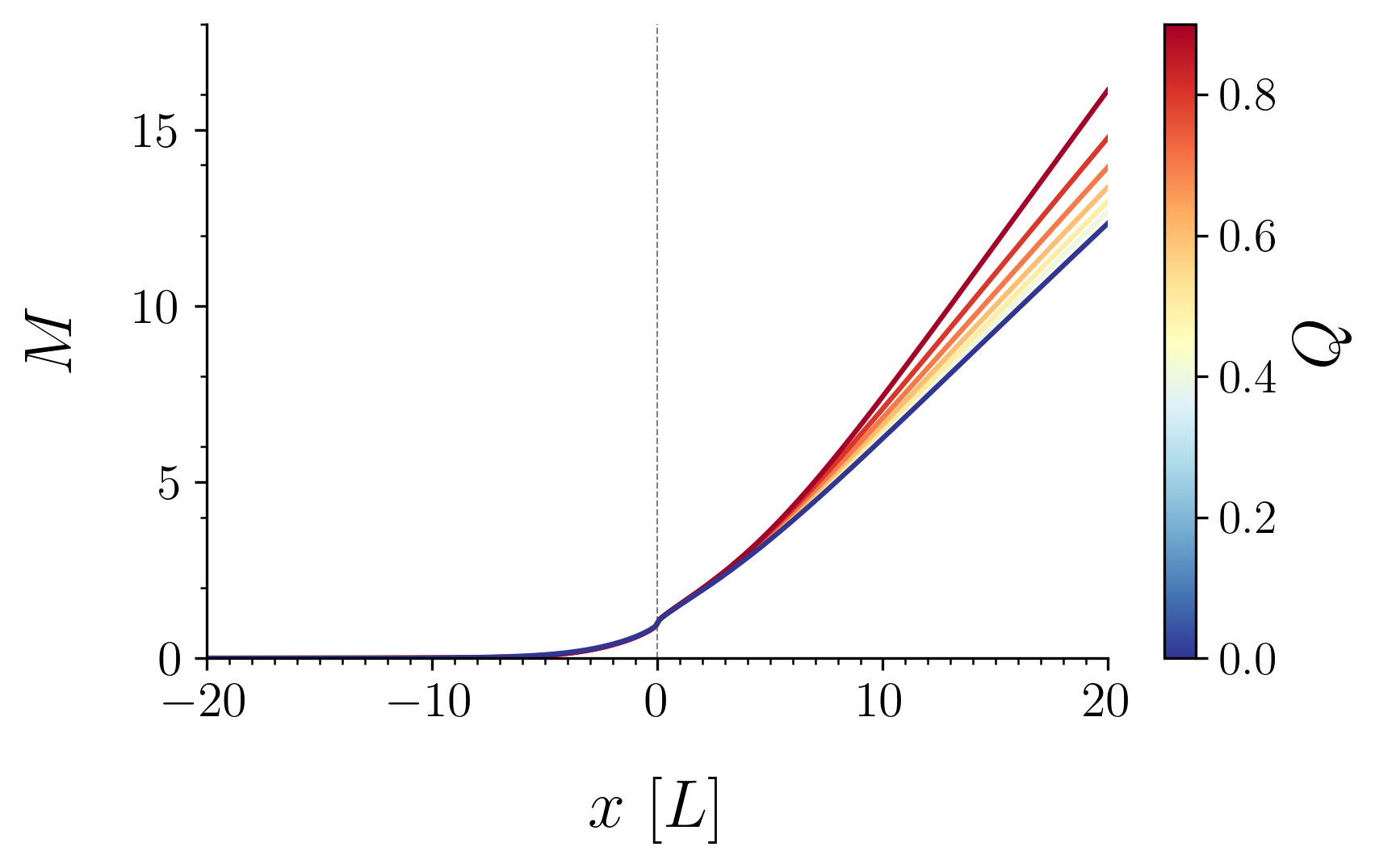}
        \caption{$s=\ell=1$.}
        \label{fig:varQ_Mach_HBHGD_c}
    \end{subfigure}
    \hfill
    \begin{subfigure}{0.48\textwidth}
        \centering
        \includegraphics[width=\linewidth]{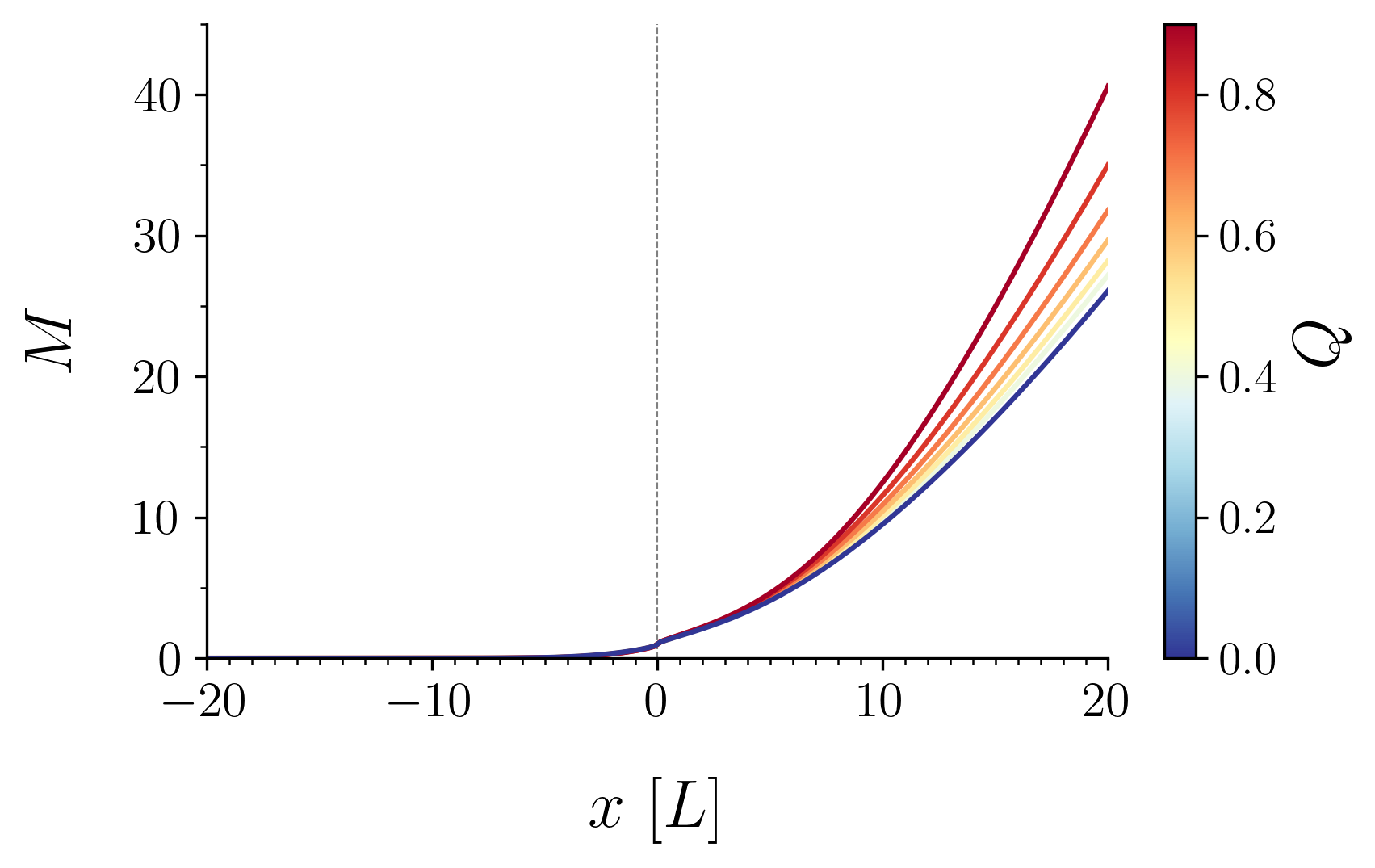}
        \caption{$s=\ell=2$.}
        \label{fig:varQ_Mach_HBHGD_d}
    \end{subfigure}
    \caption[Mach number of the hairy analogue, varying $Q$]{Mach number along the de~Laval nozzle, varying $Q$, for the hairy black hole~\eqref{eq: metric HBH} at $\alpha=\ell_h=0.2$, one panel per channel, the dashed curve being the Schwarzschild analogue. The sweep and its bounds are those of Fig.~\ref{fig:varQ_nozzle_shape_HBHGD}. As in Fig.~\ref{fig:varQ_nozzle_shape_HBHGD}, the vertical scale is proper to each panel, the exit Mach number rising with the spin. Natural units with $G_\text{N}=M=1$. First published in Ref.~\cite{Casadio:2024uwj}.}
    \label{fig:varQ_Mach_HBHGD}
\end{figure}

Figures~\ref{fig:varQ_nozzle_shape_HBHGD} and~\ref{fig:varQ_Mach_HBHGD} report the nozzle shape and the Mach-number profile at $\alpha=\ell_h=0.2$, the corner of the admissible region over which the dominant energy condition leaves most of the sweep physical. Every figure of this section carries one panel per channel, in the order $s=\ell=0$, $1/2$, $1$ and $2$, against the Schwarzschild reference of Sec.~\ref{sec:schwarzschild analogue}.

The nozzle shape responds to $Q$ in the way the effective potential of Chapter~\ref{chap:conexaoentrebnsebocais} anticipates. Raising $Q$ increases the barrier height and moves the event horizon inward, and the reconstructed duct follows. The throat is unchanged by construction, since it is the sonic point, but the divergent section opens more widely, so that a larger area ratio is needed to reproduce the same asymptotic potential. The four panels differ in scale rather than in kind, the duct opening far more widely as the spin rises, from a largest $\sqrt{\mathcal{A}_{\text{rel}}}$ of $14.3$ in the scalar channel to $1410$ in the gravitational one over the same window in $x$.

The Mach-number profile shows the asymmetry of convergent/divergent parts more directly. Upstream of the throat all curves collapse onto Schwarzschild, while downstream they separate, and the separation only becomes appreciable well past the throat. That is a practical statement about where an experiment would have to be instrumented, since the signature of the hair is not at the horizon but into the supersonic section.

What the hair changes there depends on the channel, and the change grows sharply with the spin. Read at the exit, the Mach number at $Q=0.9$ stands at $6.00$ in the scalar channel against the Schwarzschild $5.45$, at $8.99$ against $7.35$ in the Dirac one, at $16.13$ against $12.35$ in the electromagnetic one and at $40.60$ against $26.08$ in the gravitational one, a rise of $10\%$, $22\%$, $31\%$ and $56\%$ respectively. The duct opens in the same proportion, $\sqrt{\mathcal{A}_{\text{rel}}}$ at the exit going from $11.7$ to $14.3$ in the scalar channel and from $469.6$ to $1410$ in the gravitational one. Within each panel, from a quarter of the nozzle to the end, the curves are ordered by the colour bar and never cross, the scalar exit Mach number running $5.51$, $5.56$, $5.63$, $5.71$, $5.83$ and $6.00$ over the six values of $Q$, so that the response is monotone in the charge and steepest at the top of the sweep.
At the throat the hair acts on the gradient of $\mathrm{M}$, which by Eq.~\eqref{eq: temperatura hawking bocal} fixes the analogue Hawking temperature of the flow. 

\begin{figure}[htbp]
    \centering
    \begin{subfigure}{0.48\textwidth}
        \centering
        \includegraphics[width=\linewidth]{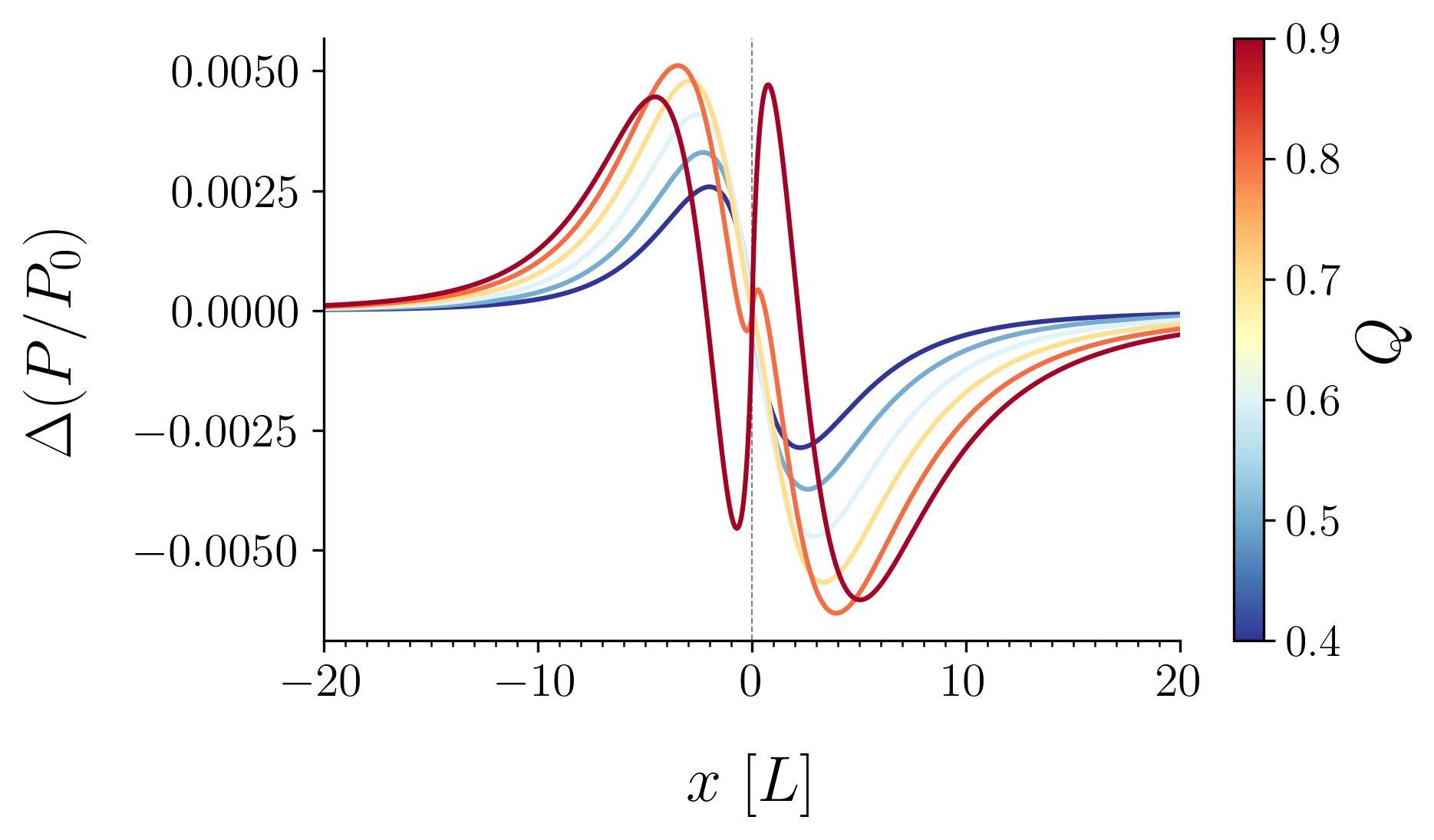}
        \caption{$s=\ell=0$.}
        \label{fig:varQ_Pressure_diff_HBHGD_a}
    \end{subfigure}
    \hfill
    \begin{subfigure}{0.48\textwidth}
        \centering
        \includegraphics[width=\linewidth]{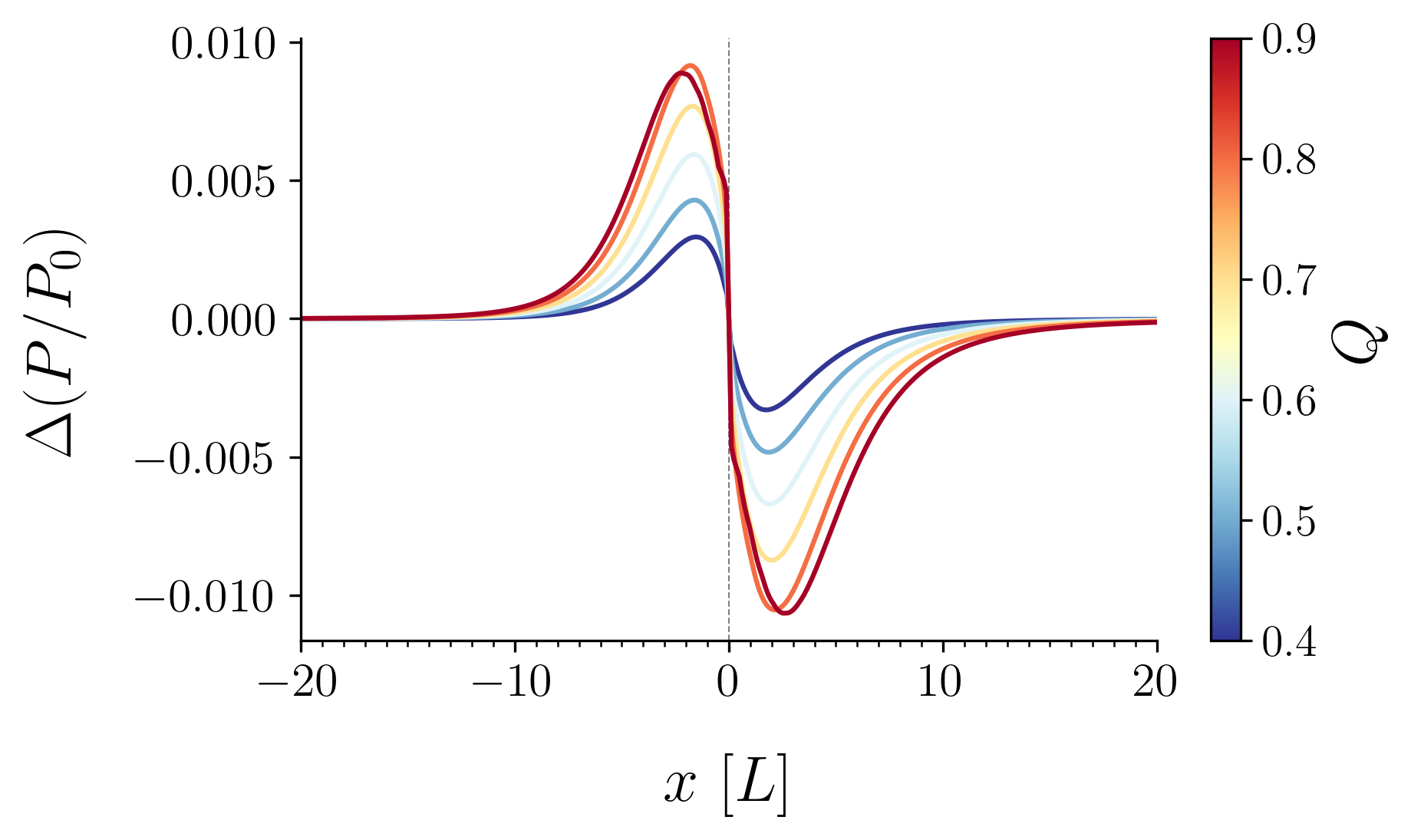}
        \caption{$s=\ell=1/2$.}
        \label{fig:varQ_Pressure_diff_HBHGD_b}
    \end{subfigure}

    \begin{subfigure}{0.48\textwidth}
        \centering
        \includegraphics[width=\linewidth]{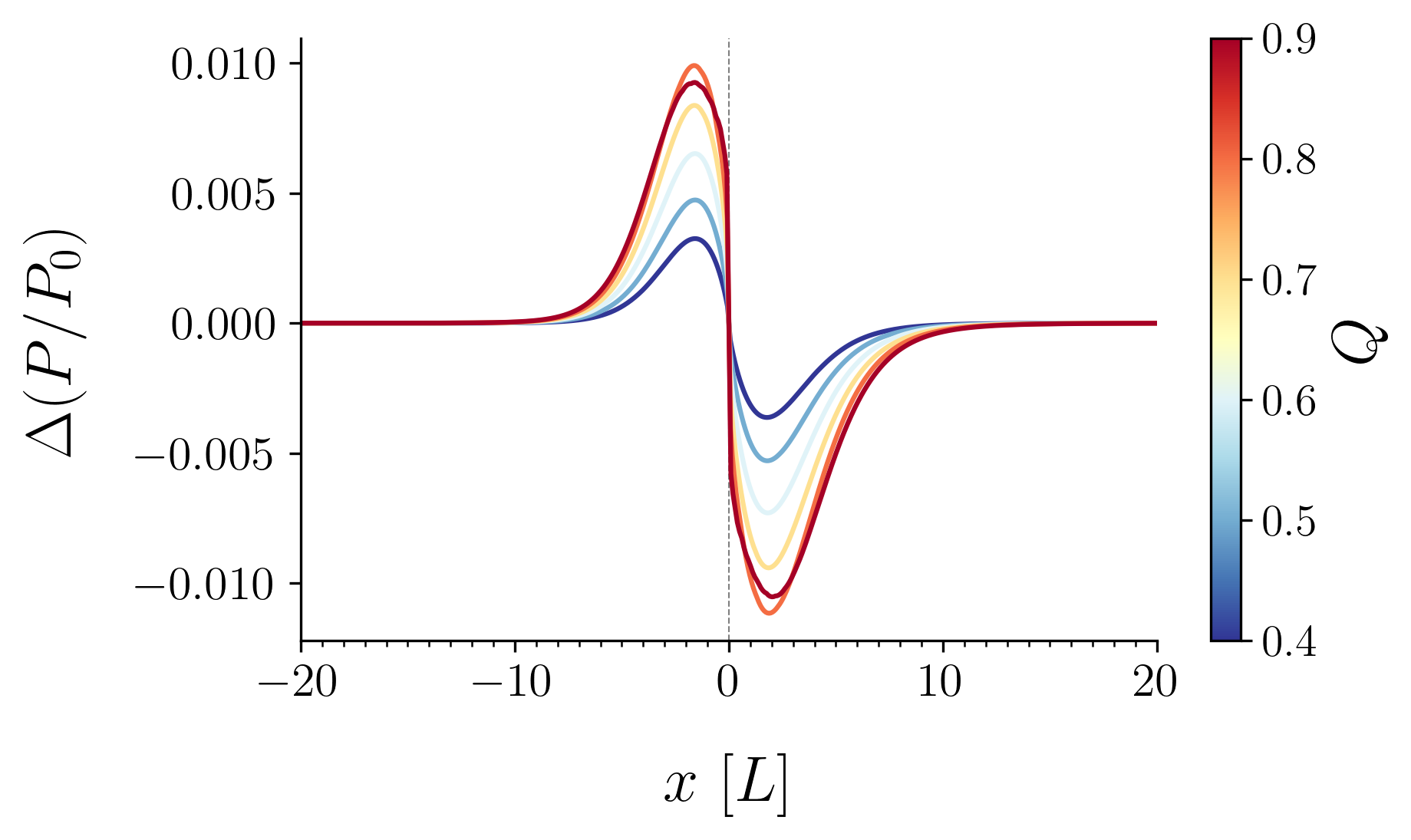}
        \caption{$s=\ell=1$.}
        \label{fig:varQ_Pressure_diff_HBHGD_c}
    \end{subfigure}
    \hfill
    \begin{subfigure}{0.48\textwidth}
        \centering
        \includegraphics[width=\linewidth]{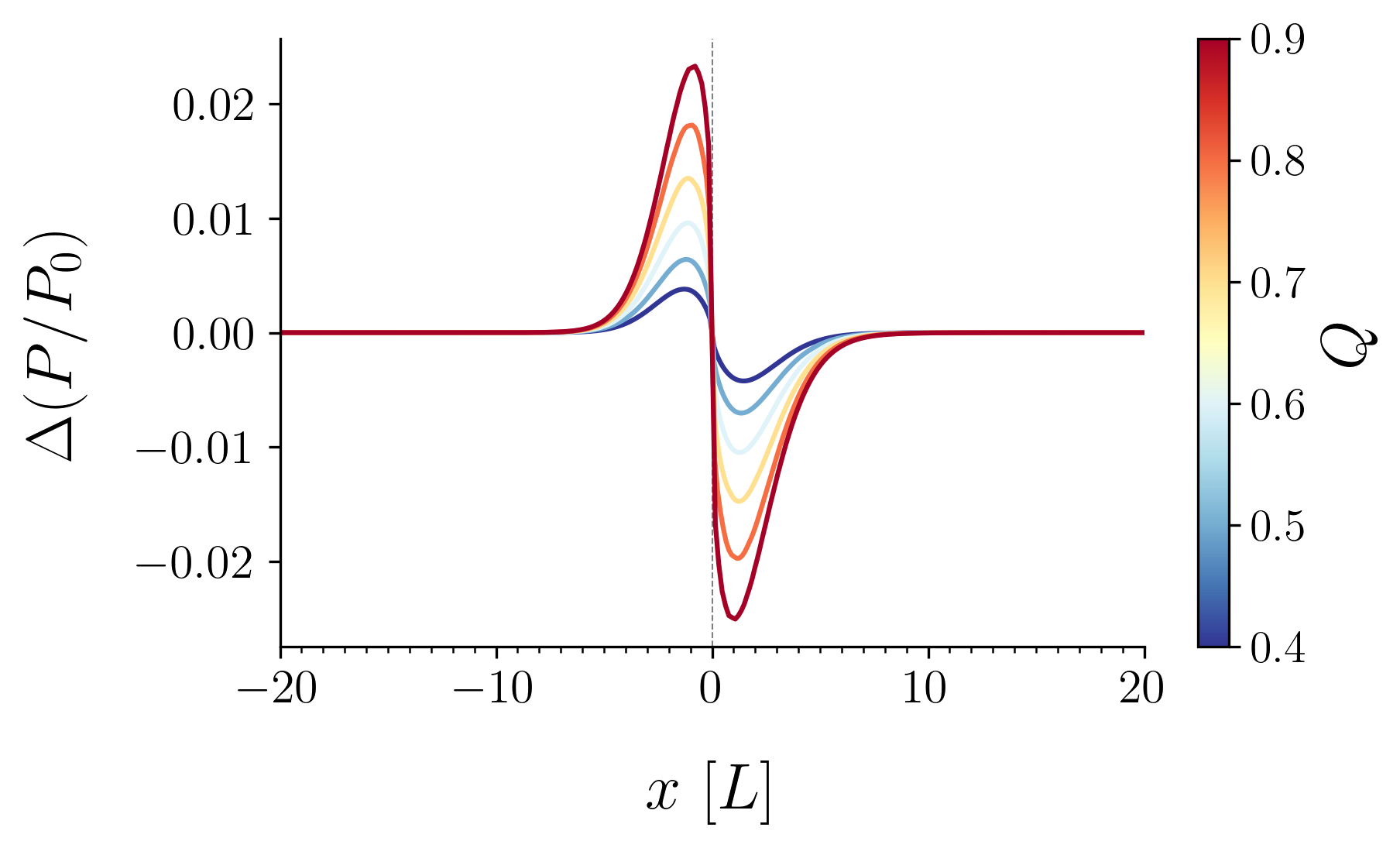}
        \caption{$s=\ell=2$.}
        \label{fig:varQ_Pressure_diff_HBHGD_d}
    \end{subfigure}
    \caption[Difference in pressure of the hairy analogue, varying $Q$]{Difference in the pressure of the flow, $\Delta({P/P_0})$, Difference from the Schwarzschild analogue of the same channel, $\Delta X(x)=X(x)-X^\text{Schw}(x)$ read station by station, for the hairy black hole~\eqref{eq: metric HBH} at $\alpha=\ell_h=0.2$ with $Q$ swept over the admissible range, from $0.4$ to $0.9$ (colour bar), one panel per channel. The vertical scale is proper to each panel. Natural units with $G_\text{N}=M=1$.}
    \label{fig:varQ_Pressure_diff_HBHGD}
\end{figure}

\begin{figure}[H]
    \centering
    \begin{subfigure}{0.48\textwidth}
        \centering
        \includegraphics[width=\linewidth]{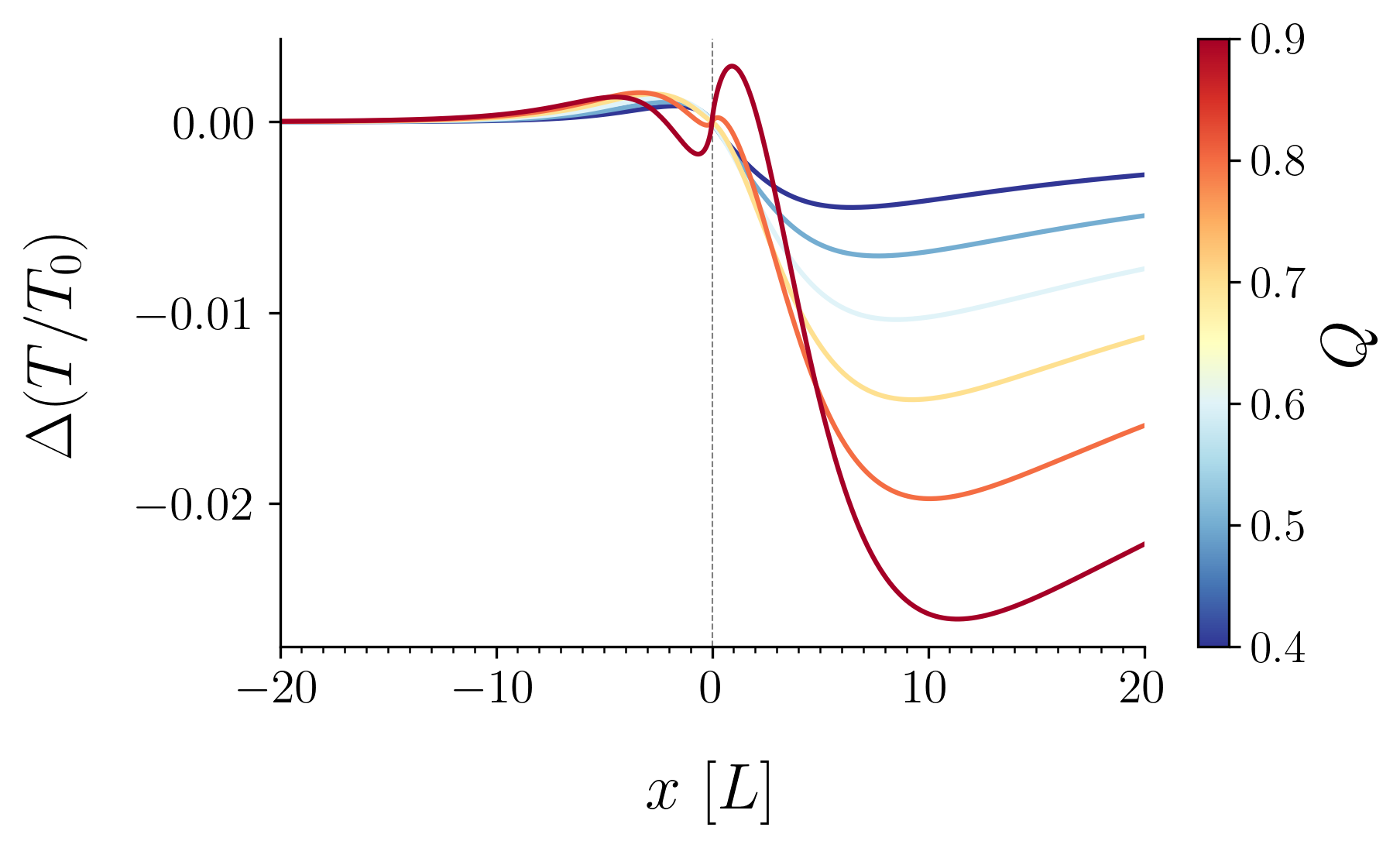}
        \caption{$s=\ell=0$.}
        \label{fig:varQ_Temperature_diff_HBHGD_a}
    \end{subfigure}
    \hfill
    \begin{subfigure}{0.48\textwidth}
        \centering
        \includegraphics[width=\linewidth]{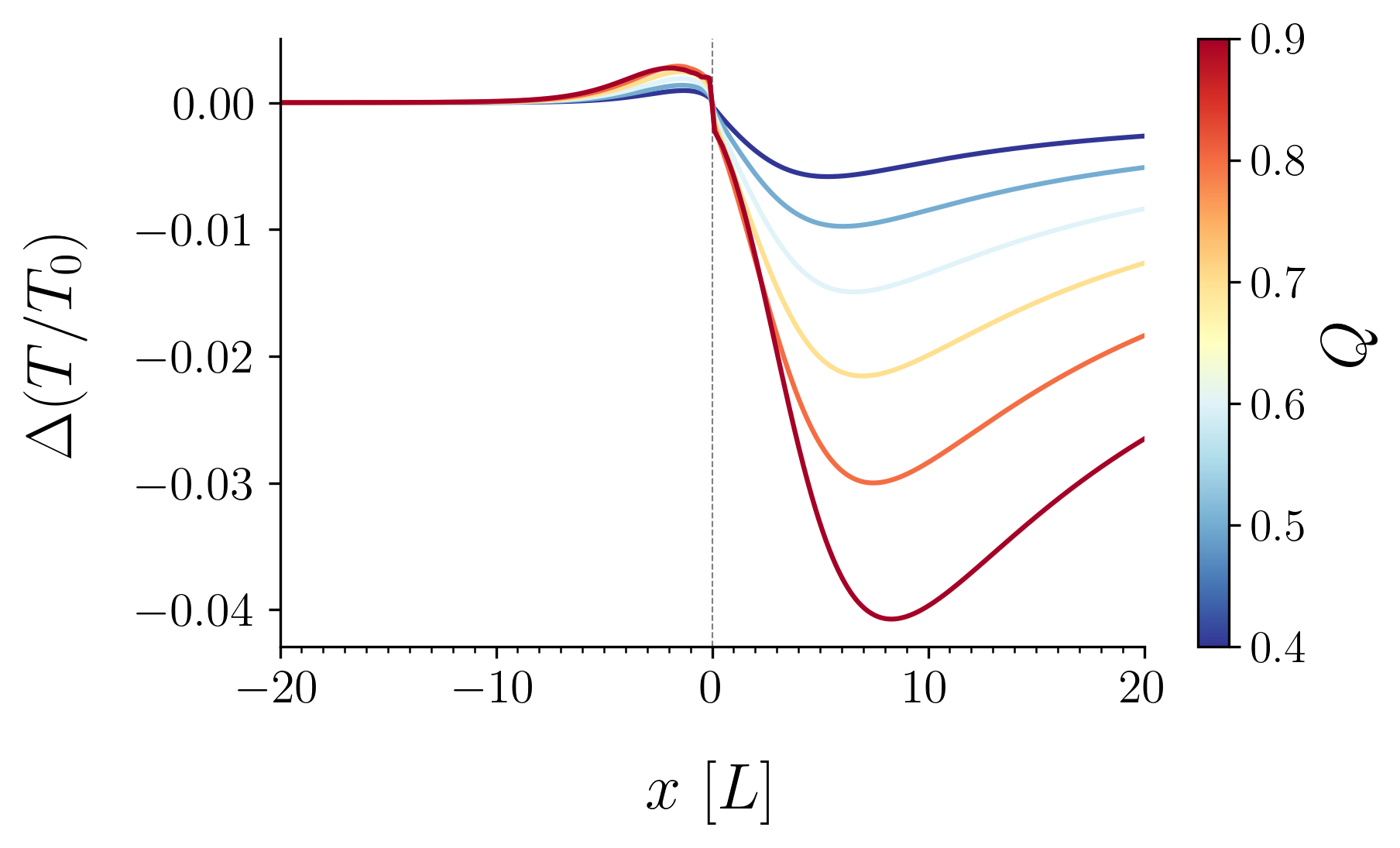}
        \caption{$s=\ell=1/2$.}
        \label{fig:varQ_Temperature_diff_HBHGD_b}
    \end{subfigure}

    \begin{subfigure}{0.48\textwidth}
        \centering
        \includegraphics[width=\linewidth]{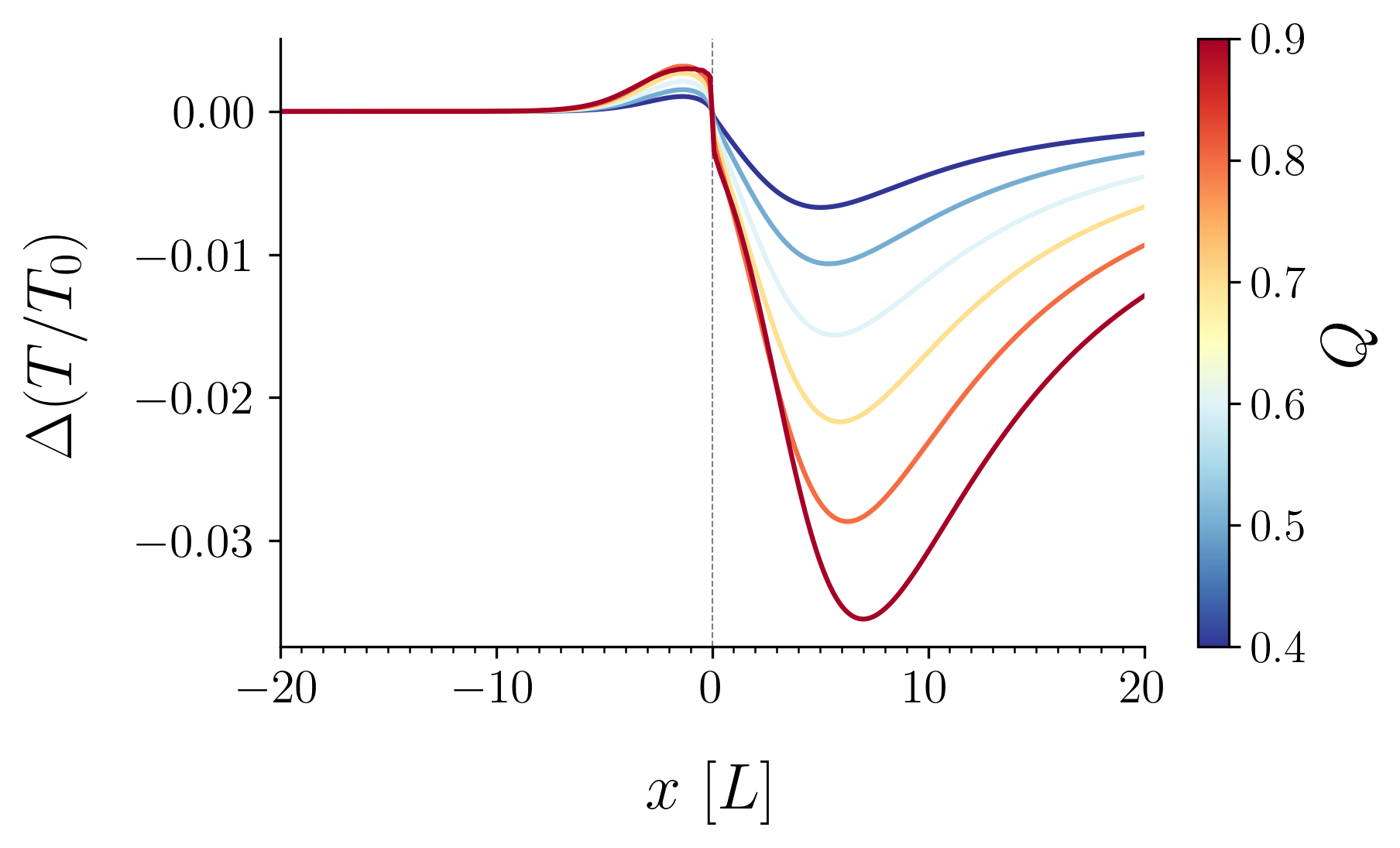}
        \caption{$s=\ell=1$.}
        \label{fig:varQ_Temperature_diff_HBHGD_c}
    \end{subfigure}
    \hfill
    \begin{subfigure}{0.48\textwidth}
        \centering
        \includegraphics[width=\linewidth]{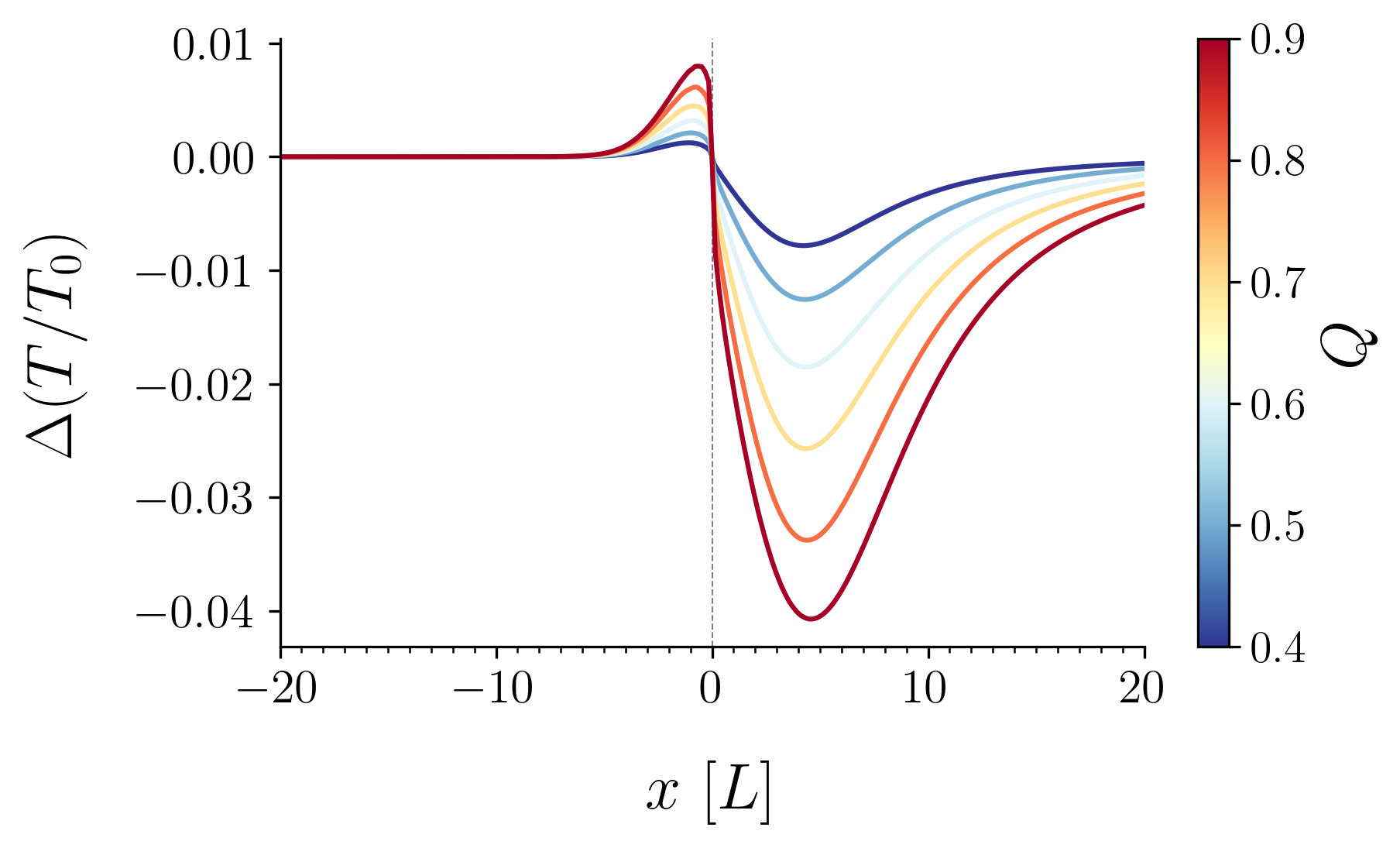}
        \caption{$s=\ell=2$.}
        \label{fig:varQ_Temperature_diff_HBHGD_d}
    \end{subfigure}
    \caption[Difference in temperature of the hairy analogue, varying $Q$]{Difference in the temperature of the flow, $\Delta({T/T_0})$, Difference from the Schwarzschild analogue of the same channel, $\Delta X(x)=X(x)-X^\text{Schw}(x)$ read station by station, for the hairy black hole~\eqref{eq: metric HBH} at $\alpha=\ell_h=0.2$ with $Q$ swept over the admissible range, from $0.4$ to $0.9$ (colour bar), one panel per channel. The vertical scale is proper to each panel. Natural units with $G_\text{N}=M=1$.}
    \label{fig:varQ_Temperature_diff_HBHGD}
\end{figure}

\begin{figure}[H]
    \centering
    \begin{subfigure}{0.48\textwidth}
        \centering
        \includegraphics[width=\linewidth]{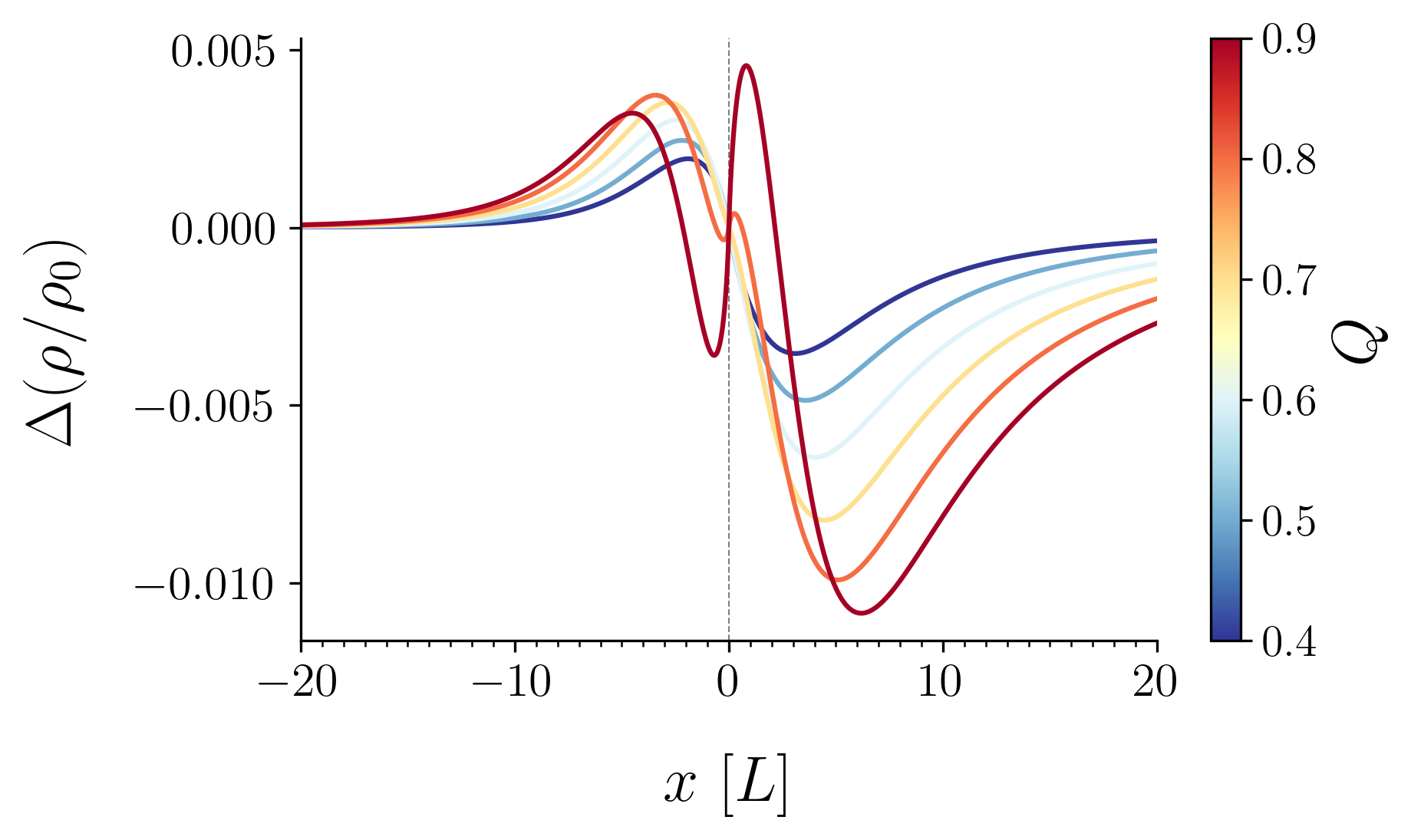}
        \caption{$s=\ell=0$.}
        \label{fig:varQ_Density_diff_HBHGD_a}
    \end{subfigure}
    \hfill
    \begin{subfigure}{0.48\textwidth}
        \centering
        \includegraphics[width=\linewidth]{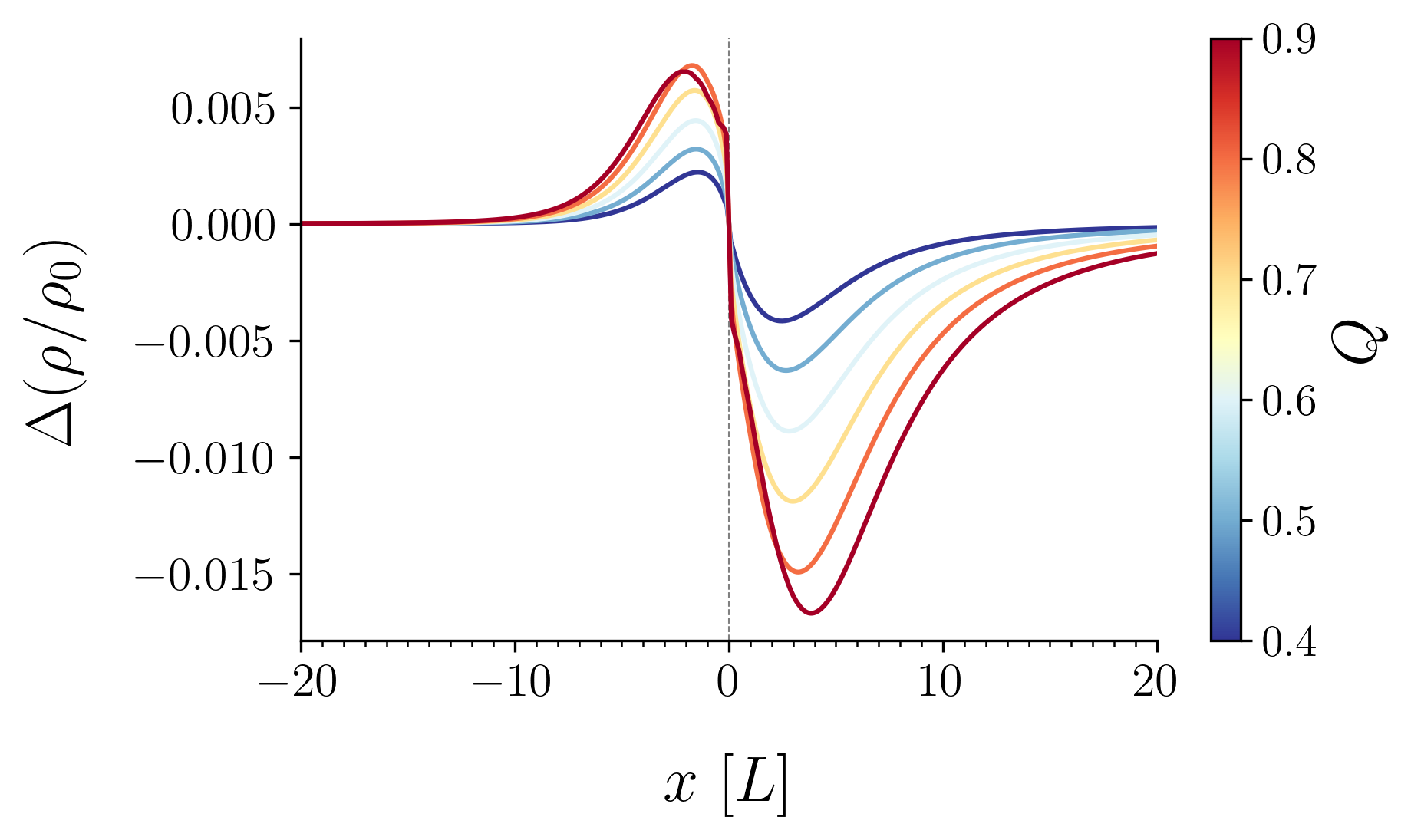}
        \caption{$s=\ell=1/2$.}
        \label{fig:varQ_Density_diff_HBHGD_b}
    \end{subfigure}

    \begin{subfigure}{0.48\textwidth}
        \centering
        \includegraphics[width=\linewidth]{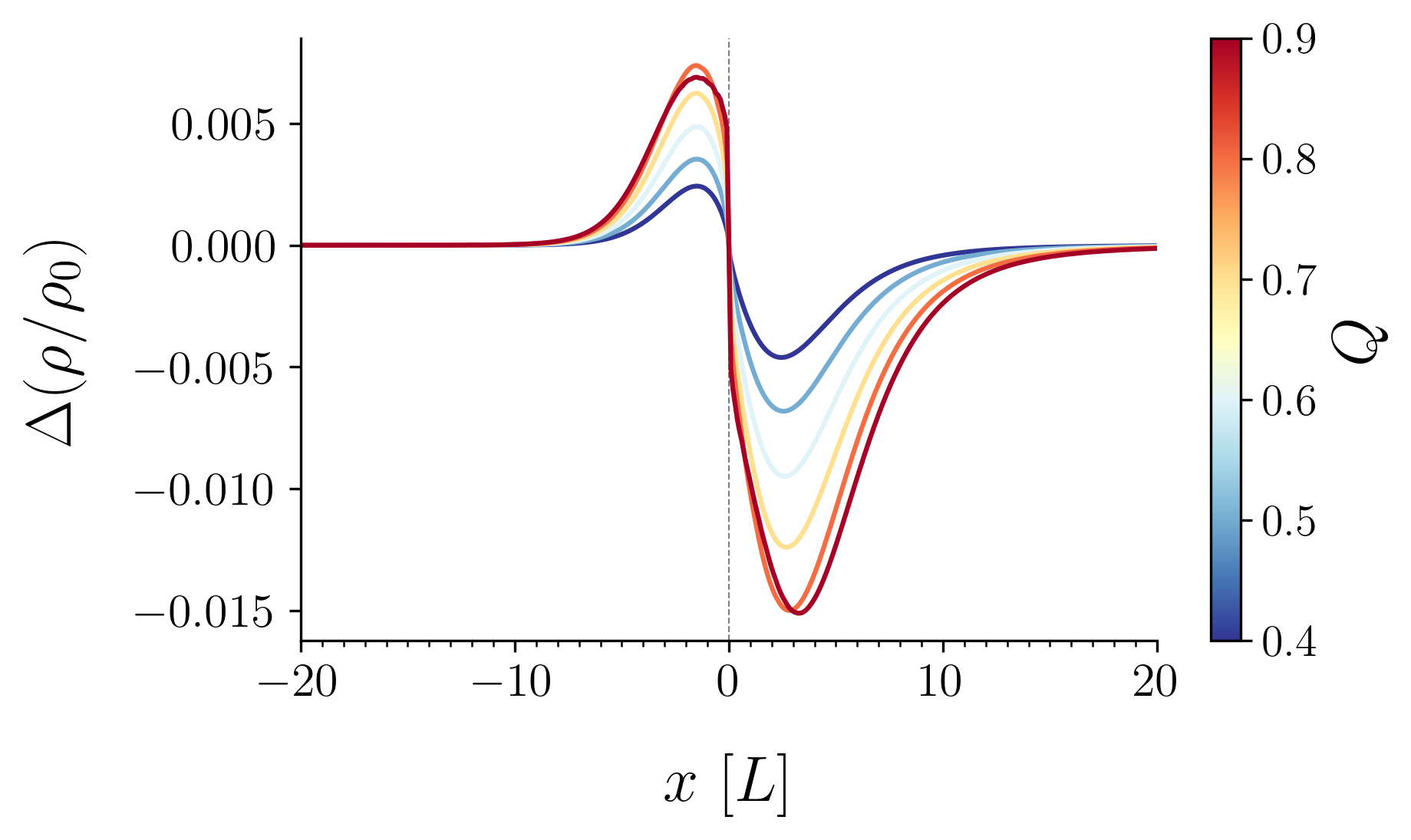}
        \caption{$s=\ell=1$.}
        \label{fig:varQ_Density_diff_HBHGD_c}
    \end{subfigure}
    \hfill
    \begin{subfigure}{0.48\textwidth}
        \centering
        \includegraphics[width=\linewidth]{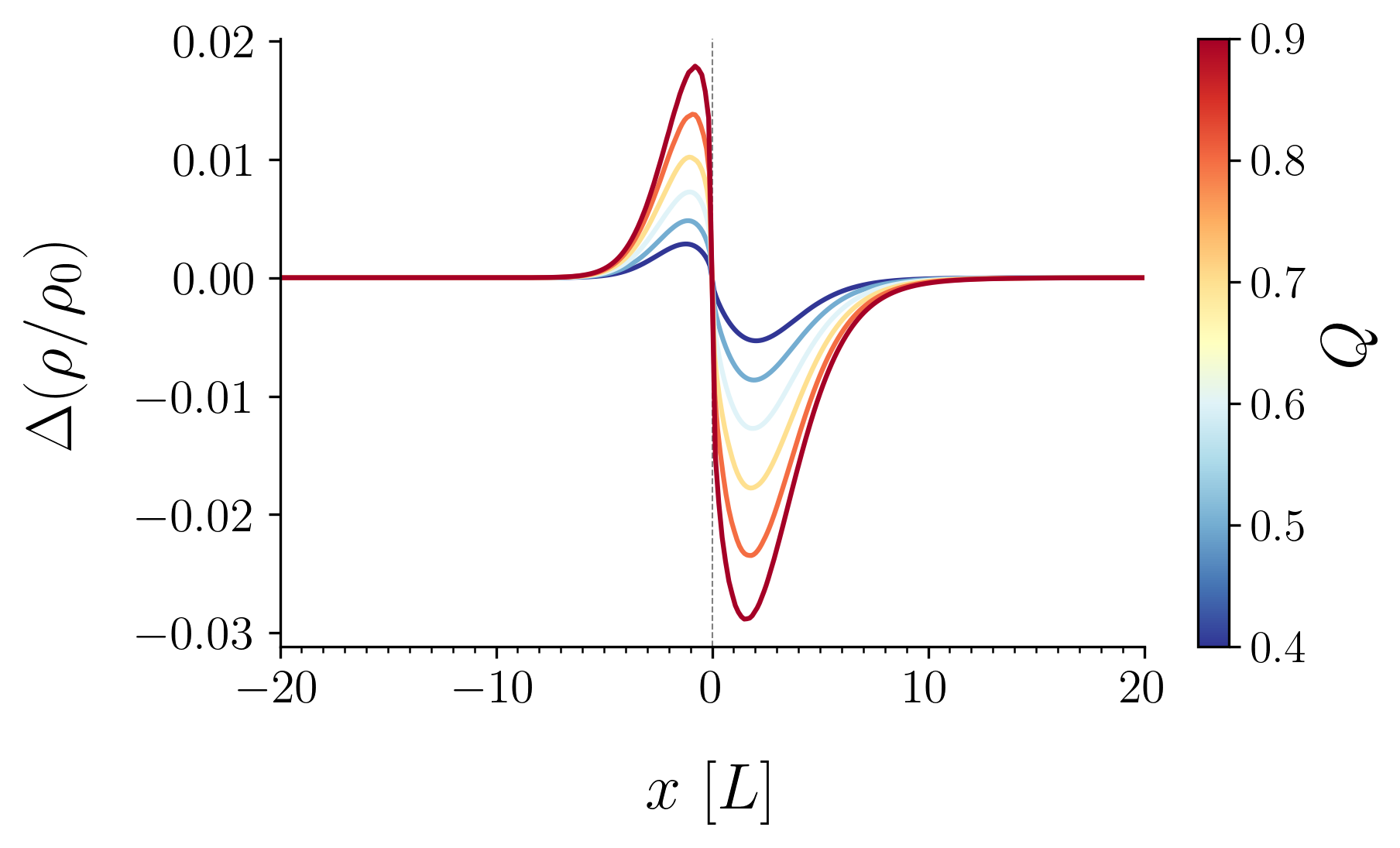}
        \caption{$s=\ell=2$.}
        \label{fig:varQ_Density_diff_HBHGD_d}
    \end{subfigure}
    \caption[Difference in density of the hairy analogue, varying $Q$]{Difference in the density of the flow, $\Delta({\rho/\rho_0})$, Difference from the Schwarzschild analogue of the same channel, $\Delta X(x)=X(x)-X^\text{Schw}(x)$ read station by station, for the hairy black hole~\eqref{eq: metric HBH} at $\alpha=\ell_h=0.2$ with $Q$ swept over the admissible range, from $0.4$ to $0.9$ (colour bar), one panel per channel. The vertical scale is proper to each panel. Natural units with $G_\text{N}=M=1$.}
    \label{fig:varQ_Density_diff_HBHGD}
\end{figure}

The pressure, temperature, and density matter because they are what an hypothetical experiment measures directly, even though the isentropic relations~\eqref{eq: rel local total} of Chapter~\ref{chap:isentropic equations} determine all three from $\mathrm{M}$ alone. Figures~\ref{fig:varQ_Pressure_diff_HBHGD} to~\ref{fig:varQ_Density_diff_HBHGD} report them as differences from the Schwarzschild analogue of the same channel, which for these three quantities is directly the departure in units of the stagnation value, that being the normalisation they already carry.

The pressure sets the pattern the other two follow. The difference vanishes at both ends of the panel and lives within a few units of the throat, positive on the convergent side and negative on the divergent one. In the Dirac, electromagnetic and gravitational channels it is a single antisymmetric deviation whose amplitude grows steadily with the charge, reaching $0.011$, $0.011$ and $0.025$ of stagnation at $Q=0.9$, and whose extremum migrates towards the throat as the spin rises, from $x=2.6$ in the Dirac channel to $x=2.0$ in the electromagnetic one and $x=1.1$ in the gravitational one, which is the same concentration at small $r$ that the $Q^2/r^2$ term produces in the metric.
Panel~\ref{fig:varQ_Pressure_diff_HBHGD_a} departs from that pattern, and the effective potential accounts for the departure. Of the four panels of Fig.~\ref{fig:hbh veff} in Chapter~\ref{chap:conexaoentrebnsebocais}, the scalar one is the only case in which the Schwarzschild barrier does not belong to the family it is compared with, its maximum standing above every hairy curve and at a smaller radius, so that $V_\text{eff}-V_\text{eff}^{\text{Schw}}$ changes sign as $r$ crosses the peak. In the gravitational panel~\ref{fig:hbh veff s2} the Schwarzschild curve lies inside the family and that difference keeps one sign throughout. The reconstruction carries the distinction straight into the duct, and from $Q=0.8$ upwards the scalar difference oscillates about zero over the two units on either side of the throat, running to $-0.004$ at $x\simeq-1$ and $+0.005$ at $x\simeq1$ before settling into its extremum of $-0.006$ at $x=5.0$. The same oscillation is visible in the temperature and in the density of that channel, and it is the one place in the whole sweep where the hairy flow crosses the Schwarzschild reference more than once.
Equation~\eqref{eq:hbh veff int explicit} says why the scalar channel is the sensitive one. At $\ell=0$ the centrifugal term $\ell(\ell+1)/r^2$ is absent and the barrier reduces to the blackening factor multiplying the bracket carried by $1-s^2$, so that the charge-like parameter reaches the potential twice over, through the $Q^2/r^2$ of the factor and through the $-2Q^2/r^3$ of the bracket, and the leading purely $Q$-dependent contribution is of order $Q^4/r^6$. At $s=\ell=1$ the bracket drops out, the charge enters only through the blackening factor and the contribution is of order $Q^2/r^4$, while at $s=\ell=2$ the centrifugal term $6/r^2$ dominates the barrier and dilutes the charge further still. The scalar channel is therefore the one in which the hair reshapes the barrier rather than merely raising it, and the nozzle inherits that.

The temperature, Fig.~\ref{fig:varQ_Density_diff_HBHGD}, has an exception worth recording. Its difference is the largest of the three, between $0.026$ and $0.041$ of stagnation depending on the channel, and above all it is the one that survives to the exit. In the scalar channel it still stands at $0.022$ there, against $0.0005$ for the pressure and $0.0027$ for the density of the same channel, and the Dirac channel carries $0.027$. The temperature is thus the only one of the three that keeps a resolvable imprint of the hair as far as the nozzle exit, which is also the station at which the hypothetical duct of Sec.~\ref{sec:constraint_equations} would be easiest to instrument, were the obstacles listed there not decisive. The advantage narrows at high spin, the difference at the exit falling to $0.013$ in the electromagnetic channel and $0.004$ in the gravitational one.

The density closes the group and behaves as the pressure does, with a difference of $0.011$ to $0.029$ of stagnation at its extremum, and at the exit $0.0027$ in the scalar channel, $0.0013$ in the Dirac one and less than $10^{-4}$ in the remaining two. Read together, the three panels say that a hypothetical measurement taken at the extremum of the difference could use any of the three, whereas one taken at the exit has only the temperature to work with. That extremum is not an arbitrary station. It falls between $x\simeq1$ and $x\simeq5$ according to the channel, which is the stretch of duct reproducing the maximum of the effective potential, and therefore the region in which the quasinormal ringing of the corresponding black hole is generated, as Chaps.~\ref{chap:conexaoentrebnsebocais} and~\ref{chap:QNM_spectrum} establish.

\begin{figure}[H]
    \centering
    \begin{subfigure}{0.48\textwidth}
        \centering
        \includegraphics[width=\linewidth]{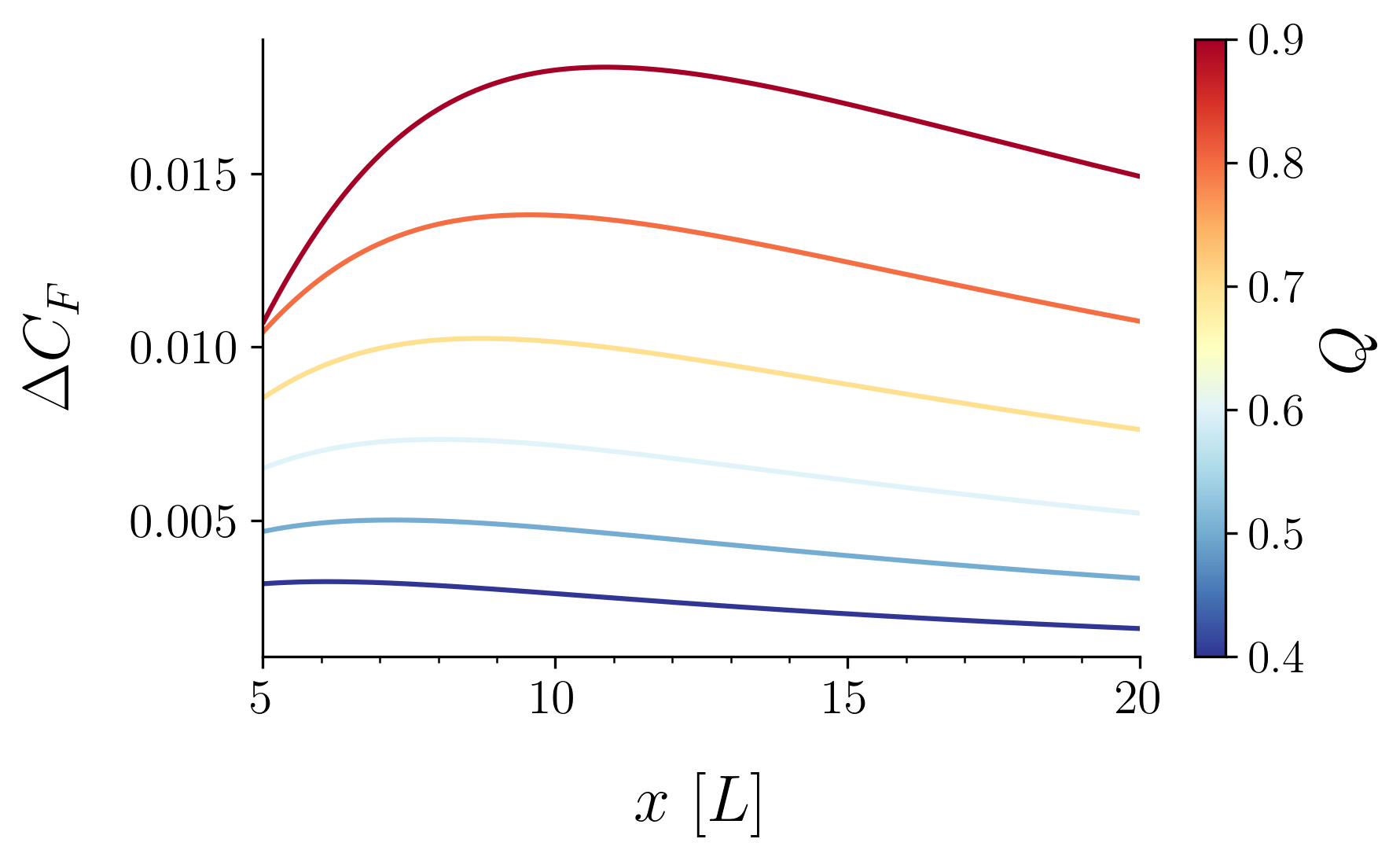}
        \caption{$s=\ell=0$.}
        \label{fig:varQ_C_F_diff_HBHGD_a}
    \end{subfigure}
    \hfill
    \begin{subfigure}{0.48\textwidth}
        \centering
        \includegraphics[width=\linewidth]{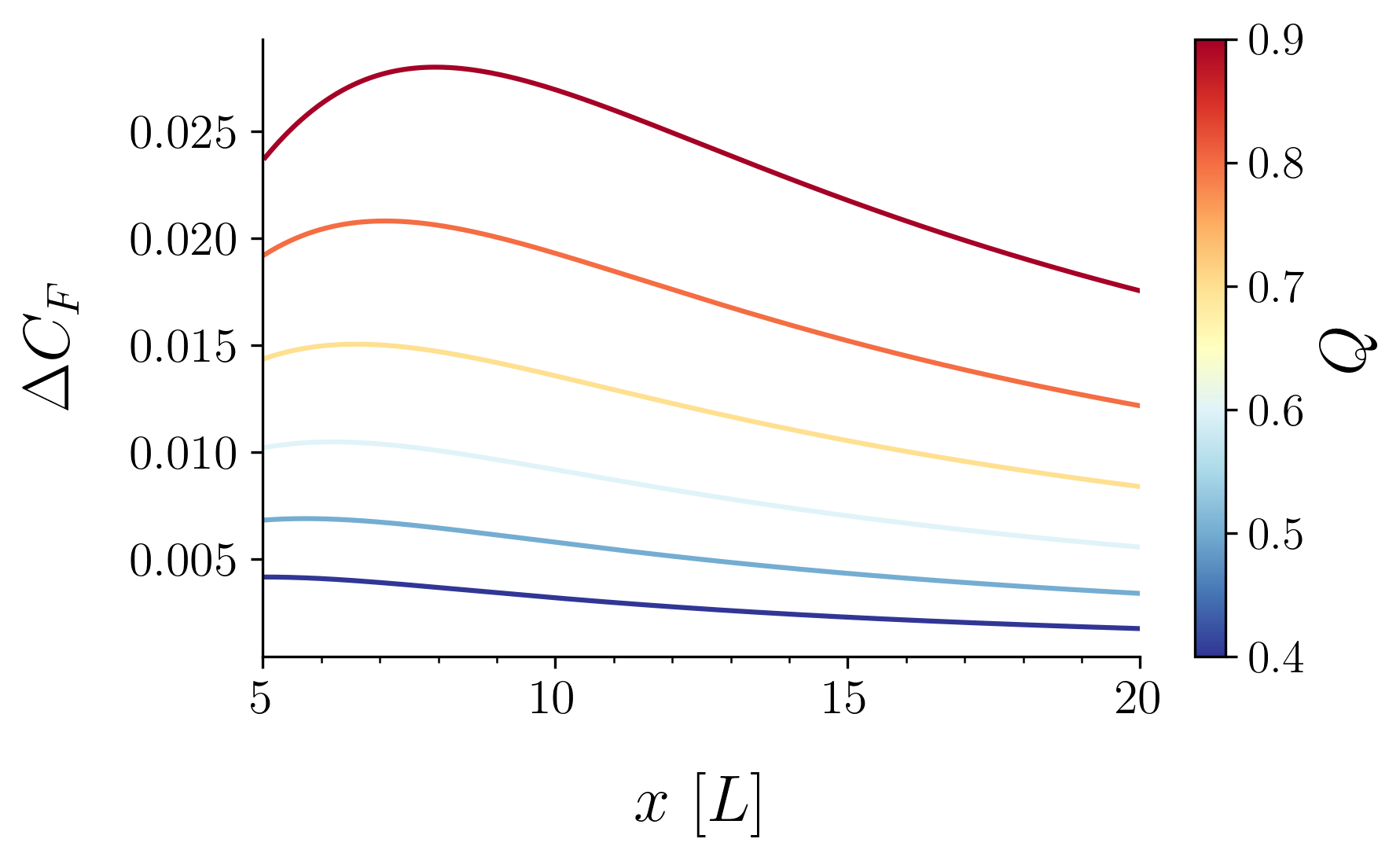}
        \caption{$s=\ell=1/2$.}
        \label{fig:varQ_C_F_diff_HBHGD_b}
    \end{subfigure}

    \begin{subfigure}{0.48\textwidth}
        \centering
        \includegraphics[width=\linewidth]{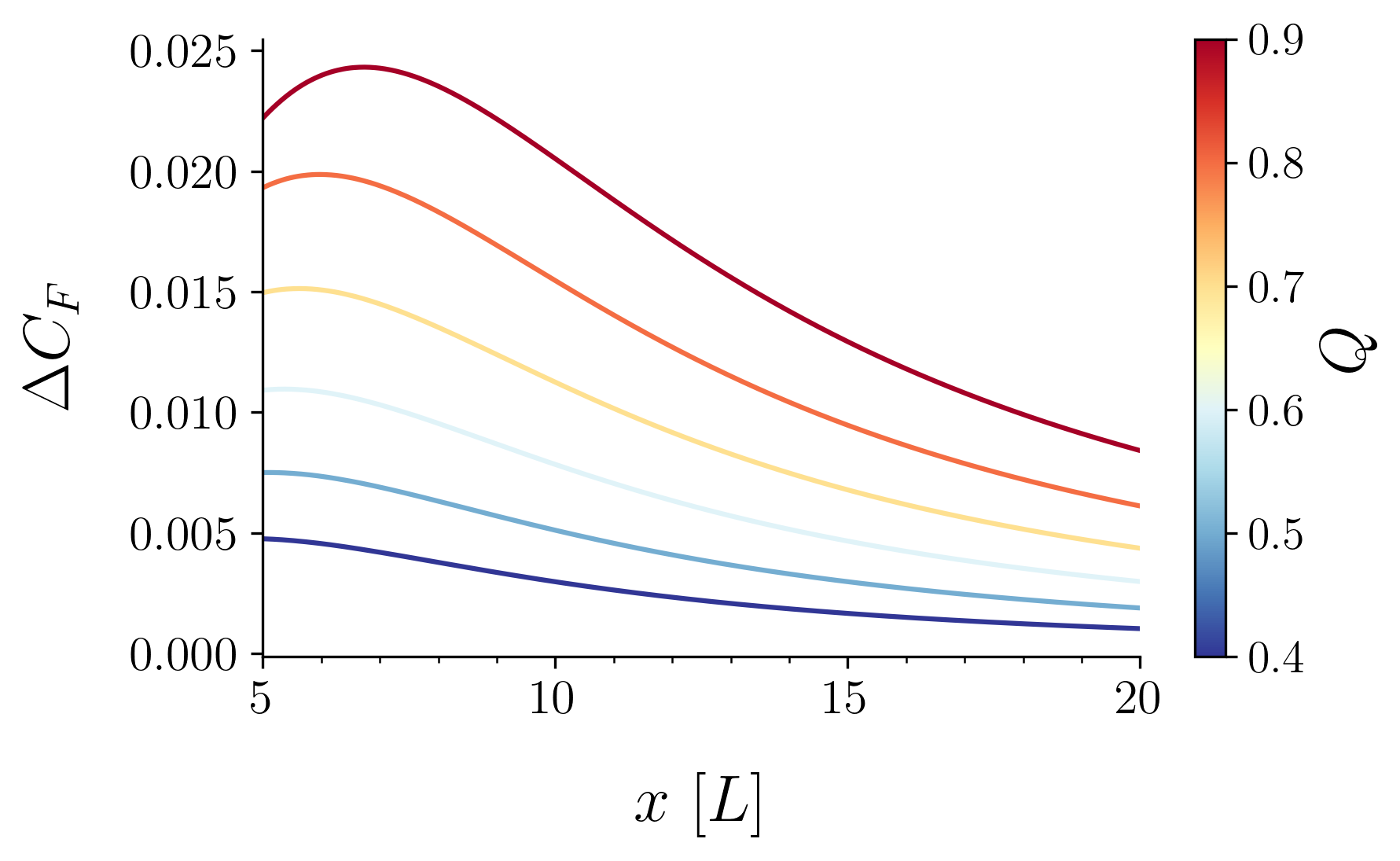}
        \caption{$s=\ell=1$.}
        \label{fig:varQ_C_F_diff_HBHGD_c}
    \end{subfigure}
    \hfill
    \begin{subfigure}{0.48\textwidth}
        \centering
        \includegraphics[width=\linewidth]{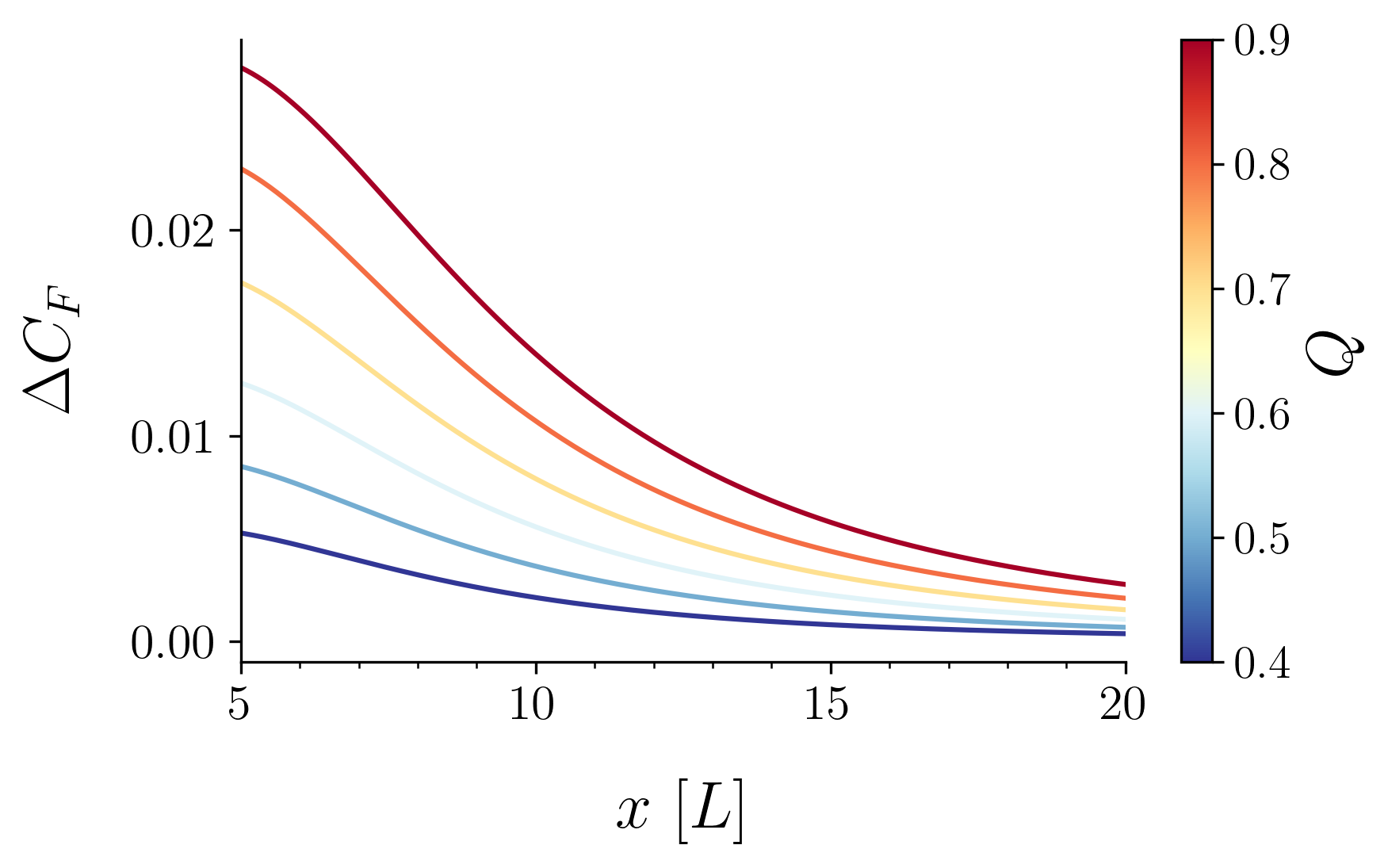}
        \caption{$s=\ell=2$.}
        \label{fig:varQ_C_F_diff_HBHGD_d}
    \end{subfigure}
    \caption[Difference in the thrust coefficient of the hairy analogue, varying $Q$]{Difference in the thrust coefficient, $\Delta C_F$, Difference from the Schwarzschild analogue of the same channel, $\Delta X(x)=X(x)-X^\text{Schw}(x)$ read station by station, for the hairy black hole~\eqref{eq: metric HBH} at $\alpha=\ell_h=0.2$ with $Q$ swept over the admissible range, from $0.4$ to $0.9$ (colour bar), one panel per channel. The vertical scale is proper to each panel. Natural units with $G_\text{N}=M=1$.}
    \label{fig:varQ_C_F_diff_HBHGD}
\end{figure}

The thrust coefficient closes the sequence. It increases with $Q$ and saturates towards its vacuum limit~\eqref{eq: coeficiente de empuxo maximo} at the exit, the scalar duct moving from $1.716$ to $1.731$ at $Q=0.9$ and the gravitational one from $1.807$ to $1.810$. Being dimensionless and routinely measured, it is the most accessible of the performance parameters, but its difference decays instead of persisting, reaching $0.018$ around $x=11$ in the scalar channel at $Q=0.9$ and falling to $0.015$ by the exit, and decaying faster the higher the spin, down to $0.003$ at the exit in the gravitational channel. The panels of Fig.~\ref{fig:varQ_C_F_diff_HBHGD} start at $x=5$, a quarter of the way along the duct, because the thrust coefficient is by construction a quantity read at the exit, being the wall pressure integrated over the whole divergent section. A value assigned to a station close to the analogue horizon, where almost none of that integral has accumulated, is therefore not a thrust that any measurement would return, and the difference plotted there would be read wrongly.

Two features of the hairy duct are worth carrying into what follows. The deformation is felt over the outer part of the duct, the deviation in the Mach profile still growing at the exit, and the parameter that drives it acts through a term of the metric that survives far outside the horizon. The other geometry of Part~\ref{part: extensions of gr} is the opposite case on both counts, its correction being short-ranged and read at the horizon itself, and the next section shows that the duct records it accordingly, close to the throat and decaying before the exit.

\section{Acoustic Analogue of the Quantum-Corrected Black Hole}
\label{sec: results quantum correction}


The reconstruction of Sec.~\ref{sec:constraint_equations} is now applied to the quantum-corrected (Calmet--Kuipers) metric of Chapter~\ref{chap:EFT}, Eq.~\eqref{eq:CK_metric}. For each spin $s$ and multipole $\ell$, the effective potential $V_{\text{eff}}(r)$ derived in Chapter~\ref{chap:conexaoentrebnsebocais} is carried into a de~Laval nozzle through the Sturm--Liouville equation~\eqref{eq: lambda}, and the nozzle relations of Sec.~\ref{sec:construcaometodocomp} then return the flow.

The tabulated solutions cover $-1.5\times10^{-2}\leq c_6\leq5\times10^{-3}$, and two bounds of Chapter~\ref{chap:EFT} delimit that interval. Perturbativity at the horizon, Eq.~\eqref{eq:c6_bound}, requires $\abs{c_6}<1/5\pi\approx6.4\times10^{-2}$ and is met throughout. The existence of the surface of infinite redshift, Eq.~\eqref{eq:B_surface_exists}, is far more restrictive on the positive branch, requiring $c_6\leq3.61\times10^{-3}$, and beyond it $B>0$ everywhere, leaving only the throat at $A^{-1}=0$. The two largest values of the sweep therefore describe a wormhole rather than a black hole, in the sense made precise in Subsec.~\ref{subsec:EFT_horizon}. They are kept in the figures, where the reconstruction uses $A^{-1}=0$ throughout and remains well defined, and are left out of the quantitative statements below, which stop at $c_6=3\times10^{-3}$.
The panels are again the four channels at their lowest admissible multipole, $\ell=s$, the same four tabulated in Part~\ref{part: quasinormal modes}, so that the acoustic results can be read directly against the gravitational spectra computed there. Differences are taken as in Sec.~\ref{sec:results HBH by GD}, against the Schwarzschild analogue of the same channel and station by station in $x$, which makes the two halves of this chapter directly comparable, the Schwarzschild duct being the same object in both.

\begin{figure}[H]
    \centering
    \begin{subfigure}{0.49\textwidth}
        \centering
        \includegraphics[width=\linewidth]{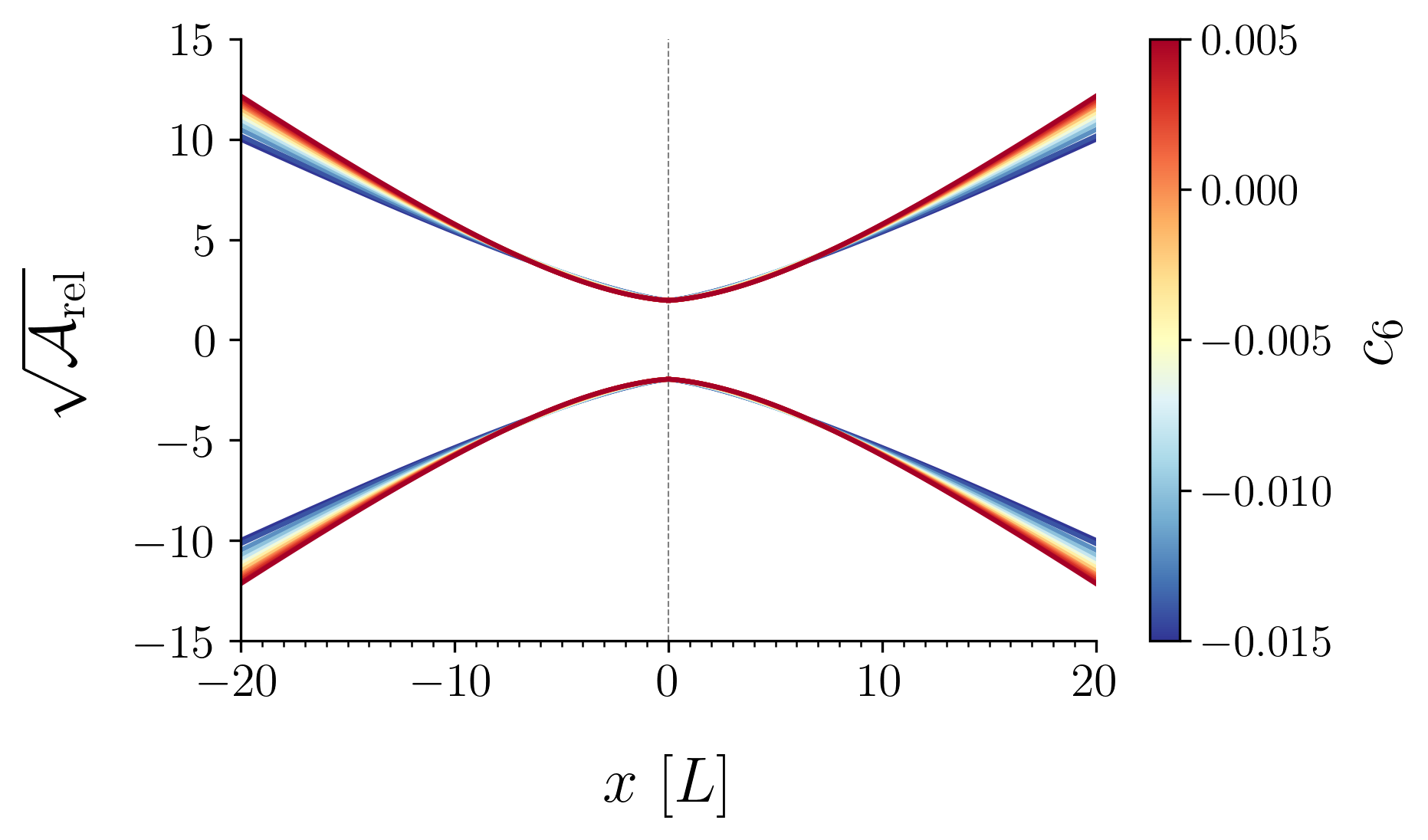}
         \caption{$s = \ell = 0$.}
        \label{fig:shapea}
    \end{subfigure}
    \hfill
    \begin{subfigure}{0.49\textwidth}
        \centering
        \includegraphics[width=\linewidth]{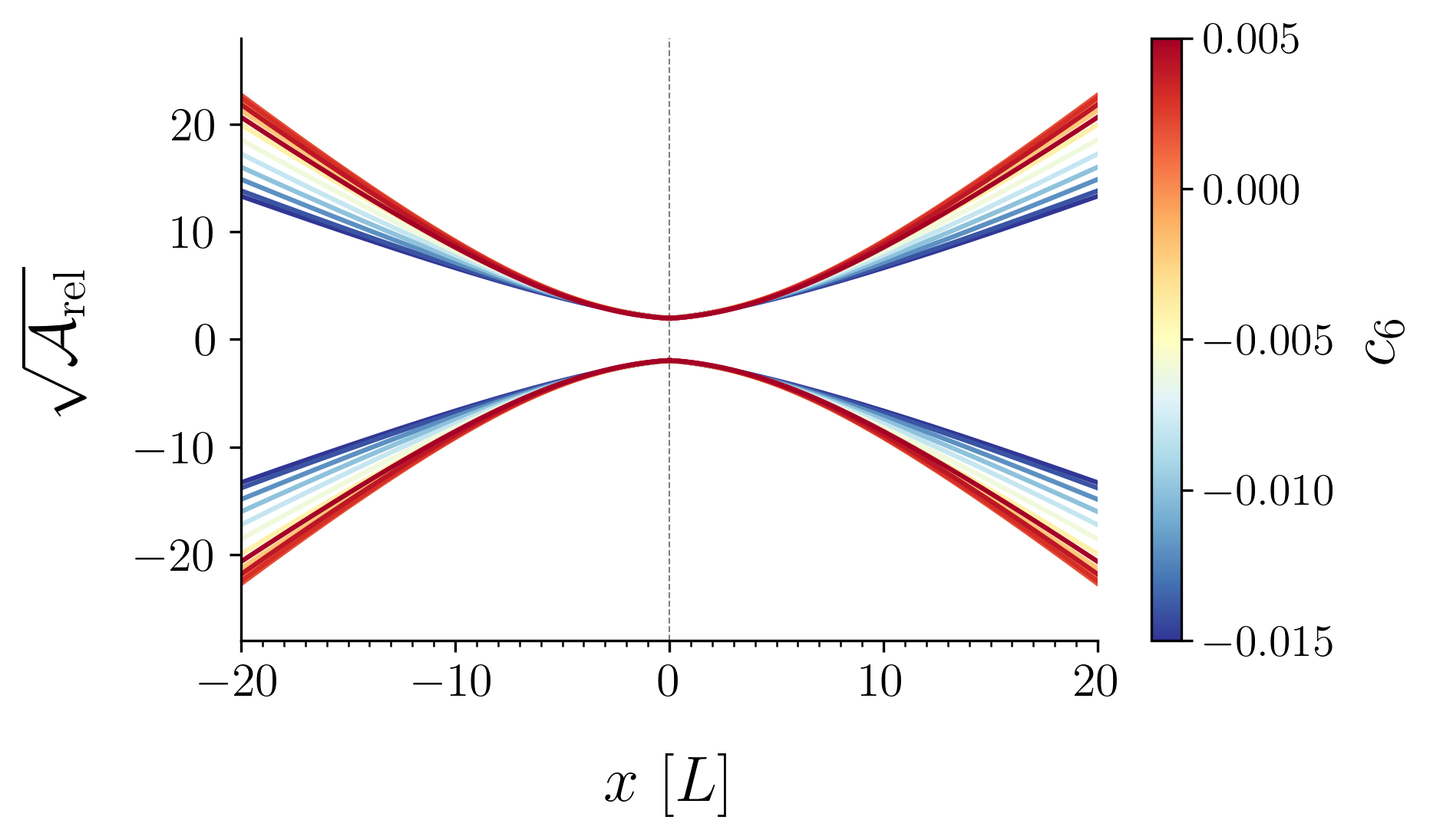}
         \caption{$s = \ell= {1}/{2}$.}
        \label{fig:shapee1}
    \end{subfigure}

    \begin{subfigure}{0.49\textwidth}
        \centering
        \includegraphics[width=\linewidth]{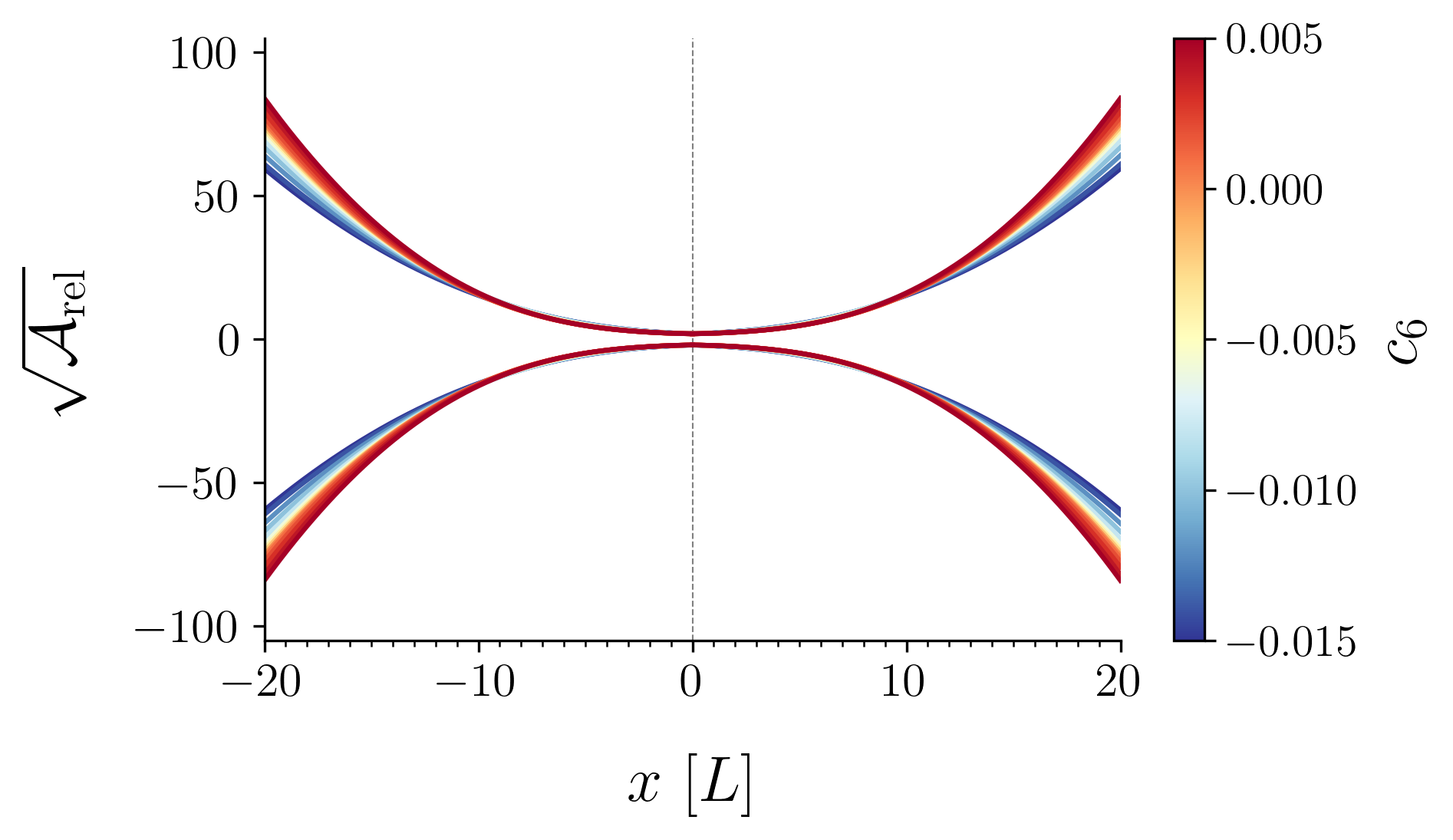}
         \caption{$s =\ell=1$.}
        \label{fig:shapeb2}
    \end{subfigure}
    \hfill
    \begin{subfigure}{0.49\textwidth}
        \centering\includegraphics[width=\linewidth]{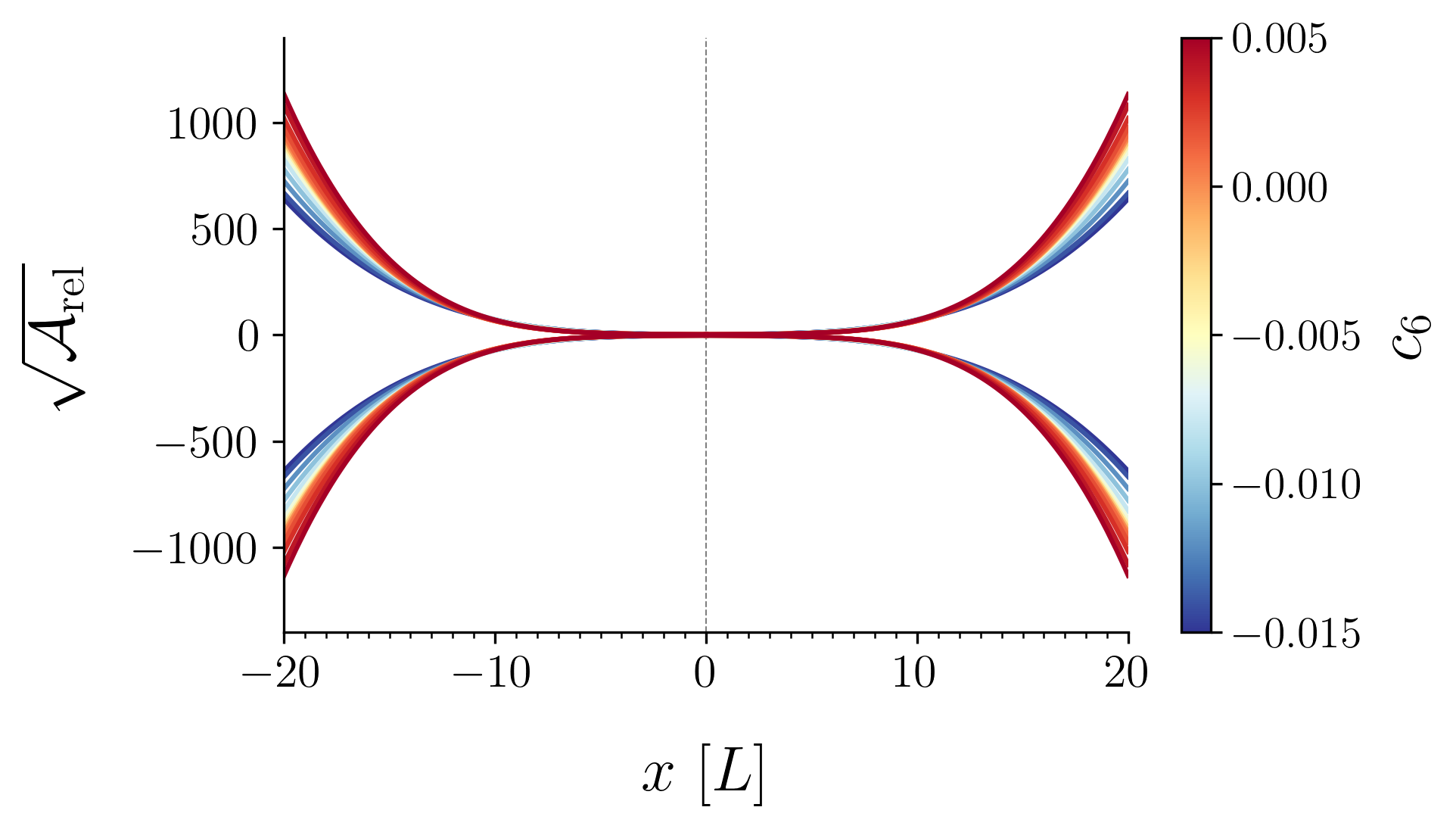}
         \caption{$s=\ell=2$.}
        \label{fig:shapec3}
    \end{subfigure}
    \caption[Nozzle shape of the quantum-corrected analogue, varying $c_6$]{Nozzle shape geometry along the de~Laval nozzle, varying $c_6$ (colour bar), for the quantum-corrected Schwarzschild metric~\eqref{eq:CK_metric}, one panel per channel at $\ell=s$, the Schwarzschild analogue being the $c_6=0$ member of the sweep. Natural units with $G_{\text{N}} = M = 1$. Computed from the tabulated solutions of Ref.~\cite{Casadio:2025lkn}.}
    \label{fig:shape}
\end{figure}

The nozzle shape, Fig.~\ref{fig:shape}, inherits the structure of the effective-potential barrier. Its throat is the analogue horizon, placed at $x=0$ by the lower limit of the integral~\eqref{eq:xemfuncaoder}, and the barrier itself, whose peak sits near the light ring at $r\simeq3G_\text{N}M$, is carried into the body of the duct on either side. The coupling acts on the duct through the corrected horizon~\eqref{eq:horizon_shift} directly. A negative coupling pushes the horizon outwards, which stretches the reconstruction over a longer radial window and leaves the wall narrower at any fixed station, $\sqrt{\mathcal{A}_{\text{rel}}}$ at the exit falling from the Schwarzschild $11.7$ to $10.0$ in the scalar channel at $c_6=-1.5\times10^{-2}$, while a positive one opens it to $12.0$ at $c_6=3\times10^{-3}$. The other three channels behave the same way and more strongly, the exit wall of the Dirac channel running from $22.3$ to $13.3$ over the negative branch.

On the positive branch the Dirac channel does something the other three do not. Its wall keeps widening only as far as $c_6\simeq2\times10^{-3}$, where $\sqrt{\mathcal{A}_{\text{rel}}}$ at the exit reaches $22.7$, and then narrows again, to $20.6$ at the top of the sweep, so that the reddest curves of panel~\ref{fig:shapee1} fall back inside the family instead of bounding it. The Mach number at the exit turns at the same place, peaking at $7.40$ against the Schwarzschild $7.35$ and dropping to $7.09$ by $c_6=5\times10^{-3}$, whereas the scalar, electromagnetic and gravitational ducts open monotonically over the whole sweep. This is the analogue counterpart of the fermionic behaviour established in Chapter~\ref{chap:conexaoentrebnsebocais}, where the height of the Dirac barrier rises for $c_6\lesssim0$ and falls beyond a global maximum at $c_6\approx5\times10^{-4}$, only marginally displaced from Schwarzschild, while the bosonic branch decreases throughout. The duct inherits the turning point, and the small offset between the two is the reconstruction reading the barrier through the whole exterior rather than at its maximum alone.

\begin{figure}[H]
    \centering
    \begin{subfigure}{0.49\textwidth}
        \centering
        \includegraphics[width=\linewidth]{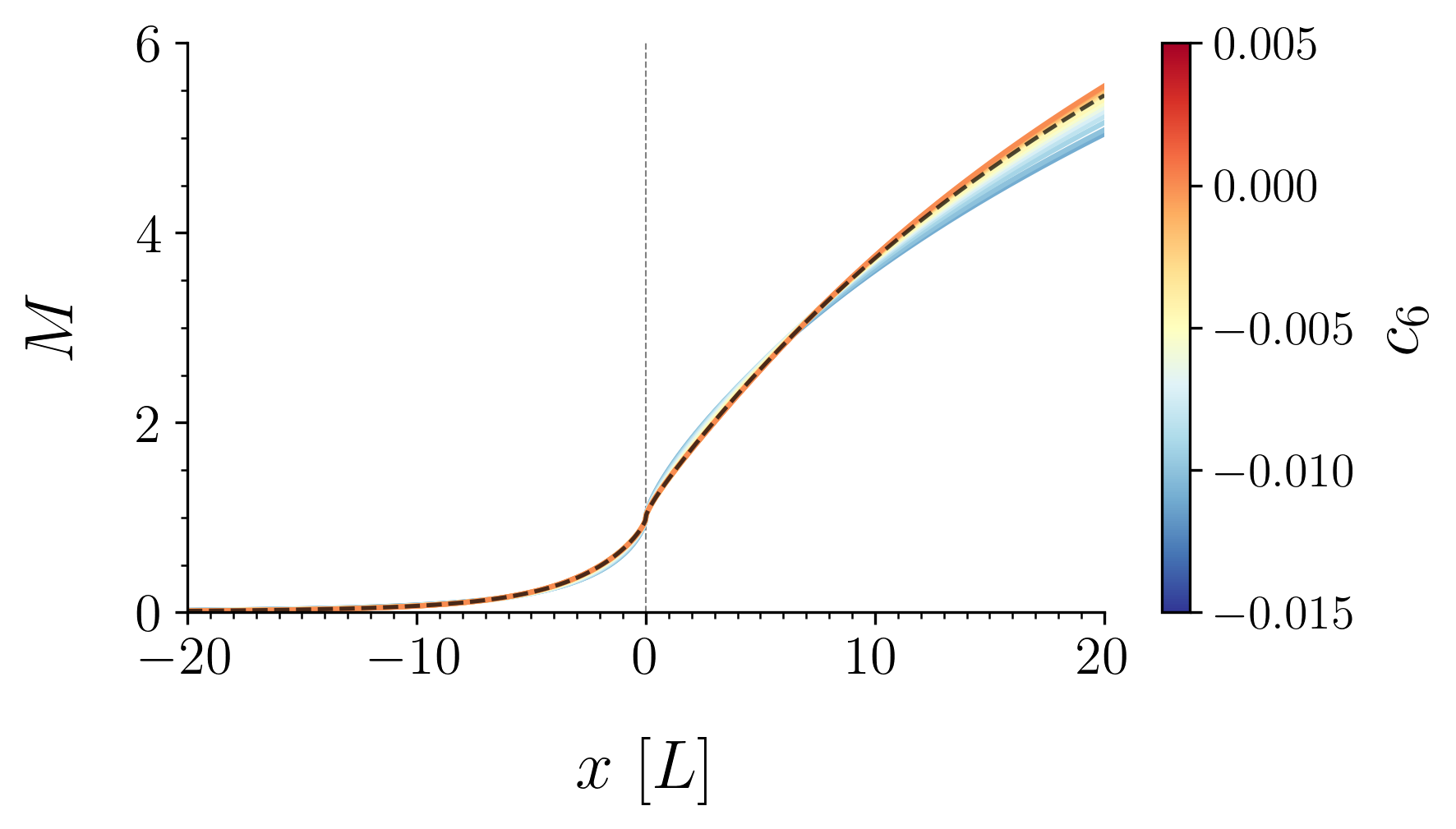}
         \caption{$s = \ell = 0$.}
        \label{fig:Macha}
    \end{subfigure}
    \hfill
    \begin{subfigure}{0.49\textwidth}
        \centering
        \includegraphics[width=\linewidth]{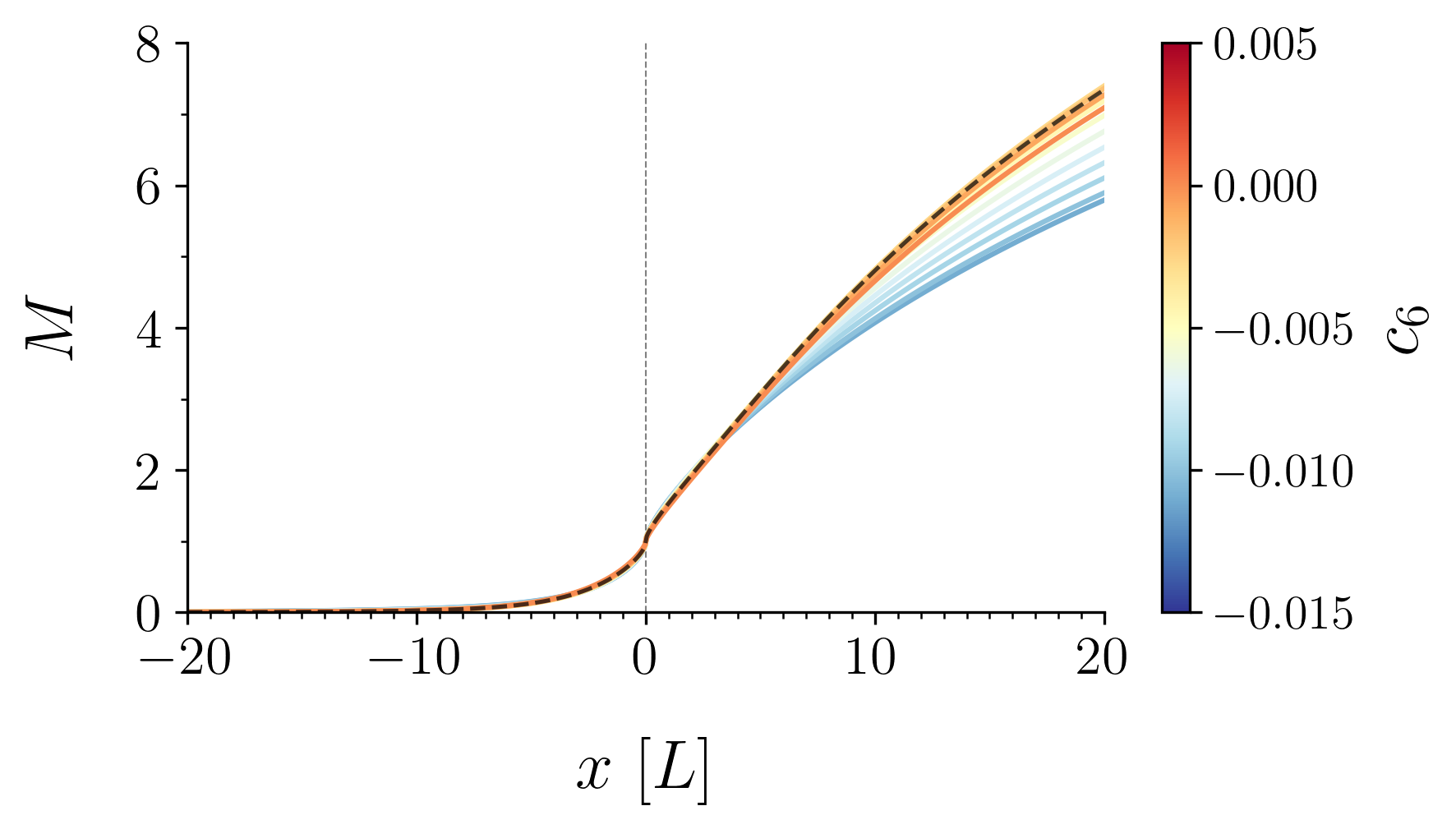}
         \caption{$s = \ell= {1}/{2}$.}
        \label{fig:Mache1}
    \end{subfigure}

    \begin{subfigure}{0.49\textwidth}
        \centering
        \includegraphics[width=\linewidth]{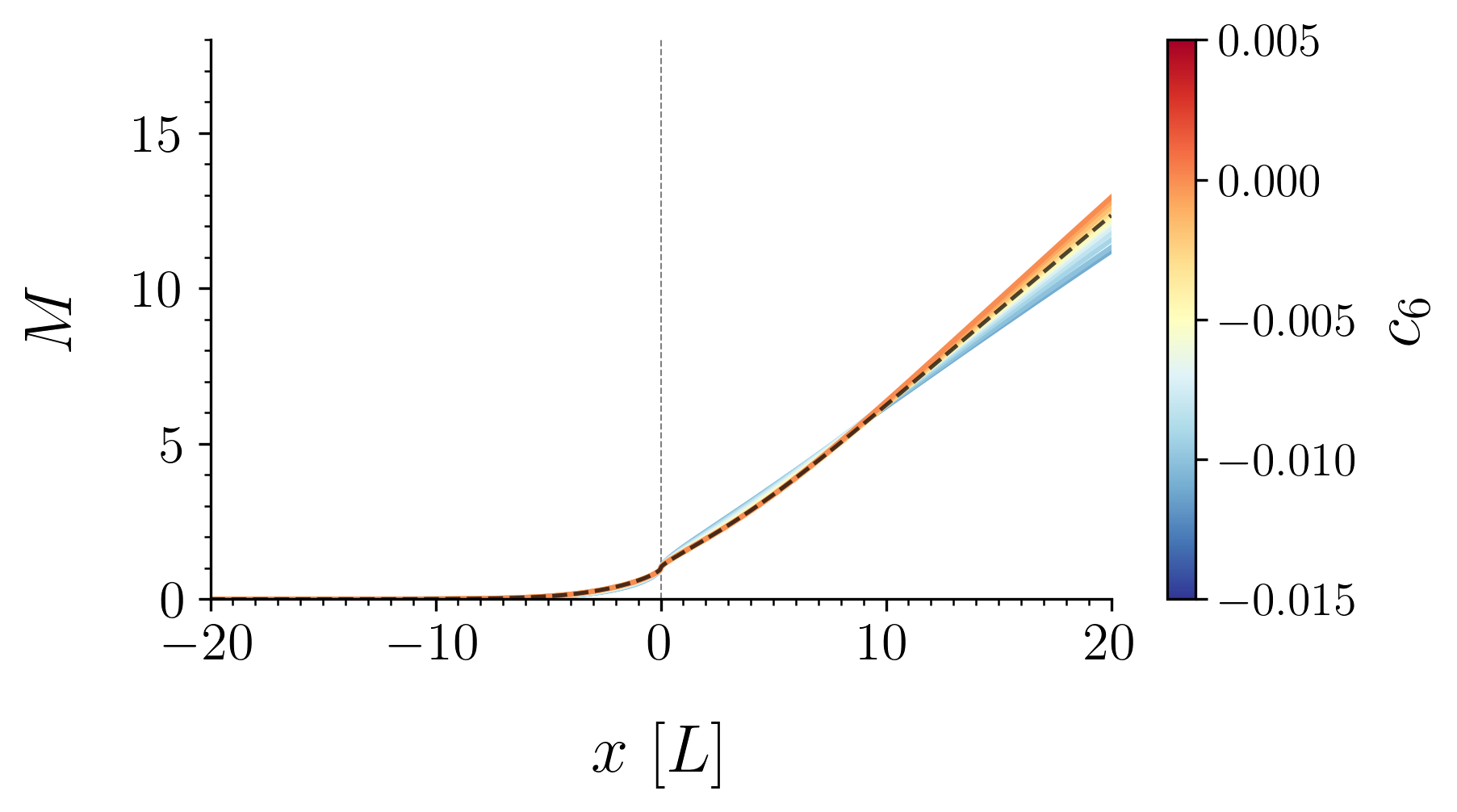}
         \caption{$s =\ell=1$.}
        \label{fig:Machb2}
    \end{subfigure}
    \hfill
    \begin{subfigure}{0.49\textwidth}
        \centering\includegraphics[width=\linewidth]{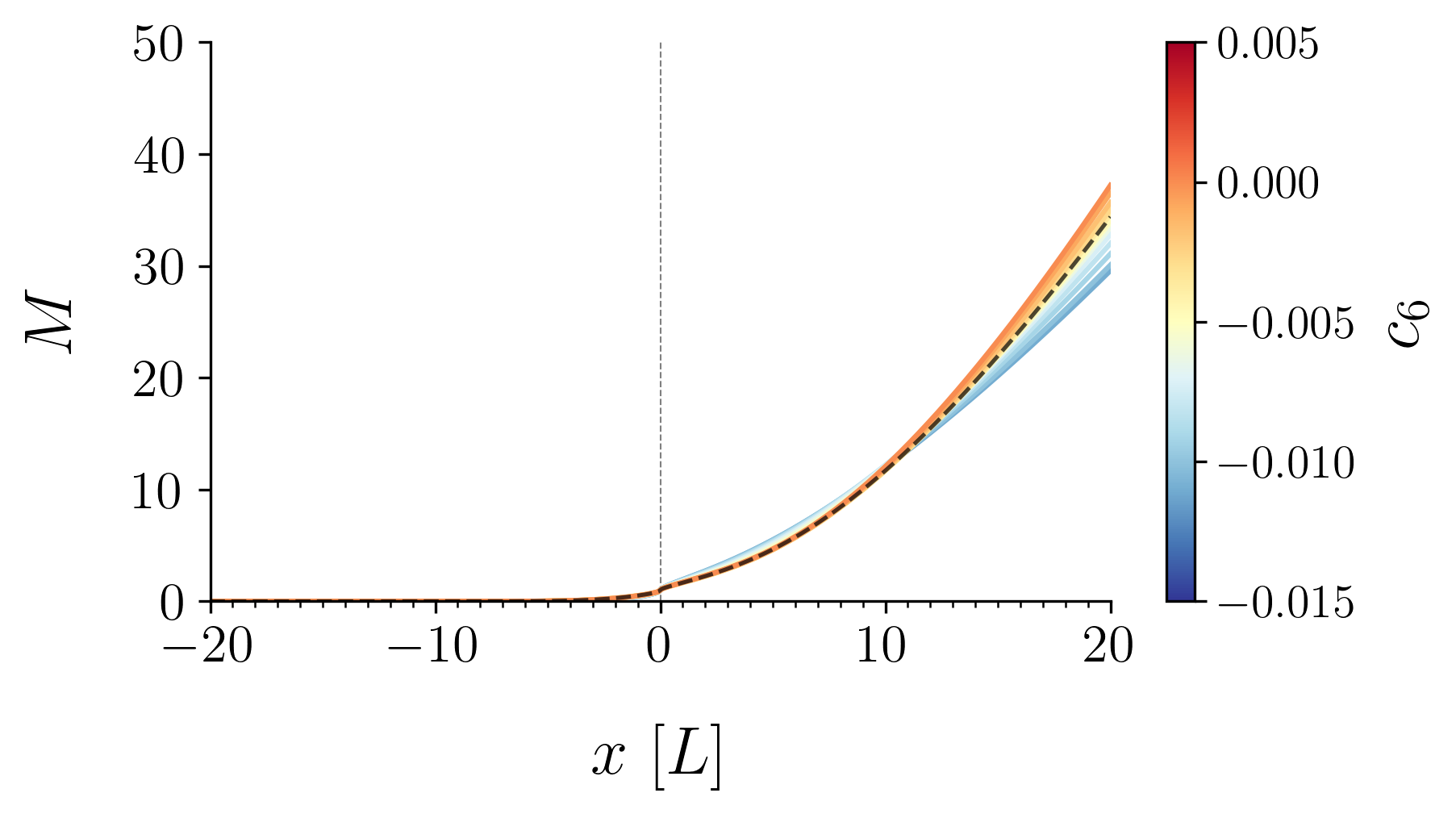}
         \caption{$s=\ell=2$.}
        \label{fig:Machc3}
    \end{subfigure}
    \caption[Mach number of the quantum-corrected analogue, varying $c_6$]{Mach number along the de~Laval nozzle, varying $c_6$, for the quantum-corrected Schwarzschild metric~\eqref{eq:CK_metric}, the dashed curve being the Schwarzschild analogue $c_6=0$. Natural units with $G_{\text{N}} = M = 1$. First published in Ref.~\cite{Casadio:2025lkn}.}
    \label{fig:Mach}
\end{figure}

The Mach-number profile, Fig.~\ref{fig:Mach}, crosses unity at the throat by construction, so what the correction changes is not the position of the transonic point but the flow on either side of it, and two features organise that change. The first is a pronounced asymmetry between the signs of $c_6$. At $c_6=-1.5\times10^{-2}$ the Mach number at the exit falls from the Schwarzschild $7.35$ to $5.79$ in the Dirac channel, a drop of $21\%$, against $7\%$, $9\%$ and $14\%$ in the scalar, electromagnetic and gravitational ones, whereas the whole positive branch admitted by Eq.~\eqref{eq:B_surface_exists} moves it by at most $4\%$ in any channel, and by only $0.3\%$ in the Dirac one. The second is that the ordering of the channels is not the one the barrier alone would suggest. In fractional terms the Dirac channel is by a wide margin the most responsive on the negative branch and the least responsive on the positive one, whereas the scalar channel, which carries the largest fractional change of barrier height according to Chapter~\ref{chap:conexaoentrebnsebocais}, moves the least of the four.
What the correction does change at the throat itself is the gradient of $\mathrm{M}$, which by Eq.~\eqref{eq: temperatura hawking bocal} fixes the analogue Hawking temperature of the flow~\cite{Casadio:2025lkn} as previously stated in Sec.~\ref{sec:results HBH by GD}.

The thermodynamic variables again follow from the Mach profile alone, and the correction concentrates their difference near the throat rather than spreading it along the duct, which is the opposite of what the hair does.

The pressure difference has its extremum within one unit of the throat in every channel, positive on the convergent side and negative on the divergent one, and has returned to zero by $\abs{x}\simeq7$. At $c_6=-1.5\times10^{-2}$ it reaches $0.040$ of stagnation in the scalar channel, $0.060$ in the electromagnetic one and $0.077$ in the gravitational one, the Dirac channel being the mild exception at $0.018$. The correction therefore concentrates its imprint at the throat, where the hair of Sec.~\ref{sec:results HBH by GD} spreads it over the first half of the divergent section, and that is the reconstruction responding to a horizon that has moved rather than to a barrier that has grown.

\begin{figure}[H]
    \centering
    \begin{subfigure}{0.49\textwidth}
        \centering
        \includegraphics[width=\linewidth]{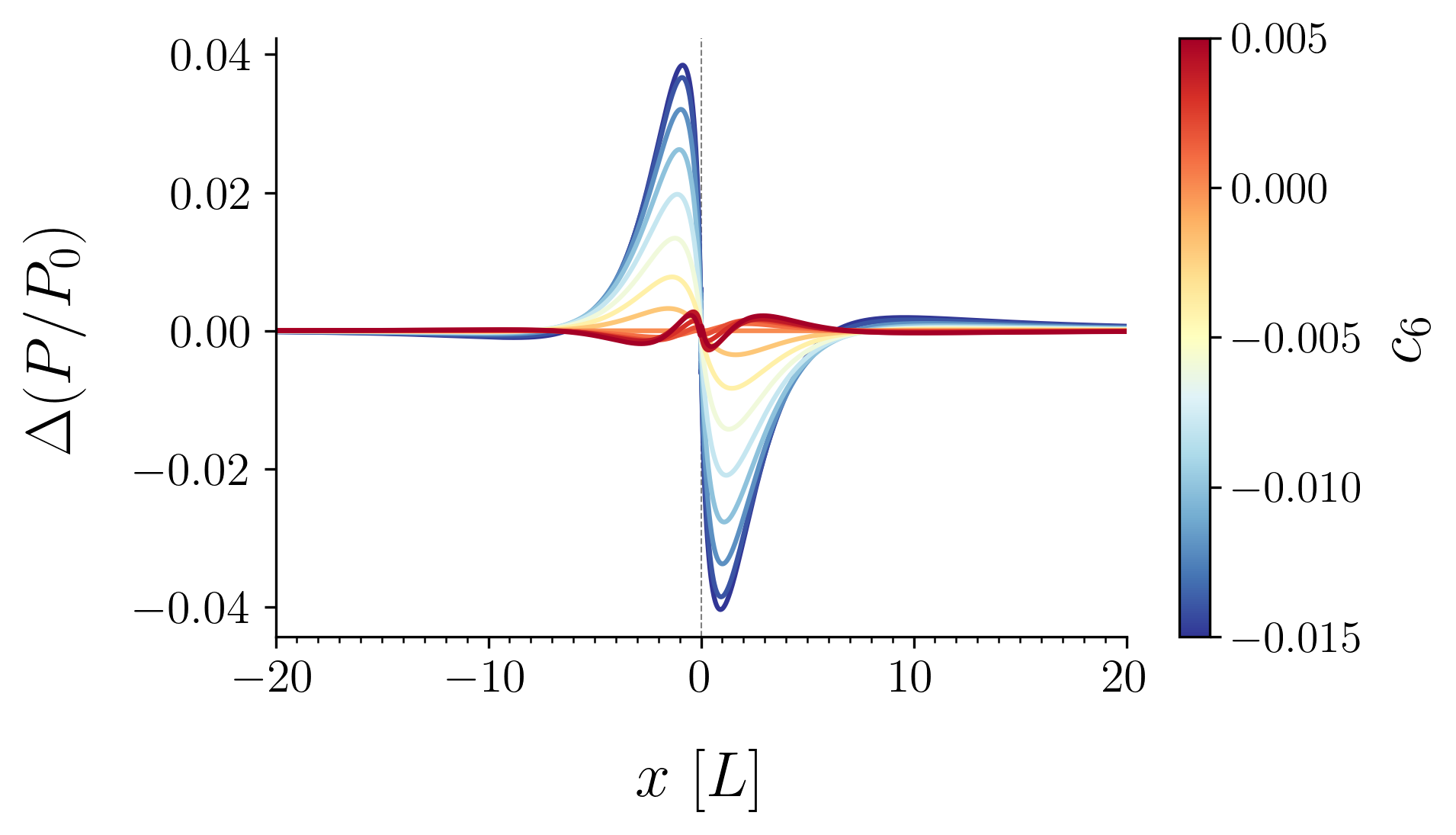}
         \caption{$s = \ell = 0$.}
        \label{fig:Pressurea}
    \end{subfigure}
    \hfill
    \begin{subfigure}{0.49\textwidth}
        \centering
        \includegraphics[width=\linewidth]{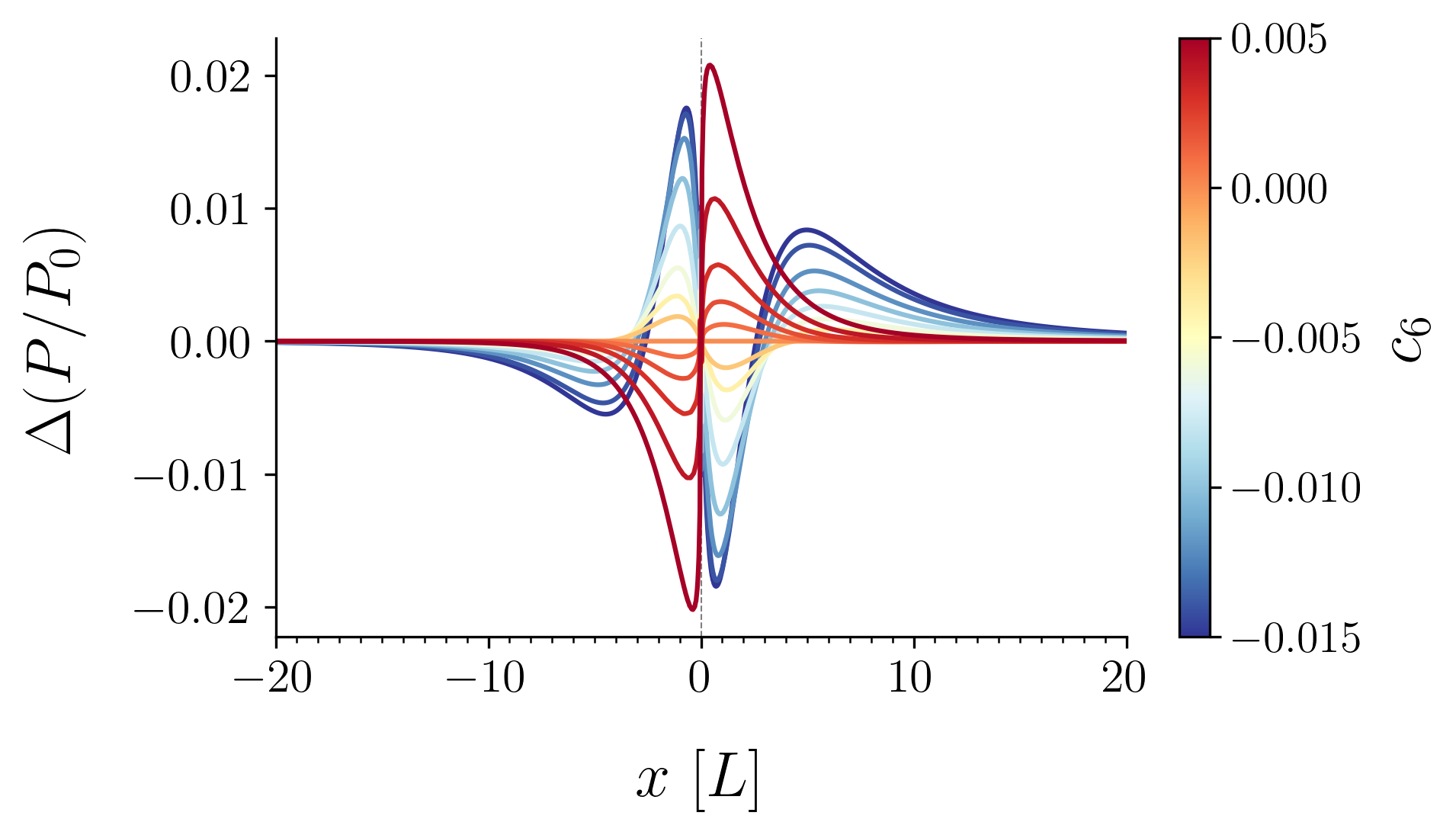}
         \caption{$s = \ell= {1}/{2}$.}
        \label{fig:Pressuree1}
    \end{subfigure}

    \begin{subfigure}{0.49\textwidth}
        \centering
        \includegraphics[width=\linewidth]{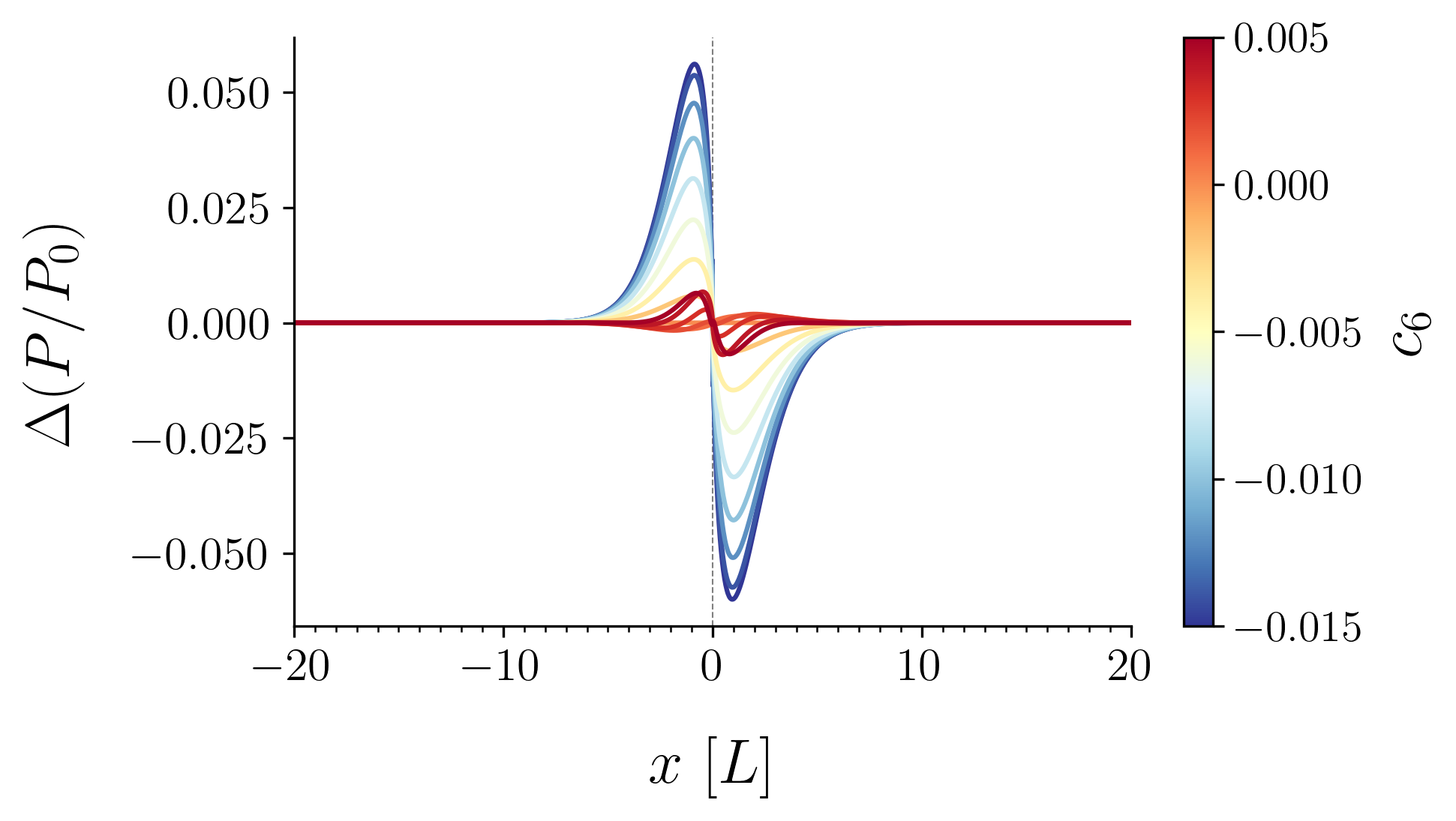}
         \caption{$s =\ell=1$.}
        \label{fig:Pressureb2}
    \end{subfigure}
    \hfill
    \begin{subfigure}{0.49\textwidth}
        \centering\includegraphics[width=\linewidth]{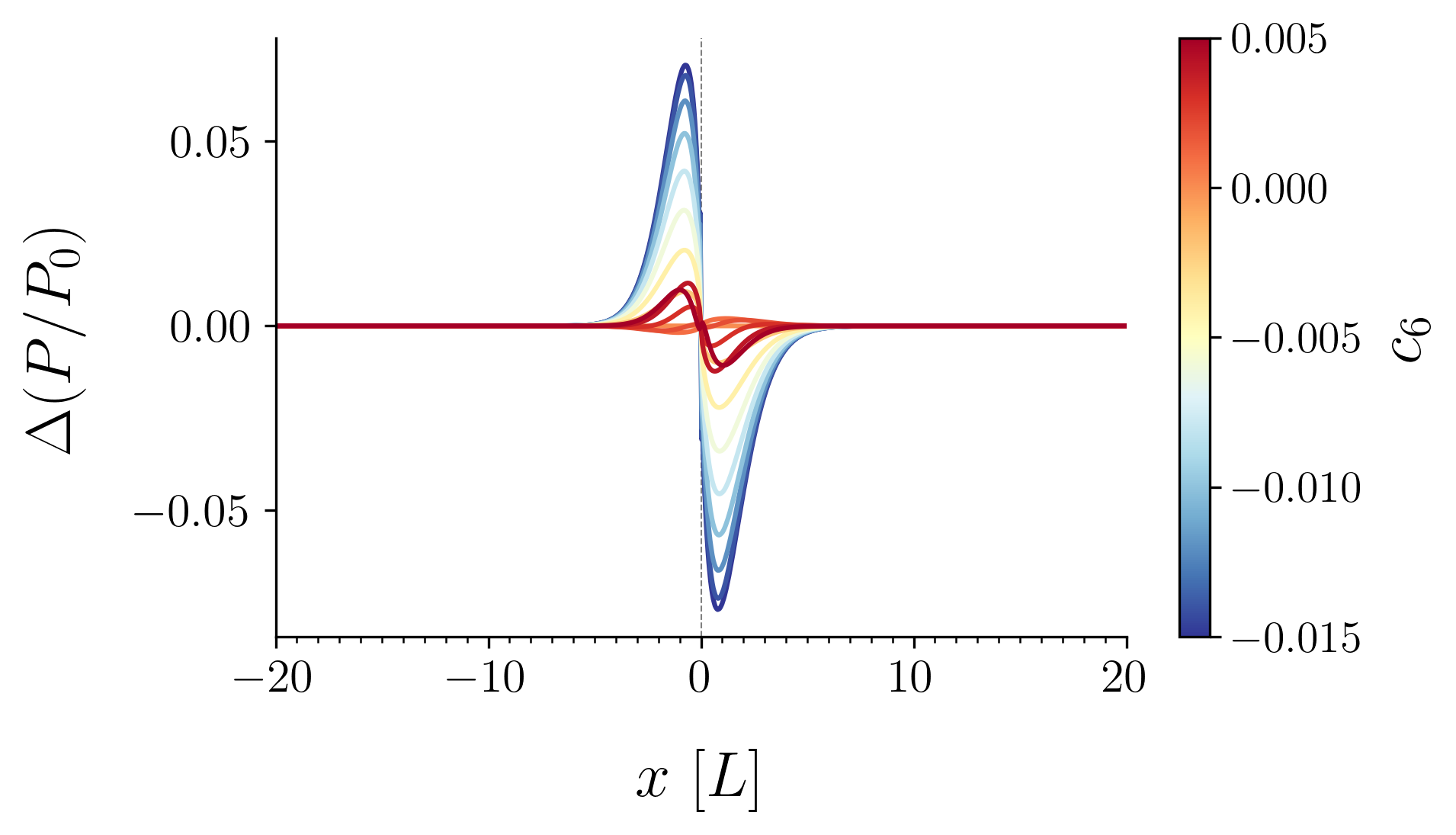}
         \caption{$s=\ell=2$.}
        \label{fig:Pressurec3}
    \end{subfigure}
    \caption[Difference in pressure of the quantum-corrected analogue, varying $c_6$]{Difference in the pressure of the flow, $\Delta (P/P_0)$. Difference from the Schwarzschild analogue of the same channel, $\Delta X(x)=X(x)-X^\text{Schw}(x)$ read station by station, for the quantum-corrected Schwarzschild metric~\eqref{eq:CK_metric}, varying $c_6$ (colour bar), one panel per channel at $\ell=s$. The vertical scale is proper to each panel. Natural units with $G_{\text{N}} = M = 1$. Computed from the tabulated solutions of Ref.~\cite{Casadio:2025lkn}.}
    \label{fig:Pressure}
\end{figure}

\begin{figure}[htbp]
    \centering
    \begin{subfigure}{0.49\textwidth}
        \centering
        \includegraphics[width=\linewidth]{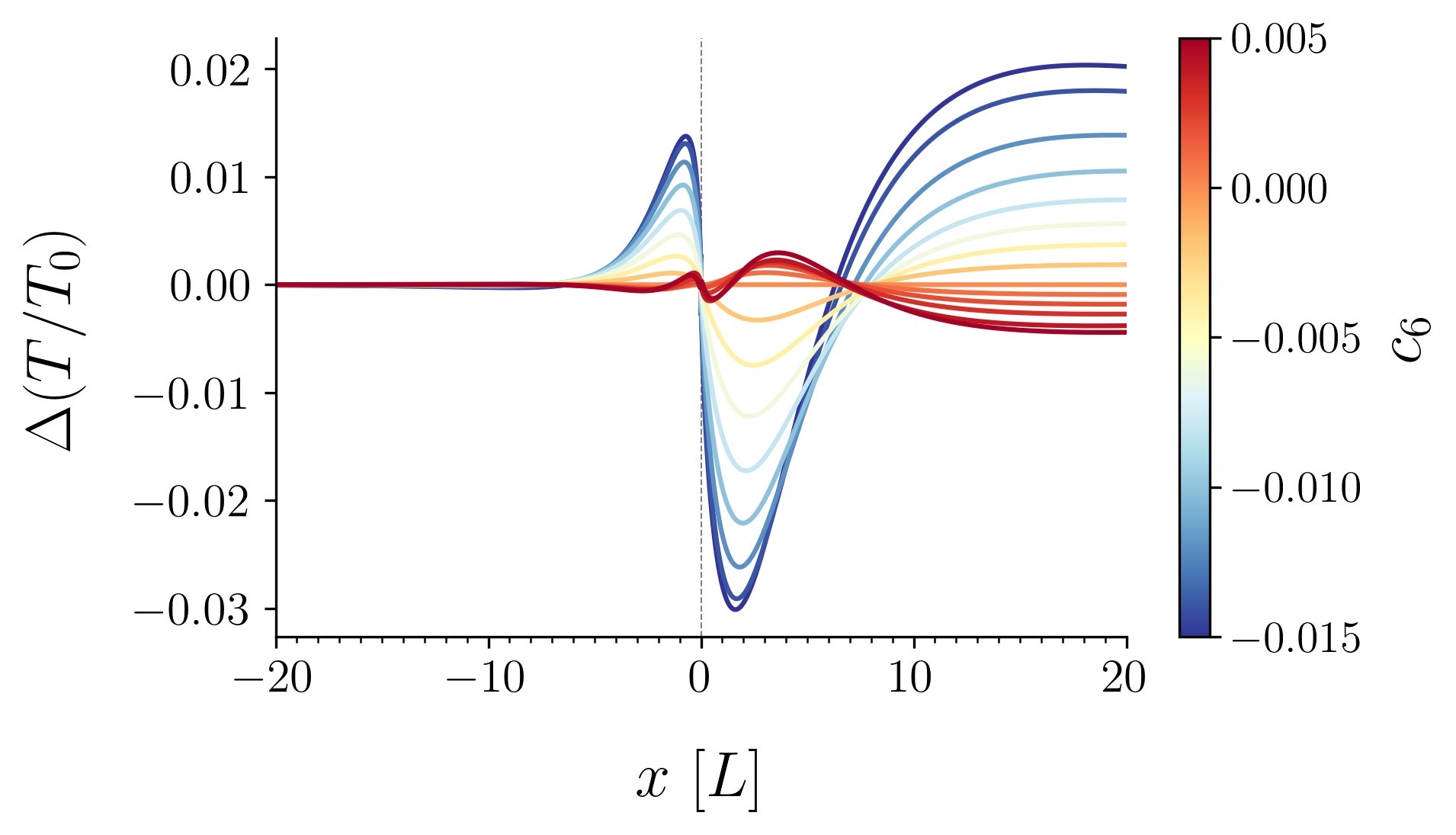}
         \caption{$s = \ell = 0$.}
        \label{fig:Temperaturea}
    \end{subfigure}
    \hfill
    \begin{subfigure}{0.49\textwidth}
        \centering
        \includegraphics[width=\linewidth]{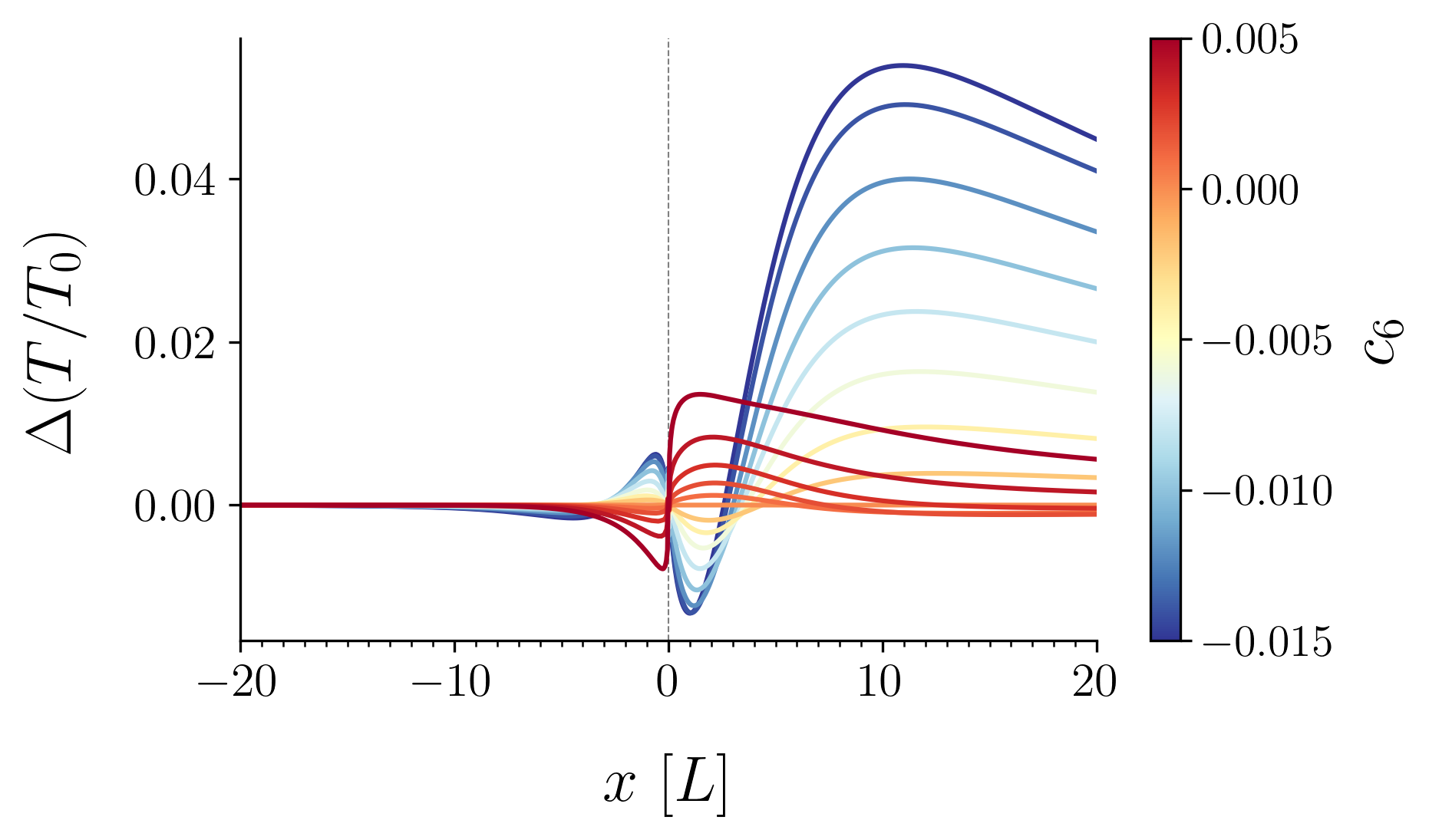}
         \caption{$s = \ell= {1}/{2}$.}
        \label{fig:Temperaturee1}
    \end{subfigure}

    \begin{subfigure}{0.49\textwidth}
        \centering
        \includegraphics[width=\linewidth]{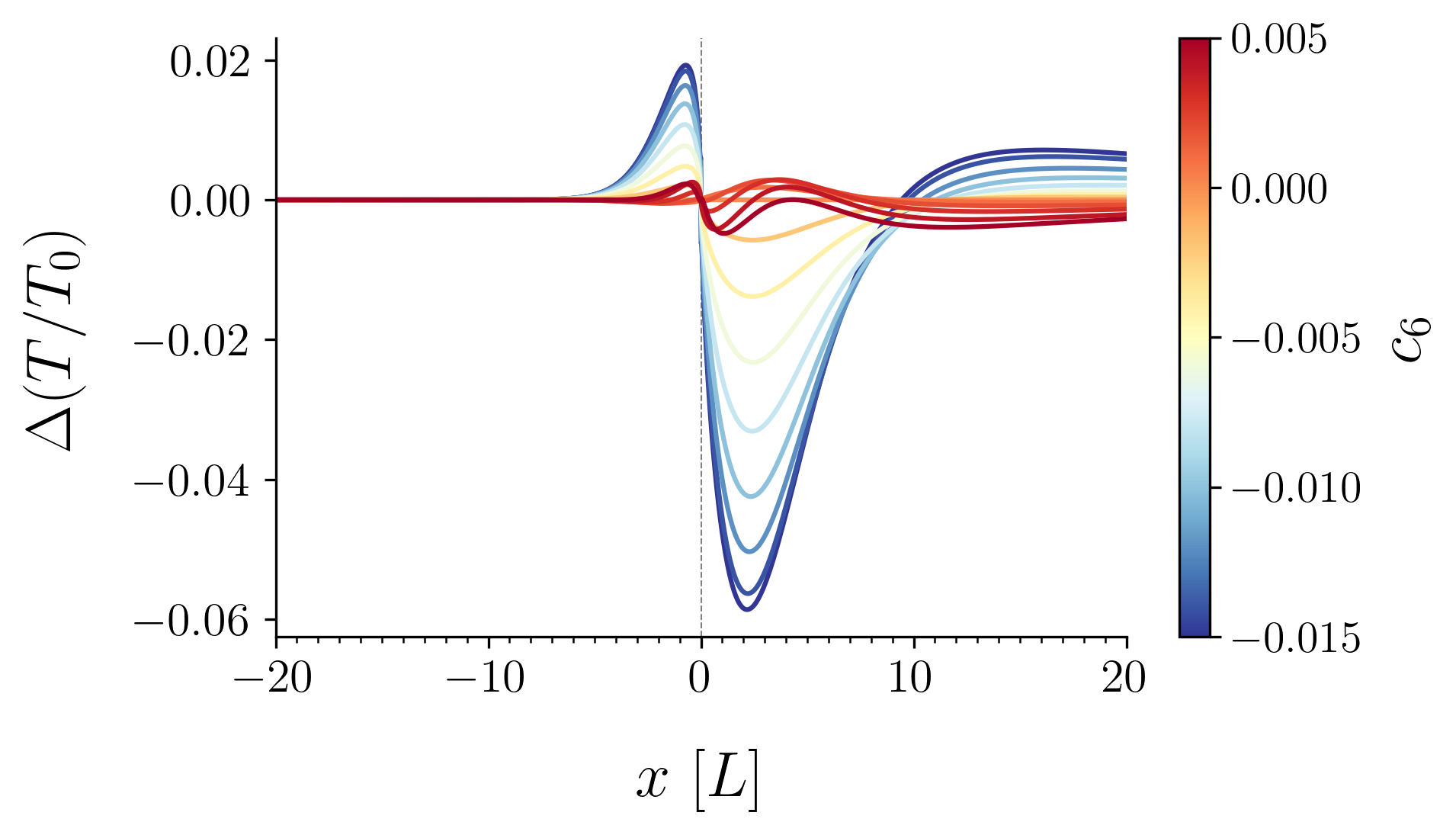}
         \caption{$s = \ell=1$.}
        \label{fig:Temperatureb2}
    \end{subfigure}
    \hfill
    \begin{subfigure}{0.49\textwidth}
        \centering\includegraphics[width=\linewidth]{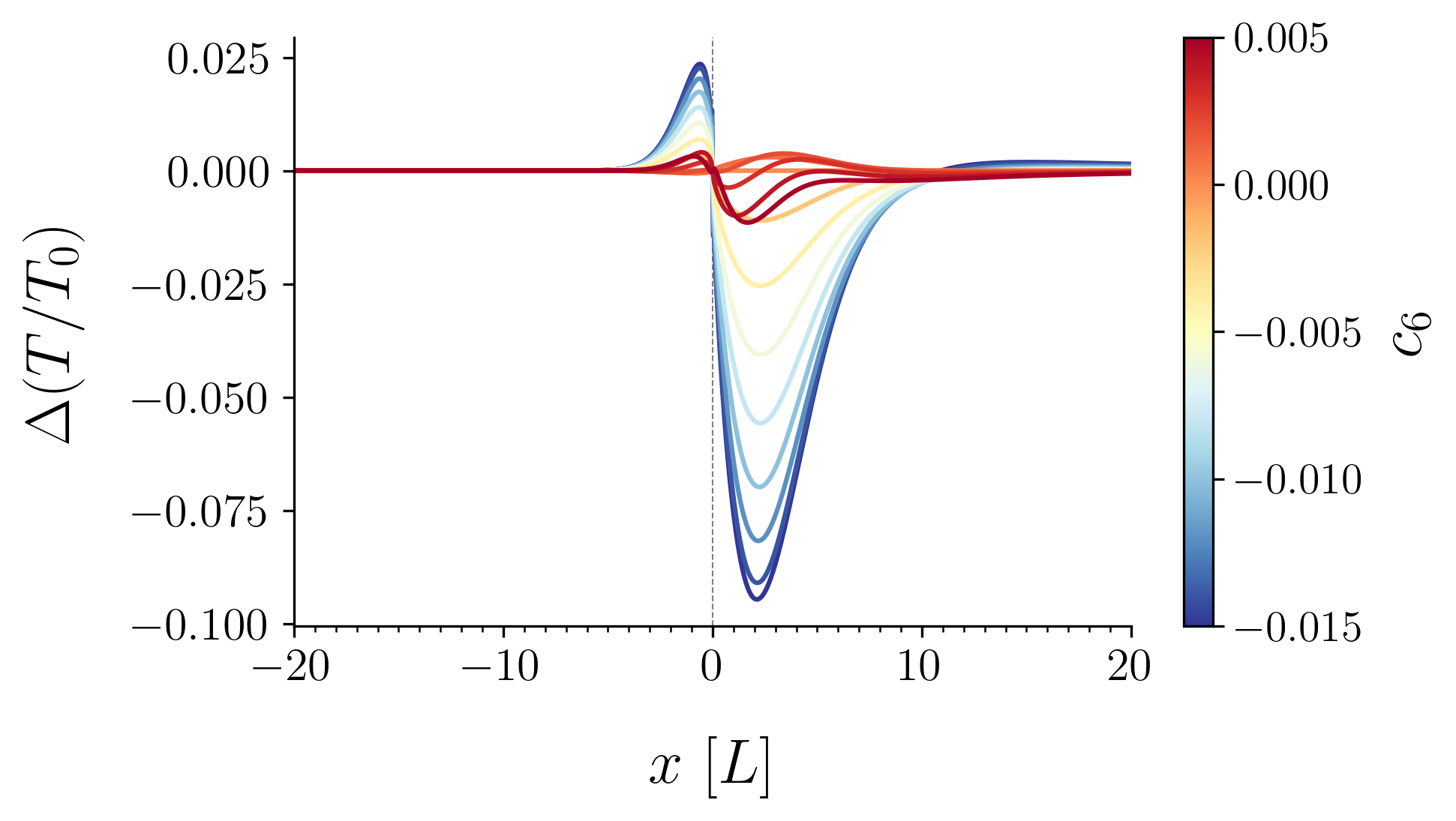}
         \caption{$s= \ell=2$.}
        \label{fig:Temperaturec3}
    \end{subfigure}
    \caption[Difference in temperature of the quantum-corrected analogue, varying $c_6$]{Difference in the temperature of the flow, $\Delta (T/T_0)$. Difference from the Schwarzschild analogue of the same channel, $\Delta X(x)=X(x)-X^\text{Schw}(x)$ read station by station, for the quantum-corrected Schwarzschild metric~\eqref{eq:CK_metric}, varying $c_6$ (colour bar), one panel per channel at $\ell=s$. The vertical scale is proper to each panel. Natural units with $G_{\text{N}} = M = 1$. Computed from the tabulated solutions of Ref.~\cite{Casadio:2025lkn}.}
    \label{fig:Temperature}
\end{figure}

The temperature repeats the role it played for the hair, being the one variable that keeps a difference as far as the exit, at $c_6=-1.5\times10^{-2}$ standing at $0.020$ of stagnation in the scalar channel and $0.045$ in the Dirac one, against $0.0007$ and $0.0006$ for the pressure of the same channels. Unlike the hair, the sign at the exit is positive, the corrected duct running hotter than Schwarzschild there.

\begin{figure}[htbp]
    \centering
    \begin{subfigure}{0.49\textwidth}
        \centering
        \includegraphics[width=7.35cm]{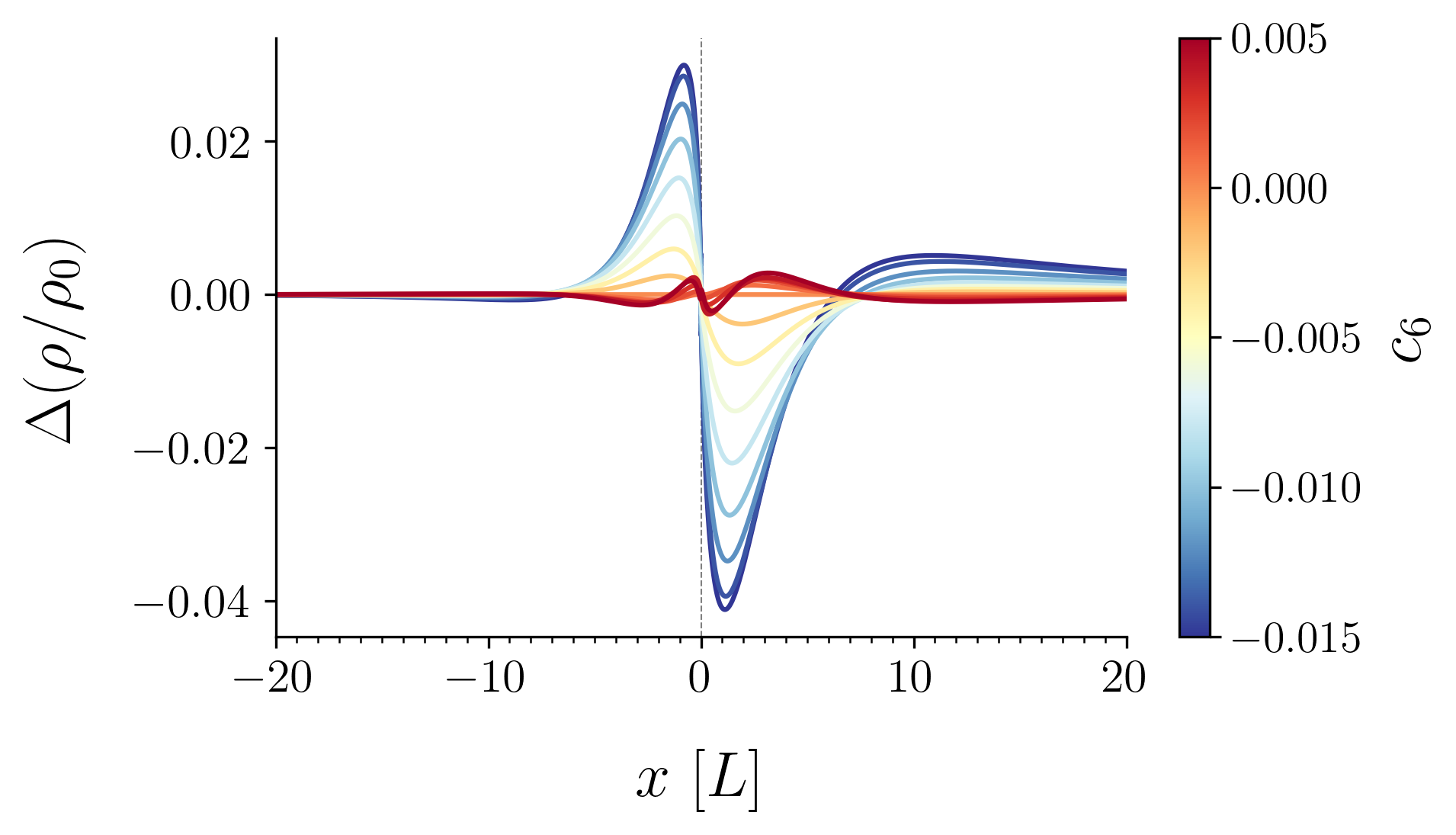}
         \caption{$s = \ell = 0$.}
        \label{fig:Densitya}
    \end{subfigure}
    \hfill
    \begin{subfigure}{0.49\textwidth}
        \centering
        \includegraphics[width=7.35cm]{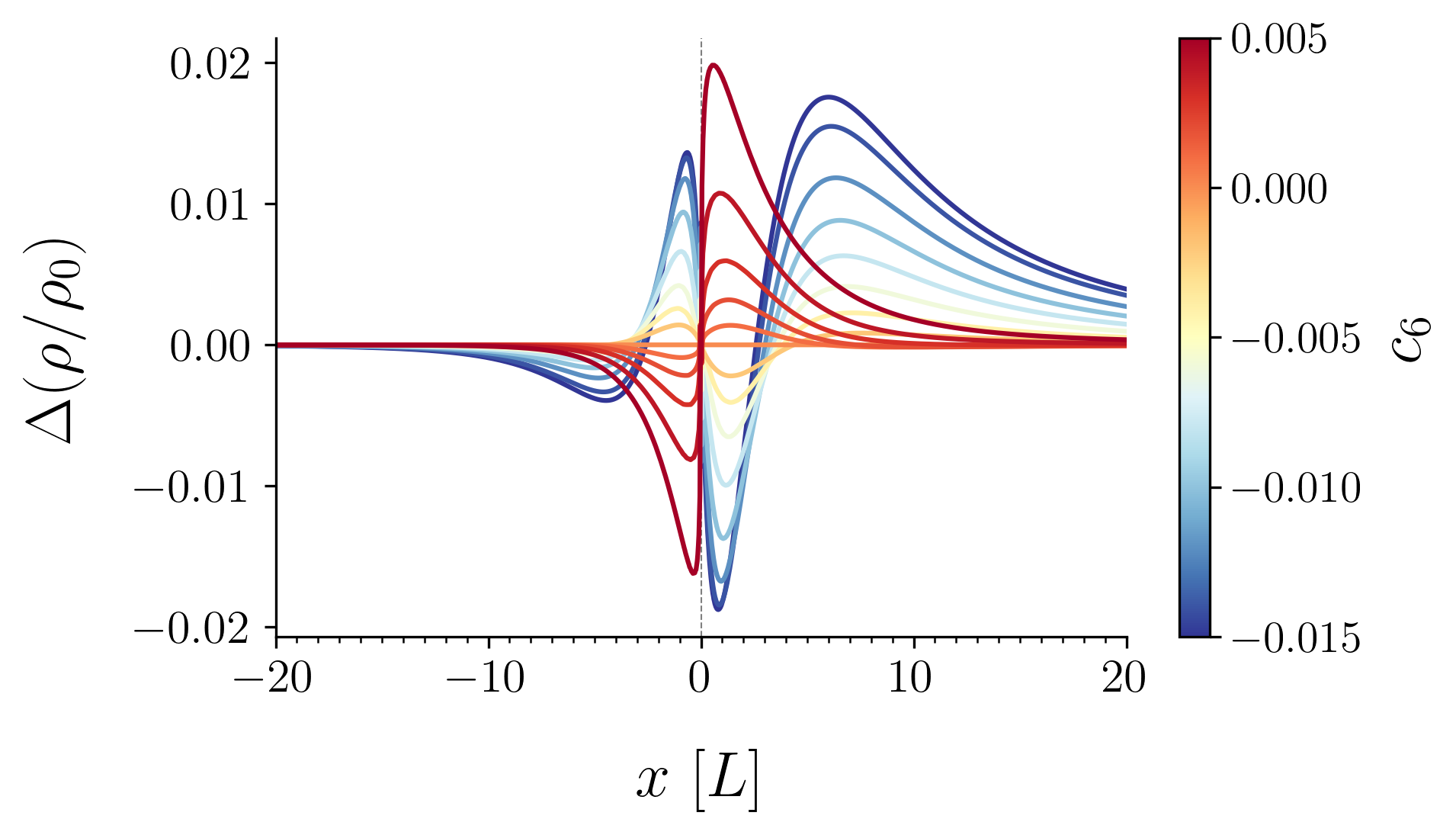}
         \caption{$s = \ell= {1}/{2}$.}
        \label{fig:Densitye1}
    \end{subfigure}

    \begin{subfigure}{0.49\textwidth}
        \centering
        \includegraphics[width=7.35cm]{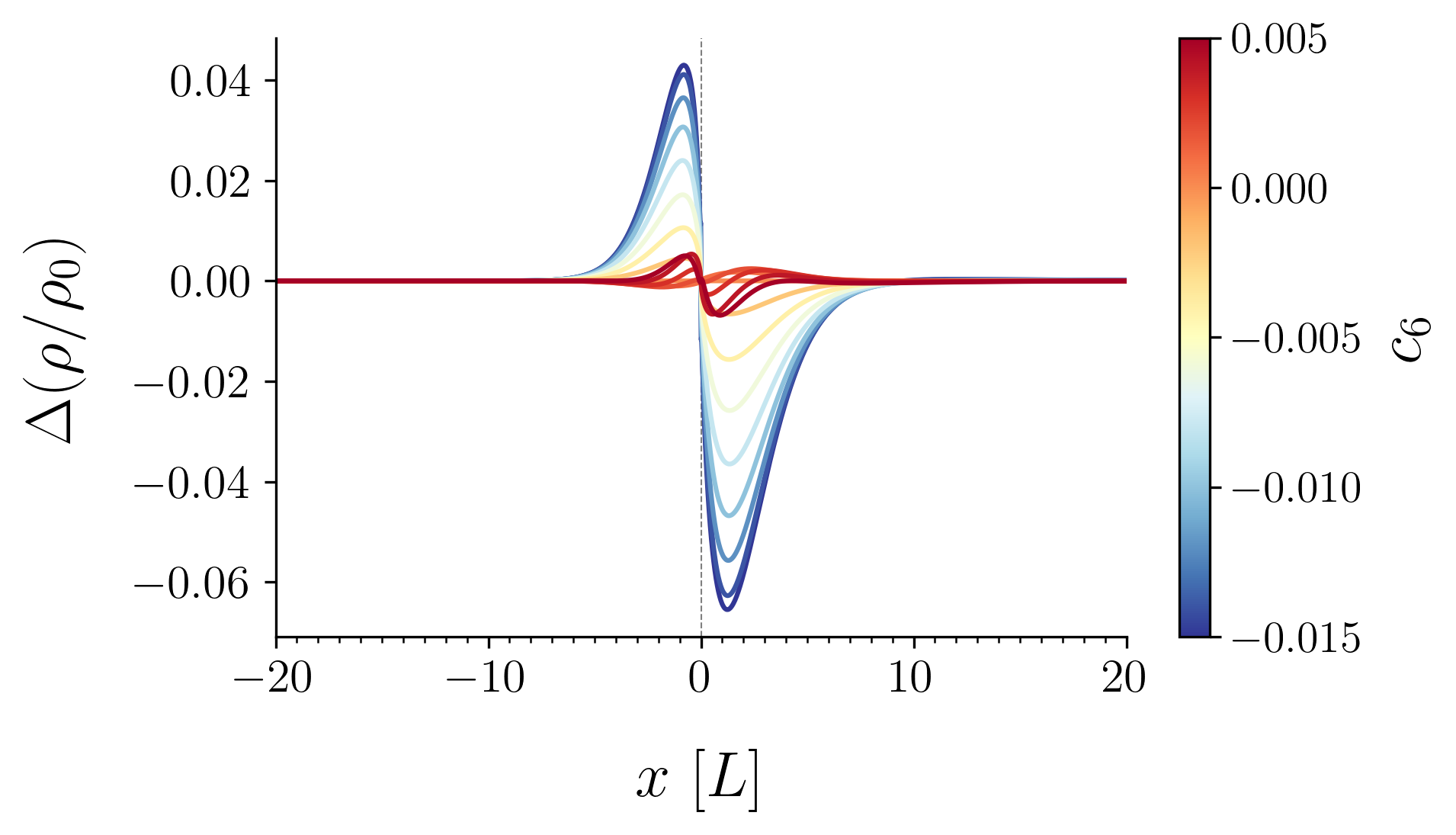}
         \caption{$s = \ell=1$.}
        \label{fig:Densityb2}
    \end{subfigure}
    \hfill
    \begin{subfigure}{0.49\textwidth}
        \centering\includegraphics[width=7.35cm]{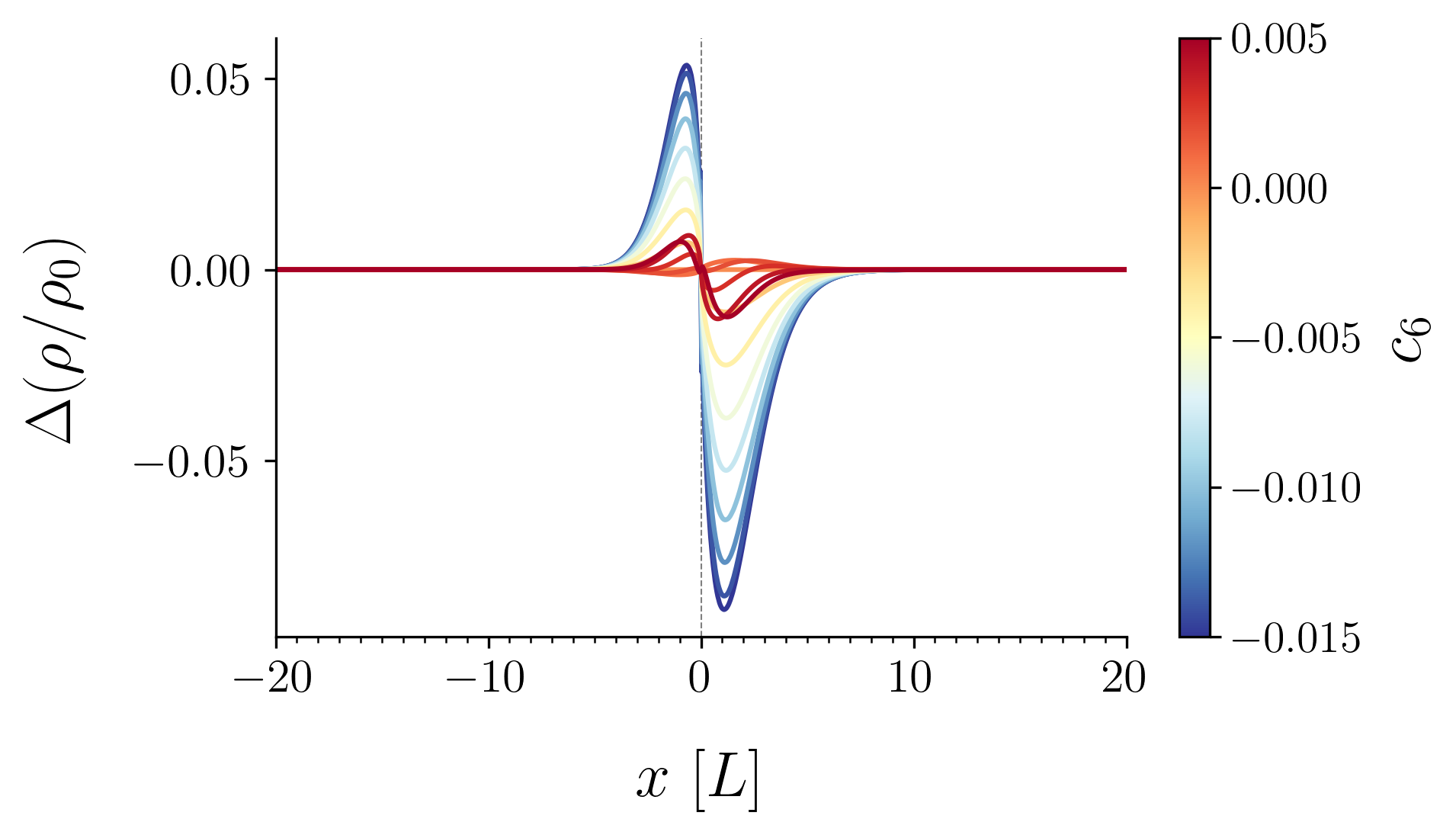}
         \caption{$s= \ell=2$.}
        \label{fig:Densityc3}
    \end{subfigure}
    \caption[Difference in density of the quantum-corrected analogue, varying $c_6$]{Difference in the density of the flow, $\Delta (\rho/\rho_0)$. Difference from the Schwarzschild analogue of the same channel, $\Delta X(x)=X(x)-X^\text{Schw}(x)$ read station by station, for the quantum-corrected Schwarzschild metric~\eqref{eq:CK_metric}, varying $c_6$ (colour bar), one panel per channel at $\ell=s$. The vertical scale is proper to each panel. Natural units with $G_{\text{N}} = M = 1$. Computed from the tabulated solutions of Ref.~\cite{Casadio:2025lkn}.}
    \label{fig:Density}
\end{figure}

The density follows the pressure closely, its extremum running from $0.019$ of stagnation in the Dirac channel to $0.089$ in the gravitational one at $c_6=-1.5\times10^{-2}$, and returning by the exit to a few parts in $10^{3}$ in the scalar and Dirac channels and to less than $10^{-4}$ in the other two.

\begin{figure}[htbp]
    \centering
    \begin{subfigure}{0.49\textwidth}
        \centering
        \includegraphics[width=\linewidth]{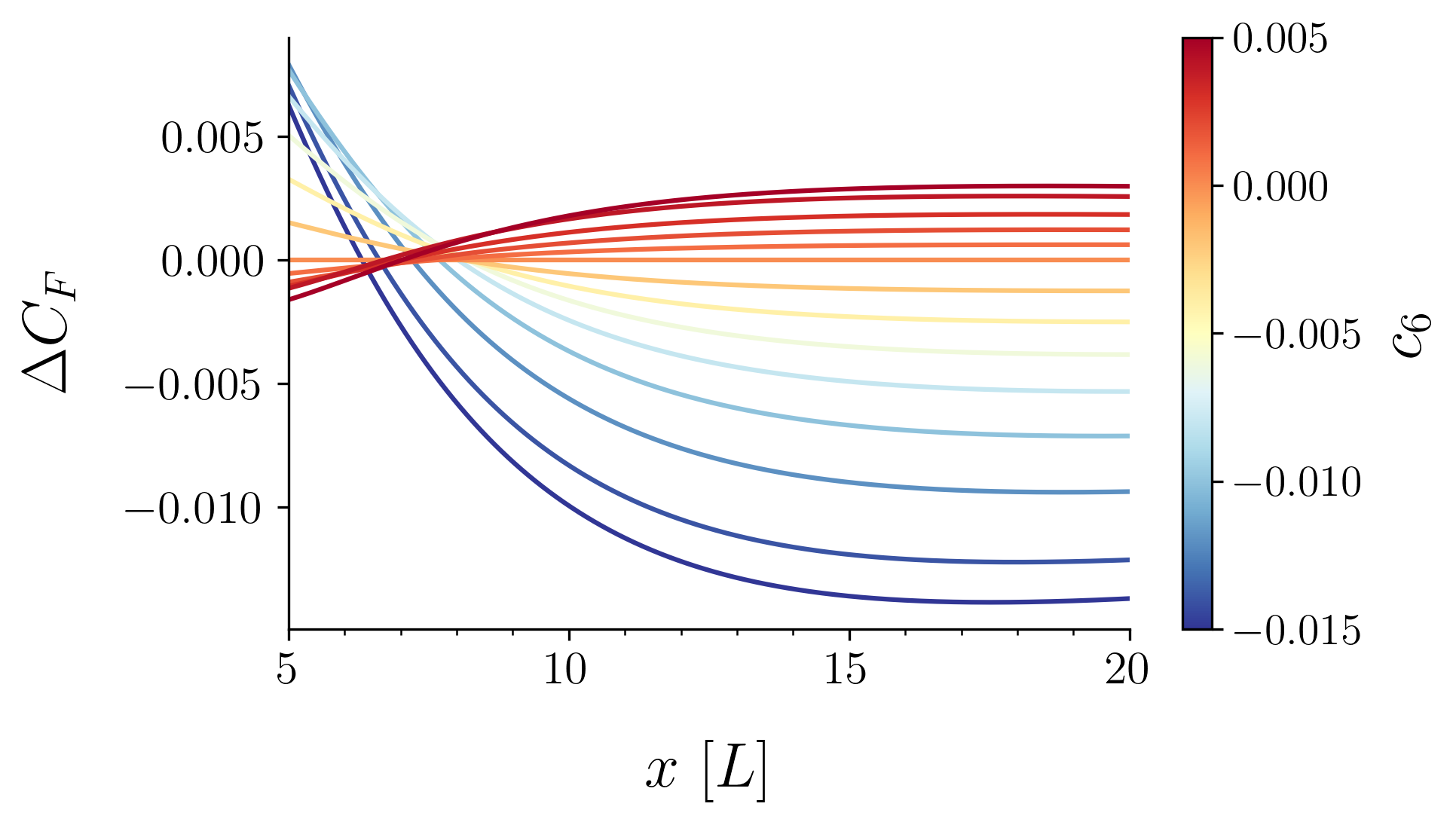}
         \caption{$s = \ell = 0$.}
        \label{fig:C_Fa}
    \end{subfigure}
    \hfill
    \begin{subfigure}{0.49\textwidth}
        \centering
        \includegraphics[width=\linewidth]{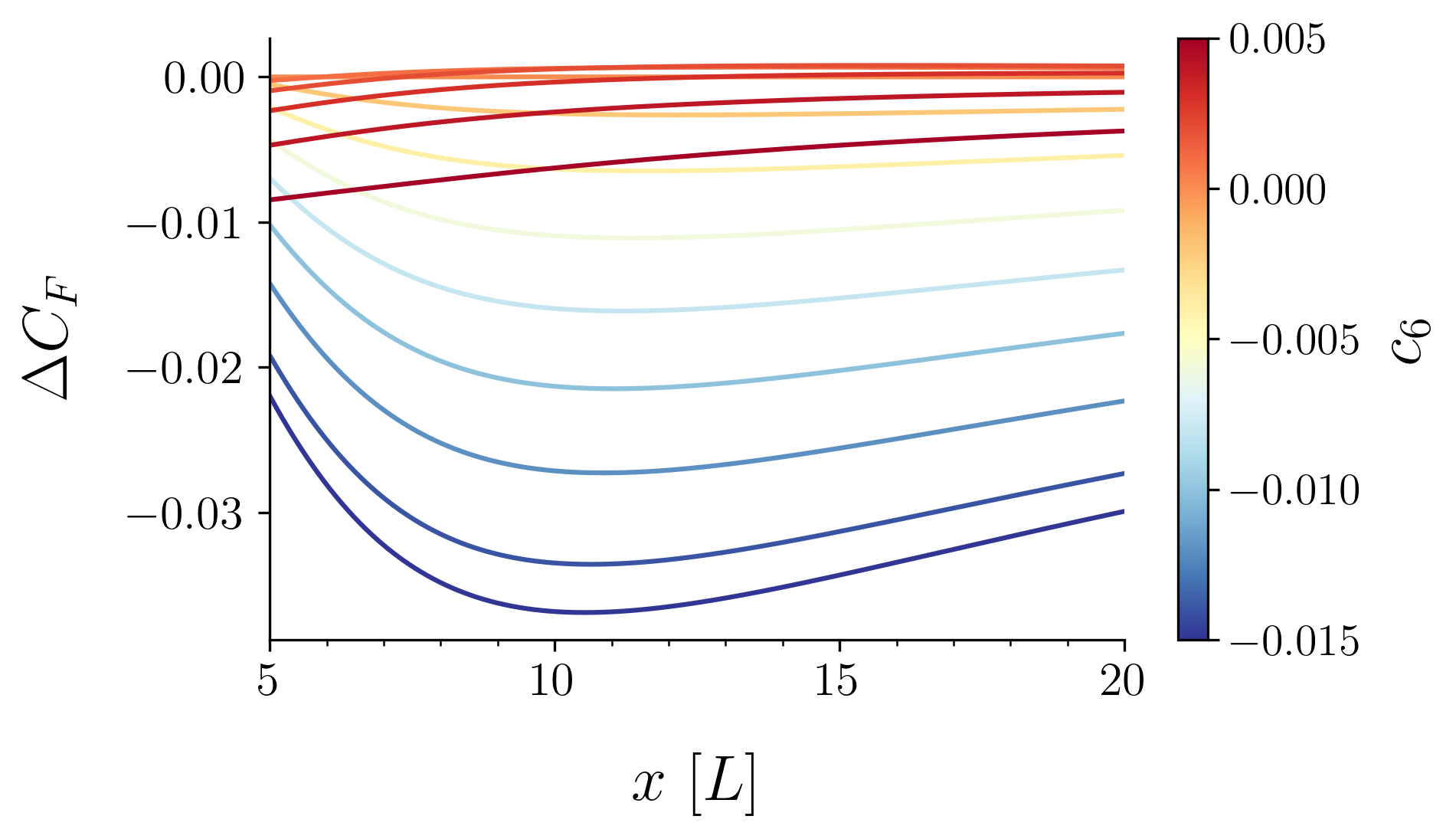}
         \caption{$s = \ell= {1}/{2}$.}
        \label{fig:C_Fe1}
    \end{subfigure}

    \begin{subfigure}{0.49\textwidth}
        \centering
        \includegraphics[width=\linewidth]{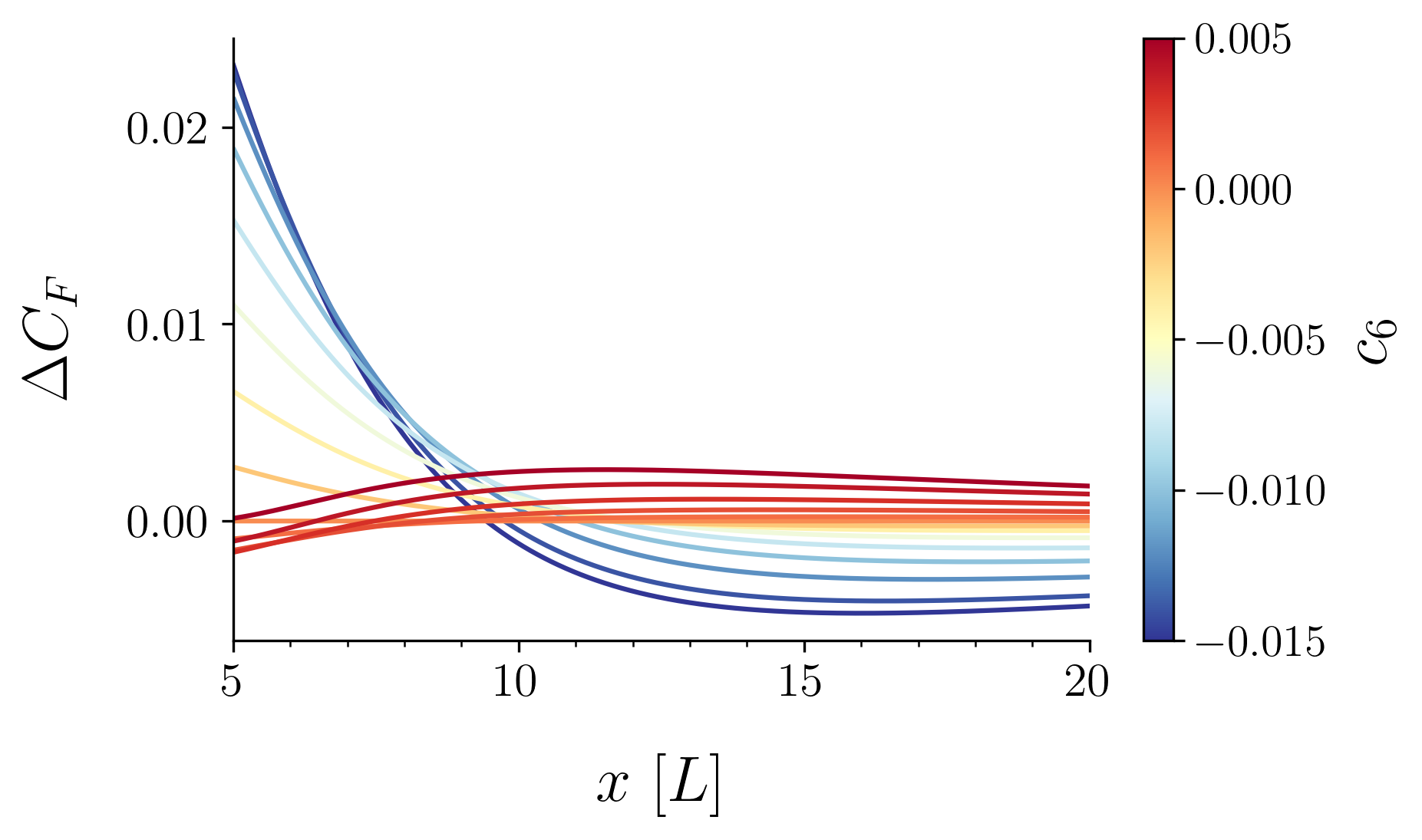}
         \caption{$s = \ell=1$.}
        \label{fig:C_Fb2}
    \end{subfigure}
    \hfill
    \begin{subfigure}{0.49\textwidth}
        \centering\includegraphics[width=\linewidth]{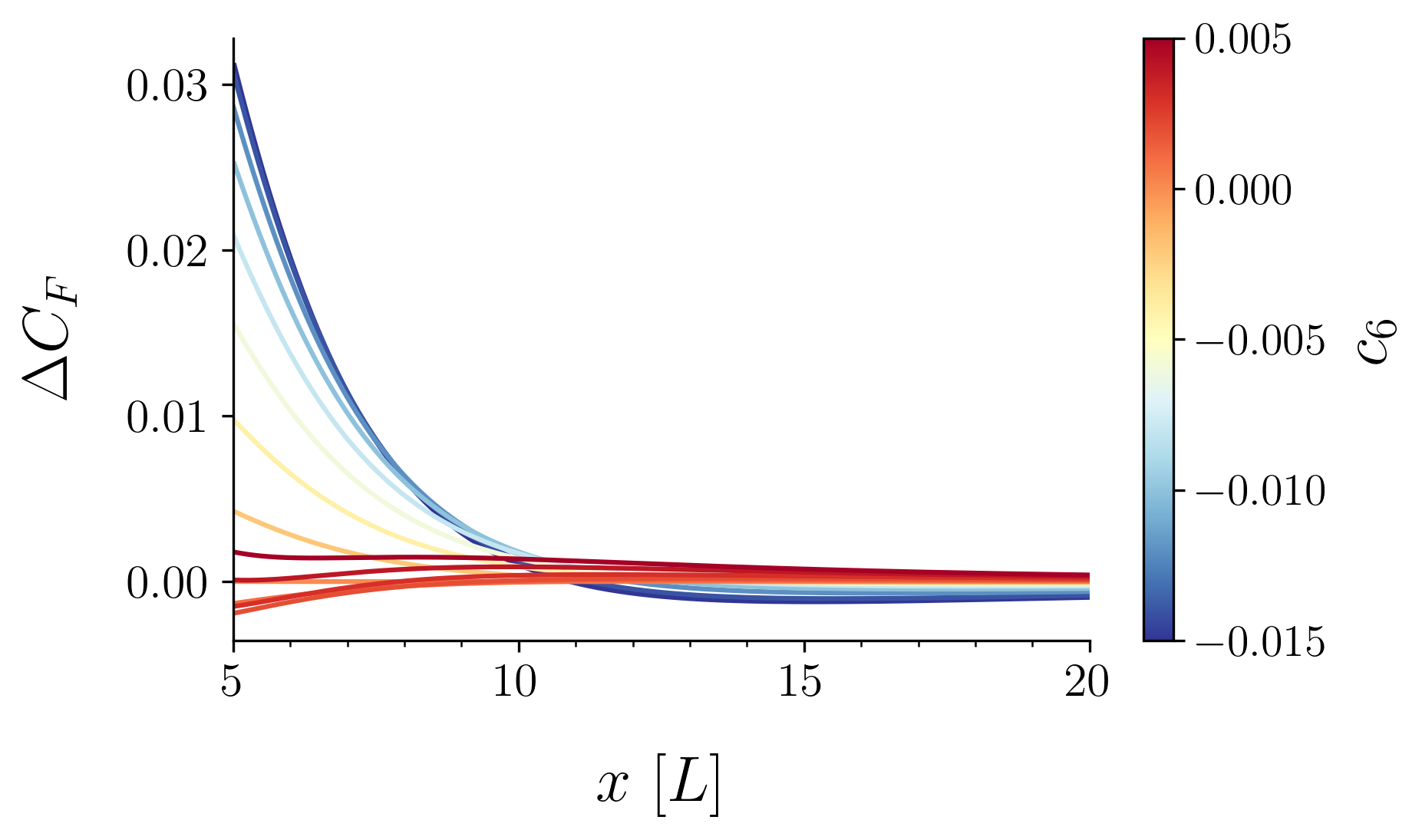}
         \caption{$s= \ell=2$.}
        \label{fig:C_Fc3}
    \end{subfigure}
    \caption[Difference in the thrust coefficient of the quantum-corrected analogue, varying $c_6$]{Difference in the thrust coefficient, $\Delta C_F$. Difference from the Schwarzschild analogue of the same channel, $\Delta X(x)=X(x)-X^\text{Schw}(x)$ read station by station, for the quantum-corrected Schwarzschild metric~\eqref{eq:CK_metric}, varying $c_6$ (colour bar), one panel per channel at $\ell=s$. The vertical scale is proper to each panel. Natural units with $G_{\text{N}} = M = 1$. Computed from the tabulated solutions of Ref.~\cite{Casadio:2025lkn}. The panels start a quarter along the duct, for the reason given in Sec.~\ref{sec:results HBH by GD}.}
    \label{fig:C_F}
\end{figure}

The thrust coefficient carries the same information in the most accessible variable, its difference staying under $0.04$ over the stretch of duct where it is meaningful and changing sign with $c_6$. Its panels start at $x=5$ for the reason given in Sec.~\ref{sec:results HBH by GD}, the coefficient being a quantity read at the exit. The effective exhaust velocity carries the same information in dimensional form and is not reproduced separately.

Two further trends complete the picture. The response grows with the multipole rather than falling with it, the deviation of the Mach profile at $c_6=-1.5\times10^{-2}$ growing along each family as $\ell$ is raised at fixed spin. This runs opposite to the barrier, whose relative sensitivity to $c_6$ falls as the centrifugal term $\ell(\ell+1)/r^2$ comes to dominate, and Fig.~\ref{fig:schw spin} of Chapter~\ref{chap:conexaoentrebnsebocais} is where that competition is displayed, one unit of $\ell$ multiplying the height of the barrier by roughly two while the whole range of integer spins changes it by less than forty per cent. The flow runs the other way because a higher multipole reconstructs a duct running to a much larger Mach number at the exit, over which the deviation accumulates along the whole divergent section. Finally, since the correction enters the metric functions~\eqref{eq:CK_metric} as $c_6\,G_{\text{N}}^{4}M^{2}/r^{6}$ and is read at the horizon, $r\sim G_\text{N}M$, its imprint on the nozzle observables grows as $M^{-4}$ towards the Planck scale. That is the scaling established for the horizon shift in Chapter~\ref{chap:EFT} and carried into the barrier in Chapter~\ref{chap:conexaoentrebnsebocais}, and it is inherited here unchanged, so that the analogue signatures are most pronounced for the lightest black holes, of potential relevance for primordial black holes and their hypothetical laboratory analogues~\cite{Hawking:1971ei,Carr:1974nx,Carr:2020gox,Casadio:2025lkn}.

\section{Summary and Outlook}
\label{sec:results summary}

This chapter closed the correspondence in the direction it was built for. Identifying the two tortoise coordinates gives the constraint~\eqref{eq:xemfuncaoder123}, and the substitution $g_c=\chi^2$ turns the condition on the potential into the Sturm--Liouville problem~\eqref{eq: lambda}, integrated outwards from the analogue horizon. Since the nozzle carries no angular sector, each duct reproduces one pair $(s,\ell)$ rather than the geometry as a whole, and Sec.~\ref{sec:schwarzschild analogue} showed that the channel already fixes the character of the undeformed duct.

For the hairy black hole the three parameters are far from equivalent, which is what allowed the section to be organised around one of them. The charge-like parameter $Q$ is the only one that provides a comparably strong and monotonic control of the nozzle deformation over the explored parameter range, opening the divergent section and raising the Mach number at the exit over the sweep by a tenth in the scalar channel and by more than a half in the gravitational one. The decoupling parameter $\alpha$ saturates as soon as the hair is introduced, and $\ell_h$ acts only through the product $\alpha\ell_h$, by the mass shift~\eqref{eq: ell_h como deslocamento de massa}, so that a measurement at this precision constrains the product and not either factor. Of the measurable quantities the temperature is the one that retains the imprint as far as the exit.
For the quantum correction the coupling acts through the corrected horizon~\eqref{eq:horizon_shift}, its two branches being far from symmetric, the negative one lowering the Mach number at the exit by more than a fifth in the Dirac channel and the whole admissible positive branch moving it by at most four per cent. Contrary to what the barrier alone would suggest, the response grows with the multipole. The whole effect is placed by the mass, since the correction falls as $M^{-4}$ and is confined to black holes far lighter than the astrophysical population.
Two limitations should be stated clearly. The profiles are those of an ideal isentropic flow, whose departures from a physical duct are listed in Chapter~\ref{chap:isentropic equations} and whose more elementary obstacles, thermal dilation of the walls and the narrowness of the shock-free window, were set out in Sec.~\ref{sec:constraint_equations}. The deviations a hypothetical measurement would have to resolve are correspondingly uneven, running from tens of per cent in the Mach number at the exit down to a few parts in $10^{4}$ of the stagnation value in the pressure at the same station. The stability of these flows, the higher overtones of the early ringdown, and the inverse problem of reading the black-hole parameters off acoustic data are collected among the perspectives of Chapter~\ref{chap:conclusion}.

\chapter{Conclusion and Future Perspectives}
\label{chap:conclusion}

\epigraph{``Now this is not the end. It is not even the beginning of the end. But it is, perhaps, the end of the beginning.''}{Winston Churchill}

This thesis built a correspondence between two exotic black-hole geometries and the transonic flow of a de~Laval nozzle, and ran it in the direction that turns a barrier into a duct. The perturbations of a static, spherically symmetric black hole and the sound waves of a nozzle flow obey the same Schr\"odinger-type equation, so a prescribed effective potential determines a wall profile and, with it, every aerodynamic quantity the flow carries. Applied to the hairy black hole of gravitational decoupling and to the quantum-corrected Schwarzschild metric of the effective field theory of gravity, the construction turns the hair $(\alpha,\ell_h,Q)$ and the coupling $c_6$ into differences in the nozzle geometry and in the aerodynamic variables, read station by station along the duct, systematising the analogue realisations established with collaborators in Refs.~\cite{Casadio:2024uwj,Casadio:2025lkn}. Beyond the two geometries treated explicitly, this shows that an idealised, isentropic transonic flow, one whose sonic and isentropic conditions would demand extreme fine-tuning to realise in practice, can, in principle, discriminate between a classical and a quantum-gravitational departure from Schwarzschild through where and how strongly the acoustic signature is registered.

The two backgrounds of Part~\ref{part: extensions of gr} depart from General Relativity in opposite ways. The extended minimal geometric deformation of Chapter~\ref{chap:GD} deforms a Schwarzschild seed by an additional source and returns the hairy metric~\eqref{eq: metric HBH}, whose admissible domain was fixed by the dominant energy condition, asymptotic flatness and the existence of a horizon, and which acts over a macroscopic range set by $G_\text{N}M$ while preserving $g_{tt}=-g_{rr}^{-1}$. In Chapter~\ref{chap:EFT} the two-loop Goroff--Sagnotti divergence instead identifies the dimension-six operator whose Riemann-cubed term produces the Calmet--Kuipers metric~\eqref{eq:CK_metric}, controlled by a single parameter bounded by $\abs{c_6}<1/(5\pi)$, decaying as $r^{-6}$ and breaking that same relation. One is long-ranged and postulated, the other short-ranged and derived, which is what makes the pair informative under a common treatment.

Part~\ref{part: quasinormal modes} reduced the perturbation problem of any static, spherically symmetric geometry to the master potentials~\eqref{final effective potential int} and~\eqref{final effective potential half}, from which the Mashhoon method returns the whole overtone tower in closed form, with $\omega_I>0$ everywhere, so neither geometry is unstable at linear order. The two corrections are distinguished by which channel responds against how strongly all of them do: the Calmet--Kuipers coupling is selective, moving the quality factor, when compared to Schwarzschild, by up to $37\%$ in the scalar and Dirac channels while the remaining two stay within about $10\%$, whereas the hair acts on every channel alike but with different intensities. The quality factor in this case is up by close to $30\%$ as $\alpha=\ell_h$ approaches unity for scalar fields, whereas is up by $54\%$ in the gravitational one. In both cases the fractional deviation of the quality factor is common to the entire overtone tower while the absolute displacement of a mode grows with $n$, so the higher overtones, which govern the early ringdown, are the sharper probes~\cite{Casadio:2025lkn,Giesler:2019uxc}.

Part~\ref{part: analogue models} assembled the acoustic side, where a linear perturbation of an ideal, irrotational and isentropic flow propagates on the acoustic metric~\eqref{elem linha geral}, whose horizon sits wherever the flow turns supersonic and is a property of the velocity field, not of the vessel~\cite{Unruh:1980cg,Barcelo:2005fc}. In a de~Laval nozzle the isentropic relations reduce the flow to the Mach number, and the acoustic perturbation to also a Schrödinger-type equation, whose potential is built from the single gravitational-acoustic mapping function $g_c=\rho\mathcal{A}/c_s$ and is invariant under any rescaling of it, so a prescribed barrier fixes the duct up to the normalisation the sonic condition supplies.

Part~\ref{part: results} carried that out, turning the equality of the potentials into the Sturm--Liouville problem~\eqref{eq: lambda}, integrated outwards from the horizon. The two wave equations differ only in the eigenvalue term, so that $\omega_\text{ac}=c_{s0}\,\omega$ exactly and the acoustic spectrum is the gravitational one rescaled. The nozzle adds observables rather than a spectrum, in a form a hypothetical laboratory instrument would read. Against a common Schwarzschild reference, the channel alone already fixes the undeformed duct, the barrier and the exit Mach number both growing by roughly an order of magnitude between the scalar and gravitational channels. Of the three hair parameters only the charge-like $Q$ acts as a dial, while $\alpha$ saturates as soon as the hair is switched on and $\ell_h$ enters only through the mass shift $M\mapsto M+\alpha\ell_h/2$, so that a measurement at this precision constrains the product $\alpha\ell_h$ and not either factor separately. The same $Q^2/r^2$ term that drives the response also bounds it, pulling the two horizons together until no black hole remains, and because that term is largest near the horizon, its imprint is carried, through the global solution of the reconstruction problem, into a departure that persists across the entire divergent section rather than staying confined to the throat. Only the temperature keeps a resolvable imprint as far as the exit, with the largest departures sitting at the image of the barrier maximum.
The quantum correction registers in the opposite place. Read at the horizon and decaying as $r^{-6}$, it concentrates its imprint within one or two units of the throat and returns to zero well before the exit. Its branches are strongly asymmetric, the negative one lowering the exit Mach number appreciably more than the entire admissible positive branch does, and the Dirac duct is the one that turns, widening only up to a bounded value of $c_6$, the analogue counterpart of the non-monotonic fermionic barrier of Chapter~\ref{chap:conexaoentrebnsebocais}. Since the correction is read at $r\sim G_\text{N}M$, its imprint grows as $M^{-4}$, so it is primordial black holes rather than the astrophysical population that such a measurement would address~\cite{Hawking:1971ei,Carr:2020gox}.

Three limitations are intrinsic to the correspondence. First, a sound wave is a scalar excitation of the gas, so the duct reproduces the scattering problem of a channel, its barrier and its frequencies, but not the tensor character of the perturbation that defines it; a realistic spin-2 gravitational wave has no counterpart in the flow~\cite{Casadio:2024uwj,Casadio:2025lkn}, and a duct has no angular sector, so each nozzle realises one pair $(s,\ell)$. Second, the analogy is kinematical, following from the continuity and Euler equations with no reference to the Einstein equations, so the duct tests how a wave scatters off a deformed geometry and not the field equations behind it, and for the same reason the thermodynamics does not transfer. Third, the distance between the reconstructed profiles and a duct that could be run is set by the walls dilating under a temperature that varies along them and by a narrow shock-free window, as Sec.~\ref{sec:constraint_equations} sets out, which makes the value of the construction numerical rather than experimental.
Several directions follow. The most natural is rotation, mapping the perturbations of a stationary, axisymmetric spacetime onto a quasi-one-dimensional flow, which is not a straightforward generalisation since the angular sector no longer decouples as it does here. Such a map would open frame dragging and the ergoregion instability to laboratory study and connect this programme to the superradiance observed in draining vortex flows~\cite{Torres:2016iee,Torres:2019sbr}, the rotating and charged quantum-corrected geometries that would serve as targets having already been derived~\cite{Casadio:2023iqt,Casadio:2022ndh}. The gravitational sector admits extension in its own right, since for charged black holes the quantum corrections appear one order lower in curvature~\cite{Calmet:2019eof,Perrucci:2024qrr}.

Of the remaining steps, the inverse problem is the one that would change the status of the programme, from illustration to measurement: everything here runs from the black hole to the duct, and recovering the potential, and with it the gravitational parameters, from acoustic data alone was not attempted here. The forward direction in the meantime is to relax the ideal-flow approximation to include viscosity and wall friction, and refining the grid about the sonic point. This would close part of that gap and settle the response of the Mach gradient at the throat, which fixes the analogue Hawking temperature and is the one quantity the figures cannot deliver. Taken together, these directions aim at consolidating the de~Laval nozzle as a platform on which extended gravity sectors can be examined in variables that a laboratory already knows how to measure.







\postextual

\renewcommand{\bibsection}{%
  \clearforchapter
  \thispagestyle{chapter}%
  \phantomsection
  \addcontentsline{toc}{chapter}{\texorpdfstring{\MakeTextUppercase{\bibname}}{\bibname}}%
  \begin{\absnamepos}\chaptitlefont\bibname\end{\absnamepos}%
  \vspace{\abstitleskip}%
  \bibmark
  \prebibhook}

\bibliographystyle{iopart-num}

\bibliography{bibliografia}



\begin{apendicesenv}

\partapendices

\chapter{Reduction of the Acoustic Wave Equation to Schr\"odinger Form}
\label{ap: reducao schrodinger}

This appendix carries out, step by step, the reduction that Sec.~\ref{sec:schrodinger_acoustic_wave} states in outline, from the time-independent equation~\eqref{eq: eq onda diferencial independente do tempo} to the Schr\"odinger form~\eqref{eq:equacao schrodinger bna} with the potential~\eqref{eq: pot eff bna}.

Three shorthands keep the expressions readable throughout,
\begin{equation}
\label{eq:ap ab k}
    \mathcal{B} \equiv \rho\mathcal{A},
    \qquad
    K \equiv \frac{\mathcal{B}}{c_s^2},
    \qquad
    a \equiv \mathcal{B}\qty(1-\mathrm{M}^2),
\end{equation}
with $\mathrm{M}=v/c_s$, and the prime denotes $\dv*{}{x}$. All four of $\rho$, $\mathcal{A}$, $c_s$ and $v$ are arbitrary functions of $x$: no relation between them is used anywhere below.

Since Eq.~\eqref{eq: eq onda diferencial independente do tempo} holds at every instant, the expression in braces vanishes for each $\omega_{\text{ac}}$ separately. Writing $\psi$ for $\delta\psi_\omega$, that leaves
\begin{equation}
\label{eq:ap ode}
    a\,\psi'' + \qty[\mathcal{B}' - \qty(v^2K)' + 2i\omega_{\text{ac}} vK]\psi'
    + \qty[\omega_{\text{ac}}^2 K + i\omega_{\text{ac}}\qty(vK)']\psi = 0.
\end{equation}
The definition of $a$ in Eq.~\eqref{eq:ap ab k} can be expanded as
\begin{equation}
\label{eq:ap identidade}
    a = \mathcal{B}\qty(1-\frac{v^2}{c_s^2}) = \mathcal{B} - v^2 K
    \qquad \Longrightarrow \qquad
    a' = \mathcal{B}' - \qty(v^2K)',
\end{equation}
so the part of the first-derivative coefficient of Eq.~\eqref{eq:ap ode} that does not involve $\omega_{\text{ac}}$ is precisely $a'$. The first two terms of Eq.~\eqref{eq:ap ode} therefore already combine into a single divergence,
\begin{equation}
\label{eq:ap ode divergencia}
    \dv{}{x}\qty(a\,\psi') + 2i\omega_{\text{ac}} vK\,\psi'
    + \qty[\omega_{\text{ac}}^2 K + i\omega_{\text{ac}}\qty(vK)']\psi = 0.
\end{equation}
This is the reason the reduction closes at all, and it is an identity of the coefficients, not a property of the flow.

The phase is introduced next. Substituting
\begin{equation}
\label{eq:ap sub fase}
    \psi = e^{-i\omega_{\text{ac}} f}\,u,
    \qquad
    \psi' = e^{-i\omega_{\text{ac}} f}\qty(u' - i\omega_{\text{ac}} f' u),
\end{equation}
\begin{equation}
\label{eq:ap sub fase 2}
    \psi'' = e^{-i\omega_{\text{ac}} f}\qty[u'' - 2i\omega_{\text{ac}} f' u'
    - \qty(i\omega_{\text{ac}} f'' + \omega_{\text{ac}}^2 f'^2)u],
\end{equation}
into Eq.~\eqref{eq:ap ode divergencia} and dividing by $e^{-i\omega_{\text{ac}} f}$, the coefficient of $u'$ becomes
\begin{equation}
\label{eq:ap coef u linha}
    a' - 2i\omega_{\text{ac}}\qty(a f' - vK).
\end{equation}
Its term linear in $\omega_{\text{ac}}$ vanishes if and only if
\begin{equation}
\label{eq:ap escolha fase}
    f' = \frac{vK}{a} = \frac{v/c_s^2}{1-v^2/c_s^2} = \frac{v}{c_s^2-v^2},
\end{equation}
which is Eq.~\eqref{eq: fase acustica}. The phase is therefore not chosen for convenience: it is fixed by the requirement itself.

Two cancellations follow from Eq.~\eqref{eq:ap escolha fase}. The coefficient of $u$ collects, at first order in $\omega_{\text{ac}}$,
\begin{equation}
\label{eq:ap coef u omega1}
    -i\omega_{\text{ac}}\qty[a f'' + a' f' - \qty(vK)']
    = -i\omega_{\text{ac}}\qty[\qty(a f')' - \qty(vK)'] = 0,
\end{equation}
since $af' = vK$ by Eq.~\eqref{eq:ap escolha fase}, so that no odd power of $\omega_{\text{ac}}$ survives. At second order it collects
\begin{equation}
\label{eq:ap coef u omega2}
    \omega_{\text{ac}}^2\qty(-a f'^2 + 2vK f' + K)
    = \omega_{\text{ac}}^2 K\qty(\frac{v^2K}{a} + 1)
    = \omega_{\text{ac}}^2\,\frac{K}{1-\mathrm{M}^2},
\end{equation}
the middle term coming from the cross product between the phase and the $2i\omega_{\text{ac}} vK\,\psi'$ of Eq.~\eqref{eq:ap ode divergencia}. Without it the factor $1/(1-\mathrm{M}^2)$ would not appear, and the final equation would not have the constant coefficient $\omega_{\text{ac}}^2/c_{s0}^2$ demanded by Eq.~\eqref{eq:equacao schrodinger bna}. Collecting Eqs.~\eqref{eq:ap coef u linha}, \eqref{eq:ap coef u omega1} and~\eqref{eq:ap coef u omega2}, the equation is left in self-adjoint form,
\begin{equation}
\label{eq:ap autoadjunta}
    \dv{}{x}\qty(a \dv{u}{x}) + \frac{\omega_{\text{ac}}^2 K}{1-\mathrm{M}^2}\,u = 0,
\end{equation}
which is Eq.~\eqref{eq: forma autoadjunta} once $a$ and $K$ are written out.

The change of variable~\eqref{eq: coordenada tartaruga bna} comes next. It gives
\begin{equation}
\label{eq:ap dxstar}
    \dv{x_\star}{x} = \frac{c_{s0}}{c_s\qty(1-\mathrm{M}^2)}
    \qquad \Longrightarrow \qquad
    \dv{}{x} = \frac{c_{s0}}{c_s\qty(1-\mathrm{M}^2)}\dv{}{x_\star},
\end{equation}
and it is here that $g_c$ appears, since
\begin{equation}
\label{eq:ap a vira gc}
    a\dv{u}{x}
    = \mathcal{B}\qty(1-\mathrm{M}^2)\,\frac{c_{s0}}{c_s\qty(1-\mathrm{M}^2)}\dv{u}{x_\star}
    = c_{s0}\,\frac{\rho\mathcal{A}}{c_s}\dv{u}{x_\star}
    = c_{s0}\,g_c \dv{u}{x_\star}.
\end{equation}
Applying Eq.~\eqref{eq:ap dxstar} once more to the outer derivative of Eq.~\eqref{eq:ap autoadjunta} and multiplying throughout by $c_s\qty(1-\mathrm{M}^2)/c_{s0}^2$,
\begin{equation}
\label{eq:ap sl}
    \dv{}{x_\star}\qty(g_c \dv{u}{x_\star})
    + \frac{\omega_{\text{ac}}^2}{c_{s0}^2}\,\frac{\rho\mathcal{A}}{c_s}\,u
    = \dv{}{x_\star}\qty(g_c \dv{u}{x_\star}) + \frac{\omega_{\text{ac}}^2}{c_{s0}^2}\,g_c\,u = 0,
\end{equation}
which is Eq.~\eqref{eq: sturm liouville acustico}. The same $g_c$ multiplies the derivative and the eigenvalue term, so the problem is a Sturm--Liouville one with $g_c$ as its weight.

The amplitude closes the reduction. Setting $u = g_c^{-1/2}H_\omega$ in Eq.~\eqref{eq:ap sl},
\begin{equation}
\label{eq:ap amplitude 1}
    g_c\dv{u}{x_\star}
    = g_c^{1/2}\dv{H_\omega}{x_\star} - \frac{1}{2}g_c^{-1/2}\dv{g_c}{x_\star}H_\omega,
\end{equation}
and differentiating once more, the two terms in $\dv*{H_\omega}{x_\star}$ cancel, leaving
\begin{equation}
\label{eq:ap amplitude 2}
    \dv{}{x_\star}\qty(g_c\dv{u}{x_\star})
    = g_c^{1/2}\dv[2]{H_\omega}{x_\star}
    + \qty[\frac{1}{4}g_c^{-3/2}\qty(\dv{g_c}{x_\star})^2
    - \frac{1}{2}g_c^{-1/2}\dv[2]{g_c}{x_\star}]H_\omega.
\end{equation}
Substituting Eq.~\eqref{eq:ap amplitude 2} into Eq.~\eqref{eq:ap sl} and dividing by $g_c^{1/2}$ gives
\begin{equation}
\label{eq:ap final}
    \dv[2]{H_\omega}{x_\star}
    + \qty[\frac{\omega_{\text{ac}}^2}{c_{s0}^2}
    - \frac{1}{2g_c}\dv[2]{g_c}{x_\star}
    + \frac{1}{4g_c^2}\qty(\dv{g_c}{x_\star})^2]H_\omega = 0,
\end{equation}
which is Eq.~\eqref{eq:equacao schrodinger bna} with the potential~\eqref{eq: pot eff bna}, and completes the reduction.

\chapter{Analytical Solution for $\chi(r)$}
\label{ap: analytical solution for lambda}

The reconstruction of a nozzle from a prescribed effective potential, developed in Chapter~\ref{chap:original results}, rests on the Sturm--Liouville problem~\eqref{eq: lambda} for the amplitude $\chi(r)$, related to the mapping function of Chapter~\ref{chap:acoustic laval} by $g_c=\chi^2$. This appendix solves it, in closed form for the Schwarzschild background and bosonic perturbations, and otherwise as a series near the acoustic horizon, which is a singular point of the equation and where the numerical integration has to be started. The closed form then measures how far that series can be trusted.

\section{Schwarzschild background and bosonic perturbations}
\label{sec: ap forma fechada}

The restriction to integer spin is not incidental, since what makes the closed form possible is the rational form of the integer-spin potential~\eqref{final effective potential int}, and the Dirac potential~\eqref{final effective potential half} carries a $\sqrt{B}$ that takes it out of this class.

On Schwarzschild the two metric functions satisfy $B=A^{-1}=1-2G_\text{N}M/r$, so that $\mathrm{F}=\sqrt{B/A}=B$ and $B/A=B^{2}$. The derivative term of Eq.~\eqref{final effective potential int} is then $\qty(1-s^{2})\,BB'/r$ with $B'=2G_\text{N}M/r^{2}$, and the potential collapses to the Regge--Wheeler form
\begin{equation}
\label{eq: ap veff schw}
    V_\text{eff}
    = B\qty[\frac{\ell\qty(\ell+1)}{r^{2}} + \frac{\qty(1-s^{2})2G_\text{N}M}{r^{3}}].
\end{equation}
Writing $z \equiv r/2G_\text{N}M$, so that the horizon sits at $z=1$ and $\mathrm{F}=1-1/z$, radial derivatives rescale as $\dv*{}{r} = \qty(2G_\text{N}M)^{-1}\dv*{}{z}$ and the two coefficients of Eq.~\eqref{eq: lambda 2} become
\begin{subequations}
\label{eq: ap coeficientes em z}
\begin{align}
    \frac{\mathrm{F}'}{\mathrm{F}}
    &= \frac{1}{2G_\text{N}M}\,\frac{z^{-2}}{1-z^{-1}}
     = \frac{1}{2G_\text{N}M}\,\frac{1}{z\qty(z-1)},
    \\[4pt]
    \frac{V_\text{eff}}{\mathrm{F}^{2}}
    &= \frac{1}{\mathrm{F}}\qty[\frac{\ell\qty(\ell+1)}{r^{2}} + \frac{\qty(1-s^{2})2G_\text{N}M}{r^{3}}]
     = \frac{1}{\qty(2G_\text{N}M)^{2}}\,
       \frac{\ell\qty(\ell+1)z + 1 - s^{2}}{z^{2}\qty(z-1)},
\end{align}
\end{subequations}
the single factor of $z/(z-1)$ from $\mathrm{F}^{-1}$ having cancelled one power of $z$ in each numerator. Multiplying Eq.~\eqref{eq: lambda 2} by $\qty(2G_\text{N}M)^{2}z^{2}\qty(z-1)$ clears both denominators at once and leaves
\begin{equation}
\label{eq: lambda em z}
    z^2\qty(z-1)\dv[2]{\chi}{z} + z\dv{\chi}{z}
    - \qty[\ell\qty(\ell+1)z + 1 - s^2]\chi = 0.
\end{equation}
This has three regular singular points, at $z=0$, $z=1$ and $z=\infty$, and is therefore of hypergeometric type. The exponents follow from keeping the leading behaviour at each point and substituting a power. Near the origin Eq.~\eqref{eq: lambda em z} reduces to $-z^{2}\ddot\chi + z\dot\chi - \qty(1-s^{2})\chi = 0$, and $\chi\sim z^{\varrho}$ requires
\begin{equation}
\label{eq: ap indicial zero}
    \varrho^{2} - 2\varrho + 1 - s^{2} = 0,
    \qquad\text{whence}\qquad
    \varrho = 1 \pm s.
\end{equation}
At the horizon, with $u \equiv z-1$, the surviving terms are $u\ddot\chi + \dot\chi = 0$, so that $\chi\sim u^{\varrho}$ gives $\varrho^{2}=0$ and the two exponents coincide at the origin of the exponent scale. At large $z$ only $z^{3}\ddot\chi - \ell\qty(\ell+1)z\chi$ survives, and $\chi\sim z^{-\varrho_\infty}$ gives $\varrho_\infty\qty(\varrho_\infty+1) = \ell\qty(\ell+1)$, with roots $\varrho_\infty = \ell$ and $\varrho_\infty = -\ell-1$. Collecting the three pairs gives the Riemann scheme
\begin{equation}
\label{eq: esquema de riemann}
    \chi = P\begin{Bmatrix}
        0 & 1 & \infty & \\
        1+s & 0 & -\ell-1 & \; z \\
        1-s & 0 & \ell & \;
    \end{Bmatrix},
\end{equation}
the exponents at infinity following the usual convention, in which $\varrho_\infty$ denotes a solution going as $z^{-\varrho_\infty}$, so that the two branches grow as $z^{\ell+1}$ and $z^{-\ell}$. They satisfy Fuchs's relation,
\begin{equation}
\label{eq: relacao de fuchs}
    \qty(1+s) + \qty(1-s) + 0 + 0 + \qty(-\ell-1) + \ell = 1,
\end{equation}
as three regular singular points require. The pair at $z=1$ decides the problem, that being the horizon. The two exponents coincide, so one solution is analytic and the other carries $\log(z-1)$, and finiteness selects the first. Expanded about $z=0$, it is the branch built on $1+s$, which is reached by factoring that exponent out. Setting $\chi = z^{s+1}w(z)$ in Eq.~\eqref{eq: lambda em z}, the terms without derivatives of $w$ cancel identically, and dividing what remains by $z^{s+2}$ and reversing the overall sign gives
\begin{equation}
\label{eq: ap hipergeometrica}
    z\qty(1-z)\dv[2]{w}{z}
    + \qty[\qty(2s+1) - \qty(2s+2)z]\dv{w}{z}
    + \qty[\ell\qty(\ell+1) - s\qty(s+1)]w = 0.
\end{equation}
This is the hypergeometric equation $z(1-z)w'' + \qty[c-(a+b+1)z]w' - ab\,w = 0$ with
\begin{equation}
\label{eq: ap abc}
    a = s-\ell, \qquad b = s+\ell+1, \qquad c = 2s+1,
\end{equation}
as is checked from $a+b = 2s+1$ and $ab = \qty(s-\ell)\qty(s+\ell+1) = s\qty(s+1)-\ell\qty(\ell+1)$. Its solution regular at the origin is $w = {}_2F_1\qty(a,b;c;z)$, so that
\begin{equation}
\label{eq: solucao geral}
    \chi(z) = C\, z^{s+1}\,
    {}_2F_1\qty(s-\ell,\, s+\ell+1;\, 2s+1;\, z),
\end{equation}
the companion branch, built on $1-s$, being degenerate in every bosonic channel, since the two exponents at $z=0$ differ by $2s$, which is an integer there and vanishes for $s=0$. Here $_{2}F_1(a,b;c;z)$ is the Hypergeometric function defined as
\begin{equation}
    _{2}F_1(a,b;c;z) = \sum_{n=0}^\infty \frac{(a)_n(b)_n}{(c)_n}\frac{z^n}{n!},
\end{equation}
and $(x)_n$ is the Pochhammer symbol:
\begin{equation} 
    (x)_n = \begin{cases}
        \displaystyle 1 , \quad \text{for }n=0, \\
        \displaystyle \prod_{k=1}^n (x+k-1) , \quad \text{for }n>0.
    \end{cases}
\end{equation}
Finiteness at the horizon constrains Eq.~\eqref{eq: solucao geral} as well, since its parameters satisfy
\begin{equation}
\label{eq: criterio de gauss}
    c-a-b = \qty(2s+1) - \qty(s-\ell) - \qty(\ell+s+1) = 0,
\end{equation}
which Gauss's criterion places exactly on the boundary of convergence. The series therefore diverges logarithmically at $z=1$ unless it terminates, and termination requires $\ell\geq s$, satisfied by every channel, a field of spin $s$ having no multipole below $\ell=s$. With $a=-m$ a non-positive integer, $m \equiv \ell-s$, the Chu--Vandermonde identity ${}_2F_1\qty(-m,b;c;1) = (c-b)_m/(c)_m$ applies. Its two Pochhammer symbols are
\begin{equation}
\label{eq: ap pochhammers}
    (c-b)_{m} = \qty(s-\ell)_{\ell-s} = (-1)^{\ell-s}\qty(\ell-s)!,
    \qquad
    (c)_{m} = \qty(2s+1)_{\ell-s} = \frac{\qty(\ell+s)!}{\qty(2s)!},
\end{equation}
the first because $(-m)_m = (-1)^{m}m!$ and the second from $(x)_n = \Gamma(x+n)/\Gamma(x)$, so that the series sums to
\begin{equation}
\label{eq: 2f1 no horizonte}
    {}_2F_1\qty(s-\ell,\, s+\ell+1;\, 2s+1;\, 1)
    = \frac{(-1)^{\ell-s}\,(\ell-s)!\,(2s)!}{(\ell+s)!}.
\end{equation}

Only $C$ is left to fix. The acoustic horizon is the sonic point of the flow, so the throat condition~\eqref{eq: cond contorno g} is imposed at $z=1$, where $g_c=\chi^{2}$ reads
\begin{equation}
\label{eq: ap normalizacao}
    \left.g_c\right|_{z=1}
    = C^{2}\qty[\frac{\qty(\ell-s)!\,\qty(2s)!}{\qty(\ell+s)!}]^{2}
    = \frac{\gamma+1}{2\sqrt{2}\sqrt{\gamma-1}},
\end{equation}
which is $3/\sqrt{5}$ for air with $\gamma\approx1.4$, and therefore
\begin{equation}
\label{eq: ap constante C}
    C = \sqrt{\frac{\gamma+1}{2\sqrt{2}\sqrt{\gamma-1}}}\;
        \frac{\Gamma\qty(1+\ell+s)}{\qty(2s)!\,\qty(\ell-s)!},
\end{equation}
the sign of the root being immaterial because only $\chi^{2}$ enters. Squaring the amplitude, the mapping function of the Schwarzschild analogue is
\begin{equation}
\label{eq: gc schwarzschild apendice}
    g_c(r)
    =
    \frac{\gamma+1}{2\sqrt{2}\sqrt{\gamma-1}}
    \qty[\frac{\Gamma(1+\ell+{s})\,{}_2F_1\qty({s}-\ell,\,{s}+\ell+1,\,1+2{s},\,\dfrac{r}{2G_\text{N} M})}{(2{s})!\,(\ell-{s})!}\qty(\frac{r}{2G_\text{N} M})^{{s}+1}]^2,
\end{equation}
which is Eq.~\eqref{gc12} of the main text. By construction the bracket equals $(-1)^{\ell-s}$ at $z=1$, so the prefactor of Eq.~\eqref{eq: gc schwarzschild apendice} can be read off directly as the value of $g_c$ at the throat, which is the check that the normalisation has been carried through correctly.

\section{Series solution near the horizon}
\label{sec: ap serie}

The three regular singular points of Eq.~\eqref{eq: lambda em z} are a property of Schwarzschild and not of the construction, so nothing of the above survives a deformation of the metric. What replaces it is the series~\eqref{eq:frob1} of Chapter~\ref{chap:original results}, whose indicial analysis and coefficients are the subject of this section.

The horizon is a singular point of Eq.~\eqref{eq: lambda 2}, since $\mathrm{F}$ vanishes there, and which kind matters. Writing $u \equiv r-r_\text{EH}$, both $\mathrm{F}$ and $V_\text{eff}$ vanish linearly there, so the two coefficients behave as
\begin{equation}
\label{eq: polos no horizonte}
    \frac{\mathrm{F}'}{\mathrm{F}} = \frac{1}{u} + \order{1},
    \qquad
    \frac{V_\text{eff}}{\mathrm{F}^2} = \frac{q_{-1}}{u} + \order{1},
    \qquad
    q_{-1} \equiv \eval{\frac{V_\text{eff}'}{\mathrm{F}'^2}}_{r_\text{EH}}.
\end{equation}
The residue of the first is unity for every background, since it follows from $\mathrm{F}$ having a simple zero and from nothing else, whereas the residue of the second carries the whole dependence on the geometry,
\begin{equation}
\label{eq: ap residuo}
    q_{-1} = \frac{\ell\qty(\ell+1)}{\qty(A^{-1})'\qty(r_\text{EH})\;r_\text{EH}^{2}}
             + \frac{1-s^{2}}{r_\text{EH}},
\end{equation}
which follows from $\qty(B/A)'' = 2\mathrm{F}'^{2}$ at the horizon and reduces to $\qty[\ell\qty(\ell+1)+1-s^{2}]/2G_\text{N}M$ on Schwarzschild. For the quantum-corrected metric Eq.~\eqref{eq: ap residuo} uses that the two surfaces of Subsec.~\ref{subsec:EFT_horizon} coincide, which they do at the order in $c_6$ to which that metric is determined. Both poles are simple, so $u^2 V_\text{eff}/\mathrm{F}^2$ vanishes at the horizon and the point is regular singular, which is what puts the problem in the reach of the Frobenius method. Substituting $\chi \sim u^{\varsigma}$ into Eq.~\eqref{eq: lambda 2} and keeping the leading order in $u$ gives the indicial equation
\begin{equation}
\label{eq: equacao indicial}
    \varsigma\qty(\varsigma-1) + \varsigma = \varsigma^2 = 0,
\end{equation}
the term in $q_{-1}$ entering only at the next order. Its two roots coincide at $\varsigma=0$, which is what makes Eq.~\eqref{eq:frob1} a series in integer powers, and what places the logarithm in the second branch, the one that finiteness at the horizon removes.

A word is needed on how Eq.~\eqref{eq: recurrence eq} is to be read. Its coefficients are functions of $r$, evaluated at the same $r$ at which the series is read, which is admissible because $\qty(V_\text{eff}/\mathrm{F}^2)u^2$ and $\qty(V_\text{eff}/\mathrm{F}^2)\qty(\mathrm{F}'/\mathrm{F})u^3$ stay finite at the horizon by Eq.~\eqref{eq: polos no horizonte}. The two boundary conditions of Chapter~\ref{chap:original results} then close the problem, the sonic condition fixing $a_0$ and smoothness at the throat fixing $a_1$.

What remains is to carry Eq.~\eqref{eq: recurrence eq} to arbitrary order, in the form in which it was first solved~\cite{Neuringer:1978},
\begin{equation}
    (n+2)(n+1) a_{n+2} = \frac{V_{\text{eff}}(r)}{\mathrm{F}(r)^2} a_n - \frac{\mathrm{F}'(r)}{\mathrm{F}(r)} (n+1)a_{n+1}.
\end{equation}
Assuming convergence, the series~\eqref{eq:frob1} is split into even and odd powers,
\begin{align}
    \chi(r) &= \sum_{k=0}^{\infty} a_k (r-r_\text{EH})^k \nonumber\\ 
    &= a_0 + a_1(r-r_\text{EH}) + \sum_{k=1}^{\infty} a_{2k} (r-r_\text{EH})^{2k} + \sum_{k=1}^{\infty} a_{2k+1} (r-r_\text{EH})^{2k+1}.
\end{align}
With the boundary data~\eqref{eq:condinilamb} the coefficients follow systematically for $k\geq2$, with the structure
\begin{equation}
a_k = 
   \frac{(-1)^k V_\text{eff} a_0}{k! \, \mathrm{F}^k} \ 
   \textcolor{lightgray}{\begin{matrix}
2 \\ 
3 \\ 
4 \\ 
5 \\ 
6 \\ 
7 \\ 
8 \\ 
9 \\ 
10 \\ 
11 \\ 
\vdots
\end{matrix}}
\begin{pmatrix}
{1} \\
{\mathrm{F}'} \\
{V_\text{eff}+\mathrm{F}'^2} \\
{\mathrm{F}'^3+2V_\text{eff}\mathrm{F}'} \\
{V_\text{eff}^2+\mathrm{F}'^4+3V_\text{eff}\mathrm{F}'^2} \\
{\mathrm{F}'^5+4V_\text{eff}\mathrm{F}'^3+3V_\text{eff}^2\mathrm{F}'} \\
{V_\text{eff}^3+\mathrm{F}'^6+5V_\text{eff}\mathrm{F}'^4+6V_\text{eff}^2\mathrm{F}'^2} \\
{\mathrm{F}'^7+6V_\text{eff}\mathrm{F}'^5+10V_\text{eff}^2\mathrm{F}'^3+4V_\text{eff}^3\mathrm{F}'} \\
{V_\text{eff}^4+\mathrm{F}'^8+7V_\text{eff}\mathrm{F}'^6+15V_\text{eff}^2\mathrm{F}'^4+10V_\text{eff}^3\mathrm{F}'^2} \\
{\mathrm{F}'^9+8V_\text{eff}\mathrm{F}'^7+21V_\text{eff}^2\mathrm{F}'^5+20V_\text{eff}^3\mathrm{F}'^3+5V_\text{eff}^4\mathrm{F}'} \\
\vdots
\end{pmatrix}_{\text{position }k-2}
\text{, for }k\geq 2.
\end{equation}

This general structure can be split according to the parity of the index. Writing the even-order coefficients as $a_{2k}$ with $k\geq1$, one finds:
\begin{equation}
a_{2k} =
   \frac{V_\text{eff} a_0}{(2k)! \, \mathrm{F}^{2k}} \
\begin{pmatrix}
{1} \\
{V_\text{eff}+\mathrm{F}'^2} \\
{V_\text{eff}^2+\mathrm{F}'^4+3V_\text{eff}\mathrm{F}'^2} \\
{V_\text{eff}^3+\mathrm{F}'^6+5V_\text{eff}\mathrm{F}'^4+6V_\text{eff}^2\mathrm{F}'^2} \\
{V_\text{eff}^4+\mathrm{F}'^8+7V_\text{eff}\mathrm{F}'^6+15V_\text{eff}^2\mathrm{F}'^4+10V_\text{eff}^3\mathrm{F}'^2} \\
\vdots
\end{pmatrix}_{\text{position }(k-1)}
\text{, for }k\geq 1.
\end{equation}
Similarly, the odd-order coefficients $a_{2k+1}$, with $k\geq1$, take the form:
\begin{equation}
a_{2k+1} =
   -\frac{a_0 V_\text{eff} \mathrm{F}'}{(2k+1)! \, \mathrm{F}^{2k+1}} \
\begin{pmatrix}
{1} \\
{\mathrm{F}'^2+2V_\text{eff}} \\
{\mathrm{F}'^4+4V_\text{eff}\mathrm{F}'^2+3V_\text{eff}^2} \\
{\mathrm{F}'^6+6V_\text{eff}\mathrm{F}'^4+10V_\text{eff}^2\mathrm{F}'^2+4V_\text{eff}^3} \\
{\mathrm{F}'^8+8V_\text{eff}\mathrm{F}'^6+21V_\text{eff}^2\mathrm{F}'^4+20V_\text{eff}^3\mathrm{F}'^2+5V_\text{eff}^4} \\
\vdots
\end{pmatrix}_{\text{position }(k-1)}
\text{, for }k\geq 1.
\end{equation}

Closed-form expressions can now be formulated for each parity. For even coefficients, the general term is given by
\begin{equation}
    a_{2k} = \frac{V_\text{eff} a_0}{(2k)! \, \mathrm{F}^{2k}} \sum_{n=0}^{k-1} V_\text{eff}^n (\mathrm{F}')^{2k-2n-2} \, \mathcal{C}^\text{even}_{n}(k-1),
\end{equation}
where the combinatorial function $\mathcal{C}^\text{even}_{n}(x)$ is defined via the Gamma function as:
\begin{align}
\mathcal{C}^\text{even}_n(x) = \lim_{m \rightarrow x}\frac{(-1)^n\Gamma(2n-2m)}{\Gamma(n+1)\Gamma(n-2m)} \text{, for }x\geq n.
\end{align}
The first few polynomials generated by this definition are:
\begin{subequations}
\begin{align}
    \mathcal{C}^\text{even}_{0}(x) &= 1 \\
    \mathcal{C}^\text{even}_{1}(x) &= (2x-1) \\
    \mathcal{C}^\text{even}_{2}(x) &= (x-1)(2x-3) \\
    \mathcal{C}^\text{even}_{3}(x) &= (x-2)(2x-5)(2x-3)/3 \\
    \mathcal{C}^\text{even}_{4}(x) &= (x-3)(x-2)(2x-7)(2x-5)/6 \\
    \mathcal{C}^\text{even}_{5}(x) &= (x-4)(x-3)(2x-9)(2x-7)(2x-5)/30 \\
    \vdots & \nonumber
\end{align}
\end{subequations}
Thus, the coefficient $a_{2k}$ can be rewritten as:
\begin{equation}
    a_{2k} = \frac{a_0}{(2k)!} \frac{V_{\text{eff}}}{\mathrm{F}'^2} \left( \frac{\mathrm{F}'}{\mathrm{F}}\right)^{2k} \  \sum_{n=0}^{k-1} \left(\frac{V_{\text{eff}}}{\mathrm{F}'^{2}}\right)^n  \mathcal{C}^\text{even}_{n}(k-1).
\end{equation}

For odd indices, following an analogous derivation, the coefficient $a_{2k+1}$ is found to be:
\begin{equation}
    a_{2k+1} =  \frac{- a_0}{(2k+1)!} \frac{V_{\text{eff}}}{\mathrm{F}'^2} \left(\frac{\mathrm{F}'}{\mathrm{F}}\right)^{2k+1} \ \sum_{n=0}^{k-1} \left(\frac{V_{\text{eff}}}{\mathrm{F}'^{2}}\right)^n \mathcal{C}^\text{odd}_{n}(k-1),
\end{equation}
where the corresponding combinatorial function is:
\begin{align}
\mathcal{C}^\text{odd}_n(x) = \lim_{m \rightarrow x}\frac{(-1)^n \Gamma(2n-2m-1)}{\Gamma(n+1)\Gamma(n-2m-1)} \text{, for }x\geq n.
\end{align}
The lowest-order terms for the odd series read:
\begin{subequations}
    \begin{align}
    \mathcal{C}^\text{odd}_{0}(x) &= 1 \\
    \mathcal{C}^\text{odd}_{1}(x) &= 2x \\
    \mathcal{C}^\text{odd}_{2}(x) &= (x-1)(2x-1) \\
    \mathcal{C}^\text{odd}_{3}(x) &= (x-2)(x-1)(2x-3) \cdot 2/3 \\
    \mathcal{C}^\text{odd}_{4}(x) &= (x-3)(x-2)(2x-5)(2x-3)/6 \\
    \mathcal{C}^\text{odd}_{5}(x) &= (x-4)(x-3)(x-2)(2x-7)(2x-5) \cdot 2/30 \\
    \vdots & \nonumber
\end{align}
\end{subequations}

Combining the results for both parities, the complete analytical solution for the normalised function $\chi(r)/a_0$ is:
\begin{align}
    \frac{\chi(r)}{a_0} = 1 
    + \sum_{k=1}^{\infty} &\left\{ \left[\frac{1}{(2k)!} \frac{V_{\text{eff}}}{\mathrm{F}'^2} \left(\frac{\mathrm{F}'}{\mathrm{F}}\right)^{2k} \  \sum_{n=0}^{k-1} \left(\frac{V_{\text{eff}}}{\mathrm{F}'^{2}}\right)^n \mathcal{C}^\text{even}_{n}(k-1)\right] (r-r_\text{EH})^{2k}- \right.\nonumber\\
    &\left.- \left[\frac{1}{(2k+1)!} \frac{V_{\text{eff}}}{\mathrm{F}'^2} \left(\frac{\mathrm{F}'}{\mathrm{F}}\right)^{2k+1} \ \sum_{n=0}^{k-1} \left(\frac{V_{\text{eff}}}{\mathrm{F}'^{2}}\right)^n \mathcal{C}^\text{odd}_{n}(k-1)\right] (r-r_\text{EH})^{2k+1}\right\}.\label{eqrt}
\end{align}

Tables~\ref{tab:Ceven} and~\ref{tab:Codd} give the values of $\mathcal{C}^\text{even}_{n}(x)$ and $\mathcal{C}^\text{odd}_{n}(x)$ needed to use Eq.~\eqref{eqrt}. Both are integer triangles whose rows sum to Fibonacci numbers, odd-indexed in the first and even-indexed in the second, which is what controls the convergence tested below.

\begin{table}[htbp]
\centering
\begin{tabular}{|c|cccccccccccc|}
\hline
\textbf{$x$↓  $n$→} & 0 & 1  & 2   & 3   & 4     & 5     & 6     & 7     & 8   & 9  & 10 & $\hdots$ \\
\hline
0               & 1 &    &     &     &       &       &       &       &     &    &   & \\
1               & 1 & 1  &     &     &       &       &       &       &     &    &   & \\
2               & 1 & 3  & 1   &     &       &       &       &       &     &    &   & \\
3               & 1 & 5  & 6   & 1   &       &       &       &       &     &    &   & \\
4               & 1 & 7  & 15  & 10  & 1     &       &       &       &     &    &   & \\
5               & 1 & 9  & 28  & 35  & 15    & 1     &       &       &     &    &   & \\
6               & 1 & 11 & 45  & 84  & 70    & 21    & 1     &       &     &    &   & \\
7               & 1 & 13 & 66  & 165 & 210   & 126   & 28    & 1     &     &    &   & \\
8               & 1 & 15 & 91  & 286 & 495   & 462   & 210   & 36    & 1   &    &   & \\
9               & 1 & 17 & 120 & 455 & 1,001 & 1,287 & 924   & 330   & 45  & 1  &   & \\
10              & 1 & 19 & 153 & 680 & 1,820 & 3,003 & 3,003 & 1,716 & 495 & 55 & 1 & \\ 
$\vdots$ &&&&&&&&&&&&\\
\hline
\end{tabular}
\caption{Values of $\mathcal{C}^\text{even}_{n}(x)$, with $x$ running down the rows and $n$ across the columns.}
\label{tab:Ceven}
\end{table}

\begin{table}[htbp]
\centering
\begin{tabular}{|c|cccccccccccc|}
\hline
\textbf{$x$↓  $n$→} & 0 & 1  & 2   & 3   & 4     & 5     & 6     & 7     & 8     & 9   & 10 & $\hdots$  \\
\hline
0               & 1 &    &     &     &       &       &       &       &       &     &  &  \\
1               & 1 & 2  &     &     &       &       &       &       &       &     & &   \\
2               & 1 & 4  & 3   &     &       &       &       &       &       &     &  &  \\
3               & 1 & 6  & 10  & 4   &       &       &       &       &       &     &   & \\
4               & 1 & 8  & 21  & 20  & 5     &       &       &       &       &     & &   \\
5               & 1 & 10 & 36  & 56  & 35    & 6     &       &       &       &     & &   \\
6               & 1 & 12 & 55  & 120 & 126   & 56    & 7     &       &       &     &  &  \\
7               & 1 & 14 & 78  & 220 & 330   & 252   & 84    & 8     &       &     &  &  \\
8               & 1 & 16 & 105 & 364 & 715   & 792   & 462   & 120   & 9     &     &  &  \\
9               & 1 & 18 & 136 & 560 & 1,365 & 2,002 & 1,716 & 792   & 165   & 10  &  &  \\
10              & 1 & 20 & 171 & 816 & 2,380 & 4,368 & 5,005 & 3,432 & 1,287 & 220 & 11 &\\
$\vdots$ &&&&&&&&&&&&\\
\hline
\end{tabular}
\caption{Values of $\mathcal{C}^\text{odd}_{n}(x)$, with $x$ running down the rows and $n$ across the columns.}
\label{tab:Codd}
\end{table}

Finally, the decay of the coefficients can be tested in the limiting case $V_\text{eff} = \mathrm{F} = \mathrm{F}' = 1$, in which the whole dependence reduces to the combinatorial sums. Writing $x=k-1$ for the argument, the two families reduce to
\begin{equation}
    \lim_{x\rightarrow \infty} \frac{1}{(2x)!} \sum_{n=0}^{x} \mathcal{C}^\text{even}_{n}(x) = 0, \label{conv1}
\end{equation}
\begin{equation}
    \lim_{x\rightarrow \infty} \frac{1}{(2x+1)!} \sum_{n=0}^{x} \mathcal{C}^\text{odd}_{n}(x) = 0.\label{conv2}
\end{equation}
Table~\ref{tab:convergencia} evaluates both, and the ratios fall by roughly two orders of magnitude per step, which is the expected factorial decay. The sums themselves are the odd- and even-indexed Fibonacci numbers, $2,5,13,34,89,\ldots$ and $1,3,8,21,55,\ldots$, so the factorial in the denominator dominates and the series converges for any finite argument. This is a test of the combinatorial factors alone, not a proof of convergence for arbitrary $V_\text{eff}$ and $\mathrm{F}$, which would require bounding the ratios $V_\text{eff}/\mathrm{F}'^2$ and $\mathrm{F}'/\mathrm{F}$ over the integration range.

\begin{table}[htbp]
\centering 
\begin{tabular}{|c|c|c|}
\hline
$x$ & $\frac{1}{(2x)!} \sum_{n=0}^{x} \mathcal{C}^\text{even}_{n}(x)$  &$ \frac{1}{(2x+1)!} \sum_{n=0}^{x} \mathcal{C}^\text{odd}_{n}(x)$  \\
\hline
0          & 1 & 1 \\
1          & 1 & $5.0\times 10^{-1}$ \\
2          & $2.1\times 10^{-1}$ & $6.7\times 10^{-2}$ \\
3          & $1.8\times 10^{-2}$ & $4.2\times 10^{-3}$ \\
4          & $8.4\times 10^{-4}$ & $1.5\times 10^{-4}$ \\
5          & $2.5\times 10^{-5}$ & $3.6\times 10^{-6}$ \\
6          & $4.9\times 10^{-7}$ & $6.1\times 10^{-8}$ \\
7          & $7.0\times 10^{-9}$ & $7.5\times 10^{-10}$ \\
8          & $7.6\times 10^{-11}$ & $7.3\times 10^{-12}$ \\
9          & $6.5\times 10^{-13}$ & $5.6\times 10^{-14}$ \\
10         & $4.5\times 10^{-15}$ & $3.5\times 10^{-16}$ \\
$\vdots$ &$\vdots$&$\vdots$ \\
\hline
\end{tabular}
\caption{Test of convergence of the combinatorial sums, Eqs.~\eqref{conv1} and~\eqref{conv2}.}
\label{tab:convergencia}
\end{table}


%

\chapter{Curvature Invariants of the Metrics Used in This Thesis}
\label{chap:curvature invariants}

The metrics of Part~\ref{part: extensions of gr} deform Schwarzschild in ways that are transparent at the level of the metric functions but not at the level of the curvature. This appendix computes the curvature invariants of each of them, in full, so that the statements made in the main text about horizons and singularities rest on something explicit. The invariants are the Ricci scalar $R$, the square of the Ricci tensor $R_{\mu\nu}R^{\mu\nu}$, the Kretschmann scalar $K = R_{\mu\nu\rho\sigma}R^{\mu\nu\rho\sigma}$, and the square of the Weyl tensor, which in four dimensions is
\begin{equation}
\label{eq: weyl squared}
    C_{\mu\nu\rho\sigma}C^{\mu\nu\rho\sigma} = K - 2R_{\mu\nu}R^{\mu\nu} + \frac{R^2}{3}.
\end{equation}
The conventions are those of the rest of the thesis, signature $(-,+,+,+)$ and $R^{\rho}_{\ \sigma\mu\nu} = \partial_\mu \Gamma^{\rho}_{\ \nu\sigma} - \partial_\nu \Gamma^{\rho}_{\ \mu\sigma} + \Gamma^{\rho}_{\ \mu\lambda}\Gamma^{\lambda}_{\ \nu\sigma} - \Gamma^{\rho}_{\ \nu\lambda}\Gamma^{\lambda}_{\ \mu\sigma}$, with $R_{\mu\nu} = R^{\rho}_{\ \mu\rho\nu}$.

\section*{The general static, spherically symmetric case}
\label{sec:app general sss}

Every metric treated here has the form
\begin{equation}
\label{eq: app metrica geral}
    \dd s^2 = -B(r)\,\dd t^2 + A(r)\,\dd r^2 + r^2\qty(\dd\theta^2 + \sin^2\theta\,\dd\varphi^2),
\end{equation}
so the invariants are computed once for arbitrary $A$ and $B$ and specialised afterwards. The inverse metric is diagonal, $g^{tt} = -1/B$, $g^{rr} = 1/A$, $g^{\theta\theta} = 1/r^2$ and $g^{\varphi\varphi} = 1/(r^2\sin^2\theta)$, and the non-vanishing Christoffel symbols are
\begin{align}
\label{eq: app christoffel}
    \Gamma^{t}_{\ tr} &= \frac{B'}{2B}, &
    \Gamma^{r}_{\ tt} &= \frac{B'}{2A}, &
    \Gamma^{r}_{\ rr} &= \frac{A'}{2A}, \nonumber \\
    \Gamma^{r}_{\ \theta\theta} &= -\frac{r}{A}, &
    \Gamma^{r}_{\ \varphi\varphi} &= -\frac{r\sin^2\theta}{A}, &
    \Gamma^{\theta}_{\ r\theta} &= \frac{1}{r}, \\
    \Gamma^{\theta}_{\ \varphi\varphi} &= -\sin\theta\cos\theta, &
    \Gamma^{\varphi}_{\ r\varphi} &= \frac{1}{r}, &
    \Gamma^{\varphi}_{\ \theta\varphi} &= \cot\theta, \nonumber
\end{align}
where a prime denotes $\dd/\dd r$. The independent non-vanishing components of the Riemann tensor follow,
\begin{align}
\label{eq: app riemann}
    R^{t}_{\ rtr} &= -\frac{B''}{2B} + \frac{(B')^2}{4B^2} + \frac{A'B'}{4AB}, &
    R^{t}_{\ \theta t\theta} &= -\frac{r B'}{2AB}, \nonumber \\
    R^{t}_{\ \varphi t\varphi} &= -\frac{r \sin^2\theta\, B'}{2AB}, &
    R^{r}_{\ \theta r\theta} &= \frac{r A'}{2A^2}, \\
    R^{r}_{\ \varphi r\varphi} &= \frac{r \sin^2\theta\, A'}{2A^2}, &
    R^{\theta}_{\ \varphi\theta\varphi} &= \frac{(A-1)\sin^2\theta}{A}, \nonumber
\end{align}
and contracting the first index with the third gives the Ricci tensor, which is diagonal,
\begin{align}
\label{eq: app ricci}
    R_{tt} &= \frac{B''}{2A} - \frac{(B')^2}{4AB} - \frac{A'B'}{4A^2} + \frac{B'}{rA}, \\
    R_{rr} &= -\frac{B''}{2B} + \frac{(B')^2}{4B^2} + \frac{A'B'}{4AB} + \frac{A'}{rA}, \\
    R_{\theta\theta} &= -\frac{r B'}{2AB} + \frac{r A'}{2A^2} + 1 - \frac{1}{A},
    \qquad R_{\varphi\varphi} = R_{\theta\theta}\sin^2\theta.
\end{align}
The Ricci scalar is therefore
\begin{equation}
\label{eq: app ricci scalar geral}
    R = -\frac{B''}{AB} + \frac{(B')^2}{2AB^2} + \frac{A'B'}{2A^2B}
        - \frac{2B'}{rAB} + \frac{2A'}{rA^2} + \frac{2}{r^2} - \frac{2}{r^2A},
\end{equation}
and the Kretschmann scalar reads
\begin{multline}
\label{eq: app kretschmann geral}
    K = \frac{(B'')^2}{A^2B^2} - \frac{(B')^2B''}{A^2B^3} + \frac{(B')^4}{4A^2B^4}
        - \frac{A'B'B''}{A^3B^2} + \frac{A'(B')^3}{2A^3B^3} + \frac{(A')^2(B')^2}{4A^4B^2} \\
        + \frac{2(B')^2}{r^2A^2B^2} + \frac{2(A')^2}{r^2A^4}
        + \frac{4}{r^4} - \frac{8}{r^4A} + \frac{4}{r^4A^2}.
\end{multline}
The corresponding expression for $R_{\mu\nu}R^{\mu\nu}$ is longer and is not displayed; it is generated by the same script. When $A = B^{-1}$, which covers every case below except the quantum-corrected metric, Eqs.~\eqref{eq: app ricci scalar geral} and~\eqref{eq: app kretschmann geral} collapse to
\begin{equation}
\label{eq: app invariantes B}
    R = -B'' - \frac{4B'}{r} - \frac{2B}{r^2} + \frac{2}{r^2},
    \qquad
    K = (B'')^2 + \frac{4(B')^2}{r^2} + \frac{4(1-B)^2}{r^4}.
\end{equation}

\section*{Validation: Schwarzschild and Reissner--Nordstr\"om}
\label{sec:app validation}

For $B = 1-2G_\text{N}M/r$ the procedure returns $R = R_{\mu\nu}R^{\mu\nu} = 0$ and
\begin{equation}
\label{eq: app K schw}
    K = C_{\mu\nu\rho\sigma}C^{\mu\nu\rho\sigma} = \frac{48 G_\text{N}^2M^2}{r^6},
\end{equation}
the familiar Schwarzschild result, finite at the horizon and divergent only at $r=0$. For the Reissner--Nordstr\"om metric, $B = 1-2G_\text{N}M/r + Q^2/r^2$, it returns $R=0$ together with
\begin{equation}
\label{eq: app K RN}
    R_{\mu\nu}R^{\mu\nu} = \frac{4Q^4}{r^8},
    \qquad
    K = \frac{48G_\text{N}^2M^2}{r^6} - \frac{96 G_\text{N} M Q^2}{r^7} + \frac{56Q^4}{r^8}.
\end{equation}
Both agree with the standard expressions, which is what licenses the results below.

\section*{Hairy black hole from gravitational decoupling}
\label{sec:app invariants hbh}

For the metric~\eqref{eq: metric HBH} of Chapter~\ref{chap:GD}, with $A = B^{-1}$ and
\begin{equation}
\label{eq: app B hbh}
    B(r) = 1 - \frac{2G_\text{N}\mathcal{M}}{r} + \frac{Q^2}{r^2}
           - \frac{\alpha G_\text{N} M}{r}\,u(r),
    \qquad u(r) \equiv e^{-r/G_\text{N}M},
\end{equation}
the Ricci scalar is remarkably compact,
\begin{equation}
\label{eq: app R hbh}
    R = \frac{\alpha\qty(r - 2G_\text{N}M)}{G_\text{N}M\,r^2}\, u(r).
\end{equation}
It depends on neither $Q$ nor $\ell_h$. The charge-like term is traceless, exactly as in Reissner--Nordstr\"om, and the primary hair enters the metric only through the shifted mass $\mathcal{M}$, which multiplies a term that is itself Ricci flat. The whole of the trace is carried by the exponential deformation, and it changes sign at $r = 2G_\text{N}M$, the Schwarzschild radius of the bare mass. Setting $\alpha = 0$ returns $R=0$, as it must.

Equation~\eqref{eq: app R hbh} is an independent check on Chapter~\ref{chap:GD}. Taking the trace of the quasi-Einstein equations gives $R = -\kappa^2 g^{\mu\nu}\tilde T_{\mu\nu} = -\kappa^2\qty(-\tilde\varrho + \tilde p_\text{rad} + 2\tilde p_\text{tan})$, and substituting the effective source~\eqref{eq:effective calc tensor}, for which $\tilde\varrho = -\tilde p_\text{rad}$, reproduces Eq.~\eqref{eq: app R hbh} identically.

Grouping the remaining invariants in powers of $u$ separates the Reissner--Nordstr\"om part from the hair. The square of the Ricci tensor is
\begin{equation}
\label{eq: app ricci2 hbh}
    R_{\mu\nu}R^{\mu\nu} = \frac{4Q^4}{r^8}
        - \frac{2Q^2\alpha\qty(2G_\text{N}M + r)}{G_\text{N}M\,r^6}\,u
        + \frac{\alpha^2\qty(4G_\text{N}^2M^2 + r^2)}{2G_\text{N}^2M^2\,r^4}\,u^2,
\end{equation}
and the Kretschmann scalar is
\begin{multline}
\label{eq: app K hbh}
    K = \qty[\frac{48G_\text{N}^2\mathcal{M}^2}{r^6} - \frac{96G_\text{N}\mathcal{M}Q^2}{r^7} + \frac{56Q^4}{r^8}] \\
    + \frac{4\alpha}{G_\text{N}M r^7}\Big[
        12G_\text{N}^3M^3 r + 6G_\text{N}^3M^2\alpha\ell_h r
        + 8G_\text{N}^2M^2r^2 + 4G_\text{N}^2M\alpha\ell_h r^2 \\
        + 2G_\text{N}Mr^3 + G_\text{N}\alpha\ell_h r^3
        - 12G_\text{N}^2M^2Q^2 - 10G_\text{N}MQ^2 r - 3Q^2r^2 \Big] u \\
    + \frac{\alpha^2\qty(2G_\text{N}^2M^2 + 2G_\text{N}Mr + r^2)\qty(6G_\text{N}^2M^2 + 2G_\text{N}Mr + r^2)}{G_\text{N}^2M^2\,r^6}\, u^2.
\end{multline}
The bracket in the first line is precisely Eq.~\eqref{eq: app K RN} with $M$ replaced by $\mathcal{M}$, so that switching the exponential off leaves the geometry indistinguishable from Reissner--Nordstr\"om with the dressed mass, as the construction of Chapter~\ref{chap:GD} requires.

Three limits summarise the singularity structure. At large radii the exponential is negligible and
\begin{equation}
\label{eq: app K hbh inf}
    K \longrightarrow \frac{48G_\text{N}^2\mathcal{M}^2}{r^6}, \qquad r \to \infty,
\end{equation}
so that $\mathcal{M}$ is the mass an observer at infinity measures, which is the statement made in Chapter~\ref{chap:GD}. At the centre the behaviour depends on whether the charge-like parameter is present. For $Q\neq0$,
\begin{equation}
\label{eq: app K hbh zero Q}
    K \longrightarrow \frac{56Q^4}{r^8}, \qquad r\to0,
\end{equation}
the Reissner--Nordstr\"om singularity, while for $Q=0$ the leading behaviour is Schwarzschild-like,
\begin{equation}
\label{eq: app K hbh zero}
    K \longrightarrow \frac{48G_\text{N}^2\mathcal{M}_\text{eff}^2}{r^6},
    \qquad
    \mathcal{M}_\text{eff} = M + \frac{\alpha\qty(M+\ell_h)}{2}, \qquad r\to0.
\end{equation}
The two masses differ, $\mathcal{M}_\text{eff} - \mathcal{M} = \alpha M/2$, because the exponential term contributes its full $\alpha G_\text{N}M/r$ at the centre and is exponentially suppressed at infinity. The hair therefore does not remove the central singularity; it dresses it, and the mass that governs the curvature there is not the one measured asymptotically.

\section*{Quantum-corrected Schwarzschild metric}
\label{sec:app invariants ck}

The metric~\eqref{eq:CK_metric} of Chapter~\ref{chap:EFT} is the only one here with $A \neq B^{-1}$, so the general expressions~\eqref{eq: app ricci scalar geral} and~\eqref{eq: app kretschmann geral} are needed in full. The exact invariants are unwieldy and of little use, since the metric is itself a first-order result in $c_6$. To that order,
\begin{align}
\label{eq: app R ck}
    R &= \frac{768\pi\, c_6\, G_\text{N}^4M^2\qty(45r - 106 G_\text{N}M)}{r^9} + \mathcal{O}(c_6^2), \\
\label{eq: app ricci2 ck}
    R_{\mu\nu}R^{\mu\nu} &= \mathcal{O}(c_6^2), \\
\label{eq: app K ck}
    K &= \frac{48G_\text{N}^2M^2}{r^6}
        - \frac{18432\pi\, c_6\, G_\text{N}^5M^3\qty(12r - 25G_\text{N}M)}{r^{12}}
        + \mathcal{O}(c_6^2).
\end{align}
That $R_{\mu\nu}R^{\mu\nu}$ starts at second order is immediate, since the Ricci tensor itself is of order $c_6$ on a Ricci-flat background. Two features of Eq.~\eqref{eq: app K ck} are worth noting. The correction falls as $r^{-12}$ against the $r^{-6}$ of the Schwarzschild term, so it is confined to the region near the horizon and dies far faster than the leading curvature, which is the covariant counterpart of the $M^{-4}$ suppression established in Chapter~\ref{chap:conexaoentrebnsebocais}. And the correction changes sign at $r = 25G_\text{N}M/12$, just outside the horizon of the uncorrected solution, so that a given sign of $c_6$ strengthens the curvature on one side of that radius and weakens it on the other.

\section*{Canonical acoustic black hole}
\label{sec:app invariants acoustic}

The acoustic geometry~\eqref{metrica bac} of Chapter~\ref{chap: analogue gravity} is not a solution of any field equation, so there is no reason for its Ricci tensor to vanish, and it does not. With the conformal factor $\rho/c_s$ held constant and set to unity, and $c_s = 1$, the blackening function is $B = 1 - r_\text{EH}^4/r^4$ and
\begin{equation}
\label{eq: app invariantes acustico}
    R = \frac{6\,r_\text{EH}^4}{r^6},
    \quad
    R_{\mu\nu}R^{\mu\nu} = \frac{90\,r_\text{EH}^8}{r^{12}},
    \quad
    K = \frac{468\,r_\text{EH}^8}{r^{12}},
    \quad
    C_{\mu\nu\rho\sigma}C^{\mu\nu\rho\sigma} = \frac{300\,r_\text{EH}^8}{r^{12}}.
\end{equation}
A constant conformal factor $\Omega = \rho/c_s$ rescales these by $\Omega^{-1}$ for $R$ and $\Omega^{-2}$ for the quadratic invariants, and changes nothing about where they are finite.

At the acoustic horizon every invariant is finite,
\begin{equation}
\label{eq: app invariantes acustico horizonte}
    \eval{R}_{r_\text{EH}} = \frac{6}{r_\text{EH}^2},
    \qquad
    \eval{K}_{r_\text{EH}} = \frac{468}{r_\text{EH}^4},
\end{equation}
which establishes what Sec.~\ref{sec: acoustic metric} asserts on the strength of the coordinate transformation alone: the divergence of the metric coefficients at $r_\text{EH}$ in the static coordinates $(\tilde t, r)$ is an artefact of those coordinates. All four invariants diverge as $r\to0$, where the velocity field~\eqref{eq: campo velocidades} itself diverges, so that the centre of the canonical model is a genuine curvature singularity of the effective geometry. In a real duct that region lies outside the domain in which the model applies, since the flow would have become supersonic without bound long before.

\chapter{Comparison of Methods for the Quasinormal Spectrum}
\label{chap:qnm methods}

Chapter~\ref{chap:QNM_spectrum} computes every quasinormal frequency by the Mashhoon method, and Sec.~\ref{sec:mashhoon} states the reason for that choice and the error it costs on a Schwarzschild background. This appendix puts the choice on a wider footing, applying four methods of different character to the same five geometries, namely direct integration, the WKB expansion, the Mashhoon substitution, and Leaver's continued fraction, which is exact but available only on Schwarzschild. The aim is to fix the range over which the spectra of Chapter~\ref{chap:QNM_spectrum} can be trusted, and to identify the backgrounds on which the methods stop answering the same question.

All of them start from the master equation~\eqref{eq:qnm_master} with the boundary conditions~\eqref{eq:qnm_bc}. It is convenient to name the combination that converts radial into tortoise derivatives,
\begin{equation}
    \label{eq:app F}
    F(r) \equiv \sqrt{\frac{B(r)}{A(r)}},
    \qquad
    \dv{}{r_\star} = F\dv{}{r},
\end{equation}
which reduces to $F=B$ whenever $A=B^{-1}$, as for Schwarzschild and the hairy solution, but not for the quantum-corrected one. Frequencies are computed with $G_\text{N}=M=1$ and reported as $\omega = \omega_R + i\,\omega_I$ with $\omega_I>0$, Eq.~\eqref{eq:qnm_freq}. With the time dependence $e^{-i\omega t}$ of Eq.~\eqref{eq:qnm_master}, a damped mode has a negative imaginary part, and that is the sign that comes out of Eqs.~\eqref{eq:app wkb} and~\eqref{eq:app leaver series} and of the direct integration. The tables follow Eq.~\eqref{eq:qnm_freq} in quoting its magnitude, the damping rate $\omega_I$. Comparisons are made in modulus, at the lowest multipole of each channel, $\ell=s$.

\section{The four methods}
\label{sec:app three methods}

\subsection{Direct integration}
\label{subsec:app direct}

This is the only method that uses the barrier as a whole rather than its neighbourhood at the peak, and the only one exact in principle on every background, so it serves as the reference wherever no continued fraction is available. It goes back to Chandrasekhar and Detweiler~\cite{Chandrasekhar:1975zza}, who integrated the perturbation equation from the horizon and from infinity and matched the two solutions at an intermediate radius. The implementation used here adapts the scheme reviewed in Ref.~\cite{Pani:2013pma}. There a near-horizon series fixes the solution at the horizon, the higher-order terms of the expansion at infinity impose the outgoing condition at a finite radius, and the quasinormal frequencies are the complex roots of the resulting function of $\omega$. The present version keeps the two-sided integration of Ref.~\cite{Chandrasekhar:1975zza}, matching the two legs through their Wronskian, and departs from both references in removing the oscillating factor before integrating and in carrying the areal radius as an integration variable.

Writing $R = e^{-i\omega r_\star}v$ near the inner surface, the derivatives are
\begin{equation}
    \label{eq:app v derivs}
    \dot R = e^{-i\omega r_\star}\qty(\dot v - i\omega v),
    \qquad
    \ddot R = e^{-i\omega r_\star}\qty(\ddot v - 2i\omega\dot v - \omega^2 v),
\end{equation}
with the dot denoting $\dv*{}{r_\star}$ as in Eq.~\eqref{eq:mama}. Substituting into $\ddot R + \qty(\omega^2 - V_{\text{eff}})R = 0$ cancels the two terms in $\omega^2 v$ and leaves an equation for the amplitude alone. Repeating with $R = e^{+i\omega r_\star}u$ at large radius only flips the sign of the first-derivative term,
\begin{align}
    \label{eq:app amplitude eqs}
    \ddot v - 2i\omega\,\dot v - V_{\text{eff}}\,v = 0,
    &&
    \ddot u + 2i\omega\,\dot u - V_{\text{eff}}\,u = 0.
\end{align}
The boundary conditions~\eqref{eq:qnm_bc} now read simply $v,u\to\text{constant}$ at their own end of the domain, and the integrator carries a slowly varying function instead of an oscillation.

A quasinormal frequency is one at which the two solutions coincide, that is, one at which their Wronskian $W = R_{\text{in}}\dot R_{\text{out}} - \dot R_{\text{in}}R_{\text{out}}$ vanishes. With the substitutions above the two exponentials cancel, so the matching radius may be chosen freely and is taken here at the peak of the barrier,
\begin{equation}
    \label{eq:app wronskian}
    W(\omega) = v\,\dot u - \dot v\,u + 2i\omega\,u\,v.
\end{equation}
The calculation then proceeds in five steps.
\begin{enumerate}
    \item Integrate $v$ outwards from $r_\text{out}+\varepsilon$, the first Frobenius coefficient being supplied analytically, and $u$ inwards from a finite cut-off $r_\infty$, both in $r_\star$, carrying the areal radius as a third variable through $\dv*{r}{r_\star}=F(r)$. No factor of $1/F$ then appears anywhere, and the surface where $F$ vanishes is a regular point of the integration rather than a stiff one.
    \item Impose the outer boundary condition through the asymptotic series
    \begin{equation}
        \label{eq:app asymptotic}
        u(r) = 1 + \sum_{k\geq1} \frac{b_k}{r^k},
    \end{equation}
    whose coefficients follow order by order in $1/r$ from Eq.~\eqref{eq:app amplitude eqs}, the first being $b_1 = i\,\ell(\ell+1)/2\omega$ for integer spin. In the Dirac channel the potential falls off as $(\ell^2-\tfrac14)/r^2$ instead, and $\ell(\ell+1)$ is replaced by $\ell^2-\tfrac14$. Twenty of them are used. Setting $u=1$ instead would not do, since $V_{\text{eff}}$ decays only as a power law. Any Yukawa term drops out here, $e^{-r}$ being smaller than every power of $1/r$. 
    \item Place the cut-off. The branch excluded by the boundary condition grows like
    \begin{equation}
        \label{eq:app contamination}
        \exp\qty[2\,\omega_I\,\qty(r_\star^{\infty} - r_\star^{\text{peak}})]
    \end{equation}
    as the outer leg is integrated inwards, so pushing $r_\infty$ out improves the boundary condition and amplifies the contamination at the same time. The two demands leave a window, not a limit, and it narrows as $\omega_I$ grows. Here $r_\infty$ sits where the exponent reaches $16$, some seven of the sixteen decimal digits carried by the arithmetic.
    \item Find the zeros of Eq.~\eqref{eq:app wronskian} by Muller's method, seeded by the WKB value of the same mode.
    \item Accept the root only if it converges. Each is sought under four combinations of $r_\infty$ and $\varepsilon$ and kept only when the four agree to two per cent in modulus, the spread among them being the uncertainty quoted. A root far from the seed is discarded, a boundary condition imposed at finite radius being able to introduce zeros of the Wronskian that belong to no mode.
\end{enumerate}

\subsection{The WKB expansion}
\label{subsec:app wkb}

Near its peak the barrier is an inverted parabola, and for that profile the connection problem between the two asymptotic WKB solutions is solved exactly by parabolic cylinder functions. Writing $Q \equiv \omega^2 - V_{\text{eff}}$, whose value at the peak is $Q_0 = \omega^2 - V_{\text{eff}}^{\max}$, the matching condition at lowest order is
\begin{equation}
    \label{eq:app wkb lowest}
    \frac{i\,Q_0}{\sqrt{2\ddot Q_0}} = n + \frac12,
    \qquad n = 0,1,2,\ldots,
\end{equation}
which is the statement that the two solutions join onto one another only for a discrete set of $\omega$. Carrying the Taylor expansion of $Q$ about the peak beyond quadratic order adds corrections built from the higher derivatives there, and through third order the condition becomes~\cite{Schutz:1985km,Iyer:1986np}
\begin{equation}
    \label{eq:app wkb}
    \omega^2 = \qty[V_{\text{eff}}^{\max} + \sqrt{-2\ddot V_0}\,\Lambda]
    - i\qty(n+\frac12)\sqrt{-2\ddot V_0}\,\qty[1 + \Omega],
\end{equation}
with every derivative taken with respect to $r_\star$ at the peak and $\ddot V_0 = -\ddot Q_0$ the second one. The second-order term $\Lambda$ is built from derivatives up to the fourth and the third-order term $\Omega$ from derivatives up to the sixth, both given in full in Ref.~\cite{Iyer:1986np}. Setting $\Lambda=\Omega=0$ recovers Eq.~\eqref{eq:app wkb lowest}, the first-order formula of Ref.~\cite{Schutz:1985km}, which is evaluated alongside the third-order one below because the separation between the two measures how well the series is converging. Extensions to sixth~\cite{Konoplya:2003ii} and thirteenth order~\cite{Matyjasek:2017psv,Konoplya:2019hlu} are standard and are not needed here.

The derivatives are obtained without finite differencing. The Taylor series of $V_{\text{eff}}(r)$ and $F(r)$ about the peak are computed in arbitrary-precision arithmetic, and the operator $F\dv*{}{r}$ of Eq.~\eqref{eq:app F} is applied to them repeatedly, which returns $\dv*[k]{V_{\text{eff}}}{r_\star}$ to machine precision.

\subsection{The Mashhoon substitution}
\label{subsec:app mashhoon}

The third method is the one used throughout Chapter~\ref{chap:QNM_spectrum} and described in Sec.~\ref{sec:mashhoon}. It keeps only $V_{\text{eff}}^{\max}$ and the curvature at the peak, replaces the barrier by the P\"oschl--Teller profile~\eqref{eq:PT_potential} reproducing both, and reads off the closed-form spectrum~\eqref{eq:PT_omega}. It is the truncation of a different expansion from the one above, keeping the exact spectrum of an approximate barrier rather than an approximate spectrum of the exact barrier, which is why its error does not fall as further derivatives are added.

\subsection{Leaver's continued fraction}
\label{subsec:app leaver}

The fourth method is exact and is used here only to calibrate the other three, since the recurrence below is specific to Schwarzschild. In the units $2G_\text{N}M=1$ of Ref.~\cite{Leaver:1985ax}, so that the horizon sits at $r=1$ and $\omega_\text{thesis}=\omega_\text{Leaver}/2$, the solution is written as
\begin{equation}
    \label{eq:app leaver series}
    R(r) = \qty(r-1)^{-i\omega} r^{2i\omega} e^{i\omega\qty(r-1)}
    \sum_{n=0}^{\infty} a_n \qty(\frac{r-1}{r})^{n},
\end{equation}
in which the prefactor already carries both boundary conditions~\eqref{eq:qnm_bc}, ingoing at the horizon and outgoing at infinity. Inserting Eq.~\eqref{eq:app leaver series} into the master equation leaves a three-term recurrence for the coefficients,
\begin{align}
    \label{eq:app leaver recurrence}
    \alpha_0 a_1 + \beta_0 a_0 = 0,
    &&
    \alpha_n a_{n+1} + \beta_n a_n + \gamma_n a_{n-1} = 0,
    \quad n\geq1,
\end{align}
with
\begin{subequations}
\label{eq:app leaver coeffs}
\begin{align}
    \alpha_n &= n^2 + \qty(2 - s - 2i\omega)n + 1 - s - 2i\omega, \\
    \beta_n &= -2n^2 + \qty(8i\omega - 2)n + 8\omega^2 + 4i\omega + s^2 - 1 - \ell\qty(\ell+1), \\
    \gamma_n &= n^2 + \qty(s - 4i\omega)n - 4\omega^2 - 2is\omega .
\end{align}
\end{subequations}
The series~\eqref{eq:app leaver series} converges, which is what the boundary conditions demand, only for those $\omega$ at which the recurrence admits a minimal solution, and that requirement is equivalent to the vanishing of a continued fraction,
\begin{equation}
    \label{eq:app leaver cf}
    \beta_0 - \cfrac{\alpha_0\gamma_1}{\beta_1 - \cfrac{\alpha_1\gamma_2}{\beta_2 - \cdots}} = 0 .
\end{equation}
The quasinormal frequencies are the roots of Eq.~\eqref{eq:app leaver cf}, obtained here by truncating the fraction at depth $1500$ in arbitrary-precision arithmetic. The recurrence~\eqref{eq:app leaver coeffs} is one of integer spin, so the Dirac channel has no exact reference in what follows.

\section{Calibration on Schwarzschild}
\label{sec:app schw}

Table~\ref{tab:app schw} sets the three approximate methods against Leaver on the one background whose spectrum is known exactly. Direct integration reproduces it to better than $0.01\%$ in every channel, which validates the implementation of Subsec.~\ref{subsec:app direct} and licenses its use as the reference where no continued fraction exists.

\begin{table}[H]
\centering
\caption[Quasinormal methods compared on Schwarzschild]{Fundamental mode $\omega_0$ on a Schwarzschild background at $\ell=s$, computed with $G_\text{N}=M=1$. Panel~\subref{tab:app schw a} lists the frequencies and panel~\subref{tab:app schw b} the departures, that of the direct integration measured against Leaver and those of the two approximations against the direct integration, with the Mashhoon error also given part by part for comparison with Sec.~\ref{sec:mashhoon}. The Dirac channel has no continued-fraction entry, the recurrence~\eqref{eq:app leaver coeffs} being one of integer spin.}
\label{tab:app schw}
\footnotesize
\setlength{\tabcolsep}{4pt}
\renewcommand{\arraystretch}{1.2}
\begin{subtable}{\linewidth}
\centering
\caption{Fundamental mode $\omega_0$.}\label{tab:app schw a}
\begin{tabular}{||c||c|c|c|c||}
\hline\hline
$s$ & Leaver & Direct & WKB & Mashhoon \\
\hline\hline
$0$ & $0.1105+0.1049\,i$ & $0.1104+0.1049\,i$ & $0.1046+0.1152\,i$ & $0.1148+0.1148\,i$ \\ \hline
$1/2$ & --- & $0.1872+0.0986\,i$ & $0.1765+0.1001\,i$ & $0.1890+0.1048\,i$ \\ \hline
$1$ & $0.2483+0.0925\,i$ & $0.2483+0.0925\,i$ & $0.2459+0.0931\,i$ & $0.2546+0.0962\,i$ \\ \hline
$2$ & $0.3737+0.0890\,i$ & $0.3737+0.0890\,i$ & $0.3732+0.0892\,i$ & $0.3783+0.0905\,i$ \\ \hline
\hline
\end{tabular}
\end{subtable}

\vspace{1.2em}

\begin{subtable}{\linewidth}
\centering
\caption{Departures (\%). The first three columns are taken in modulus, the last two part by part, which is the metric of Sec.~\ref{sec:mashhoon}.}\label{tab:app schw b}
\begin{tabular}{||c||c|c|c||c|c||}
\hline\hline
 & \multicolumn{3}{c||}{In modulus} & \multicolumn{2}{c||}{Mashhoon vs.\ Leaver} \\
\cline{2-6}
$s$ & \makecell{Direct\\vs.\ Leaver} & \makecell{WKB\\vs.\ Direct} & \makecell{Mashhoon\\vs.\ Direct} & $\omega_R$ & $\omega_I$ \\
\hline\hline
$0$ & $0.008$ & $7.8$ & $7.1$ & $3.9$ & $9.4$ \\ \hline
$1/2$ & --- & $5.1$ & $3.0$ & --- & --- \\ \hline
$1$ & $0.000$ & $0.9$ & $2.8$ & $2.5$ & $4.0$ \\ \hline
$2$ & $0.000$ & $0.1$ & $1.3$ & $1.2$ & $1.7$ \\ \hline
\hline
\end{tabular}
\end{subtable}
\end{table}

Panel~\subref{tab:app schw a} agrees with Chapter~\ref{chap:QNM_spectrum} mode by mode: its Mashhoon column reproduces the Schwarzschild columns of Tables~\ref{tab:QNM_GD} and~\ref{tab:QNM_QC} to the last digit in all four channels, as it must. Section~\ref{sec:mashhoon} quotes the error part by part, within $4\%$ of the exact $\omega_R$ and within $10\%$ of the exact $\omega_I$, and the last two columns of panel~\subref{tab:app schw b} show the scalar channel saturating those bounds at $3.9\%$ and $9.4\%$. The other columns are departures in modulus, the metric used throughout this appendix because it survives where no exact reference exists, and they combine the two parts into $7.1\%$ for that same channel.

The accuracy of both approximations improves monotonically along $\ell=s=0,\tfrac12,1,2$, from $7.1\%$ to $1.3\%$ for the Mashhoon method, since both keep only the neighbourhood of the peak and the barrier grows taller and sharper as $\ell$ rises, Fig.~\ref{fig:schw spin}. This ordering recurs on every background below, the single exception being the third-order WKB at $c_6=+0.003$, where the gravitational departure of $0.9\%$ slightly exceeds the electromagnetic one of $0.7\%$. On Schwarzschild the two methods also fail in opposite directions, the Mashhoon frequency lying above the exact one in both parts, its P\"oschl--Teller profile being the wider barrier, and the WKB one undershooting the real part and overshooting the damping rate. This is a tendency rather than a rule, and it fails on several of the backgrounds below.

The separation between the first- and third-order WKB values measures whether that series is converging at all. On Schwarzschild it runs from $56\%$ in the scalar channel to $6.7\%$ in the gravitational one, so the series is asymptotic rather than convergent at low multipole, and the $7.8\%$ departure of the third-order value from the direct integration there is not a controlled error bar.

\section{Hairy black hole}
\label{sec:app hairy}

The hairy solution~\eqref{eq: metric HBH} is examined at the strongest hair of Table~\ref{tab:QNM_GD}, $\alpha=\ell_h=1$, at the two ends of the admissible charge range. The lower end is the dominant-energy-condition threshold $Q_\text{DEC}=2\sqrt{\alpha}\,G_\text{N}M/e\simeq0.7358$, Eq.~\eqref{eq: limit Q alpha M}. At the upper end, $Q=Q_\text{max}\simeq1.6086$, the two roots of $B(r)$ merge and the configuration is extremal, so the surface gravity vanishes, the tortoise coordinate diverges as $1/(r-r_\text{EH})$ rather than logarithmically, and the barrier decays there as a power of $r_\star$ instead of exponentially. Direct integration loses its footing, the gravitational channel giving $\omega_R$ anywhere between $0.34$ and $0.41$ as the starting offset is moved over two decades. The whole extremal block of Table~\ref{tab:app exotic} is therefore rejected by step 5 of Subsec.~\ref{subsec:app direct}, and the comparison is made at $Q=0.99\,Q_\text{max}$, where the same test passes in three of the four channels with a spread of at most three parts in $10^{4}$. The Dirac channel fails it even there. Its potential vanishes towards the horizon only as $\sqrt{B}$, more slowly than the others, so it is the first to feel the small surface gravity of a near-extremal configuration. The two local methods notice none of this, a vanishing surface gravity leaving the peak almost unchanged.

Away from extremality the methods agree as well as on Schwarzschild. In the gravitational channel the Mashhoon method lies within $1.3\%$ of the direct integration at $Q_\text{DEC}$ and within $0.8\%$ at $Q=0.99\,Q_\text{max}$, the third-order WKB within $0.2\%$ and $0.1\%$. The scalar channel is again the worst, at $7.2\%$ and $8.9\%$. The hair therefore does not degrade the approximation, which is what matters for Chapter~\ref{chap:QNM_spectrum}, since Table~\ref{tab:QNM_GD} reports a shift measured between two geometries treated by one and the same method.

\section{Quantum-corrected black hole}
\label{sec:app qc}

The quantum-corrected metric~\eqref{eq:CK_metric} is examined at one coupling of each sign. The negative one, $c_6=-0.010$, lies somewhat beyond the most negative column of Table~\ref{tab:QNM_QC}. The positive one, $c_6=+0.003$, lies just below the threshold $c_6\simeq3.6\times10^{-3}$ of Eq.~\eqref{eq:B_surface_exists}, above which the surface $B=0$ ceases to exist and the truncated metric no longer describes a black hole.

Here the methods stop answering the same question, and the reason is the inner boundary. The surfaces $B=0$ and $A^{-1}=0$ coincide at linear order in $c_6$, which is why Chapter~\ref{chap:EFT} can define the horizon through $A^{-1}$, but at the couplings needed here they no longer do. The root of $A^{-1}$ lies outside that of $B$ by $7\%$ at $c_6=-0.010$ and by $4\%$ at $c_6=+0.003$, so the outer surface of the domain is one where the radial metric diverges while the redshift stays finite. It is a throat in the sense of Subsec.~\ref{subsec:veff quantum} even on the negative branch, where the surface of infinite redshift lies further in. It sits at finite tortoise distance, and neither the blackening factor nor the effective potential vanishes there. Measured against the height of the barrier,
\begin{equation}
    \label{eq:app Vout}
    \begin{split}
    \left.\frac{V_{\text{eff}}(r_\text{out})}{V_{\text{eff}}^{\max}}\right|_{c_6=-0.010}
        &= 0.37,\ 0.39,\ 0.48,\ 0.51, \\[4pt]
    \left.\frac{V_{\text{eff}}(r_\text{out})}{V_{\text{eff}}^{\max}}\right|_{c_6=+0.003}
        &= 0.18,\ 0.14,\ 0.17,\ 0.16,
    \end{split}
\end{equation}
along the four channels in the order of Table~\ref{tab:app exotic}. The barrier is cut off at between a seventh and half of its own height, whereas the boundary condition~\eqref{eq:qnm_bc} assumed by all four methods is stated where the potential vanishes. Direct integration, the only method that reaches the inner boundary, registers the difficulty by failing the convergence test in the scalar channel at $c_6=-0.010$, the only fundamental mode it rejects outside the hairy configurations at and near extremality.

The departures follow that pattern. In the gravitational channel the Mashhoon method stays within $4.4\%$ of the direct integration at $c_6=-0.010$ and within $1.3\%$ at $c_6=+0.003$, so the frequencies of Table~\ref{tab:QNM_QC} are as sound there as those of Schwarzschild. In the Dirac channel at $c_6=-0.010$ the same comparison gives $35\%$, and in the electromagnetic one $10\%$. The WKB series behaves likewise, its first- to third-order separation reaching $104\%$ in the scalar channel and $53\%$ in the Dirac one at $c_6=-0.010$, against $56\%$ and $30\%$ on Schwarzschild. On every measure available here, the quantum-corrected background at negative coupling is the hardest of the five.

\begin{table}[H]
\centering
\caption[Quasinormal methods compared on the exotic backgrounds]{Fundamental mode $\omega_0$ at $\ell=s$ on the four exotic backgrounds, computed with $G_\text{N}=M=1$, with the departure in modulus of each approximation from the direct integration. An entry marked $\ast$ is one that the direct integration returned but step 5 of Subsec.~\ref{subsec:app direct} rejected. The whole extremal block is rejected, for the reason given in Sec.~\ref{sec:app hairy}. Both hairy blocks are at $\alpha=\ell_h=1$.}
\label{tab:app exotic}
\footnotesize
\setlength{\tabcolsep}{4pt}
\renewcommand{\arraystretch}{1.2}
\begin{tabular}{||c||c|c|c||c|c||}
\hline\hline
 & \multicolumn{3}{c||}{Fundamental mode $\omega_0$} & \multicolumn{2}{c||}{Departure (\%)} \\
\cline{2-6}
$s$ & Direct & WKB & Mashhoon & WKB & Mashhoon \\
\hline\hline
\multicolumn{6}{||c||}{Hairy black hole, $Q=Q_{\rm DEC}\simeq0.7358$} \\ \hline
$0$ & $0.0761+0.0718\,i$ & $0.0703+0.0783\,i$ & $0.0790+0.0788\,i$ & $8.3$ & $7.2$ \\ \hline
$1/2$ & $0.1296+0.0669\,i$ & $0.1219+0.0685\,i$ & $0.1310+0.0721\,i$ & $5.4$ & $3.7$ \\ \hline
$1$ & $0.1722+0.0636\,i$ & $0.1706+0.0640\,i$ & $0.1765+0.0661\,i$ & $0.9$ & $2.7$ \\ \hline
$2$ & $0.2598+0.0614\,i$ & $0.2594+0.0615\,i$ & $0.2630+0.0625\,i$ & $0.2$ & $1.3$ \\ \hline
\multicolumn{6}{||c||}{Hairy black hole, $Q=0.99\,Q_{\max}\simeq1.5925$} \\ \hline
$0$ & $0.0924+0.0659\,i$ & $0.0807+0.0705\,i$ & $0.0962+0.0752\,i$ & $11.1$ & $8.9$ \\ \hline
$1/2$ & $\ast$ & $0.1615+0.0630\,i$ & $0.1693+0.0662\,i$ & --- & --- \\ \hline
$1$ & $0.2303+0.0603\,i$ & $0.2282+0.0601\,i$ & $0.2346+0.0631\,i$ & $0.9$ & $2.1$ \\ \hline
$2$ & $0.3563+0.0574\,i$ & $0.3560+0.0575\,i$ & $0.3591+0.0584\,i$ & $0.1$ & $0.8$ \\ \hline
\multicolumn{6}{||c||}{Hairy black hole, $Q=Q_{\max}\simeq1.6086$ (extremal)} \\ \hline
$0$ & $\ast$ & $0.0807+0.0706\,i$ & $0.0963+0.0752\,i$ & --- & --- \\ \hline
$1/2$ & $\ast$ & $0.1628+0.0620\,i$ & $0.1711+0.0652\,i$ & --- & --- \\ \hline
$1$ & $\ast$ & $0.2311+0.0587\,i$ & $0.2380+0.0618\,i$ & --- & --- \\ \hline
$2$ & $\ast$ & $0.3617+0.0558\,i$ & $0.3653+0.0568\,i$ & --- & --- \\ \hline
\multicolumn{6}{||c||}{Quantum-corrected, $c_6=-0.010$} \\ \hline
$0$ & $\ast$ & $0.0494+0.1437\,i$ & $0.1014+0.1362\,i$ & --- & --- \\ \hline
$1/2$ & $0.1182+0.1060\,i$ & $0.1438+0.1017\,i$ & $0.1711+0.1231\,i$ & $16.4$ & $35.0$ \\ \hline
$1$ & $0.2265+0.0900\,i$ & $0.2332+0.0972\,i$ & $0.2507+0.0973\,i$ & $4.1$ & $10.4$ \\ \hline
$2$ & $0.3520+0.0845\,i$ & $0.3497+0.0938\,i$ & $0.3656+0.0926\,i$ & $2.7$ & $4.4$ \\ \hline
\multicolumn{6}{||c||}{Quantum-corrected, $c_6=+0.003$} \\ \hline
$0$ & $0.0930+0.1044\,i$ & $0.1515+0.1267\,i$ & $0.1221+0.1024\,i$ & $44.7$ & $20.9$ \\ \hline
$1/2$ & $0.1788+0.1208\,i$ & $0.1829+0.1082\,i$ & $0.1869+0.1069\,i$ & $6.1$ & $7.4$ \\ \hline
$1$ & $0.2486+0.0951\,i$ & $0.2478+0.0935\,i$ & $0.2564+0.0948\,i$ & $0.7$ & $2.9$ \\ \hline
$2$ & $0.3783+0.0912\,i$ & $0.3815+0.0928\,i$ & $0.3831+0.0898\,i$ & $0.9$ & $1.3$ \\ \hline
\hline
\end{tabular}
\end{table}

\section{The first overtone}
\label{sec:app overtone}

The overtones separate the methods more sharply than the fundamental mode, and in the opposite direction. Table~\ref{tab:app overtone} collects the cases at $n=1$ in which the direct integration converges. The window~\eqref{eq:app contamination} closes as $\omega_I$ grows, so it is narrow already at $n=1$ and the method returns nothing at $n=2$ on any of the five backgrounds, which is the quantitative content of the remark in Sec.~\ref{sec:mashhoon} that direct integration is unstable at large $n$. The near-extremal hairy configuration is absent as well. Its gravitational overtone passes step 5, but moves by $2\%$ as $\varepsilon$ goes from $10^{-4}$ to $10^{-2}$ and returns no root at $\varepsilon=10^{-6}$, the symptom of the extremal case in milder form, whereas on Schwarzschild and at $Q_\text{DEC}$ the same test moves the overtone by $0.3\%$ at most. The integer-spin fundamental modes at $Q=0.99\,Q_\text{max}$ do not move at all over $10^{-6}\leq\varepsilon\leq10^{-3}$, the range from which their entries in Table~\ref{tab:app exotic} are taken.

Where all three are available the ordering of the fundamental mode reverses, and the third-order WKB outperforms the Mashhoon substitution, by $0.3\%$ against $7.2\%$ on Schwarzschild and by $0.4\%$ against $7.2\%$ at $Q_\text{DEC}$, both in the gravitational channel. Equation~\eqref{eq:PT_omega} gives the same real part for every overtone, whereas Eq.~\eqref{eq:app wkb} carries the overtone number into both parts through $\Lambda$ and $\Omega$. This is why the overtone rows of Tables~\ref{tab:QNM_GD} and~\ref{tab:QNM_QC} carry a warning, which, as item 3 of Sec.~\ref{sec:app conclusion} notes, reaches further than the $\dagger$ marks there.

\begin{table}[H]
\centering
\caption[Quasinormal methods compared at the first overtone]{First overtone $\omega_1$, in the cases where the direct integration passes step 5 of Subsec.~\ref{subsec:app direct}, with the departure in modulus of each approximation from it. No case at $n=2$ passes that test on any of the five backgrounds. Computed with $G_\text{N}=M=1$ at $\ell=s$.}
\label{tab:app overtone}
\footnotesize
\setlength{\tabcolsep}{3pt}
\renewcommand{\arraystretch}{1.2}
\begin{tabular}{||c|c||c|c|c||c|c||}
\hline\hline
Background & $s$ & Direct & WKB & Mashhoon & \makecell{WKB\\(\%)} & \makecell{Mashhoon\\(\%)} \\
\hline\hline
Schwarzschild & $1$ & $0.2149+0.2926\,i$ & $0.2113+0.2958\,i$ & $0.2546+0.2887\,i$ & $1.3$ & $11.0$ \\ \hline
Schwarzschild & $2$ & $0.3467+0.2739\,i$ & $0.3460+0.2749\,i$ & $0.3783+0.2716\,i$ & $0.3$ & $7.2$ \\ \hline
Hairy, $Q_{\rm DEC}$ & $1$ & $0.1495+0.2008\,i$ & $0.1464+0.2030\,i$ & $0.1765+0.1983\,i$ & $1.5$ & $10.9$ \\ \hline
Hairy, $Q_{\rm DEC}$ & $2$ & $0.2411+0.1886\,i$ & $0.2402+0.1894\,i$ & $0.2630+0.1874\,i$ & $0.4$ & $7.2$ \\ \hline
QC, $c_6=-0.010$ & $2$ & $0.3014+0.2618\,i$ & $0.2940+0.3043\,i$ & $0.3656+0.2778\,i$ & $10.8$ & $16.6$ \\ \hline
QC, $c_6=+0.003$ & $2$ & $0.3644+0.2879\,i$ & $0.3742+0.2879\,i$ & $0.3831+0.2695\,i$ & $2.1$ & $5.6$ \\ \hline
\hline
\end{tabular}
\end{table}

\section{What the comparison shows}
\label{sec:app conclusion}

Four statements survive, and the first three support the use made of the Mashhoon method in Chapter~\ref{chap:QNM_spectrum}.

\begin{enumerate}
    \item \textbf{The accuracy improves with the multipole on every background.} Along $\ell=s=0,\tfrac12,1,2$ the departure of the Mashhoon fundamental mode from the direct integration falls monotonically on all five geometries, and in the gravitational channel it stays within $1.3\%$ everywhere except at $c_6=-0.010$, where it reaches $4.4\%$. That channel, the one a ringdown measurement would supply, is thus the best determined of the four, and the error quoted in Sec.~\ref{sec:mashhoon} is not degraded by the hair or by the quantum correction.
    \item \textbf{The scalar and Dirac channels are a different matter.} On the quantum-corrected background at $c_6=-0.010$ the Mashhoon frequency departs from the direct integration by $35\%$ in the Dirac channel, against $3.0\%$ on Schwarzschild, so the figures of Sec.~\ref{sec:mashhoon} understate the uncertainty there. The sensitivity of these two channels to $c_6$ found in Sec.~\ref{sec:QNM_QC} is a shift between backgrounds computed by a single method and is not in doubt. The absolute frequencies of the scalar and Dirac rows of Table~\ref{tab:QNM_QC}, however, carry a method uncertainty an order of magnitude larger than the gravitational ones.
    \item \textbf{At the first overtone the ordering reverses}, and the WKB expansion becomes the more accurate approximation in every case where the direct integration converges. In the gravitational channel the gain exceeds an order of magnitude on Schwarzschild and on the hairy background at $Q_\text{DEC}$, but amounts only to a factor between $1.5$ and $3$ on the quantum-corrected one. Nothing in Chapter~\ref{chap:QNM_spectrum} rests on the absolute value of an overtone, but the caution attached there to the rows with $n\geq\ell$ should be read as extending to every overtone, since the unmarked rows at $n=1$, $\ell=2$ carry a Mashhoon error between $5.6\%$ and $16.6\%$, above the $4\%$ quoted for the fundamental mode. The shifts between backgrounds in those rows remain sound, for the reason given in item 2.
    \item \textbf{Two geometries lie outside the comparison}, and the local methods cannot detect it on their own. At the extremal hairy configuration the surface gravity vanishes and the inner boundary recedes to an infinite tortoise distance, while the truncated quantum-corrected metric ends, once $\abs{c_6}$ separates the roots of $B$ and $A^{-1}$ appreciably, at a throat where the potential still stands at between a seventh and half the height of the barrier, Eq.~\eqref{eq:app Vout}. In both, the boundary-value problem of Eqs.~\eqref{eq:qnm_master} and~\eqref{eq:qnm_bc} is not the one being solved, and the direct integration says so by refusing to converge. WKB and the Mashhoon substitution return a frequency regardless, a smooth barrier peak existing in both cases, so a frequency from them alone is no evidence about the spectrum. This is the clearest reason for having run the comparison.
\end{enumerate}


\end{apendicesenv}


\clearpage


\end{document}